\documentclass[notitlepage,12pt]{report}

\usepackage[square,numbers]{natbib}

\usepackage{makeidx}
\usepackage{graphics}

\newcommand{\bm}[1]{\mbox{\boldmath $#1$}}

\title{\large{\textbf{NEGATIVE-ENERGY MATTER}}\\
\large{\textbf{AND THE DIRECTION OF TIME\footnote{Cite as: arXiv:1608.01531v12}}}}

\author{J.C. Lindner\footnote{Email address: research@jclindner.ca}\\
Department of Physics, Universit\'{e} de Montr\'{e}al\\
Montr\'{e}al, QC, Canada}

\date{17 June 2016 (updated 27 May 2026)}

\makeindex

\begin{document}

\maketitle

\begin{abstract}
This report offers a modern perspective on the problem of negative energy, based on a re-examination of the concept of time direction as it arises in a classical and quantum-mechanical context. From this analysis emerges an improved understanding of the general-relativistic stress-energy of matter as being a manifestation of local variations in the energy density of zero-point vacuum fluctuations. Based on those developments, a set of axioms is proposed from which is derived a generalized classical theory of gravitation that preserves the basic mathematical structure of relativity, but that simplifies the equations of the theory in the presence of negative-energy matter and vacuum energy. Those results are then applied to provide original solutions to several long-standing problems in cosmology, including the problem of the nature of dark matter and dark energy, that of the origin of thermodynamic time asymmetry, and a few other issues which are usually approached using inflation theory. Important progress is also achieved concerning quantitative aspects of black hole entropy and gravitational information. Finally, we draw on those developments to provide significant new insights into the foundations of quantum theory, regarding, in particular, the problem of quantum non-locality, that of the emergence of time in quantum cosmology, as well as the question of the persistence of quasiclassicality following decoherence.
\end{abstract}

\newpage

\chapter*{Note to the reader}
\markboth{}{}

This is the extended version of a report which was also released in abridged form as an update of preprint \cite{Lindner-1} by the present author. It contains a more complete discussion of the subjects which are covered in the abridged version, along with significant additional developments concerning the relationship between discrete symmetries and black hole thermodynamics, quantum-mechanical aspects of gravitation theory and cosmology, and the problem of the interpretation of quantum theory itself. For reading convenience, adapted versions of both the extended and the abridged edition of this report have been published in print under the title \textit{Negative-Energy Matter and the Direction of Time}.

The present update provides many significant improvements and corrections to all chapters, but more particularly concerning the problem of the origin of repulsive gravitational forces between opposite-energy objects and the analogy of voids in an expanding matter distribution from section \ref{sec:2.7}, as well as regarding the question of energy conservation from section \ref{sec:2.10} and the formulation of the generalized gravitational field equations from section \ref{sec:2.14}. Section \ref{sec:3.10} concerning the degrees of freedom of matter which give rise to black hole entropy has also been completely revamped, along with parts of section \ref{sec:4.7} about gravitational entropy.

Even if you have been in contact with the author's earlier publications, it is highly recommended to read those sections again in order to fully appreciate the value of the many profound changes they introduce to our understanding of the matters discussed.

\newpage

\tableofcontents
\newpage

\listoffigures
\newpage
\listoftables
\newpage

\raggedbottom



\pagestyle{headings}

\chapter{Introduction\label{chap:1}}


The reflection which gave rise to the developments that will be introduced in this report started with a very simple question: could gravitation be a repulsive force\index{repulsive gravitational force} under certain circumstances and what would it mean for gravitational mass\index{negative gravitational mass} to be negative? Even though there appears to be important difficulties associated with the possibility that a gravitationally repulsive object may exist, particularly in the context of a general-relativistic theory, the idea of a symmetry which would have to do with the sign of mass or energy is certainly quite appealing aesthetically. Indeed, if the electric charge and all the other charges turning up in particle physics are allowed to be both positive and negative, why should mass or energy be restricted to positive values?

What I came to realize, through a careful analysis of the assumptions behind the common idea that gravitationally repulsive matter\index{gravitationally repulsive matter} does not exist, is that there is actually a general misunderstanding surrounding the whole concept of negative energy in modern physical theory and that this is the single most important stumbling block that is preventing necessary progress from being achieved in several fields of fundamental theoretical physics\index{fundamental theoretical physics}. The objective of this essay is to clear up this misunderstanding and to provide a detailed account of the most crucial advances which are made possible by adopting a more consistent approach for the integration of negative-energy matter to gravitation theory that draws on the analogy provided by the gravitational dynamics of voids in an expanding matter distribution\index{void in expanding matter distribution}.

I will begin by identifying the difficulties which are associated with the conventional concept of negative mass\index{negative mass!conventional concept}. I will show that conventional expectations, regarding the interaction of negative-mass matter with itself and with positive-mass matter, are inappropriate, because they violate the requirement that all physical properties be defined in a relational way. From this analysis will emerge an improved understanding of the notion of gravitational repulsion\index{gravitational repulsion} involving negative-energy matter as a form of dark matter\index{dark matter} whose existence must, under certain circumstances, be considered unavoidable from both a theoretical and an empirical viewpoint. An alternative set of axioms, which allows an appropriate and at last, consistent integration of negative energy states\index{negative energy states} to physical theory will then be proposed. I will conclude this portion of my analysis with a reformulation of the relativistic gravitational field equations\index{gravitational field equations} that provides the foundation for the first-ever bi-metric theory\index{bi-metric theory} of gravitation that is truly symmetric under exchange of positive and negative energy states and which actually simplifies the original theory in the presence of a non-zero cosmological constant\index{cosmological constant}.

What allowed me to achieve a better understanding of the concept of negative-energy matter is the acknowledgment that there exists a fundamental time-direction degree of freedom\index{time-direction degree of freedom!fundamental}, independent from the thermodynamic concept of time direction\index{time direction!thermodynamic concept}. In such a context, it emerges that only the sign of energy defined in relation to a given direction of propagation in time\index{direction of propagation in time} is significant from a gravitational viewpoint. Once the significance of this insight was properly assimilated, it became possible to develop an alternative concept of time reversal\index{time reversal!alternative concept} that allows a reformulation of the discrete symmetry operations\index{discrete symmetry operations} and a more consistent description of the changes occurring under a reversal of space- and time-related parameters. In order to achieve full consistency, it was necessary to introduce an additional set of discrete symmetry operations of a kind which had never been considered and which transforms a positive energy state into various negative energy states. Those developments then allowed the derivation of an exact value for the number of discrete degrees of freedom\index{discrete degrees of freedom!particle inside black hole} of matter particles which are under the influence of a black hole, that reflects the measure of missing information encoded on the event horizon of both ordinary and extremal black holes\index{extremal black hole!missing information}.

As a consequence of the relatively long period of gestation during which the mere intuitive insights from which this work originates evolved into a revised, classical theory of gravitation, I was able to explore the consequences of some of the most decisive results which were reached in the course of that process for a rather large number of questions of fundamental interest. Thus, I can now provide a complete account of the implications of this improved understanding of gravitational physics for classical cosmology\index{classical cosmology} theory. I will, in particular, provide significant new insight regarding the whole question of dark energy\index{dark energy} and dark matter\index{dark matter} and the related problem of the formation of large-scale structures\index{large-scale structures!formation}. By making use of the results derived in the first portion of this report, I will then propose alternative solutions to a few other problems which were originally addressed using inflation theory\index{inflation theory}. I will conclude this part of my analysis by providing a definitive solution to the problem of the origin of thermodynamic time asymmetry\index{thermodynamic time asymmetry} which relies on a more accurate assessment of the gravitational entropy associated with homogeneous and inhomogeneous matter distributions.

In the fifth chapter of the report, I will then offer a fresh perspective on several aspects of the problem of the interpretation of quantum theory\index{quantum theory!interpretation} which centers around a reconsideration of the significance of the requirement of time reversal symmetry\index{time reversal symmetry}. Following a critical review of some early time-symmetric formulations of quantum mechanics\index{quantum mechanics!early time-symmetric formulations}, I will argue that a more consistent approach must overcome the contradictions of the orthodox interpretation of quantum theory\index{quantum theory!orthodox interpretation} that follow from its rejection of scientific realism\index{scientific realism}. I will also show that the condition of time reversal invariance\index{time reversal invariance} provides strong enough a constraint to allow a realistic interpretation of quantum theory\index{quantum theory!realistic interpretation} to satisfy the principle of local causality\index{principle of local causality} in the face of quantum entanglement\index{quantum entanglement}. Finally, in the second portion of my discussion concerning the foundations of quantum theory, I will explain that the existence of a maximum quasiclassical domain\index{maximum quasiclassical domain} can only be predicted to arise and to persist, following decoherence\index{decoherence}, once we consider the problem of the emergence of time\index{emergence of time} in quantum cosmology\index{quantum cosmology} from the perspective of the solution provided in the first portion of the report to the problem of the origin of thermodynamic time asymmetry.

\section{Motivations\label{sec:1.1}}

Even though I became interested in the basic idea underlying the developments discussed in this report based on mostly aesthetic motives, the actual reasons that later fueled my interest in developing a viable model around it were of a more pragmatic nature. In particular, I saw the need that existed, but that few authors recognized, to reformulate the current classical theory of gravitation\index{classical gravitation theory} in a way that would be consistent with the possibility for elementary particles to be found in the negative energy states\index{negative energy states} which are allowed by special relativistic quantum theory\index{special relativistic quantum theory}. I had also come to understand that the current interpretation of negative energy states as merely being those of particles propagating backward in time (the antiparticles), whose behavior is identical to that of ordinary matter from a gravitational viewpoint, amounts to assume that only some of those energy states are allowed.

In other words, we had solved the puzzling problem of the prediction of negative energy states\index{negative energy states!problem} by postulating that those states were not allowed, without providing any \textit{theoretical} justification for this empirically convenient hypothesis. But if we recognize that the whole spectrum of energy states predicted to exist by quantum theory can actually be occupied, even if transitions between positive and negative energy states may not be allowed, then we need a classical theory of gravitation that is consistent with this requirement. However, further considerations indicated that the general theory of relativity\index{general relativity theory} is not entirely compatible with an appropriate notion of negative energy obeying certain theoretical requirements which must be imposed in order to achieve consistency.

Despite those difficulties, I believe that the imperative to provide an appropriate description of negative-energy matter should prevail over our willingness to leave untouched the current theory of gravitation, because I have recognized the inadequacy of the arguments against the physical nature of negative energy states, while I also understand that quantum theory constitutes a more appropriate basis to decide what states are allowed for elementary particles. Thus, I persisted in seeking to achieve this integration and as it turned out, this insistence was vindicated, given that I was able to develop an alternative framework that merely generalizes relativity theory in a very elegant manner, without affecting its basic mathematical structure, while allowing an appropriate description of negative-energy matter.

But I was also motivated by the desire to obtain a better agreement between theoretical predictions and astronomical observations concerning certain aspects of the gravitational dynamics of the universe. In particular, there was the exceptionally severe disagreement between most theoretical derivations of the expected average value of vacuum energy density\index{average vacuum energy density} and observational constraints on the upper (positive or negative) value of the cosmological constant\index{cosmological constant}. Very early on, I saw that the hypothesis that there should exist a usually ignored portion of zero-point vacuum fluctuations\index{zero-point vacuum fluctuations} that would interact (other than gravitationally) only with matter in a negative energy state could potentially provide a whole new class of contributions to vacuum energy density, which would naturally allow an overall cancellation of all contributions, if some level of symmetry exists between the viewpoint of positive-energy observers and that of negative-energy observers. Thus, despite the apparent difficulties, perceived or real, associated with the existence of negative-energy matter, it had become very clear to me that this is a hypothesis we can no longer reject without serious motives.

Finally, I also wanted to bring some much-needed clarity to the theoretical context in which we are to address the problem of the elaboration of a theory of the gravitational interaction compatible with the basic principles of quantum theory. Here I will show the essential role played by the discrete spacetime and energy-momentum symmetry operations\index{discrete symmetry operations} (appropriately redefined and extended to comply with an improved concept of time reversal\index{time reversal!improved concept}) in characterizing states of matter particles on the spatial scale and energy level at which we can expect the gravitational interaction to be as strong as the other known interactions. This will be achieved by demonstrating the relevance of those symmetry operations for a determination of the microscopic degrees of freedom\index{microscopic degrees of freedom} of matter that must be taken into consideration in order to provide an appropriate measure of black hole\index{black hole!information and entropy} information and entropy. But I will also explain that one of the main consequences of the solution I have developed to explain the origin of thermodynamic time asymmetry\index{thermodynamic time asymmetry} is that it allows one to understand how a uniformly flowing time variable\index{time variable!uniform flow} can emerge from the timeless equations of quantum gravity\index{quantum gravity}, thereby imparting the metric of spacetime\index{metric of spacetime!signature} with its unique signature.

\section{Approach\label{sec:1.2}}

Basically, the approach I will follow in this report consists in explaining how some specific aspect of the quantum world, namely the ignored possibility for both positive and negative energy states to propagate forward and backward in time, changes our understanding of the \textit{classical} theory of gravitation\index{classical gravitation theory} and allows to actually improve and simplify its formulation in a way that will have decisive consequences for the description of certain phenomena which are taking place on the cosmological scale. Once there, I will go the opposite way and show how those original insights regarding cosmology and the thermodynamic arrow of time\index{thermodynamic arrow of time} shall affect our understanding of \textit{quantum} physics and open up the way to a more pragmatic approach toward a quantum theory of gravitation\index{quantum gravitation theories}.

Concerning the methodology which is reflected in the style of this treatise, I must emphasize that I have been introduced to the mathematical methods of theoretical physics very early on, but that I later came to realize that in the context where all the really useful mathematical developments that could be carried out in the field of fundamental theoretical physics\index{fundamental theoretical physics} have already been performed over and over again by competent people, real progress can only arise at the level of interpretation. Indeed, a fully consistent interpretation of the existing frameworks is currently missing, perhaps because the vast majority of competent researchers prefer to dedicate their efforts to more technical aspects, and this is restraining our ability to distinguish between what are viable developments and what is logically and empirically inappropriate.

Thus, I was led to concentrate my efforts on interpretative issues, but without sacrificing precision and accuracy. As a result it has become both possible and necessary to state most of the results I have obtained in plain language, while referring to already developed mathematical frameworks and by providing very few mathematical developments. But if mathematical developments do not dominate the content of this report, it is also simply a consequence of the fact that, while I have achieved a crucial revision of the mathematical framework of relativity theory\index{relativity theory!mathematical framework} and a necessary improvement of the interpretation of quantum theory\index{quantum theory!interpretation}, I nevertheless ended up confirming the validity of the basic mathematical structures of both theories within a certain limit, so that practically no further mathematical developments were required.

The reader must be warned, however, that given that the broadness of scope of the following discussion was not achieved at the expense of precision, the density of information that is to be found in this document is very high. It may also be difficult, at first, to appreciate the value of some of the most radical solutions which are proposed to various problems, because, due to the unusually large number of disciplines which are touched by the developments which will be introduced in this report, a good portion of the arguments which are formulated in the earlier portions of the document only become fully understandable after reading its latter portions. This, however, does not mean that the present essay is difficult to read, to the contrary. In fact, I tend to follow a rather educational approach and I do not avoid making statements and providing explanations that may appear obvious to some readers, because I think that it is better to make too many unnecessary statements than to more or less willingly avoid making some which would have been useful. This approach should not be considered as condescending or as an indication that this treatise is not suitable for professional researchers.

Now, I do recognize that the approach I followed in order to achieve the valuable results that will be described and justified here really \textit{is} different from that which is usually followed in theoretical physics. Indeed, very early on in my career, I learned to rely on the expertise of specialists concerning certain technical aspects which are not essential to an accurate understanding of the issues on which I was working. Thus, instead of assimilating all the complex machinery that allows to solve specific problems in various fields of theoretical physics\index{theoretical physics}, I was satisfied with studying problems of a more general nature that still required careful reasoning and analysis, but that were not considered serious work by most conventional researchers. I'm convinced that, if I had insisted on following a more conventional approach, I simply would never have been able to derive all of the important results that figure in this report.

What I'm trying to say is that the kind of work I have accomplished requires specialization, but while most researchers develop very elaborate technical skills in one specific field of study, my specialization consisted in developing skills in analyzing certain general aspects common to several different fields of fundamental theoretical physics\index{fundamental theoretical physics} which all have to do with time directionality\index{time directionality}. If I had not focused my attention on questions of interpretation and had rather tried to develop elaborate technical skills in every field I studied, as I thought to be necessary when I began studying physics in a conventional academic environment, I would certainly have failed to contribute to our understanding of the physical world. The truth is that a certain level of technical expertise was required to achieve those results, but I was fortunate enough that, when I first began to work at a more qualitative level, I had already developed most of that mathematical proficiency.

But the very fact that, for many researchers, the preceding comments will merely reflect incompetence indicates that, at the present epoch, theoretical physics\index{theoretical physics!evolution} has reached a point in its evolution which is similar in many regards to that in which natural philosophy\index{natural philosophy} ended up when it began deviating into mathematical idealism\index{mathematical idealism} during antiquity. Indeed, it has recently been emphasized that the absence of philosophical underpinning that characterizes some currently favored approaches and the excessive recourse to advanced mathematics in formulating physical theories (which is often achieved even at the expense of clarity or usefulness), has driven the field of fundamental theoretical physics into a state of stagnation. But this overly technical strategy is not a requirement of the scientific method and there is no need to use complex mathematics at \textit{every} level of discussion and under \textit{all} circumstances, especially when language allows sufficient or better clarity and references to existing mathematical constructs are provided.

In fact, I believe that there is a trend in the evolution of scientific research, from the first theoreticians who invented their own mathematics, to later physicists who made use of existing mathematical developments to build their models, and on to some present day physicists using already existing mathematical physics frameworks to produce further original insights, still building on what had previously been achieved. It must be clear, however, that I'm not trying to deny the effectiveness, or the usefulness, and certainly not the necessity of a quantitative approach to physics, but simply to emphasize that, in order to develop a globally consistent understanding of so many different aspects of fundamental theoretical physics\index{fundamental theoretical physics}, I had no choice but to follow an unconventional approach and to adjoin to mathematical reasoning the benefits, nowadays somewhat forgotten, of rigorous philosophical analysis.

But, even though I would not myself have believed that one could achieve significant results by concentrating on interpretative issues when I started studying physics, which I did the usual way by learning about the mathematics of quantum theory, statistical mechanics and relativity, it is through experience and by force of circumstance (although not as a result of mere inability), after having slowly and partly unwillingly deviated from the conventional path, that I began to understand that there is real value in such an approach, which I developed by making systematic a learning process that initially appeared to merely be a faithful, but irresponsible time-wasting improvisation. If the reader has enough courage to immerse herself in a similar experience and to loosen her grip on more traditional ways of achieving deep understanding, while nevertheless being ready to spend a minimum amount of effort to follow simple logical arguments, I can assure her that she will not be deceived and will learn useful physics, which is not so bad already by today's standards.

\section{Historical context\label{sec:1.3}}

There are many similarities between the current state of theoretical physics\index{theoretical physics!turning points} and those it went through at other crucial turning points in its history. Indeed, the situation we have now arrived at is characterized by an accumulation of unanswered questions which creates an impasse that prevents further progress from being achieved. It is my belief that answering just a few key questions among those will greatly facilitate future theoretical research. When we examine the present situation in physics it becomes clear, in effect, that if there are questions which we are justified in not being able to answer right now, because they are related to what may be said of reality under conditions which we cannot yet reproduce in experiments (think of trying to explain the origin of the free parameters of the standard model of particle physics\index{standard model of particle physics!free parameters}), there are also questions which have to do with known difficulties which we have puzzled about for a long time and which we have no reason to believe further experiments may be particularly useful in helping resolve. But those are problems whose existence is often simply unknown to most people or which are sometimes considered to have already been solved, while careful examination shows that this is not always entirely the case.

Most research programs in fundamental theoretical physics\index{fundamental theoretical physics} are focused on trying to solve the problems raised by questions of the first type and this is unfortunate, because here is precisely the domain in which progress is limited by technological constraints of a practical nature and the cost of achieving the required experiments. Very early on I recognized that, if I was going to enable progress to be made in physics, I had to concentrate my efforts on trying to answer questions of the latter type, where progress could occur not only in my lifetime, but also as a consequence of the success or failure of my own enterprise.

Among the questions we may have hope to answer using our current knowledge is the question I mentioned earlier on as having being that which launched the reflection process from which this report emerged. It is, in effect, one of those unsolved questions whose very existence is usually unrecognized or which is considered to have already been solved, while this is clearly not the case, as I will explain later. You will not see it mentioned in most accounts as being one of today's open questions\index{open questions!physics} in physics, but it is one of the most important categories of question regarding classical physics and a field most people currently consider to be free of major difficulties. This problem of negative energy states\index{negative energy states!problem} could actually be called the `classical gravitation theory problem'\index{classical gravitation theory!problem} or the `general relativity problem', because properly answering that question requires introducing slight modifications to that theory, which actually consist in a generalization of its own founding principles. This is the first question I will address in this report and satisfactory answers will be provided to the mostly unrecognized issues it currently raises.

An additional category of questions, which is also related to classical gravitation theory, can be collectively described as the `cosmology problem'\index{cosmology problem}. It asks what is the origin of the constants of the standard model of cosmology\index{standard model of cosmology}, what is dark matter\index{dark matter} and what is dark energy\index{dark energy}, how are we to resolve the flatness\index{flatness problem} and horizon problems\index{horizon problem}, and what explains the existence of certain visible astronomical structures which appear to have developed too early or on too large a scale to be explainable by conventional theories? It also asks why it is that the energy which is contained in zero-point vacuum fluctuations\index{zero-point vacuum fluctuations} is so low in comparison with the very large value that is provided by most theoretical estimates? Finally, it asks whether there was a beginning to time in the past and why it is that matter was present in the first instants of the Big Bang?

While it is often considered that some of those questions have already been answered by developments like inflation theory\index{inflation theory}, I will explain that there remain important unresolved issues in this context and that we are justified in seeking alternative answers, which I will show do exist. In fact, even though the objectives I had in mind when I started this research project were quite humble, in the end I was able to provide original solutions to nearly all aspects of the cosmology problem.

But I will also address a further category of questions that is usually considered to regard classical physics, but which actually sits right at the interface between the classical theory of gravitation\index{classical gravitation theory} and quantum theory\index{quantum theory}. This is the question of the origin of the thermodynamic arrow of time\index{thermodynamic arrow of time!problem of origin}. Given that this problem can be traced back to the peculiar properties of the distribution of matter energy which existed during the first instants of the Big Bang (as I will explain), it follows that the question of the origin of the unidirectionality of thermodynamic processes is, in effect, also a question for cosmology and as such, it will hugely benefit from the insights I have gained while solving other aspects of the cosmology problem\index{cosmology problem}.

What was somewhat unexpected to me, though, was the realization that answering those questions actually constitutes an essential condition for addressing an additional and apparently unrelated category of questions. Indeed, as I have mentioned above, the solution I will propose to the problem of the origin of time irreversibility\index{time irreversibility} turned out to be essential for developing a proper understanding of quantum reality\index{quantum reality} and for providing a satisfactory explanation to the emergence of a quasiclassical world\index{quasiclassical world!emergence} and this is why I will discuss the problem of the interpretation of quantum theory\index{quantum theory!interpretation} as part of my analysis of the question of time directionality\index{time directionality}.

Richard Feynman\index{Feynman, Richard} has emphasized the fact that, acquiring knowledge about one physical law, or getting insight into one important problem, and being confident in the validity of those developments, often allows us to find other laws. I have been able to experience the validity of this remark while my understanding of physics progressed. Indeed, by carefully applying the knowledge I had gained by solving the problems I was originally interested in, I was allowed to achieve further improvements to our understanding of distinct and apparently unrelated issues, always based on an unassailable confidence in the validity of what I had already been able to understand.

But on several occasions I also had to revise the solutions I had tentatively provided to certain problems in a given field of theoretical physics, following improvements in my understanding of a distinct, but not totally unrelated field, and sometimes this revision itself had an impact on the validity of other results in other fields, because, at this level, nothing can really be conceived independently of anything else. It takes time, however, to get a proper understanding of the whole picture in which everything agrees with everything else and with observations.

There may, thus, still remain insights which have not been fully integrated into the global picture I developed, simply due to the fact that I did not had the chance to rethink their significance in the context of all the other advances. This is perhaps unavoidable given the considerably large scope of the subject of time directionality\index{time directionality}, which is relevant to so many different fields of fundamental theoretical physics. Given that I wanted to publish the results of my research within my lifetime, I had no choice but to eventually let the outcome of my reflection go out for others to benefit, but also to criticize for what it may still contain of imperfections.

I have done my best, however, at providing the most exhaustive account of the progress achieved and at identifying the various relationships between the many insights that form the substance of this report which, therefore, provides a complete account of the subject it covers. It is my hope that by reading about what I have found, some young and not yet indoctrinated mind will be inspired to explore even more remote territories and bring forth a significant shift in our understanding of reality that will prove, again, that it is only by wandering far from the beaten track that one can gain the perspective necessary to see the vast landscape that goes unnoticed to those who do not dare to deviate from the normal course of research imposed by the practices which are of common use at a given epoch.

\section{Organizing principles\label{sec:1.4}}

Every successful venture into unknown territory requires relying on the appropriate beacons and guidelines and this is particularly true when the voyage takes you to the boundaries of conventional certainties and brings you to question some essential aspects of what had previously appeared to constitute a fixed background for scientific exploration. I would, therefore, like to briefly describe what were the essential principles that guided me while I was elaborating the improved understanding of physical reality that will be discussed in this report. It must be understood that those principles were not given as preconditions imposed on any vision of the world, but actually developed alongside improvements in my and other people's knowledge and understanding of physical reality.

My awareness of the importance of the first of those principles developed mainly in conjunction with my appreciation of the requirements imposed by the classical theory of gravitation\index{classical gravitation theory}. Indeed, it is while I was tackling the problem of negative energy that I realized the importance (emphasized by others in a different context) of a relational definition\index{requirement of relational definition!physical attributes} of the physical attributes of objects and that I understood the real significance of the requirement of relativistic invariance\index{relativistic invariance}. This allowed me to perceive the true meaning of Einstein's\index{Einstein, Albert} insistence that the objects of physics must be conceived of only in relation to the spacetime structure to which they belong, because I came to understand that the metric properties\index{metric properties of spacetime} of space and time are not merely dependent on the state of motion of an object, but also on the sign of its energy (as will be explained later), in contrast with what one would expect from a conventional viewpoint.

Thus, if a determination of the relationships between physical objects in different states of motion is possible only when we determine the common spacetime structure shared by those objects, then the fact that the spacetime structure itself is dependent on the nature of the objects means that the relationships between them are dependent on their nature and in particular, their energy signs. It therefore appeared to me that it is not merely the state of motion of an object which constitutes a relative notion, but that any physical attribute must always be defined or characterized only in relation to similar attributes of other objects present in the same universe (the physical attributes of a system enable to characterize it merely in relation to the similar attributes of other systems and those relationships are determined through the use of reference systems).

When I tried to understand what could logically impose such a requirement, I slowly came to realize that it is the very fact that it would be meaningless to relate some physical attribute, in order to define its value, to a reference point not part of the same physical universe. Indeed, in the absence of a well-defined, continuous network of causal relationships\index{causal relationships} that would extend to those immaterial reference systems, there can be no meaningful definition of the physical attributes involved, because the attributes of an object cannot be determined without comparing them to those of another object that is part of the same causally related ensemble (the universe\index{universe!causally related ensemble}) to which it belongs. This requirement of a relational definition of physical attributes\index{requirement of relational definition!physical attributes} will have enormously important consequences on many aspects of the developments to be discussed in the following chapters.

This remark illustrates the importance of another broad requirement that slowly emerged as being essential for solving the problems which will be discussed in the fifth chapter of this report. There is, in effect, a tendency, nowadays, to designate as metaphysical every aspect of reality which may be impossible to probe through direct observation and to conclude that such aspects are not worth the attention of the scientific community. What I have come to understand is that the self-imposed requirement of systematically characterizing as metaphysical any notion that refers to aspects of physical reality which may not be directly accessible to observation is actually a mild form of solipsism\index{solipsism} and constitutes one of the most serious obstacle on the way to developing more accurate models in fundamental theoretical physics.

In fact, I think that the greatest challenge with which science is currently faced may well be that of surmounting the obstinate refusal to accept as a legitimate object of scientific inquiry what cannot be \textit{directly} observed by the means of measuring instruments and as physically meaningful what lies outside the limits of observation of a given observer (think of the reality behind event horizons for example). In this particular sense, the success of science might, in the end, depend on our willingness to adopt a position analogue to scientific realism\index{scientific realism} and opposite to instrumentalism\index{instrumentalism}, concerning ultimately the idea that something really exists outside our immediate domain of perception of reality. What justifies this position is the fact that the opposite approach is sterile, because one never really knows for sure whether something exists outside our immediate domain of perception of reality.

This requirement is not so different from the original condition of objective reality\index{objective reality} which was advocated by Einstein\index{Einstein, Albert} and which he proposed in an attempt to demonstrate the validity of the hypothesis that reality actually exists, in a quantum mechanical context, even when it is not subjected to direct observation. But given that, in the physical sciences, objectivity has rather come to characterize any conception of reality that is derived solely from empirical knowledge and observation, then it would not be appropriate to use the term `objective reality' in order to refer specifically to a reality that is not directly observable under all conditions, even if the nature of this reality is still \textit{derived} from experimental facts. Thus, I cannot avoid having to speak about a \textit{realistic} conception of reality\index{realistic conception of reality}, even if that may appear tautological, as there does not exist a more appropriate term to denote this kind of approach.

Such a requirement, however, should not be confused with a belief in the validity of theoretical constructs that have no experimental justification, which doesn't constitute a desirable position to hold on to and which would actually consist in the exact opposite of the viewpoint I'm defending here. What I'm suggesting, in effect, is that it may sometimes be appropriate to extend the validity of what we know to be true with absolute certainty to a larger domain of reality where this validity may not be \textit{directly} assessed and not that it would be right to try to extend the domain of validity of a description for which there does not yet exist any empirical evidence. In other words, if we are justified in extrapolating beyond the domain of direct observation, as may be found necessary, principles and notions which we have good reasons to believe are indeed valid, it would be wrong to take advantage of the absence of observational data to try to justify the validity of hypotheses which cannot yet be independently corroborated and which may, therefore, have absolutely no value from a scientific viewpoint.

Those considerations will have decisive consequences for the formulation of an interpretation of quantum theory\index{quantum theory!interpretation} that contains no contradiction when considered in the broader context of the representation of reality that emerges from the progress which will be achieved, in the first portion of this report, in solving other long-standing problems in the fields of gravitational physics\index{gravitational physics}, cosmology, and statistical mechanics\index{statistical mechanics}.

\chapter{Negative-energy matter and gravitation\label{chap:2}}

\section{The negative-energy problem\label{sec:2.1}}

Regarding the question of negative energy, the current situation has much in common with that in which we were at the turn of the previous century with regard to the quantization hypothesis\index{quantization hypothesis}. There was, in effect, some reluctance, initially, to recognize the validity of the original suggestion by Max Planck\index{Planck, Max} that energy is quantized, despite the fact that this proposal would have solved the problem of black body radiation\index{black body radiation problem}. The trouble was, of course, that recognizing the validity of the quantization hypothesis required abandoning classical physics. There is a similar dilemma with negative energy today because, as I will show, the hypothesis that matter can be found in such a state has the potential to solve many important problems facing theoretical physics, but those benefits come at a price which may, at first, appear to be too high, because the introduction of negative-energy matter as a concept somewhat distinct from that which is currently favored (which I believe is required in order to allow it to be consistent from a theoretical viewpoint) seems to imply that general relativity theory\index{general relativity theory} must be abandoned.

But rejecting a theory so well established and so beautifully simple as general relativity is not something that most people would do without very good motives. Yet, while the currently favored assumptions concerning the rules governing negative-energy matter (if it was to actually exist) may appear to better agree with relativity, they actually contradict some of the basic principles on which this theory is founded, therefore making it just as untenable. We must then either abandon the idea that negative-energy matter can exist, or else provide a better interpretation of negative energy states\index{negative energy states} which may force a reformulation of relativity theory itself. But I will show that the conclusion that the latter alternative is the only viable one is not necessarily as dramatic in terms of its consequences as one would expect, because what is required, in this context, is mainly a reinterpretation of the equivalence principle\index{equivalence principle} and not a rejection of the whole mathematical framework of relativity theory\index{relativity theory!mathematical framework}.

There is, however, an additional problem associated with negative energy states, which is that there appears to be no observational evidence for matter in such a state. But I think that it would be incorrect to conclude, on the basis of this difficulty alone, that matter cannot exist in a negative-energy state, because there are also strong, but usually ignored, arguments for recognizing that this conclusion is not unavoidable. In fact, I will later explain that there are very good reasons to expect that it won't be easy to confirm the existence of negative-energy matter, because, as I have come to understand, even the portion of baryonic negative-energy matter\index{baryonic negative-energy matter} that may still be present in the universe today should not be directly observable, just as the more common, hypothetical dark matter\index{dark matter}. Thus, if I'm right, the implicit assumption that negative-energy states are forbidden would be just one of those `reasonable' assumptions which we should be careful not taking too seriously.

With regards to the problem of negative-energy states\index{negative energy states!problem}, we are currently in a situation where we have built into the very formalism of our most fundamental theory of elementary particle interactions (which is quantum field theory\index{quantum field theory}) the apparently necessary, but theoretically unjustified hypothesis that only positive frequencies\index{positive frequencies} (associated with positive energies) are allowed to propagate in the future (the constraint on negative frequencies\index{negative frequencies} being merely that they must propagate toward the past). However, I think that the fact that this restriction appears to be valid does not imply that positive frequencies cannot propagate backward in time or that negative frequencies cannot propagate forward in time, but merely that if such negative-energy matter exists then, for some reason, it can only interact with matter having the same sign of energy, at least electromagnetically.

This absence of interaction is what really justifies the hypothesis that quantum field theory only deals with matter of one energy sign under most circumstances. But given that I will suggest that energy sign is a relatively defined physical property, so that there is no absolute (non-relational) distinction between positive- and negative-energy matter, then it becomes necessary to conclude that there cannot exist a constraint that would impose that negative-energy matter, and only matter with such an energy sign, cannot exist under any circumstances, if positive-energy matter itself is allowed to exist, as required, because it is not even possible to identify the distinguishing property, specific to negative-energy matter, that would justify that its existence be ruled out in such a way.

In the context where we would recognize that there is no \textit{theoretical} motive to reject the possibility that negative-energy matter may be present in our universe it would become apparent that one often mentioned argument that must definitely be rejected concerning the nature of the gravitational interaction is the idea that the predominance of gravitation on the largest scales is a consequence of the `fact' that this interaction is always attractive. This is a conclusion which is usually assumed to follow from the observation that negative-energy matter does not exist and therefore cannot cancel out the gravitational forces exerted by positive-energy matter. Yet, what actually explains the fact that gravitation is a dominant force on larger scales (in addition to its long-range property) is not the absence of matter in a negative energy state, but the simple fact that gravity is attractive between objects with the same positive sign of energy.

Thus, if gravitation dominates over electrical forces on astronomical scales, this is really a consequence of the fact that while identical electric charges tend to disperse under mutual electrostatic repulsion\index{electrostatic repulsion}, positive energies have a tendency to coalesce and to accumulate under mutual gravitational attraction\index{mutual gravitational attraction!positive energies} and the fact that electromagnetism is known to have both positive and negative charges has nothing to do with the fact that those charges do not so readily accumulate, because even if there were only positive electric charges they would not cluster, because identical electric charges mutually repel one another and the possibility for such opposite charges to cancel out actually facilitates an \textit{accumulation} of those charges, but only in neutral configurations and under the influence of gravitation.

Even if it was found that there actually exist negative-energy particles, the possibility for energy to cancel out would not necessarily prevent the accumulation of matter with one or another energy sign, because negative-energy matter may also be gravitationally attracted to itself (despite what is usually assumed) and could therefore also be subject to accumulation. But it is also true that the apparent absence of large accumulations of \textit{negative-energy} matter would not necessarily mean that such matter cannot exist, even if we were to assume that this matter gravitationally attracts matter of the same kind, because it may turn out that this matter is dark and given that it may also be repelled by positive-energy matter (even if this is not what we usually assume), then we might be justified to expect that it should be located mainly in regions of the universe where the density of positive-energy matter is the lowest and that it should be virtually absent from regions where positive-energy matter is more abundant, like that in which we are located, and this would help explain that we have never noticed its existence\footnote{Later in this chapter I will explain that it can also be expected that, even on the cosmological scale, where negative-energy matter is necessarily present, there is no compensating gravitational force attributable to the presence of a homogeneous distribution of such matter.}.

All that the frequently encountered remark, to the effect that gravitation is attractive for all particles, should be understood to mean, therefore, is that it is attractive for all currently known forms of matter. What needs to be recognized is that the commonly held view that the occurrence of negative energy in a theory is necessarily always indicative of a problem is not rationally motivated and that it is not true that all traces of negative energy must be eradicated, at all costs, whenever they are encountered. Dirac\index{Dirac, Paul}, at least, understood that the prediction of negative energy states\index{negative energy states} could not be ignored and tried to provide an explanation for the absence of transitions to such states \cite{Dirac-1}. His solution, based on the idea that negative energy states are already all occupied, was not satisfactory, but at least he did not simply reject the possibility that negative-energy matter might have to be considered real. There is no motive to argue, as people often do, that negative energy is totally unacceptable, other than the difficulty to find an appropriate interpretation that would be compatible with empirical facts for this logically unavoidable counterpart to positive energy.

In the absence of a theoretical justification for the absence of negative-energy matter I think that the only appropriate approach would be to seek to find out why it is that we never observe matter in such a state, rather than try to build that assumption into a then necessarily incomplete theory of quantum fields. In this particular sense, it is significant that the prediction that there must exist antiparticles emerged as a by-product of Dirac's original interpretation of negative energy states, because this contributed to the belief that the discovery of antiparticles constitutes a definitive solution to the problem of negative energy. But, given that Dirac's interpretation was later found to be inappropriate, I think that we need to recognize that, in fact, antiparticles are only one particular aspect of a complete solution to the problem of negative energy, which therefore remains unsolved.

In any case, even if we were to succeed in justifying that it should be imposed that there cannot be transitions from a positive energy state to a negative energy state, we would not have solved the problem of negative energy. This is because such a restriction would merely impose that no positive-energy particle can turn into a negative-energy particle (and vice versa maybe), but there would be nothing in that constraint to forbid a particle from already being in a negative energy state, in which case we would still need to provide a consistent description of the properties of matter in such a state and to justify that we do not observe those negative-energy particles under most conditions.

I do understand, of course, that there are a number of issues associated with the possibility that matter may occupy negative energy states\index{negative energy states}. Of particular concern would be the issue of `vacuum decay\index{vacuum decay problem}' or the apparent problem that all positive-energy particles should fall within a very short interval of time into the available negative energy states by releasing a compensating amount of positive-energy radiation, if those states are not assumed to be forbidden. This is of course the difficulty that motivated Dirac's\index{Dirac, Paul} problematic proposal that those energy states should already be nearly completely filled, so that no further decay should occur once a particle looses all of its positive energy. But I will show, in later portions of this chapter, that this problem, as well a few others which may seem to arise when we recognize that it should be possible for negative-energy states to be occupied, are merely a consequence of the inappropriateness of the conventional concept of negative-energy matter.

I also recognize that the tentative interpretation of negative energy states that came to replace Dirac's solution does, in effect, provide some level of relief in that it at least allows to take into account those negative energy states that cannot be ignored as they actually interfere with processes involving ordinary matter. This is because we are indeed allowed to consider that antiparticles\index{antiparticles!backward-in-time-propagating particles} are negative-energy particles propagating their non-gravitational charges backward in time. But according to that particular interpretation, antiparticles can still be conceived as ordinary particles from the viewpoint of the gravitational interaction, given that it is merely their charges which appear to be reversed from the forward-in-time\index{forward-in-time!perspective} perspective, while their energy remains positive. Therefore, the existence of antiparticles cannot be considered to allow a satisfactory interpretation of negative energy states, because the negative energy of antiparticles is not truly significant from a gravitational viewpoint.

What must be clear it that there is no consistency principle behind the exclusion of negative energy states as states of particles propagating forward in time. It is therefore quite amazing that so many otherwise well-informed authors suggest that no negative-energy, or negative-mass particle can exist, as if this was an obvious and unavoidable conclusion. It must be clear that I'm not complaining about this situation, I merely want it to be recognized for what it is, because I will take a different course and it should be understood that I'm not doing this without good motives or out of a fondness for hopeless, exotic or eccentric ideas.

I must therefore mention that I'm aware that the originators of the steady state model of cosmology\index{steady state model of cosmology} once also criticized (based on distinct motives) the conventional position according to which the existence of negative-energy matter is forbidden. But if I do find this criticism to be valid and appropriate, I do not, however, find suitable the whole concept of negative-energy matter (which is actually very conventional) proposed by these authors, nor do I agree with the objectives they unsuccessfully (given the failure of steady state cosmology) sought to achieve by using this otherwise interesting idea.

I think that the fact that the hypothesis that negative-energy matter may exist was historically associated with such failed theoretical models and was also developed into many different inconsistent formulations lacking any epistemological support is more than anything else responsible for the state of suspicion and confusion that currently surrounds the whole idea of negative-energy matter. The objective I will try to achieve in this chapter will therefore be to clarify the situation regarding what should be expected of the properties of matter in a negative energy state and to demonstrate the validity of the concept itself, in the context where it is properly defined and justified.

\section{The time-direction degree of freedom\label{sec:2.2}}

One very important and often unrecognized outcome of the application of special relativity\index{special relativity!application to quantum theory} to quantum theory is that there exists a fundamental degree of freedom for the direction of propagation in time\index{direction of propagation in time!fundamental degree of freedom} of particles that is distinct and independent from the direction of time which is associated with thermodynamic processes\index{thermodynamic processes!direction of time} and the growth of entropy\index{entropy!growth} that is responsible for the formation of memories. What the degree of freedom associated with time direction embodies is simply the sum of all relationships of time directionality\index{time directionality!relationships} between a given particle and all the other particles in the universe. The existence of such a degree of freedom means that a positive charge can, in effect, be positive either in relation to the positive direction of time or in relation to the negative direction of time. It is not possible, therefore, to completely specify the physical properties of a particle at a given instant by simply providing the sign of its charges and that of its energy independently from the direction of propagation in time of the particle.

This is what the relativity\index{relativity!of simultaneity} of simultaneity implies, when it is considered in the context of a quantum mechanical description of particle interactions. Indeed, what makes the notion of particles propagating backward in time unavoidable is the fact that the emission and absorption events of any process during which a virtual interaction boson\index{virtual interaction boson} is exchanged are space-like separated, which implies that their order of occurrence in time\index{order of occurrence in time!observer dependence} is dependent on the state of motion of an observer. In such a context one cannot avoid having to conclude that what would appear, for one observer, to be a consequence of the emission of some particle carrying a negative charge, would, for another observer, appear to be a consequence of the absorption of a similar particle carrying a positive charge, which certainly requires the sign of charge to be dependent on the perceived direction of propagation in time.

But it would also be incorrect to assume that the direction of propagation in time\index{direction of propagation in time} of a given type of particle, carrying a unit of electric charge with a given positive or negative sign, is definitely the future direction, say, while the direction of propagation of the antiparticle of the same type is definitely the past, or even that there exists an absolutely defined character of being an antiparticle by opposition to being a particle. The only physical property that can be objectively defined by referring exclusively to quantitative attributes of objects which are present in our universe is the relative direction of propagation in time\index{direction of propagation in time!relative attribute} of two particles. Two particles with the same charge may be both propagating in the same direction of time or they may be propagating in opposite directions of time and this is all we can ascertain through physical means.

Thus, what appears to be a positively-charged particle, in relation to another particle propagating forward in time, would actually appear to be a negatively-charged particle in relation to yet another particle propagating backward in time and the same would be true of energy sign. Those relative alterations of the sign of charges occurring as a consequence of a reversal of time are manifested merely in the fact that what is found to be a repulsive non-gravitational interaction between two identical particles propagating in the same direction of time, would upon a reversal of the direction of propagation in time of one of the particles become an attractive interaction, or vice versa, as a result of the equivalent reversal of the sign of charge that occurs when a particle reverses its direction of propagation in time without actually reversing its charge.

This is an unavoidable consequence of the fact that the departure of a positively-charged particle from a region of space would from a reversed-time viewpoint necessarily appear as the arrival of a particle of opposite (negative) charge, therefore implying that there is a relationship between the relative direction of propagation in time\index{direction of propagation in time!relative attribute} and the relative sign of any conserved physical attribute. We do not even need to know what an electric charge is or what energy is, from an exact mathematical viewpoint, to draw that conclusion. The reversal of charges\index{charge reversal!reversal of time} associated with a reversal of time simply illustrates the subtlety of the requirement of relational definition\index{requirement of relational definition!sign of conserved quantities} of the sign of conserved (time-invariant) physical quantities in the context where there exists a fundamental degree of freedom associated with the direction of propagation in time\index{direction of propagation in time!fundamental degree of freedom}. But given that, from the unidirectional-time viewpoint\index{unidirectional-time viewpoint}, the sign of a charge or that of an energy depends on whether it is propagating forward or backward in time, then it is possible to distinguish between the situation in which a particle is propagating a positive charge forward in time and that in which it is propagating the same charge in the opposite direction of time.

It is merely the fact that we are used to experience time as a unidirectional phenomenon that makes us feel suspicious of the possibility that some elementary particles could be propagating effects backward in time. But thermodynamic time asymmetry\index{thermodynamic time asymmetry} only applies to the flow\index{information flow} of information as it occurs through the formation of records\index{formation of records} and in no way forbids individual particles from propagating backward in time as long as they are not involved in processes which (collectively) would allow information to be transferred from the future to the past (I will better explain what motivates this distinction in the first portion of chapter \ref{chap:5}).

Thus, even independently from the argument based on the relativity\index{relativity!of simultaneity} of simultaneity, we may consider that the sign of charges and that of energy are defined only in relation to the state of motion of the particle carrying those charges and that energy. But here `motion' is relative to time instead of space. If we may, in addition, assume that the attribution of a direction of propagation in time\index{direction of propagation in time!convention} is merely a matter of convention, because all that can be asserted is whether any two particles are propagating in the same direction of time or in opposite directions, then it would appear that the sign of energy itself would become a relative notion, dependent on which direction of time is chosen as that in which a given particle propagates. In this particular sense we would have to recognize that associated with the relativity of `motion' in time there is also a relativity of the sign of energy\index{sign of energy!relative attribute}.

Acknowledgment that the sign of energy is a relative property actually allows one to reject the validity of the constraint usually imposed that all energy must be positive, because it means that, even what appears to be positive energy according to one particular convention for the direction of propagation in time, is actually negative energy according to an alternative choice for the same time-direction parameter. Negative energies must be considered to be possible states of matter, even if only for particles propagating in the backward direction of time. This dependence of energy sign on the assumed direction of propagation in time is what allows antiparticles\index{antiparticles!backward-in-time-propagating particles} to be described as particles propagating backward in time with negative energies and non-reversed charges (as Ernst St\"{u}ckelberg\index{St\"{u}ckelberg, Ernst} \cite{Stuckelberg-1} and Richard Feynman\index{Feynman, Richard} \cite{Feynman-1} once suggested), even if we can also consider those particles to be positive-energy particles with reversed charges propagating in the usual forward-in-time direction.

It needs to be recognized, however, that if the energy of an electron is by convention considered positive relative to the future direction of time in which it is, again by convention, assumed to propagate, then the energy of an anti-electron \textit{must} be considered negative relative to the past direction of time in which it must, according to the same convention, be assumed to propagate. It is merely because we ignore the requirement to describe the positron as propagating backward in time that we can attribute to it a positive energy (and a positive electric charge). The fact that energy appears to be a positive-definite physical attribute is thus a mere consequence of this implicit choice of the positive direction of time as that relative to which energy is always measured.

But despite the enormous simplification of our world-view that is made possible by the hypothesis of the existence of a fundamental degree of freedom related to time direction\index{time direction!fundamental degree of freedom}, it is still often suggested that the interpretation of antiparticles\index{antiparticles!backward-in-time-propagating particles} as particles propagating backward in time with negative energy is merely a mathematical artifact and corresponds to nothing real. I think that this attitude is similar to that of nineteenth century philosophers and scientists rejecting the hypothesis of the existence of atoms\index{atoms!existence hypothesis}, even in face of the overwhelming evidence in favor of this concept, supposedly because the atoms could not be seen directly, but actually because of an unjustified prejudice in favor of a continuous description of matter. Given the above discussion concerning the relative nature of the sign of energy\index{sign of energy!relative attribute}, I think that it is clear that there is no basis for assuming, as is often done, that the negative energy\index{negative energy!antiparticles} of antiparticles, as particles propagating backward in time, is not real and that those particles are merely `ordinary' particles which happen to carry opposite non-gravitational charges. If we are allowed to describe antiparticles as particles propagating backward in time, then we must recognize the existence of negative energy states\index{negative energy states}.

It is true, though, that if it was not for the fact that the charges carried by a particle are not reversed when the particle reverses its direction of propagation in time\index{direction of propagation in time}, then it would be impossible to distinguish between the case of a particle propagating a positive energy forward in time and that of a particle propagating a negative energy backward in time, just as it would be impossible to distinguish between the case of a particle propagating a negative energy forward in time and that of a particle propagating a positive energy backward in time. But there is no reason to assume that there would be no distinction between particles propagating positive and negative energies in the same direction of time and therefore the truly significant measure, concerning the sign of energy, is the sign of action\index{sign of action}, which is obtained by multiplying the positive or negative energy of a particle by the positive or negative time interval during which it propagates.

If the hypothesis that energy must necessarily be positive has always appeared valid it is merely as a consequence of the fact that we always measure energy relative to the positive or forward direction of time and for all known particles action remains positive. As I suggested above, however, this does not mean that energy really is always positive, but merely that the sign of action, or the sign of energy relative to that of time intervals, is, in effect, always positive for all currently known particles, independently from the true sign of energy of those particles.

Ultimately, it is not only the sign of energy\index{sign of energy!relative attribute} that must be viewed as a relative physical attribute, but also the sign of action\index{sign of action!relative attribute} itself, because there could never exist a generally agreed, absolutely defined, positive or negative value for the sign of action of a particle. In this context, not only would the sign of energy be dependent on the direction of time in which a particle is assumed to propagate, but the sign of action would itself depend on the choice of what direction of time is to be that in which what we assume to be positive-energy particles propagate, or what is the sign of energy of those particles which are considered to propagate forward in time.

Here all that matters is that once you define one particle as having positive action, because you assume that it is this particle that propagates positive energy forward in time, then the particles that you must assume to be carrying negative energies forward in time or positive energies backward in time, as a consequence of this \textit{choice}, are those which will have negative action. But it must be clear that you are always free to describe the first particle as propagating negative energy forward in time and therefore as having negative action, as all by itself this choice is arbitrary, but in this case the other particles would necessarily have to be assumed to carry positive action instead of negative action, because their \textit{relationships} of time directionality\index{time directionality!relationships} and energy sign with the first particle (the difference or the identity of the signs of time intervals and energy) would remain unchanged.

It must also be remarked that the fact that what we would currently define as negative-action particles are related to ordinary matter through a simple convention regarding the direction of propagation in time\index{direction of propagation in time!convention} means that the motive for rejecting the possibility that negative-action matter may exist is no stronger than that which would consists in arguing that ordinary matter itself is not allowed to exist. There is absolutely no rational motive for rejecting the viewpoint described here and many reasons to recognize its validity. In any case, the fact that the sign of action\index{sign of action!relative concept} is a purely relative concept, which can vary as a consequence of assumptions regarding the direction of propagation in time, means that if the direction of the gravitational acceleration produced by a local matter distribution depends on the sign of action of its source, then it should also vary as a function of the assumptions made concerning the direction of propagation in time of the objects submitted to it (which determine their own action signs in relation to that of the source) and therefore the gravitational field\index{gravitational field!relative concept} must itself be considered a relative concept dependent on the conventions used by an observer.

\bigskip

\noindent Regarding the relation between the sign of charges in general and the direction of propagation in time\index{direction of propagation in time} it must be noted that energy actually distinguishes itself from non-gravitational charges by the fact that it is naturally reversed when a particle reverses its direction of propagation in time. Indeed, in the context where a particle-antiparticle annihilation\index{particle-antiparticle annihilation!bifurcation in time} process must be considered as an event during which a particle bifurcates in time to begin propagating the same non-gravitational charges backward in time (which would effect the same kind of change as reversing the charges and keeping the direction of propagation in time unchanged), it must be assumed that the energy of the particle is reversed, along with the direction of time intervals, when the bifurcation occurs, given that the particle now propagates backward in time while its energy remains positive from the conventional forward-in-time\index{forward-in-time!viewpoint} viewpoint.

In fact, we have no choice but to consider that only non-gravitational charges are left unchanged (relative to the true direction of propagation in time) when the particle begins propagating backward in time during what appears to be a particle-antiparticle annihilation process, because energy is always released by such a process and if the sign of energy had remained unchanged along with that of non-gravitational charges when the direction of propagation in time of the particle reversed, then an antiparticle's energy would be opposite that of its particle counterpart with respect to the forward direction of time and therefore the annihilation of such a pair could occur without any energy at all being released. Thus, energy must actually reverse along the `true' direction of propagation in time of a particle, when the particle reverses its direction of propagation in time during a pair-annihilation process, just like momentum naturally reverses when a particle reverses its direction of motion in space.

If this relational interpretation of the energy signs of particles involved in pair-annihilation processes\index{pair-annihilation processes} is valid, then, based on the fact that we also have many reasons to believe that the gravitational properties of antiparticles\index{antiparticles!gravitational properties} are the same as those of particles, I can deduce that, from a gravitational viewpoint, the sign of energy is physically significant merely in relation to the direction in which a particle with that sign of energy is propagating in time. In other words, to produce an anomalous gravitational field\index{anomalous gravitational field}, or to respond anomalously to a gravitational field\index{gravitational field!anomalous response}, a particle would have to propagate its negative energy forward in time rather than backward, as does an ordinary antiparticle. This is a simple, but very significant result whose consequences will be developed in the following sections.

Concerning the gravitational properties of antimatter\index{antimatter!gravitational properties}, it appears that it is unnecessary to appeal to any independent constraint, like the equivalence principle\index{equivalence principle} (which seems to require all matter to have the same acceleration in a gravitational field), to justify that antimatter should not `fall' up in the gravitational field of a positive-energy planet like the Earth, as was often proposed before experiments began to rule out such a possibility, because all the arguments usually provided to rule out the possibility of an anomalous gravitational behavior of antimatter become unnecessary once it is understood that it is actually only matter propagating its negative energy forward in time that could experience gravitation distinctively from ordinary matter. There is, thus, a very good reason to assume that antimatter falls down in the gravitational field of the Earth, but this is not an argument that we could use to rule out the possibility that some matter that would not be ordinary antimatter could perhaps experience anomalous gravitational interaction\index{anomalous gravitational interaction} with ordinary matter, because there is no \textit{a priori} motive for assuming that there cannot exist particles propagating negative energy forward in time.

Now, it must be understood that while the relationship between the direction of propagation in time\index{direction of propagation in time} and the sign of a given charge is a matter of coordinative definition\index{coordinative definition} (a definition that must be applied similarly to all processes in the whole universe on the basis of their relationships to one particular process for which an arbitrary choice of properties is assumed), once such a definition is applied, the difference between the sign of time intervals and the sign of charges is an objective physical property that is not dependent on a particular viewpoint. But it is not just the relationship between the sign of charge and the direction of propagation in time of a particle which can be given clear meaning through the use of a coordinative definition, because once we define one kind of particle as actually propagating a positive charge forward in time, then it should also be possible to differentiate such a particle from an otherwise identical particle propagating a negative charge in the opposite direction of time.

It must be clear, therefore, that once we assume an ordinary electron to be propagating a negative charge forward in time, it is not possible to consider another \textit{ordinary} electron as perhaps propagating backward in time while carrying a positive electric charge in this direction of time (so that the electron would still appear to be propagating a negative charge relative to the forward direction of time). Indeed, if a certain condition of continuity of the flow of time\index{condition of continuity of flow of time} on which I will elaborate in section \ref{sec:4.3} is assumed to apply, such a backward-in-time-propagating ordinary electron could only annihilate with an anti-electron which would be propagating the same positive charge forward in time (instead of propagating a negative charge backward in time). But this would actually mean that certain positrons cannot annihilate with certain electrons, while no constraint of this kind is observed to apply, as all known electrons have the same unique probability of annihilating with any positron.

Thus, if a constraint of continuity of the flow of time does indeed apply along an elementary particle world-line\index{elementary particle world-line}, then an ordinary electron must be assumed to always propagate a negative charge in the future direction of time, while its antimatter counterpart must similarly be assumed to always propagate this same negative charge in the opposite direction of time. Perhaps that this restriction is a consequence of the fact that there actually exists only one electron, in the sense that all electrons are `the same particle' propagating forward and backward in spacetime, as John Wheeler\index{Wheeler, John} once argued, but the condition of continuity of the flow of time does not specifically require the validity of this hypothesis.

Concerning the properties of negative-action matter, what the preceding considerations, regarding the requirement of relational definition\index{requirement of relational definition!physical attributes} of physical attributes, would mean is that a negative-action particle cannot possibly be considered to have physical properties that would qualify it as responding to the gravitational field\index{gravitational field!anomalous response} of a positive-action object in an anomalous fashion that would not also be shared by an ordinary positive-action matter particle submitted to the gravitational field of a negative-action object. This must be considered an unavoidable conclusion in the context where one can physically distinguish only a difference or an equality in the signs of action of any two particles and cannot attribute absolute meaning to the sign of action itself. That does not mean that there would actually be no anomalous response, only that, in a configuration where all `anomalously' gravitating matter is replaced by ordinary matter and all ordinary matter is replaced by anomalously gravitating matter, we should observe no difference (attributable merely to the gravitational interaction).

Thus, a particle defined as having negative energy relative to the positive direction of time and which would be located in the gravitational field of a planet having opposite energy relative to the positive direction of time should behave in the same way as a positive-energy particle in the gravitational field of a negative-energy planet and similarly for any combination of energy signs of particle and planet, because only the relative difference in forward-propagated energy signs can be considered significant. Given the preceding discussion, this should be crystal-clear. But that is not what people considering the possibility that matter could be found in a negative-energy state usually assume would occur.

What is usually assumed is that a positive-energy or positive-mass object would attract all objects, regardless of whether those objects have positive or negative energy or mass, while a negative-mass object would repel all objects, again regardless of whether those objects have positive or negative mass. It is currently believed that this is the consequence of taking inertial mass\index{inertial mass} to be reversed along with gravitational mass\index{gravitational mass}, as would appear to be required by the equivalence principle\index{equivalence principle}. It must be clear, however, that those are not results which are `derived' from relativity theory, as is sometimes suggested, but merely the consequence of a choice that is implicitly made regarding what properties should be associated with negative inertial mass\index{negative inertial mass}, while trying to be as accommodating as possible with the conventional formulation of the equivalence principle.

But if I find it appropriate and indeed necessary to consider, as most people do, that inertial mass is reversed along with gravitational mass when we are considering an object with negative energy (which would normally allow the equivalence principle to apply), I cannot agree with the conclusion that is usually drawn from such an assumption. Indeed, for the response of various masses to the presence of a negative mass\index{negative mass} to be in line with common expectations, it must be possible to determine the sign of mass, or the sign of action\index{sign of action} of particles in an absolute, non-relational manner, because we are assigning the attractive or repulsive character of the gravitational field in precisely such an absolute manner (the field is either repulsive for everything or attractive for everything), which I believe could never be justified.

It cannot be assumed that a negative mass is repulsive in an absolute, invariant way, because it would not be possible to tell in relation to which aspect of physical reality the distinctiveness of this character is defined, given that positive mass cannot be used as a reference if its gravitationally-attractive nature is itself absolutely defined (does not vary merely in relation to a variation of the sign of mass of the object with which it is interacting). I will explain, in section \ref{sec:2.4}, why it is that the assumption that a negative inertial mass\index{negative inertial mass} is associated with a reversal of the sign of action, far from having the undesirable consequence of allowing absolutely defined physical attributes\index{absolutely defined physical attributes} into physical theory (if there could ever be such a theory), actually gives rise to a description of the gravitational interaction between positive and negative-mass objects that is in perfect agreement with the requirement of relational definition\index{requirement of relational definition!sign of mass} of the sign of mass or energy (once the inertial properties of negative-mass matter are well understood). All that would then remain to understand is how the equivalence principle\index{equivalence principle} can still be satisfied by such a description.

For that purpose, I will provide arguments to the effect that a simple reconsideration of the true significance of the principle of equivalence, and a better understanding of its motivation in the principle of relativity\index{principle of relativity!accelerated motion} of accelerated motion, allows its foundations to be preserved while enabling the more consistent, relational viewpoint\index{relational viewpoint!sign of mass} on the sign of mass to be retained and to actually be integrated into the core mathematical framework of relativity theory\index{relativity theory!mathematical framework} by introducing a slight modification to this classical theory of gravitation that is actually a simple generalization of it. In order to further justify this approach, I will first try to identify what the true properties of negative-action matter actually are and explain why we should not expect such matter to behave in ways that would make it undesirable, not only from the viewpoint of the requirement of a relational description of physical attributes, but with respect to other constraints and other physical principles which we can be confident must also be obeyed.

\section{Our current understanding\label{sec:2.3}}

Before addressing the question of how a negative-energy particle would actually behave, it is essential to explore what the current situation is regarding the notion, or indeed the problem of negative energy. For this purpose, it should first of all be noted that for many reasons no one seems to like the idea that there could exist negative-energy particles. Thus, it is no surprise that one of the most basic and often implicit assumption that enters our description of physical reality is that energy must always be positive. There are many different mathematical formulations of that requirement which impose various degrees of conformity to the hypothesis that matter cannot find itself in a negative energy state (for a technical review of those conditions see Ref. \cite{Pfenning-3}).

In its least restrictive form this condition is called the weak energy condition\index{weak energy condition} and merely constitutes a statement about the positivity of the contribution of the stress-energy tensor\index{stress-energy tensors} of matter to the curvature of spacetime\index{curvature of spacetime}. More constraining conditions have also been proposed, among which is the appropriately named strong energy condition\index{strong energy condition} which, if obeyed under all circumstances, would mean that gravity must always be attractive (between all forms of matter which would then be allowed to exist). Those conditions are used as rigorously defined hypotheses in various theorems dealing with the behavior of matter under the influence of the gravitational interaction.

The problem is that it was found, at some point, that configurations involving negative energy densities\index{negative energy densities!quantum field theory} are allowed to occur, under certain conditions, in quantum field theory \cite{Epstein-1}. This does not mean that negative-energy particles are explicitly allowed by current theories, but merely that unlike what we would expect from a classical viewpoint, where the vacuum is described as a total absence of matter, quantum field theory allows for the local density of energy to not always be positive-definite, even in the context where only positive-energy matter is present.

A well-known experiment illustrates the kind of phenomena involved. It requires placing two parallel mirrors\index{parallel mirrors in vacuum} a very small distance apart in a vacuum, so as to forbid some states, which would normally exist in the vacuum, from being present in the space between the mirrors, as a consequence of the incompatibility of their characteristic wavelengths with the spatial constraints imposed by the presence of the mirrors. The predicted result, which is actually observed, is that there should arise a small negative pressure\index{negative pressure} between the mirrors that can be attributed to the restriction which is imposed on the presence of certain virtual particles in this volume. This negative pressure is too small to exert a measurable repulsive gravitational force, but it does result in the two mirrors being pulled together as a consequence of the comparatively larger pressure exerted from the outside. This is of course the phenomenon known as the Casimir effect\index{Casimir effect} \cite{Casimir-1}.

It is clear though that we are not directly measuring a negative energy density in such an experiment, but merely the indirect effects of an absence of some positive contribution to vacuum energy, which is then assumed to imply that the energy density in the small volume between the mirrors is smaller than that which exists even in the absence of any matter and which is usually considered to be null. But even this particular occurrence of negative energy is assumed to be so serious a problem by some theorists that they suggested that the description of the vacuum as involving virtual particles\index{virtual particles} coming in and out of existence is actually only a mathematical trick and does not reflect what is really going on in the absence of `real' matter.

However, this aversion for whatever is negative of energy is not shared by all authors and some more open-minded specialists have tried to address the issue of negative energies\index{negative energy!quantum field theory} as they occur in quantum field theory and in so doing gained some significant insights into what exactly is allowed by a quantized description of the vacuum\footnote{
It was found out \cite{Ford-1} \cite{Ford-2} \cite{Pfenning-1} \cite{Pfenning-2} that quantum field theory places strong limits on the values of negative energy density\index{negative energy densities!quantum field theory} that can be observed over finite periods of time under various conditions. Thus, there appears to be a constraint on the magnitude of negative energy that can be observed which requires it to be merely as large as the time interval during which it is measured is short.}.
 A modified version of the weak energy condition\index{weak energy condition} was thus proposed that allows to take into account the fluctuations of energy which arise in the quantum realm. This condition, which is appropriately called the averaged, weak energy condition\index{weak energy condition!averaged}, involves only quantum\index{quantum!expectation value} expectation values of the stress-energy tensor\index{stress-energy tensors} averaged over some period of time during which the observations are assumed to occur, rather than idealized measurements at a spacetime point.

A feature of the constraint provided by this condition is that it allows for the presence of large negative energies over relatively large regions of space if there is a compensation by the presence of a larger amount of positive energy during the time period over which the observations are made. I believe that this is indicative of the fact that while negative energy states cannot be ruled out as strictly forbidden, they should also clearly not be expected to materialize in stable form in the context where we are dealing with ordinary matter configurations, for which the particles are already predominantly in positive energy states.

A similar limitation can also be observed to restrain another form of negative energy that occurs in the presence of an attractive force field\index{attractive force field!negative energy}, even in a classical context. Indeed, the energy contained in the force field between two positive-energy particles submitted to an attractive interaction must be considered negative. This is because positive energy must be provided to separate two positive-energy particles attracted to one another in such a way and given that the attractive force field responsible for this interaction would contain no energy at all when the particles are separated by a distance that tends to infinity (given that the strength of the field would then itself be null), then we must conclude that the energy contained in the same attractive force field, when the particles were near one another, was actually negative (so that adding positive energy can produce a null final value).

This is unavoidable, because a negative contribution is required to reduce the energy of a bound system\index{bound systems!energy}, which is lower than the sum of the positive energies of its component particles. The additional energy that was present before the formation of a bound system is in fact released (through the emission of radiation for example) when the system is created, but except for the additional negative energy contained in the attractive force field, the system is identical, in terms of its matter particle content, to what it was initially and therefore we definitely need the negative energy. This is made more obvious when we consider larger systems like those bound by the gravitational interaction. It was shown, in effect, that even a system as large as the Earth-Moon system\index{Earth-Moon system} has an asymptotically-defined total mass (providing a measure of its total energy) which is smaller than that of its constituent planets (when it is possible to neglect any contribution which would normally be attributed to the presence of dark matter\index{dark matter}), which means that the energy contained in the gravitational field maintaining the two planets in orbit around their center of mass must be negative.

What is crucial to understand regarding the situation described here, however, is that even if we must acknowledge the existence of a well-defined negative contribution to the energy of some physical systems that diminishes their total energy, it is again impossible to measure that energy directly and it can merely be deduced to occur from the behavior of the positive-energy subsystems which are submitted to the attractive interaction. Here also, the negative energy must be associated with virtual particles\index{virtual particles}, namely the unobserved bosons that mediate the interaction, and cannot be measured independently from the total energy of the bound systems, which usually remains positive. It is simply not possible to isolate the attractive force field of a bound system from its positive-energy sources and this is true for systems of any size.

It would, nevertheless, certainly be a concern if the negative binding energy\index{negative binding energy} of a system made of positive-energy components could become so negative as to make the total energy of the bound system\index{bound systems!energy} itself negative. Once again, however, it was shown that there are unavoidable theoretical constraints on the values that observable total energy can take\footnote{
It was shown \cite{Brill-1} \cite{Brill-2} \cite{Deser-1} \cite{Brill-3} \cite{Schoen-1} \cite{Schoen-2} \cite{Schoen-3} \cite{Schoen-4} \cite{Schoen-5}, concerning the gravitational interaction in particular, that the energy of matter (everything except gravitation) plus that of gravitation is always positive when the dominant energy condition\index{dominant energy condition|nn} is assumed to be valid, which actually amounts to assume that the energy of the component particles is itself positive.}.
 If we compress a positive-energy object too tightly, it simply collapses into a black hole before its surface area is allowed to become so small and its energy density so large that the magnitude of its negative gravitational potential energy\index{gravitational potential energy!negative} would be larger than the positive energy of the matter. Thus, positive-energy matter cannot turn into negative-energy matter through an increase of negative gravitational potential energy.

What must be retained from the preceding considerations, therefore, is that despite the fact that they are usually considered forbidden, negative energy states\index{negative energy states} are nevertheless quite commonplace, even from a conventional viewpoint. But it appears that it is the very conditions necessary for the existence of those negative-energy states which are responsible for the fact that it is not possible to observe them as stable states of matter particles propagating forward in time, which would be similar to those occupied by positive-energy matter particles. But we still have no argument to rule out the possibility that there may exist configurations where the matter particles which allow those conditions to exist would themselves have negative energies and for which there would exist constraints, similar to those unveiled here, enforcing the \textit{negativity} of energy.

\bigskip

\noindent It was discovered, a long time ago, by Paul Dirac\index{Dirac, Paul} (when he achieved his unification of special relativity\index{special relativity!unification with quantum theory} and quantum theory) that there actually exists a mathematical requirement for the existence of negative energy states. Indeed, it turned out that in order to obtain Lorentz-invariant equations\index{Lorentz-invariant equations!wave function} for the wave function one had to sacrifice the positivity of energy\index{positivity of energy}. After having considered various possible interpretations for what in nature could possibly correspond to those negative energy states, Dirac concluded that it required the existence of a new category of particles, the antiparticles\index{antiparticles!holes in filled matter distribution}, which would consist of holes in a filled distribution of such negative-energy matter. But despite the fact that it was later found that antiparticles do exist, as he predicted, Dirac's solution to the problem of negative energy states\index{negative energy states!Dirac's solution} was never considered fully satisfactory.

Antiparticles\index{antiparticles!backward-in-time-propagating particles} were eventually described by Feynman\index{Feynman, Richard} (following St\"{u}ckel\-berg\index{St\"{u}ckelberg, Ernst}) as particles propagating negative energies backward in time, which allowed to fulfill the mathematical requirements imposed by the existence of the negative energy states (by providing an interpretation for those transitions which were predicted to involve a reversal of energy) without requiring the presence of the filled, negative energy continuum. But in the process, it seems that the discovery that particles could actually occupy negative energy states, which appeared to be implied by the original developments, was somehow forgotten and lost in the details of the proposed solution. This indifference was probably justified by the fact that antiparticles could still be considered to have positive energy, from a practical viewpoint, given that, if they really are propagating a negative energy backward in time, then relative to the future direction of time in which observations are made, they would still appear to have a positive energy.

But what is usually unrecognized is that while attributing a positive energy to antiparticles may appear more `reasonable' than assuming that those particles propagate negative energy backward in time, such a choice would actually imply that it is the particles themselves (by opposition to antiparticles) which must then be considered to carry negative energy backward in time, because it must be either that or the opposite. This is what the subtleties of the quantum-mechanical definition of energy\index{energy!quantum-mechanical definition} seems to require that was not apparent classically. The reluctance to recognize the existence of negative energy states\index{negative energy states} is probably also in part a consequence of the apparently insurmountable difficulties which would be associated with the possibility for particles to occupy those physically allowed states.

First of all, it is certainly not desirable from a theoretical viewpoint to assume that antiparticles would be submitted to anomalous gravitational interaction\index{anomalous gravitational interaction!antiparticles} as a consequence of propagating negative energy backward in time, because it was demonstrated some time ago \cite{Nieto-1} that if, for any reason, antimatter was to be found experiencing repulsive gravitational interactions with ordinary matter, we would run into a number of problems ranging from violations of the conservation of energy\index{conservation of energy!violation} and up to the undesirable and unlikely (from a theoretical perspective) possibility of producing perpetual motion\index{perpetual motion} machines.

But an analysis of the arguments presented against the possibility of anomalously gravitating antimatter has led me to conclude (for reasons which will be explained later) that the problem really has to do merely with the possibility for antimatter `as we know it' to experience gravitational repulsion. It cannot be considered to mean that matter propagating negative energy in the future direction of time could not exist and experience anomalous gravitational interactions with ordinary matter without violating the principle of conservation of energy or the second law of thermodynamics\index{second law of thermodynamics}, because matter in such a negative energy state may also, by necessity, have properties different from those which are known to characterize antimatter, in particular with what regards non-gravitational interactions.

Nevertheless, most people today seem to consider that the developments that followed the introduction of the early theory of relativistic quantum mechanics\index{relativistic quantum mechanics} and which gave rise to modern quantum field theory\index{quantum field theory} have eliminated the problem of negative energy states\index{negative energy states!problem}, which can now be considered a mere artifact of the former single-particle theory\index{single-particle theory}. Thus, the predicted negative energy states would simply be nonphysical solutions that must be discarded as irrelevant to physical reality. But it must be clear that this is indeed what we are doing here. We are rejecting the possibility that a particle could be found in a whole set of states that are allowed by the most basic equations without providing any justification as to why those states should be forbidden.

Upon closer examination it becomes clear that if `true' negative energy states do not explicitly arise in quantum field theory it is not because the structure of the theory forbids them, but simply because we \textit{choose} to ignore those solutions to start with and then integrate that choice into the formalism. More specifically, it turns out that what prevents negative-action particles from showing up in quantum field theory is merely a choice of boundary conditions\index{boundary conditions!path integrals} for the path integrals that provide the probability amplitude for transitions involving particle trajectories in spacetime.

There are several possible choices for expanding those integrals which all constitute valid solutions of the equations of the theory, but only those solutions propagating positive frequencies\index{positive frequencies} forward in time and negative frequencies\index{negative frequencies} backward in time are usually considered to be physically significant, while the solutions propagating negative frequencies forward in time and positive frequencies backward in time, which are also valid from a mathematical viewpoint, are systematically rejected. But this actually amounts to retain only the positive-action portion of the theory, while ignoring all transitions involving negative-action (although not negative-energy) particles. There is no other origin for the often-mentioned conclusion that quantum field theory\index{quantum field theory} does not involve negative-energy matter. It is our own arbitrary decision to reject all transitions involving negative-action particles.

In order to make the choice of boundary conditions\index{boundary conditions!absence of negative-action particles} responsible for the absence of negative-action particles in quantum field theory more acceptable it is sometimes suggested that the negative energies predicted by the single-particle, relativistic equations\index{relativistic equations!single-particle} are simply transition energies\index{transition energies!positive or negative}, or differences between two positive energy states and there is obviously no reason why those variations could not be negative if they can be positive. But no explanation has ever been provided for why the same reasoning could not be applied to the energy states themselves, which are also energy differences, given that the energy of a particle is always defined in relation to the zero level of energy\index{zero-energy level} associated with the vacuum in which it propagates. There is no justification for this arbitrary distinction between transition energies and particle energies, except for the satisfaction that is obtained by the physicist in having easily disposed of an embarrassing problem.

It may of course be argued that there is nothing wrong with those methods, given that they appear to be validated by experimental results. Indeed, we have never observed interferences by negative-action particles into the outcome of any experiment conducted at any level of energy and to any degree of precision. But I would like to emphasize that this still doesn't constitute an explanation for the absence of negative-action particles. Thus, the problem I have with the modern approach to quantum field theory\index{quantum field theory!modern approach} is that the formalism is generally introduced in a way that encourages us to believe that, after all, no particle is actually propagating backward in time with negative energy and that a positron is really just another particle, identical to the electron, but with an opposite electrical charge. However, this viewpoint does not only complicate things unnecessarily as a consequence of rejecting the possibility that electrons and positrons and all other particles and their related antiparticles could actually consist in the same particles observed from different perspectives, it is also completely ignorant of the requirement of a relational definition\index{requirement of relational definition!physical attributes} of any physical attribute dependent on the fundamental time-direction degree of freedom\index{time-direction degree of freedom!fundamental}.

But if we choose to recognize the validity and the greater value of the viewpoint defended here and according to which antiparticles\index{antiparticles!backward-in-time-propagating particles} are really just ordinary particles propagating backward in time, then we must accept that there definitely exist in nature particles which carry negative energies and if the arguments provided above concerning the arbitrariness of the current restrictions imposed on the propagation of those negative energy states are valid, then we would have to conclude that there should necessarily also exist particles propagating such energies forward in time and which could be submitted to anomalous gravitational interactions\index{anomalous gravitational interaction} in the presence of ordinary matter. To better understand the nature of the difficulties which would appear to emerge if such particles were allowed to exist it will be necessary to first investigate the conventional concept of negative mass\index{negative mass!conventional concept}, which is usually associated with gravitational repulsion\index{gravitational repulsion}.

\section{The negative-mass concept\label{sec:2.4}}

When discussing the issue of negative mass\index{negative mass}, what must first of all be understood is that, if the physical property of mass is to have any polarity associated with it, such that we could attribute to mass either a positive or a negative sign, then this polarity must be directly related to the sign of action\index{sign of action}, that is, to the sign of energy relative to the positive direction of time. This is because, as I previously emphasized, the sign of action is the only physical property from which the attractive or repulsive character of the gravitational interaction between two objects could depend. We may, thus, attribute positive mass to a positive-action particle and negative mass to a negative-action particle.

Mass being a Newtonian concept, its polarity must be determined in relation to a particular Newtonian gravitational field\index{Newtonian gravitational field}. Thus, the sign of mass\index{sign of mass} of a given particle must be understood as determining the response to the gravitational field of a given source, in the sense that it determines the \textit{direction} of the gravitational force exerted on such a particle. If we may consider the gravitational field of the source (represented by a vector in Newtonian mechanics\index{Newtonian mechanics}) to be uniform, then only its own direction or polarity (which we may assume to be dependent merely on the sign of mass of the source, when its position is assumed to be fixed) would be decisive in determining the kind of response experienced by a given type of mass submitted to it.

Equipped with such a definition, we can meaningfully discuss the problem of the gravitational interaction of negative-action particles with positive-action particles and with themselves as the problem of the gravitational interaction of positive and negative masses\index{negative mass}. This will allow us to better grasp the significance of the assumptions that will form the basis of the new interpretation of negative-energy matter which I shall propose and therefore, also, to gain better confidence in their validity, even in the more appropriate context of a general-relativistic theory\index{general-relativistic theory}.

If we may agree on those requirements, then I think that what must emerge is that, if it is indeed important to have a well-defined concept of negative mass, then it also seems that such a negative mass must be negative in all respects. That there could be a difference between the sign of gravitational mass\index{gravitational mass} and the sign of inertial mass\index{inertial mass} is usually considered to be forbidden merely by the general theory of relativity\index{general relativity theory} which is, in effect, founded on the principle of equivalence\index{equivalence principle} which requires the equality of gravitational and inertial masses. However, I think that if this hypothesis is justified, it is not because our negative-mass concept must comply with some perceived requirement from general relativity theory, but because it would not be acceptable to attribute mutually exclusive values to one single physical attribute.

Thus, I do believe that the mass of any particle or bound object should be either definitely positive, or definitely negative (but still in a relational way), regardless of whether we are considering gravitational mass or inertial mass, if the concept itself is to have any consistent physical meaning. But unlike most theorists, I do not consider that this requirement must be assumed to imply the kind of behavior that is usually attributed to negative-mass matter, where gravitational repulsion\index{gravitational repulsion!intrinsic property} is an intrinsic property of this type of matter itself, independently from the sign of mass of the matter with which it is interacting. This is indeed the conclusion I was able to draw based on the outcome of the previously discussed analysis of the constraints imposed by the requirement of relational definition\index{requirement of relational definition!sign of energy} of the sign of energy, for reasons I will now explain.

The difficulty I originally met when I first began to explore the possibility that inertial mass\index{inertial mass} could be reversed along with gravitational mass\index{gravitational mass} when we are dealing with negative-mass matter is that, if both the gravitational mass and the inertial mass are to be negative at once, then it seems that there could occur situations where the principle of inertia\index{principle of inertia!violation} would be violated (I will explain what motivates this conclusion below). I was eventually able to understand, however, that this is merely a consequence of the inappropriateness of current assumptions regarding what we should expect to be the behavior of matter with both a negative gravitational mass\index{negative gravitational mass} and a negative inertial mass\index{negative inertial mass}.

Currently, it is assumed that given that positive-mass matter gravitationally attracts all matter and resist the action of any force exerted on it, then this must be an intrinsic property of such positive masses. On the other hand, it is usually assumed that two choices exist for what could possibly characterize the behavior of matter with a negative mass. First of all, we must assume that \textit{gravitational} mass is indeed negative when mass is reversed. This would give rise to gravitational repulsion\index{gravitational repulsion} when only the mass of the source (the active gravitational mass\index{active gravitational mass}) is negative, because it reverses the polarity of the Newtonian gravitational field\index{Newtonian gravitational field} to which any passive gravitational mass\index{passive gravitational mass} is submitted and therefore should at least reverse the force exerted on positive-mass objects. But once this is recognized two possibilities would appear to exist for a negative-mass particle submitted to a given gravitational field, depending on whether \textit{inertial} mass is assumed to remain positive or is itself also negative.

The inertial mass\index{inertial mass!sign} sign is usually assumed to determine the direction of the acceleration of a particle submitted to a given force, including a gravitational force, while the gravitational mass\index{gravitational mass!sign} sign of a particle is assumed to determine both the polarity\index{polarity!gravitational field} of the gravitational field it produces and its response to a given gravitational force field. If we were to agree with those assumptions, then we would have to conclude that a negative-gravitational-mass\index{negative gravitational mass} particle with a negative inertial mass\index{negative inertial mass}, should actually respond `normally' to any gravitational force field (because the nature of its response is changed twice, once by the reversal of its inertial mass and once by that of its gravitational mass) while its response to non-gravitational forces would be reversed (same force, opposite acceleration), as current assumptions concerning the effects of a reversal of inertial mass would imply.

But we must keep in mind that the fact that such negative-mass particles would respond normally to gravitational force fields would mean that they are repelled by other negative masses, because the gravitational fields produced by those masses are themselves reversed. Thus, if the above stated commonly accepted assumptions are valid, negative masses should repel masses of all signs, be repelled by other negative masses and be attracted to positive masses. Given that it is usually considered that, in a general-relativistic context, all mass (gravitational and inertial) must be negative, this is the choice that is usually retained as defining the behavior of negative-mass matter if it could exist.

But despite the support that is usually granted to such a conception of negative-mass or negative-energy matter, I think that enormous problems would arise if it was retained as a valid proposal. Some of those problems, involving black holes and the second law of thermodynamics\index{second law of thermodynamics}, will be discussed later, but even if we remain at the level of classical Newtonian dynamics\index{classical Newtonian dynamics} we can readily identify one very serious problem, which is that the existence of such matter would allow violations of the principle of inertia\index{principle of inertia!violation} (considered as a generalization of Newton's first law\index{Newton's first law}) or the hypothesis that no physical system can accelerate without work being done on it by an external force. This is a consequence of assuming that a negative-mass object would both repel positive-mass objects and be attracted to them, because such a combination of features would give rise to unlikely phenomena, like that where a pair of opposite-mass objects\index{pair of opposite-mass objects!self-acceleration} accelerates itself to arbitrarily large velocities, the negative-mass object being gravitationally attracted toward the positive-mass object it repels, without any external force being applied on the two particles, as first described in Ref. \cite{Bondi-1}.

But this is usually considered appropriate and necessary, given that energy would appear to be conserved under such conditions (because the energy gained by one of the objects would be opposite that of the other), while it may seem that this would not be the case for a pair of opposite-energy objects that would interact with one another through mutually repulsive gravitational forces\index{mutually repulsive gravitational forces}. But, as I will explain in section \ref{sec:2.10}, this is not really an advantage of the conventional approach, because in the context of a fully consistent theory of the gravitational interaction, other energy variations must be taken into considerations, which already allow energy and momentum to be conserved in the presence of negative-energy matter. In fact, the possibility that would seem to exist, from a conventional viewpoint, for both positive and negative kinetic energies\index{kinetic energy!creation out of nothing} to be created out of nothing in the way described above would be very problematic from a theoretical viewpoint, given that it would allow perpetual motion\index{perpetual motion} to be achieved.

But an even more significant problem is that there would be no equal and opposite force to balance that which is applied on one of the two objects that could be attributed to its interaction with the other object and this would be a violation of the principle of action and reaction\index{principle of action and reaction} (Newton's third law\index{Newton's third law} in a classical context), while this is one requirement that in all fairness we should recognize as being as essential as that of conservation of energy, because if it does not rigorously apply then absolutely anything could occur and under such conditions we could not give much of even the principle of conservation of energy\index{principle of conservation of energy}.

However, I think that what those observations show is not the nonphysical nature of negative mass\index{negative mass}, but merely the ineffectiveness of the conventional concept of negative-mass matter. It is important to mention, by the way, that even though this hypothetical situation of a pair of self-accelerating opposite-mass objects\index{pair of opposite-mass objects!self-acceleration} has been described by other authors in the past, none of them has ever recognized that what it actually demonstrates is the inconsistency of the currently favored notion of negative mass, which I believe is illustrative of the state of denial in which most people remain concerning the possibility that there could actually exist negative-mass matter.

What is also significant concerning the unlikely phenomenon described above is that it would necessarily be the positive-mass objects that would be chased in this way, while the negative-mass objects would inevitably be those trailing them. But isn't it strange that there should be such a clear and decisive distinction between what constitutes the role of positive masses and what constitutes that of negative masses? Doesn't it seem like there is something wrong with such a possibility? Shouldn't we only be allowed to define the property of gravitational attraction and repulsion in such a way that we could not observe such mass-sign-distinguishing behavior?

I believe that the unease we experience in face of the strangeness of such phenomena is in fact justified, because it does not just seem like there is something wrong here. What we have just described is actually the perfect example of an attempt to distinguish a physical property (the positivity of mass or the attractiveness of gravitation) despite the absence of any reference in the physical universe to which that arbitrary distinction could be related, which violates the basic requirement of relational determination\index{requirement of relational determination!physical attributes} of physical attributes discussed above. The mistake which is made by assuming the validity of the conventional viewpoint is that we suppose that we can define attraction and repulsion in an absolute (non-relative) manner such that one kind of mass always attracts all kinds of masses regardless of their polarities and another always repels all masses, still regardless of their polarities, as if attractiveness and repulsiveness were intrinsic aspects of one and the other type of mass.

However, if the sign of mass\index{sign of mass} is to be considered a meaningful physical property of elementary particles, then it must be taken to indicate that there can be a reversed or opposite value to a given mass and this reversed value can be considered to be reversed merely in relation to a non-reversed mass and to nothing else. A mass cannot be considered to be reversed with respect to an absolute point of reference lacking any physical significance in the universe. Therefore, if a gravitational field is to be assumed repulsive as a consequence of the reversed (negative) sign of the mass of the matter that is the source of the field, then this gravitational field should be repulsive only for an unchanged (positive) mass particle and not with respect to other negative masses. It would be incorrect to assume that the attractive or repulsive nature of gravitational fields depends solely on the sign of mass of the source itself, because no distinction exists for the sign of a mass other than its sameness or oppositeness compared to that of another mass.

That does not mean that the field itself must be assumed to change as a consequence of the reversal of the sign of mass of the particle experiencing it (even though that may be one way to describe things if other conventions are adopted for the sign of mass itself as we will see later), but merely that the response of a negative-mass particle to a given gravitational field must be reversed in comparison to the response we would expect from a positive-mass particle submitted to the same field, despite the associated reversal of the inertial mass\index{inertial mass!reversal} of such a particle. If that was not the case, then I think that we would have to conclude that negative mass\index{negative mass} is, in effect, forbidden.

If the incorrect hypothesis on which the conventional approach is based, regarding the effect of a reversal of inertial mass, nevertheless allows to successfully (from my viewpoint) predict that a positive mass would be repelled in the gravitational field of a negative mass, it is simply because we assume the right inertial properties for the positive-mass matter\index{positive-mass matter!inertial properties} submitted to the gravitational force of the negative mass. Thus, the positive mass responds in the appropriate way to the gravitational force exerted by the negative mass, which is correctly assumed to be a repulsive force, given that the gravitational field produced by the negative mass is necessarily opposite that which would be produced by a positive mass of similar magnitude located in the same position.

The problem is that, given that it seems that we cannot expect the same kind of behavior from a negative mass\index{negative mass} submitted to the gravitational field of a positive mass, then it would appear that the behavior of both positive and negative masses is the consequence of some predetermined properties of absolute attractiveness and repulsiveness\index{absolute attractiveness and repulsiveness} associated with the gravitational fields emanating from positive and negative masses, respectively, and which do not depend merely on a property of the source defined with respect to a property of the matter with which it interacts. The difficulty to which the conventional interpretation gives rise is also made apparent when we consider the case of a negative mass in the gravitational field of another negative mass, given that now the negative mass would be repelled by the same negative-mass matter (because the gravitational force is unchanged, but the response to this force would be reversed), while, on the basis of the requirement of relational definition\index{requirement of relational definition!sign of mass} of the sign of mass, there should be no difference between this case and that of a positive mass in the gravitational field of another positive mass (which is symmetric to the other case under exchange of mass signs).

The appropriate outcome could only be obtained if, in addition to the assumption regarding the nature of the gravitational force between two negative-mass objects, it was also assumed that the reversal of the inertial mass\index{inertial mass!reversal} of the negative-mass object submitted to this force actually changes nothing to the response of that object to the force that is exerted by the other negative-mass object. Thus, the problem of the absoluteness of the attractive or repulsive nature of the gravitational field arises as a direct consequence of current assumptions regarding the effect of a reversal of inertial mass. It is only in this context that the direction of the Newtonian gravitational force\index{Newtonian gravitational force!absolute direction} associated with a concentration of matter of positive or negative mass sign acquires an absolute meaning and is not merely dependent on the identity or the difference between the sign of mass\index{sign of mass} of the matter submitted to the gravitational field and that of the matter that is the source of this field.

Even if merely as a consequence of the previously discussed considerations regarding the relative nature of the sign of energy\index{sign of energy!relative attribute} (as dependent on the direction of propagation in time\index{direction of propagation in time} of a particle) and the purely conventional (subject to an arbitrary coordinative definition\index{coordinative definition}) nature of the sign of action\index{sign of action!conventional nature}, it would appear that a consistent concept of negative mass\index{negative mass!consistent concept} would require that it is the relative difference or absence of difference between the mass signs of two gravitationally interacting objects that determines the attractive or repulsive character of this interaction, so that two negative-mass objects should be submitted to the same mutual gravitational attraction\index{mutual gravitational attraction!negative-mass objects} that is experienced by two positive-mass objects, while the same negative-mass objects would also repel ordinary positive-mass objects and be repelled by them, unlike is usually assumed.

To be fair, I must acknowledge that some authors did suggest in the past that the gravitational interaction should perhaps be repulsive between two objects with opposite mass signs, while it would be attractive between two negative-mass objects (just as it is between two positive-mass objects), but simply on the basis of the fact that the sign of the gravitational force that is obtained by reversing the sign of one of the masses in Newton's equation for universal gravitation\index{Newton's equation for universal gravitation} would itself be reversed, while it would be unchanged if the signs of the two masses were together reversed. But even though it is not necessarily wrong to suggest that the repulsive or attractive nature of the gravitational interaction is determined by the signs of mass in Newton's equation for universal gravitation, it is only when we realize that the sign of mass must be related to the sign of action that we can begin to understand why it is that there should be a symmetry under exchange\index{exchange symmetry!positive and negative masses} of positive and negative masses.

I previously mentioned, in effect, that positive action states are related to negative action states by a simple convention regarding the sign of energy and that of time intervals, so that the sign of action\index{sign of action!relative concept} is itself a purely relative concept. There must consequently be a symmetry under exchange\index{exchange symmetry!positive- and negative-action matter} of positive- and negative-action matter, which would then require the behavior of positive masses in relation to themselves and in relation to negative masses to be similar to that of negative masses in relation to themselves and in relation to positive masses.

The contradictions of the conventional concept of negative-mass matter\index{negative-mass matter!conventional concept} can be illustrated by using a rarely discussed thought experiment. It has, in effect, been proposed that the sign of energy of a negative-mass object could be determined by measuring the energy lost or gained while raising or lowering the object in the gravitational field of a large mass. Now, according to the conventional approach, if we were to raise a negative-mass object in the gravitational field of a positive-mass object, like a planet, we would have to produce work and exert a force directed downward, because the inertial mass\index{inertial mass!negative} of the object is negative, which according to the conventional viewpoint means that it responds perversely to the applied force. But then, it is also the case, according to this same viewpoint, that the gravitational force exerted by the planet on the object should be attractive, because the planet has positive mass. Thus, we would be in the situation where we would have to exert a force downward to raise a negative-mass object in the gravitational field of a planet that exerts an \textit{attractive} force on that object.

I do not know to what extent people actually believe in the validity of such a conclusion, but I think that, faced with such absurdities, one has to come to realize that the contradictions involved are a clear indication that the conventional assumptions regarding the behavior of negative-mass or negative-action matter are incorrect and that a better interpretation of the whole concept of mass sign is required.

\bigskip

\noindent Despite the fact that the question of the validity of the conventional concept of negative-mass matter had never been clearly analyzed before, it is no doubt the general feeling that there is something wrong with the possibility of observing phenomena of the type described above which is responsible for having transformed the idea of negative-energy or negative-mass matter into the synonym of nonsense it has become in the minds of so many researchers. But, is negative mass really to blame here, or could it be that we are not attributing to it the right physical properties?

There is, of course, even under the conventional assumptions regarding the response of negative-mass particles to applied forces, another possibility, which is that when gravitational mass\index{negative gravitational mass} is negative, inertial mass\index{inertial mass!positive} may remain positive for some reason. Of course that would not only appear to contradict the equivalence principle\index{equivalence principle}, as is already understood, it would also itself be nonsense, as we would have to assume that one single physical attribute of one single particle (the mass of that particle) is at once both positive and negative, for the same observer. The latter problem has never been discussed, but I think that it is actually the strongest argument one can make against this second possibility. We may nevertheless begin by exploring the consequences of such a choice.

Under the same commonly held assumption to the effect that the response of an object to any force is dependent on the sign of its inertial mass, we would have to conclude that a negative-gravitational-mass object, to which a positive inertial mass would be attributed, would respond anomalously (in comparison to the response expected of a positive mass) to any gravitational force field (because the nature of the response is changed only once by the reversal of its gravitational mass), while its response to non-gravitational forces would be unchanged (same force, same acceleration), because the inertial mass remains positive or unchanged in comparison with that of positive-mass objects.

Therefore, if material objects were to exist that would be made of such negative-mass matter they should, from the conventional viewpoint, gravitationally attract one another (as do positive masses), repel positive-mass objects and also be repelled by those same positive-mass objects. As a consequence, we would observe no violation of the principle of inertia\index{principle of inertia!absence of violation} in this case and also no acceleration without work\index{acceleration without work}. If this behavior was to be observed, it would in fact be possible to exchange all positive-mass objects by negative-mass objects and vice versa and no apparent change in the phenomenology of the gravitational interaction\index{gravitational interaction!phenomenology} would be detectable, because gravitational repulsion\index{gravitational repulsion} would only occur when there is a difference in the signs of the \textit{gravitational} masses\index{gravitational mass!sign} which are interacting. Thus, from a purely phenomenological viewpoint there would be equivalence between positive and negative-mass objects.

Given the previous discussion regarding the necessity of a relational determination\index{requirement of relational determination!sign of energy} of the sign of energy, which would here be a requirement for the relational determination\index{requirement of relational definition!sign of mass} of the sign of mass, this situation would appear more appropriate, because, indeed, it would be impossible in principle to differentiate any intrinsic property of gravitational attraction or repulsion\index{gravitational attraction and repulsion!intrinsic property} and only the difference or the equality of the signs of \textit{gravitational} mass of two particles would be physically significant. The problem that most people would have with this possibility, however, is that it would explicitly violate the equivalence principle\index{equivalence principle}, because positive and negative gravitational masses\index{gravitational mass} would respond differently to a given gravitational field, produced by a given matter distribution, even if they are located in the same local inertial reference system\index{local inertial reference system}.

But even before we consider the issue of the apparent incompatibility with the principle of equivalence, we must first ask whether it is reasonable to assume, in order to avoid the above discussed difficulties, that the inertial mass\index{inertial mass!positive} would remain positive when the gravitational mass is reversed? The truth is that it would be inappropriate to assume that mass could be at once positive and negative, not because this would forbid all masses from always having the same acceleration in a gravitational field, thereby allowing violations of the principle of equivalence\index{equivalence principle!violation}, but simply because such a hypothesis would involve a contradiction. An electric charge is either positive or negative and mass, appropriately defined as the charge associated with the gravitational interaction, must also be either positive or negative. It would be incorrect to pretend that there are multiple attributes of mass and that each of those independent attributes can have a different sign. Clearly there is still something wrong, even with the second possibility that is normally considered for assigning physical properties to negative-mass objects.

Arguing that the problem here is with the notion that there exists only one single attribute of mass\index{mass!single attribute}, while the difficulty can be avoided when the appropriate distinction is made between what we would call the inertial mass, which always remains invariant, and what constitutes the `real mass', which we would call the gravitational mass\index{gravitational mass} and which may alone be reversed, would in my opinion not just be confused, it would be nonsense. What is positive cannot also at the same time be negative, if this polarity is to have any meaningful physical significance. Mass is not an abstruse, complicated property, with multiple independent and yet interrelated aspects, it is the gravitational charge and even though the stress-energy tensor\index{stress-energy tensors} replaces mass as the source of gravitational fields in a general-relativistic context, the lessons learned here are still valid and significant even in the context of the modern theory of gravitation.

It took me some time to realize that the problems we are dealing with here (if we are willing to recognize that the whole question of identifying the properties of negative-energy matter is not itself insignificant) originate from what is usually assumed concerning the response to any force field in the case of an object with negative inertial mass\index{negative inertial mass}. It is only after a rather long process of getting to understand the meaning of the phenomenon of inertia\index{inertia} that I was finally able to gain the insight required to solve the problem of identifying the actual properties of negative-mass matter, in the context where we consider it a consistency requirement to impose on such matter that it should have both a negative gravitational mass and a negative inertial mass.

What one must acknowledge, first of all, is that in a general-relativistic context there must exist a gravitational field exerting a gravitational force on a positive-mass object which is accelerating relative to a local inertial reference system\index{local inertial reference system}, even far from any large mass. The existence of the inertial force\index{inertial force} associated with this equivalent gravitational field\index{equivalent gravitational field} is what allows a dynamic (by opposition to static) equilibrium\index{dynamic equilibrium!external force} to occur when an external force is applied on an object, which gives rise to an acceleration. Indeed, in the accelerated reference system\index{accelerated reference system} relative to which a positive-mass object submitted to an external force does not accelerate, a gravitational force is present which balances the applied external force and this is what explains that there is no acceleration of the object relative to this particular (accelerated) reference system. In fact, the equivalent gravitational field is a general feature of acceleration and is present in any accelerated reference system, but in the absence of an external force to balance the associated inertial force the equivalent gravitational field only serves to determine the local inertial reference system associated with free-fall motion\index{free-fall motion}.

Given that the force associated with the equivalent gravitational field is a gravitational force, it follows that when the external force responsible for the acceleration is itself gravitational, we are actually in a situation where there would appear to be no force at all. It is therefore possible to assume that what determines the local inertial reference systems relative to which a positive mass experiences no gravitational force is the local matter distribution which is the source of the applied gravitational forces which are balanced by the inertial force which would otherwise be present relative to those reference systems (this is the essence of the insight that led to the general theory of relativity\index{general relativity theory}).

For a positive-mass object, the inertial force\index{inertial force} attributable to the equivalent gravitational field\index{equivalent gravitational field} is always directed opposite the direction of the external force which gives rise to its acceleration and this means that the direction of the equivalent gravitational field experienced by such an object is opposite the direction of its acceleration, that is, opposite the direction of acceleration of the reference system relative to which this equivalent gravitational field exists.

Now, it must be clear that the gravitational force $\bm{F}_g=m\bm{g}$ exerted on a particle of mass $m$ by a given matter distribution would be reversed if the mass of the particle was reversed, because the Newtonian gravitational field vector\index{Newtonian gravitational field!vector} $\bm{g}$ at the particle's position would be left unchanged (because the matter distribution that is the source of the field does not change), while the sign of mass\index{sign of mass!reversed} of the particle experiencing the field would be reversed. The problem, however, is that when we want to determine the response of a particle to some gravitational force $\bm{F}$ using Newton's second law\index{Newton's second law} $\bm{F}=m\bm{a}$, if the mass of the particle is reversed (negative), then it would appear that the resulting acceleration $\bm{a}$ needs to be opposite that experienced by a positive mass submitted to the same force (the acceleration would be in the direction opposite that of the applied force). This is the conventional concept of negative mass\index{negative mass!conventional concept}.

But if we consider things in a more general context, where Newton's second law would be an equation expressing the dynamic equilibrium\index{dynamic equilibrium!external force} between external forces $\bm{F}_{ext}$ and the inertial force\index{inertial force} $\bm{F}_i=m\bm{g}_{eq}$ produced by the equivalent gravitational field\index{equivalent gravitational field} $\bm{g}_{eq}$ associated with a given acceleration, then we may write $\bm{F}_{ext}=-\bm{F}_i$, so that for example if the external force is gravitational $\bm{F}_{ext}=\bm{F}_g=m\bm{g}$ then we would have $m\bm{g}=-m\bm{g}_{eq}$ and this means that the equivalent gravitational field $\bm{g}_{eq}$ is usually opposite both the applied gravitational field and the acceleration, because in the present case we also have $\bm{F}_{ext}=m\bm{a}$, which means that $m\bm{g}_{eq}=-m\bm{a}$ for the considered positive mass $m$ at least.

But would the equivalent gravitational field experienced by a negative-mass object really be directed opposite the direction of its acceleration, as is the case for a positive-mass object? To that question I think that, contrarily to what is usually assumed implicitly, we would have to answer that this cannot be the case. In fact, the equivalent gravitational field $\bm{g}_{eq}^-$ experienced by a negative-mass object accelerating in a given direction, far from any local matter inhomogeneity, would have to be opposite the equivalent gravitational field $\bm{g}_{eq}^+$ experienced by a similar positive-mass object with the same acceleration under the same conditions, so that we have $\bm{g}_{eq}^-=-\bm{g}_{eq}^+=-(-\bm{a})=\bm{a}$ for a negative-mass particle and given that we still have $\bm{F}_{ext}=-\bm{F}_i=-m\bm{g}_{eq}^-$ it means that $\bm{F}_{ext}=-m\bm{a}$ when the mass $m$ is negative.

The fact that Newton's second law\index{Newton's second law} was always observed to work in its original form, that is, when the equivalent gravitational field\index{equivalent gravitational field} is implicitly considered to be opposite the acceleration, is merely a consequence of the fact that it has only been verified to apply using positive-mass matter. But what is it indeed that might allow one to assume that the equivalent gravitational field would be reversed (would be directed in the same sense as the acceleration) for an accelerating negative-mass particle in comparison to what it would be for a similarly accelerating positive-mass particle? To understand what is going on we may consider the example of Einstein's accelerated elevator experiment\index{accelerated elevator experiment}.

We are allowed by the equivalence principle\index{equivalence principle} to assume that the effects observed inside an elevator accelerated in space, away from any local matter inhomogeneity, could also be explained by assuming that the elevator is not accelerating relative to the local inertial reference system\index{local inertial reference system} which would exist in the absence of any local matter inhomogeneity, but that it is instead maintained in place in the gravitational field of a large mass (located beneath the elevator) by the same external force which was originally causing it to accelerate. Thus, it seems that acceleration relative to a local inertial reference system always gives rise to an equivalent gravitational field similar to that which we would normally attribute to the presence of a local concentration of matter. We may then define an \textit{equivalent source}\index{equivalent source} to be the matter distribution which would give rise to the equivalent gravitational field experienced by an accelerated object if the presence of this field was not merely the consequence of acceleration.

Now, if the equivalent gravitational \textit{field}\index{equivalent gravitational field} associated with the inertial gravitational force\index{inertial gravitational force} reverses when the mass of the accelerated object reverses, it is simply because the sign of mass of the equivalent source associated with the equivalent gravitational field experienced by a negative-mass object must itself reverse. There should be no question, in effect, that if an accelerating positive-mass observer\index{accelerating positive-mass observer} is allowed to assume that the equivalent gravitational field she experiences is actually attributable to the presence of an equivalent source with \textit{positive} mass located in the direction opposite her acceleration, then a similarly accelerating negative-mass observer\index{accelerating negative-mass observer} should himself be allowed to attribute the equivalent gravitational field that he would experience to the presence of some equivalent source with \textit{negative} mass also located in the direction opposite his acceleration, otherwise we would have a way to determine in an absolute fashion, the positivity of mass.

If it was always an equivalent source\index{equivalent source} with positive mass (located in the same position relative to the accelerating object) that gave rise to the equivalent gravitational field, we could simply accelerate an object of any mass sign and measure the equivalent gravitational field experienced by this object, which could then be identified as the gravitational field attributable to a positive mass in the assumed position. Therefore, any gravitational field exerting on a given object a force such as that which was observed could be identified as the gravitational field of a positive mass, independently from the mere difference or equality between the polarity of the mass producing the field and that of the particle experiencing it. But this would constitute a violation of the above discussed requirement of relational definition\index{requirement of relational definition!sign of mass} of the sign of mass.

As it turns out, an additional difficulty arises when we try to assess the response of negative-mass matter to applied forces if we insist on assuming that the equivalent gravitational field\index{equivalent gravitational field} associated with acceleration is an invariant property of the acceleration itself, because it is not only in the presence of an external force that the inertial force\index{inertial force} on a negative-mass object would have to point in the direction of its presumed acceleration when it is assumed that the equivalent gravitational field is opposite this acceleration (as is the case for a positive-mass object). The truth is that, in the context of the generalized form of Newton's second law\index{Newton's second law!generalized form}, if we assume that it is an equivalent source with positive mass that gives rise to the inertial force experienced by a negative-mass object in an accelerated reference system, then it follows that even in the absence of external forces the inertial force would have the same direction as the acceleration, which means that the negative-mass object would actually accelerate in the same direction as the \textit{accelerated} reference\index{accelerated reference system} system itself.

Thus, a negative-mass object would not merely accelerate in the direction opposite the applied external force, as is usually assumed, it would rather accelerate in the direction of the applied force, but with twice the acceleration of a similar positive mass object, as if a net force existed relative to the accelerated reference system itself. This is due to the fact that, under such conditions, the inertial force would not balance the applied force, but would rather add to it, given that the inertial mass\index{inertial mass} is merely the mass that responds to the equivalent gravitational field and when it is reversed it responds anomalously to any equivalent gravitational field, whether this field is reversed or not. As a consequence, there would no longer be a dynamic equilibrium\index{dynamic equilibrium} between the applied forces and the inertial force that is experienced by a negative-mass object due to its acceleration, which is certainly not a desirable outcome. Thus, even if the equivalent gravitational field experienced by an accelerating negative-mass object was the same as that experienced by a similarly accelerating positive-mass object, this would not give rise to the kind of motion which is conventionally expected from a negative-mass object.

In order to avoid those difficulties, the condition that must be imposed is that there should always be an equilibrium between the applied external forces and the inertial force and under such conditions the acceleration to which an object with a given mass sign is submitted is determined solely by the requirement that the inertial force\index{inertial force} it experiences actually balances the applied forces. Thus, once the direction of an applied force is known, the acceleration of the object submitted to this force is determined only by the condition that it does, in effect, give rise to an inertial force which balances the applied force. But if the equivalent gravitational field\index{equivalent gravitational field} which gives rise to the inertial force is dependent on both the direction of acceleration and the sign of mass\index{sign of mass} of the accelerated object, then the fact that the sign of mass would be reversed would not affect the direction of the acceleration, because the equivalent gravitational field would also be reversed, which allows the inertial force associated with this acceleration to remain invariant under a reversal of the sign of mass.

What the preceding argument shows is that it would be a mistake to assume that the conventional formulation of Newton's second law\index{Newton's second law!conventional formulation} also applies when the mass is negative. This equation does not apply when the mass is negative simply because the formula was not derived under the assumption that mass can be negative and was never intended to apply under such circumstances. It would not be appropriate, therefore, to assume that it is the sign of mass\index{sign of mass} itself which determines the direction of the acceleration, because in fact the acceleration of an object submitted to a given force is determined merely by the requirement that the inertial force experienced by the object balances the applied force in the accelerated reference system\index{accelerated reference system} relative to which this inertial force is present. There is no \textit{a priori} justification for considering that a negative-mass object with negative inertial mass\index{negative inertial mass} should experience an acceleration opposite the applied force. This would be an incorrect interpretation of the conventional equation between force and acceleration, which must be assumed to be valid only when the mass is positive.

If we are willing to recognize that it would be a serious inconsistency to allow for the same equivalent source\index{equivalent source} (with the same mass sign) to give rise to both the equivalent gravitational field\index{equivalent gravitational field} experienced by positive-mass particles and that experienced by negative-mass particles, then we must also recognize that similarly accelerating positive and negative-mass objects would experience opposite equivalent gravitational fields, because those gravitational fields would arise from equivalent sources with opposite mass signs. But given that a negative mass\index{negative mass} must experience a force opposite that experienced by a positive mass of similar magnitude in response to any gravitational field, it follows that the inertial \textit{force}\index{inertial force} actually has the same direction for positive- and negative-mass objects accelerating in the same direction, as a consequence of being submitted to the same external force (which is more constraining than requiring the same applied force \textit{field}), even if we consider inertial mass\index{inertial mass!reversal} to be reversed along with gravitational mass\index{gravitational mass}, as I previously argued to be necessary.

In such a context we have no choice but to recognize that the response of a negative-mass object to any applied force would be that which we ordinarily (but inappropriately) attribute to a negative gravitational mass\index{negative gravitational mass} whose inertial mass\index{inertial mass!positive} would remain positive. But it is now possible to understand why it is that assigning a positive inertial mass to a negative gravitational mass object would seem to produce phenomena that would not violate the independently motivated requirement of a relational definition\index{requirement of relational definition!sign of mass} of mass sign. It is simply because, in such a case, instead of appropriately reversing the equivalent gravitational field for a negative mass accelerating in a given direction, we would reverse the sign of inertial mass (which must be negative for a negative-mass particle) a second time, from negative to positive again (while keeping the gravitational mass negative), which, superficially, would be equivalent to simply reversing the direction of the equivalent gravitational field while keeping the mass negative as required.

As later developments will illustrate, it appears that, in fact, the reversal of the equivalent gravitational field\index{equivalent gravitational field!reversal} is the trade-off we have to accept for keeping the value of the gravitational field attributable to a local matter inhomogeneity generally invariant while assuming that it is actually the mass experiencing it that can be reversed. But if, instead, we considered that the motion of an object must always be determined using the measure of gravitational field experienced by an observer made of matter with an invariant sign of energy, then it would be natural to assume that the sign of mass of the object (both inertial and gravitational) is positive-definite, while it is the gravitational field\index{gravitational field!observer-dependent property} attributable to a given matter inhomogeneity that is an observer-dependent property.

From this viewpoint, the equivalent gravitational field due to acceleration far from any local matter inhomogeneities would no longer be dependent on the sign of mass of the accelerating object (because the mass itself would not change), while the gravitational field due to the presence of a local matter inhomogeneity would depend on the perceived sign of energy\index{sign of energy!observer dependence} of its sources, which would become an observer-dependent property (again because the mass or energy of the object experiencing the fields would actually be considered positive-definite). In this context there would then still be a practical (although not fundamental) distinction between an equivalent gravitational field due to acceleration far from any local mass concentration (which wouldn't depend on the nature of the accelerating object) and the gravitational field due to the presence of a local matter inhomogeneity (which would depend on the nature of the object submitted to it).

Now, if we do consider the mass (both gravitational and inertial) of the particle experiencing a gravitational field to always be positive-definite, so that that it is the direction of the gravitational field itself which varies as a function of the \textit{relative} difference between the observer-dependent sign of mass\index{sign of mass!observer dependence} of the source (which can still be either positive or negative) and that of the particle experiencing the field (which would be assumed to always be positive) then we obtain a framework that can be more easily generalized to a relativistic theory. But it must be clear that the two approaches discussed here are equivalent in the Newtonian context and still require all mass (gravitational and inertial) to be either positive or negative and when the direction of the gravitational field\index{gravitational field!observer-dependent property} due to a local matter inhomogeneity is not considered to be an observer-dependent property we must consider the \textit{equivalent} gravitational field\index{equivalent gravitational field!mass-sign dependence} to itself be dependent on the sign of the accelerated mass (which is no longer positive-definite), otherwise the equivalence between the two viewpoints breaks down.

It should then be clear that in the context of an approach according to which the particles experiencing a gravitational field would be assumed to always have a positive mass, the crucial assumption is that while the gravitational fields attributable to local matter concentrations are dependent on the nature of the object experiencing their effects, the equivalent gravitational field associated with acceleration away from local masses would, for its part, remain invariant, regardless of how the object experiencing it perceives the gravitational fields attributable to local matter inhomogeneities. This hypothesis can be considered to be equivalent to that which in the above described approach consists in assuming that the equivalent gravitational field must actually be reversed for a negative mass, because this is indeed what allows the inertial properties of an object to be independent from its mass sign.

I believe that this observation clearly shows that I'm justified in analyzing the problem of negative mass from a conventional perspective, according to which the mass experiencing a gravitational field is explicitly assumed to be reversed, because in such a context the underlying assumptions are made more apparent and it is also easier to explain what I'm referring to when discussing the case of anomalously-gravitating matter\index{anomalously-gravitating matter}. In a Newtonian context I will therefore continue to use the first viewpoint, according to which it is possible for the mass experiencing a gravitational field to be negative.

Now, we may want to dig a little deeper and ask why it is exactly that we are allowed to assume that the direction of the equivalent gravitational field\index{equivalent gravitational field!mass-sign dependence} is dependent on the sign of mass of the object experiencing it? I have tried very hard to develop a better understanding of the whole phenomenon of inertia and what I have learned has actually helped me to derive the above discussed results. Indeed, this investigation has enabled me to realize that the assumption that the equivalent gravitational field is reversed, when the mass which is subject to acceleration is itself reversed, is not just a requirement of the necessary relational definition\index{requirement of relational definition!sign of mass} of the sign of mass, but must be imposed in order to allow a relational description of the phenomenon of inertia\index{inertia!relational description} itself, in the sense that inertia should be conceived as arising from purely relative motions between matter particles, as suggested by Ernst Mach\index{Mach, Ernst} a long time ago.

In this context, it appears necessary to assume that the inertial forces\index{inertial force} acting on a particle must arise as a consequence of an imbalance, caused by acceleration relative to the global inertial reference system\index{global inertial reference system} (associated with the distribution of matter on the largest scale), in the sum of forces attributable to the interaction of the accelerating particle with each and every other particle in the universe. But this imbalance of gravitational forces must be the same for similarly accelerating positive- and negative-mass objects, because the imbalance responsible for the existence of the inertial gravitational force\index{inertial gravitational force!imbalance of gravitational forces} is similar to a skewed mass distribution\index{skewed mass distribution} and if the actual large-scale matter distribution responsible for those effects is roughly the same from the viewpoint of both positive and negative masses, in the absence of local matter inhomogeneities, then the imbalance should develop in a similar way for both positive and negative masses from the viewpoint of their own mass sign.

What must be retained of this investigation is that the equivalent gravitational field\index{equivalent gravitational field} which applies on a negative-mass object is required to be the opposite of that which would be experienced by a positive-mass object with the same acceleration that is located within the same matter distribution, even if simply due to the fact that for a reversed mass the same motion relative to the same matter distribution should give rise to a similar imbalance in the sum of \textit{forces} attributable to interaction with all the matter in the (observable) universe. This may be considered to actually explain why it is appropriate to assume that it is the inertial force itself, instead of merely the product of mass and acceleration, that would be opposite the direction of the applied external force for a negative-mass object, as the generalization of Newton's second law\index{Newton's second law!generalized form} I have proposed allows to express.

But it must be clear that if there is a requirement for inertial mass\index{inertial mass!reversal} to be reversed, along with gravitational mass\index{gravitational mass}, it does not follow from imposing the validity of the equivalence principle\index{equivalence principle} as a condition that all matter should have the same acceleration in the absence of any interaction other than gravitation, as is usually considered. Indeed, as the previous analysis allows to understand, even a negative-mass object for which both the gravitational and the inertial masses are negative cannot be expected to follow the same trajectory as a positive-mass object in the presence of a local positive or negative mass concentration (despite what is usually assumed). What I have tried to explain is precisely that, even when inertial mass is assumed to be reversed along with gravitational mass, it is not possible to preserve the validity of the equivalence principle integrally.

Thus, a local inertial reference system\index{local inertial reference system!mass-sign dependence} cannot be defined independently from the sign of mass of the object experiencing it, given that the direction of the gravitational force resulting from a particular matter distribution depends on the sign of mass of this object. What I will explain in the following section is that the requirement that all matter with the \textit{same mass sign}, in the same location, experiences the same acceleration is in fact restrictive enough for the equivalence between gravitation and acceleration to apply in a certain way, that allows a metric theory of the gravitational field\index{gravitational field!metric theory} to emerge which merely relativizes the curvature of spacetime\index{curvature of spacetime!observer dependence} by making it an observer-dependent aspect of reality.

\section{The equivalence principle with negative mass\label{sec:2.5}}

It is not a well-known fact that the general theory of relativity\index{general relativity theory!two postulates} is actually based on two postulates, because only the first postulate, which concerns the equivalence between the effects of acceleration and those of a Newtonian gravitational field\index{Newtonian gravitational field}, is explicitly taken into account. But actually, a second postulate is required to obtain the current formulation of the theory and is implicitly assumed to be valid without justification. It is the hypothesis that the sign of energy is not relativistically defined. This second assumption appears to be necessary in order to preserve the validity of the first postulate under conditions where the presence of negative-energy matter would, in effect, need to be taken into account.

But even though the postulate that the sign of energy is not relationally defined may be considered problematic in the context of the preceding analysis, it remains to be shown whether it is possible to provide a consistent classical theory of the gravitational field\index{gravitational field!classical theory} in which only this second postulate would be rejected. Thus, I will try to show, in this section and later on when discussing the mathematical aspects of a generalized theory of gravitation\index{generalized gravitation theory}, that it is, in effect, still possible and necessary to maintain the validity of the equivalence principle\index{equivalence principle} in a certain form, while nevertheless rejecting the assumption of an absolute significance of the sign of mass or energy.

First of all, it must be emphasized that the true motivation behind the equivalence principle is to be found in a requirement which we may call the principle of relativity\index{principle of relativity} and which is actually one particular expression of the requirement of relational definition\index{requirement of relational definition!physical attributes} of all physical attributes. This relativity principle imposes that the state of motion of an object, and in particular its rate of acceleration, is to be determined merely in relation to the state of motion of other physical systems, so that there is no absolute state of acceleration\index{absolute acceleration} relative to an arbitrarily-chosen, unique reference system. This consistency requirement is what motivated Einstein\index{Einstein, Albert} to propose that there is an equivalence between a Newtonian gravitational field\index{Newtonian gravitational field} and an acceleration, because it is only under such conditions that what might have otherwise appeared to be an acceleration relative to absolute space\index{absolute space} may sometimes merely consist in a state of rest in the vicinity of a local mass concentration not accelerating relative to the same `absolute' space.

I think that it must be recognized that, in fact, the only essential implication of the equivalence principle\index{equivalence principle} is that there is no longer any motive for arguing that because acceleration is felt (unlike velocity) it must be absolute. Thus, it may appear problematic that even if we can find generally covariant equations\index{generally covariant equations} for the gravitational field in the presence of negative-energy matter, the fact that, according to the previous analysis, such matter would not share the same accelerated motion as positive-energy matter in the presence of a local matter inhomogeneity (while it should in the absence of such a perturbation) would appear to allow the effects of acceleration far from any local mass to be distinguished from those attributable to the gravitational field of a local mass.

There is, in effect, a tension between the principle of relativity\index{principle of relativity} and the previously discussed requirements concerning negative-mass matter which we may illustrate by once again using Einstein's accelerated elevator experiment\index{accelerated elevator experiment}. This is because when appropriately behaving negative-energy matter is present it seems that we can differentiate an acceleration of the elevator occurring far from any local mass from an acceleration of the elevator occurring while it is at rest near such a large mass, given that near a planet or another large matter inhomogeneity, positive- and negative-mass objects would accelerate in opposite directions, one upward, the other downward, while in the elevator which is simply accelerating far from any large mass, positive- and negative-energy objects would share the same downward acceleration, apparently betraying the fact that the acceleration is `real'.

It would, therefore, appear that an observer in the elevator would be able to tell when it is that she is simply standing still in the gravitational field of a planet and when it is that she is actually accelerating far from any big mass. The `true' acceleration would have been revealed to the occupants of the elevator as that for which both the positive- and the negative-mass objects have the same acceleration. Consequently, we would seem to be justified to conclude that the notion that the effects of acceleration are totally equivalent to those of a gravitational field (which is the essence of the principle of equivalence\index{equivalence principle}) is no longer valid when we introduce negative-mass matter with properties otherwise required to make it a consistent concept (according to the preceding analysis).

I made it clear before that it is not feasible to try to save the principle of equivalence by simply allowing negative-mass matter to react anomalously to applied forces (as if this was required when inertial mass\index{inertial mass!negative} is negative), so that it can accelerate in the same way positive-mass matter does in the presence of local matter inhomogeneities, because this would mean that the principle of inertia\index{principle of inertia} no longer applies in general and in such a case there is no guarantee that even the alternative situation we expect to observe under such conditions would really exist. Clearly, there must be something wrong with certain assumptions we take for granted concerning the equivalence principle itself. The fact that this is the principle upon which relativity theory\index{relativity theory} and our modern concept of gravitation is founded should not prevent us from re-examining some of the implicit assumptions surrounding it. Failing to do so would mean that we have to give up trying to provide a consistent interpretation of negative energy states\index{negative energy states}.

It is important to note, at this point, that it would be inappropriate to suggest that it may be possible to accommodate the requirement that the principle of equivalence\index{equivalence principle} also applies in the presence of negative-mass matter by assuming that opposite-mass objects always share \textit{opposite} accelerations instead of always sharing the same acceleration, as is conventionally believed. It is certainly true that, under such circumstances, it would still be impossible to distinguish a true acceleration given that opposite-mass objects would always accelerate in opposite directions, whether those accelerations are the result of the presence of a local concentration of matter or the result of the presence of an equivalent gravitational field\index{equivalent gravitational field} far from any large mass. But this situation could only occur, in the context of an appropriate conception of the phenomenon of inertia\index{inertia} based on the previously discussed generalized formulation of Newton's second law\index{Newton's second law!generalized form}, if it was assumed that the equivalent gravitational field associated with acceleration is not reversed despite the reversal of the mass of the accelerated object experiencing it.

From such a viewpoint we should actually expect that one of two opposite-mass objects would fall down while the other would fall up in the accelerating elevator, far from any local mass, even when no force is applied on any of the two masses independently. However, this kind of behavior would constitute an even more severe violation of the principle of inertia\index{principle of inertia!violation} than that which would occur in the case of the pair of self-accelerating opposite-mass objects\index{pair of opposite-mass objects!self-acceleration} described before, given that, in this case, there wouldn't even exist any identifiable cause for the upward acceleration of one of the two objects, because the elevator does not even interact with any of the masses and merely constitutes a reference system.

In fact, this situation is so devoid of plausibility that it clearly means that it is not possible to try to salvage the equivalence principle\index{equivalence principle} by assuming that the equivalent gravitational field\index{equivalent gravitational field} is not reversed for an accelerating negative-mass object. The fact that the kind of invariance of the equivalent gravitational field that is involved here would also violate the requirement of relational definition\index{requirement of relational definition!sign of mass} of the sign of mass, as I explained in the previous section, only contributes to confirm the validity of this conclusion. We must therefore accept that while the local inertial reference systems\index{local inertial reference system!mass-sign dependence} can differ for positive- and negative-mass objects near some local matter inhomogeneities, they must nevertheless be identical for opposite-mass objects far from local mass concentrations.

It is important to point out that in the case of the elevator suspended in the gravitational field of a local mass we are, in effect, considering an inhomogeneous matter distribution for which positive- and negative-energy matter concentrations are \textit{not} superposed in space (in the classical sense) and therefore do not produce mutually compensating local gravitational fields. If such compensations between the effects of \textit{local} matter inhomogeneities were to occur, as would be the case for example in the presence of two superposed gas clouds\index{superposed gas clouds} of opposite energy signs with the same overall motion or rotation, then the acceleration of positive- and negative-energy objects located near or within those matter distributions would be the same despite the presence of local inhomogeneities in the configuration of positive- and negative-energy matter.

What this means is that such a matter distribution could not influence the determination of the local inertial reference system\index{local inertial reference system} relative to which matter with a given mass sign does not accelerate, because its total energy would be null everywhere within the inhomogeneities. But in such a context it may seem that one would also need to assume that if the positive- and negative-energy matter distributions are homogeneous and are at rest with respect to one another on a global scale (a hypothesis which appears necessary, as I will explain in sections \ref{sec:4.5} and \ref{sec:4.9}), then they could not influence the determination of the global inertial reference system\index{global inertial reference system} relative to which matter of \textit{any} mass sign does not accelerate, if the average density of positive-energy matter equals that of negative-energy matter, at least in the initial Big Bang state (a hypothesis that also appears necessary, for reasons which will be discussed in section \ref{sec:4.5}).

It would, therefore, appear that there is something wrong with one or more of the implicit assumptions entering this deduction, because observations do show that the global inertial reference system is determined by the very-large-scale matter distribution\index{very-large-scale matter distribution} and its average state of (absence of) motion and rotation. Of course, the idea that there simply never was any negative-energy matter in the universe, so that the matter energy of the universe was never null, may be tempting, because after all we do not observe any such matter. But keep in mind that it will later be explained that this hypothesis is not required and that, in any case, it would again amount to simply reject the possibility that such matter may exist, without providing any justification for this very convenient hypothesis.

Thus, while it may seem that we have no choice but to conclude that there needs to exist something like absolute acceleration\index{absolute acceleration} relative to an arbitrarily-chosen, unique reference system lacking any physical underpinning, I have come to understand (based on independent motives which will be discussed later in this chapter) that, in fact, what constitutes the incorrect assumption, which appears to invalidate the hypothesis that all motion (including accelerated motion) is relative, even in the presence of negative-energy matter, is the hypothesis that both the positive- and the negative-energy matter distributions contribute to determine the global inertial reference system\index{global inertial reference system} relative to which an object with a given sign of mass does not accelerate in the absence of a local force.

If we drop the assumption that a \textit{negative-energy} matter distribution that is uniform on the cosmological scale can exert a gravitational force on \textit{positive-energy} objects (and vice versa for the effects of positive-energy matter on negative-energy objects), then it seems that we can explain the correspondence between the reference system relative to which both the large-scale positive- and negative-energy matter distributions do not accelerate on the average and the global inertial reference system relative to which objects of any mass sign do not accelerate in the absence of local forces, because, in such a case, the gravitational field produced by the positive-energy matter distribution is not canceled out by that which would be produced by the negative-energy matter distribution (and vice versa) on the largest scale.

I believe that this is due merely to the fact that, in the absence of matter inhomogeneities, particles with one energy sign interact only with the matter distribution carrying the same sign of energy. I'm particularly confident in the validity of this hypothesis, given that I had actually understood the requirement of absence of interaction between a positive-energy object and the uniform, large-scale distribution of negative-energy matter before I even realized that it was required to solve the problem of the relativity of motion, in the context where negative-energy matter is allowed to exist. I will explain what independently justifies this conclusion in section \ref{sec:2.7}.

Thus, what differentiates the situation of the elevator suspended in the gravitational field of a large mass of positive \textit{or} negative sign and the situation we have in the elevator accelerating far from any such local mass is that, in the first case, the gravitational force responsible for the observed acceleration is the result of an imbalance that is caused by unequally distributed inhomogeneities in the positive- and negative-energy matter distributions and this imbalance is dependent on the sign of energy of the object experiencing it (as there are two possibilities for both the sign of mass of the source and that of the accelerated object), while, in the latter case, the observed inertial gravitational force\index{inertial gravitational force!imbalance of gravitational forces} responsible for the acceleration is the result of an imbalance of gravitational forces that is always caused by the motion of an object with a given mass sign relative to a uniform matter distribution with the same mass sign (necessarily and invariably), so that it is not dependent on the sign of energy or mass of the object experiencing it, as long as the distributions of positive- and negative-energy matter are both homogeneous and are not accelerating or rotating relative to one another on the largest scale.

All accelerations are therefore measured relative to reference systems which are determined by the gravitational fields attributable to the matter that is present in the universe, and there exists no state of absolute acceleration\index{absolute acceleration}. This is true even if there does exist one unique reference system (actually two unique, but corresponding reference systems) which is singled out as that relative to which both positive- and negative-mass objects are not accelerating in the absence of local disturbances, as a result of the fact that the positive- and negative-energy matter distributions are not accelerating with respect to one another on the largest scales.

In light of those developments, it appears that what the previously discussed insight concerning the nature of the equilibrium involved in determining local inertial reference systems\index{local inertial reference system} should be understood to mean is that free-fall motion\index{free-fall motion}, instead of involving a total absence of forces, as is usually assumed in a general-relativistic context, must be considered to be the consequence of an acceleration-dependent equilibrium in the sum of gravitational forces\index{gravitational forces!equilibrium} attributable to interaction with both local masses and the very-large-scale matter distribution\index{very-large-scale matter distribution}. This interpretation appears to be required in the context where negative-energy matter must be recognized to exist, given that, in such a case, there cannot even be a unique local inertial, or free-fall reference system\index{local inertial reference system} dictated by the geometry of spacetime, so that we are forced to consider the reality of the general-relativistic gravitational field\index{general-relativistic gravitational field!physical interaction} as being associated with such a physical interaction.

It is only when we are dealing with a universal force\index{universal force}, defined precisely as a force that affects all objects in the same way, that we can \textit{choose} (as a mere convention) to include this force in our definition of the metric properties of space and time\index{metric properties of space and time}, given that, in principle, geometry must be shared by all objects present in the related space. What remains to decide is whether this convenient choice is still appropriate for gravitation, in the context where the force in question can no longer be assumed to affect all objects similarly (therefore betraying its material nature).

Einstein\index{Einstein, Albert} himself insisted that once we recognize the validity of a principle of general relativity\index{principle of general relativity}, then the speed of light can no longer be assumed to be constant (even though it is left invariant locally, along a geodesic\index{geodesics}), given that, in the accelerated elevator experiment\index{accelerated elevator experiment}, light rays\index{light rays!curved paths} may follow curved paths. But, from this viewpoint, the curvature of spacetime\index{curvature of spacetime} should naturally be expected to arise as the consequence of a local perturbation in the equilibrium of gravitational forces attributable to the interaction of the objects experiencing it with all the matter in the universe (except the large-scale matter distribution with opposite mass sign), otherwise it would be impossible to determine what affects the trajectory of light in an accelerated reference system far from any local matter inhomogeneity.

Even in a flat space, far from any local matter concentration, the motion of light in a straight line, which is usually considered to be a consequence of geometry itself, would be a consequence of the equilibrium of forces arising from the gravitational interaction with the rest of matter in the universe. This does not mean, however, that the geometrical interpretation of gravitation\index{geometrical interpretation of gravitation} is incorrect, but merely that the geometrical properties of space must definitely be conceived as arising from those interactions and more precisely, from an equilibrium in the sum of gravitational forces\index{gravitational forces!equilibrium} that can be altered by the presence of local matter inhomogeneities. As I will explain in section \ref{sec:5.13}, such a viewpoint has the added benefit of being more easy to generalize to a theory where the gravitational interaction\index{gravitational interaction!quantum mechanical} must not only be described as an interaction mediated by quantum particles, as is already recognized to be necessary, but must really be integrated into the quantum mechanical framework in the manner I shall propose.

In any case, I think that it is clear that statements to the effect that relativity theory\index{relativity theory} has made the concept of gravitational interaction obsolete and replaced it with that of spacetime curvature\index{spacetime curvature!relatively defined property} (so that gravitation is merely a manifestation of the geometry of spacetime\index{geometry of spacetime}) can no longer be assumed meaningful, if curvature is itself a relatively defined property which arises as a consequence of an equilibrium of local and inertial gravitational forces which depend on the sign of energy of the objects involved.

The situation we are facing here is similar to that in which electromagnetic theory\index{electromagnetic theory} was before the quantization of energy\index{quantization of energy} and the photon concept were proposed, because spacetime is now viewed as a continuous medium that directly takes part in determining the motion of objects, just like the electromagnetic field\index{electromagnetic field!continuous wavelike phenomenon} was originally considered to be a continuous wavelike phenomenon, directly influencing the motion of charged particles. When it was shown that light\index{light!corpuscular phenomenon} is a corpuscular phenomenon, the whole notion of electromagnetic wave\index{electromagnetic waves} was not abandoned, of course, because there was something real about the wavelike character of electromagnetic phenomena and this is the element which came to be integrated into quantum mechanics. Similarly, I think that the concept of spacetime curvature cannot and need not be abandoned when gravitation is described as an interaction mediated by the exchange of particles that would allow mass-sign-dependent local inertial reference systems\index{local inertial reference system!mass-sign dependence} to emerge, only, the curvature of spacetime can no longer be considered as actually \textit{being} gravitation itself.

As Hans Reichenbach\index{Reichenbach, Hans} once emphasized \cite{Reichenbach-1} (ch. 3, sec. 40), if we choose to integrate the gravitational force into our definition of spacetime we may no longer need to explicitly take the force into consideration to explain the motion of objects, but we must still invoke a force as the cause of the geometry itself. Thus, it is not gravitation which was replaced by curved geometry, but all of geometry that became a manifestation of the universality of the gravitational interaction, and I think that this remark is particularly appropriate in the context of a theory of gravitation that allows to take into account the possibility of the existence of negative-energy matter.

Actually, the commonly made remark to the effect that relativity allowed to eliminate gravitation as a real force appears to be motivated by the fact that the gravitational force arising from local mass concentrations was given the status of inertial force\index{inertial force} (similar in kind to the Coriolis force\index{Coriolis force}) by relativity and given that inertial forces were never seen as real forces, then it is believed that gravitation\index{gravitation!fictitious force} can now be considered a fictitious force under all circumstances. But I believe that it is rather the contrary that is true and that it is the inertial forces which can be considered as real gravitational forces in a general-relativistic context. The fact that inertial forces are involved in giving rise to the dynamic equilibrium which determines the mass-sign-dependent local inertial reference systems\index{local inertial reference system!mass-sign dependence} would then be a further indication that the geometry of spacetime\index{geometry of spacetime} is the product of an equilibrium of real gravitational forces\index{gravitational forces!equilibrium} arising from the interaction of local masses with the rest of matter in the universe.

Now, the fact that one particular reference system appears to be singled out as being that relative to which both positive- and negative-mass objects do not accelerate in the absence of local perturbations is not a unique feature of the approach described here. Actually, in a general-relativistic context, even in the absence of negative-energy matter, this feature of our description of the motion of objects should appear all the more natural given that all inertial reference systems are an outcome of the gravitational interaction and are therefore determined by the surrounding matter distribution. There exists, in effect, one very particular reference system in our universe, which I call the global inertial reference system\index{global inertial reference system} and which is that which is determined by the average motion of all masses together and relative to which no mass in the universe accelerates in the absence of a local force. That there may be such a unique reference system (or at least a unique set of such reference systems) does not mean that it is not relationally defined.

Relativity theory allows to explain the existence of this particular reference system as being a result of the combined gravitational interactions of a local object with all the other masses in the universe (with the same mass sign). Indeed, even far from any big mass, there remains the gravitational effect of the universe as a whole, which can never be ignored. Thus, the situation we usually refer to as corresponding to an absence of gravitational field and which we expect to be experienced far from any local mass concentration, is not really different from that which is occurring in the presence of such a local mass, only it is characterized by the fact that the gravitational field is then attributable to the average state of motion of either positive- or negative-mass matter on the cosmic scale and cannot be compensated by the presence of matter with an opposite mass sign, as long as all matter is uniformly distributed and positive-mass matter is not accelerating or rotating relative to negative-mass matter on such a scale.

The fact that inertial reference systems are always determined by the average state of motion of matter in the universe becomes particularly obvious when we consider the reference system associated with a felt motion of rotation which, as experiments have revealed, must be one that takes place relative to the most distant galaxies and therefore relative to the largest ensemble of matter in the universe. The reference system relative to which a positive-mass observer feels no rotation must then be determined by the gravitational field attributable to all matter particles with the same mass sign which are present in the visible universe, in a way that is dependent on the average state of motion of those particles. Such a reference system, therefore, is definitely unique, even though its description involves only relationally defined properties. We may still consider the average matter distribution on the largest scale to be rotating, but then its gravitational field would give rise to a rotating inertial reference system which, through relativistic frame dragging\index{relativistic frame dragging}, would put the whole matter content of the universe in rotation with it.

Since Einstein\index{Einstein, Albert}, there is no longer any mystery with the existence of such a preferred reference system\index{reference system!preferred} and what I'm trying to explain is that there is also no problem with the fact that there is a unique inertial reference system relative to which both positive- and negative-mass objects have no acceleration when free from external non-gravitational forces. We are not faced here with a metaphysical reference system associated with absolute acceleration\index{absolute acceleration}, but merely with an ordinary reference system relative to which the sum of gravitational interactions exerted on a local mass by the ensemble of matter present on the largest scale imposes an absence of acceleration that is shared by positive- and negative-mass objects.

The situation we are dealing with here, concerning the relativity\index{relativity!of acceleration} of acceleration in the presence of negative-energy matter, is similar to that regarding the relativity\index{relativity!of velocity} of velocity, because there also exists a preferred reference system relative to which the temperature of the cosmic microwave background\index{cosmic microwave background!temperature uniformity} is mostly uniform and which may appear to define a state of absolute rest\index{absolute rest}. But this unique reference system is merely that which is not moving relative to the average state of motion (not acceleration) of matter on the largest scale. If there is no conflict with the principle of relativity in such a case, then there need not be a problem in the case of the global inertial reference system\index{global inertial reference system} singled out as being that relative to which there is no difference between the states of acceleration of freely falling positive- and negative-mass objects.

There would then be no substance to the argument that the distinction between acceleration and gravitation, which appears to be revealed by the distinct accelerations of positive- and negative-energy objects in the standing still elevator near a local mass (in the context where negative-energy matter does not respond perversely to applied forces), allows absolute acceleration\index{absolute acceleration} (or absolute absence of acceleration) to be determined. The local gravitational fields and the associated local inertial reference systems\index{local inertial reference system!relatively determined} are always determined in a relative fashion as dependent on the presence of the local masses which are the source of the fields, while the reference system where the states of acceleration of positive- and negative-energy objects are identical is determined as that relative to which the large-scale matter distribution itself is not accelerating.

This all follows from the fact that positive- and negative-energy objects interact only with the homogeneous matter distribution with the same sign of energy as their own on the cosmological scale, so that motions relative to those matter distributions must be treated differently from motions relative to local matter inhomogeneities, although they are still relative motions. In fact, as I will explain later in this chapter, the large-scale distribution of negative-energy matter may exert a gravitational force on positive-energy objects, but only when inhomogeneities are present in this matter distribution. The nature of those interactions is such, however, that there is necessarily a cancellation in the sum of the effects involved on the largest scale, so that there can be no overall gravitational force and the same is true for the effects exerted by positive-energy matter on negative-energy objects.

Based on the above discussed considerations, I have come to the conclusion that the principle of relativity\index{principle of relativity} is not really threatened by the introduction of negative-energy matter obeying the requirement of relational definition\index{requirement of relational definition!sign of mass} of its mass sign. But clearly the equivalence principle\index{equivalence principle} itself (which allows accelerated motion to be treated relativistically) is no longer to be considered valid in the sense it is conventionally believed to be and if it need not and indeed cannot be abandoned it must, however, be generalized or somewhat relativized. In fact, we already know for sure that the equivalence principle always applies only in local reference systems whose state of acceleration may differ with position. We can tell, in effect, that a gravitational field is attributable to the presence of local masses instead of being the consequence of an acceleration, even in the total absence of negative-energy matter, when we consider a portion of space that is sufficiently large.

For example, if we consider two elevators suspended on opposite sides of a planet, instead of a single elevator, it is obvious that even though observers in each of those elevators could assume that they are accelerating far from any local mass, from the global viewpoint, where we would be observing oppositely-directed gravitational fields and an absence of relative motion of the elevators, we would have to conclude that those fields are due to the presence of a local mass and not to acceleration relative to the homogeneous large-scale matter distribution, even in the absence of negative-mass objects inside the elevators. In fact, even in a single elevator standing still on the surface of a small planet, freely falling positive-mass particles would have a tendency to slightly converge toward one another, therefore betraying the fact that the observed acceleration is an effect of the presence of a nearby mass attracting the particles toward its center. Yet we do not consider the equivalence principle to be violated under such conditions.

What I'm suggesting, therefore, is that, in addition to assuming that the equivalence of gravitation and acceleration applies only for particles in the same location, we also have to recognize that it applies only for particles with the same mass sign. For all particles that share a given sign of energy, there would never be a difference between acceleration and a gravitational field. It is only when we consider opposite-energy particles together, or more precisely in relation to one another, in the presence of a gravitational field attributable to a local matter inhomogeneity (when there is no compensation between the gravitational fields attributable to the local positive- and negative-energy matter distributions) that we can tell the difference between acceleration relative to the large-scale matter distribution and such a gravitational field. But this distinction would become somewhat irrelevant if there was reason to believe that negative-energy matter\index{negative-energy matter!dark matter} is dark from the viewpoint of positive-energy observers, due to the fact that opposite-energy particles cannot interact with one another non-gravitationally, as I will suggest in section \ref{sec:2.7}.

It is generally recognized, however, that what makes gravitation different from other interactions is the fact that the motion of objects in a gravitational field does not depend on the physical properties of those objects (when no other force field is present globally). But even though this characteristic would appear to be violated in the presence of negative-energy matter obeying the consistency conditions I have identified above, this does not make gravitation any less distinct, because from the previously discussed viewpoint where it is the direction of the gravitational field attributable to a given matter distribution which varies upon a reversal of the mass of the particle submitted to it (which would actually be considered positive-definite), the equivalence principle\index{equivalence principle!relativized} would merely be relativized by the presence of such negative-energy matter, given that the difference between the motion of positive-energy objects and that of negative-energy objects would actually be a consequence of the different measures of spacetime curvature\index{spacetime curvature} which (as I will explain later) can be associated with those two measures of the Newtonian gravitational field\index{Newtonian gravitational field}.

But under such conditions it is natural to expect that opposite-mass objects should not be restricted to share the same local inertial reference systems\index{local inertial reference system}, because, in fact, they do not even evolve in the same spacetime, but in spaces characterized by different metric properties. The fact that the gravitational field\index{gravitational field!observer dependence} can be conceived in such an observer-dependent way means that, in the case of gravitation, it is not the force that varies when the charge (here the mass) which is submitted to the field is reversed, but the field itself (to which is associated a given spacetime curvature). What remains true, therefore, is that, in any given situation, all objects sharing the same measure of gravitational field follow the same motion, as acceleration does not depend on the detailed characteristics of the objects experiencing the same gravitational field.

The equivalence principle\index{equivalence principle} can be assumed to still be valid in the presence of negative-energy matter, only it applies separately for positive- and negative-energy objects (just as it applies separately for separate portions of space), because each of those two kinds of matter particle is to be attributed its own free-fall reference system\index{free-fall reference system!mass-sign dependent} defined in relation to its mass sign. What's significant is that those two sets of free-fall reference systems and the two distinct measures of space and time intervals with which they are associated, can be related to one another by a simple, unique transformation, as all particles of one type share the same free-fall motion.

Those observations become even more significant when it is recognized that opposite-energy objects cannot interact other than gravitationally with one another, as I will suggest below, while one can only tell the difference between the case of a uniform acceleration and that of a local gravitational field by comparing the states of acceleration of two particles with opposite signs of action. In any case, all particles with the same sign of energy still share the same local inertial reference system and this is all that is truly required for a general-relativistic gravitational field\index{general-relativistic gravitational field!theory} theory to apply that allows the sign of energy itself to be relativistically defined.

\section{Six problems for negative-energy matter\label{sec:2.6}}

The preceding discussion may already make us feel more comfortable with the possibility that there could exist negative-energy matter, despite the conventional reluctance to accept the reality of negative energy states\index{negative energy states}. But at the current stage of my account, this confidence would not yet be totally appropriate. Even in the context of the new understanding unveiled in the previous sections, there remain many problems associated with the possibility that negative-energy matter may exist in our universe. First of all, we do not observe in the universe any matter or astronomical object which would clearly appear to be involved in repulsive gravitational interaction\index{repulsive gravitational interaction} with other astronomical objects. This is a very basic, but also very constraining fact.

Associated with this problem is the fact that the current predictions of quantum field theory\index{quantum field theory} are based on a systematic rejection of the possibility of a transition to negative action states\index{negative action states} (as states of negative energy propagated forward in time or positive energy propagated backward in time) and yet they appear to produce results which agree very well with observations in all situations where the nature of the interactions involved is well understood and computational methods\index{computational methods} are sufficiently well developed to allow the derivation of such verifiable predictions. This could provide an additional motive for arguing against the possibility of the existence of negative-energy matter. Such pieces of evidence certainly cannot be dismissed without very good reasons. Any theory involving particles propagating negative energies forward in time must explain why it is that we can safely ignore the existence of those particles in formulating a quantum theory\index{quantum theory!elementary particles} of elementary particles and their interactions, even while we would presumably have to take some of their effects into account in an astronomical context, where gravitational forces are not negligible.

A second category of difficulties has to do with the possibility that seems to be allowed, in the context where negative-energy particles would exist, for the annihilation of particle-antiparticle pairs to occur in which one of the particles would have negative action, therefore permitting matter to vanish, leaving absolutely nothing behind. This would, of course, require the annihilating opposite-action particles\index{opposite-action particles!annihilation} to also have opposite electric and other non-gravitational charges, because charge must still be conserved. We have no reason, however, to assume that negative-action matter does not also come in two varieties, one propagating negative energy and all non-gravitational charges forward in time and the other propagating positive energy and the same charges backward in time (so that we have opposite charges from the forward-time viewpoint).

Therefore, the possibility that such annihilations could take place cannot be ignored. But that is a much worse problem than may perhaps appear to be, because if such annihilations were possible there would then be no reason why the time-reverse processes could not also take place and if that was the case it would mean that pairs of opposite-action particles\index{opposite-action pairs!creation out of nothing} could be spontaneously created out of nothing without immediately returning to the vacuum like ordinary particle-antiparticle pairs, given that the process could occur without requiring a violation of energy conservation.

Even though one cannot entirely dismiss the possibility that pairs of opposite-action particles\index{opposite-action particles!annihilation to nothing} may be prevented from annihilating to nothing because they gravitationally repel one another, the fact that gravitational repulsion would not affect the possibility for the associated \textit{creation} processes to occur means that the problem is real. It may, therefore, seem like positive-energy matter particles could be created out of nothing abundantly, even under ordinary circumstances, or else annihilate to nothing at an arbitrarily large rate upon encounter with negative-energy particles, under appropriate circumstances, while both kinds of phenomena would clearly violate observational constraints, which provide no evidence at all that such events are taking place. This category of difficulties may then appropriately be called the energy-out-of-nothing problem\index{energy-out-of-nothing problem}.

A third potential problem has to do with the possibility that appears to be offered, as a consequence of the existence of negative energy states, for ordinary positive-energy matter particles or even any pre-existing negative-energy matter particles to `fall' into the allowed negative energy states in a continuous, unstoppable process during which they would either release positive radiation or absorb negative-energy radiation and reach ever `lower' energies. This is a difficulty which would also affect negative-energy matter as it is usually conceived and which is known as the vacuum decay problem\index{vacuum decay problem}. It would arise from the fact that the zero-energy level\index{zero-energy level} would no longer constitute a minimum level of energy (the ground state\index{ground state}) from which there can be no transition to lower energies.

Here we appear to have a situation where the existence of negative energy states raises the specter of allowing an arbitrarily large amount of work to be generated out of nearly nothing (by letting matter fall into the negative energy states and using the energy difference to produce work), as if energy conservation alone was not enough to restrict the evolution to negative energy states. This is clearly another issue of incompatibility with observation, because such decays are not observed to occur, even under the previously discussed conditions where negative energy densities\index{negative energy densities!quantum field theory} are allowed to occur in a limited way by ordinary quantum field theory. If negative-energy states\index{negative-energy states} can be occupied, what is it exactly that prevents positive-energy particles from falling into those lower energy levels? This is, all by itself, a legitimate question which has remained unanswered. Even from the viewpoint of the conventional interpretation of negative energy states this situation looks like a deep mystery.

But what is probably the most serious problem which one must face upon recognizing the necessity of introducing a notion of negative-energy matter obeying the requirement of a relational definition\index{requirement of relational definition!physical attributes} of physical attributes (which imply that opposite-energy objects must gravitationally repel one another) is that the existence of such matter may appear to allow violations of the principle of conservation of energy\index{principle of conservation of energy!violation}. This issue arises as a consequence of the fact that it seems possible for energy and momentum to be exchanged between positive- and negative-energy systems in a way that is similar to that by which positive-energy systems exchange energy among themselves.

Basically, it appears that the positive energy of a positive-energy object could be turned into an equal amount of negative energy belonging to a negative-energy object (or vice versa) when a `collision' between two such opposite-energy objects\index{opposite-energy objects!collision} would occur. For example, positive kinetic energy could be lost by a positive-energy object colliding with a negative-energy object initially at rest, while negative kinetic energy would be gained by the negative-energy object with which the first object has interacted. This would give rise to a net variation in the total energy of the two objects that would be equal to twice the individual energy changes (rather than allowing a cancellation of changes, as is observed when two positive-energy objects collide). The solution to that problem will have to arise from a proper understanding of the fact that it is not possible, in a general-relativistic context, for the energy of matter to be conserved independently from that of the gravitational field.

A further difficulty could arise in the context where the inertial force\index{inertial force} on a negative-mass object has the same direction as that which applies on a similarly accelerating positive-mass object, despite the reversal of inertial mass\index{inertial mass!reversal} which I have argued must occur when gravitational mass\index{gravitational mass!reversal} itself reverses. Indeed, from the viewpoint of an improved conception of the phenomenon of inertia\index{inertia} based on a generalized formulation of Newton's second law\index{Newton's second law!generalized formulation}, it is no longer possible to consider that acceleration would take place in the direction opposite the applied force for a negative-mass object, given that the equivalent gravitational field\index{equivalent gravitational field} due to acceleration would be reversed for such an object, which means that the inertial force it would experience is identical to that which is experienced by a similar positive-mass object. It would therefore appear that while the presence of a negative-mass object could contribute to reduce the gravitational mass in a region of space in which positive-mass matter is also present, it would still provide the same resistance to acceleration, despite the fact that it would also provide a negative contribution to the inertial mass contained in this volume.

This may not be a problem when we are dealing with independent physical systems with opposite masses, but as I previously mentioned, when a bound system\index{bound systems} is involved, the energy contained in the field of interaction\index{interaction field!energy} between its constituent particles would be opposite that of the system as a whole and in such a case it would seem that while the energy of the field should reduce the gravitational mass\index{gravitational mass} of the system, it should nevertheless contribute to increase its resistance to acceleration. Given that bound systems with differing force field configurations are quite common, it would seem that objects made of different materials should experience distinct accelerations when submitted to a gravitational force, but no such variations are observed. Some further clarification is required here, if the concept of negative mass which I have proposed is to be considered viable from an observation viewpoint.

One last potential category of arguments which one might believe could disprove the validity of the idea of gravitationally repulsive, negative-energy matter does not actually have to do with the concept of negative-energy matter developed here, but merely with more conventional concepts of anti-gravity\index{anti-gravity!conventional concept} and gravitational repulsion\index{gravitational repulsion!conventional concept}. The problems involved would undermine the validity of a theory according to which ordinary antimatter is gravitationally repulsive. They would also constitute a challenge for the usually favored interpretation of negative energy states\index{negative energy states!favored interpretation}, according to which gravitational repulsion is an absolute property of negative-energy matter itself, while gravitational attraction is an absolute property of positive-energy matter (so that negative-energy matter should repel positive-energy matter and be attracted to it). If such conceptions were to be retained as valid, they would allow paradoxical phenomena such as perpetual motion\index{perpetual motion} and time travel\index{time travel} to arise. Given that, for most people, those difficulties are associated with the general concept of negative energy matter, it is important to explain why the issues involved here would not affect a more consistent theory of gravitationally repulsive, negative-energy matter such as that which will emerge from the developments I introduced in the preceding sections.

We are then faced with six categories of problems, which appear to undermine a conception of physical reality according to which matter would be allowed to occupy energy levels below zero. I have wrestled with the questions raised by those difficulties for a long time and on many occasions I had nearly given up on the possibility to ever be able to find appropriate answers that would perhaps explain why negative energy is not an inappropriate concept for physical theory. But, gradually, I came to understand that each of those problems really has to do with one or more incorrect, implicit assumptions\index{implicit assumptions!negative energy states} we make when considering the expected behavior of matter, in the context where those negative energy states are actually allowed to be occupied. In the next six sections I will explain the nature of the insights required to appropriately deal with those severe problems.

\section{The origin of repulsive gravitational forces\label{sec:2.7}}

In order to explain the absence of observational evidence for the existence of negative-action matter\index{negative-action matter!absence of observational evidence} it will be necessary to first examine how repulsive gravitational forces\index{repulsive gravitational force!origin} may arise between objects of opposite energy signs. For this purpose it is useful to consider the analogy of voids in an expanding matter distribution\index{void in expanding matter distribution}. There is, in effect, a kind of equivalence between the gravitational forces attributable to the presence of an underdensity in an expanding, uniform distribution of positive-energy matter and the gravitational forces which we can expect to be produced by a negative-energy object with a mass of similar magnitude. This is because, in an expanding universe, we can expect underdensities in the positive-energy matter distribution to produce a local acceleration of the rate of expansion\index{expansion rate!local acceleration}, as if the universe had a negative space curvature\index{negative space curvature} locally. But while positive-energy matter underdensities would seem to exert a repulsive gravitational force\index{repulsive gravitational force} on the surrounding positive-energy matter, the same structures would have a tendency to merge into ever larger spherical voids\index{spherical voids} \cite{Peebles-1}, as if they were attracted to one another by the force of gravity, just like we can expect to be the case with astronomical objects made of negative-energy matter.

Thus, in this particular case at least, it seems that the underdense structures\index{underdense structures!mutual gravitational attraction} would actually be submitted to mutual gravitational attraction, even while they would seem to repel the surrounding matter and any overdense structure. But it is usually considered that there is nothing more than an accidental analogy between the case of those underdense structures and any gravitationally repulsive matter\index{gravitationally repulsive matter}, because if they give rise to the kind of phenomena described above, then, according to conventional understanding, those gravitationally repulsive voids\index{gravitationally repulsive voids} would need to have not only negative gravitational mass\index{negative gravitational mass}, but also positive inertial mass\index{inertial mass!positive} \cite{Piran-1} and as everyone `knows', this kind of negative mass is forbidden by the equivalence principle\index{equivalence principle} and relativity theory\index{relativity theory}, which require the equality of gravitational and inertial masses. Therefore, what we would observe to be happening, in this particular case, is not what most people would consider should occur if we were actually dealing with gravitationally repulsive matter.

What is usually assumed, as I previously explained, is that gravitational repulsion\index{gravitational repulsion!absolute property} is an absolute property of negative-energy matter and that this kind of matter would, therefore, itself also be repelled by matter of the same type. But given the clarifications which were provided in the previous sections of this chapter regarding what would be a consistent concept of negative-mass or negative-energy matter, it should be clear that we would not be justified to argue that the gravitational dynamics between positive-energy matter and voids in an expanding, uniform distribution of positive-energy matter are not similar to the gravitational dynamics\index{gravitational dynamics!positive- and negative-mass objects} between positive- and negative-mass objects. Therefore, we cannot so easily reject the possibility that the discussed phenomenon is actually telling us something important about the nature of negative-energy matter.

I do believe that there is something very profound which we need to understand concerning the analogy between voids in an expanding, uniform distribution of positive-energy matter\index{void in expanding matter distribution} and negative-energy matter. It must be clear, however, that the discussed equivalence is only valid in the case of voids in an expanding matter distribution and that it does not mean that voids in a uniform distribution of positive-energy matter actually exert a repulsive gravitational force\index{repulsive gravitational force} on the surrounding positive-energy matter, that would exist even in absence of universal expansion. The forces exerted by voids in an expanding, uniform distribution of positive-energy matter on the surrounding positive-energy particles are merely a result of the fact that the variation of the rate of change of the distance between two particles located on the boundary of such a void depends on the density of matter \textit{inside} the void and given that this density is lower than average, then the rate of expansion\index{expansion rate!acceleration} of space may actually be made to accelerate if it is large enough, even if no repulsive gravitational force is actually exerted on the surrounding matter.

An underdense region in a uniform positive-energy matter distribution, therefore, does not produce a gravitational field opposite that which would be produced by a quantity of positive-energy matter equal to that which is missing in this underdense region, because it does not contribute negatively to the local density of energy. The conclusion that a void in a uniform positive-energy matter distribution does not produce a repulsive gravitational field\index{repulsive gravitational field} from the viewpoint of positive-energy particles, even though it does exert a smaller than average gravitational attraction on the surrounding positive-energy matter, is supported by Birkhoff's theorem\index{Birkhoff's theorem} \cite{Birkhoff-1}. What Birkhoff's theorem implies is that there can be no net gravitational force on matter located inside a spherically symmetric region in a globally uniform matter distribution from matter located outside that region, because for a spherically symmetric matter distribution\index{matter distribution!spherical symmetry}, the only possible solution to the general-relativistic gravitational field equations\index{gravitational field equations!solution} is the Schwarzschild solution\index{Schwarzschild solution}, which means that if there is no matter inside a spherical region, then there will be no gravitational field on its boundary, as the mass of the object is null.

But in fact, it is not even necessary to appeal to Birkhoff's theorem\index{Birkhoff's theorem} to understand why it is that a void in a uniform matter distribution cannot be the source of a repulsive gravitational field\index{repulsive gravitational field}. Once one recognizes that for a local void to form in an originally uniform positive-energy matter distribution a compensating overdensity, such as a spherical shell of positive-energy matter\index{spherical shell of positive-energy matter}, would have to be produced in the surroundings of the void, then it follows that if the void was really giving rise to a repulsive gravitational field, then given that the surrounding overdense shell would, for its part, still exert the same attractive gravitational field (from the viewpoint of positive-energy particles), then, for a particle located outside this overdense shell, the strength of the gravitational field originating from the whole structure would have to decrease as a result of its formation, even though the same amount of matter would still be present within the outer boundary of the surrounding shell, and this would require assuming that the energy that is the source of this gravitational field can vanish due to the formation of an inhomogeneity in the matter distribution.

Therefore, if energy is to be conserved as inhomogeneities develop in a matter distribution, the gravitational field attributable to the presence of an underdense region in the matter distribution can only be reduced until it is null everywhere inside the created void and its polarity cannot reverse as if matter with an opposite energy sign was present inside the structure. But that does not necessarily mean that there is no substance to the analogy which appears to exist between the gravitational dynamics\index{gravitational dynamics!positive- and negative-mass objects} of positive- and negative-mass objects and that of overdense and underdense structures\index{gravitational dynamics!overdense and underdense structures} in an expanding, uniform distribution of positive-energy matter. To explain what motivates that conclusion, it is necessary to recall the previous discussion concerning the possibility for negative densities of energy to arise under certain unusual conditions which are predicted to exist by ordinary quantum field theory\index{quantum field theory}.

In section \ref{sec:2.3} I pointed out that the absence of some positive energy from the vacuum, in a limited region of space (between the plates of two parallel mirrors\index{parallel mirrors in vacuum} for example), can actually give rise to a vacuum with negative energy density\index{negative energy densities} in the volume considered, because removing positive energy from a vacuum state whose energy is already minimum is like decreasing the energy below its zero point, into negative territory. The fact that the vacuum is known to have a relatively small average energy density should not be considered an obstacle to the occurrence of large negative energies in such a way, because, as I will explain later in this chapter and in section \ref{sec:4.2}, this small energy density appears to arise from the imperfect cancellation of two arbitrarily large (actually maximum) opposite contributions which can each be reduced to any extent by the conditions which are responsible for locally decreasing (under particular circumstances) the energy of the vacuum\index{vacuum energy!equilibrium value} below its equilibrium value.

But if we may, in effect, measure energy to be negative in a certain region of space where positive energy is missing from the vacuum, then there is no reason why we could not consider that negative energy states\index{negative energy states} in general are equivalent, in certain ways, to a local absence of positive energy from the vacuum, if from a phenomenological viewpoint there is no distinction between those two situations. In such a context, it would appear appropriate to assume that the presence of negative-energy matter\index{negative-energy matter!absence of positive vacuum energy} itself is equivalent to an absence of positive energy from the vacuum.

I must again mention, in this regard, that many authors have expressed doubts concerning the validity of the concept that energy should exist in the vacuum that would be the outcome of the presence of zero-point fluctuations\index{zero-point vacuum fluctuations} involving virtual particles\index{virtual particles} and have suggested that there may be nothing real with the processes so described outside of the context where they are occurring as part of processes involving `real' particles\index{real particles}. But I think that what really motivates this mistrust is precisely the fact that the existence of those processes would imply the reality of negative energy states, because it is no secret that, for most physicists, the theoretical possibility of the existence of negative energy states is not well-viewed. However, I believe that this aversion is merely a consequence of the fact that the conventional concept of negative-energy matter\index{negative-energy matter!conventional concept} is, in effect, not viable and that it has not yet been realized that a better description of negative-energy matter is possible and even necessary, as I emphasized above.

In any case, the idea that virtual processes\index{virtual processes} only occur as part of processes involving real particles, thus explaining why we must nevertheless consider the effects of such fluctuations when calculating transition probabilities\index{transition probabilities}, is meaningless, because even in empty space, far from any `real' matter, the virtual processes of particle creation and annihilation characteristic of the quantum-mechanical vacuum\index{vacuum!quantum-mechanical} are causally influenced by nearby processes of the same kind, which may themselves be influenced by other such processes, which would be causally influenced by processes involving real particles present in their surroundings. Therefore, the argument that the negative energy states\index{negative energy states} which are predicted to occur in the vacuum under appropriate conditions are not real, because our description of the vacuum is inadequate, cannot be retained. Also, the fact that it has been confirmed that the cosmological constant\index{cosmological constant!non-zero value} is not absolutely null is a strong motive to conclude that it is not possible to deny the reality of vacuum fluctuations as essential aspects of our description of empty space and therefore that negative energy states are a real possibility.

Now, I have already explained why we should expect to observe mutual gravitational attraction between two objects with the same sign of mass or energy and gravitational repulsion between opposite-energy objects. But given the analogous nature of the gravitational dynamics\index{gravitational dynamics!overdense and underdense structures} of overdense and underdense structures in an expanding matter distribution and those of positive- and negative-mass objects, it becomes possible to assert what would be the effects of missing positive vacuum energy\index{missing positive vacuum energy}.

As I will explain later in this chapter, we can expect the density of vacuum energy to be the outcome of two opposite contributions of equal, maximum magnitudes which, under appropriate conditions, add up to zero. The situation we would have in the presence of a local underdensity in the positive-energy portion of zero-point vacuum fluctuations\index{zero-point vacuum fluctuations} would, therefore, be very similar to that which exists in the presence of a local underdensity in the expanding, uniform distribution of positive matter energy. What's different between the case of a void in the expanding distribution of positive matter energy and that of a void in the positive portion of vacuum energy\index{void in positive vacuum energy}, however, is that, for reasons I will discuss below, while only the positive portion of the uniform distribution of matter energy contributes to determine the gravitational field experienced by positive-energy particles, both the positive and the negative portions of the uniform distribution of vacuum energy\index{vacuum energy!uniform distribution} contribute to determine the gravitational field experienced by the same positive-energy particles.

As a consequence, in the absence of any matter, the gravitational field attributable to the uniform distribution of \textit{vacuum} energy may well be null, if the density of vacuum energy\index{vacuum energy density!zero value} itself happens to be zero. Under such conditions, the energy density would not merely be null, in the presence of a local void in the positive portion of vacuum energy, but would actually be lower than the zero density of energy that exists in the absence of any local perturbation. Thus, if Birkhoff's theorem\index{Birkhoff's theorem} is valid under such conditions as well, one would expect the gravitational field around a local void in the positive portion of vacuum energy\index{vacuum energy!positive portion} to not merely be null, but to instead have a polarity opposite that which would otherwise be produced by the positive vacuum energy which is missing in the void, because the density of vacuum energy is negative in this particular location.

In the case of a void in the positive portion of vacuum energy\index{void in positive vacuum energy!gravitational repulsion} we are, therefore, in a situation where there does exist a gravitational field that grows with the distance to the center of the void and that would \textit{not} be null on the boundary of the structure. It may therefore seem that such a void in the uniform distribution of positive vacuum energy could give rise to a repulsive gravitational force\index{repulsive gravitational force} that would be present even inside the structure, from the viewpoint of positive-energy particles, because under such conditions the energy is lower, locally, than it would be even in the total absence of any positive-energy matter, while it is only when the total measure of vacuum energy is null in a region of space (due to the existence of mutually compensating opposite contributions) that the gravitational field on a spherical surface surrounding that region needs to be null.

But even when one recognizes the adequacy of those assumptions, a significant difficulty remains for any theory of the gravitational interaction between positive- and negative-action particles that obeys the requirement of relational definition\index{requirement of relational definition!physical attributes} of physical attributes. So, before exploring the consequences and the profound significance of the above description of matter as consisting of voids in vacuum energy, let me explain what the subtleties of the gravitational interaction of opposite-energy particles are, which turned out to be essential to make the whole concept of negative-energy matter proposed here viable from an observational viewpoint.

\bigskip

\noindent When, as a young man, I first started to contemplate the possibility that there could exist matter in a state of negative energy, I soon realized that if such matter was to attract matter of the same type while it would repel ordinary matter and be repelled by it (as I had intuitively assumed should occur, ignorant of the dominant paradigm), then this matter would have to be dark, because no observation had ever revealed the existence of planets, stars, or galaxies exerting a gravitational repulsion\index{gravitational repulsion} on other visible astronomical objects. While I was working on improving my understanding of physics in general and trying to develop a theory incorporating the concept of negative mass\index{negative mass}, I simply assumed that negative-mass particles where such that they would interact with ordinary matter only through gravitation.

I remember that I had read that Feynman\index{Feynman, Richard} once said that we must not question \textit{why} things are the way they are, but simply try to describe in the most accurate way possible \textit{how} they behave. Thus, for a while, I was comfortable with the idea that negative-energy matter simply doesn't interact, other than through the gravitational force, with ordinary matter (although it could interact with itself through the whole spectrum of forces), even if I had no idea why that should be the case and had to assume that this is just the way things are. The only concern I had, regarding this situation, is that it appeared odd that negative-energy matter should not interact with ordinary positive-energy matter through the same forces by which positive-energy particles were interacting among themselves, given that negative-energy matter can be assumed to be composed of the exact same particles as form positive-energy matter.

While as was ruminating this question, I was also trying to figure out what determined the repulsive or attractive nature of an interaction, which clearly depends on the signs of the charges of the interacting particles, and had slowly come to realize that this property seems to be related to the sign of energy of the field of interaction\index{interaction field!sign of energy}, not yet fully aware that it is actually rather the attractive or repulsive nature of an interaction (determined by the sign of the charges involved) that determines the sign of energy of the field and not the opposite. Anyhow, I eventually understood that the energy of a field associated with a repulsive interaction between positive-energy particles, for example the energy of the electromagnetic field\index{electromagnetic field!energy} between two electrons, is always positive, while the energy of a field associated with an attractive interaction between positive-energy particles, for example the energy of the electromagnetic field between an electron and a positron, is always negative.

But it also had to be the case (as I will explain below) that the energy of a field associated with a repulsive interaction between negative-energy particles is always \textit{negative}, while the energy of a field associated with an attractive interaction between negative-energy particles is always \textit{positive}. What this means is that when two negative-action particles are attracted toward one another or bound together in a single system\index{bound systems!energy}, the contribution of the attractive field mediating the interaction to the energy of the whole system should be positive, while for positive-action particles it would be negative.

As I was trying to make sense of this observation in the context where the interaction involved would be that between a positive- and a negative-energy object, I suddenly realized that there was a problem, because if this relation between the sign of energy of the field and the attractive or repulsive nature of the related interaction is right in general, then it means that any gravitational interaction between positive- and negative-energy objects should be either repulsive for positive-energy matter and attractive for negative-energy matter (if the field is attributed positive energy) or repulsive for negative-energy matter and attractive for positive-energy matter (if the field is attributed negative energy), but never repulsive for both the positive- and the negative-energy objects involved in the interaction.

In other words, a repulsive interaction field\index{interaction field!sign of energy} would need to have positive energy for positive-energy matter particles, while a field with the same energy would need to exert an attractive force, from the viewpoint of negative-energy matter particles for which the same relation would exist in general between the \textit{difference} between the signs of energy of the matter particles and their interaction field on the one hand and the repulsive or attractive nature of the interaction on the other (the problem is not restricted to gravitation). This is again a consequence of the requirement of relational definition\index{requirement of relational determination!sign of energy} of the sign of energy which implies that the energy sign of an interaction field cannot be determined merely by the attractive or repulsive nature of this interaction, but must also depend on whether this attractive or repulsive interaction involves positive- or negative-energy particles.

But it was just nonsense to conclude that an interaction could be both attractive and repulsive at the same time, or that a field of interaction could actually contribute both positively and negatively to the energy of a system, and it is even more so now, in the context where we must recognize that the hypothesis that positive- and negative-energy matter particles are submitted to mutual gravitational repulsion is also necessary for a relational description of the gravitational interaction between those two types of objects. The conclusion I had to draw was thus very clear: no definite energy sign could be attributed to the fields of interaction between positive- and negative-energy particles (as must be the case for any interaction involving particles with the same sign of energy) and therefore those two types of particles simply cannot interact with one another, not even gravitationally. This appeared to be a fatal blow, because if there are no interactions between positive- and negative-energy matter, then how could negative-energy matter have any relevance to the world we experience?

When I realized the existence of this difficulty for a theory of negative-energy matter, I had already come to appreciate the many advantages that there would be if such matter was allowed to exist (if it could indeed gravitationally interact in some way with ordinary matter). This is because I had been able to solve important problems even while just using the incomplete description I had by then managed to develop and it seemed improbable to me that the whole approach could simply be wrong. I know that this may look like it was more a hopeful wish than a rational conclusion, but in fact it was actually both hope and reason.

We had struggled with the problems I was able to solve for a very long time and there really appeared to be no viable alternative solutions to those problems, while, theoretically, the hypothesis that there could exist matter in a negative energy state\index{negative energy states} had a lot of appeal. It is as a consequence of the fact that I had so much confidence in the validity of the basic concept of a symmetry between positive and negative energy states that I didn't stop working on developing the idea when I encountered the difficulties discussed here. And as it turned out, the problems encountered became just another challenge on the way to a satisfactory solution to the problem of negative energy states.

So, I went from having to explain why there appears to be no electromagnetic interactions between positive- and negative-action matter to having to explain why there can be any interaction at all between the same two kinds of matter. Of course, I was glad that, at least, I now had an explanation for why there appears to be no electromagnetic or other non-gravitational interactions between opposite-energy particles, because on the basis of the above discussed considerations, it is clear that we have no choice but to recognize that there cannot exist any direct interactions (mediated by the exchange of interaction bosons\index{interaction bosons}) between such particles.

But gravitation is different, because no fully satisfactory description of this interaction exists that is compatible with quantum theory and I had hope that what would allow the existence of \textit{some kind} of interaction between positive- and negative-energy objects is the classical character of this interaction. It must be clear, however, that the problem described above is very real and unavoidable and its significance should not be underestimated, as it actually means that there can be no direct interaction between positive- and negative-action particles. It must also be understood that this is not a hypothesis, as no logically consistent theory could describe such an interaction and therefore the only conclusion we can draw is that those hypothetical interactions are, in effect, nonexistent\footnote{
These conclusions are the reason why I did not argue in this report that the gravitational interaction between opposite-energy particles must be considered repulsive merely on the basis of the fact that gravitation is mediated by a spin-two interaction boson (the graviton\index{graviton!spin-two interaction boson|nn}), because, obviously, if there cannot be any direct interaction between opposite-energy matter particles, then it is pointless to argue that it is the spin of the particles they exchange that determines the repulsive nature of such an interaction.}.

What I have since come to understand is that, if negative-energy matter appears to exert a repulsive gravitational force\index{repulsive gravitational force} on surrounding positive-energy matter, this can only be a consequence of the fact that, the positive vacuum energy that goes missing\index{missing positive vacuum energy!negative-energy matter} locally, as a result of the presence of negative-energy matter, no longer contributes to balance the gravitational attraction\index{gravitational attraction!surrounding positive vacuum energy} exerted on this positive-energy matter by the surrounding positive portion of vacuum energy.

When two maximum contributions with opposite signs exist for vacuum energy\index{vacuum energy!maximum positive and negative contributions}, it is necessary to assume, in effect, that in the absence of any positive or negative energy object, that is to say, in the absence of \textit{local} voids in the negative and positive portions of the uniform distribution of vacuum energy (respectively), no gravitational force is exerted by vacuum energy on a positive-energy particle, as the forces attributable to positive vacuum energy must then cancel out those which are attributable to negative vacuum energy. This is true, even though positive-energy matter doesn't interact with negative-energy \textit{matter}, because it is necessary to assume (as I explain below) that positive-energy matter itself is equivalent to missing negative vacuum energy\index{missing negative vacuum energy!positive-energy matter} and the negative portion of the \textit{uniform} distribution of vacuum energy does interact with itself.

The equilibrium of gravitational forces\index{gravitational forces!equilibrium} that exists under such conditions, however, is not attributable merely to the presence of two opposite contributions of equal maximum magnitudes to the energy of the vacuum, but also to the absence of local inhomogeneities in one or another of those two opposite-energy distributions. When a void is present locally, in the positive portion of vacuum energy, this equilibrium is broken, not because the gravitational force exerted by the negative vacuum energy that is still present in the void is no longer compensated by that which would otherwise be exerted by the missing positive vacuum energy, but because the gravitational forces exerted on a positive-energy particle by the uniform portion of the surrounding \textit{positive} vacuum energy distribution no longer cancel out locally, while the gravitational forces exerted by the uniform portion of the surrounding negative vacuum energy distribution still cancel out locally, as if they were actually inexistent.

Thus, when positive vacuum energy is missing\index{missing positive vacuum energy} locally, it is as if the portion of the surrounding uniform distribution of positive vacuum energy above that which is present in the void and which would otherwise exert no gravitational force on either positive- or negative-energy particles, can now exert some kind of attractive gravitational force on positive-energy particles, even when the average density of vacuum energy remains null outside the void.

It is important, therefore, to differentiate between the equilibrium of gravitational forces that is attributable to the presence of two uniformly distributed, opposite contributions to the energy of the vacuum\index{vacuum energy!uniformly distributed opposite contributions}, which are allowed to cancel out on the cosmic scale (thereby possibly giving rise to a zero cosmological constant\index{cosmological constant!null value} that does not affect the rate of expansion of matter), and the local equilibrium of gravitational forces\index{gravitational forces!equilibrium} exerted on positive-energy particles, which arises from the uniformity of the \textit{spatial} distribution of positive vacuum energy and which may be affected by the absence of positive vacuum energy in a limited portion of space (while the gravitational force exerted on positive-energy particles as a result of a local absence of negative vacuum energy is merely the outcome of an ordinary interaction between particles with the same positive sign of energy).

What we can actually expect, therefore, is that a local void in the uniform, positive portion of vacuum energy, which is equivalent to the presence of a local concentration of negative-energy matter, would gravitationally repel a positive-energy particle, not because the positive-energy particle is gravitationally repelled by the negative vacuum energy that is present in the void, but because the positive vacuum energy that is missing\index{missing positive vacuum energy} does not exert the gravitational attraction it would otherwise exert in the direction of the void, while the surrounding positive portion of vacuum energy still exerts an attractive gravitational force\index{attractive gravitational force!outward-directed} directed outward from the center of the void. This is possible, despite the fact that the uniform distribution of vacuum energy may have a null density outside the void, because when positive vacuum energy is missing locally, from a uniform distribution of vacuum energy that is already null, the gravitational attraction exerted on a positive-energy particle in the direction of the void is weaker than it would be even in the total absence of any energy, which means that it is actually pointing in a direction opposite that in which it would be pointing if it was attributable to the presence of positive vacuum energy in the void.

Given that there would be no gravitational force on positive-energy particles, arising from a void in vacuum energy\index{void in vacuum energy}, if equal amounts of positive and negative vacuum energies were missing at the same time in this void, while such a situation would only differ from the case where only positive vacuum energy is missing in that there would be no negative vacuum energy in the void, then it is necessary to assume that it is the surrounding uniform distribution of positive vacuum energy\index{positive vacuum energy!uniform distribution} that is exerting a gravitational force on a positive energy particle when the only energy that is missing is positive vacuum energy, because when the negative vacuum energy is missing\index{missing negative vacuum energy!positive-energy matter} as well in the void, the absence of forces is not due to an absence of gravitational repulsion\index{gravitational repulsion} by the negative vacuum energy that is missing, but to the presence of an attractive gravitational force by the positive-energy matter which is located \textit{inside} the void and which exists as a result of this local absence of negative vacuum energy.

Despite what one may be tempted to argue, Birkhoff's\index{Birkhoff's theorem} theorem doesn't imply that it is impossible for the surrounding uniform distribution of positive vacuum energy\index{vacuum energy!uniform distribution} to be the cause of the gravitational force that pulls a positive-energy particle away from a region where positive vacuum energy is missing, because even though the spherical symmetry that characterizes the distribution of vacuum energy that exists in the presence of a single spherical void in the positive portion of vacuum energy\index{void in positive vacuum energy} does imply that the only solution to the gravitational field equations is the Schwarzschild solution\index{Schwarzschild solution}, as when a void in the uniform positive-energy \textit{matter} distribution is considered, in the present case the density of energy inside the spherical void is not null, but rather negative and under such conditions it is possible for a repulsive gravitational field to be present on the boundary of the void.

Once again, however, what this means is not that the negative vacuum energy that is present inside such a void in positive vacuum energy is allowed to exert a gravitational repulsion\index{gravitational repulsion} on positive-energy particles, but that, in such a situation, the positive energy that is located \textit{outside} the void can actually exert on those particles an uncompensated attractive gravitational force\index{uncompensated gravitational attraction} that is oriented outward from the center of the structure. All that is necessary for such a phenomenon to occur is that the energy inside the void be smaller than it is outside the structure, despite the fact that this energy is already null in the absence of such a void. When this condition is satisfied a net gravitational force can exist that arises from the fact that the gravitational attraction attributable to the uniform distribution of positive vacuum energy is no longer compensated in the direction of the void, as it would if this positive energy actually was homogeneously distributed.

It must be clear that the situation we have in the presence of a single void in the uniform distribution of positive vacuum energy\index{vacuum energy!uniform distribution} is different from that of a hollow sphere\index{hollow sphere} of positive-energy matter of finite size located in an otherwise empty universe, because in such a case we are dealing with a spherical symmetry\index{spherical symmetry!dependence on observer position} that is dependent on the position of the observer, unlike for the hollow sphere. Given that the uniform distribution of vacuum energy can be assumed to remain symmetric around \textit{any} location in the absence of a void in the positive portion of it, then one must conclude that the presence of such a void would necessarily alter the spatial equilibrium of gravitational forces\index{gravitational forces!spatial equilibrium} in its surroundings.

When we are considering a spherical void in the uniformly distributed positive portion of vacuum energy\index{void in positive vacuum energy}, it is not possible to assume that the rest of the distribution of positive vacuum energy surrounding that region constitutes a hollow sphere, based merely on the fact that this energy is uniformly distributed. In fact, such a spherical region in a uniform distribution of positive vacuum energy would be free of uncompensated gravitational forces\index{uncompensated gravitational force} only if it was itself filled with positive vacuum energy as uniformly distributed as the vacuum energy that is found outside the region, because it is only in such a case that the condition of spherical symmetry\index{spherical symmetry!condition} would apply to any point inside the spherical region.

Here, again, there would only be a problem if we were to fail to apply the requirement of relational definition\index{requirement of relational definition!physical attributes} of physical attributes when we are dealing with the gravitational force attributable to the entire distribution of positive vacuum energy. When we are dealing with a chosen spherical region of the universe, we cannot assume that the surrounding positive vacuum energy, which may exert a gravitational force on positive-energy particles located inside that region is spherically distributed around the center of the spherical region considered, as if the location of the center of mass of the universe\index{center of mass of universe} was an intrinsic invariable feature of the whole configuration. The center of mass of a uniform distribution of vacuum energy\index{vacuum energy!uniform distribution}, in a physical universe without boundary, cannot be chosen arbitrarily and must rather be defined in a relational manner as any other property, if we are to be able to determine the consequences on a given object of being located in such a position.

When we are dealing with a uniform distribution of positive matter and vacuum energy in a universe without spatial boundary, in which the local inertial reference systems\index{local inertial reference system} are determined by this entire distribution of matter and vacuum energy (following Mach's principle\index{Mach's principle}), the true center of mass, defined in terms of the influences exerted on a given positive-energy particle, is \textit{always} located right where that particle is to be found, whatever its position actually is. Thus, a positive-energy particle located \textit{at the center} of a void in a uniform distribution of positive vacuum energy could actually be considered to be in the situation of a particle in a hollow sphere\index{hollow sphere}, because for this particle the entire distribution of vacuum energy is centered on the void (in this situation the surrounding positive vacuum energy actually is a hollow sphere centered on the particle's position). Therefore, such a particle would feel no uncompensated gravitational force\index{uncompensated gravitational force} from the whole positive vacuum energy distribution, as required.

But if this particle moves to one side or another in the void, the distribution of vacuum energy that is exerting a gravitational force on the particle in its new position would be centered on the new position and this means that the void in the previous hollow sphere is shifted to the opposite side, just as the sphere itself is shifted in the direction of the particle's new position. The symmetry of the initial configuration would therefore no longer be present and the spatial equilibrium of gravitational forces\index{gravitational forces!spatial equilibrium} would no longer apply. In the new configuration, a whole layer of positive vacuum energy must be `removed' on one side of the external surface of the imaginary hollow sphere (in the direction opposite the particle's displacement) and added on the other (this is easier to visualize in a closed universe\index{closed universe}) which, given the distances involved, means that an enormous amount of positive energy has changed position from the viewpoint of the particle.

It must therefore be recognized that, in the final configuration, the void in the imaginary sphere\index{imaginary sphere!center of mass} is no longer centered on the center of mass of the sphere, but is actually located away from the center of the sphere. As a consequence, the condition of spherical symmetry\index{spherical symmetry!condition}, from which depended the conclusion that there would be no net gravitational force on a positive-energy particle inside the sphere, arising from the distribution of positive vacuum energy, is no longer satisfied in the final configuration and therefore it can be expected that there would be an attractive gravitational force\index{attractive gravitational force!outward-directed} on this positive-energy particle that would necessarily be directed away from the center of the void. But such a force would be completely equivalent to a repulsive gravitational force\index{repulsive gravitational force} arising from the void itself.

It is important to understand that, however large you consider the imaginary sphere encompassing the whole distribution of vacuum energy (the size of the universe) to be when you are trying to determine the gravitational force it would exert on a local particle, if the center of the sphere is shifted to one side, there would be a non-negligible effect from the displacement of its center of mass. This is true even if the distance to the periphery of the sphere (where the changes occur) is very large and the strength of the gravitational interaction decreases with the square of the distance, because the larger the distances (the larger the sphere) considered, the larger the quantity of energy that is shifted from one side to the other and thus the larger the changes produced on the local gravitational field. We should not be surprised, then, that, even the retarded interaction\index{retarded interaction} with energy so distant could have an effect similar in magnitude to the effect that would be exerted by the positive vacuum energy that is missing from a void located near some positive-energy particle experiencing those forces.

Now, even if, under certain circumstances, there may be an equivalence between a local imbalance in the sum of attractive gravitational forces\index{imbalance of attractive gravitational forces} attributable to the uniform distribution of positive vacuum energy and what would appear to be a gravitational repulsion\index{gravitational repulsion} exerted on a positive-energy particle by negative-energy matter, it must be clear that we are nevertheless always dealing with gravitational attraction. There is no question that it is the uncompensated gravitational attraction\index{uncompensated gravitational attraction} of surrounding positive vacuum energy that is responsible for the apparent gravitational repulsion which would be exerted on a positive-energy particle by a void in the otherwise uniform distribution of positive vacuum energy\index{vacuum energy!uniform distribution}. As I explained above, it is clearly as a consequence of the fact that positive vacuum energy is missing in the direction where the void is located, while the positive vacuum energy present in the opposite direction still exerts its gravitational pull, that there exists a net force directed away from the void.

Thus, what looks like a repulsive gravitational force\index{repulsive gravitational force} exerted in a given direction by some negative-energy matter and which could, from a certain viewpoint, be equivalent to it, would actually be the product of a gravitational attraction arising from an absence of positive vacuum energy exerting a compensating gravitational attraction\index{compensating gravitational attraction} in the opposite direction. This conclusion is particularly significant in the context where local inertial reference systems\index{local inertial reference system} are to be considered as always arising from a perturbation of the equilibrium of large-scale gravitational forces\index{equilibrium of gravitational forces!perturbation} by the gravitational forces attributable to local matter concentrations, as I have emphasized in section \ref{sec:2.5}.

In the presence of a void in the uniform distribution of positive vacuum energy\index{vacuum energy!uniform distribution} one can, therefore, expect that a positive-energy particle would only experience an absence of repulsive gravitational force \textit{at the center} of a spherical void\index{void in positive vacuum energy!gravitational repulsion} in the distribution of positive vacuum energy, while there would actually exist a net force everywhere else inside the void, as would be the case if we were considering the gravitational attraction existing inside an isolated sphere filled with positive-energy matter (like a planet or a spherical gas cloud\index{spherical gas cloud}). This is a necessary condition if missing positive vacuum energy\index{missing positive vacuum energy!negative-energy matter} is to be considered completely equivalent to negative-energy matter, in the presence of which this property must necessarily apply, given that it does in the presence of positive-energy matter itself.

One assumption that will be crucial for my derivation of the modified general-relativistic gravitational field equations\index{gravitational field equations!modified} is that a phenomenon of gravitational repulsion\index{gravitational repulsion} similar to that described here would apply from the viewpoint of negative-energy particles in the presence of voids in the negative portion of vacuum energy\index{void in negative vacuum energy!positive-energy matter}. Using the same logic that allowed me to derive the consequences of the presence of a void in the uniform distribution of positive vacuum energy, it is possible, in effect, to deduce that a local absence of negative vacuum energy would be equivalent, from a gravitational viewpoint, to the presence of positive-energy matter. The equivalence between matter and missing vacuum energy is therefore valid both ways, so that the proposed description is invariant under an exchange of positive- and negative-energy matter\index{positive- and negative-energy matter!invariance under exchange} and does not depend on the energy sign of an observer.

But concerning the effects which I'm suggesting should be attributed to missing positive vacuum energy\index{missing positive vacuum energy}, we may ask to what extent such a void may actually be considered physically significant, in the sense of being merely an anomaly in an otherwise uniform distribution of vacuum energy\index{vacuum energy!uniform distribution}? If we examine the situation carefully it becomes clear, in effect, that, given that it must be the surrounding positive vacuum energy that exerts the \textit{outward}-directed attractive gravitational force\index{attractive gravitational force!outward-directed} that would be experienced as a gravitational repulsion by positive-energy particles, then it follows that, as we consider voids of increasingly larger sizes, there may come a point when there would be no positive vacuum energy left outside the void to produce the uncompensated gravitational attraction\index{uncompensated gravitational attraction} that must exist to produce the equivalent repulsion.

If the equivalence between missing positive vacuum energy\index{missing positive vacuum energy!negative-energy matter} and the presence of negative-energy matter is considered to be generally valid, then it follows that the presence of a uniform negative-energy matter distribution would imply the existence of a void in the positive-energy portion of the vacuum\index{void in positive vacuum energy!whole universe extent} which would actually extend to the whole universe. This void would have been present in the vacuum from the very beginning of the universe's history and would not have developed through the production of a compensating overdensity of positive vacuum energy in its environment. In such a case it would no longer be possible to assume the existence of an uncompensated gravitational pull\index{uncompensated gravitational pull} on positive-energy particles from the surrounding positive vacuum energy, because indeed there would be no vacuum energy with higher positive density outside the void to generate the attraction. I'm therefore allowed to conclude that, under such conditions, no outward-directed gravitational attraction, which would be equivalent to a gravitational repulsion\index{gravitational repulsion} exerted by the void itself, could exist, that would be experienced by positive-energy particles.

But if the indirect gravitational force\index{indirect gravitational force} exerted on positive-energy particles by voids in the positive-energy portion of the vacuum\index{void in positive vacuum energy} actually constitutes the only form of gravitational interaction between positive- and negative-energy matter, it would appear that the preceding conclusion imposes very strong limitations on such an interaction. Indeed, it transpires that the absence of gravitational force on positive-energy matter, originating from a globally homogeneous distribution of negative-energy matter, is a very general and unavoidable feature of the description of the gravitational interaction between positive- and negative-energy matter. This is because such a limitation would also be verified in the case of a uniform distribution of positive-energy matter from the viewpoint of negative-energy particles, if the gravitational repulsion exerted on those objects by positive-energy matter can be attributed to an absence of \textit{negative} energy from the vacuum.

Thus, if opposite-energy objects interact only through their respective same-energy-sign portions of the vacuum, we must conclude that positive-energy matter interacts with negative-energy matter only in the presence of inhomogeneities in any of the two matter distributions. But given that we can expect the matter distribution to be homogeneous on the largest scale (for reasons that will be clarified in section \ref{sec:4.9}), then it follows that the presence of negative-energy matter has absolutely no effect on the gravitational dynamics of positive-energy matter (and vice versa) on such a scale. This would mean, in particular, that the rate of expansion\index{expansion rate!positive-energy observers} experienced by positive-energy observers is not influenced by the presence of negative-energy matter on a global scale and similarly that the rate of expansion\index{expansion rate!negative-energy observers} experienced by negative-energy observers is not affected by the presence of positive-energy matter on a global scale. This is a very significant result, which will have an impact on many aspects of cosmology theory\index{cosmology theory} and whose implications will be developed in chapter \ref{chap:4}.

The conclusion discussed here is the one on which is founded the hypothesis, discussed in section \ref{sec:2.5}, that allowed a relational description of the phenomenon of inertia\index{inertia!relational description}. There, I explained that if both the large-scale positive- and negative-energy matter distributions were to exert a gravitational force on positive-energy objects, then the hypothesis that acceleration is relative\index{relativity!of acceleration} would be invalidated in the presence of negative-energy matter, on a cosmological scale, because under such circumstances there would be no reason to assume that the global inertial reference system\index{global inertial reference system} relative to which both positive- and negative energy objects are not accelerating in the absence of local forces should be that relative to which matter is not accelerating or rotating on the average, on the largest scale, because the sum of gravitational forces exerted by the large-scale matter distribution on local particles would cancel out whenever the total energy of matter\index{energy of matter!null total value} is null globally.

But if only matter with a positive energy sign exerts a gravitational force on positive-energy objects on a cosmic scale, then the global inertial reference system experienced by a positive-energy object could actually be determined by the average state of motion of positive-energy matter, given that it would only be the outcome of its gravitational interaction with the large-scale distribution of positive-energy matter. Thus, we can now see why the assumption that a uniform, cosmic-scale distribution of negative-energy matter cannot exert a force on positive-energy matter (and vice versa), which appears to be required in order to arrive at a relational explanation of the phenomenon of inertia based on the principle of relativity\index{principle of relativity}, was, in effect, justified. The preceding discussion actually shows that this is not just a desirable hypothesis, but that it actually constitutes an unavoidable requirement of the description of negative-energy matter as missing positive vacuum energy\index{missing positive vacuum energy!negative-energy matter}.

It should be clear, however, that, even in the context where we need to assume that positive-energy matter only interacts (gravitationally) with negative-energy matter in the presence of inhomogeneities in the positive- and negative-energy matter distributions, it would be wrong to consider that positive-energy matter gravitationally interacts only with the uniform, positive-energy portion of the \textit{vacuum}\index{uniform portion of vacuum energy} and not with the uniform, negative-energy portion of it, because, as I have explained above, positive-energy matter must itself be assumed to consist of voids in the negative-energy portion of the vacuum\index{void in negative vacuum energy} and as such, certainly cannot be considered to evolve independently from this negative portion of vacuum energy. Yet it should be clear that we are not really dealing with an interaction between opposite-energy particles here, but merely with the gravitational interaction of this negative-energy portion of the vacuum with itself.

Such a phenomenon is somewhat similar to that which arises when voids are present in an expanding, uniform distribution\index{void in expanding matter distribution} of positive-energy matter which would appear to exert attractive gravitational forces on each other and repulsive forces on the rest of the positive-energy matter. In such a case it is clear that even if the voids seem to be equivalent to the presence of negative-energy matter, their effects are actually the outcome of a gravitational interaction of positive-energy particles among themselves. The conclusion that both the positive and the negative portions of the uniform distribution of vacuum energy\index{vacuum energy!uniform distribution} exert gravitational forces on both positive- and negative-energy particles will have important consequences for the description of the cosmological effects of vacuum energy\index{vacuum energy!cosmological effects} that will be discussed in chapter \ref{chap:4}.

Now, even though this may appear obvious, it is important to mention that in the context where positive-energy matter is equivalent to the presence of voids in the negative-energy portion of the vacuum\index{void in negative vacuum energy}, it must also be the case that a uniform distribution of positive-energy matter that extends to the whole universe still exerts an attractive gravitational force on positive-energy matter, despite the fact that a similar distribution of missing positive vacuum energy\index{missing positive vacuum energy!negative-energy matter}, which is equivalent to the presence of a uniform distribution of negative-energy matter, would have no effect on positive-energy matter, because positive-energy matter does interact with itself.

In other words, the fact that a void in the uniform negative-energy portion of the vacuum, which is equivalent to the presence of positive-energy matter, could leave no outside, surrounding negative vacuum energy to exert a gravitational force on negative-energy matter (if this void is itself uniformly distributed over the entire volume of the universe) would not affect the ability for such a void to gravitationally attract \textit{positive-energy} matter, that is to say, any \textit{local} void in this uniform, negative portion of vacuum energy, because in such a case the interaction is actually occurring between the matter particles themselves and not between a particle and the surrounding vacuum with the same energy sign.

When one appropriately assumes an absence of gravitational interaction of positive-energy particles with the uniform portion of the negative-energy matter distribution, one can also expect that a local absence of \textit{missing} positive vacuum energy\index{missing positive vacuum energy}, that is to say, an underdensity in the uniform negative-energy matter distribution, would exert on positive-energy particles an attractive gravitational force equivalent to that which we may attribute to the presence of positive-energy matter, given that under such conditions the density of gravitationally attractive, positive vacuum energy\index{positive vacuum energy!larger than average density} would be larger than its average value locally, while there would be no gravitational force on positive-energy particles from the presence of an infinitely large void in the positive portion of vacuum energy\index{void in positive vacuum energy!infinitely large} which is equivalent to the presence of a homogeneous distribution of negative-energy matter.

A local excess of \textit{missing} positive vacuum energy, that is to say, a negative-energy matter overdensity, for its part, would appear to exert a repulsive gravitational force\index{repulsive gravitational force} on positive-energy particles, given that under such conditions the density of gravitationally-attractive, positive vacuum energy would be smaller than its average value locally, thereby allowing the surrounding positive vacuum energy to exert an uncompensated gravitational attraction\index{uncompensated gravitational attraction} directed away from the center of the structure.

If it was not the case that a local absence of missing positive vacuum energy\index{missing positive vacuum energy!local absence}, which is equivalent to a void in the uniform distribution of negative-energy matter, was giving rise to an attractive gravitational field, from the viewpoint of the surrounding \textit{positive-energy} matter, then difficulties would emerge, because as we consider voids of increasingly larger sizes in the negative-energy matter distribution, there would come a point when such a void would no longer be a mere anomaly, as it would extend to the entire observable universe\index{observable universe}. But such a void in negative matter energy would (under those inadequate assumptions) still be prevented from raising the density of gravitationally-attractive, positive vacuum energy, while it must be assumed that a uniform distribution of positive vacuum energy\index{vacuum energy!uniform distribution} whose density is maximum (as a result of the absence of negative-energy matter) does exert a gravitational force on positive-energy matter. Therefore, it is absolutely necessary to assume that a local absence of missing positive vacuum energy produces an attractive gravitational field from the viewpoint of a positive-energy observer.

What makes unavoidable the conclusion that, from the viewpoint of positive-energy particles, a local \textit{excess} of missing positive vacuum energy\index{missing positive vacuum energy!local excess}, which is equivalent to the presence of a negative-energy matter overdensity, would produce a repulsive gravitational field\index{repulsive gravitational field}, on the other hand, is the fact that if this was not the case, then the formation of a void in the negative-energy matter distribution would give rise to a violation of the conservation of energy\index{conservation of energy!violation}, given that this void in negative \textit{matter} energy must itself be the source of an attractive gravitational field, for positive-energy particles, so that if the local excess of negative-energy matter which must be produced in the surroundings of the void, as a result of its formation, was not giving rise to a repulsive gravitational field that did not exist originally, then no compensation would occur, as if positive matter energy\index{matter energy!creation out of nothing} could be created out of nothing.

But even if those conclusions would also apply from the viewpoint of negative-energy particles, in the sense that, for those particles, a local excess of missing negative vacuum energy\index{missing negative vacuum energy!positive-energy matter} (which is equivalent to the presence of a positive-energy matter overdensity) would produce a repulsive gravitational field, while a local absence of missing negative vacuum energy (which is equivalent to the presence of a positive-energy matter underdensity) would produce an attractive gravitational field, it must be clear that this does not mean that \textit{positive-energy} particles themselves would experience a gravitational field in the presence of a positive-energy matter underdensity.

It is merely the fact that the average density of negative \textit{matter} energy produces no gravitational force, from the viewpoint of a positive-energy observer, that requires one to conclude that a void in the negative-energy matter distribution must give rise to gravitational attraction for positive-energy particles. But the average density of positive matter energy does exert a gravitational force on positive-energy matter and from such a viewpoint, therefore, the only possible consequence of the presence of a void in the positive-energy matter distribution is an absence of attractive gravitational field, as required by Birkhoff's theorem\index{Birkhoff's theorem}.

In any case, it must be clear that the gravitational repulsion\index{gravitational repulsion} exerted on positive-energy particles by a negative-energy matter overdensity would not be the outcome of a direct interaction with negative-energy matter, but would rather arise from a local absence of interaction of those positive-energy particles with the positive portion of \textit{vacuum} energy which is missing as a result of the presence of this negative-energy matter. It is merely the fact that the force arising from this portion of vacuum energy would no longer compensate that which arises from the surrounding uniform distribution of positive vacuum energy that would allow gravitational repulsion to be produced.

There is no contradiction here, because, even in the context where negative-energy matter is equivalent to missing positive vacuum energy\index{missing positive vacuum energy!negative-energy matter}, the gravitational repulsion exerted on positive-energy matter, in the presence of a local void in the uniform, positive portion of vacuum energy\index{void in positive vacuum energy}, does not arise from an absence of compensation (by the positive vacuum energy that is missing in the void) of the repulsive gravitational force\index{repulsive gravitational force} that would be exerted by the negative vacuum energy that is still present in the void, but rather from an absence of compensation of the attractive gravitational force exerted by the surrounding positive portion of vacuum energy, as I explained above.

But given that a force of similar origin would gravitationally repel negative-energy particles in the presence of a positive-energy matter overdensity, as a result of a local absence of interaction of those negative-energy particles with the negative portion of vacuum energy that is missing\index{missing negative vacuum energy!positive-energy matter} as a result of the presence of this positive-energy matter, then it is possible to deduce that the gravitational forces that would exist between positive- and negative-energy objects, as a result of their interaction with the uniform, positive and negative portions of vacuum energy\index{vacuum energy!positive and negative portions}, are precisely such as would be required in order to obtain a relational description of the gravitational dynamics\index{gravitational dynamics!positive- and negative-energy objects} of positive- and negative-energy objects. This would appear to confirm the validity of the hypothesis that, from a phenomenological viewpoint, the presence of matter with a given sign of energy is equivalent to an absence of energy with opposite sign from the vacuum.

It may, therefore, seem appropriate to assume, as I mentioned above, that if states of negative vacuum energy are allowed by current theories, then negative-energy matter itself must be allowed to exist. But if the presence of negative-energy matter was always constrained by the limitations which are observed to apply in the above discussed experiments (such as that involving two parallel mirrors\index{parallel mirrors in vacuum} in a vacuum), in which negative energies arise as a consequence of a suppression of positive energy from the portion of vacuum fluctuations\index{vacuum fluctuations!direct interaction} that directly interacts with positive-energy matter, then it would be clearly impossible for negative-energy matter to exist with physical properties similar to those which characterize positive-energy matter.

As I will explain in section \ref{sec:2.14}, however, the portion of vacuum fluctuations that directly interacts, other than gravitationally, with positive-energy matter is \textit{not} the portion of zero-point vacuum fluctuations\index{zero-point vacuum fluctuations!maximum contribution} that contributes a maximum \textit{positive} value to the energy of the vacuum in which voids may be present that would actually be equivalent to the presence of negative-energy matter particles, but rather the portion that contributes a maximum negative value. It is only when one recognizes that negative-energy matter consists of voids in the portion of zero-point vacuum fluctuations that directly interacts with itself, and that contributes a maximum positive value to the density of vacuum energy, that one can explain how it is possible for negative-energy matter to exist without being constrained by the above mentioned limitations (such as the averaged, weak energy condition\index{weak energy condition!averaged}) which are known to apply to the magnitude of those negative densities of vacuum energy which arise from an absence of positive energy from that portion of vacuum fluctuations that directly interacts with positive-energy matter and whose \textit{overall} contribution to vacuum energy density\index{vacuum energy density!overall negative contribution} is negative.

But it must also be recognized that if the presence of negative-energy matter in a region of space is equivalent, for positive-energy particles, to an absence of positive energy from the vacuum, this is simply because in general, for an equilibrium state\index{equilibrium state} of any kind, the presence of a negative contribution is equivalent to the absence of a positive contribution of equal magnitude and it just happens that, in the absence of real matter\index{real matter}, the gravitational forces produced by the uniform portion of vacuum energy\index{uniform portion of vacuum energy} are precisely in such an equilibrium state.

Thus, if voids in an \textit{expanding}, uniform distribution\index{void in expanding matter distribution} of positive-energy matter may appear to produce effects similar, from a gravitational viewpoint, to those we can attribute to the presence of local voids in the positive portion of vacuum energy\index{void in positive vacuum energy}, which are equivalent to the presence of negative-energy matter overdensities, it is because, in the case of an expanding matter distribution, an equilibrium exists, which is similar to that which exists, in the absence of any matter, between the gravitational forces attributable to the maximum positive and negative contributions to the energy of the vacuum\index{vacuum energy!maximum positive and negative contributions}, and that opposes the expansion of space to the gravitational pull of positive-energy matter, while this equilibrium is altered locally by the presence of a void in the otherwise uniform matter distribution, because, under such conditions, the attractive gravitational forces which would normally be exerted by the positive-energy matter that is absent locally no longer balance the expansion of space, as if there existed an equivalent gravitational force opposite that which would normally be exerted by the positive-energy matter that is missing.

It must be clear, however, that I'm not suggesting that matter particles are equivalent to voids in a filled matter distribution\index{void in filled matter distribution} of opposite energy sign, even if the gravitational dynamics\index{gravitational dynamics!voids in expanding matter distribution} of voids in an expanding, uniform distribution of positive-energy matter appear to be analogous to those of negative-energy matter overdensities, as local voids in the uniform distribution of positive vacuum energy. I must emphasize, once again, that a void in a uniform \textit{matter} distribution remains clearly distinct from a void in the average distribution of \textit{vacuum} energy. This means that my proposal is distinct from Dirac's failed, hole theory\index{Dirac's hole theory} (proposed as an attempt to solve the problem of negative energy states\index{negative energy states!problem}), in particular because what I'm suggesting is that \textit{all} positive-energy matter particles (and not just antimatter particles) are actually equivalent to voids in the negative-energy portion of the vacuum\index{void in negative vacuum energy}, rather than in a filled continuum of negative-energy matter.

What Dirac\index{Dirac, Paul} proposed is that all negative energy states are already occupied, so that positive-energy fermions, at least, should not be expected to make transitions to those negative energy states. But even if, the existence of such a filled, uniform continuum of negative-energy matter\index{negative-energy matter continuum!filled} was to have no effect on positive-energy matter (perhaps due to its uniformity), the fact that, from my viewpoint, there would be no reason to assume that positive energy states are not completely filled in the same way means that this hypothesis would not agree with observations. Indeed, it is not possible to assume, in a theory that respects the requirement of a relational definition\index{requirement of a relational definition!sign of energy} of the sign of energy, that positive-energy antiparticles are merely voids in a completely filled negative-energy matter continuum, as Dirac proposed, without also assuming that negative-energy antiparticles would be voids in a completely filled positive-energy matter continuum\index{positive-energy matter continuum!filled}, if one recognizes that negative-energy particles should have a tendency to fall toward less negative energy states, as I will propose in section \ref{sec:2.9}. But, given that positive energy states are obviously not all occupied by matter particles, it appears that this requirement cannot be satisfied.

We may then, instead, assume that \textit{all} positive-energy particles are voids in a filled negative-energy matter continuum, but again, in such a case, we would have no reason not to assume as well that all negative-energy particles are voids in a filled positive-energy matter continuum. The problem, however, is that it seems impossible to assume that we could have a completely filled distribution of negative-energy matter and at the same time a completely filled distribution of positive-energy matter (as required in order to avoid having to assume that there exists an absolute distinction between positive- and negative-energy matter), because so many voids in the positive-energy matter distribution as would be necessary to describe the filled negative-energy matter distribution would leave no possibility for the positive-energy matter distribution to itself be nearly completely filled.

But given that, as I have explained above, even a void in an \textit{expanding}, uniform matter distribution\index{void in expanding matter distribution}, which may appear to exert an equivalent gravitational repulsion\index{gravitational repulsion} on the surrounding matter with the same sign of energy, cannot really produce a repulsive gravitational field\index{repulsive gravitational field} that would exist on the boundary of the structure, but merely a local absence of attractive gravitational field, then it is not even necessary to take into consideration the necessity of a relational definition\index{requirement of a relational definition!sign of energy} of the sign of energy to rule out an interpretation of negative-energy matter as consisting of voids in a uniform distribution of positive matter energy, because a negative-energy matter overdensity must be allowed to actually produce a repulsive gravitational field in and around itself, from the viewpoint of positive-energy particles, if the negative-energy matter concept is to have any meaning.

When matter particles are merely equivalent (from a gravitational viewpoint) to a local absence of vacuum energy\index{vacuum energy!local absence} of opposite sign, it becomes possible for both positive- and negative-energy particles to actually exist as real observable particles\index{real observable particles} independently from the presence of one another. Thus, if the voids in the negative portion of vacuum energy\index{void in negative vacuum energy} which are equivalent to the presence of positive-energy matter are not equivalent to voids in a hypothetical, filled distribution of negative-energy matter\index{negative-energy matter distribution!filled} it is simply because, in fact, voids in vacuum energy cannot be equivalent to an absence of voids in vacuum energy.

I may add that, from the viewpoint of a consistent interpretation of negative-energy matter, there would also be a problem with Dirac's\index{Dirac, Paul} original proposal that a void in the filled negative-energy matter continuum\index{negative-energy matter continuum!filled} could be created, along with a positive-energy particle (as would a particle-antiparticle pair), when positive-energy photons provide enough energy to raise a negative-energy particle to a positive energy level. Indeed, as I explained above, negative-energy matter\index{negative-energy matter!dark matter} must be dark from the viewpoint of positive-energy observers, given that there can be no electromagnetic interactions between opposite-action particles, and this means that a positive-energy photon could not even interact with a negative-energy electron (propagating both a negative charge and a negative energy forward in time) to provide it with the positive energy it needs in order to reach a positive energy state. Thus, even if we insist on assuming the existence of a filled negative-energy matter continuum, we could not use this hypothesis to explain the existence of antimatter.

It is essential to understand, therefore, that the situation we would have if all negative energy states were filled is different from that we would have when dealing with a vacuum in which there would be a maximum negative contribution to the average energy density of zero-point fluctuations\index{zero-point vacuum fluctuations!maximum contribution}, because in contrast with the vacuum, a negative-energy matter distribution which would be filled at one particular epoch would no longer be filled at a later time, in the context where space is expanding. This is reflected in the fact that the uniform portion of vacuum energy obeys an equation of state\index{equation of state!vacuum energy} which is different from that of a homogeneous matter distribution.

Also, even if there is a large negative contribution to the energy of the fluctuating vacuum, there is no reason to expect that it gives rise to a situation similar, from a gravitational viewpoint, to that which would occur if space was filled with negative-energy matter and almost completely devoid of positive-energy matter (as Dirac's\index{Dirac, Paul} solution would require assuming), because a large positive contribution to the energy of empty space\index{energy of empty space!positive contribution} must always be present as well (the motives behind this conclusion will be clarified in section \ref{sec:4.2}). A space filled with positive- or negative-energy matter would be as different from the true vacuum as the primordial soup\index{primordial soup} which existed in the first instants of the Big Bang is different from the space nearly devoid of matter particles that currently exists between galaxies. Thus, if a theory of voids\index{voids theory} is to have any relevance in a gravitational context, it must involve a description of matter of any energy sign as consisting of voids in the opposite-energy portion of the vacuum, so that the presence of matter with a given energy sign does not imply an absence of \textit{matter} with opposite energy sign.

When it is understood that all positive- and negative-energy particles are actually equivalent to voids in their respective opposite-energy portions of the \textit{vacuum}, as I propose, then it follows that the unsatisfactory, categorical distinction between matter and vacuum becomes meaningless, because, in such a context, all matter can actually be considered to consist in a particular aspect of zero-point vacuum fluctuations\index{zero-point vacuum fluctuations}. It is by building on this insight that I will be able to provide a mathematically consistent and totally symmetric description of the gravitational dynamics\index{gravitational dynamics!positive- and negative-energy matter} of positive- and negative-energy matter, according to which the measure of matter energy is significant merely in relation to an energy scale associated with objective properties of the vacuum.

\bigskip

\noindent What must be retained from this discussion is that even if we assume an absence of interaction between positive- and negative-energy particles, we can nevertheless expect gravitational forces to be exerted on matter with a given sign of energy by inhomogeneities in the distribution of matter with an opposite energy sign. But under such conditions the above discussed problem of the impossibility of direct interactions of either gravitational or non-gravitational type between positive- and negative-energy particles is turned into an advantage, because even though it forbids any interactions from occurring between opposite-energy particles themselves, it allows repulsive gravitational forces\index{repulsive gravitational force} to be exerted on positive- and negative-energy particles, as a result of a local absence of interaction of those particles with the positive and negative portions of vacuum energy (respectively) which are missing as a result of the presence of matter with an opposite energy sign.

It must be clear, in effect, that while negative-energy particles, conceived as missing positive vacuum energy\index{missing positive vacuum energy!negative-energy particles}, must have charges opposite those of the virtual particles\index{virtual particles!missing from vacuum} which are missing from the vacuum as a result of this local absence of energy, this does not mean that negative-energy objects\index{negative-energy objects!electric charges} carrying positive or negative electric charges are allowed to exert indirect electromagnetic forces\index{indirect electromagnetic forces} on charged positive-energy particles. The conclusion that some indirect interactions\index{indirect interactions} should exist between opposite-energy objects only applies to gravitation, because even if there must be an equivalence between non-gravitational charges missing\index{missing non-gravitational charge!opposite-sign charge} from a neutral charge distribution\index{neutral charge distribution} and the presence of opposite-sign charges and even if it may be necessary to assume that the presence of matter is in fact equivalent to both missing energy and missing non-gravitational charges from zero-point vacuum fluctuations\index{zero-point vacuum fluctuations!missing non-gravitational charges} (in order to avoid having to conclude that ordinary positive-action particles themselves, as missing negative vacuum energy\index{missing negative vacuum energy!positive-action particles}, have no charges), an additional difficulty would emerge if we were to assume that inhomogeneities in the distribution of negative-energy matter \textit{charges}\index{negative-energy matter charge distribution!inhomogeneities} can exert electromagnetic and other non-gravitational forces on electrically charged positive-energy matter particles.

As I mentioned is section \ref{sec:2.5} and as will be further explained in section \ref{sec:2.14}, opposite-energy objects do not share the same metric properties of spacetime\index{metric properties of spacetime} and this means that it is not possible to define the strength of an electromagnetic force\index{electromagnetic force strength!opposite-action particles} in a unique and consistent manner when it is assumed that this force is attributable to an interaction between opposite-action particles, just like it is not possible to define the sign of energy of the field that would need to mediate such an interaction, while those differences would give rise to genuinely distinct effects (of a non-metric kind) from the viewpoints of opposite-energy observers. Therefore we have no choice but to recognize that such forces cannot exist, even in the presence of inhomogeneities in the distribution of positive- and negative-energy matter \textit{charges}\index{positive- and negative-energy matter charges!inhomogeneous distribution}. It is merely the fact that gravitation is the interaction that determines the very metric properties of spacetime that allows the strength of the indirect gravitational forces\index{indirect gravitational force!opposite-energy objects} between opposite-energy objects to be well-defined, unlike is the case for non-metric interactions\index{non-metric interactions}.

Thus, even if a positive-energy particle may be gravitationally repelled by a negative-energy matter overdensity, as a result of experiencing less interaction with the positive-energy portion of the vacuum, it would experience no external electromagnetic or other non-gravitational force of either attractive or repulsive kind as a result of the absence of any positive or negative charges arising from the presence of this void in positive vacuum energy\index{void in positive vacuum energy!absence of charges}. It is necessary to assume, as a basic consistency requirement, that only particles with the same sign of energy can interact through non-gravitational forces and that even the kind of effect that allows inhomogeneities in the negative-energy matter distribution to exert gravitational forces on positive-energy particles cannot exist for non-gravitational forces\index{non-gravitational force}. Positive-energy matter particles\index{positive-energy matter!missing charges from vacuum}, as missing charges from the negative-energy portion of vacuum fluctuations, interact non-gravitationally only with other missing charges from the same portion of vacuum fluctuations and never with missing charges from the positive-energy portion of vacuum fluctuations.

Those results should be encouraging, as the category of problems they allow to solve was the most basic and the most serious of those which I have identified above as facing a theory of negative-energy matter\index{negative-energy matter!theory}. Thus, it is now possible to explain why it is that we have never observed gravitationally repulsive matter\index{gravitationally repulsive matter!invisible}, because such matter, if it exists, should be invisible, as it would not interact with ordinary positive-energy matter through the long-range electromagnetic interaction.

It is also possible to explain why it is that the predictions of quantum field theory\index{quantum field theory!predictions}, made under the hypothesis that negative energy states are not allowed in the formalism, produce accurate results which correspond with observations to a very high degree of precision. If only the indirect gravitational interactions\index{indirect gravitational interactions} just described can exist as a kind of influence of negative-energy matter on the processes, involving positive-energy particles, which are described by conventional quantum field theory\index{quantum field theory!conventional}, then, given the weakness of the gravitational interaction\index{gravitational interaction!weakness}, there should only be a marginal impact from the existence of this negative-energy matter on estimations of physical observables, which are currently obtained under the assumption that negative-energy particles do not exist. If we don't need to take into account the effects of the attractive gravitational interaction between ordinary \textit{positive-energy} matter particles in such calculations, then we certainly don't have to take into account any effects arising from the indirect gravitational interactions of positive-energy particles with the very sparse amount of negative-energy particles that could perhaps be found to wander around apparatuses located on Earth.

Thus, if I'm right, we would have here the solutions to two quite serious problems which were never addressed by any of the authors that previously discussed the possibility of gravitationally repulsive matter\index{gravitationally repulsive matter!unaddressed problems}, because it can now be understood, at once, why negative-energy matter\index{negative-energy matter!dark matter} is dark and why it would nevertheless appear to interact gravitationally with positive-energy matter under appropriate conditions.

\bigskip

\noindent Before concluding this section, it may be of interest to recall the above mentioned conclusion to the effect that, from the viewpoint of a positive-energy particle, a local void in a uniform negative-energy \textit{matter} distribution should give rise to gravitational attraction. This can be predicted to occur due to the fact that a uniform distribution of negative-energy matter exerts no gravitational force on positive-energy particles. Indeed, if such a uniform matter distribution is equivalent to the presence of an infinitely large void\index{void in positive vacuum energy!infinitely large} in the positive-energy portion of the vacuum and you remove negative-energy matter in a portion of this void, then what you obtain is a local absence of \textit{absence} of energy, which is really just the same as a local presence of energy and if the energy that was absent (when negative-energy matter was present) was positive, then the energy that is now present, locally, should be positive. This local absence of negative-energy matter is therefore equivalent to the presence of a similar amount of positive-energy matter and should produce on positive-energy particles (even those present inside the structure) a gravitational attraction directed toward the center of the void in the negative-energy matter distribution.

It is important to understand that this conclusion is not invalidated by the fact that a void in a uniform distribution of negative-energy matter cannot produce a repulsive gravitational field\index{repulsive gravitational field} from the viewpoint of negative-energy observers, but merely a missing attractive gravitational field, because the formation of such a void must take place along with the formation of an overdensity of negative-energy matter in its surroundings and given that, from the viewpoint of positive-energy observers, this overdensity produces a repulsive gravitational field, while there was absolutely no gravitational field (attributable to the presence of negative-energy matter) prior to the formation of the void (as a uniform distribution of negative-energy matter exerts no influence on positive-energy particles), then, from the viewpoint of positive-energy observers, no matter energy\index{matter energy!creation out of nothing} is created out of nothing, even if the void actually produces an attractive gravitational field that was not present prior to its formation, because from this particular viewpoint the configuration of energy that is the source of the field produced by the void is opposite that which is the source of the field produced by the surrounding overdensity, while no gravitational field was present initially.

But from the viewpoint of negative-energy observers the energy of matter is unchanged as well, overall, because an attractive gravitational field was present in the region where the void is located, before it was created, and while this gravitational field no longer exists in the void, following its formation, there is now an additional attractive gravitational field in the surroundings of the void, where the negative-energy matter was displaced, so that, overall, the energy which is the source of the observed gravitational fields does not vary, as required. In this particular case, no mutually compensating, opposite variations actually take place as a result of the formation of the void, given that the initially uniform distribution of negative-energy matter does produce an attractive gravitational field from the viewpoint of negative-energy observers. All that must be understood, under such conditions, is that even the magnitude of the gravitational field\index{magnitude of gravitational field!observer dependence} attributable to one and the same matter distribution must be allowed to vary depending on the sign of energy of the particles or the observer experiencing it.

The fact that, in an initially smooth matter distribution, an overdensity in the negative-energy matter distribution can only form if a compensating underdensity also forms in its surroundings, while from the viewpoint of a positive-energy observer those two types of inhomogeneities should produce oppositely-directed gravitational fields, is what allows me to conclude that the presence of negative-energy matter has absolutely no effect on the gravitational dynamics of positive-energy matter on the cosmic scale, as required for a relational definition of the phenomenon of inertia\index{inertia!relational description}. Even though the inhomogeneities which are present in the negative-energy matter distribution do exert gravitational forces on positive-energy matter, those forces necessarily cancel out on the largest scale.

Now, one may be tempted to argue that no attractive gravitational field could exist in the presence of a void in an otherwise uniform negative-energy matter distribution, from the viewpoint of positive-energy particles, because in the absence of \textit{missing} positive vacuum energy\index{missing positive vacuum energy!absence} (in the absence of negative-energy matter), if no positive-energy matter is present, vacuum energy is merely null in the void and therefore it should exert no gravitational force at all, not just from the viewpoint of negative-energy particles, but for positive-energy particles as well. This deduction, however, would be incorrect in the context where it is necessary to assume that the uniform portion of negative matter energy exerts no gravitational force on positive-energy particles, due to the fact that such a matter distribution is equivalent to the presence of a void of universal proportion in the positive portion of vacuum energy\index{void in positive vacuum energy!universal proportion}, which leaves no surrounding positive vacuum energy to produce uncompensated attractive gravitational forces\index{uncompensated gravitational attraction}.

In such a case, it is as if the uniform portion of the distribution of negative-energy matter just wasn't there, from the viewpoint of positive-energy particles, and therefore when less positive vacuum energy is missing in such a way, in a certain region of space, it is just as if a positive-energy matter overdensity was present in this region, because in the absence of this local contribution, no gravitational force is exerted on positive-energy particles by the negative-energy matter, so that if the density of positive vacuum energy is larger, in comparison to what it would be in the absence of the local negative-energy matter underdensity, then the density of matter energy actually is positive in this region of space and therefore it must be the source of an attractive gravitational field, but only from the viewpoint of positive-energy particles (how this is possible will be clarified in section \ref{sec:2.14}). However, if such a void in the uniform portion of the negative-energy matter distribution really is equivalent, from a gravitational viewpoint, to the presence of positive matter energy, for positive-energy observers, then the presence of a gravitational field inside the void would not violate Birkhoff's theorem\index{Birkhoff's theorem}, as long as the right positive measure of matter energy is considered for the spherically-symmetric Schwarzschild metric\index{Schwarzschild metric!spherically-symmetric} of spacetime.

The effect described here will have interesting consequences on the cosmological scale, in situations where variations in the density of negative-energy matter have a magnitude that is comparable with the average density of matter itself. I will explore the practical consequences of this important result in section \ref{sec:4.3}. But for now, let me mention that the effectiveness of the preceding description is a further confirmation of the existence of a close relationship between vacuum energy and matter energy, while the invariance of the described phenomenon under exchange of positive- and negative-energy matter\index{positive- and negative-energy matter!invariance under exchange} also indicates that the concept of negative-energy matter proposed above fully agrees with the requirement of a relational definition\index{requirement of relational definition!sign of energy} of the physical attribute of energy sign.

\section{No energy out of nothing\label{sec:2.8}}

Before we can conclude that there should be no interference with current predictions, made using quantum field theory\index{quantum field theory}, from allowing the existence of negative-energy particles in stable states, we must first explain why it is that there should be no creation or annihilation processes involving pairs of opposite-action particles\index{pairs of opposite-action particles!creation or annihilation} with opposite charges, as such a phenomenon could also disrupt current predictions. This is the second category of problems I previously identified as potentially affecting the viability of the negative-energy matter hypothesis.

Given the plausibility of the hypothesis that negative-action particles should be very rare in our region of the universe, it may seem that the problem of the annihilation of opposite-action particles does not constitute a decisive issue. But, as I previously mentioned, we cannot avoid having to face the related problem of the creation of pairs of opposite-action particles, because in such a case it would appear that no favorable initial conditions are required for the discussed processes to occur. Thus, an explanation must be provided for why matter is not being created out of the vacuum in massive amounts, even under normal conditions, despite the fact that the processes involved could occur without violating the principle of conservation of energy, because this prediction clearly disagrees with observations which indicate a complete absence of such processes.

One may perhaps suggest that it is the fact that the opposite-action particles emerging from a creation event in opposite directions would have their momenta both pointing in the same direction (because it must be assumed that a negative-action particle\index{negative-action particle!momentum} would have momentum opposite the direction of its velocity) that prevents the creation of such pairs, when we require momentum to be conserved. But it does not seem that this would constitute a strong enough constraint, under all circumstances, because the pairs could be created without much momentum, or with their velocities pointing in the same direction, or even through an input of momentum from the environment, as is the case for ordinary particle-antiparticle creation\index{particle-antiparticle!creation} processes arising from the disintegration of a single boson. It is not possible, therefore, to conclude that it is the requirement of momentum conservation\index{conservation of momentum} which prevents pair-creation processes\index{pair-creation processes!opposite-action particles} involving particles with opposite action signs from occurring.

The fact that the kind of creation (or annihilation) processes which would require no energy input (or output) could be described as processes during which a particle reverses its direction of propagation in time while retaining the sign of its energy, may suggest another explanation for why such events would be forbidden. Indeed, we may ask why it is that when a particle changes its direction of propagation in time, in the course of all those particle-antiparticle annihilation\index{particle-antiparticle!annihilation} processes which do occur under the right conditions, the energy is invariably reversed relative to the new direction of propagation in time (so that it appears to be unchanged from the forward-time perspective)?

Why must it be imposed that a reversal of the direction of propagation in time\index{direction of propagation in time!reversal} be combined with a reversal of energy\index{reversal of energy} which leaves the sign of action\index{sign of action!invariance} invariant, so that the energy of the annihilating pair needs to be compensated by the emission of photons carrying away the energy? Could it be that it is a requirement of continuity of physical properties along the world-lines of elementary particles\index{elementary particle world-line} that prevents a positive-action particle from turning into a negative-action particle? Such a change would, in effect, involve the transformation of a particle experiencing the gravitational interaction in a given way into a particle experiencing it in a different way, but perhaps that a particle cannot change the way it gravitationally interacts with the rest of the universe along a continuous world-line.

But in fact, if the \textit{charge} of a particle can remain unchanged when the particle reverses its direction of propagation in time in a continuous fashion (during an ordinary particle-antiparticle annihilation process), then there is no \textit{a priori} reason why the energy of a particle could not also remain unchanged when a particle reverses its direction of propagation in time. But such a transformation is precisely what is never observed to occur. Must one, then, conclude that there exists an inexplicable decree, simply banning negative-action particles (carrying positive energy backward in time) from existing? This would again be the easy way out: there is a difficulty, so let's just forget about the whole thing. But if we recognize that the existence of particles carrying positive energies backward in time is theoretically inevitable, then a satisfactory explanation for the absence of spontaneous matter creation out of nothing\index{matter creation out of nothing} is required.

In fact, the problem of the creation of pairs of opposite-action particles\index{opposite-action pairs!creation out of nothing} from nothing and the related problem of the annihilation of pairs of opposite-action particles\index{opposite-action pairs!annihilation to nothing} to nothing turned out to be much simpler to solve than I had originally envisaged. To understand what imposes a limit on the creation and annihilation of pairs of opposite-action particles, we simply need to take into account the results obtained in the preceding section. Indeed, one may ask how it is supposed to occur that a positive-action particle with positive charge, say, could annihilate with a negative-action particle with negative charge, when annihilation is to be considered a kind of interaction (that normally involves a photon or another interaction boson) and there can be no direct interaction between opposite-action particles?

It appears that what prevents the creation and the annihilation of particles with opposite action signs is simply the absence of direct interaction between such particles, because given that two opposite-action particles\index{opposite-action particles!absence of interaction} with opposite charges cannot interact with one another, even if they find themselves arbitrarily close to one another, then it is certainly not possible for their trajectories to merge. But this does not only prevent opposite-action pair annihilation, it forbids opposite-action pair creation as well, because it would also be impossible for the trajectories of opposite-action particles to merge while the particles are propagating backward-in-time, from the future toward the present, while from a forward-in-time perspective, such an annihilation process would be equivalent to a process of pair creation. Thus, it seems that it is not possible for a positive-action particle to transform into a negative-action particle on a continuous spacetime path, even while reversing its direction of propagation in time\index{direction of propagation in time}. The absence of interaction between opposite-action particles would therefore appear to constitute a necessary and sufficient condition to prevent opposite-action particles from being created out of nothing or from annihilating to nothing.

It is important to realize that the solution proposed here to the problem of creation out of nothing\index{creation out of nothing!problem} does, in effect, also apply to hypothetical processes of annihilation to nothing, because, even though it may seem that the problem of the annihilation of pairs of opposite-action particles\index{opposite-action pairs!annihilation to nothing} does not constitute a decisive issue in the context where there cannot be very many negative-action particles and antiparticles in our region of the universe (so that encounters between opposite-action particles\index{opposite-action particle encounters} should be rare), the situation was different in the very first instants of the Big Bang. When the magnitudes of the densities of positive and negative matter energy are both arbitrarily large, or actually maximum, and the matter is very homogeneously distributed in space (as must have been the case in the initial singularity\index{initial singularity}, for reasons that will be discussed in sections \ref{sec:4.5} and \ref{sec:4.9}), it follows that if the annihilation of opposite-action particles to nothing was not forbidden, then most of the matter, as well as most of the energy contained in radiation, would vanish within a very short instant, leaving absolutely nothing behind.

This is certainly not an outcome that would agree with astronomical observations (unless creation out of nothing\index{creation out of nothing} was allowed to compensate this annihilation) and therefore one must recognize that the validity of the explanation proposed here for the absence of creation of energy out of nothing is also confirmed by the observation that matter does exist in our universe at the present time. In fact, once this is understood, it transpires that it may not even be necessary for all matter to be created out of nothing during the Big Bang, if it is possible to assume that time extends past the initial singularity following a hypothetical quantum bounce\index{quantum bounce}, as the most promising, tentative quantum gravitation theories\index{quantum gravitation theories} seem to imply. The plausibility of the results discussed above would therefore merely provide one more reason to acknowledge the validity of this prediction.

\section{The problem of vacuum decay\label{sec:2.9}}

There is an unavoidable question that arises whenever one proposes that negative energy states may be physically allowed. What is it, in effect, that prevents particles from falling into those `lower' energy states? It is often argued that what would prevent positive-energy matter particles from falling into ever lower, negative energy states is the fact that their positive energy cannot be reduced below the minimum limit of energy contained in their rest mass\index{rest mass!energy}. This constraint, however, would not prevent a particle that is already in a negative energy state from being submitted to such a decay, under conditions where negative energy states\index{negative energy states} are not completely filled. Neither would it prevent massless, positive energy particles from reaching those below-zero energy states, if they are allowed to release a compensating amount of positive energy in some way.

Under such conditions it would seem that if even a small amount of matter was to ever find itself in one of the available negative energy states this would give rise to a catastrophic process of creation of negative matter energy and positive radiation energy, because the matter would radiate energy in going from `higher' energy states (with negative values nearer to zero) toward `lower' energy states (with larger negative values) without ever reaching a minimum energy\index{minimum energy} in which it could settle down. Thus, as I mentioned before, it would seem that if negative-energy matter can exist, we could produce an infinite amount of work by simply harvesting the positive-energy radiation produced when negative-energy particles fall into lower negative energy states.

But given that quantum field theory\index{quantum field theory!negative energy states} already allows for states of negative energy to occur in limited portions of space, it would seem that we have a very serious problem, even in a conventional theoretical context, because if negative energy can be made to exist under such conditions (which have already been produced in the laboratory) it should immediately collapse to even lower negative energies and in the process produce an arbitrarily large amount of positive-energy radiation, while, of course, no such a phenomenon has ever been observed. It may, therefore, appear preferable to assume that negative energy states\index{negative energy states} are simply forbidden, because there is no evidence that the current density of negative-energy matter is already maximum, as would be the case if negative-energy states were all filled.

The insights gained while studying the problem of energy creation out of nothing\index{energy creation out of nothing!problem} discussed in the preceding section, however, provide the elements needed to tackle this additional difficulty from a different angle. As I have explained, an important consequence of the absence of direct interactions between opposite-action particles\index{opposite-action particles!absence of direct interactions} is that it is impossible for a particle to annihilate with one of its opposite-action antiparticle counterpart, which is another way to say that a particle cannot reverse its direction of propagation in time\index{direction of propagation in time!reversal} without also reversing its energy sign (relative to its new direction of propagation in time), therefore describing an ordinary particle-antiparticle\index{particle-antiparticle!annihilation} annihilation process. But another, perhaps less obvious, consequence of the absence of direct interaction between opposite-action particles is that a negative-energy particle cannot emit a real (by opposition to virtual), positive-energy interaction boson\index{interaction bosons!real} regardless of what energy changes the original particle goes through, because the positive-energy boson is not even allowed to interact with the negative-energy particle it is assumed to get its energy from.

Therefore, a negative-energy particle could not `lose' energy (gain negative energy) through the production of a compensating amount of positive radiation energy and the same limitation also implies that a positive-energy particle couldn't absorb negative-energy radiation and diminish its own positive energy in the process. This constraint must apply even if such processes could occur without violating conservation laws\index{conservation laws}, when the change in the energy of the matter particle involved would be compensated by the emission or the absorption of an opposite amount of radiation energy\index{radiation energy!emission or absorption}. But this means that even the emission of \textit{positive-energy} radiation by a positive-energy matter particle could not occur in such a way that the positive-energy particle could turn into a negative-energy particle, given that this would imply that there would have been an interaction between the matter particle that now has negative energy and the positive-energy radiation it would have released (at the particular point in spacetime where the reversal of energy would have taken place), while according to my analysis this must be considered impossible.

This conclusion would be valid even for a massless positive-energy particle, because once such a particle would reach a null energy, by releasing positive energy, it would have to continue releasing positive energy if it is to reach a negative energy state, but once it crosses the zero-energy limit\index{zero-energy limit} it would, in effect, be forbidden from emitting positive-energy radiation, with which it cannot interact. Thus, the same constraint whose existence allowed me to conclude that a particle cannot change its direction of propagation in time without reversing its energy sign\index{sign of energy!reversal}, also implies that it is impossible for a particle to reverse its energy without reversing its direction of propagation in time\index{direction of propagation in time!reversal} (in which case the particle would not continue to exist with opposite energy in the future).

The existence of such a limitation suggests that no interaction vertex\index{interaction vertex!mixed action signs} involving matter particles with mixed action signs needs to be taken into account in determining the transition probabilities\index{transition probabilities!quantum processes} associated with quantum processes. A certain limitation against the possibility of transitions to negative energy states\index{negative energy states}, therefore, actually exists, because a positive-energy particle cannot `fall' into a negative energy state by releasing positive-energy radiation. The only reversal of energy which may occur on a continuous particle world-line\index{particle world-line!continuous} would have to involve a reversal of the direction of propagation in time, in which case the energy of the particle in its final state would not be reversed relative to the forward direction of time\index{time!forward direction} and we would merely observe a conventional antiparticle in a positive action state annihilating with the `original' positive-action particle.

The limitation imposed on interaction vertexes that they cannot involve real particles with mixed action signs would therefore actually prevent a particle that is already in a negative energy state from falling into even `lower' energy states by releasing positive-energy radiation, because such a negative-energy particle could never have interacted with the positive-energy radiation it is assumed to emit. In fact, this explanation works both ways, as it is also true that a particle in a negative energy state could not `gain' energy and turn into a positive-energy particle by releasing a compensating amount of negative-energy radiation\index{negative-energy radiation}, because the interaction bosons\index{interaction bosons} so released could not have been emitted by the particle that now has positive energy at the particular point in spacetime where the reversal would have taken place, given that they cannot interact with such a particle.

What must be understood, again, is that while the requirement of energy conservation\index{conservation of energy!requirement} may not, alone, forbid transitions involving a reversal of the sign of energy\index{sign of energy!reversal}, the fact that those transitions would involve the emission or the absorption of radiation\index{radiation energy!emission or absorption} with an energy sign opposite that of the final or original particle (respectively) actually prevents them from occurring in the context where a negative-energy particle (be it matter or radiation) is not allowed to interact with a positive-energy particle. But the constraint described here would not prevent a negative-energy particle from absorbing negative-energy radiation and `falling' into ever more negative energy states, if such an evolution is favored from a thermodynamic viewpoint. In this particular sense it may therefore appear that a certain aspect of the problem of vacuum decay\index{vacuum decay problem} remains unsolved.

I believe that the situation we have here is analogous to that which was faced upon the introduction of the Rutherford atom model\index{Rutherford atom model}, which was initially rejected despite its apparent empirical inevitability, because it was assumed that the electrons in orbit around the nucleus would lose energy in the form of electromagnetic radiation and end up collapsing into the nucleus, while no such catastrophe was observed. But just like the Rutherford model, it appears that negative energy states\index{negative energy states} are unavoidable and thus a solution to the problem of vacuum decay that does not simply amount to reject the physical nature of those energy states must be provided. Based on the results achieved in the preceding sections, I would suggest that the difficulties described here arise, again, from the fact that we ignore the requirements imposed by the necessary relational definition\index{requirement of relational definition!physical attributes} of physical attributes.

What is happening is that we are attributing a preferred direction to energy variations\index{energy variations!preferred direction} without referring to a physical aspect of our universe relative to which that direction could be compared. In other words, we use an absolutely defined direction on the energy scale\index{energy scale!absolutely defined direction} which we arbitrarily define as `lower' and we attribute distinctive physical properties to energy variations occurring along that absolutely defined direction, despite the fact that it actually has no objective significance. This conventional assumption seems to be justified by the observation that, for positive energy states at least, there does exist a singled-out direction on the energy scale which is related to the natural tendency for matter to disintegrate and to reach thermal equilibrium\index{thermal equilibrium}. This direction can be associated with a well-defined physical aspect of our universe, which is the direction of time in which entropy is growing. In the absence of such a relationship, we would have no motive to assume the existence of a preferred direction on the positive energy scale, that would not necessarily be opposite any such direction on the negative energy scale.

However, when I examined what the motives are, exactly, which allow us to consider the existence of this objectively defined `lower' direction on the positive energy scale\index{positive energy scale!objectively defined lower direction}, arising in relation to the direction of time in which entropy grows, I realized that there is absolutely no reason to assume that this direction on the energy scale can be extended into negative-energy territory without being subjected to a reversal like energy itself. It rather seems that the objectively defined `low' direction on the energy scale\index{energy scale!objectively defined lower direction} would actually be effective toward smaller, \textit{less negative} energy states (toward the zero-energy ground state\index{zero-energy ground state}) for negative-energy matter. What allows me to conclude that the low-energy direction, for negative-energy matter, is toward the zero energy, as is the case for positive-energy matter, is that the singled out, objectively defined direction on the energy scale is simply that relative to which energy\index{energy!tendency to dissociate} tends to dissociate itself and to become less concentrated, so as to spread into a larger number of independent particles which thus necessarily have smaller (nearer to zero) energy as time goes.

What explains this tendency is the fact that such a final configuration is associated with a larger number of microscopic degrees of freedom\index{microscopic degrees of freedom} and a higher entropy (when gravitation can be neglected) and therefore is more likely to be reached in this direction of time in which entropy is actually allowed to grow. Therefore, the direction that would emerge as the low direction on the negative energy scale must be the opposite of that which constitutes the equivalent, objectively, or relationally defined low direction on the positive energy scale, because the spreading of energy into a larger number of particles with smaller \textit{negative} energies, which is necessarily associated with a higher entropy, occurs in the direction on the energy scale which is opposite that in which smaller positive energies are reached.

Thus, what we conventionally called `low' energies, far below the zero level of energy, are in fact high energies for negative-energy matter and what we called `higher' energies, nearer to the zero level on the negative energy scale, are actually lower energies for negative-energy matter. This is in perfect agreement with the previously discussed requirement to the effect that there should be a symmetry under exchange\index{exchange symmetry!positive- and negative-energy matter} of positive- and negative-energy matter, so that the sign of energy\index{sign of energy!relative attribute} can be defined as a relative physical attribute.

Such a conclusion is significant, because it allows one to deduce that it is not to be expected that matter should have a tendency (arising from a thermodynamic necessity) to decay into larger negative energy states past the zero-energy level\index{zero-energy level}. Negative-energy matter must be expected to have the same tendency as positive-energy matter to decay to energy states which from the perspective of an observer made of such matter would be lower energies and therefore to produce a larger number of particles with smaller negative energies and reach for the vacuum ground state\index{vacuum ground state} in the future direction of time. If matter was found in a negative energy state\index{negative energy states}, it would not have a natural tendency to decay in a direction on the energy scale which is actually upward for a negative-energy observer.

It would be incorrect to assume that negative-energy particles have a tendency to decay by spontaneously gaining negative energy through absorption of negative-energy radiation\index{negative-energy radiation!absorption}, as time goes, because such configurations are not thermodynamically favored, but are actually less likely to be reached, for the same reason that positive-energy matter particles are not likely to reach states where energy would become more concentrated into fewer particles as a result of the absorption of positive-energy radiation\index{positive-energy radiation!absorption}. As a consequence, regardless of its present energy, a positive-energy particle can only release radiation until it reaches the energy contained in its rest mass\index{rest mass!energy} and if it disintegrates and loses its mass it cannot be expected that it would continue to decay past the zero level of energy, by gaining more negative energy through absorption of negative-energy radiation.

The unavoidable character of the conclusion that there is no preference for states of larger negative energy means that there should be no continuous decay to more concentrated, negative energy states, especially in the context where there already exists a constraint on the emission of positive radiation energy by matter entering a negative energy state. It would not be possible, therefore, to produce a large amount of work by making use of processes during which particles would gain larger negative energies, either by releasing positive-energy radiation\index{positive-energy radiation!emission}, or by spontaneously absorbing negative-energy radiation, despite the fact that matter is actually allowed to occupy those negative energy states\index{negative energy states}. I should, finally, mention that the fact that we observe no catastrophic collapse to larger negative energies, under conditions where small negative energy densities are produced in a limited way (as when a negative pressure\index{negative pressure} is observed between two parallel mirrors\index{parallel mirrors in vacuum} in a vacuum), is a confirmation of the validity of those conclusions.

The outcome of the progress achieved in the last two sections is that a fully consistent interpretation of negative energy states now exists that allows to at least preserve the validity of the current mathematical framework of quantum field theory\index{quantum field theory!current mathematical framework}. What we obtain are two more or less independent frameworks, describing two more or less independently evolving categories of systems with opposite energies, which interfere with one another only under conditions where it is possible for an observer made of matter with one energy sign to measure densities of vacuum energy with an opposite sign, in the context where constraints are imposed which forbid the presence of certain states which would otherwise be present in that portion of the vacuum which is directly experienced (other than through the gravitational interaction) by such an observer. This particularity allows the near perfect agreement between the predictions and the observations related to the small-scale realm of quantum theory to naturally be maintained, despite the fact that it is possible for matter to occupy the available negative energy states.

\section{Energy and momentum conservation\label{sec:2.10}}

I would now like to discuss the case of that most serious of problems, which could have proved fatal to the alternative concept of negative energy developed here and which I have identified above as being that raised by the apparent possibility of a violation of the law of conservation of energy\index{conservation of energy!violation}, under conditions where indirect gravitational interactions\index{indirect gravitational interactions} are allowed to occur that would appear to involve positive- and negative-energy matter objects. The nature of the issue can be illustrated through the use of a simple thought experiment.

I briefly discussed, in a previous section, the problem that would seem to arise in the case where a collision would occur between two opposite-energy objects\index{opposite-energy objects!collision}, due to the fact that they both repel one another. I explained that such a collision would involve a loss or gain of positive kinetic energy\index{positive kinetic energy!gain or loss} by the positive-energy object that would not be compensated, but instead be made worse by the associated gain or loss (respectively) of negative kinetic energy\index{negative kinetic energy!gain or loss} by the negative-energy object. This is because, instead of witnessing a loss of energy by one particle that would be gained by another, as when two particles with the same energy sign collide, we would here seem to have equal variations of kinetic energy, either both positive or both negative, depending on which particle accelerates and which decelerates as a result of the collision. For example, a negative-action object could lose negative kinetic energy, while the positive-action object it repels would gain positive kinetic energy, resulting in a net overall increase of energy twice as large as the individual changes. It may then seem that energy conservation is not possible under such circumstances.

The problem discussed here is also apparent when we consider the variations of momentum involved in such a process. Indeed, if action is to be assumed negative for a particle propagating negative energy forward in time, then it means that the sign of its momentum\index{sign of momentum!relative to direction of propagation in space} relative to its direction of propagation in \textit{space} must be negative, that is, momentum must be opposite the direction of the motion for a negative-energy particle (because action\index{action!dimensions} has the dimension of an energy multiplied by a time or that of a momentum multiplied by a distance). In such a context, it is easy to deduce that the variation of momentum occurring upon a collision between two opposite-energy objects would be twice as large as the absolute values of the changes in each object's momentum rather than be zero, as when two positive-energy objects collide.

This is a problem that does not exist in the context of the conventional conception of negative-energy matter\index{negative-energy matter!conventional concept}, according to which positive-energy objects attract negative-energy objects, which repel them (if we assume that only gravitational forces exist between opposite-energy objects) and therefore the existence of such a difficulty could be used as an argument in favor of this conventional approach, despite the fact that it raises other problems of its own, as I previously explained.

But given that we now understand that there are no direct interactions between opposite-energy particles, we have to recognize that the only way a collision between opposite-energy objects\index{opposite-energy objects!collision} could occur would be through the indirect gravitational interactions\index{indirect gravitational interactions} which can be expected to arise as a result of the local absence of positive or negative vacuum energy which is equivalent to the presence of matter with an opposite energy sign. Under such conditions, it would appear unlikely that there could occur violations of energy conservation\index{conservation of energy!violation} arising from a collision between a positive- and a negative-energy object, if indeed there are no direct interactions between those objects. Mathematically at least, it certainly seems that a general-relativistic theory\index{general-relativistic theory} of negative-energy matter involving only gravitational interactions should not give rise to violations of the law of conservation of energy, given that energy conservation, in such a context, is actually a constraint regarding the exchange of energy between matter and the gravitational field.

Thus, if positive-energy objects interact only with one another and with vacuum energy, then it means that, from the viewpoint of a general-relativistic description of those interactions, any variation in the energy of matter that would occur in the course of an indirect gravitational interaction between a positive- and a negative-energy object would need to come from a variation in the energy of the gravitational field\index{gravitational field energy!variation} attributable to the variation of vacuum energy\index{vacuum energy!variation} which is produced by such an interaction. The problem I initially had, however, is that I was not able to figure out how this could come about in the more intuitive context of a Newtonian description of such interactions and I'm always suspicious of conclusions drawn solely on the basis of mathematical deductions, which often conceal totally inappropriate assumptions. So, where exactly does the positive energy go, which is lost by a fast-moving positive-energy object colliding with a negative-energy object initially at rest and where does the negative energy come from, which is gained by the negative-energy object during such a collision?

I was allowed to understand what is really going on when a positive-energy object appears to interact with a negative-energy object only when I became aware of the possibility that the energy of matter and its gravitational field may be null for the universe as a whole. Indeed, as certain specialists now recognize, it seems that when positive-energy matter collapses into a spacetime singularity\index{spacetime singularity} its negative gravitational potential energy\index{gravitational potential energy!negative} becomes equal in magnitude to the energy of the matter itself (even though, in the case of a future singularity\index{future singularity}, this can only happen after an event horizon\index{event horizon} has already formed).

Thus, if the initial Big Bang state must be considered to consist of a spacetime singularity (which is required even in the presence of negative-energy matter, for reasons I will discuss in chapter \ref{chap:4}), then it means that the gravitational potential energy of positive-energy matter was initially the exact opposite of the energy of this matter. As space expanded this potential energy immediately began to decrease (toward the zero value) along with the positive kinetic energy of expansion\index{kinetic energy of expansion}, but it remains that under such circumstances there naturally occurs a compensation between the energy of matter and its gravitational potential energy (although it is actually the kinetic energy of expansion that must compensate the gravitational potential energy at all times, as I will explain in section \ref{sec:4.5}).

What happens, therefore, when a positive-energy object loses some of its kinetic energy as a result of an indirect gravitational interaction\index{indirect gravitational interactions} with a negative-energy object is that the gravitational force exerted on this negative-energy object by the \textit{negative} vacuum energy that surrounds the void in vacuum energy\index{void in vacuum energy} that is equivalent to the presence of the positive-energy object allows the magnitude of its energy to grow larger, which means that, following the exchange, there is more gravitational interaction between all negative-energy matter in the universe and the negative-energy object whose presence is equivalent to this absence of positive vacuum energy, but given that the energy of the field associated with this additional gravitational attraction is positive, then there must be a positive change in gravitational energy\index{gravitational energy!positive change} that exactly balances the negative change in positive matter energy.

It must be clear, however, that the interaction of the negative-energy object is with the distribution of negative vacuum energy and that the described change in positive gravitational potential energy\index{gravitational potential energy!variation} is a result of the gravitational force exerted by the void in the negative portion of vacuum energy\index{void in negative vacuum energy} on that \textit{distinct} negative-energy object. Thus, it is not the loss of negative gravitational potential energy attributable to the interaction of the positive-energy object with all the matter with the same energy sign in the universe that compensates its own loss of positive energy, because if that was the case then it would mean that no interaction would be required to trigger those changes, which could then occur without any identifiable cause (for both positive- and negative-energy matter). It is merely the gain in positive gravitational potential energy which is attributable to the influence exerted by a positive-energy object on a negative-energy object \textit{in its environment}, as a result of their indirect gravitational interaction, which can compensate a loss in the positive energy of matter, while it is only the loss which is taking place in the negative gravitational potential energy of positive-energy matter that can compensate the gain in negative matter energy attributable to the same interaction.

What allows me to conclude that a local increase in the amount of negative matter energy will give rise to a variation of gravitational potential energy that is attributable only to the increased amount of interaction of this negative-energy matter with the ensemble of \textit{negative-energy} matter in the universe is the fact that this additional negative matter energy only interacts with the ensemble of matter in the universe that shares the same sign of energy and cannot directly interact with positive-energy matter, as I explained in section \ref{sec:2.7}.

If that was not the case, then one would need to conclude that the variation of gravitational potential energy discussed above is compensated by an opposite variation that would arise from the increased amount of interaction of this negative-energy matter with the ensemble of \textit{positive-energy} matter in the universe, which means that the variation of positive matter energy itself would remain uncompensated. The fact that only as much energy as is present in the gravitational field attributable to the interaction of an object with the ensemble of matter with the same sign of energy in the universe can actually be exchanged between opposite-energy objects means that those energies are relatively small, but they are not completely negligible and it is possible to understand how it is exactly that they are compensated, when one takes into account the variations of gravitational potential energy\index{gravitational potential energy!variation} which are occurring as a result those indirect interactions.

It must be understood, though, that we are not dealing here with the gravitational potential energy\index{gravitational potential energy!repulsive force} that could be associated with a repulsive force field mediating an interaction between the opposite-energy objects themselves, which cannot exist, in fact, as I explained before, but merely with \textit{independent} measures of gravitational potential energy associated with the interaction of each of two opposite-energy objects with all the matter in the universe that shares the same sign of energy.

Following any indirect gravitational interaction\index{indirect gravitational interactions} between opposite-energy objects, there is actually a variation in the total energy of matter\index{total energy of matter!variation}, but this is only half of the equation, as to any such change there must be a compensating change in gravitational potential energy\index{gravitational potential energy!variation} that is attributable to the local variations which take place in the energy of matter. But it must be clear that this only reflects the exchange of energy between positive-energy matter and the gravitational field produced by negative-energy matter, on the one hand, and between negative-energy matter and the gravitational field produced by positive-energy matter, on the other, because there is no actual exchange of energy between opposite-action particles\index{opposite-action particles!absence of energy exchanges} themselves.

It is, therefore, possible to conclude that kinetic energy\index{kinetic energy!positive-definite quantity} is exchanged between positive- and negative-energy objects as if it was a positive-definite quantity, from the viewpoint of positive-energy observers (or as a negative-definite quantity from the viewpoint of negative-energy observers), which means that if the magnitude of the energy of an object with a given energy sign varies as a result of its indirect gravitational interaction with an object of opposite energy sign, then the magnitude of the energy of that second object should necessarily vary in the opposite way, except while the interaction is under way and changes in the kinetic energy of matter\index{kinetic energy!matter} are compensated by changes in the gravitational potential energy associated with the interaction of the two objects with the portion of vacuum fluctuations with the same sign of energy.

It must also be mentioned that, in a general-relativistic context, one can expect that a similar conclusion would apply for those momentum variations which take place as a result of an indirect gravitational interaction\index{indirect gravitational interactions} between opposite energy objects, which would be compensated by opposite variations in the momentum of the gravitational field\index{gravitational field momentum} (actually the equivalent components of spacetime curvature), which would occur as a consequence of the changes which are produced in the momentum of matter by those interactions. The fact that the gravitational interaction is very weak means that the energy flow between matter and gravitational field that occurs in the course of any indirect gravitational interaction between opposite-energy objects is relatively small, but it nevertheless exists and it appears to be what allows energy to be conserved despite the mutually repulsive nature of those interactions.

\section{Absolute inertial mass\label{sec:2.11}}

One last objection which could be raised against the interpretation of negative energy states\index{negative energy states!interpretation} I have proposed has to do with the fact that, from my viewpoint, negative-energy matter would offer the same resistance to acceleration as does positive-energy matter. From a conventional viewpoint, this would occur whenever we would assume that inertial mass\index{inertial mass!positive} is positive, even for negative-energy matter otherwise characterized as having a negative gravitational mass\index{negative gravitational mass}. Of course, as I already explained, the inertial mass\index{inertial mass!reversal} must be considered to actually be reversed, along with the gravitational mass, from the viewpoint of a consistent description of the gravitational dynamics\index{gravitational dynamics!positive- and negative-energy matter} of positive- and negative-energy matter. But in the context of the previously discussed, improved conception of the phenomenon of inertia\index{inertia!improved concept} that emerged from my generalized formulation of Newton's second law\index{Newton's second law!generalized formulation}, it was shown that acceleration would not occur in the direction opposite the applied force for a negative-mass object, even when inertial mass is assumed to be reversed along with gravitational mass.

Once it is recognized that the equivalent gravitational field\index{equivalent gravitational field} experienced by a negative-mass object as a result of acceleration must be opposite that experienced by a positive-mass object, it is necessary to conclude that negative-mass matter would actually experience the same resistance to acceleration as positive-mass matter, when submitted to the same forces, despite the reversal of its inertial mass. Thus, negative-mass or negative-energy matter would appear to violate the principle of equivalence\index{equivalence principle!violation} as it is usually conceived. In fact, there could also be situations where the mass in a volume of space would be relatively small or even zero, despite the presence of a potentially large amount of matter in this volume, as when two opposite-mass objects are present all at once in the same location (which would be allowed in the absence of strong interactions between them). Yet such configurations would not be equivalent, from an inertial viewpoint, to the case of a system with nearly vanishing total mass, because the matter that is present would be more difficult to accelerate than if it actually had such a small mass.

To better describe such vanishing energy configurations, which are clearly different from the vacuum, we may define a measure of inertial mass that would be related to the physically significant properties with which it is usually associated and that would correspond to the true amount of matter present under such circumstances, independently from the total amount of mass, which may partially or totally cancel out. The \textit{absolute inertial mass}\index{absolute inertial mass}, obtained by adding the absolute values of the masses of all material objects present in some volume of space (or by adding all masses as negative from the viewpoint of a negative-mass observer), would constitute such a measure of the true amount of matter present.

Now, while it is clear that the acceleration of a negative-energy object in the gravitational field of a local matter inhomogeneity (such as the gravitational field which exists on the surface of the Earth) would not be that which is shared by all objects made of positive-energy matter, experiments provide very strong constraints on the degree of violation of the equivalence principle\index{equivalence principle!violation} and to date there is, in fact, no evidence at all that any such a violation has ever occurred when objects of various different compositions are utilized. However, I did say, in a previous section, that negative energy is as common as bound systems\index{bound systems} of particles, such as atomic nuclei and molecules, due to the negative energy of their attractive force field\index{attractive force field!negative energy}. Why, then, do we never observe an altered level of resistance to gravitational acceleration?

We may, for example, consider atomic nuclei\index{atomic nuclei!bound systems} formed of many protons and neutrons bound together by the strong nuclear interaction\index{strong nuclear interaction}, with various measures of negative force-field energy, associated with various configurations, involving a variable number of component particles. It may then seem that the gravitational mass\index{gravitational mass} of such bound systems should be reduced by the negative value of the energy of the field, while their inertial resistance would be proportionately larger, as the absolute inertial masses\index{absolute inertial mass} attributable to the component particles and the force field would not cancel out like the gravitational masses. If we measured the acceleration of a whole object composed of one such type of nucleus on the surface of the Earth and compared it with the acceleration of another object made of another kind of nucleus, containing a lesser proportion of such negative energy, we may then expect to discern a difference.

But it appears that this is precisely what the experiments discussed above rule out to a very good degree of precision, because objects formed of various substances, with different amounts of negative binding energy\index{negative binding energy}, always have exactly the same acceleration in the gravitational field of the Earth. Shall we then once again abandon everything and conclude that negative energy, even though it is definitely present in bound systems\index{bound systems!negative energy}, must be described in a non-relational manner (so that the sum of forces necessary to accelerate positive and negative inertial masses is allowed to cancel out like those produced by opposite gravitational masses)? It must be understood that, in fact, this conclusion would constitute a theoretical problem as grave as apparently is the empirical difficulty revealed by the absence of differences in the acceleration of various bound systems. But can we ever hope to solve a problem by creating a `new' one and assume that, despite all indications to the contrary, the latter difficulty is not real, simply because it only affects consistency on a more general level?

This is not the path I chose to follow, because I realized that, despite what is often suggested, there is simply no reason to expect the kind of violations of the principle of equivalence\index{equivalence principle!violation} which are described here, even if inertial forces do not cancel out when we consider two masses with opposite signs. What is wrong, I believe, with conventional assumptions is that, when we are considering a bound system and its force field, we assume that we have two masses with opposite signs, while what we really have is one single mass with one overall magnitude and one polarity, both from the viewpoint of inertia and from that of the response to local gravitational fields. What motive do we have, then, for considering that there could be independent contributions to the total mass of a bound system\index{bound systems!independent mass contributions} (inertial or otherwise) when, in fact, the energy of the subsystems forming it (in particular the particles mediating the attractive force fields) could not be measured independently, given that they arise from virtual processes\index{virtual processes} which do not even have classically well-defined physical properties?

It is a fact that the particles mediating an interaction are virtual and as such, exist merely by virtue of quantum uncertainty\index{quantum uncertainty}, which allows them to carry energy, but only for a time that is short enough that this energy cannot be determined. The energies of the virtual particles\index{virtual particles!unobservable energies} involved in giving rise to interactions must then be considered unobservable, even if only because, to establish their value in any one particular instance would require a time length greater than the duration of the exchange process. But under such circumstances, how could we be talking about an \textit{independent} contribution of those particles to the energy or the mass of the bound systems in which they materialize? I think that this would, in effect, be non-sense and that it must be recognized that any physical attribute of a component of a bound system\index{bound systems!physical attributes} which cannot be directly and independently measured (such as the energy of an attractive force field) does not contribute \textit{independently} to determine the value of that same physical attribute of the system as a whole, when it is actually measured.

There simply cannot exist an independent contribution, by the negative force field energy, to the inertial mass of a bound system\index{bound systems!inertial mass} and therefore we have no choice but to recognize that the measured value of the inertial mass of the system is the outcome of both positive and negative contributions and that only this total value is physically significant, even if it is possible to \textit{deduce} from a knowledge of the mass of the constituent subsystems and an observation of the total mass of the bound system, what the value of this unmeasured negative contribution actually is, on the average, for such a system. There is no contradiction here, because there is definitely a negative contribution to the energy of bound systems\index{bound systems!negative energy contribution}, only this energy contribution cannot be independently measured in any specific case and this is the crucial distinction we must take into account when estimating the absolute inertial mass\index{absolute inertial mass} of such a system.

Failure to understand this decisive requirement would mean that we again allow an inconsistency to obscure our conception of negative energy in a way that could only be made acceptable by rejecting one or another of the fundamental constraints identified above. In the present context, this could not even be avoided by assuming that negative energy does not exist at all, because the issue is no longer merely about deciding if negative energy exists, but about determining its properties in a context where we must definitely accept that it is occurring.

Thus, the difference between the situation described above of the two superposed opposite-mass objects with large absolute inertial masses\index{absolute inertial mass} and that of a composite system\index{composite system} with absolute inertial mass smaller than that of its constituent particles is that, in the former case we are actually dealing with two independent systems\index{independent systems}, which may be interacting only negligibly with one another, while in the latter case we have one single bound system\index{bound systems}, which is physically different from the sum of its parts and to which must therefore be associated one single combined measure of mass, gravitational and inertial. In any case, the fact that we do not observe violations of the principle of equivalence\index{equivalence principle!violation} for bound systems whose observable total energy is positive confirms that this conclusion is appropriate.

\section{A few other misconceptions\label{sec:2.12}}

Before finishing this discussion concerning the potential problems facing a theory of negative-energy matter\index{negative-energy matter!theory} I would like to provide arguments to the effect that a few other problems which are often associated with the possibility that there could exist gravitationally repulsive matter\index{gravitationally repulsive matter!potential problems} are actually of no concern, because they are significant only in the context of a conventional conception of negative-energy matter\index{negative-energy matter!conventional concept} and gravitational repulsion\index{gravitational repulsion!conventional concept}\footnote{
It is not possible to provide a detailed, critical review of all the arguments which were put forward in various papers which claim to offer a proof that gravitationally repulsive, negative-energy matter\index{negative-energy matter!non-existence proofs|nn} cannot exist in our universe, but I can assure the reader that, even though I have carefully analyzed many of the so called `theorems' concerning the positivity of energy\index{positivity of energy!theorems|nn}, I have never found any that does not contain one or another implicit or explicit assumption which would not apply to the kind of approach developed in this report and which invalidates them as theoretical arguments against the possibility of developing a consistent model based on the assumption that matter is allowed to occupy the available negative energy states.}.
 It is nevertheless important for me to discuss those issues, because I have come to realize that the perception of negative energy as being associated with all sorts of strange phenomena that defy common sense is responsible, more than anything else, for having transformed the perfectly acceptable idea of negative-energy matter\index{negative-energy matter!pseudo-scientific concept} into a pseudo-scientific concept which appears to have no relevance to physical reality. I will thus try to make clear that what is wrong is not the hypothesis of matter in a negative energy state, but merely the current assumptions regarding what would be the properties of such matter.

One of the problems I would like to discuss arose as an outcome of the first attempts at finding an interpretation for the negative energy states\index{negative energy states!interpretation} which were predicted to occur by special relativistic quantum theory\index{special relativistic quantum theory}. Indeed, when the existence of antimatter\index{antimatter!anti-gravity} was experimentally confirmed, it was suggested that this kind of matter may perhaps actually give rise to anti-gravity, in the sense that antimatter would experience repulsive gravitational forces in the presence of ordinary matter. But only theoretical arguments, such as those provided in Ref. \cite{Nieto-1}, could be given to disprove this possibility when it was first suggested, because no experiment had yet been performed to demonstrate that antimatter doesn't fall upward in the gravitational field of the Earth.

One of those arguments was based on the recognition that if antimatter was to gravitationally repel ordinary matter or to be gravitationally repelled by it, this would allow perpetual motion\index{perpetual motion} machines to be build that would extract more energy from a process than was initially available. Indeed, under such circumstances, it would take no energy to slowly raise a particle-antiparticle pair in the gravitational field of our planet (because there would be as much gravitational repulsion as there would be attraction). But once this would be accomplished, the pair could be made to annihilate and the positive energy of the photons so produced could fall back to a detector on the ground where it would be measured as larger than the energy the pair initially had, as a consequence of the frequency increase to which the positive-energy photons would be submitted on their way down (this would be allowed in the context where the energy of the gravitationally repelled antiparticle would nevertheless be assumed to be positive relative to the forward direction of time, so that the pair annihilation\index{pair annihilation!positive-energy radiation} process is allowed to produce positive-energy radiation). It would then seem that energy\index{energy!free production} can be freely produced if antimatter `falls' upward.

I think that this argument is perfectly valid, only it cannot be used to justify the rejection of anomalous gravitational interactions\index{anomalous gravitational interaction} in general, but rather simply means that, given that antimatter does not have negative energy (as observed in the forward direction of time), then it should not be expected to be submitted to anomalous gravitational forces. Now, could the same experiment be performed with negative-energy (actually negative-action) antimatter\index{negative-action antimatter} and then what would it mean for energy conservation\index{conservation of energy}? The answer to that question is to be found in the developments introduced in previous sections of this chapter, while solving other aspects of the problem of negative energy states\index{negative energy states!problem}.

First of all, it must be understood that, given that there are no direct interactions between positive- and negative-energy particles, then it seems that it would be much more difficult to raise a pair of opposite-energy particles together in the gravitational field of a planet. Yet, this may not constitute an insurmountable difficulty, because it is possible to imagine arrangements which would allow a negative-energy object to achieve the task of raising a positive-energy object in the gravitational field of a positive-energy planet by making use of the indirect, repulsive gravitational forces\index{repulsive gravitational force} which must exist between the two objects (which could be composed of matter with opposite charges). But, in fact, the same limitation, concerning the absence of any direct interaction between opposite-energy particles, would also imply that it is impossible to make a pair of opposite-action particles\index{opposite-action particles!annihilation} to annihilate, as I explained in section \ref{sec:2.8}. However, other means would probably exist for harvesting the energy contained in those particles, so that this limitation does not really constitute a decisive constraint that would allow to rule out the kind of processes discussed here.

For example, one could arrange things so that the positive-energy particles annihilate with positive-energy antiparticles already in place at the destination point, while the negative-energy antiparticles would annihilate with negative-energy particles already in place. But if the positive-energy photons produced by the annihilation of the positive-energy particles could actually gain positive energy while falling back to a detector on the ground, the negative-energy photons produced by the annihilation of the negative-energy particles, for their part, would lose negative energy while reaching the same detector and would therefore end up with less negative energy than they would have had if the negative-energy particles had been submitted to annihilation before rising to a higher altitude. Thus, while positive radiation energy would be gained during such a process, negative radiation energy would be lost and this means that no work can be performed in such a way.

In order to better understand the significance of the changes involved, we can consider the variations occurring in the potential energy of two opposite-energy objects\index{opposite-energy objects!potential energy variations} as they are raised in the gravitational field of a positive-energy planet. From this more general perspective what would be observed, in effect, is that any potential energy that would be gained by one of the two objects (the one that was actually lifted by the other) would necessarily be lost by the other object, thereby preventing any useful energy\index{useful energy!production} from being produced in the course of such a process. In the case at hand, what would happen is that, while the positive-energy object would gain positive potential energy, the negative-energy object would lose negative potential energy.

Now, this may seem to imply that a forbidden net increase of (positive) energy\index{energy increase!without work} can be obtained despite the fact that no work would have been done to take the system to its final state. But, as I have explained in section \ref{sec:2.10}, this variation is not significant, because any variation in the total energy of matter\index{total energy of matter!variation} resulting from the indirect gravitational interaction\index{indirect gravitational interactions} of a positive-energy object with a negative-energy object is compensated by an opposite variation in the energy of the gravitational field\index{gravitational field energy!variation} associated with the interaction of each of those objects with the rest of the matter in the universe with the same sign of energy. What is significant is that even if a positive change may occur in the potential energy of matter, this would not mean that we have gained the ability to perform more work, as would be required to produce perpetual motion\index{perpetual motion}, because what the loss of negative potential energy by the negative-energy object means is precisely that there was a loss of useful negative energy\index{useful negative energy!loss} (energy that could be used to do work) for that object during the process by which it would have performed work to raise the positive-energy object and increase the ability of this positive-energy object to perform work\index{work!ability to perform}.

In other words, despite the net gain in potential energy for the pair as a whole, the ability to do work would not have increased, because the negative-energy object, having been raised by the repulsive gravitational field it experiences, would have exhausted its ability to perform work (even though its kinetic energy would remain unchanged). The gain in useful energy by the positive-energy object would actually have been provided by the negative-energy object which would have lost its own useful energy and in fact, if the usual friction and other degradation of energy\index{friction and degradation of energy} had been taken into consideration, it should be observed that the positive-energy object would have gained less useful energy than the negative-energy object would have lost, thereby precluding any perpetual motion from being achieved. This line of arguments against gravitational repulsion\index{gravitational repulsion!arguments against}, therefore, cannot be considered significant, other than as an argument against the possibility of an anomalous gravitational interaction\index{anomalous gravitational interaction!matter and antimatter} between ordinary matter and ordinary antimatter.

A more exotic and hypothetical phenomenon, which according to certain accounts could have interesting practical applications, but which would raise serious problems from a theoretical viewpoint, given that it may provide the means of achieving faster-than-light space travel\index{faster-than-light space travel} and therefore, also, time travel\index{time travel}, is that of wormholes\index{wormhole}. It is often thought that wormholes would naturally occur in the presence of some types of black hole singularities\index{black hole!singularity} and may allow remote regions of space to be directly connected in some way, so that traveling through such wormholes would enable to bypass the limitations associated with the passage of time experienced under normal circumstances when traveling over such long distances at slower-than-light velocity\index{slower-than-light velocity}. It is not clear exactly what regions of space could be connected in such a way, or if we are really talking about connecting regions of our own universe, but if we leave aside those uncertainties, then it would seem that all that is required for unlocking the potential of faster-than-light space travel is the existence of traversable versions of such hypothetical shortcuts through space and time.

What must be provided, therefore, is a means to maintain the `throat' of a wormhole\index{wormhole!throat} open for a long enough period of time that space travelers can safely traverse it, despite the tendency for the matter configurations involved here to collapse under the effect of the gravitational attraction exerted by the singularity. The idea is that gravitationally repulsive, negative-energy matter (often called exotic matter\index{exotic matter!negative-energy matter}) may allow to achieve that goal, given that it could be used to exert a gravitational repulsion that would compensate the attraction exerted by the spacetime singularity at the center of the black hole. But again, when we look at the details of such proposals, it becomes clear that the conditions necessary for achieving the desired results are incompatible with a consistent notion of negative-energy matter. This may not be good news for science fiction\index{science fiction} lovers, but if I'm right, negative-energy matter could never be used to achieve such a goal.

To help identify what's wrong with current expectations, I would suggest that we ask how it is exactly that negative-energy matter could be brought, not just inside some positive-energy black hole, but toward the point of maximum density of positive-energy matter (the singularity), despite the enormous gravitational repulsion that this positive-energy matter would exert on the exotic matter? It should be clear that it is merely because we usually assume that negative-energy matter would be attracted by a positive-energy black hole\index{black hole!singularity} and its singularity, even while it would repel it, that this appears to constitute an achievable goal. But the truth is that any negative-energy matter approaching a large concentration of positive-energy matter, such as an ordinary black hole, would be submitted to repulsive forces as large as those maintaining positive-energy matter trapped inside the same black hole.

In this context, the only way by which negative-energy matter could find itself inside the event horizon of a positive-energy black hole would be by having already been present inside the region destined to collapse into that positive-mass black hole, before it formed. But even if that was to happen, there is no way that the negative-energy matter could be made to remain near the black hole singularity, where repulsive forces would be the strongest. This situation is simply unstable and given that stability is precisely what is required for a traversable wormhole\index{wormhole!traversable} to exist, we must recognize that negative-energy matter could not provide the necessary element for allowing spacetime singularities to be used for faster-than-light space travel\index{faster-than-light space travel} and time travel\index{time travel}.

The possibility that the kind of phenomenon discussed here could actually have been used for achieving theoretically problematic, causality-violating processes\index{causality-violating process} may seem far-fetched, but I think that it is nevertheless important to show that, even under such extreme conditions, there is no reason to expect that the existence of negative-energy matter could facilitate such an outcome (in section \ref{sec:5.11} I will explain why it is exactly that closed time-like curves\index{closed time-like curves}, of the kind that could have been allowed by the existence of traversable wormholes, must be considered problematic and it will become clear that the difficulty is not that they may allow a time traveler to alter his or her own past).

The same argument I have used to rule out the possibility of engineering traversable wormholes\index{wormhole!traversable} can also be utilized to solve a more down-to-earth problem that is not often discussed, but which would contradict one of the most unavoidable constraint applying to the evolution of irreversibly evolving physical systems, such as black holes. The problem is that negative-energy matter, as it is usually conceived, could be used to reduce the mass of a black hole and therefore, also, the area of its event horizon. This could be achieved by simply throwing negative-energy matter into a black hole, which would presumably absorb it, given that negative-energy matter is usually assumed to be gravitationally attracted by a positive-energy black hole. This would be possible, even if negative-energy matter repels a positive-mass black hole, because we could throw negative-energy particles in small amounts and their gravitational fields would be too small to resist the much larger gravitational attraction of the black hole.

But the surface area of a black hole\index{black hole!surface area} has been shown to constitute a measure of the entropy of such an object, so that reducing the area of the black hole\index{black hole!entropy reduction} is similar to reducing its entropy. Again, however, if we reject the conventional conception of negative-energy matter\index{negative-energy matter!conventional concept}, the problem does not exist, because a negative-energy particle cannot even get near a positive-energy black hole without experiencing extreme gravitational repulsion\index{gravitational repulsion}, so that it certainly cannot be absorbed by the object, as would be necessary for reducing its mass and the area of its event horizon. If negative energy states are to be considered a true possibility, then the fact that the conventional concept of negative-energy matter would allow such violations of the second law of thermodynamics\index{second law of thermodynamics!violation}, while the alternative approach developed in this report would forbid them, constitutes a strong indication to the effect that this latter proposal is more appropriate.

In fact, we are dealing with a much more general problem in this case, because, from a conventional viewpoint, it is usually assumed that when negative-energy matter comes into contact with positive-energy matter (not necessarily a black hole), its negative thermal energy\index{negative thermal energy} can be used to withdraw positive thermal energy\index{positive thermal energy!withdrawal} from this positive-energy matter (as if it was providing negative heat\index{negative heat}), while positive thermal energy could presumably be used to withdraw negative thermal energy from the same negative-energy matter, therefore allowing entropy\index{entropy!decrease} to decrease in both the positive-energy system and the negative-energy system with which it is interacting, without any heat being released in their environment, which would again violate the second law of thermodynamics. Of course, given that, from my viewpoint, negative-energy radiation cannot even interact with positive-energy matter, the possibility raised here appears to be mostly irrelevant from a practical viewpoint. We may nevertheless examine the situation which would arise following an exchange of thermal energy between positive- and negative-energy systems occurring as a consequence of the indirect, repulsive gravitational forces\index{indirect repulsive gravitational force} they are allowed to exert on one another (under appropriate conditions).

The conclusion we must draw, in such a case, is that negative energy is not equivalent to negative heat for a positive-energy system. Indeed, according to my conception of negative-energy matter, from the viewpoint of a positive-energy observer, kinetic energy\index{kinetic energy!positive-definite quantity} is exchanged between opposite-energy particles as if it was a positive-definite quantity. This is allowed given that the energy of interacting opposite-energy particles is not conserved independently from certain opposite contributions to gravitational potential energy\index{gravitational potential energy} associated with their interaction with all matter in the universe with the same sign of energy, as I explained in section \ref{sec:2.10}. But the fact that the kinetic energy of matter appears to be conserved as if the interacting particles all had the same sign of energy means that thermal energy\index{thermal energy!positive-definite quantity} itself can only be gained as a positive-definite quantity by positive-energy systems, or equivalently as a negative-definite quantity by negative-energy systems, even when the exchange involves opposite-energy systems.

Thus, when heat is provided by a negative-energy system it can only raise the positive temperature of a positive-energy system (as if positive thermal energy was provided) and the same is true for the heat provided by a positive-energy system to a negative-energy system, which can only raise the negative temperature\index{negative temperature} of the negative-energy system toward more negative values (as if negative thermal energy\index{negative thermal energy} was provided by the positive-energy system). It is necessary to assume, in effect, that temperature\index{temperature!local intensity of thermal energy}, as a measure of the local intensity of thermal energy, is negative for negative-energy matter, even under normal circumstances (when the number of possible microscopic states of a system cannot decrease as it reaches higher negative energies), given that when the energy of matter rises (into positive or negative territory), the entropy of matter\index{entropy of matter} itself rises, so that if such a change takes place as a result of the absorption of negative heat (as may be the case for a negative-energy system), then it can only mean that the temperature of the system in which those changes are taking place is negative.

We have no reason, therefore, to expect that even the indirect gravitational interactions\index{indirect gravitational interactions} between opposite-energy systems could be used to transform useless forms of energy into more useful forms and in such a way reduce the entropy of matter globally. The absorption of negative thermal energy cannot reduce the energy of a positive-energy system any more than the absorption of positive thermal energy could diminish the magnitude of the energy of a negative-energy system. The thermal energy of a positive-energy system can only be reduced through the emission of positive heat, just like the thermal energy of a negative-energy system can only be reduced (toward less negative values) when it releases negative heat\index{negative heat}. For a positive-energy system to lose thermal energy at the benefit of a negative-energy system, the magnitude of its temperature\index{temperature!magnitude} must be larger than that of the negative-energy system it is interacting with, and under such conditions the magnitude of the temperature of the negative-energy system would be raised by an amount proportional to that which is lost by the positive-energy system, as when all temperatures are positive.

What must be understood is that transferring heat from a negative energy source to a positive-energy system is not equivalent to removing positive heat from that system (despite what most people considering the possibility of the existence negative-energy matter usually assume). In fact the positive thermal energy\index{positive thermal energy} of a gas of positive-energy particles can actually be raised through contact with a gas of negative-energy particles, as long as the magnitude of the negative temperature\index{negative temperature!negative-energy gas} of the negative-energy gas is larger than the positive temperature\index{positive temperature!positive-energy gas} of the positive-energy gas. This is a consequence of the fact that thermal energy\index{thermal energy!average kinetic energy} is a measure of the average kinetic energy of gas molecules, while this energy would become more evenly distributed between the two gases (independently from energy signs), if they could come into contact by interacting gravitationally with one another, given that negative kinetic energy\index{negative kinetic energy} can be turned into positive kinetic energy\index{positive kinetic energy} and vice versa, even when energy is assumed to be conserved, as I previously explained.

In this context, it transpires that all that matters from a thermodynamic viewpoint, for a positive-energy system which interacts gravitationally with a negative-energy system, is whether negative thermal energy\index{negative thermal energy!gain or loss} is actually gained or lost by the negative-energy system and not whether the sign of this energy is positive or negative. The rule that emerges is that when negative heat\index{negative heat} is lost by a negative-energy system in contact with a positive-energy system, it is gained as positive heat\index{positive heat} by the positive-energy system, while when heat is lost by a positive-energy system in the same situation, it is gained as negative heat by the negative-energy system. One must keep in mind, though, that this conclusion is valid only when the strength of gravitational fields can be neglected and gravitational entropy\index{gravitational entropy} is not involved (as I will explain in section \ref{sec:3.10}).

Once again, the conventional expectation can be seen to arise from a misconception. You should take note, however, that I'm not just trying to debunk myths here. The opposite conclusion, that a low temperature gas made of positive-energy particles would be cooled even further upon contact with heat from a negative-energy gas, regardless of the magnitude of the temperature of this negative-energy gas, and the above discussed assumption that the mass of a positive-energy black hole could be reduced through the absorption of negative-energy matter, would constitute serious problems for a gravitational theory integrating the concept of negative-energy matter. There are very strong motives behind my desire to demonstrate that the possibility of such entropy-decreasing processes\index{entropy-decreasing processes} can be rejected and they are actually related to those which one might raise against the above discussed possibility of causality-violating processes\index{causality-violating process}. I will explain what is the profound significance of the results discussed here in the multiple sections of chapter \ref{chap:4} that deal with the problem of time irreversibility\index{time irreversibility!problem}.

\section{An axiomatic formulation\label{sec:2.13}}

Before I introduce the generalized gravitational field equations\index{generalized gravitational field equations} and complete this integration of negative-energy matter to classical gravitation theory\index{classical gravitation theory}, I would like to provide formal statements of each of the significant rules which were derived in the preceding sections of the present chapter, concerning the concept of negative-energy matter. Basically, there are eleven fundamental rules which clarify the situation regarding the nature and the behavior of negative-energy matter itself, as well as the behavior of positive-energy matter in the presence of negative-energy matter. Those rules actually constitute the axioms on which a generalized theory of gravitation\index{generalized gravitation theory!axioms} can be based that integrates a relational concept of negative-energy matter\index{negative-energy matter!relational concept} compatible with the principle of relativity\index{principle of relativity}. The axioms are legitimized by the fact that they have been shown to be necessary on the basis of both logical consistency\index{logical consistency} and agreement with experimental facts and thus we may appropriately refer to them as principles.

The first principle is the most fundamental and a recognition of its validity opens the way for a derivation of all the other results. The formal statement of this principle goes like this:
\begin{quote}
\textbf{Principle 1}: The distinction between a positive-action particle and a negative-action particle (propagating negative energy forward in time or positive energy backward in time) can only be defined by referring to the difference or the equality of the sign of action of one particle in comparison with that of another, so that this sign of action or energy has no absolute meaning.
\end{quote}
From a gravitational viewpoint, this principle is satisfied when positive-energy objects are submitted to mutual gravitational attraction among themselves (as we observe), while negative-energy objects (formed of negative-action particles) are also attracted to one another gravitationally and positive- and negative-energy objects repel one another (at least when the average matter densities can be neglected). Compliance with this rule means that for a positive-energy particle, a negative-energy particle should be physically equivalent to what a positive-energy particle is for a negative-energy particle.

Another rule applies only in the classical Newtonian context where mass is a significant concept, but given that it allows to derive the rules which must also be obeyed in a general-relativistic context it is necessary to mention it as a basic result. It simply amounts to recognize that
\begin{quote}
\textbf{Principle 2}: When mass is reversed from its conventional positive value (as a result of a reversal of the sign of action), both gravitational mass\index{gravitational mass} and inertial mass\index{inertial mass} are reversed and together become negative.
\end{quote}
This is actually equivalent to assume that there is indeed only one physical property to which we may refer to as being that of mass and that there cannot be any arbitrary distinction between gravitational and inertial mass. But this requirement does not have the consequences one usually expect it would have, as it actually means that negative-mass objects do not respond perversely to applied forces, as I already explained.

While principles 1 and 2 are for the most part theoretically motivated, the third principle is both theoretically and observationally motivated. Indeed, principle 3 arose as the unavoidable consequence of an analysis of the relationship between the attractive or repulsive nature of a field of interaction\index{interaction field!energy} and the sign of the energy classically contained in this field, but it is also a necessary requirement of the fact that we do not observe any negative-energy matter, despite the fact that the existence of such matter appears to be allowed from a theoretical viewpoint. The third principle consists in the following statement:
\begin{quote}
\textbf{Principle 3}: There can be no direct interactions, either gravitational or non-gravitational, mediated by the exchange of interaction bosons\index{interaction bosons}, between positive- and negative-action particles.
\end{quote}
Compliance with this principle means that positive-energy observers cannot directly observe negative-energy matter (and vice versa).

Another important result was discussed at length in section \ref{sec:2.7}, where its validity was shown to be unavoidable in the context where it must be assumed that there exist two opposite contributions of maximum magnitude to the energy of zero-point vacuum fluctuations\index{zero-point vacuum fluctuations!energy}. This result simply states that
\begin{quote}
\textbf{Principle 4}: The distribution of vacuum energy that surrounds a local void in an otherwise uniform distribution of positive vacuum energy\index{positive vacuum energy!uniform distribution} must give rise to uncompensated gravitational forces\index{uncompensated gravitational force} which attract positive-action particles away from the center of mass of the void, as if the void itself was producing a repulsive gravitational force.
\end{quote}
The effect it describes is the consequence of an alteration (caused by the presence of the local void in the uniform distribution of positive vacuum energy\index{positive vacuum energy!uniform distribution}) in the equilibrium of gravitational forces\index{equilibrium of gravitational forces!alteration} exerted on a positive-action particle as a result of its interaction with the portion of the surrounding uniform distribution of positive vacuum energy above that which is still present in the void. The importance of this principle becomes clear when it is recognized that principle 5 below applies.

The following principle is probably the most decisive after principle 1, given that it is the result that allows the whole concept of negative-energy matter to have a significance despite the validity of principle 3 and the absence of direct interactions between positive- and negative-action particles. It states that
\begin{quote}
\textbf{Principle 5}: The presence of negative matter energy in a given location is equivalent to the absence of an amount of energy of equal magnitude from the portion of zero-point vacuum fluctuations that provides a maximum positive contribution to the uniform distribution of vacuum energy\index{vacuum energy!maximum positive contribution}.
\end{quote}
And by principle 1 it would also follow that the presence of positive matter energy is equivalent to the absence of an amount of energy of equal magnitude from the portion of zero-point vacuum fluctuations that provides a maximum negative contribution to the uniform distribution of vacuum energy\index{vacuum energy!maximum negative contribution}. As I explained in section \ref{sec:2.7}, those equivalences constitute the particularity that allows opposite-energy objects to exert gravitational forces on one another despite the absence of direct interactions between them, simply because, according to principle 4, voids in a uniform distribution of positive vacuum energy do give rise to uncompensated gravitational forces\index{uncompensated gravitational force}, which arise from the gravitational interaction of positive-action particles with the surrounding positive vacuum energy.

Now, even in the context where we assume the existence of a symmetry under exchange\index{exchange symmetry!positive- and negative-energy matter} of positive- and negative-energy matter, principle 5 would require that it is, in fact, only the inhomogeneities (either overdensities or underdensities) present in the negative-energy matter distribution which can affect the gravitational dynamics of positive-energy matter, while it is only the inhomogeneities present in the positive-energy matter distribution which can affect the gravitational dynamics of negative-energy matter. This is because, as I explained in section \ref{sec:2.7}, the void in the positive portion of vacuum energy\index{void in positive vacuum energy!negative-energy matter} that is equivalent to a homogeneous distribution of negative-energy matter leaves no surrounding positive vacuum energy to produce an uncompensated gravitational attraction\index{uncompensated gravitational attraction} equivalent (according to principle 4) to a gravitational repulsion\index{gravitational repulsion} arising from the void itself and the same is true concerning a homogeneous distribution of positive-energy matter from the viewpoint of negative-energy matter. An additional principle thus emerges that expresses this limitation applying on principle 4. It amounts to assume that
\begin{quote}
\textbf{Principle 6}: The void of universal proportion\index{void in positive vacuum energy!universal proportion} in the uniform distribution of positive vacuum energy that is equivalent to a uniform distribution of negative-energy matter exerts no gravitational force on positive-energy matter and does not contribute to determine the curvature of space experienced by positive-energy observers.
\end{quote}
Of course, a similar limitation also applies, which expresses the absence of gravitational forces on negative-energy matter arising from the void of universal proportion\index{void in negative vacuum energy!universal proportion} in the uniform distribution of negative vacuum energy that is equivalent to a uniform positive-energy matter distribution.

A further particularity could be derived from the already stated principles, but I will provide it as an additional specific rule, because it is of particular importance for cosmology and it may not be obvious that it must apply in the context where principles 3 and 6 are assumed to constrain the interaction between positive- and negative-energy matter. This ordinance states that
\begin{quote}
\textbf{Principle 7}: Despite its negative sign the uniform, negative portion of vacuum energy\index{vacuum energy!negative portion} does exert a gravitational force on positive-energy matter.
\end{quote}
As I previously explained, this deduction (which would also apply to the uniform, positive portion of vacuum energy\index{vacuum energy!positive portion} from the viewpoint of negative-energy matter) follows from the fact that the restriction that applies on the interaction of positive- and negative-energy matter cannot prevent the negative portion of the uniform distribution of vacuum energy\index{negative vacuum energy!gravitational self-interaction} from exerting a gravitational force on positive-energy matter, when this matter is conceived as voids in this very same portion of the vacuum, just like voids in an expanding uniform distribution\index{void in expanding matter distribution} of negative-energy matter cannot be assumed to evolve independently from this matter distribution. In the present case, we are not dealing with an interaction between positive-energy matter and a void of universal proportion in the positive-energy portion of the vacuum, but rather with an interaction between voids in the negative portion of vacuum energy and that very same portion of vacuum energy, or with a gravitational interaction of the negative portion of vacuum energy with itself.

Now, even though it is appropriate to assume that the presence of matter is in fact equivalent to both missing energy and missing non-gravitational charges\index{missing non-gravitational charge} from zero-point vacuum fluctuations\index{zero-point vacuum fluctuations}, as I explained in section \ref{sec:2.7}, and while such a local absence of charges is equivalent to the presence of opposite-sign charges, this does not allow negative-energy matter to exert indirect non-gravitational forces\index{indirect non-gravitational forces} on positive-energy particles, because not only is it impossible to define the sign of energy of an interaction field\index{interaction field!sign of energy} between opposite-action particles, it is also impossible to define the strength of any non-gravitational force\index{non-gravitational force!strength} that would arise between them, and therefore it must be considered a consistency requirement to assume that
\begin{quote}
\textbf{Principle 8}: Even in the presence of inhomogeneities in the distribution of negative-energy matter charges\index{negative-energy matter charge distribution!inhomogeneities}, no \textit{indirect} forces\index{indirect forces!non-gravitational interactions} can be exerted by negative-energy matter on positive-action particles which would result from a non-gravitational interaction between those particles and the portion of zero-point vacuum fluctuations\index{zero-point vacuum fluctuations!maximum positive energy contribution} that provides a maximum positive contribution to the uniform distribution of vacuum energy.
\end{quote}
And the same conclusion would apply for negative-action particles in the presence of inhomogeneities in the distribution of positive-energy matter charges.

In a previous section I have explained that a consequence of principle 1, in the context where principle 2 (regarding the negativity of the inertial mass\index{inertial mass!negative} of a negative gravitational mass\index{negative gravitational mass}) is considered to apply, is that the usual assumption that reversing both gravitational and inertial mass would allow to maintain agreement with the equivalence principle\index{equivalence principle!usual concept} (as it is usually conceived) is wrong. Therefore, only an altered principle of equivalence\index{equivalence principle!alteration} between acceleration and a gravitational field remains valid. The additional condition applying on the equivalence principle\index{equivalence principle!additional condition} would be the following:
\begin{quote}
\textbf{Principle 9}: The equivalence of the effects of acceleration and a gravitational field does not apply merely for matter in the same location, but only for matter with the same sign of mass or energy in the same location.
\end{quote}
What remains true, in this context, is that the motion of objects in a gravitational field does not depend on any physical properties of those objects other than the sign of their mass\index{sign of mass} or energy and this is what will allow the essence of the current theory of gravitation to be retained, while accommodating a consistent, relational concept of negative-energy matter\index{negative-energy matter!relational concept}.

Another rule must be obeyed in the context where all matter is governed by principle 1 above and where the inertial properties\index{inertial properties!negative-energy object} of negative-energy objects are determined by principles 2 and 6 (which actually means that the inertial response\index{inertial response!negative-mass object} of a negative-mass or negative-energy object to a given force is the same as that which is experienced by positive-energy objects, as I explained before). This rule would not be required to apply if the conventional assumptions regarding the inertial response of negative-energy or negative-mass objects were valid. But given that I have argued that those assumptions are problematic and cannot be justified, then it seems that we have no choice but to take the following experimentally motivated principle into account.
\begin{quote}
\textbf{Principle 10}: As the negative contribution of a field of interaction to the energy of a bound physical system\index{bound systems!negative energy contribution} with overall positive energy cannot be independently and directly observed, only the diminished total energy of the bound system contributes to its (previously defined) absolute inertial mass\index{absolute inertial mass}.
\end{quote}
Again, this is also valid for bound physical systems\index{bound systems!negative energy} with overall negative energy, for which we may say that, when the positive contribution of a field of interaction to the energy of the bound system\index{bound systems!positive energy contribution} cannot be independently and directly observed, only the diminished (less negative) total energy of the bound system contributes to its absolute inertial mass.

It must be remarked that the validity of this rule does not mean that the opposite contribution to the total energy of a bound system by the field of interaction responsible for the mutual attraction of its component particles cannot be well-defined, only that if it cannot be isolated and independently measured then it also does not independently contribute to the inertial mass of the whole system.

One last constraint is observed to apply when negative energy states\index{negative energy states} are allowed to be occupied (can be propagated forward in time). While this rule is theoretically motivated, I originally derived it based on purely phenomenological arguments\index{phenomenological arguments}. It is the following:
\begin{quote}
\textbf{Principle 11}: A particle cannot reverse its direction of propagation in time\index{direction of propagation in time} on a continuous particle world-line\index{particle world-line!continuous} without also reversing its energy and equivalently, a particle cannot reverse its energy on a continuous particle world-line without also reversing its direction of propagation in time.
\end{quote}
Here by `negative energy' I mean negative energy\index{negative energy!direction of propagation in time} relative to the true (even though relationally defined) direction of propagation in time, as in the case of the positron as a negative-energy electron propagating its negative electric charge backward in time. This rule is equivalent to assume that it is impossible for pairs of opposite-action particles\index{opposite-action pairs!creation out of nothing} to be created out of nothing, or to annihilate to nothing\index{opposite-action pairs!annihilation to nothing}, which is an indirect consequence of principle 3 (regarding the necessary absence of direct interactions between opposite-action particles).

The eleven principles enunciated above embody the essence of the insights I have gained through an analysis of the problem of negative energy states\index{negative energy states!problem} in light of the requirement of relational definition\index{requirement of relational definition!sign of energy} of the physical properties of mass and energy signs. They will now be used to help derive a generalized formulation of the gravitational field equations\index{gravitational field equations!generalized formulation} that allows to describe the motion of particles with a given sign of energy in the gravitational field of an object with an opposite sign of energy.

\section{Generalized gravitational field equations\label{sec:2.14}}

I previously indicated that equations would be scarce in this report. But the point has now been reached where it is absolutely necessary to provide some level of quantitative detail regarding the manner by which the concept of negative energy that was developed in the preceding sections of the current chapter is to be integrated into a classical theory of gravitation\index{classical gravitation theory}. The objective I'm seeking here, though, is not to provide a complete treatise on the subject, but merely to introduce the modified gravitational field equations\index{gravitational field equations!modified} which constitute the core mathematical structure of the generalized theory\index{generalized gravitation theory!core mathematical structure} that emerges from the alternative set of axioms\index{alternative axioms} introduced in the preceding section.

The essential requirement that must be imposed on a formulation of the gravitational field equations is that the gravitational field\index{gravitational field!attractive or repulsive nature} attributable to a local source shall not be attractive or repulsive depending merely on the sign of energy of this source (regardless of the sign of energy of the particle that is submitted to it). This can be satisfied by assuming that the gravitational field\index{gravitational field!observer dependence} experienced by a negative-energy particle (which, in a general-relativistic theory, determines the metric properties of space and time\index{metric properties of spacetime} experienced by a negative-energy observer) is different from the one experienced by a positive-energy particle in the same conditions. For this purpose it is necessary to assume that the sign of energy\index{sign of energy!positive-definiteness} or action of the particle submitted to the gravitational field is an invariant property that may be chosen to be positive-definite, while it is the sign of energy of the source\index{sign of energy of source!observer dependence} that constitutes the variable, observer-dependent physical property.

When this convention is adopted, we can write a set of observer-dependent equations\index{gravitational field equations!observer dependence} in replacement of the original gravitational field equations. The motion of matter with a given energy sign is, therefore, determined by a gravitational field that may differ from that which is experienced by matter with an opposite energy sign, because the gravitational field varies as a function of both the energy sign of the sources and the energy sign of the particles submitted to it. This means that only the difference or the identity between the energy sign of the source and that of the matter submitted to the observer-dependent gravitational field\index{gravitational field!attractive or repulsive nature} determines the repulsive or attractive nature of the gravitational interaction.

In a Newtonian context this would mean that both the gravitational mass\index{gravitational mass!positive definiteness} and the inertial mass\index{inertial mass!positive definiteness} of the particle experiencing the field would be kept positive, while the equivalent gravitational field\index{equivalent gravitational field!invariant property} due to acceleration far from any local source would be an invariant property, as required for a relational description of the phenomenon of inertia\index{inertia!relational description}. The crucial assumption here is that, while the gravitational field\index{gravitational field!observer-dependent property} attributable to a local source is an observer-dependent physical property, the equivalent gravitational field associated with acceleration far from local sources is, for its part, an observer independent property.

In a general-relativistic context, on the other hand, the observer dependence of the gravitational field of a local source means that observers of opposite energy signs actually experience space and time in a different way. But despite the awkwardness of this possibility from the perspective of our conventional perception of spatial relationships, from a mathematical viewpoint this requirement does not constitute an insurmountable difficulty. We merely have to assume two spacetimes, related to one another by the fact that the same unique set of events is taking place in both of them, but which may nevertheless have distinct metric properties\index{metric properties of spacetime}, in the sense that the events which are taking place in the universe are separated by space and time intervals\index{space and time intervals!observer dependence} which are dependent on the energy sign of the observer.

As I mentioned before, the equations which will be proposed here merely constitute a generalization of the existing mathematical framework of relativity theory\index{relativity theory!mathematical framework} and we will, therefore, be in familiar territory. I'm, in effect, assuming that the reader already has a proper understanding of the current general-relativistic theory\index{general-relativistic theory!mathematical objects} of gravitation and of the physical significance of the various mathematical objects which are relevant to the conventional formulation of this theory. Also, given that attempts at formulating a relativistic theory of gravitation\index{gravitation!relativistic theory} that would allow for the existence of observer-dependent space curvature\index{space curvature!observer dependence} were the subject of earlier publications by various authors and since it would be pointless to simply reproduce what has already been discussed elsewhere, I will here concentrate on describing the essential, distinctive mathematical features unique to the theory I'm proposing. This choice is appropriate, despite the fact that the approach I favor involves several distinctive aspects, because the most general features of the kind of framework involved are not dependent on the specific assumptions of the model considered.

But keep in mind that even the most suitable of the currently available mathematical frameworks which were developed in an attempt to integrate the concept of negative-energy matter to physical theory still involve theoretical constructs and assumptions which I consider inappropriate for the formulation of a fully consistent classical theory of gravitation\index{classical gravitation theory} compatible with the requirement of relational definition\index{requirement of relational definition!physical attributes} of physical attributes and therefore only the general structure provided by those developments must be retained. I will here provide an interpretation of such bi-metric theories\index{bi-metric theories!interpretation} that is different from those which were tentatively proposed by the few authors that preceded me and this will have significant consequences which will be reflected in the fact that the final equations\index{final equations} at which I have arrived are actually distinct from those which had been proposed until now.

To my knowledge, the first observer-dependent gravitational field equations\index{gravitational field equations!observer dependence} which were proposed \cite{Petit-1} in order to accommodate the concept of negative-energy matter simply amounted to allow for negative contributions to the stress-energy tensor of matter\index{stress-energy tensor of matter!negative contributions}, while tentatively (but unsatisfactorily) trying to conform to the requirement of symmetry under exchange\index{exchange symmetry!positive and negative energy} of positive and negative energy signs. But even in more recent publications, no justifications were ever provided in support of the axioms on which are implicitly based the emerging frameworks and the only experimental consequences which were derived from those developments actually disagreed with observations, in particular those which allow to determine the rate of expansion\index{expansion rate!primordial universe} of space in the primordial universe.

Meaningful equations were, nevertheless, proposed, which happened to be compatible with the simplest of the conditions I have identified above as characterizing a consistent theory of negative-energy matter\index{negative-energy matter!consistent theory}. Those equations were a step forward in deriving a quantitative model for the gravitational dynamics\index{gravitational dynamics!positive- and negative-energy matter} of positive- and negative-energy matter, even if they failed to provide a totally appropriate framework and had to be assumed to apply only under particular circumstances, as they were clearly inappropriate to describe the early phases of cosmic evolution\index{early phases of cosmic evolution}. In any case, the equations which were initially proposed were of the following form:
\begin{eqnarray}
R_{\mu\nu}-\frac{1}{2}g_{\mu\nu}R=\frac{8\pi G}{c^4} (T_{\mu\nu}-T^-_{\mu\nu})\label{eq:2.1}\\
R^-_{\mu\nu}-\frac{1}{2}g_{\mu\nu}R^-=\frac{8\pi G}{c^4} (T^-_{\mu\nu}-T_{\mu\nu})\label{eq:2.2}
\end{eqnarray}
Here and in what follows $G$ is Newton's constant, $c$ is the speed of light in a vacuum, and the Greek indexes $\mu$ and $\nu$ run over the four general coordinate system labels (assuming a metric with diagonal elements $+1$, $-1$, $-1$, $-1$ in an inertial coordinate system\index{inertial coordinate system}).

In those equations, the usual notation is used for the curvature tensors\index{curvature tensors} $R_{\mu\nu}$ and $R$ experienced by positive-energy observers and for the stress-energy tensor\index{stress-energy tensors} $T_{\mu\nu}$ of what we conventionally consider to be positive-energy matter, as measured by a positive-energy observer, while $-T^-_{\mu\nu}$ is the stress-energy tensor of what we usually consider to be negative-energy matter, as determined by a positive-energy observer. The curvature tensors\index{curvature tensors!negative-energy observers} experienced by negative-energy observers are for their part denoted $R^-_{\mu\nu}$ and $R^-$, while the stress-energy tensor of what we would conventionally consider to be negative-energy matter, as measured by a negative-energy observer, is here denoted $T^-_{\mu\nu}$ and of course $-T_{\mu\nu}$ is the stress-energy tensor of what we usually consider to be positive-energy matter, as determined by a negative-energy observer (who would measure a negative contribution by positive-energy matter to the density of matter energy).

The first of those two equations can be used to determine the geodesics\index{geodesics} followed by positive-energy particles, while the second determines the geodesics followed by negative-energy particles. Here, all stress-energy tensors\index{stress-energy tensors!positive-definite energy densities} would correspond with positive-definite energy densities if it was not for the negative sign in front of the second stress-energy tensor on the right-hand side of each equation, which allows for a negative contribution to the total stress-energy tensor of matter\index{stress-energy tensor of matter!negative contributions} that is dependent on the particular measure of the sign of energy associated with one or the other type of observer. The negative sign of stress-energies can thus be attributed alternatively to what we would usually consider to be negative-energy matter and to what we usually consider to be positive-energy matter.

This means that what appears to be negative-energy matter to a conventional positive-energy observer would really be positive-energy matter for an observer we would normally consider to be a negative-energy observer, while what appears to be positive-energy matter to a positive-energy observer would really be negative-energy matter for an observer usually considered to be made of negative-energy matter. Therefore, all energy signs\index{sign of energy!observer dependence} must now be assumed to depend on the energy sign of the observer\index{observer!positive-definite energy}, which is itself assumed positive as a matter of convention. This means that we adopt the viewpoint I previously identified as being equivalent to assuming an observer-dependent sign of mass\index{sign of mass!observer dependence} for the local sources of gravitational field\index{gravitational field!local sources} and according to which it is the gravitational field itself (represented here by the curvature tensors\index{curvature tensors}) which actually varies, while the sign of mass\index{sign of mass!observer} (replaced here by the sign of energy\index{sign of energy!observer}) of the observer which experiences that gravitational field is to be considered positive-definite. This is certainly appropriate given that it gives rise to equations of the simplest form.

It is because there are two different measures for the gravitational field\index{gravitational field!two different measures}, associated with the two different ways by which the positive and negative contributions to the total energy of matter can be attributed, that there are two different equations for the gravitational field\index{gravitational field!two different equations}, instead of the single one that is usually considered. Otherwise, however, those equations are fairly conventional and were certainly the most straightforward that one could derive for a bi-metric theory\index{bi-metric theory}, as they were the closest to Einstein's own equation\index{Einstein's equation} that one could propose.

The fact that, in the context of those equations, the sign of energy\index{sign of energy!observer dependence} contributed by a given mass must now be assumed to depend on the sign of energy which we would normally attribute to the observer determining the associated gravitational field has important consequences. Indeed, if variations in the gravitational field (which is represented by the curvature tensors\index{curvature tensors}) are to compensate variations in the stress-energy of matter\index{stress-energy!of matter}, as the general covariance\index{general covariance} of the equations require, then it means that the gravitational field\index{gravitational field!observer dependence} attributed to the presence of some matter can actually be either attractive or repulsive depending on the sign of energy of the observer that measures the energy of this matter. This is certainly appropriate from the viewpoint of the principles identified in the preceding section. But given the insights I had already obtained when I first learned about the exact form of the equations that can be used to articulate those requirements, it nevertheless appeared to me that what the available framework provided was, at best, an incomplete formulation of the gravitational field equations\index{gravitational field equations!incomplete formulation} to associate with a theory of negative-energy matter.

To try to address those shortcomings I thus proposed, in the first version (released in early 2006) of
 what became the abridged edition of this report \cite{Lindner-1},
 the following equations which allowed to express the particularities of the indirect gravitational interaction\index{indirect gravitational interactions} of positive- and negative-energy mater that I had come to consider as unavoidable:
\begin{eqnarray}
R^+_{\mu\nu}-\frac{1}{2}g_{\mu\nu}R^+=\frac{8\pi G}{c^4} T^+_{\mu\nu}\label{eq:2.3}\\
R^-_{\mu\nu}-\frac{1}{2}g_{\mu\nu}R^-=\frac{8\pi G}{c^4} T^-_{\mu\nu}\label{eq:2.4}
\end{eqnarray}
Here $R^+_{\mu\nu}$ and $R^+$ are simply the curvature tensors\index{curvature tensors} experienced by positive-energy observers, while $R^-_{\mu\nu}$ and $R^-$ are the curvature tensors experienced by negative-energy observers. But the stress-energy tensors\index{stress-energy tensors} figuring in the equations I proposed are actually different from those entering the previously mentioned set of equations, despite the similar notation I adopted here, because the $T^+_{\mu\nu}$ tensor encompasses all contributions to the energy and momentum experienced by positive-energy observers, while the $T^-_{\mu\nu}$ tensor encompasses all contributions to the energy and momentum experienced by negative-energy observers and I did assume contributions to those stress-energy tensors which were different from those which had previously been considered.

Thus, when written in a more explicit form, with all the components actually entering the stress-energy tensors on the right-hand side, the equations I had proposed were the following:
\begin{eqnarray}
R^+_{\mu\nu}-\frac{1}{2}g_{\mu\nu}R^+=\frac{8\pi G}{c^4} (T^+_{\mu\nu}+\check{T}^-_{\mu\nu}-\hat{T}^-_{\mu\nu})\label{eq:2.5}\\
R^-_{\mu\nu}-\frac{1}{2}g_{\mu\nu}R^-=\frac{8\pi G}{c^4} (T^-_{\mu\nu}+\check{T}^+_{\mu\nu}-\hat{T}^+_{\mu\nu})\label{eq:2.6}
\end{eqnarray}
In this notation $T^+_{\mu\nu}$ is the stress-energy tensor of what is usually considered to be positive-energy matter, as measured by a positive-energy observer, while $\check{T}^-_{\mu\nu}$ is the stress-energy tensor associated with the (positive) measure of energy of negative-energy matter\index{negative-energy matter!average cosmic density} below its average cosmic density (as determined by a positive-energy observer) and $-\hat{T}^-_{\mu\nu}$ is the stress-energy tensor associated with the (negative) measure of energy of negative-energy matter above its average cosmic density (as determined by a positive-energy observer). Similarly, $T^-_{\mu\nu}$ is the stress-energy tensor of what we would usually consider to be negative-energy matter, as measured by a negative-energy observer (and which actually provides a positive contribution to the energy of matter) while $\check{T}^+_{\mu\nu}$ is the stress-energy tensor associated with the (positive) measure of energy of positive-energy matter\index{positive-energy matter!average cosmic density} below its average cosmic density (as determined by a negative-energy observer) and $-\hat{T}^+_{\mu\nu}$ is the stress-energy tensor associated with the (negative) measure of energy of positive-energy matter above its average cosmic density (as determined by a negative-energy observer).

This alternative formulation of the generalized gravitational field equations\index{generalized gravitational field equations} allowed me to take into account the fact that there are two distinct categories of contributions to the total energy density experienced by positive-energy observers, one positive-definite for all densities of positive-energy matter and one that can be either positive or negative depending on the value of energy density of negative-energy matter relative, not to the zero-energy ground state\index{zero-energy ground state}, but to the density of this negative-energy matter averaged over the entire volume of the observable universe. Basically, what this means is that the energy measures of the second category of contributions experienced by a positive-energy observer are shifted from the conventional zero level of energy\index{zero level of energy!conventional} toward a lower (more negative) energy level below which energies are negative and above which energies are positive, up to a maximum value which is reached when no negative-energy matter is present at all in the considered location.

This redefinition of the measures of energy\index{energy measures!redefinition} associated with what we conventionally assume to be negative-energy matter simply amounts to subtract the true, average density of negative matter energy\index{negative matter energy!average density} (add the absolute value of this density) from every measure of its energy density that contributes to determine the gravitational field experienced by what we conventionally assume to be positive-energy matter, that is, the gravitational field measured by positive-energy observers. But at the present epoch, the required shift in the origin of the measures of energy\index{energy measures!shift in origin}, for matter with an energy sign opposite that of the observer, becomes significant only on a very large scale, because in the case of stars and planets it doesn't make much difference if we instead simply consider the true density of positive- or negative-action matter, given that the typical densities which are involved are much larger than the mean cosmic energy density\index{mean cosmic energy density}, which can thus be neglected.

The refinement discussed here is justified by the fact that, in the context where negative-energy matter is understood to consist of voids in the positive-energy portion of the vacuum, it must be assumed that the uniform portion of the distribution of negative-energy matter exerts no gravitational force on positive-energy matter (because no surrounding positive vacuum energy is present to produce an uncompensated gravitational attraction\index{uncompensated gravitational attraction}), which requires considering the contributions of negative-energy matter to the stress-energy tensor\index{stress-energy tensors!negative-energy matter contributions} experienced by positive-energy observers as being significant merely relative to the average density of negative matter energy (and therefore to actually be positive in the presence of underdensities in the otherwise uniform distribution of negative-energy matter), as I explained in section \ref{sec:2.7}.

The equations I initially proposed were also a reflection of the fact that a similar requirement exists for the contributions of positive-energy matter to the total stress-energy tensor\index{stress-energy tensors!positive-energy matter contributions} experienced by negative-energy observers. But, still, I did not find the set of equations I had proposed completely satisfactory. I thought that the right solution should bring about a simplification of the gravitational field equations\index{gravitational field equations!simplification} (particularly those involving a non-zero cosmological constant\index{cosmological constant!non-zero value}), while, visibly, the equations I had derived were not even as simple and elegant as the equations originally proposed by Einstein\index{Einstein, Albert}, despite the fact that, in their compact form, they were similar.

As I now understand, however, the equations I had proposed also fell short of meeting a certain mathematical requirement which I have come to appreciate as being essential to a consistent bi-metric theory\index{bi-metric theory} of gravitation of the kind I sought to develop. This became clear when a paper \cite{Hossenfelder-1} was published (whose existence came to my attention as a result of the fact that its author cited the original version of the
 abridged edition of this report)
 in which new equations were proposed, which introduced a further refinement to bi-metric theories, by not assuming that there is a unique predefined relationship between the metric properties of spacetime\index{metric properties of spacetime} experienced by positive-energy observers and those experienced by negative-energy observers in the same situation (even though, for some reason, the author of this paper preferred not to consider that the matter contributing a negative measure to the total stress-energy tensor experienced by positive-energy matter actually constitutes negative-energy matter). As a consequence of this revised assumption, additional variables had to be considered that affected the contribution of negative-energy matter to the total stress-energy tensor experienced by positive-energy observers, or the contribution of what we usually consider to be positive-energy matter to the total stress-energy tensor experienced by negative-energy observers.

The equations proposed were the following, in which the additional factors are written in their explicit form, using my notation, and the quantities are now expressed in units where $c=1$ and $G=1/8\pi$:
\begin{eqnarray}
R^+_{\mu\nu}-\frac{1}{2}g_{\mu\nu}R^+=T^+_{\mu\nu}-\sqrt{\frac{g^{-+}}{g^{++}}}a_{\nu}^{\;\underline{\nu}}a_{\mu}^{\;\underline{\mu}} T^-_{\underline{\nu\mu}}\label{eq:2.7}\\
R^-_{\nu\mu}-\frac{1}{2}g_{\nu\mu}R^-=T^-_{\underline{\nu\mu}}-\sqrt{\frac{g^{+-}}{g^{--}}}a^{\mu}_{\;\underline{\mu}}a^{\nu}_{\;\underline{\nu}} T^+_{\mu\nu}\label{eq:2.8}
\end{eqnarray}
In this notation, tensors which refer to positive or negative stress-energies, as determined from the viewpoint of positive-energy observers, are given a plus or minus upper right index, respectively. Tensors which refer to the measures of spacetime curvature\index{spacetime curvature}, as observed by positive-energy observers, are also given a plus upper right index, while tensors which refer to the same kind of measures, as observed by negative-energy observers, are given a minus upper right index. Also, when the distinct, ordinary or underlined Greek letter indexes used in the original paper are not explicitly present to show the nature of the tensor considered, I simply add another plus or minus index to the right of that which already characterizes this tensor to define it as an object associated with physical properties as they are experienced by positive- or negative-energy observers, respectively, and associated with their own specific metric. For all such tensors, therefore, the first plus or minus index refers to the matter or gravitational field that is observed while the second plus or minus index (to the right) refers to the matter that is observing.

In those equations, the decisive additional factors are the determinants of what the author calls the pull-overs\index{pull-overs}, which are the maps $g^-_{\nu\mu}$ and $g^+_{\underline{\mu\nu}}$ (originally denoted $h_{\nu\mu}$ and $g_{\underline{\mu\nu}}$), which we may also write as $\bm{g}^{-+}$ and $\bm{g}^{+-}$ in tensor form. Those determinants are written here as $g^{-+}=\det(g^-_{\nu\mu})$ and $g^{+-}=\det(g^+_{\underline{\mu\nu}})$, while $g^{++}=\det(g^+_{\nu\mu})$ is the determinant of the usual metric tensor\index{metric tensor} related to properties of positive-energy matter as observed by positive-energy observers and $g^{--}=\det(g^-_{\underline{\mu\nu}})$ is the determinant of the metric tensor related to properties of negative-energy matter as observed by negative-energy observers (the map $\bm{a}$ is simply used as a means to transform the metric $\bm{g}^{++}$ into the $\bm{g}^{-+}$ pull-over or the metric $\bm{g}^{--}$ into the $\bm{g}^{+-}$ pull-over). It is clear, therefore, that the pull-over $\bm{g}^{-+}$ is the map which allows to describe the metric properties\index{metric properties of spacetime} obeyed by negative-energy matter as they are observed by positive-energy observers, while the pull-over $\bm{g}^{+-}$ is the map which allows to describe the metric properties obeyed by positive-energy matter as they are observed by negative-energy observers (which justifies my notation).

To better illustrate the relationships involved we may rewrite those equations as:
\begin{eqnarray}
R^+_{\mu\nu}-\frac{1}{2}g_{\mu\nu}R^+=T^+_{\mu\nu}-\gamma^{-+}\sqrt{\frac{g^{--}}{g^{++}}}a_{\nu}^{\;\underline{\nu}}a_{\mu}^{\;\underline{\mu}} T^-_{\underline{\nu\mu}}\label{eq:2.9}\\
R^-_{\nu\mu}-\frac{1}{2}g_{\nu\mu}R^-=T^-_{\underline{\nu\mu}}-\gamma^{+-}\sqrt{\frac{g^{++}}{g^{--}}}a^{\mu}_{\;\underline{\mu}}a^{\nu}_{\;\underline{\nu}} T^+_{\mu\nu}\label{eq:2.10}
\end{eqnarray}
where $\gamma^{-+}$ is the absolute value of the determinant of the previously considered map of the metric properties of spacetime experienced by negative-energy matter as negative-energy observers measure them, to the metric properties of spacetime experienced by negative-energy matter as positive-energy observers measure them, and vice versa for $\gamma^{+-}$. We can then rewrite those gravitational field equations\index{gravitational field equations!compact tensor form} in compact tensor form by making use of those \textit{metric conversion factors}\index{metric conversion factors} as:
\begin{eqnarray}
\bm{G}^+=\bm{T}^{++}-\gamma^{-+}\bm{T}^{-+}\label{eq:2.11}\\
\bm{G}^-=\bm{T}^{--}-\gamma^{+-}\bm{T}^{+-}\label{eq:2.12}
\end{eqnarray}
where $\bm{G}^+$ is the Einstein tensor\index{Einstein tensor} $G^+_{\mu\nu}=R^+_{\mu\nu}-\frac{1}{2}g_{\mu\nu}R^+$ related to positive-energy observers, $\bm{G}^-$ is the similar Einstein tensor related to negative-energy observers, $\bm{T}^{++}$ is the stress-energy tensor\index{stress-energy tensors} of positive-energy matter as measured by positive-energy observers, $-\gamma^{-+}\bm{T}^{-+}$ is the stress-energy tensor of negative-energy matter as measured by positive-energy observers, $\bm{T}^{--}$ is the stress-energy tensor of negative-energy matter as measured by negative-energy observers and finally $-\gamma^{+-}\bm{T}^{+-}$ is the stress-energy tensor of positive-energy matter as measured by negative-energy observers.

As is apparent, however, the proposed equations were still of the conventional kind, in the sense that they did not allow to take into account the requirement that negative-energy matter be experienced as voids in the positive-energy portion of the vacuum\index{void in positive vacuum energy}, or the fact that a uniform distribution of negative-energy matter exerts no gravitational force on positive-energy particles (and vice versa for a uniform distribution of positive-energy matter, from the viewpoint of negative-energy particles). The complexity of those equations and their lack of symmetry under exchange\index{exchange symmetry!positive- and negative-energy matter} of positive- and negative-energy matter can be made more apparent by explicitly adding a term for the observed positive value of (average) vacuum energy density\index{average vacuum energy density}:
\begin{eqnarray}
\bm{G}^+=\bm{T}^{++}+\bm{T}_{\Lambda}^{++}-\gamma^{-+}\bm{T}^{-+}\label{eq:2.13}\\
\bm{G}^-=\bm{T}^{--}-\bm{T}_{\Lambda}^{+-}-\gamma^{+-}\bm{T}^{+-}\label{eq:2.14}
\end{eqnarray}
This cosmological term\index{cosmological term} $\bm{T}_{\Lambda}^{++}=\Lambda\bm{g}^{++}$ would provide the value of stress-energy associated with the positive density of vacuum energy $\rho_{\Lambda}^{++}=\Lambda$ measured on a global scale by a positive-energy observer (with $\Lambda$ as the positive cosmological constant\index{cosmological constant} experienced by such an observer), while an additional cosmological term $-\bm{T}_{\Lambda}^{+-}=-\Lambda\bm{g}^{--}$ would provide the value of stress-energy associated with the same vacuum energy, as measured by a negative-energy observer. The density of vacuum energy measured by a negative-energy observer must be the opposite of that measured by a positive-energy observer if the sign of energy\index{sign of energy!observer dependence} is to remain an observer-dependent physical property (which justifies the presence of a minus sign in front of the $\bm{T}_{\Lambda}^{+-}$ tensor that enters the gravitational field equations\index{gravitational field equations} for negative-energy observers).

But given that we are indeed dealing with vacuum energy, it would seem inappropriate to assign to this tensor the same metric conversion factor\index{metric conversion factors} $\gamma^{+-}$ as apply to measures of positive-energy matter density performed by negative-energy observers, even if the sum of all positive and negative contributions to the energy of the vacuum is a positive number, because, in principle, the magnitude of all such contributions is the same for positive- and negative-energy observers on the cosmological scale. Anyhow, it is apparent that once all relevant contributions to the stress-energy tensors\index{stress-energy tensors} are considered, the symmetry of the original equations is lost, as their form becomes dependent on the actual sign of the average energy density of vacuum fluctuations. To me at least, it is obvious that those equations cannot be considered to embody a simplification of Einstein's theory\index{Einstein's theory!simplification} that could be considered a substantial improvement over the original equations, in the presence of negative-energy matter.

In order that such a formulation of bi-metric theory\index{bi-metric theory} be allowed to at least meet the requirements concerning the contribution of negative-energy matter to the measure of stress-energy experienced by a positive-energy observer which I had already identified and which were not taken into account by the preceding authors, it is necessary, first of all, to replace the usual stress-energy tensors associated with the measures of energy of negative- and positive-energy matter made by observers of opposite energy sign with the following \textit{irregular stress-energy tensors}\index{irregular stress-energy tensors}, which provide the observed densities of energy of negative- and positive-energy matter relative to their average cosmic densities:
\begin{eqnarray}
-\gamma^{-+}\bm{\widetilde{T}}^{-+}=-\gamma^{-+}(\bm{T}^{-+}-\bm{\bar{T}}^{-+})\label{eq:2.15}\\
-\gamma^{+-}\bm{\widetilde{T}}^{+-}=-\gamma^{+-}(\bm{T}^{+-}-\bm{\bar{T}}^{+-})\label{eq:2.16}
\end{eqnarray}
where $-\gamma^{-+}\bm{T}^{-+}$ and $-\gamma^{+-}\bm{T}^{+-}$ are the usual measures of stress-energy of negative- and positive-energy matter experienced by observers of opposite energy signs (relative to the conventional zero level of energy\index{zero-energy level}) and $-\gamma^{-+}\bm{\bar{T}}^{-+}$ and $-\gamma^{+-}\bm{\bar{T}}^{+-}$ are the \textit{average} values of stress-energy\index{stress-energy!average values} of negative- and positive-energy matter determined by observers with an opposite energy sign, based on their own measures of spatial volume.

In such a context, it appears that negative-energy matter would contribute negatively to the total measure of stress-energy experienced by a positive-energy observer only when the magnitude of its local energy density is larger than the magnitude of its average energy density. Otherwise negative-energy matter would actually contribute positively to the total measure of stress-energy experienced by a positive-energy observer, up to a maximum level which is fixed by the average density of negative-energy matter that is measured by such a positive-energy observer. The same remark would apply for the contribution of what is usually considered to be positive-energy matter to the total measure of stress-energy experienced by a negative-energy observer, which would be opposite the energy contribution of negative-energy matter only when the magnitude of the local density of positive-energy matter is larger than the magnitude of its average cosmic density.

It must be noted, however, that even though positive contributions to the energy density measured by positive-energy observers may occur which would be attributable to the presence of underdensities in the negative-energy matter distribution, we must nevertheless apply the metric conversion factor\index{metric conversion factors} $\gamma^{-+}$ to such energy measures, because they still relate to measurements regarding the density of negative-energy matter, which are subject to the same mapping relationships as apply to other (truly negative) measures of negative-energy matter density determined by a positive-energy observer. Of course, this is also true concerning below average measures of the energy density of what we would usually consider to be positive-energy matter made by negative-energy observers.

Even when the second category of contributions to the energy density of matter is of the same sign as the energy of the matter experiencing the gravitational field, it is still undetermined to the same extent as negative contributions, because what is unknown (due to the impossibility to directly compare the measures of distances experienced by positive- and negative-energy observers) is the exact true density of negative-energy matter (in comparison with that of positive-energy matter) and this indefiniteness also affects the positive value of such contributions. Therefore, positive energy contributions arising from underdensities of negative-energy matter are contained in the same irregular stress-energy tensor\index{irregular stress-energy tensors} as negative energy contributions.

A more appropriate set of gravitational field equations\index{gravitational field equations!more appropriate set} would, therefore, take into account the shifted origin of the measures of stress-energy\index{stress-energy!shifted origin} related to positive- and negative-energy matter as they are experienced by observers of opposite energy signs:
\begin{eqnarray}
\bm{G}^+=\bm{T}^{++}+\bm{T}_{\Lambda}^{++}-\gamma^{-+}\bm{\widetilde{T}}^{-+}\label{eq:2.17}\\
\bm{G}^-=\bm{T}^{--}-\bm{T}_{\Lambda}^{+-}-\gamma^{+-}\bm{\widetilde{T}}^{+-}\label{eq:2.18}
\end{eqnarray}
But clearly, for what regards simplicity, we appear to be no better off than with the previous set of equations. Something is still missing from those equations. At this point I suggest that we take a bold step forward and instead of trying to derive the gravitational field equations from a variational principle\index{variational principle}, as is usually done, we rather follow Einstein's\index{Einstein, Albert} way and simply guess what the final form of the equations should be that would generalize the set of equations (\ref{eq:2.17}) and (\ref{eq:2.18}) I have just proposed, which would otherwise constitute the most accurate description of the gravitational dynamics\index{gravitational dynamics!positive- and negative-energy matter} of positive- and negative-energy matter.

As I have been able to understand, the crucial step in this process consists in reconsidering the meaning of the cosmological terms\index{cosmological term} whose contributions I had long suspected were inappropriately attributed, in the context of bi-metric theories\index{bi-metric theory}. Indeed, I always thought that the cosmological term should arise from an imbalance between some positive and some negative contributions to the energy budget, while in the current set of equations it occurs only as an additional term, which must merely be attributed the appropriate energy sign depending on whether it is observed by a positive-energy observer or a negative-energy observer, which I do not find satisfactory.

It is only when I recognized the profound significance of my description of positive- and negative-energy matter as voids in their respective opposite-energy portions of the vacuum, that I was able to achieve the breakthrough that allowed me to guess what the appropriate generalized gravitational field equations\index{generalized gravitational field equations} are that allow the concept of negative-energy matter to be integrated into a general-relativistic framework\index{general-relativistic framework} in a way that actually simplifies Einstein's\index{Einstein's theory} theory rather than further complicate things. What I realized, basically, is that if the results of the above described analysis is right, then all energy is vacuum energy, either present or missing.

An additional insight was then necessary, which consisted in recognizing that the magnitude of the natural, positive and negative values of vacuum energy density relative to which are measured the missing energies which are equivalent to the presence of negative- and positive-energy matter (respectively) is actually provided by the Planck energy\index{Planck energy}. What must be understood is that when we remove energy from the vacuum, we decrease its energy density from a maximum (positive or negative) value which is fluctuating quantum mechanically (upon measurement) in just the same measure as does the energy of matter itself. Therefore, if the presence of negative-energy matter is to be considered as equivalent to the presence of a void in the positive portion of vacuum energy\index{void in positive vacuum energy}, then locally we should observe a value of fluctuating vacuum energy density\index{vacuum energy density!natural maximum value} that would be decreased from its natural maximum value in just the same measure as that of the energy of the matter that is present.

But given that the level of fluctuation of vacuum energy involved would be as large as the void considered is small, it is possible to assume that there is an exact correspondence between the missing vacuum energy\index{missing vacuum energy} and the energy of the matter that we ordinarily expect to be present, which is known to be fluctuating (even if it is actually the measure of momentum that is involved) in proportion with the level of precision of the measurement of spatial position to which the matter is submitted. The natural level of energy\index{natural energy level!Planck energy} involved would thus correspond to that which is known to be associated with the highest possible magnitude of energy fluctuation, which is the Planck energy\footnote{
The validity of this assumption could be the subject of controversy, but given that the most advanced and least speculative theoretical developments toward a theory of quantum gravitation\index{quantum gravitation theories|nn} indicate that this is an appropriate and unavoidable constraint, I will nevertheless consider it to be universally valid. However, even if the existence of such a limit to the energy associated with quantum fluctuations was to be found irrelevant, there is no \textit{a priori} reason why the following results would have to be considered invalid. I believe that the situation we have here is once again similar to that which existed at the turn of the twentieth century concerning the hypothesis of the existence of atoms\index{hypothesis of existence of atoms|nn}, which was often rejected on the basis of an absence of direct observational evidence, despite the fact that this assumption had actually become unavoidable theoretically.}.
 Therefore, any missing vacuum energy attributable to the presence of matter with an energy sign opposite that of the portion of vacuum energy in which it arises may be considered to actually be a local decrease over the maximum energy density determined by the Planck scale\index{Planck scale}.

Let me thus introduce the generalized gravitational field equations\index{generalized gravitational field equations} which allow to fulfill all the requirements I have identified as being essential aspects of a classical theory of gravitation\index{classical gravitation theory} that solves the problem of negative energy states\index{negative energy states!problem}. The formula, in all its beauty and simplicity, is the following:
\begin{equation}\label{eq:2.19}
\bm{G}^{\pm}=\bm{V}^{\pm}
\end{equation}
where $\bm{G}^{\pm}$ is the Einstein tensor\index{Einstein tensor} associated with the metric properties experienced by what we would usually consider to be positive- and negative-energy observers and $\bm{V}^{\pm}$ is the \textit{vacuum stress-energy tensor}\index{vacuum stress-energy tensor} associated with the measures of vacuum energy effected by those same positive- and negative-energy observers. The similarity with the compact form of Einstein's own equation\index{Einstein's equation} is very clear, but it is also somewhat misleading, as the right-hand side of the equation proposed here is a more general object than the stress-energy tensor\index{stress-energy tensors} of matter which appeared in the original theory. I will now define it with various levels of precision and generality.

If we first consider the significance of the equation for a positive-energy observer, we would obtain the following equation:
\begin{equation}\label{eq:2.20}
\bm{G}^+=\gamma^{-+}\bm{V}^{++}-\bm{V}^{-+}
\end{equation}
in which $\bm{G}^+$ is, again, the Einstein tensor associated with the gravitational field experienced by positive-energy observers, but now the vacuum stress-energy tensor is decomposed into its positive- and negative-energy portions $\gamma^{-+}\bm{V}^{++}$ and $-\bm{V}^{-+}$ as they would be measured by such positive-energy observers, either directly or based on the curvature of space\index{curvature of space} they produce. This is the most basic form of the proposed generalized gravitational field equations\index{generalized gravitational field equations!most basic form} for a positive-energy observer.

In accordance with what was explained above we would then obtain the next level of decomposition of the equations, in which the two opposite contributions to the energy of vacuum fluctuations determined by positive-energy observers are given their explicit form:
\begin{equation}\label{eq:2.21}
\bm{G}^+=(\gamma^{-+}\bm{V}_P^{++}-\gamma^{-+}\bm{T}^{-+})-(\bm{V}_P^{-+}-\bm{T}^{++})
\end{equation}
where $\gamma^{-+}\bm{V}_P^{++}$ and $-\bm{V}_P^{-+}$ are the \textit{natural vacuum-stress-energy tensors}\index{natural vacuum-stress-energy tensors} associated with the maximum, positive and negative contributions to the energy density of zero-point vacuum fluctuations\index{zero-point vacuum fluctuations!maximum contributions} set by the Planck scale\index{Planck scale} (as determined by positive-energy observers) and from which are subtracted the missing vacuum energies\index{missing vacuum energy} $\gamma^{-+}\bm{T}^{-+}$ and $\bm{T}^{++}$ which are equivalent to the presence of negative- and positive-energy matter, respectively. What justifies the attribution of the previously introduced metric conversion factor\index{metric conversion factors} $\gamma^{-+}$ to the positive measure of vacuum stress-energy\index{vacuum stress-energy} in equation (\ref{eq:2.20}) and therefore, also, to the maximum \textit{positive} contribution to the energy of zero-point vacuum fluctuations\index{zero-point vacuum fluctuations!maximum contribution} in equation (\ref{eq:2.21}) is precisely the fact that this is the portion of vacuum energy relative to which the negative measure of matter energy $-\gamma^{-+}\bm{T}^{-+}$ is determined and which we can therefore expect to be directly experienced (other than through the gravitational interaction) only by this negative-energy matter, even though it does exert an observer-dependent gravitational force\index{gravitational forces!observer dependence} on positive-energy matter as well.

Given that the previously introduced metric conversion factors are made necessary as a result of the absence of predetermined relationships between the metric properties of spacetime\index{metric properties of spacetime} experienced by negative-energy matter and those experienced by positive-energy matter, it is natural to assume that if the density of negative-energy matter itself cannot be directly observed by a positive-energy observer, then the positive measure of vacuum energy density relative to which this matter energy is defined cannot be directly determined either, because if this was not true, then by directly measuring the density of energy contained in this positive portion of vacuum energy\index{vacuum energy!positive portion}, a positive-energy observer could determine the density of negative-energy matter which is experienced by negative-energy observers. What must be understood is that the fact that this portion of vacuum energy density is positive should not be assumed to invalidate the conclusion that it cannot be directly experienced by positive-energy observers \textit{other than through the gravitational interaction}\footnote{
Yet this is not what I had originally proposed, because it once seemed to me (for reasons that will be explained in section \ref{sec:4.2}) that this hypothesis would be ruled-out from an observational (astronomical) perspective. But I have since realized that there are very good reasons to believe that this is not the case, after all, and that consistency requires that it is the portion of zero-point vacuum fluctuations\index{zero-point vacuum fluctuations!maximum positive energy contribution|nn} that gives rise to a maximum positive contribution to the density of vacuum energy that cannot be directly observed by a positive-energy observer, even though it does contribute to determine the gravitational field experienced by positive-energy matter (while the portion of vacuum fluctuations\index{zero-point vacuum fluctuations!maximum negative energy contribution|nn} that gives rise to a maximum negative contribution is the one that cannot be directly observed by a negative-energy observer, even though it does contribute to determine the gravitational field experienced by negative-energy matter).}.

The preceding equation can then be rewritten in the following form, by making use of the previously defined, irregular stress-energy tensor\index{irregular stress-energy tensors} $-\gamma^{-+}\bm{\widetilde{T}}^{-+}$ (equation (\ref{eq:2.15})) that provides the actual measure of stress-energy of negative-energy matter experienced by positive-energy observers, which are only affected by local \textit{variations} in the density of negative-energy matter:
\begin{equation}\label{eq:2.22}
\bm{G}^+=\bm{T}^{++}-\gamma^{-+}\bm{\widetilde{T}}^{-+}+(\gamma^{-+}\bm{V}_P^{++}-\bm{V}_P^{-+})
\end{equation}
This allows one to isolate a term, in the generalized gravitational field equations\index{generalized gravitational field equations}, that can be associated with pure vacuum energy\index{pure vacuum energy} and that would be provided by the following tensor:
\begin{equation}\label{eq:2.23}
\bm{T}_{V}^+=\gamma^{-+}\bm{V}_P^{++}-\bm{V}_P^{-+}
\end{equation}
where the positive index attributed to this \textit{vacuum-energy term}\index{vacuum-energy term} (associated with the energy that is present in the vacuum independently from the contribution of ordinary matter) now merely denotes the purely conventional energy sign of the observer experiencing it, without referring to an actual energy sign of the vacuum fluctuations themselves, which could in principle be either positive or negative (without affecting the form of the equations) and which is determined solely by the metric conversion factor\index{metric conversion factors} provided by the previously discussed map of the metric properties of spacetime\index{metric properties of spacetime} experienced by negative-energy observers onto those experienced by positive-energy observers.

Given the invariant nature of the maximum positive and negative contributions to the density of vacuum energy\index{vacuum energy!maximum positive and negative contributions} associated with the Planck scale\index{Planck scale}, for an observer having an energy sign opposite that of the contribution considered, the above equation means that a non-zero value of vacuum energy density\index{vacuum energy density!non-zero value} can only be measured by positive-energy observers when there exists a difference between the metric properties of space\index{metric properties of space} they experience and those which are experienced by negative-energy observers.

But as I will explain in section \ref{sec:4.3}, it must be assumed that positive-energy matter would remain unaffected by the average portion of a locally varying component of negative vacuum energy\index{negative vacuum energy!locally varying component}. Therefore, only a redefined measure of vacuum-energy would actually contribute to determine the gravitational field experienced by a positive-energy observer, which is given by the following \textit{irregular vacuum-energy term}\index{irregular vacuum-energy term}:
\begin{equation}\label{eq:2.24}
\bm{\widetilde{T}}_{V}^+=\bm{T}_{V}^{+}-(-\bm{\bar{T}}_{VM}^{-+})
\end{equation}
where $-\bm{\bar{T}}_{VM}^{-+}$ is the average value of stress-energy arising from all locally variable negative contributions to the energy of zero-point vacuum fluctuations\index{zero-point vacuum fluctuations!locally variable negative energy contributions}
 (which can be associated with the presence of dark matter),
 as measured by a positive-energy observer on a global scale, and which would not include the contribution provided by a negative cosmological constant\index{cosmological constant!negative} (as the uniform portion of a negative density of vacuum energy).

It is now possible to write the generalized gravitational field equations\index{generalized gravitational field equations!most explicit form} associated with positive-energy observers in their most explicit form as:
\begin{equation}\label{eq:2.25}
\bm{G}^+=\bm{T}^{++}-\gamma^{-+}\bm{\widetilde{T}}^{-+}+\bm{\widetilde{T}}_{V}^+
\end{equation}
The formal equivalence of this formula with equation (\ref{eq:2.17}) above, at which I had arrived on the basis of considerations of a physical nature, is quite clear. But while one may be tempted to deduce from this that the irregular vacuum-energy term $\bm{\widetilde{T}}_{V}^+$ is equivalent to the cosmological term\index{cosmological term} $\bm{T}_{\Lambda}^{++}$ which is present in the original version of the gravitational field equations\index{gravitational field equations!original version}, this would not be entirely appropriate, because contrarily to the cosmological term (associated with the cosmological constant\index{cosmological constant!uniform and invariant energy} $\Lambda$), which must by necessity provide a uniform and invariant energy contribution, the vacuum-energy term\index{vacuum-energy term!variable in space and time} can vary in space and incidentally also with time, given that it is determined by the locally variable, metric conversion factor\index{metric conversion factors} $\gamma^{-+}$.

Only the contribution associated with the uniformly distributed portion of vacuum energy\index{vacuum energy!uniformly distributed portion} contained in the irregular vacuum-energy term\index{irregular vacuum-energy term!average value at one particular time} at one particular time can be expected to be equivalent to the original cosmological term associated with the positive cosmological constant\index{cosmological constant!positive}. In sections \ref{sec:4.2} and \ref{sec:4.3} I will explain how one must interpret the variable nature of the vacuum-energy term and why it is still appropriate to consider that, in general, the density of vacuum energy does not vary with position, in the absence of local inhomogeneities in the positive- and negative-energy matter distributions. Anyhow, given that we know that on the cosmic scale, at least, the vacuum-energy term $\bm{T}_{V}^+=\gamma^{-+}\bm{V}_P^{++}-\bm{V}_P^{-+}$ is very small, compared with the natural energy scale provided by the Planck energy\index{Planck energy!natural energy scale}, then it is possible to conclude that the correction provided by the $\gamma^{-+}$ conversion factor is itself actually very small on such a scale. This observation, therefore, indicates that there is a near perfect level of symmetry between the metric properties of space\index{metric properties of space} experienced by positive-energy observers and those experienced by negative-energy observers at the present epoch, on a global scale.

Now, if we consider the above equation in a cosmological context, the irregular stress-energy tensor\index{irregular stress-energy tensors!zero average value} $-\gamma^{-+}\bm{\widetilde{T}}^{-+}$ presumably reduces to zero on average (as the overdensities of negative-energy matter cancel out the underdensities present in the same matter distribution), so that the relevant equations, for positive-energy observers, take the following form:
\begin{equation}\label{eq:2.26}
\bm{G}^+=\bm{T}^{++}+\bm{\widetilde{T}}_{V}^+
\end{equation}
which is similar to their conventional form, except for the fact that the cosmological term\index{cosmological term} $\bm{T}_{\Lambda}^{++}$ is here replaced by the irregular vacuum-energy term\index{irregular vacuum-energy term} $\bm{\widetilde{T}}_{V}^+$ that may vary with position. But given that \textit{local} variations would presumably cancel out for the variable component of negative vacuum energy\index{negative vacuum energy!locally varying component} as well, on a very large scale, and given the (relative) success of current cosmological models for predicting the relevant features of our universe's history, then this outcome would appear appropriate from an observational viewpoint.

We may then also write the following set of equations, which would provide the various levels of decomposition of the general equation (\ref{eq:2.19}) which apply from the viewpoint of negative-energy observers:
\begin{eqnarray}
\bm{G}^- & = & \gamma^{+-}\bm{V}^{--}-\bm{V}^{+-}\label{eq:2.27}\\
\bm{G}^- & = & (\gamma^{+-}\bm{V}_P^{--}-\gamma^{+-}\bm{T}^{+-})-(\bm{V}_P^{+-}-\bm{T}^{--})\label{eq:2.28}\\
\bm{G}^- & = & \bm{T}^{--}-\gamma^{+-}\bm{\widetilde{T}}^{+-}+\bm{\widetilde{T}}_{V}^-\label{eq:2.29}
\end{eqnarray}
where $\gamma^{+-}\bm{V}_P^{--}$ and $-\bm{V}_P^{+-}$ are the natural vacuum-stress-energy tensors\index{natural vacuum-stress-energy tensors} associated with the maximum, negative and positive contributions to the energy density of zero-point vacuum fluctuations\index{zero-point vacuum fluctuations!maximum contributions} set by the Planck scale\index{Planck scale} (as determined by negative-energy observers) and $\bm{\widetilde{T}}_{V}^-=\bm{T}_{V}^{-}-(-\bm{\bar{T}}_{VM}^{+-})$ is the irregular vacuum-energy term associated with a negative-energy observer, with $\bm{T}_{V}^-=\gamma^{+-}\bm{V}_P^{--}-\bm{V}_P^{+-}$ as the locally variable (positive or negative) value of vacuum energy density which would be observed by such an observer in the absence of any ordinary positive- or negative-energy matter and from which is subtracted the average cosmological value of stress-energy $-\bm{\bar{T}}_{VM}^{+-}$ arising from all locally variable negative contributions to the energy of zero-point vacuum fluctuations\index{zero-point vacuum fluctuations!locally variable negative energy contributions} which are experienced by a negative-energy observer (who measures a negative contribution from what would be positive vacuum energy to a positive-energy observer) and which would not include the contribution provided by what would appear to such an observer as a negative cosmological constant\index{cosmological constant!negative} (as the uniform portion of a negative density of vacuum energy which would appear positive to a conventional positive-energy observer).

The last equation, as well the other two, are now manifestly symmetric with the corresponding equations associated with positive-energy observers under a reversal of the sign of energy, as I have argued should be required. But the most remarkable feature of those equations and the related equations for the gravitational field experienced by a positive-energy observer is that they are actually obtained from a very simple expression (the first of the three equations) according to which the gravitational field experienced by an observer with a given sign of energy is determined merely by the appropriate measures of (positive and negative) vacuum energy density. This equation alone allows to embody the essence of the emerging framework.

Indeed, it turns out that for the general equation (\ref{eq:2.19}) to give rise to the decomposition of energy contributions exhibited in the first and second of the observer-specific gravitational field equations\index{gravitational field equations!observer-specific} (in which the metric conversion factors\index{metric conversion factors} are present), all that is required is that the portion of zero-point vacuum fluctuations\index{zero-point vacuum fluctuations} which directly interacts (other than through the gravitational interaction) with positive-energy matter produces a maximum value of energy density that is measured to be negative by a positive-energy observer, while the portion of zero-point fluctuations which directly interacts (other than gravitationally) with what we would normally consider to be negative-energy matter produces a maximum value of energy density that is also measured to be negative by what we would usually consider to be a negative-energy observer (in the sense that the sign of this energy must be opposite that of the observer, which from a conventional Newtonian viewpoint would mean that it is positive).

\bigskip

\noindent The quantitative aspects of the proposed integration of negative energy states to classical gravitation theory\index{classical gravitation theory} having being properly introduced, it is now possible to look back and examine whether the equations obtained can actually provide the structure of an alternative model, which would conform to all the principles enunciated in the preceding section. As I previously remarked, the basic structure of the proposed bi-metric theory\index{bi-metric theory} was adopted precisely because it allows the kind of arbitrariness of the attribution of the sign of energy that is required for this physical property to be defined in a relational manner. But the ultimate confirmation that the proposed framework is compatible with the fundamental requirement expressed by principle 1 is the fact that, even in the presence of a non-vanishing value for the cosmological constant\index{cosmological constant!non-vanishing value}, the set of equations (\ref{eq:2.27}), (\ref{eq:2.28}), and (\ref{eq:2.29}) describing the motion of matter experienced by a negative-energy observer is now symmetric with the corresponding set of equations describing the motion of matter experienced by a positive-energy observer under a reversal of the sign of energy.

Furthermore, the requirement set by principle 2, that inertial mass\index{inertial mass!reversal} be reversed along with gravitational mass\index{gravitational mass!reversal}, is also fulfilled by the proposed gravitational field equations\index{gravitational field equations}, given that my analysis of the phenomenon of inertia\index{inertia} has shown that imposing such a condition should give rise to gravitational attraction between masses of the same sign (whatever this sign is assumed to be) and to gravitational repulsion\index{gravitational repulsion} between masses of opposite signs and this is precisely what we obtain with the proposed equations, even if the sign of energy\index{sign of energy!arbitrariness} that replaces the sign of mass\index{sign of mass} is here arbitrary and the gravitational field\index{gravitational field!relative concept} is a relative concept, dependent on the nature of the matter submitted to it.

On the other hand, the validity of principle 3 and the absence of direct interaction between positive- and negative-energy matter particles may seem to be threatened by the fact that the stress-energy tensor\index{stress-energy tensors} associated with negative-energy matter contributes to determine the gravitational field experienced by positive-energy matter. But again, in the context of the more refined set of equations I have proposed, it is explicit that the negative contribution that enters the total measure of the stress-energy of matter that determines a gravitational field and which we associate with the presence of a negative-energy matter overdensity is actually a measure of the amount of stress-energy missing from the positive portion of vacuum energy\index{missing positive vacuum energy}.

The gravitational forces exerted on positive-energy matter, which must be taken into account in the presence of negative-energy matter, cannot, therefore, be attributable to an interaction with negative-energy matter (whose presence is not directly felt by a positive-energy observer), but must necessarily come from an absence of interaction with that portion of positive vacuum energy that is missing as a consequence of the presence of negative matter energy, which allows the surrounding positive vacuum energy to exert an uncompensated gravitational attraction\index{uncompensated gravitational attraction} directed away from the region where this positive vacuum energy is missing. As a result, the equations naturally require that there be no direct interactions between particles with opposite energy signs.

The new equations are also the perfect embodiment of the requirements set by principles 4 and 5, because they allow voids in the positive-energy portion of the vacuum to actually provide a negative contribution to the total stress-energy tensor\index{stress-energy tensors} of matter and in a general-relativistic context a negative contribution to the stress-energy of matter must be matched by a contribution to the gravitational field that is opposite that which is produced by positive stress-energy, so that if positive-energy matter produces an attractive gravitational field from the viewpoint of positive-energy matter, then negative-energy matter must produce a repulsive gravitational field\index{repulsive gravitational field} from the same viewpoint. The presence of voids in an otherwise uniform distribution of positive vacuum energy\index{positive vacuum energy!uniform distribution} should therefore give rise to uncompensated gravitational forces\index{uncompensated gravitational force} opposite those attributable to the presence of an equivalent amount of positive-energy matter.

We can now understand why it would be inappropriate to assume that the energy of the gravitationally repulsive matter\index{gravitationally repulsive matter!positive energy} whose behavior is described by conventional bi-metric theories\index{bi-metric theories!conventional} is positive, even for an observer that measures a negative contribution from it to the total stress-energy of matter (so that the difficulties usually associated with the presence of negative-energy matter could perhaps be avoided). Indeed, according to the above proposed equations, such matter would produce a gravitational field\index{gravitational field!energy} that would itself have an energy content (to the extent that a definite energy could actually be associated with the gravitational field) opposite that of the gravitational field which is produced by particles contributing positively to the total stress-energy of matter. But this means that if matter was assumed to always have positive energy, then, when energy is exchanged between the two types of matter, the variation of total gravitational energy\index{total gravitational energy!variation} (which would occur because opposite variations of \textit{opposite} gravitational energies are involved) would not be compensated by a variation of the total energy of matter\index{total energy of matter!variation} (which would involve opposite variations of \textit{positive} energy).

Therefore, in the case of our two colliding opposite-energy objects\index{colliding opposite-energy objects} exerting a gravitational repulsion\index{gravitational repulsion} on one another, it would be impossible for the decrease in the kinetic energy of the positive-energy object to be compensated by an increase in the positive gravitational potential energy\index{gravitational potential energy!variation} of the negative-energy object (associated with its interaction with the rest of the negative-energy matter in the universe) that would arise from the corresponding variation of its kinetic energy, despite the fact that this must be considered necessary if energy is to be conserved, as I previously explained. Those problems can be avoided, however, when real negative energy states are allowed for matter, because changes in the energy of the gravitational field can then actually balance the changes occurring in the stress-energy of the two interacting matter components.

I think that this is a clear indication that the tentative solution to the problem of vacuum decay\index{vacuum decay problem} (the collapse of matter to ever more negative energy states) through the contradictory proposal of a gravitationally repulsive matter\index{gravitationally repulsive matter!positive energy} that would have positive energy (from all viewpoints) is misguided and ineffective. Thus, if an observer is allowed to attribute a positive energy to matter of his own kind, regardless of which matter he is made of, it should be clear that, once this choice is made, the energy sign of the matter which from the viewpoint of this same observer provides a negative contribution to the stress-energy tensor\index{stress-energy tensors} of matter must be assumed negative.

I must mention again that, from a cosmological viewpoint, the growth of negative-energy matter overdensities occurring in an initially homogeneous distribution of such matter will always be compensated by an opposite growth of underdensities in the surrounding environment. But given that from my viewpoint those two kinds of inhomogeneities provide opposite contributions to the total stress-energy tensor of matter experienced by a positive-energy observer, then it follows that there exists an additional constraint regarding the conservation of that portion of energy contributed by negative-energy matter and this is a further confirmation of the viability of the proposed equations.

Returning to the criteria imposed by the principles enunciated in the preceding section, we can readily assess that the additional condition set by principle 6 (which implies that only density variations over and below the average cosmic density of negative-energy matter\index{negative-energy matter!average cosmic density} have an effect on positive-energy matter), is also reflected in the equations proposed above. Indeed, the modified measure of negative stress-energy provided by the irregular stress-energy tensor\index{irregular stress-energy tensors} $-\gamma^{-+}\bm{\widetilde{T}}^{-+}$ which naturally enters the gravitational field equations\index{gravitational field equations} associated with a positive-energy observer (given that the presence of negative-energy matter is here explicitly equivalent to an absence of positive energy from the vacuum) actually allows to fulfill the requirement set by principle 6, given that it provides a measure of stress-energy from which is subtracted the average stress-energy of negative-energy matter.

This compliance of the proposed gravitational field equations may perhaps appear to be of secondary concern, given how negligible the average cosmic density of positive-energy matter\index{positive-energy matter!average cosmic density} (and even more so, that of negative-energy matter) really is in comparison with the density variations encountered under most circumstances when we are dealing with astronomical objects of interest, like stars or even galaxies. But, if it was not for the modified measure of negative stress-energy provided by the second term of equation (\ref{eq:2.25}), or the corresponding term from equation (\ref{eq:2.29}), serious problems would occur.

In section \ref{sec:2.5} (in which was elaborated the alternative concept of negative mass\index{negative mass!alternative concept} on which is based the mathematical framework developed here) I mentioned, in effect, that if an object with a given mass sign was to interact with all matter of both positive and negative mass that is present on the cosmological scale, then the classical phenomenon of inertia\index{inertia} itself could not even exist, because the average density of matter energy\index{average density of matter energy!null} may be null on a global scale and under such conditions if both positive- and negative energy matter contribute to determine the gravitational field experienced by matter with a given sign of energy, then the global inertial reference system\index{global inertial reference system} could not be determined by the average state of rest of the matter distribution. However, a Newtonian model is all about inertia, so that if inertial reference systems cannot be appropriately determined in the presence of negative-energy matter, then a reduction of the relativistic equations to a Newtonian gravitation theory\index{Newtonian gravitation theory!reduction to} with gravitationally repulsive, negative mass densities would actually be impossible, even as an approximation.

I believe that ignorance of the requirement to impose a suitable, modified measure of negative stress-energy for the generalized gravitational field equations\index{generalized gravitational field equations!negative stress-energy} is, in fact, the ultimate source of the difficulties which, according to certain authors, are encountered in trying to obtain an appropriate Newtonian limit from conventional bi-metric theories\index{bi-metric theories!Newtonian limit}. This is in addition to the fact that, without the appropriate measure of negative stress-energy, complex hypotheses would have to be introduced concerning the variation in time of the ratio of the average cosmic densities\index{average cosmic matter densities!ratio} of positive- and negative-energy matter in order to try to maintain the agreement of the proposed models with astronomical observations, regarding the rate of expansion of ordinary, positive-energy matter, which is already predicted with (relatively) good accuracy by conventional cosmological models, when no negative-energy matter is assumed to be present initially.

Finally, the fact that two maximum contributions of opposite signs to the energy density of the vacuum\index{vacuum energy density!maximum opposite contributions} are now explicitly present in the most general form of each of the gravitational field equations\index{gravitational field equations} means that both positive and negative contributions to the energy of the vacuum itself (ignoring voids) are allowed to contribute to the gravitational field experienced by positive- or negative-energy matter on the cosmological scale, as required by principle 7. From this alternative viewpoint, what allows one to appropriately ignore most of the effects that the vacuum would have on the gravitational field or the curvature of spacetime\index{curvature of spacetime!vacuum energy} experienced by positive- or negative-energy observers is merely the fact that, in the absence of any matter, those opposite energy contributions nearly cancel each other out at the present epoch.

I may also mention that the condition set by principle 9 (that the equivalence principle\index{equivalence principle} be valid, not merely for matter in a given location, but really for matter with a given sign of energy in a given location) is implicitly contained in the structure of the proposed equations at the most basic level, because they describe gravitational fields which are themselves dependent, not merely on the location, but also on the sign of energy of the particles submitted to them. On the other hand, principles 8, 10 and 11, which identify requirements that have to do with the properties of matter particles (namely the absence of indirect non-gravitational forces\index{indirect non-gravitational forces!absence} between opposite-action particles, the absence of independent contributions to the energy of bound systems\index{bound systems!energy} and the impossibility of a reversal of action\index{reversal of action!continuous particle world-line} on a continuous particle world-line), are not explicitly contained in the gravitational field equations proposed here, but if we assume the validity of those equations, then experimental facts make those constraints unavoidable.

\chapter{Time Reversal and Information\label{chap:3}}

\section{The problem of discrete symmetries\label{sec:3.1}}

In this chapter I would like to explain how a more consistent and adequate formulation of the discrete $P$, $T$, and $C$ symmetry operations, involving a revised concept of time reversal\index{time reversal!revised concept}, can be obtained that integrates the insights gained while studying the problem of negative energy and that offers a better understanding of why and how such symmetries can, under certain circumstances, be violated. 

Discrete symmetry operations\index{discrete symmetry operations} are usually assumed to be relevant only in the context of quantum field theory\index{quantum field theory}, but in fact they can also be examined from a semi-classical standpoint. Their level of application is actually right at the interface between the classical world of gravitation theory and that of quantum theory and it should not come as a surprise, therefore, that some of the results I have obtained will allow progress to be achieved concerning the problem of identifying the nature of the microscopic degrees of freedom\index{microscopic degrees of freedom!matter in black hole} of matter which give rise to black hole entropy\index{black hole entropy} as a semi-classical phenomenon. In order to do so, it will be necessary to introduce two additional categories of discrete symmetry operations. The first of those two categories of symmetry operations relates positive- and negative-action matter particles in a way that is similar in many respects to that by which the charge conjugation\index{charge conjugation} symmetry operation relates ordinary matter and antimatter particles, while the other symmetry operation reverses the non-gravitational charge sign of particles independently from their direction of propagation in time\index{direction of propagation in time}.

I had long ago realized that it is necessary to revise our conception of space and time reversals, because the current formulation of those symmetry operations is based on unreasonable assumptions regarding the significance of time reversal\index{time reversal!significance} and its relationship with the sign of energy and that of non-gravitational charges. It is presently believed that the charge conjugation or $C$ symmetry operation is not a discrete space or time symmetry operation, but simply an additional symmetry having to do with charge as an independent concept. But I came to suspect that the relationships which are known to exist between this charge reversal operation\index{charge reversal operation} and the discrete $P$ and $T$ symmetry operations associated with space and time reversals are an indication that $C$ should be conceived and explicitly defined as a particular instance of discrete spacetime symmetry operation\index{discrete spacetime symmetry operation}. What constitutes the underlying basis of those considerations is the acknowledgment that the sign of certain physical attributes (including charge) are dependent on their direction of propagation in time. From that viewpoint it would seem that both the $T$ and the $C$ symmetry operations should be assumed to involve some form of time reversal and this is reason enough to suspect that they may also both give rise to a reversal of charge.

The problem, however, does not really have to do with our current concept of charge reversal operation as such. What is truly inappropriate is the simple, kinematic representation of time reversal\index{time reversal!kinematic representation} as involving a backward motion of all particles and their angular momenta, which I believe is too rudimentary to characterize a reversal of the fundamental time-direction degree of freedom\index{time-direction degree of freedom!fundamental}. I also think that if $T$ is to be assumed to actually reverse time, then it should leave momentum unchanged (despite common expectations) as this is a physical attribute that should rather be reversed independently, along with the direction of space intervals. In this context, if some reversal of momentum may still be of relevance to $T$ it would clearly have to be due to the fact that it is actually equivalent to the effects we should expect to obtain from an appropriate reversal of time, when we insist on measuring physical attributes against the perceived, rather than the actual direction of the flow of time\index{direction of flow of time}.

In any case, it must be understood that what we observe from our classical historical perspective is not representative of the true evolution that takes place when we are dealing with the propagation of elementary particles. The subtleties of what is going on at the microscopic level are not directly apparent from the superficial viewpoint associated with a global representation of events `after the fact' that provides a historical picture of the spacetime paths\index{spacetime paths!historical picture} followed by elementary particles. Therefore, it is not appropriate to define a reversal of the fundamental (non-thermodynamic) time-direction degree of freedom based merely on narrative aspects of phenomena which are all directly discernible at this superficial level of description. Better formulations of the discrete spacetime symmetry operations are required which would reflect the actual and sometimes unrecognized variations, or absence of variation of physical parameters associated with each of those reversals of the fundamental space- and time-direction degrees of freedom.

\section{The constraint of relational definition\label{sec:3.2}}

To begin this discussion, I must first of all mention that, once again, the most significant constraint which we need to consider and upon which our understanding of the discrete symmetry operations\index{discrete symmetry operations} must be developed is that of the necessary relational definition\index{constraint of relational definition!physical attributes} of physical attributes and their changes. Those attributes are here the sign of space and time intervals\index{space and time intervals!sign}, the direction of momentum and angular momentum\index{direction of momentum and angular momentum} and the sign of energy and non-gravitational charges\index{sign of energy and non-gravitational charges}. The main point I want to emphasize is that there can be no meaning in considering a change of any one of those attributes (to its opposite value) that does not occur relatively to some remaining, unchanged parameter of the same kind. Breaking that rule is to be considered logically impossible, simply because if it was allowed it would mean that we can define an absolute (metaphysical) direction or polarity\index{absolute direction or polarity} (in the general sense), which would not be related to any reference point of a physical nature in our universe.

What I'm suggesting is that the profound reason why a certain level of lopsidedness\index{lopsidedness}, such as the observed breaking of $P$ symmetry by the weak interaction\index{weak interaction}, can exist is that such asymmetries merely occur when one or two physical parameters are reversed \textit{relative} to a fixed background of unchanged directional parameters of a similar kind. In other words, what makes those violations of discrete symmetry\index{discrete symmetry!violations} possible is simply the fact that application of a reversal operation to a single attribute leaves some other parameters unchanged, which allows the asymmetry to occur as a real feature characterized by a measurable change relative to a distinct physical attribute. In the case of $P$ symmetry, the reversal of space intervals involved occurs relative to the direction of time intervals, which remain unchanged by such an operation and therefore it should be expected that violations of $P$ can be observed, given that the reversal of physical parameters associated with this operation can be measured against the unchanged attributes.

But those asymmetries cannot imply the existence of an absolute lopsidedness\index{absolute lopsidedness or directionality} or directionality at the most fundamental level, for the universe as a whole, because they can be compensated by an appropriate reversal of the unchanged attributes relative to which the original transformation took place. This is what explains that despite the violation of $P$ symmetry by the weak interaction, it remains impossible to provide an absolute definition of left and right\index{left and right!absolute definition}, because indeed reversing the sign of charges allows to regain invariance. Thus, contrarily to what is sometimes assumed, the preferred handedness unveiled by the weak interaction\index{weak interaction!preferred handedness} is not more profound than that we observe in certain complex biological structures\index{biological structures!preferred handedness} and which exists as a result of accidental events that took place in the far past.

As long as invariance under a more general discrete symmetry operation like $CP$ is observed to hold, it is impossible to communicate the significance of right and left without knowing which of two $C$-related particles is to be considered as having positive electric charge. But if it is impossible to distinguish an absolute (non-relational) difference between positive and negative charges themselves, as I previously suggested, then only observers from the same universe and which are allowed to directly compare physical attributes, could differentiate between left and right.

This is a very general feature which I think would always be observed to apply, given that it is actually required by the constraint of relational definition\index{constraint of relational definition!physical attributes} of physical attributes, which is relevant to any change of direction or polarity (such as a reversal of the sign of charges). The directions of space and time which are singled out by any process which appears to violate a discrete symmetry are significant only in relation to other aspects of reality which must be identifiable from within the universe in which those processes take place. If, in one particular instance, it was found that no combination of discrete symmetry operations allows invariance to be regained, then it would mean that there can exist physical attributes whose state must be established by referring to elements of reality which are not part of our universe. In other words, if directional asymmetries\index{directional asymmetries} not occurring merely in relation to unchanged attributes (not defined as mere relative physical properties) were allowed, it would, in effect, be impossible to describe the polarities so revealed by referring only to measurable attributes of physical reality\index{physical reality!measurable attributes}.

The problem that there would be if such violations of discrete symmetry\index{discrete symmetry!violations} were possible is that completeness and self-determination are the defining characteristics of the universe\index{universe!completeness and self-determination} concept, in the sense that the universe consists precisely in that ensemble of physical components which are all causally related to one another and to nothing else. Thus, if we were to find that the description of our universe can refer to absolute and immaterial notions of direction, not defined merely as relationships between physical attributes of particles which are present in that universe, then the only logically valid conclusion would have to be that there exists a \textit{causally related} reality outside what we consider to be the universe\index{universe!causally related outside reality} (this has nothing to do with the multiverse\index{multiverse} concept, whose component universes must be considered causally independent from one another), relative to which the otherwise metaphysical polarities\index{metaphysical polarities} would be properly defined.

As a consequence, there is definitely no way our universe\index{universe!lopsided} could be considered lopsided if it is actually the whole universe and I believe that the fact that it can be shown that the existence of such an irreducible asymmetry\index{asymmetry!irreducible} would imply that some physical attributes may not be conserved for the universe as a whole, is a confirmation of the validity of this conclusion. It must be understood, however, that the identified requisite does not mean that symmetry could never be preserved following a reversal of one single parameter, like space direction alone, which can be defined in a relational way, but simply that such invariance is not absolutely required to apply under all circumstances.

Given those considerations, we can be totally confident that there is no such thing as an absolute direction of space and time intervals\index{space and time intervals!absolute direction}, because, indeed, this would imply a violation of the principle of relativity\index{principle of relativity!violation} (in its most general form, which predates relativity theory) and the validity of this criterion is necessary for the consistency of any model of physical reality. Even without going into elaborate mathematical arguments, such as those entering the CPT theorem\index{CPT theorem}, it is therefore possible to appreciate that the only problem there could be in relation to the observation of an asymmetry under a properly defined discrete symmetry operation\index{discrete symmetry operations!proper definition}, would have to involve a violation of invariance under a combined operation that reverses all parameters and leaves absolutely none unchanged.

I will later explain why an appropriately defined $PTC$ transformation must be considered as one instance of such a discrete symmetry operation\index{discrete symmetry operations!inviolable} that reverses all parameters and leaves nothing unchanged (by actually reversing all space- and time-related parameters twice) and which we are thus justified to categorize as inviolable. But I believe that the fact that it would be impossible to provide a mathematical framework for quantum field theory\index{quantum field theory!compliance with special relativity} that would satisfy the requirements set by special relativity if the equations of the theory were not invariant under $PTC$ (which constitute the substance of the argument behind the conventional CPT theorem), confirms that relativistic imperatives (all measures of space and time intervals are relative) are the true constraints which impose invariance under the most general, combined, discrete symmetry operation.

The fact that this simple, but most unavoidable requirement has never been considered as a means to restrict allowed violations of discrete symmetry illustrates the fact that our treatment of space and time reversals\index{space and time reversals} is incomplete and inadequate, due to multiple misconceptions which do not concern only the aspect discussed here. The often met remarks to the effect that there is no \textit{a priori} reason why the universe could not be asymmetric in a fundamental way and that it is only the above mentioned mathematical requirements, arising from the CPT theorem\index{CPT theorem!mathematical requirements}, that motivate the conclusion that some overall symmetry must nevertheless be obeyed under all circumstances, are therefore inappropriate and misleading. But it should also not come as a surprise that the discrete symmetry operations\index{discrete symmetry operations}, when performed independently from one another, may not produce invariance.

What explains the unexpectedness of the violations of $P$ and $CP$ symmetries, when they were first observed, is the intuitive belief that absolute directionality\index{absolute directionality} should not be allowed, while, as I just explained, this is rather the argument that would apply to a more general symmetry operation like $PTC$ whose required conservation, ironically, is usually not believed to be intuitively explainable. The truth is that, for an imbalance under reflection\index{imbalance under reflection} to exist, all that is required is that the world be unbalanced with respect to something. This conclusion is the outcome of the most unequivocal interpretation of the requirement of relational definition of physical attributes\index{requirement of relational definition!physical attributes}, which itself constitutes the one rule we can be most confident need to apply to the physical world we experience. In fact, the argument against the possibility of a violation of symmetry\index{violation of symmetry!reversal of space and time parameters} under a combined reversal of all space- and time-related parameters is probably the strongest kind of argument which can be proposed from a theoretical viewpoint.

Regarding time reversal\index{time reversal}, in particular, and the question of what it would mean to assume that the whole universe is running backward in time and whether there can be any objective meaning to such a reversal operation, I think that, given the preceding discussion, we would have to recognize that such a reversal could, in effect, be physically significant, if it is defined as a reversal that leaves other parameters, such as the direction of space intervals\index{direction of space intervals} unchanged. But this means that such a time reversal operation cannot consist in a mere reversal of the motions and rotations\index{reversal of motions and rotations} of objects taking place in a reverse chronological order\index{chronological order!reversal}.

A reversal of time\index{time reversal!relational definition} that would be relationally defined would have to be meaningful both globally and locally, as it would allow a distinction between a physical system with unchanged time direction and one with reversed time. This difference could be determined by directly comparing the physical attributes of one of the systems with those of the other, if the two systems are part of the same universe. But a difference could also be identified as occurring for the whole universe in relation to the unchanged direction of space intervals. In any case, the above discussed constraint would require that such a relative backward-in-time evolution\index{backward-in-time evolution} be clearly identifiable from the physical attributes of the particles involved, precisely because it is only under such conditions that the change of direction in time could be objectively determined by comparing it with that of the unchanged parameters. But given that those differences would then actually be determined in relation to the value of parameters which are themselves reversible, it follows that no absolutely characterized notion of asymmetry would be involved.

In the context where absolute lopsidedness\index{absolute lopsidedness} is to be considered impossible, it follows that it is of primordial importance to identify all the physical attributes which can be related to one another and which could be affected by transformations of the kind that involve a reversal of space and time directions\index{reversal of space and time directions!fundamental level} at the fundamental level. Indeed, if we are to be allowed to determine whether there remain physical attributes not reversed when a certain discrete symmetry operation\index{discrete symmetry operations} is performed, we must certainly be able to determine which attributes are actually affected by the operation involved. It is my belief that some of the violations of discrete symmetries\index{discrete symmetry!violations} which are usually assumed to have been observationally confirmed are actually a consequence of the fact that the effect of the considered reversals on certain physical attributes are not taken into account, while invariance would actually be inferred if all attributes dependent on the parameters which are assumed to be reversed were appropriately transformed.

I already mentioned the fact that there are indications to the effect that we may, in particular, need to assume that the sign of charges is dependent on the sign of time intervals experienced by the particles carrying them. Yet the conventional definition of the time reversal\index{time reversal!conventional definition} operation $T$ does not involve any reversal of charges (from whatever viewpoint) and thus we could observe violations of such a $T$ symmetry that would occur simply because we do not appropriately reverse the sign of charges when we try to verify invariance under a reversal of time (from a certain viewpoint). We must, therefore, first take care of identifying all unaccounted dependencies which may confuse our assessment of symmetry violations, before we can truly appreciate under which conditions they are actually allowed to occur.

\section{The concept of bidirectional time\label{sec:3.3}}

Concerning the problem of discrete symmetries, another essential aspect must be recognized, in addition to that regarding the necessity of a relational definition of all such symmetry operations. Awareness of what it involves is of the highest importance for a proper resolution of all matters associated with time directionality\index{time directionality} and given that this is the central problem with which this report is concerned, it is crucial to grasp the significance and the implications of the notions involved. Basically, what must be understood is that a distinction is to be made between the conventional concept of time direction\index{time direction!conventional concept} associated with changes occurring at a statistically significant level, where the notion of entropy is meaningful, and a more fundamental concept of time direction\index{time direction!fundamental concept} associated with the existence of a fundamental time-direction degree of freedom\index{time-direction degree of freedom!fundamental}, independent from the constraints related to entropy variation.

The conventional concept of time direction related to statistically significant changes and the growth of entropy gives rise to what I call the unidirectional-\index{unidirectional-time viewpoint} or thermodynamic-time viewpoint\index{thermodynamic-time viewpoint}, while the alternative concept of time direction, related to the existence of a fundamental time-direction degree of freedom independent from statistical constraints, gives rise to what I call the time-symmetric\index{time-symmetric viewpoint}, or bidirectional-time viewpoint\index{bidirectional-time viewpoint}. In chapter \ref{chap:5} the refinement of the concept of time direction associated with the bidirectional viewpoint will be shown to allow the formulation of a principle of causality\index{principle of causality} that is different from the conventional one and which no longer requires the existence of an absolute distinction between causes and effects.

But, associated with this alternative concept of time direction\index{time direction!alternative concept}, is also a different notion of time reversal, not limited by the constraints imposed on our description of physical processes by the second law of thermodynamics\index{second law of thermodynamics}. The conventional concept of time reversal\index{time reversal!conventional concept}, associated with the thermodynamic-time viewpoint\index{thermodynamic-time viewpoint}, merely consists in assuming a reversal of the motion\index{reversal of motion} of all particles involved in a process, so as to give rise to the same events as observed in the original process, but in the reverse order. However, those events would still be described from the same unique and immutable forward direction of time associated with entropy growth. This is a consequence of the fact that the unidirectional-time viewpoint\index{unidirectional-time viewpoint} involves considering that there can only be one direction in time at once for the propagation of all particles, indiscriminately, which actually amounts to ignore the existence of a fundamental time-direction degree of freedom\index{time-direction degree of freedom!fundamental}.

From a conventional viewpoint, if time was reversed, all particles would have to propagate backward, not relative to some fundamental time-direction parameter, but in comparison with the direction of motion which they were all \textit{observed} to have originally. Thus, the time-reverse of a process would simply be the equivalent process for which the same observations are made, but in the reverse order. The time-symmetric\index{time-symmetric viewpoint} or bidirectional-time viewpoint\index{bidirectional-time viewpoint}, on the other hand, is at once less restrictive and more distinctive, in that it actually recognizes the existence of a fundamental time-direction degree of freedom, distinct from the observed direction of motion of particles apparent to an observer constrained by the law of entropy increase. This time-direction parameter must be allowed to vary from one particle to another, even between those of an otherwise identical nature which are involved in the same process at the same time.

Now, of course, I have already discussed the significance of the existence of a fundamental time-direction degree of freedom\index{time-direction degree of freedom!fundamental} as being that property which allows to explain the distinction that exists between a particle and its antiparticle, despite the fact that from an observational viewpoint both objects appear to be ordinary particles propagating forward in time, but which merely happen to carry opposite non-gravitational charges. However, I previously made clear that, in fact, the sign of charge is \textit{not} affected by a reversal of the direction of propagation in time\index{direction of propagation in time!reversal} which transforms a particle into its antiparticle and therefore, if it nevertheless appears to be reversed, it can only mean that the direction of time relative to which we measure the charge is not the true direction in which the particle is propagating in time, because an observer measuring the same physical attribute while following the true direction of propagation in time\index{direction of propagation in time!true direction} of the particle would not observe any change\footnote{
I will henceforth use the term `propagation' in place of `motion' to designate the true direction in which a particle is traversing space and time intervals, as occurs from a bidirectional-time viewpoint\index{bidirectional-time viewpoint|nn}. This allows to explicitly refer to those aspects associated with the fundamental time-direction degree of freedom\index{time-direction degree of freedom!fundamental|nn} which are ignored from the viewpoint of unidirectional time\index{unidirectional-time viewpoint|nn}, relative to which all changes refer to a particle's observed (semi-classical) trajectory.}.

It is merely the fact that a backward-in-time observation\index{backward-in-time observation!impossibility} is impossible that justifies assuming an apparent reversal of charges for a particle propagating toward the past. Indeed, measuring apparatuses always record changes as they occur in the future direction of time due to the fact that the processes involved in the amplification of the signal which gives rise to a measurement can only take place in this direction of time in a universe where a thermodynamic arrow of time\index{thermodynamic arrow of time} governs the evolution of processes involving a large number of independently evolving particles. This constraint is therefore what justifies the use of a unidirectional-time viewpoint\index{unidirectional-time viewpoint}, relative to which all physical properties are given as they would appear relative to the future direction of time, even when the true direction of time in which the processes involved occur is the past direction. Non-gravitational charges, therefore, actually remain unchanged from the bidirectional-time viewpoint\index{bidirectional-time viewpoint} when the fundamental time-direction degree of freedom is reversed, but this is the very reason why they appear to be reversed from the unidirectional-time viewpoint.

A rule, thus, emerges which is that, for any particle propagating in the past direction of time, a time-direction-dependent physical attribute of that particle which would be positive when considered from the bidirectional-time viewpoint (relative to the true direction of propagation of that attribute in time\index{direction of propagation in time!true direction}), would appear as negative from the unidirectional-time viewpoint. But this reversal of observed attributes from their true value is not restricted to charge or energy, which I had already identified as attributes dependent on the direction of propagation in time, but would also have to apply to the direction of space intervals\index{direction of space intervals} associated with the motion of particles (which is always given in relation to that of time intervals\index{direction of time intervals}) and thus, also, to momentum (even if the time intervals entering the conventional definition of momentum\index{momentum!conventional definition} were assumed positive-definite as a consequence of adopting a unidirectional-time viewpoint).

What this means is that if momentum was assumed to be left unchanged by a properly defined reversal of time\index{time reversal!proper definition}, it would nevertheless appear to be reversed in comparison with its actual value, from the unidirectional-time viewpoint\index{unidirectional-time viewpoint}. But as the direction of momentum\index{momentum direction} is not fixed for a given type of particle, propagating in a given direction of time (it also changes when the direction of propagation of the particle in \textit{space}\index{direction of propagation in space} is reversed), then it cannot be taken as a clear indicator of the true direction of propagation in time of a particle. That, however, is not the case with charge, which from the bidirectional-time viewpoint\index{bidirectional-time viewpoint} remains unchanged, even as a particle reverses its direction of propagation in time\index{direction of propagation in time!reversal} (while also reversing its energy sign), and this is why it is possible, from the unidirectional-time viewpoint, to identify the true (even if merely conventionally-defined) direction of propagation in time of a particle, based on the observed value of its non-gravitational charges (in relation to those of an otherwise identical particle)\footnote{
In fact, even if this relationship between time direction and observable charge was valid only for ordinary particles and antiparticles\index{ordinary particles and antiparticles|nn}, while it would be possible to conceive of a distinct operation of charge reversal\index{charge reversal operation!direction of propagation in time|nn} that would reverse charge independently and not merely as result of a reversal of the direction of propagation in time of particles, this conclusion would still be valid, because, as I will explain in section \ref{sec:4.3}, particles carrying such a reversed bidirectional charge\index{reversed bidirectional charge|nn} would remain clearly distinguishable from ordinary particles and antiparticles, regardless of the direction of time in which they are propagating.}.

What is important to understand is this interdependence of space and time intervals\index{space and time intervals!interdependence}, even as they would be separately and independently transformed by their respective, discrete symmetry operations\index{discrete symmetry operations}. Thus, when we reverse the direction of the motion of a particle in space, we reverse the sign of the space intervals associated with this motion, not merely relative to the position axes, but also relative to time intervals (same time interval, opposite space interval). The sign of space intervals associated with the propagation of a particle submitted to a reversal of space directions would be reversed not merely from what it previously was (or relative to the space intervals associated with the motion of a particle not subject to the reversal), but also relative to the direction of time intervals in which the particle is still propagating. A particle which was propagating to the right, relative to the future direction of time, will now be propagating to the left, relative to the same future direction of time, which was not affected by the reversal of space directions (this is illustrated in Figure \ref{fig:3.1} where I consider the effects of the various discrete symmetry operations as they will be defined below). In other words, the particle is not just propagating left, it is propagating left, forward in time, because indeed we are always concerned with the physical properties of processes involving particles propagating in space and time and not just with the properties of space or time themselves.

\begin{figure}
\begin{center}
\begin{picture}(290,290)
\put(140,0){\vector(0,1){280}}
\put(140,282){$t$}
\put(0,140){\vector(1,0){280}}
\put(282,140){$x$}
\put(140,140){\vector(1,1){60}}
\put(200,200){\line(1,1){60}}
\put(262,262){$I$}
\put(210,210){\circle*{3}}
\put(210,210){\vector(0,1){30}}
\put(200,242){$\Delta t$/$E$}
\put(210,210){\vector(1,0){30}}
\put(242,210){$\Delta x$/$p$}
\put(140,140){\vector(-1,1){60}}
\put(80,200){\line(-1,1){60}}
\put(10,262){$P$}
\put(70,210){\circle*{3}}
\put(70,210){\vector(0,1){30}}
\put(60,242){$\Delta t$/$E$}
\put(70,210){\vector(-1,0){30}}
\put(10,210){$\Delta x$/$p$}
\put(140,140){\vector(1,-1){60}}
\put(200,80){\line(1,-1){60}}
\put(262,10){$T$}
\put(210,70){\circle*{3}}
\put(210,70){\vector(0,-1){30}}
\put(200,30){$\Delta t$/$E$}
\put(210,70){\vector(1,0){30}}
\put(242,70){$\Delta x$/$p$}
\put(140,140){\vector(-1,-1){60}}
\put(80,80){\line(-1,-1){60}}
\put(10,10){$C$}
\put(70,70){\circle*{3}}
\put(70,70){\vector(0,-1){30}}
\put(60,30){$\Delta t$/$E$}
\put(70,70){\vector(-1,0){30}}
\put(10,70){$\Delta x$/$p$}
\end{picture}
\end{center}
\caption[Variation of physical parameters under the proposed alternative definition of $P$, $T$, and $C$, as described from the bidirectional-time viewpoint]{Variation of physical parameters under the proposed alternative definition of $P$, $T$, and $C$, as described from the bidirectional-time viewpoint\index{bidirectional-time viewpoint}. In this figure and in the related ones from the current chapter, $I$ represents the original state and the diagonal lines correspond to particle trajectories. The space and time intervals $\Delta x$ and $\Delta t$ are indicated by vectors whose lengths correspond to the magnitude of the intervals and whose directions indicate the sign of the intervals relative to the space and time coordinates. The direction of the vectors associated with the momentum $p$ and energy $E$ of particles corresponds with the sign of momentum and energy relative to the direction of the space and time coordinates.}\label{fig:3.1}
\end{figure}

What matters, therefore, is not just the direction of space intervals\index{direction of space intervals} associated with some arbitrarily-fixed spatial coordinate system\index{spatial coordinate system}, but the direction of space intervals for a particle propagating in a given direction of time, as asserted from a fundamental bidirectional viewpoint\index{bidirectional viewpoint}. Similarly, when time is assumed to be reversed, it must be considered that the time intervals\index{time intervals!reversal} are reversed relative to the unchanged direction of space intervals in which a particle submitted to the reversal is propagating, so that the same positive space intervals are now traveled in the opposite direction of time. This does not mean that a reversal of both space and time cannot have clear meaning, however, because, as I will explain later, even in such a case there would still remain unchanged physical parameters relative to which the transformation could be characterized.

This interdependence between space and time intervals\index{space and time intervals!interdependence} is what gives a true physical meaning to the notion of time reversal, when it is to be considered as a symmetry operation clearly distinct from space reversal and which should, therefore, leave momentum unaffected (from the bidirectional viewpoint at least). In fact, it is what allows the very notion of a fundamental degree of freedom associated with direction in time\index{time-direction degree of freedom!fundamental} to have a definite meaning, because it allows to distinguish (as a theoretical possibility) the process by which a particle is going through a given spacetime trajectory\index{spacetime trajectories} forward in time from the similar process by which an identical particle would be going through the exact same spacetime trajectory, only now backward in time. Such a distinction is crucial, given that if we were to ignore it, then from a unidirectional-time viewpoint\index{unidirectional-time viewpoint} there would be no meaning to assume that it may be possible for a trajectory to be traversed backward in time, given that from such a viewpoint we always observe particles as if they were necessarily going forward in time.

But given that charge can be assumed to be left unchanged by a reversal of time (from the bidirectional viewpoint\index{bidirectional viewpoint}), we are actually allowed to differentiate between the situation of a particle going through a spacetime trajectory forward in time and that of an identical particle going through the same trajectory backward in time, even in the context where all particle trajectories are necessarily followed as if they were occurring in the `normal' chronological order\index{chronological order!normal} (forward in time) associated with the growth of entropy, regardless of the true direction of propagation in time\index{direction of propagation in time!true direction} of the particles. It is, therefore, the interdependence of space intervals and time intervals that allows to distinguish backward-in-time propagation\index{backward-in-time propagation} from forward-in-time\index{forward-in-time!propagation} propagation, and the fact that the observed value of the sign of charge\index{sign of charge!observed value} is dependent on that distinction simply confirms that it is appropriate to consider the existence of such a directionality parameter for the time dimension at the fundamental, elementary-particle level.

It must be clear, however, that the coordinate systems for space and time still have a physical significance, because it is possible to reverse the direction of the space intervals\index{direction of space intervals!reversal} traveled by particles in the forward direction of time, as well as the associated momenta, while keeping the positions of the particles in space unchanged (not reversed as they would under a conventional space reversal operation). Indeed, as a comparison of Figure \ref{fig:3.1} and Figure \ref{fig:3.2} allows to reveal, it is only from the bidirectional-time viewpoint\index{bidirectional-time viewpoint} that the sign of space and time \textit{intervals}\index{space and time intervals!sign}, corresponding to the directions of propagation of particles, always change in association with the sign of space and time \textit{positions}\index{space and time positions!sign} on the coordinate axes, while from the unidirectional-time viewpoint\index{unidirectional-time viewpoint} that need not be the case. Under such conditions, attributes like angular momentum\index{angular momentum}, which depend on both the position in space and the direction of space intervals, may not always be left invariant, as they would when a complete space reversal operation is performed.

\begin{figure}
\begin{center}
\begin{picture}(290,290)
\put(140,0){\vector(0,1){280}}
\put(140,282){$t$}
\put(0,140){\vector(1,0){280}}
\put(282,140){$x$}
\put(140,140){\vector(1,1){60}}
\put(200,200){\line(1,1){60}}
\put(262,262){$I$}
\put(210,210){\circle*{3}}
\put(210,210){\vector(0,1){30}}
\put(200,242){$\Delta t$/$E$}
\put(210,210){\vector(1,0){30}}
\put(242,210){$\Delta x$/$p$}
\put(140,140){\vector(-1,1){60}}
\put(80,200){\line(-1,1){60}}
\put(10,262){$P$}
\put(70,210){\circle*{3}}
\put(70,210){\vector(0,1){30}}
\put(60,242){$\Delta t$/$E$}
\put(70,210){\vector(-1,0){30}}
\put(10,210){$\Delta x$/$p$}
\put(140,140){\line(1,-1){80}}
\put(260,20){\vector(-1,1){40}}
\put(262,10){$T$}
\put(210,70){\circle*{3}}
\put(210,70){\vector(0,1){30}}
\put(200,102){$\Delta t$/$E$}
\put(210,70){\vector(-1,0){30}}
\put(150,70){$\Delta x$/$p$}
\put(140,140){\line(-1,-1){80}}
\put(20,20){\vector(1,1){40}}
\put(10,10){$C$}
\put(70,70){\circle*{3}}
\put(70,70){\vector(0,1){30}}
\put(60,102){$\Delta t$/$E$}
\put(70,70){\vector(1,0){30}}
\put(102,70){$\Delta x$/$p$}
\end{picture}
\end{center}
\caption[Variation of physical parameters under the proposed alternative definition of $P$, $T$, and $C$, as apparent from the unidirectional-time viewpoint]{Variation of physical parameters under the proposed alternative definition of $P$, $T$, and $C$, as apparent from the unidirectional-time viewpoint\index{unidirectional-time viewpoint}. We can see that, from this viewpoint, the only difference between the original process and the $T$-reversed process is that the space intervals are traversed in the opposite direction, just as would be expected according to the conventional definition of backward-in-time motion\index{backward-in-time motion!conventional definition}. The case of the $C$-reversed process is also quite in line with conventional expectations, given that such a process should not be different from the original process except for a reversal of the sign of charges (which is not illustrated here) which would in fact also occur for the $T$-reversed process, despite conventional expectations.}\label{fig:3.2}
\end{figure}

Such a reversal would occur for processes submitted to a reversal of time, when they are described from the unidirectional viewpoint\index{unidirectional viewpoint} in which time is maintained positive, even for backward-in-time-propagating particles\index{backward-in-time-propagating particles}, and all time-direction-dependent attributes like the direction of space intervals\index{direction of space intervals!reversal} and the momentum of a particle consequently appear to be reversed, while the positions are left unchanged (which implies that spin\index{spin!reversal} would appear to be reversed). In this context, it seems that space intervals, as quantities defined in relation to the direction of propagation in time\index{direction of propagation in time}, can actually be reversed in two different ways. They may be reversed because space directions are reversed (which also reverses positions), or they may be reversed because the direction in which they are assumed to be traversed in time is reversed (which leaves positions unchanged). This distinction is what allows the conventional concept of time reversal\index{time reversal!conventional concept}, as affecting the directions of momentum and angular momentum, to still be relevant, even in the context of the existence of a fundamental time-direction degree of freedom\index{time-direction degree of freedom!fundamental}, when those directions should in fact be left invariant (from a bidirectional viewpoint\index{bidirectional viewpoint}) by a properly defined time reversal operation.

Another point must be emphasized regarding the kind of time reversal operation which can be developed in the above described context. Indeed, if we no longer consider appropriate the picture of time reversal\index{time reversal!reversal of observed motion} as consisting in a simple reversal of the observed motion of each and every particle, then it must also be recognized that a properly defined time reversal operation could never give rise to a reversal of the thermodynamic arrow of time\index{thermodynamic arrow of time!reversal} for the physical systems involved. In fact, I think that we should already suspect that there is something wrong with the often-met suggestion that a reversal of the motion of every particle in a region of space would give rise to entropy decreasing evolution\index{entropy decreasing evolution} (in the absence of any external perturbation). For such a proposal to be valid it would have to be shown that the origin of the observed time asymmetry\index{time asymmetry!thermodynamic processes} of thermodynamic processes in our universe is to be found in a very precise adjustment of the motion of every single particle in the universe at the present time, which would occur in just such a way as to allow a state of minimum entropy to be reached as time unfolds in the past, right back to the initial Big Bang state\index{initial Big Bang state!minimum entropy}.

However, given the inherently random nature of quantum processes and the extreme sensitivity to initial conditions\index{initial conditions!sensitivity to} (here the `final' conditions giving rise to a given past evolution) which are known to exist, even in a classical context, this hypothesis appears highly implausible (I will address this question more thoroughly in section \ref{sec:4.6}). But if, in addition, we admit the existence of a fundamental time-direction degree of freedom\index{time-direction degree of freedom!fundamental}, distinct from the observed motion of particles, then we clearly have to reject the possibility that a reversal of time may produce anti-thermodynamic evolution\index{anti-thermodynamic evolution}, because backward-in-time propagation\index{backward-in-time propagation} is in fact already taking place in the course of processes for which there is no apparent change to the direction of the thermodynamic arrow of time\index{thermodynamic arrow of time}. This means that the direction of propagation in time\index{direction of propagation in time} of particles (the sign of time intervals\index{time intervals!sign} associated with a bidirectional viewpoint\index{bidirectional viewpoint}) is not necessarily that relative to which entropy increases, despite the fact that it may appear unnatural that evolution could proceed in a direction of time other than that in which we do observe time to be `flowing' (as a thermodynamic necessity). The thermodynamic arrow of time and the notion of time directionality occurring from a bidirectional viewpoint are two completely independent concepts.

\section{Alternative definition of $C$, $P$, and $T$\label{sec:3.4}}

One last remark is necessary before I can provide a full description of exactly how the fundamental physical attributes of matter particles should be considered to vary under an alternative set of discrete symmetry operations\index{discrete symmetry operations!alternative set} formulated so as to allow the above discussed requirements to be satisfied.

I previously hinted at the fact that the direction of momentum should be considered as independent from the direction of time, at least from the most consistent viewpoint, which is provided by a bidirectional perspective on time. I believe, in effect, that momentum\index{momentum!conjugate attribute to position}, as the physical attribute conjugate to spatial position, should only be considered to reverse along with space and not along with time, just as energy\index{energy!conjugate attribute to time}, being the physical attribute conjugate to time, should necessarily reverse when time reverses and only then. There is, however, an additional motivation for requiring this kind of joint variation of all space- or time-related attributes (independently), besides the fact that consistency may require that it be imposed when what we seek to assert is precisely the dependence of various parameters under reversal operations which are defined after the attributes they are assumed to reverse. This, perhaps more unavoidable, justification for the joint reversal of conjugate attributes\index{conjugate attributes!joint reversal} is to be found in the requirement that the considered discrete symmetry operations\index{discrete symmetry operations!invariance of action sign} should leave invariant the sign of action of the physical systems on which they operate.

It is my understanding of the true physical significance of a reversal of the sign of action\index{sign of action!reversal} that allows me to recognize the necessity to define the discrete symmetry operations in such a way that momentum would necessarily reverse as a consequence of a reversal of space coordinates, while energy would necessarily reverse as a consequence of a reversal of the time coordinate. Indeed, in the context where a reversal of space coordinates would necessarily give rise to a reversal of space intervals\index{space intervals!reversal}, while a reversal of the time coordinate would necessarily give rise to a reversal of time intervals\index{time intervals!reversal}, if the sign of action\index{sign of action!invariance} itself is to remain invariant, then it means that a reversal of space must also involve a reversal of momentum\index{momentum!reversal} and a reversal of time must also involve a reversal of energy\index{energy reversal}. In fact, we always implicitly assume that the $P$, $T$, and $C$ reversal operations do not relate physical processes in which the particles involved would have opposite action signs or energies (as measured from the forward direction of time). But the implications this should have for the dependence (under application of conventional discrete symmetry operations\index{discrete symmetry operations!conventional}) of the signs of momentum and energy on those of space and time intervals is not always recognized.

I believe that this lack of clarity is responsible for a good part of the misunderstanding regarding what parameters should really be affected by any symmetry operation involving a reversal of time\index{time reversal}. In Table \ref{tab:3.1}, Table \ref{tab:3.2}, and Table \ref{tab:3.3} I will therefore provide an explicit account of the dependence of the signs of momentum and energy, along with those of space and time intervals, under all relevant discrete symmetry operations\index{discrete symmetry operations}. It will be apparent from this account that clear distinctions exist between the conventional and the redefined time reversal and charge conjugation\index{charge conjugation} symmetry operations. Yet, given that the original definitions need to be replaced and cannot even be considered meaningful anymore, I think that it will not be necessary to relabel those operations and associate them with new symbols or letters, so that I will continue to use the $T$ and $C$ notation when referring to those redefined discrete symmetry operations\index{discrete symmetry operations!redefined}.

In the following tables and in the corresponding diagrams (Figure \ref{fig:3.1} corresponds to Table \ref{tab:3.2} and the bidirectional-time viewpoint\index{bidirectional-time viewpoint}, while Figure \ref{fig:3.2} corresponds to Table \ref{tab:3.3} and the unidirectional-time viewpoint\index{unidirectional-time viewpoint}) the position along the space and time axes are denoted $x$ and $t$ (I'm assuming a one-dimensional space for simplicity), while the space and time intervals\index{space and time intervals} corresponding to the motion, or the propagation of the particles involved in the processes which are transformed by the symmetry operations are denoted $\Delta x$ and $\Delta t$ respectively. The energy of the particles involved in the same processes is denoted $E$ and can actually vary in sign, while the momentum of those particles along the $x$ axis is simply denoted $p$.

The sign of non-gravitational charges (which allows to distinguish between the state of a particle and that of its antimatter counterpart), even though it should be understood not to be reversed by any of the conventional discrete symmetry operations\index{discrete symmetry operations!conventional} (including $C$) from the bidirectional viewpoint (which provides the most accurate description of the transformations involved), is nevertheless included in the tables and denoted $q$, as it may actually appear to be reversed from the unidirectional viewpoint by some of those symmetry operations. The direction of angular momentum\index{angular momentum!direction}, related to the motion of the particles involved in the processes transformed by the $P$, $T$, and $C$ operations, as well as the spin direction\index{spin!direction} of elementary particles, which again should be understood not to be affected by those operations from a bidirectional-time viewpoint, are together denoted by the letter $s$, while the associated parameter of handedness\index{handedness} (the direction of spin along the axis associated with the momentum of a particle) is here denoted $h$ and should be expected to vary, even from a bidirectional-time viewpoint.

\begin{table}
\begin{center}
\begin{tabular}{c||c|c|c||c|c|c||c|c|c}
Conv. & $t$ & $\Delta t$ & $E$ & $x$ & $\Delta x$ & $p$ & $q$ & $s$ & $h$ \\ \hline\hline
$I$ & $t$ & $\Delta t$ & $E$ & $x$ & $\Delta x$ & $p$ & $q$ & $s$ & $h$ \\ \hline
$P$ & $t$ & $\Delta t$ & $E$ & $-x$ & $-\Delta x$ & $-p$ & $q$ & $s$ & $-h$ \\ \hline
$T$ & $-t$ & $\Delta t$ & $E$ & $x$ & $-\Delta x$ & $-p$ & $q$ & $-s$ & $h$ \\ \hline
$C$ & $t$ & $\Delta t$ & $E$ & $x$ & $\Delta x$ & $p$ & $-q$ & $s$ & $h$ 
\end{tabular}
\end{center}
\caption[Variation of the physical parameters associated with a process transformed by the discrete $P$, $T$, and $C$ symmetry operations, as they are conventionally defined]{Variation of the physical parameters associated with a process transformed by the discrete $P$, $T$, and $C$ symmetry operations\index{discrete symmetry operations!conventional definition}, as they are conventionally defined. The variations of the $\Delta t$ and $\Delta x$ parameters indicated here are only implicitly assumed from a conventional viewpoint. The absence of reversal of $\Delta t$ when time is assumed to be reversed can be noted. The variation of the direction of angular momentum $s$, as well as that of the handedness $h$, can be derived from those of the other fundamental parameters, but they are nevertheless indicated here and in the other tables, because in certain cases they differ from what is conventionally expected. The identity operation $I$ which corresponds to an absence of reversal is shown for reference purpose.}\label{tab:3.1}
\end{table}

\begin{table}
\begin{center}
\begin{tabular}{c||c|c|c||c|c|c||c|c|c}
Bidir. & $t$ & $\Delta t$ & $E$ & $x$ & $\Delta x$ & $p$ & $q$ & $s$ & $h$ \\ \hline\hline
$I$ & $t$ & $\Delta t$ & $E$ & $x$ & $\Delta x$ & $p$ & $q$ & $s$ & $h$ \\ \hline
$P$ & $t$ & $\Delta t$ & $E$ & $-x$ & $-\Delta x$ & $-p$ & $q$ & $s$ & $-h$ \\ \hline
$T$ & $-t$ & $-\Delta t$ & $-E$ & $x$ & $\Delta x$ & $p$ & $q$ & $s$ & $h$ \\ \hline
$C$ & $-t$ & $-\Delta t$ & $-E$ & $-x$ & $-\Delta x$ & $-p$ & $q$ & $s$ & $-h$ 
\end{tabular}
\end{center}
\caption[Variation of physical parameters under the redefined discrete $P$, $T$, and $C$ symmetry operations, as described from the bidirectional-time viewpoint]{Variation of physical parameters under the redefined discrete $P$, $T$, and $C$ symmetry operations\index{discrete symmetry operations!bidirectional-time viewpoint}, as described from the bidirectional-time viewpoint. The necessary reversal of $\Delta t$ with $E$, as well as that of $\Delta x$ with $p$, can be noted, as also the necessary reversal of $t$ with $\Delta t$ and that of $x$ with $\Delta x$. This is the variation of physical parameters which would be produced by the most appropriately defined discrete symmetry operations that can be formulated in a semi-classical context. Here, all reversals of physical attributes are seen to occur twice or not at all, as required for explicit invariance under a joint $PTC$ operation.}\label{tab:3.2}
\end{table}

\begin{table}
\begin{center}
\begin{tabular}{c||c|c|c||c|c|c||c|c|c}
Unidir. & $t$ & $\Delta t$ & $E$ & $x$ & $\Delta x$ & $p$ & $q$ & $s$ & $h$ \\ \hline\hline
$I$ & $t$ & $\Delta t$ & $E$ & $x$ & $\Delta x$ & $p$ & $q$ & $s$ & $h$ \\ \hline
$P$ & $t$ & $\Delta t$ & $E$ & $-x$ & $-\Delta x$ & $-p$ & $q$ & $s$ & $-h$ \\ \hline
$T$ & $-t$ & $\Delta t$ & $E$ & $x$ & $-\Delta x$ & $-p$ & $-q$ & $-s$ & $h$ \\ \hline
$C$ & $-t$ & $\Delta t$ & $E$ & $-x$ & $\Delta x$ & $p$ & $-q$ & $-s$ & $-h$ 
\end{tabular}
\end{center}
\caption[Variation of physical parameters under the redefined discrete $P$, $T$, and $C$ symmetry operations, as described from the unidirectional-time viewpoint]{Variation of physical parameters under the redefined discrete $P$, $T$, and $C$ symmetry operations\index{discrete symmetry operations!unidirectional-time viewpoint}, as described from the unidirectional-time viewpoint. Again, all attributes are reversed either twice or never by a combination of all operations, which guarantees explicit invariance under $PTC$. The equivalent reversal of charge $q$ by both $T$ and $C$, as well as the apparent absence of any variation of $\Delta t$ and $E$, and the absence of joint variation of $x$ and $\Delta x$ when $t$ is reversed, can be noted.}\label{tab:3.3}
\end{table}

From a semi-classical viewpoint, the displayed tables, giving the variations of the space- and time-related physical parameters\index{space- and time-related parameters} under the conventional and redefined discrete symmetry operations\index{discrete symmetry operations!conventional and redefined}, along with the assumptions which are made concerning the variation of the sign of charge, provide the most precise definitions that can be achieved of the operations involved. Using those definitions, one can rebuild the quantum\index{quantum!operators} operators which are needed to transform the state vectors\index{state vectors} or the propagators\index{propagators} corresponding to specific quantum states or processes\index{quantum states or processes}. It must be clear that quantum field theory\index{quantum field theory} itself does not dictate how the discrete symmetry operations should be defined and it is merely the assumptions used while formulating the related operators (to achieve transformations that match our expectations regarding which parameters should be affected by a given operation) that provide the necessary constraints on which depend their precise mathematical formulation. What I'm providing, therefore, is an improved understanding of the constraints that must apply to those transformations, based on a re-examination of the meaning of space and time reversals\index{space and time reversals}, as they would occur in a semi-classical context.

It is important to recognize, indeed, that despite the apparent freedom, the discrete symmetry operations cannot be arbitrarily defined, but must be the outcome of the most unavoidable consistency requirements (formulated in an empirically motivated context), which I believe are those I have identified in the above discussion. The fact that greater simplicity has been achieved while redefining those symmetry operations is only a further confirmation of the appropriateness of the alternative viewpoint that emerged from the preceding analysis. Indeed, the pattern of variations of physical parameters which is illustrated in Figure \ref{fig:3.1} is strikingly simple in comparison with that we would have according to the conventional definition of the discrete symmetry operations and this simplification was actually one of the objectives I sought to achieve while redefining them. Let me then describe what the elegance of this proposal really embodies.

Looking at the tables in which the outcomes of the various discrete symmetry operations are displayed, one thing we may first remark is that the parity operation\index{parity operation} $P$ remains as it was originally defined, even in the context of the proposed alternative formulation of those transformations and this regardless of whether we consider things from the bidirectional-\index{bidirectional-time viewpoint} or the unidirectional-time viewpoint\index{unidirectional-time viewpoint}. Of course, the reversal of space intervals\index{space intervals!reversal} associated with the propagation of particles (which must occur as a result of the reversal of space coordinates) is now explicitly stated, but, otherwise, the conventional definition of space reversal\index{space reversal!conventional definition} remains unchanged. There is one good reason for that, which is that the revision I'm operating regards the concept of time direction, essentially, and the $P$ operation is unique for being the only one that does not involve any time reversal, regardless of the viewpoint favored. This is what explains that this operation was properly defined already, in the form it originally was, despite the inadequacy of the conventional approach in general.

What $P$ produces is a reversal of space coordinates that reverses the positions, the space intervals, and naturally, also, the momentum (as a requirement of action sign invariance\index{sign of action!invariance}), while it leaves unchanged (now as a matter of definition) the position in time, the time intervals and the sign of energy. No reversal of charge is to be observed in this case (particles are not replaced by antiparticles), from any perspective, because there is no time reversal involved from a bidirectional viewpoint\index{bidirectional viewpoint} and thus no change to be associated with the adoption of a unidirectional viewpoint\index{unidirectional viewpoint}. There is no reversal of angular momentum either (because both momentum and position are together reversed), which is appropriate given that if angular momentum or spin were reversed, a forbidden reversal of action\index{reversal of action} would occur from the bidirectional viewpoint (because spin has the dimension of an action) that would not be associated merely with the shift to a unidirectional-time viewpoint. But again, this is in perfect agreement with conventional expectations regarding the effects of $P$. Handedness\index{handedness!reversal} is to be assumed reversed by such a reversal of space, however, because momentum is reversed while spin is left invariant from all viewpoints.

It should be noted that the explicit mention of a reversal of space intervals\index{space intervals!reversal} $\Delta x$ under a symmetry operation like $P$ does not mean that a reversal of space intervals must be assumed to occur in addition to that produced by the reversal of space coordinates\index{space coordinates!reversal}. In other words, if the space intervals are indeed reversed, it is merely as a consequence of the reversal of space coordinates, as otherwise there would be no real change in the direction of space intervals, that is to say, no change relative to the new coordinates. We may, in fact, consider it more appropriate to assume that it is the intervals themselves which are reversed along with the position of particles while the coordinates remain unchanged, which would still be equivalent to reversing the coordinates themselves. If I choose to explicitly mention a reversal of space intervals, along with the assumed reversal of positions, it is because there may be situations where the intervals would be reversed independently from the positions on the coordinate axes and we must be able to distinguish between the two situations.

What the explicit statement of a reversal of $\Delta x$ should be understood to imply is that there must occur a reversal of the sign of space intervals traversed by the particles involved in the $P$-reversed processes, in comparison with the sign of space intervals experienced by particles involved in processes which would not be submitted to this reversal. We must, therefore, assume that those reversed intervals are the space intervals which are traversed during unchanged time intervals and which we may ordinarily associate with the directions of the momenta of the particles involved. Indeed, the reversal of space intervals associated with the motion of particles is usually assumed to be implied by the reversal of momentum\index{momentum!reversal} itself, but given that I will later suggest that momentum can be reversed without space intervals being equally reversed (when action is reversed\index{action reversal}), then it becomes necessary to explicitly define the variation of space intervals under $P$ and to recognize that momentum direction\index{momentum direction} is an independent parameter, whose specification is not sufficient to determine the sign of space intervals\index{sign of space intervals} spanned during a given time interval (except if the action sign\index{sign of action!invariance} is, in effect, required to be invariant).

It must be recognized, therefore, that the reversal of $\Delta x$ is not \textit{merely} a reflection of the reversal of space coordinates\index{space coordinates!reversal}, but that it also allows to denote the physical changes that occur when a particle reverses its direction of propagation in space\index{direction of propagation in space!reversal}, while retaining its direction of propagation in time\index{direction of propagation in time} and those changes would be significant even if the position in space was to itself remain unchanged. Likewise, what the specific statement about the reversal of momentum $p$ under space reversal $P$ is intended to mean is that the direction of momentum is now the opposite of what it was, not merely relative to the new coordinates, but also relative to the directions of the momenta of particles which would not be subject to the symmetry operation. I may add that the same remarks would apply to time intervals $\Delta t$ and the sign of energy, because if the reversal of those physical parameters under the $T$ and $C$ operations (from a bidirectional viewpoint) can be understood to occur as a consequence of the reversal of the time coordinate, it is clear that it also arises in relation to the time intervals experienced by particles which would be left unaffected by the reversal.

\section{The time reversal operation\label{sec:3.5}}

Despite a concordance of the rules from which are derived the variation of physical parameters under any one of the redefined discrete symmetry operations\index{discrete symmetry operations!redefined}, there are important differences between the case of time reversal $T$ or charge conjugation $C$ and that of space reversal $P$ and this is reflected in the fact that those two symmetry operations would produce results which are unexpected from a conventional viewpoint. In the case of $T$, it must be required, in effect, that the physical time intervals $\Delta t$ associated with the propagation of elementary particles and the energy $E$ be together reversed when the time coordinate is reversed (if action is to remain positive when it already is), while it is conventionally assumed (even if only implicitly) that both energy signs and \textit{bidirectional} time intervals\index{bidirectional time intervals} are in fact unchanged by $T$ despite the reversal of the time coordinate.

But it must now also be assumed that there is no \textit{a priori} reversal of the space intervals $\Delta x$ and momentum $p$ when time is reversed (which is allowed when those parameters are recognized to be independent from the time-related parameters $\Delta t$ and $E$). This is required, despite the fact that, conventionally, momentum is assumed to be dependent on time intervals (I will explain below how this apparent contradiction is to be resolved). In fact, the conventional assumption that $p$ would be reversed by $T$, while the position $x$ on the space axis would remain unchanged, would be problematic if, in this context, we did not again implicitly assume an independent reversal of physical space intervals $\Delta x$, by presuming an invariance of the sign of action\index{sign of action!invariance}.

What must be recognized, therefore, is that from a consistent, bidirectional viewpoint\index{bidirectional viewpoint}, when the time coordinate is reversed, it must be assumed that the time intervals of propagating particles (associated with the fundamental time-direction degree of freedom\index{time-direction degree of freedom!fundamental}) are reversed along with their energies (as defined relative to the true direction of propagation in time\index{direction of propagation in time!true direction}), while momentum and space intervals are left unchanged, just like a reversal of space coordinates is assumed to imply a reversal of the space intervals and momenta, but no change to energy sign and no reversal of time intervals. This independence of space- and time-related physical attributes\index{space- and time-related attributes!independence} (from one another) under reversal is a requirement of the constraint of relational definition\index{constraint of relational definition!physical attributes} of those attributes, which imposes that something remains unchanged when $T$ or $P$ is applied, and those invariant parameters are in fact the spatial directions themselves (when the direction of time is reversed) or the direction of time itself (when space directions are reversed).

Now, if we appropriately assume that the spatial positions, the space intervals, and the momenta remain unchanged under a properly defined time reversal operation, it follows that the spin and the handedness must also remain invariant. Those relationships may appear unnatural (spin is usually considered to be reversed under a reversal of time), but from a bidirectional-time viewpoint\index{bidirectional-time viewpoint} they are perfectly acceptable and in the context where we want to define time reversal as really affecting time-related parameters\index{time-related parameters} in a specific way, they are actually unavoidable.

The discussed invariance, however, is observed from the bidirectional-time viewpoint, according to which the values of physical parameters are such as they would appear to an observer following the direction of propagation in time\index{direction of propagation in time} of the particles involved in the processes submitted to such a time reversal operation. From a unidirectional-time viewpoint\index{unidirectional-time viewpoint} (of the kind that is required from a practical perspective), the only parameters which would appear to be left unchanged when time is reversed would actually be the time intervals $\Delta t$ and the energies $E$, because they would be submitted to twice the same reversal, once as time-related attributes, and once as a consequence of the additional reversal occurring when we are forcing a forward-in-time\index{forward-in-time!perspective} perspective. This is what justifies the validity of the assumption that energy would appear to not be reversed from the conventional forward-in-time viewpoint and it means that if energy wasn't, in effect, reversed from the time-symmetric viewpoint\index{time-symmetric viewpoint}, then from the unidirectional viewpoint it would actually appear to be reversed by $T$, which is certainly not desirable.

On the other hand, the physical space intervals and the momenta associated with the propagation of particles do need to be reversed (once) when time is reversed, if we insist on describing the motion of particles as it appears to take place from the conventional forward-in-time viewpoint and this despite the fact that only the physical time intervals experienced by the particles should actually be reversed by $T$. Indeed, given that the direction of space intervals\index{direction of space intervals} is defined in relation to the direction of time intervals\index{direction of time intervals}, if time intervals are followed in the wrong direction, then space intervals are also traversed in the wrong direction, so that the observed direction of motion of particles\index{direction of motion of particles!observed} are opposite the true direction of their motions\index{direction of motion of particles!true}, which means that those directions are actually reversed under a properly defined $T$ operation, when the outcome of this operation is considered from a unidirectional-time viewpoint (this is made apparent when we reverse the direction of the arrows associated with the time-reversed states in Figure \ref{fig:3.1} to produce those in Figure \ref{fig:3.2}).

Thus, when the direction of time is reversed, but the time intervals in which the particles propagate are kept unchanged, as a consequence of practical limitations imposed by the thermodynamic nature of the observation process, the associated space intervals actually appear to be reversed (they are the negative of those really experienced by the particles), even though the spatial positions remain unchanged. This is true, again, despite the fact that at the most fundamental level of description, which is that of bidirectional time\index{bidirectional time}, the direction of space intervals is to be considered unchanged by a reversal of time. As a consequence, we obtain results which comply with the conventional definition of time reversal\index{time reversal!conventional definition}, according to which momentum (and implicitly also space intervals) should, in effect, be reversed by $T$, along with angular momentum or spin, because given that momentum is here reversed independently from the position parameter $x$ it follows that angular momentum would also appear to be reversed.

From the unidirectional viewpoint\index{unidirectional viewpoint} it may, in effect, seem like the conventional concept of time reversal, as involving a reversal of motion\index{reversal of motion} which simply allows the particles to follow a trajectory backward, could be valid. We must recognize, however, that just as there is no reason to assume that momentum is affected by a reversal of time from a bidirectional viewpoint\index{bidirectional viewpoint} (which explains that it is reversed from a unidirectional viewpoint), there is also no reason to assume that the sign of charge, as distinct from that of energy (the gravitational charge), would be affected from this same viewpoint when $T$ is applied, because charge is not constrained to reverse by the requirement of action sign\index{sign of action!invariance} invariance, when the direction of propagation in time\index{direction of propagation in time!reversal} reverses. This may also appear to comply with conventional expectations, but in fact (as I previously remarked) it rather constitutes the one aspect which introduces a radical departure from what is normally assumed concerning time reversal.

What the invariance of the sign of charge under a reversal of time really means is that the same reversal that does apply to momentum from the unidirectional-time viewpoint, would have to apply to non-gravitational charges as well, because if the direction of propagation in time of the charges is actually reversed as required, then the fact that time is followed in the same forward direction relative to which the charges were originally propagating means that the charges would now appear to be reversed. We must, therefore, consider a reversal of charges\index{charge reversal!reversal of time} to be associated with a reversal of time, as a result of the fact that this physical attribute is not experienced along the true direction of time in which it is being propagated. This is a very important result, which is definitely not expected from a conventional viewpoint, given that it asserts that a physical parameter which was previously assumed to be unaffected by a reversal of time (namely the sign of charge), would actually appear to be reversed under such a transformation, and if the preceding argument is valid then this conclusion would have to be considered unavoidable.

Thus, it seems that considering a reversal of time without assuming a consequent reversal of charge is incorrect and may give rise to violations of symmetry which are a simple artifact of the inappropriateness of conventional assumptions concerning which attributes are reversed along with the time coordinates, from the unidirectional viewpoint\index{unidirectional viewpoint}. To be meaningful, the experiments which seek to verify invariance under $T$ would actually have to assume a reversal of momentum and spin retracing a process backward, but combined with a reversal of charge (a permutation of particle and antiparticle). In other words, to test the invariance of physical laws under time reversal, we would have to use antimatter, which may explain why a violation of $T$ symmetry is so difficult to observe despite the fact that violations of the combined $CP$ symmetry were actually observed (which implies that $T$ should also be violated, given that $CPT$ is inviolable).

It appears that we are simply not using the right kind of matter to probe for $T$ violation. It is not the invariance of a process relative to the thermodynamic arrow of time\index{thermodynamic arrow of time} which must be probed, but invariance under a reversal of the true directions of propagation in time\index{direction of propagation in time!true direction} of elementary particles. I believe that the improved consistency of the interpretation suggested here, from both an observational and a theoretical viewpoint, confirms that the conventional definition of time reversal\index{time reversal!conventional definition} as involving nothing more than a reversal of the directions of motion and rotation\index{directions of motion and rotation!reversal} of particles can no longer be considered appropriate.

It may also be noted that, from a unidirectional viewpoint\index{unidirectional viewpoint}, the reversal of charge\index{charge reversal!reversal of time} and the reversal of spin\index{spin reversal!reversal of time} under a properly defined time reversal operation are now the only aspects that differentiate this $T$ operation from the $P$ operation, apart from the respective reversals of the time and space coordinates themselves. But given that spin can also vary independently from the direction of propagation in time of a particle, this means that the only unmistakable distinction between the time-reverse of a given state and the space-reverse of the same state is, in effect, the sign of charge, which again emphasizes the importance of recognizing the dependence of this parameter on the direction of time.

In such a context, it seems possible that the violations of $T$ which may have been observed despite all the previously mentioned experimental difficulties could actually be violations of combined symmetries under which charge is left invariant by being reversed twice, because indeed those experiments do not compare matter and antimatter processes. Yet it might be considered that, despite what is commonly believed, violations of time reversal symmetry\index{time reversal symmetry!violation} had already been observed, even before the violations of conventional $T$ symmetry were reported, because, as I will explain below, the $C$ operation also involves some time reversal and violations of charge conjugation symmetry\index{charge conjugation symmetry!violations} do occur. In any case, it is clear that a violation of the time reversal symmetry operation $T$, as it was here redefined, would not provide us with an absolute direction of time\index{absolute direction of time} at a fundamental level, but merely with a preferred direction of time relative to some arbitrarily-chosen direction in space.

Another particularity of the alternative definition of time reversal\index{time reversal!alternative definition} proposed here is that it implies that it would now be electric fields\index{electric field!reversal} which would reverse under application of the $T$ operation, instead of magnetic fields, because electric fields depend only on the sign of charge of the source particles and charge must be assumed to reverse under time reversal. Magnetic fields, on the other hand, would now remain unchanged under time reversal, because from the unidirectional viewpoint\index{unidirectional viewpoint} the direction of motion\index{direction of motion of particles!reversal} of the source particles would reverse, as is currently understood, but charge would also reverse, despite what is currently assumed, so that electric currents\index{electric currents!invariance} (which are the source of magnetic fields) would remain unchanged as a consequence of being submitted to this additional reversal. We must, therefore, assume that a relative change between the direction of an electric field and that of a magnetic field does, in effect, take place under a properly defined time reversal operation, only it is not attributable to a reversal of the magnetic field, but rather to a reversal of the electric field.

Failure to recognize the dependence of the sign of charge on the direction of propagation in time\index{direction of propagation in time} of elementary particles gives rise to an incorrect appraisal of the response of electromagnetic fields to a reversal of time\index{reversal of time!response of electromagnetic fields}. A more consistent definition of the operation of time reversal, on the other hand, allows to avoid the troubling conclusion that certain phenomena involving electromagnetic fields would actually constitute a challenge to the necessary relational definition of discrete symmetry operations\index{discrete symmetry operations!relational definition}.

Violations of time symmetry\index{time symmetry!violations} could arise, for example, in the case where neutrons would be observed to have an electric dipole moment\index{electric dipole moment!neutron} and as such could effect a movement of precession\index{precession movement} around the direction of an external electric field, because this movement would appear to vary depending on the direction of time, but independently from the direction of the field and the direction of the electric dipole\index{electric dipole!direction}. However, while the direction of the dipole is not affected by the reversal of a neutron's spin angular momentum\index{spin angular momentum!reversal} occurring as a consequence of the reversal of time, according to my proposal it would nevertheless be reversed together with it, because it depends on the sign of the constituent particles' electrical charges, which we must now also assume to be reversed as a consequence of applying the $T$ operation.

In this context, it is not possible to assume that a reversal of time\index{time reversal!change in precession motion of neutrons} would allow a change in the precession motion of the neutron (associated with the direction of the neutron's spin) to occur \textit{independently} from the direction of its electrical dipole\index{electric dipole!reversal} in the presence of an invariant external electric field, because in fact both the spin and the dipole must be assumed to be reversed by $T$, along with the external electric field\index{electric field!reversal}. In other words, it is no longer possible to assume that while we should observe the precession movement\index{precession movement!reversal} to occur in reverse upon reversing time, the same dipole would nevertheless be interacting with the same electric field, as would happen if applying $T$ actually reversed spin, but left the direction of the dipole and the external electric field unchanged. When the appropriate time reversal symmetry operation\index{time reversal symmetry!appropriate operation} is considered, only \textit{relative} differences can occur between the direction associated with the precession motion and the direction of the dipole.

Still concerning the $T$ operation, it must be clear that it is not possible to assume that what the conventional definition of this transformation involves is a reversal of the time coordinate that reverses physical time intervals and leaves energy unchanged, even if that may not appear to disagree with the explicit definition of $T$ as it is usually conceived. Such a definition of time reversal\index{time reversal!reversal of action} would be inapplicable, simply because it would reverse the sign of action of the physical systems involved and this is certainly not desirable knowing that negative-action matter (propagating positive energies backward in time) would be an entirely different kind of matter from a gravitational viewpoint and therefore certainly cannot be involved in those processes which we currently assume to be the time-reverse of processes involving positive-action matter. This has nothing to do with the fact that a unidirectional viewpoint\index{unidirectional viewpoint} is used conventionally. It is a different problem that would be unique to the $T$ operation, despite the fact that I'm here assuming that $C$ also involves some time reversal, because charge conjugation\index{charge conjugation!space and time reversal} is simply not assumed to involve any space or time reversal conventionally and as such cannot be mistaken to involve action sign reversal.

From the viewpoint of unidirectional time, we can, therefore, only assume that the space intervals are reversed by $T$, along with the momenta, and that the time intervals, along with the energies, are left unchanged by the same operation, despite the reversal of the time coordinate. In other words, an appropriate (action-sign-preserving) time reversal symmetry\index{time reversal symmetry!action-sign-preserving operation} operation needs to reverse both momentum and space intervals together (from a unidirectional viewpoint) or leave them unchanged together (from the time-symmetric viewpoint\index{time-symmetric viewpoint}) and those constraints must be explicitly stated in the definition of the symmetry operation. This, again, illustrates how important it is to identify the variability of all physical parameters under any discrete symmetry operation, in particular for what regards the sign of charge and that of energy, in relation to the direction of propagation in time\index{direction of propagation in time}, as otherwise we may misinterpret ordinary phenomena for potentially forbidden, symmetry violating occurrences.

\section{The charge conjugation operation\label{sec:3.6}}

In the context of the preceding analysis, it becomes clear that the common assumption that time reversal amounts to simple motion (including rotation) reversal\index{reversal of motions and rotations} is what prevents a proper understanding of the nature of the charge conjugation symmetry operation\index{charge conjugation symmetry operation} from being achieved. The problem is that if we ignore the dependence of the observed sign of charges on the true direction of propagation in time\index{direction of propagation in time!true direction} of the particles carrying them, then this direction of propagation becomes impossible to assert, which explains that the existence of such a degree of freedom has conventionally been ignored altogether. Thus, I believe that the mistake we do when we consider time reversal\index{time reversal!conventional definition} as it is conventionally defined (even if we can now recognize that this error is not \textit{only} a consequence of using a unidirectional viewpoint\index{unidirectional viewpoint}), is that we do not consider an evolution according to which the direction of propagation in time\index{direction of propagation in time!reversal} of particles is really reversed, but instead consider processes for which a series of events occur forward in time, merely in the reverse order to that in which they would otherwise be observed to occur.

But given that non-gravitational charges are not affected by a reversal of the direction of propagation in time of the particles carrying them (which is distinct from the observed direction of their motion), it follows that we have a means to determine the direction of propagation in time of particles, which therefore becomes a meaningful, well-defined concept\footnote{
This conclusion is also justified by the fact that if an observer was `following' the actual direction of propagation in time of an antiparticle, then this antiparticle would appear to have the same charge as its particle counterpart, but then it would be all the other particles in the universe which would appear to have a reversed charge, which is certainly a significant change.}.
 It would be incorrect to argue that only thermodynamic phenomena\index{thermodynamic phenomena} allow to distinguish a direction of time (even in the absence of violations of $T$ symmetry), because the sign of charge\index{sign of charge!relative to direction of time} is always dependent on the direction of time relative to which it is measured. It is simply the fact that the sign of charge itself cannot be characterized in an absolute manner that prevents a direction of time from being singled out as objectively distinct, in the way thermodynamic processes may appear to allow.

Now, what makes the acknowledgment of the existence of a relationship between direction of time and sign of charge unavoidable is the recognized validity of the interpretation of antiparticles\index{antiparticles!backward-in-time-propagating particles} as particles propagating backward in time, which allows to identify reversal of time as the very cause of the apparent reversal of charge\index{charge reversal!reversal of time} occurring from the unidirectional-time viewpoint\index{unidirectional-time viewpoint}. I believe, in effect, that despite what is often suggested, the interpretation of antiparticles as particles propagating in the opposite direction of time is not merely a helpful analogy with no real significance. Given the absence of a rational motive for rejecting the existence of a fundamental time-direction degree of freedom\index{time-direction degree of freedom!fundamental} equivalent to the space direction\index{space direction!degree of freedom} degree of freedom and given the simplification made possible by the discussed interpretation of antimatter in a relativistic context, I think that we must recognize that there definitely exists a relationship between the direction of time and the sign of charge.

It must also be clear, however, that despite what is sometimes proposed, there is no equivalence between a reversal of \textit{space} directions and a reversal of the sign of charge (which could imply that antiparticles\index{antiparticles!enantiomorphic equivalent particles} are merely the enantiomorphic equivalent of their corresponding particles), even if there do occur situations when reversing the space coordinates may appear to counteract asymmetries associated with the sign of charge, because the relationship between space direction and sign of charge is, in fact, always a consequence of the existence of a relationship between the direction of space intervals and that of time intervals. In any case, if the relationship between time reversal and charge reversal which is suggested by the above-mentioned interpretation, is considered valid, then it would mean that the charge conjugation symmetry operation must actually be understood as itself involving some time reversal.

What I'm proposing, therefore, is that we should recognize that the charge conjugation\index{charge conjugation!space and time reversal} symmetry operation $C$ must actually be conceived as a combined space and time reversal operation that leaves the sign of non-gravitational charges invariant relative to the direction of time in which particles would be propagating following such a reversal. Thus, $C$ must be understood to reverse the time parameter $t$ (associated with the `position' in time), along with the physical time intervals $\Delta t$ associated with the propagation of particles, and the sign of the energy $E$ of those particles (which is reversed as a requirement of action sign invariance). But it must also reverse the space position parameter $x$, the physical space intervals $\Delta x$ associated with the propagation of particles, and the momentum $p$ of those particles (which is also reversed as a requirement of action sign invariance).

Here, again, we must recognize that the charge $q$ is actually left unchanged, along with the spin of elementary particles, from a fundamental viewpoint, even by this reversal operation we call charge conjugation. Yet it still makes sense to consider $C$ as a reversal of charge, given that, from the viewpoint of unidirectional time\index{unidirectional-time viewpoint}, non-gravitational charges would appear to be one of the few physical attributes of elementary particles which would actually be reversed by this symmetry operation, while the space and time intervals, along with the energies and the momenta would appear to remain unchanged.

This must happen for the same reasons that justified assuming that momentum and space intervals are reversed by $T$ from a unidirectional-time perspective, even though they are left invariant by this symmetry operation from the bidirectional viewpoint. Indeed, upon applying $C$, we are in a situation where all intervals and their conjugate attributes are reversed from a fundamental time-symmetric viewpoint\index{time-symmetric viewpoint}, which means that to satisfy the needs of a unidirectional perspective we must reverse the time-related parameters $\Delta t$ and $E$ again, but given the relationships that exist between the physical time intervals and the space intervals, this means that the space-related parameters $\Delta x$ and $p$ must also be reversed a second time, just as they were shown to be reversed (once) by $T$ from this unidirectional viewpoint.

If the physical time intervals and the energies must be reversed from what they really are (what they have become as a result of applying the operation in the first place) it is, therefore, due to the fact that from the unidirectional viewpoint\index{unidirectional viewpoint} we use the wrong direction of time, but given that following time in the wrong direction also implies that the space intervals are followed in the wrong direction (the relational aspect), then this actually means that the space intervals must also be reversed from what they really are (what they have become), along with the momenta. As a result, there appears to be no change to space and time intervals from applying $C$, even though it is here defined as a space and time reversal operation.

But given that charge is not a spacetime-related physical attribute\index{spacetime-related physical attribute}, in the sense that it is associated with interactions distinct from gravitation (unlike energy or momentum, which can be conceived as the charges determining the metric properties of spacetime\index{metric properties of spacetime}), it should actually remain unchanged, from the fundamental bidirectional viewpoint\index{bidirectional viewpoint}, under a space and time reversal operation, such as the properly defined $C$, which means that it would appear to be reversed, as we would normally expect, from the unidirectional-time viewpoint (because time is then followed in the wrong direction).

There is a slight difference, however, between the outcome of a properly defined $C$ operation and the expected outcome of a conventionally defined charge conjugation operation\index{charge conjugation operation!conventional definition}, because the reversal of the space and time position parameters $x$ and $t$ themselves (which now occurs from both the bidirectional- and the unidirectional-time viewpoint\index{unidirectional-time viewpoint}), even if it is without any effect on the sign of the space and time intervals\index{space and time intervals!sign} associated with the propagation of particles from a unidirectional viewpoint (given that those intervals must then be reversed a second time), actually implies that angular momentum\index{angular momentum!reversal} would appear to be reversed by $C$ (because momentum is unchanged, while the position in space is reversed). Thus, despite common expectations, a $C$-reversed process would also appear to involve reversed angular momentum or spin, which means that contrarily to what is sometimes suggested, the behavior of spin under charge conjugation\index{charge conjugation!reversal of spin} is not a mere matter of convention and its reversal (apparent from a unidirectional-time perspective) must be considered an unavoidable outcome of applying this symmetry operation.

The reversal of spin under $C$ is certainly unexpected according to the conventional approach, but from my perspective it appears natural, given that $C$ involves a reversal of time. It must be clear, though, that this reversal of spin is only apparent and does not occur at the most fundamental level of description, in accordance with the requirement that an action-sign-preserving symmetry operation\index{action-sign-preserving operation!charge conjugation} like $C$ should not reverse the sign of action\index{sign of action!angular momentum} associated with angular momentum. This must be required, even if, in general, the direction of spin is not uniquely tied to the sign of action associated with energy and momentum, because the only way spin can reverse is when either the position in space or the momentum are independently reversed and an action-sign-preserving reversal operation that reverses momentum would necessarily also reverse spatial position given that it must reverse space intervals (which is not required from the unidirectional-time viewpoint, relative to which momentum can be made to vary independently from the sign of space position, even when action is to remain positive).

We are now therefore in a situation where we must recognize that, from a certain viewpoint, charges are reversed\index{charge reversal!reversal of time} by a properly defined time reversal operation $T$, while spin and angular momentum are reversed\index{spin reversal!charge conjugation} by a properly defined charge reversal operation $C$, despite what had conventionally appeared to be required from such discrete symmetry operations. Another distinction of the proposed approach is that handedness\index{handedness!reversal} is now also reversed by $C$ from whatever viewpoint, because either momentum is reversed and spin is invariant (as from the bidirectional viewpoint\index{bidirectional viewpoint}), or momentum is invariant and spin is reversed (as from the unidirectional viewpoint\index{unidirectional viewpoint}), so that there is always a relative change between the direction of spin and that of momentum. The outcome of the proposed charge reversal operation $C$, as it was here redefined, would therefore differ from that of a properly defined $T$ operation mainly through the fact that, unlike $C$, $T$ would reverse the momentum and space intervals\index{momentum and space intervals!reversal} (from a unidirectional viewpoint), but would not reverse the handedness of particles, just as we would also expect conventionally.

Thus, both the $P$ operation and the redefined $C$ operation would alone and from any viewpoint reverse the handedness. In this context, the fact that under certain circumstances, such as when the weak interaction\index{weak interaction!particle handedness} is involved, particles of a given handedness seem to be naturally related to antiparticles with opposite handedness, could be understood to follow from the fact that the handedness is reversed by a properly defined charge conjugation operation\index{charge conjugation operation!reversal of handedness} (which still relates particles to antiparticles), so that if there can be invariance under such a symmetry operation, then reversing both charge and handedness should not be expected to produce any change. This is an important result which confirms that the suggestion, usually made on the basis of purely phenomenological considerations, that charge conjugation should perhaps involve a reversal of handedness, was in fact justified from a theoretical viewpoint.

\section{Invariance under combined reversals\label{sec:3.7}}

I think that I have appropriately justified the inevitability of the above discussed conclusions regarding which parameters should be expected to reverse under the various discrete symmetry operations\index{discrete symmetry operations!invariance of action sign} (in particular when I discussed the requirement of action sign invariance and the constraint of relational definition\index{constraint of relational definition!reversal operations} of the reversal operations), but I must nevertheless mention how remarkable it is that the described variations of physical parameters under the redefined $P$, $T$, and $C$ operations happen to be just such that they \textit{explicitly} require invariance to occur under a combined $PTC$ operation. This happens because all the parameters which are independently reversed by any of the symmetry operations are actually reversed twice when the operations are combined and this regardless of whether we are considering a unidirectional-\index{unidirectional-time viewpoint} or a bidirectional-time viewpoint\index{bidirectional-time viewpoint} (a look at Table \ref{tab:3.2} and Table \ref{tab:3.3} allows to quickly confirm this fact). Either a parameter such as $\Delta t$ is reversed twice, or else it is not reversed a single time by a properly defined $PTC$, and this actually guarantees that there is invariance under a combination of the three discrete symmetry operations\index{discrete symmetry operations!combinations}, because anything that may be reversed is reversed again and only once.

As I will explain below, what we really need in order to necessarily obtain invariance is twice a reversal of \textit{all} fundamental space- and time-related parameters (that is both the time-related parameters\index{time-related parameters} $t$, $\Delta t$, and $E$, and the space-related parameters\index{space-related parameters} $x$, $\Delta x$, and $p$) and this actually occurs following application of a properly defined $PTC$, when the appropriate bidirectional-time viewpoint is considered. Charge and spin, on the other hand, need not reverse at all from such a viewpoint, under a $PTC$ operation, as they necessarily transform independently from the action-sign-preserving discrete symmetry operations\index{discrete symmetry operations!action-sign-preserving} and only reverse as a consequence of adopting a unidirectional viewpoint and in such a case they do reverse twice, as required. This is in contrast with the conventional definition of the discrete symmetry operations\index{discrete symmetry operations!conventional definition} (described in Table \ref{tab:3.1}) according to which some parameters, like the space and time coordinates, the charge, and the spin, can be reversed a single time only by the combined $PTC$ operation.

One can understand, however, why it is that this combined symmetry operation\index{combined symmetry operation} should be expected to produce invariance, even as it is conventionally defined (as required by the CPT theorem\index{CPT theorem}). This is possible simply because, according to the conventional approach, while charge would be reversed only once (by $C$), spin would also be reversed only once (by $T$), but as one can show, there is a kind of equivalence, at least for fermions, between a reversal of the polarization state\index{polarization state!reversal} associated with spin and a reversal of charge (conceived as arising from a reversal of time) and this is why, even as it is conventionally defined, the combined $PTC$ symmetry operation would have to leave physical states invariant (although it would seem to alter the direction of space and time coordinates, which could turn out to be physically significant under particular circumstances).

It is also interesting to observe that, in the context of my revised definitions of the discrete symmetry operations, any two operations applied together is \textit{explicitly} equivalent to the remaining operation, so that applying $PT$, for example, is totally equivalent to applying $C$, which again demonstrates that charge conjugation\index{charge conjugation!space and time reversal} must really be conceived as a space and time reversal operation and that time reversal must involve a reversal of charge\index{charge reversal!reversal of time} from a certain viewpoint. What those relationships really show is that the discrete symmetry operations\index{discrete symmetry operations!redefined}, as they have been redefined, are all necessary and together sufficient to provide a complete account of the possible transformations involving a reversal of any of the fundamental attributes of matter, aside from the sign of action (in fact, charge\index{charge!independent reversal} can also be reversed independently from the direction of propagation in time\index{direction of propagation in time} of particles, as I will explain in sections \ref{sec:3.10} and \ref{sec:4.3}, but the states of matter so obtained usually do not interfere with the processes involving ordinary matter and antimatter particles).

In this regard, I must also mention that it is not possible to assume that applying either $P$ or $T$ alone, but twice, should necessarily produce invariance (in the sense that it would leave any system with no discernible change that could be related to unchanged physical parameters), despite the fact that it would appear to leave all parameters unchanged, because such a combined transformation may not leave the phase of the wave function\index{phase of wave function!fermion processes} of processes involving fermions unchanged, given that it would only be equivalent to a rotation in space\index{rotation in space} by $2\pi$ radiant (as a single space reversal introduces a $\pi$ radiant rotation and a single time reversal introduces an equivalent additional $\pi$ radiant rotation in space) and only twice such a complete rotation would necessarily produce invariance in the presence of fermions. Of course, applying $P$ or $T$ alone, twice, would already be more likely to produce invariance than applying $P$ alone or $P$ combined with $T$ only once, because at least some of the effects of applying $P$ or $T$ once would indeed be neutralized by a second application of the same operation, but the point is that, in such a case, invariance would not \textit{necessarily} follow.

The case of $C$ is different, however, given that this operation involves a reversal of both space- and time-related parameters\index{space- and time-related parameters!combined reversal}, all at once, which produces an equivalent $2\pi$ radiant rotation with only one application (therefore allowing the changes involved to be related to the incomplete transformation of fermion wave functions\index{fermion wave function!incomplete transformation}), so that applying $C$ twice reverses \textit{all} parameters twice and introduces twice a $2\pi$ rotation, that must leave even the phase of the wave function\index{phase of wave function!fermion processes} of fermion processes invariant. The $C$ operation, as I redefined it, is thus unique, because it is the only one of the three relationally distinct discrete symmetry operations\index{discrete symmetry operations!relational distinction} that reverses both space- and time-related parameters together and given its alternative definition, it follows that applying $C$ is actually and explicitly equivalent to applying a combined $PT$ operation. In this context applying $PTC$ could be considered equivalent to applying $PT$ twice, which clearly shows that the $PTC$ operation involves a reversal of all parameters twice and is also equivalent to two complete rotations, which can only produce invariance.

In fact, any one of the three basic discrete symmetry operations can be considered as equivalent to a combination of the other two, so that $T$, for example, would here be equivalent to $CP$ and $P$ would be equivalent to the combined $CT$. Therefore, applying $T$ twice would be equivalent to applying $CP$ twice, which would amount to reverse both space- and time-related parameters twice (which considered alone would have to produce invariance) and then also reverse space-related parameters\index{space-related parameters} twice (the order of application of the discrete symmetry operations\index{discrete symmetry operations!order of application} in a combined operation has no importance and only the number of times a parameter is reversed is significant). But such a combined symmetry operation\index{combined symmetry operation} would not leave fermion wave functions\index{fermion wave function!invariance} invariant, for the same reason that applying $P$ alone twice should not be expected to necessarily leave things invariant.

It remains, however, that the fact that some combinations of basic discrete symmetry operations\index{discrete symmetry operations!combinations}, which are not required to necessarily produce invariance do involve twice a reversal of some specific physical parameters, allows one to expect that an invariance which was lost when one of those fundamental operations was applied alone, can sometimes be regained by application of such combined operations. This should indeed be expected to occur given that, as I mentioned above, reversing one physical parameter twice, even if it is not guaranteed to leave all processes invariant, still allows the possibility of neutralizing some asymmetries which would occur as a consequence of a reversal of this single parameter.

What must be retained here is that there may be a difference between applying a symmetry operation twice and applying the \textit{outcome} of this combined operation\index{combined symmetry operation} only once (which would amount to effect no change), even if in certain cases, as when the operation considered is the $C$ symmetry operation, we would necessarily observe no change when the same operation is applied twice. This particularity of the $C$ operation is merely a consequence of the fact that it reverses more individual parameters all at once, so that applying it in combination with itself actually allows to leave no parameter unchanged, relative to which an asymmetry could be properly defined.

It must be understood, however, that despite their equivalence with combinations of distinct symmetry operations, each of the three basic discrete symmetry operations\index{discrete symmetry operations!three basic operations} defined above is essential to produce a reversal of one or another combination of physical attributes of elementary particles and none is more fundamental than any other. Indeed, two operations are distinct from a relational viewpoint, when one of them reverses one category of parameters, say space-related parameters\index{space-related parameters}, relative to the other category, say time-related parameters\index{time-related parameters}, while the other reverses another category of parameters, say time-related parameters, relative to the previous one, say space-related parameters, and each one of those operations is relationally distinct from yet another one that reverses both categories of parameters together and which constitutes the necessary complement to the other two operations.

\section{The significance of classical equations\label{sec:3.8}}

We can now return to the problem of understanding how it is possible for the momentum $p$ to be left unchanged by a properly defined time reversal operation $T$ which, from the most fundamental viewpoint, must be assumed to reverse time intervals $dt$, but to leave space intervals $dx$ unchanged. A problem would, in effect, appear to arise from the fact that, according to the classical equation\index{classical equations!momentum} that defines the momentum of a particle with mass $m$, we should have $p=m\,dx/dt$, which would clearly imply that if $dt$ is reversed or negative, while $dx$ is invariant or positive, then $p$ should be negative, which is contrary to my proposal that both space intervals and momentum are unaffected by a reversal of time\index{reversal of time!invariance of momentum}. But I believe that this contradiction is only apparent and a result of the fact that the classical equation for momentum is actually valid only from a unidirectional-time viewpoint\index{unidirectional-time viewpoint}, because it was originally introduced under the implicit assumption that physical attributes are always measured in the conventional forward direction of time\index{time!forward direction}.

What the classical equation is telling us is merely that, from the unidirectional viewpoint of an observer always following events in the unique direction of time associated with entropy increase and providing an account of physical attributes like momentum and space intervals in relation to that unique direction of time, relative to which time intervals $dt$ are, in effect, positive-definite, independently from the true direction of propagation in time\index{direction of propagation in time!true direction} of the particles involved, some parameters, like $dx$, which we might assume not to be reversed by $T$, are actually observed to be reversed, while $dt$ itself is kept unchanged.

Thus, if we consider things from the unidirectional-time viewpoint\index{unidirectional-time viewpoint} relative to which we are allowed to assume that the above equation is valid, then $dt$ would actually remain positive-definite, despite the reversal of time, but $dx$ would have to be assumed reversed (for reasons I have already explained), which according to this classical equation\index{classical equations!momentum} would imply that momentum is also reversed, a conclusion that agrees with the definitions I provided for the time reversal operation $T$ in section \ref{sec:3.5}, and this is certainly appropriate, given that particles submitted to such an action-sign-preserving time reversal symmetry\index{time reversal symmetry!action-sign-preserving operation} operation must have unchanged momentum relative to the apparent (but false) direction of their motion, which is satisfied when both the momenta and the physical space intervals are together reversed.

There is no contradiction here, despite the fact that we must assume that the true signs of conjugate physical attributes\index{conjugate physical attributes}, such as the space intervals and the momenta, are together invariant under a reversal of time\index{reversal of time!invariance of space intervals and momenta} from the alternative time-symmetric viewpoint\index{time-symmetric viewpoint} (according to which the sign of time intervals is itself reversed), because in such a case the classical equation no longer applies, simply because, as a conventional formula, it \textit{never} really applied to such situations. The classical relation between momentum and the space and time intervals was deduced on the basis of the validity of a thermodynamic viewpoint of time\index{thermodynamic-time viewpoint} and therefore does not apply in a context where time intervals are allowed to change sign. The classical equations are logical deductions, dependent on a certain viewpoint of time which must be considered inappropriate at the most fundamental level of description.

It is not the validity of classical equations\index{classical equations!limited validity} in a limited context that implies that the hypotheses made from a time-symmetric viewpoint (concerning the variation of the signs of physical attributes) are contrary to experimental evidence, but really the limited domain of validity of the classical equations which implies that the hypotheses which apply from a unidirectional viewpoint\index{unidirectional viewpoint} are not generally valid. We must recognize that the deductions which are made from the more appropriate time-symmetric viewpoint, regarding the variations of space- and time-related parameters\index{space- and time-related parameters} under a reversal of time, are not just theoretically well-founded, but that, under the right interpretation, they are fully supported by observations, while the variations deduced from a unidirectional-time viewpoint are explainable merely in the context where they are assumed to derive from the more fundamental bidirectional description\index{bidirectional description}.

It must be clear that, in this context, we would also be unjustified to make use of the classical equation\index{classical equations!angular momentum} for angular momentum $\bm{L}$, to which the spin of elementary particles is related, to decide what would happen to spin, from a fundamental viewpoint, under a reversal of time generated by a properly defined $T$ or $C$ operation. Indeed, the classical formula defines the angular momentum $\bm{L}=\bm{r}\times\bm{p}$ in terms of the position vector $\bm{r}$ and the momentum $\bm{p}=m\,(dx/dt)\,\bm{i}$ (where $\bm{i}$ is a unit vector pointing in the direction of this momentum), and if we assume a reversal of time intervals $dt$ to follow from both a $T$ and a $C$ reversal operation then according to this equation it would seem that $\bm{L}$ should reverse under both types of time reversal, because, either $dt$ reverses alone (as under a properly defined $T$), or else it reverses along with $\bm{r}$ and $dx$ (as under a properly defined $C$). But, as I already mentioned, and for reasons I have previously discussed, it would be incorrect to assume that angular momentum reverses under either $T$ or $C$, from the bidirectional-time viewpoint\index{bidirectional-time viewpoint!reversal of angular momentum} relative to which $dt$ does, in effect, reverse.

Yet there is no problem here, because the classical equation is only right when we consider things from the unidirectional viewpoint\index{unidirectional viewpoint}, according to which $dt$ is positive-definite, but under such conditions either $dx$ and $\bm{p}$ reverse together with unchanged $\bm{r}$ (as occurs when $T$ is applied), or else $dx$ and $\bm{p}$ are unchanged and $\bm{r}$ is reversed (as occurs when $C$ is applied and only space positions are reversed), so that in both cases spin angular momentum should actually reverse, as expected. Again, it must be emphasized that the incompatibility of the classical equation for angular momentum with the proposed definition of time reversal, as it occurs from a fundamental, bidirectional-time viewpoint, cannot be considered to imply that the proposed fundamental definition is inapplicable, because all that it means is that the classical equation\index{classical equations!limited scope} itself is of limited scope, having been developed in the context of a unidirectional perception of the evolution of physical systems, when it had not yet even been realized that there exists a fundamental degree of freedom associated with the direction of propagation in time\index{direction of propagation in time!fundamental degree of freedom}.

\section{Reversal of action\label{sec:3.9}}

The clarification of the situation which was achieved in the preceding sections, regarding the interdependence of fundamental physical attributes\index{physical attributes!fundamental} as they vary under application of any of the three discrete symmetry operations\index{discrete symmetry operations}, has allowed to establish that none of the usually considered discrete symmetry operations engenders a reversal of the sign of action\index{reversal of action}. This is of course a consequence of the fact that, regardless of the viewpoint we adopt, those symmetry operations always reverse the sign of energy in combination with the sign of time intervals associated with the propagation of particles, just as they always reverse the direction of momentum in combination with the direction of space intervals. Thus, the $T$ operation in particular, despite the ambiguity of its conventional definition, cannot be assumed to reverse the action, because, while it reverses the time coordinates and leaves the sign of energy unchanged from the unidirectional-time viewpoint\index{unidirectional-time viewpoint}, it is also implicitly assumed to leave the sign of time intervals\index{time intervals!sign} associated with the propagation of elementary particles unchanged. The role of reversing the sign of action must therefore be attributed to some symmetry operations distinct from all of those which are usually considered.

I have come to understand that there is not a unique single operation relating positive and negative action states, but that there are basically four different ways by which action can be reversed, which give rise to four different action reversal symmetry operations\index{action reversal symmetry operations!four operations}, whose four different outcomes are each related to phenomenologically distinct states of negative-action matter. If any one of those operations is applied independently from the others, it may not necessarily produce invariance. I will collectively denote those operations by the letter $M$ to emphasize the fact that they constitute a different category of reversal transformations which are unlike those already studied. The states produced by those four distinct operations can be transformed into one another by individually applying each of the three action-sign-preserving symmetry operations\index{action-sign-preserving symmetry operations} $P$, $T$, and $C$, and therefore I will denote the various action reversal operations by applying the appropriate indexes corresponding to the operations which relate the states they generate to the state which is produced by one of those action reversal operations\index{action reversal operations!basic operation} chosen arbitrarily as the basic operation, which will itself be denoted $M_I$.

The four discrete symmetry operations so defined are thus the $M_I$, $M_P$, $M_T$, and $M_C$ operations displayed in Table \ref{tab:3.4}. It must be clear, however, that the choice of which action reversal transformation must be associated with the basic operation $M_I$ is completely arbitrary and we could, for example, have defined the operation originally denoted $M_C$ to be the basic operation, which we would instead denote $M'_I$ and we would then obtain the states produced by the other three operations by applying $P$, $T$, and $C$ to the state generated by $M'_I$. That way it would appear that it is the redefined $M'_C$ which would be equivalent to the original $M_I$, while $M'_P$ would be equivalent to $M_T$, and of course $M'_T$ would be equivalent to $M_P$ and therefore we see that attribution of the indexes is purely a matter of convention. I chose the letter $M$ to denote action reversal\index{action reversal}, because the operations it represents actually alter the gravitational attributes of the matter submitted to such reversals and mass (which is usually denoted $m$) is the attribute that was traditionally associated with the gravitational interaction.

\begin{table}
\begin{center}
\begin{tabular}{c||c|c|c||c|c|c||c|c|c}
Bidir. & $t$ & $\Delta t$ & $E$ & $x$ & $\Delta x$ & $p$ & $q$ & $s$ & $h$ \\ \hline\hline
$M_I$ & $t$ & $\Delta t$ & $-E$ & $x$ & $\Delta x$ & $-p$ & $q$ & $-s$ & $h$ \\ \hline
$M_P$ & $t$ & $\Delta t$ & $-E$ & $-x$ & $-\Delta x$ & $p$ & $q$ & $-s$ & $-h$ \\ \hline
$M_T$ & $-t$ & $-\Delta t$ & $E$ & $x$ & $\Delta x$ & $-p$ & $q$ & $-s$ & $h$ \\ \hline
$M_C$ & $-t$ & $-\Delta t$ & $E$ & $-x$ & $-\Delta x$ & $p$ & $q$ & $-s$ & $-h$ 
\end{tabular}
\end{center}
\caption[Variations of physical parameters under the four relationally distinct action reversal symmetry operations, as described from the bidirectional-time viewpoint]{Variations of physical parameters under the four relationally distinct action reversal symmetry operations\index{action reversal symmetries!bidirectional-time viewpoint}, as described from the bidirectional-time viewpoint. Here I chose the basic action reversal operation $M_I$ to be that which reverses energy $E$ independently from time intervals $\Delta t$, and momentum $p$ independently from space intervals $\Delta x$. Under an equivalent definition it would be the time intervals $\Delta t$ and the space intervals $\Delta x$ which would be reversed by the basic action reversal operation, while the energy $E$ and the momentum $p$ would be kept invariant.}\label{tab:3.4}
\end{table}

From Table \ref{tab:3.4} it is possible to see that there are two different ways by which a given type of fundamental physical parameter\index{fundamental space- or time-related parameters}, either space- or time-related, can be reversed in such a way that the sign of action\index{sign of action!reversal} is reversed. We can either assume a reversal of the signs of momenta and energies relative to unchanged space and time intervals, or we can assume a reversal of the space and time intervals associated with the propagation of particles that would occur while keeping the signs of momenta and energies invariant. But given that those two different kinds of action reversal\index{action reversal!two kinds of operation} symmetry operations can be applied differently to space- and time-related parameters (you can apply one kind of reversal to space and the other to time, or vice versa, as long as you do apply any one type of reversal to each type of parameter), it follows that four different operations can reverse the sign of action.

What the $M_I$, $M_P$, $M_T$, and $M_C$ operations really involve is the reversal of an additional degree of freedom, relationally distinct from those already affected by the $P$, $T$, and $C$ operations, because, even the state obtained by applying the basic $M_I$ operation actually involves a reversal of action, which means that all possible states related by application of $P$, $T$, and $C$, including the original state obtained by application of the identity operation $I$, have their counterpart as $M$-reversed states, and under such conditions we can only conclude that we are actually dealing with a transformation that applies to a distinct property of matter. The illustration of the effects of the various action reversal operations depicted in Figure \ref{fig:3.3} allows to clearly identify this degree of freedom as the relative orientation of momentum $p$ compared to space intervals $\Delta x$ or equivalently that of energy $E$ compared to time intervals $\Delta t$, which for negative action states is the opposite of what it is for positive action states.

\begin{figure}
\begin{center}
\begin{picture}(290,290)
\put(140,0){\vector(0,1){280}}
\put(140,282){$t$}
\put(0,140){\vector(1,0){280}}
\put(282,140){$x$}
\put(140,140){\vector(1,1){60}}
\put(200,200){\line(1,1){60}}
\put(262,262){$M_I$}
\put(210,210){\circle*{3}}
\put(210,210){\vector(0,1){30}}
\put(205,242){$\Delta t$}
\put(210,210){\vector(0,-1){30}}
\put(205,170){$E$}
\put(210,210){\vector(1,0){30}}
\put(242,210){$\Delta x$}
\put(210,210){\vector(-1,0){30}}
\put(170,210){$p$}
\put(140,140){\vector(-1,1){60}}
\put(80,200){\line(-1,1){60}}
\put(10,262){$M_P$}
\put(70,210){\circle*{3}}
\put(70,210){\vector(0,1){30}}
\put(65,242){$\Delta t$}
\put(70,210){\vector(0,-1){30}}
\put(65,170){$E$}
\put(70,210){\vector(-1,0){30}}
\put(20,210){$\Delta x$}
\put(70,210){\vector(1,0){30}}
\put(102,210){$p$}
\put(140,140){\vector(1,-1){60}}
\put(200,80){\line(1,-1){60}}
\put(262,10){$M_T$}
\put(210,70){\circle*{3}}
\put(210,70){\vector(0,-1){30}}
\put(205,30){$\Delta t$}
\put(210,70){\vector(0,1){30}}
\put(205,102){$E$}
\put(210,70){\vector(1,0){30}}
\put(242,70){$\Delta x$}
\put(210,70){\vector(-1,0){30}}
\put(170,70){$p$}
\put(140,140){\vector(-1,-1){60}}
\put(80,80){\line(-1,-1){60}}
\put(10,10){$M_C$}
\put(70,70){\circle*{3}}
\put(70,70){\vector(0,-1){30}}
\put(65,30){$\Delta t$}
\put(70,70){\vector(0,1){30}}
\put(65,102){$E$}
\put(70,70){\vector(-1,0){30}}
\put(20,70){$\Delta x$}
\put(70,70){\vector(1,0){30}}
\put(102,70){$p$}
\end{picture}
\end{center}
\caption[Four different outcomes of applying each of the relationally distinct action reversal symmetry operations, as described from the bidirectional-time viewpoint]{Four different outcomes of applying each of the relationally distinct action reversal symmetry operations\index{action reversal symmetry operations}, as described from the bidirectional-time viewpoint\index{bidirectional-time viewpoint}. Here we notice that the orientation of the vectors which correspond to the signs of space and time intervals is always opposite that corresponding to the signs of momentum and energy, as we should expect to observe when action is negative. If we were to consider a unidirectional-time viewpoint\index{unidirectional-time viewpoint}, we would have to reverse all space and time intervals and all momentum and energy signs for the processes obtained by application of both the $M_T$ and $M_C$ operations, which means that all four operations would give rise to the propagation of negative energies forward in time.}\label{fig:3.3}
\end{figure}

The $C$, $P$, and $T$ operations, therefore, do not together operate a reversal of all fundamental physical parameters, because they merely reverse all parameters while leaving the sign of action\index{sign of action!invariance} invariant. The four action reversal symmetry operations\index{action reversal symmetry operations!four operations} proposed here are then the additional operations which are required to complete the set of discrete space- and time-related symmetry operations\index{discrete space- and time-related symmetry operations}, because they perform the only remaining possible changes that the conventional operations do not produce, by actually reversing the sign of momentum and energy relative to the direction of space and time intervals\index{direction of space and time intervals}.

From that viewpoint, it appears that even though they are usually ignored, the $M_I$, $M_P$, $M_T$, and $M_C$ operations cannot in fact be avoided. The fact that there are actually four distinct operations that can perform a reversal of action, on the other hand, simply means that it is not possible to associate a unique state of momentum or energy, or of propagation in either space or time, to negative-action matter and that all the different action-sign-preserving variations of the direction of fundamental physical parameters which can be applied to positive-action matter, can also be applied to negative-action matter. There must, for example, be a charge conjugation symmetry operation\index{charge conjugation symmetry operation!negative-action matter} $C$ applying independently to negative-action matter, which would therefore have its own antimatter particles, distinct from ordinary antiparticles.

In this context, it appears that the distinction that exists between matter and antimatter must be attributed essentially to the true direction of propagation in time\index{direction of propagation in time!true direction} of particles, independently from their sign of action. An antiparticle is therefore always just a particle which reversed its energy while changing its direction of propagation in time, which is not very different from the situation of a particle which reverses its momentum while changing its direction of motion in space\index{direction of motion in space}. Indeed, by reversing its momentum\index{momentum!reversal} when it changes its direction of propagation in \textit{space}, a particle is allowed to keep the sign of its momentum relative to the direction of its motion unchanged, so that its action sign\index{sign of action!invariance} is also unchanged, just like a positron retains the sign of action of the electron with which it annihilates, because the electron reverses its energy when it starts propagating backward in time (which is viewed as the annihilation process forward in time).

But a negative-action particle would be clearly distinct in this respect, as a consequence of the fact that it would not merely carry negative energy forward in time (or positive energy backward in time, which is equivalent from a unidirectional-time viewpoint\index{unidirectional-time viewpoint}, when the sign of charge can be ignored), but also negative momentum in the observed direction of its propagation in space (the momentum would point in the direction opposite the observed velocity of the particle), unlike any ordinary matter particle (including antiparticles). It must be clear, however, that according to the proposed definition of action reversal symmetry operations\index{action reversal symmetry operations} which is described in Table \ref{tab:3.4}, non-gravitational charges are assumed to be unaffected be a reversal of action\index{action reversal!invariance of charge}, just as they were left invariant by the action-sign-preserving reversal operations. Only the practical necessity of a forward-in-time\index{forward-in-time!viewpoint} viewpoint would, for negative-action matter, also imply that charges appear to be reversed when a process is submitted to an action-sign-preserving reversal of time.

Another particularity of the operations of action reversal\index{action reversal!reversal of angular momentum and spin} defined above is that angular momentum and spin are deduced to be reversed under all such relationally distinct operations, when their effects are considered from the bidirectional-time viewpoint\index{bidirectional-time viewpoint}. This is certainly just as appropriate as the invariance of angular momentum that is produced by all action-sign-preserving discrete symmetry operations\index{discrete symmetry operations!action-sign-preserving} from the same bidirectional-time viewpoint, because, as I previously mentioned, angular momentum has the dimension of an action and should therefore be reversed by an action reversal symmetry operation. Yet this doesn't mean that spin must necessarily reverse along with the action sign of elementary particles associated with their energy and momentum, because while there clearly exists a constraint that requires the momentum to be reversed independently from space intervals, when the energy is reversed independently from time intervals (due to the fact that a particle cannot have, at the same time, both positive and negative action), there is no necessity for spin to always be up for positive-action particles and down for negative-action particles (or vice versa).

The requirement that energy be reversed independently from time intervals when the momentum is reversed independently from space intervals is a simple matter of consistency, because a physical system cannot have, at once, both the gravitational attributes of positive-action matter and those of negative-action matter if, as I suggested in the previous chapter, the attractive or repulsive nature of the gravitational interaction\index{gravitational interaction!attractive or repulsive nature} between two particles actually depends on the difference or identity of their action signs. But no such a constraint exists that would forbid spin from varying independently from the sign of action associated with energy and momentum under application of the appropriate action-sign-preserving discrete symmetry operation, when the effects of such an operation are considered from the unidirectional-time viewpoint\index{unidirectional-time viewpoint}.

It may also be noted that just as is the case for the action-sign-preserving discrete symmetry operations\index{discrete symmetry operations!action-sign-preserving}, some combinations of two of the four symmetry operations describing a reversal of action\index{reversal of action!combinations of symmetry operations} are equivalent to a combination of the other two operations (in the case of the action-sign-preserving operations, one operation, which is that of charge conjugation $C$, was equivalent to the other two, but in fact this single operation was implicitly combined with the invariant operation $I$, which produced no additional change and could thus be ignored). Here, a combination of $M_I$ and $M_C$ or a $M_{I}M_{C}$ operation, would be equivalent to a combination of $M_P$ and $M_T$ and this is what allows a combination of all action reversal symmetry operations\index{discrete symmetry operations!action reversal} (or a $M_{I}M_{P}M_{T}M_{C}$ operation) to necessarily produce invariance, given that all relevant parameters are actually reversed twice by such a combined operation.

In fact, it turns out that combining any of $M_P$, $M_T$, or $M_C$ with $M_I$ produces an operation equivalent to the above defined $P$, $T$, or $C$, respectively (while a combination of $M_I$ with itself produces an operation equivalent to the identity operation $I$), so that a combination of the other two remaining action reversal operations would also be equivalent to those action-sign-preserving operations. For example, the combined $M_{P}M_{T}$ operation is mathematically equivalent to a $C$ operation, because it reverses both space- and time-related parameters once and reverses the action twice, which is equivalent to leave action unchanged.

One must understand, however, that even though applying any one action reversal operation\index{action reversal operations} twice would be equivalent to applying the identity operation $I$, such a combined operation would not necessarily produce invariance and this for the same reason that applying $P$ or $T$ twice would not necessarily leave everything invariant, despite the fact that it would also appear to be equivalent to applying the $I$ operation, which produces no change. This is, again, because applying an operation that does not reverse all physical parameters twice, even if it may appear to return a system to its original state, may still produce a change which can be characterized in a relational way, because some parameters would be reversed relative to other parameters which remain unaffected by the transformation and this may not leave the processes involved invariant.

Still regarding the conditions necessary to ensure invariance, it should be clear that simply combining a $PTC$ operation with the basic $M_I$ or any other action reversal operation, as a way to try to regain invariance which may be lost upon reversing the action (in the way we would apply $T$ to a $CP$ violating process), cannot be expected to produce invariance, given that the action sign degree of freedom\index{action sign degree of freedom} would then be reversed only once. Thus, a violation of any of the $M$ symmetries would not imply that there must be a violation of $PTC$ symmetry, as we may understand to be independently required on the basis of the fact that invariance under $PTC$ alone must itself be considered unavoidable. The appropriate generalization of the $PTC$ (or really $IPTC$) symmetry must then be recognized to be the $M_{I}M_{P}M_{T}M_{C}$ symmetry, which combines all the relationally distinct action reversal symmetry operations and which must therefore be equivalent to no change at all (because there would remain no unchanged physical parameter relative to which a change could be determined). Indeed, as indicated in Table \ref{tab:3.4}, a physical parameter may either not be reversed by any of the action reversal operations, or else be reversed by two or all of those symmetry operations, which explicitly guarantees invariance under a combination of the four operations.

Now, in order to avoid confusion, it is important to understand that the action reversal symmetry operations\index{action reversal symmetry operations} must be considered as operations distinct from one another that apply to an identical state, rather than as an identical operation that applies to different states. In such a context, it transpires that the fact that the $M_I$, $M_P$, $M_T$, and $M_C$ operations are related to one another through application of the various action-sign-preserving discrete symmetry operations\index{discrete symmetry operations!action-sign-preserving}, merely shows that the states obtained by applying the four action reversal symmetry operations are themselves related to one another through the same action-sign-preserving reversal operations that transform states with unchanged action sign into one another. What must be clear, then, is that no action reversal symmetry operation can be identified as \textit{the} action reversal operation and under such circumstances it is not possible to avoid having to consider the many operations as distinct from one another, despite the fact that all such operations can be obtained by combining, in turn, each of the action-sign-preserving symmetry operations with just one single action reversal operation.

Action reversal symmetry can, therefore, be violated to different degrees when one transforms a state of positive-energy matter into the different states of negative-energy matter which are related to one another by the redefined action-sign-preserving reversal operations\index{action-sign-preserving reversal operations} $P$, $T$, and $C$, because each of those states is related to a corresponding state of positive-energy matter by a specific action reversal symmetry operation and these operations do not necessarily produce invariance when applied separately. Thus, the $P$, $T$, and $C$ operations can be violated to different degrees by negative-energy matter (compared to how they are violated by positive-energy matter), when applied independently from one another, and this precisely because $M_I$, $M_P$, $M_T$, and $M_C$ can themselves be violated to different degrees in comparison with one another, so that they relate the different asymmetric states\index{asymmetric states!positive-energy matter} of positive-energy matter to corresponding states\index{asymmetric states!negative-energy matter} of negative-energy matter which can be asymmetric in different ways relative to one another.

The only requirement, here, is that the different states of negative-energy matter which are related to the different states of positive-energy matter by the various action reversal symmetry operations\index{action reversal symmetry operations} be subject to the same invariance under a combined $PTC$ transformation as are states of positive-energy matter, even if $P$, $T$, and $C$ are violated to different degrees by negative-energy matter, in comparison with the violations occurring for posi\-tive-energy matter. The four action reversal symmetry operations, therefore, simply allow to relate all the positive energy states, which are transformed into one another by the action-sign-preserving symmetry operations, to all the negative energy states which are transformed into one another by similar operations. Thus, despite the existence of four distinct action reversal symmetry operations\index{action reversal symmetry operations!four operations}, action reversal must really be conceived as transforming one single degree of freedom and this means that I'm justified in referring to the action reversal operations collectively as the $M$ symmetry.

At this point it must be emphasized that there is something wrong with the commonly met remark, to the effect that gravitation\index{gravitation!time reversal invariance} is invariant under a reversal of time. What I mean is that, while it is certainly true that there would be no change to the attractive or repulsive nature of the gravitational interaction\index{gravitational interaction!attractive or repulsive nature} if time was locally reversed for some physical system by a time reversal operation such as $T$, we should certainly expect a reversal of time\index{time reversal!energy sign independence} independent from the sign of energy, such as that produced by an $M_T$ operation, to exert a change on the nature of the interaction of the affected system with the rest of the universe, because such a transformation would reverse the sign of action and as I previously explained, the repulsive or attractive nature of the gravitational force between two objects depends on the relative value of their action signs (because gravitation is always attractive only for particles with the same sign of action).

But, even if we consider the reversal of time produced by an action reversal operation\index{action reversal operations!time reversal} like $M_T$ to apply to the whole universe (in which case we would have to use negative-energy matter in place of positive-energy matter when testing for invariance), the preceding discussion made clear that we should not expect to necessarily observe phenomena which would be completely similar with those of the original universe, because $M_T$ applied alone could be violated, just as any symmetry operation which is not reversing all physical parameters twice. This would also be true of $M_P$, for example, because, just as the change in the sign of time intervals produced by an $M_T$ operation can be related to an unchanged sign of energy, so the change in the direction of space intervals produced by an $M_P$ operation can be related to an unchanged direction of momentum.

Yet, the fact is that there \textit{could} be invariance under a reversal of time that does not preserve the sign of action, if the operation is applied to all particles in the universe, because in such a case the difference or the identity of the signs of action of the various particles would not be affected and this is the only aspect that would be significant from a gravitational viewpoint. This invariance, however, would apply only to the extent that there is, in effect, no violation of symmetry under exchange of positive and negative action states. Anyhow, as I have explained in the preceding sections of this chapter, simply reversing the direction of motion\index{direction of motion of particles!reversal} of particles cannot be considered to consist in a true time reversal operation in any meaningful way, so that assuming that such a transformation would leave all processes unaffected, even when gravitation is involved, could not be understood to mean that gravitation\index{gravitation!time reversal invariance} is invariant under time reversal.

Finally, I may add that even if one might be tempted to conclude, based on a certain interpretation of the generalized gravitational field equations\index{generalized gravitational field equations!interpretation} which were proposed in section \ref{sec:2.14}, that the minute imbalance responsible for the observed small, but non-vanishing positive value of the cosmological constant\index{cosmological constant!non-vanishing value} actually arises from a violation of $M$ symmetry, this would not be a valid conclusion, because, as I will explain in section \ref{sec:4.2}, this imbalance rather develops as a consequence of the fact that the rates of expansion\index{expansion rate!opposite-energy observers} of space experienced by observers of opposite energy signs are allowed to differ as time goes, even if they were initially the same and this can occur even in the absence of a violation of $M$ symmetry.

\section{Black hole entropy\label{sec:3.10}}

In this section I will show that based on certain plausible hypotheses concerning the constraints that must apply on matter collapsing into a spacetime singularity\index{spacetime singularity}, it is possible to deduce that a finite number of discrete degrees of freedom\index{discrete degrees of freedom} characterizes the state of the elementary particles which were captured by the gravitational field of a black hole\index{black hole!state of elementary particles}. As a consequence, it becomes possible to confirm the existence of an exact relationship between those matter degrees of freedom\index{matter degrees of freedom} and the binary measure of missing information or entropy\index{missing information or entropy!event horizon} which, according to the semi-classical theory of black hole thermodynamics\index{black hole thermodynamics!semi-classical theory}, should be distinctive of those situations in which an event horizon is present. Those developments will be based on the results achieved in the previous sections while deriving an improved formulation of the discrete symmetry operations\index{discrete symmetry operations!improved formulation}, as well as on a better assessment of the fundamental degrees of freedom\index{fundamental degrees of freedom!matter} of matter which must be taken into account in the derivation of the semi-classical formula for black hole entropy\index{black hole entropy!semi-classical formula}.

I will be working here under the hypothesis, now usually recognized as appropriate, that black hole entropy\index{black hole entropy!origins} originates from the fact that a certain amount of information is missing from a description of a black hole in terms of its macroscopic physical parameters of mass, angular momentum, and charge, that is encoded in the detailed configuration of certain degrees of freedom associated with microscopic, quantum-gravitational units of area\index{quantum-gravitational unit of area!black hole event horizon} on the event horizon of the object about the state of the particles it contains. It is merely the classical nature of the general-relativistic description of the event horizon of a black hole which makes it incompatible with the hypothesis that information must be encoded in the microscopic quantum-gravitational degrees of freedom\index{quantum-gravitational degrees of freedom!black hole event horizon} on such a surface.

What is not fully understood presently is how it can be that matter particles appear to be characterized by physical attributes that vary in a continuous fashion, while all the information concerning the state of any one matter or interaction field particle inside a black hole must be encoded in only a few elementary units of area\index{elementary units of area!black hole event horizon} on the event horizon of the object, each of which is known to provide only one binary unit of information\index{binary unit of information}. What is the exact nature of the microscopic degrees of freedom\index{microscopic degrees of freedom!matter in black hole} of matter which would correspond with the missing information\index{missing information!black hole event horizon} encoded in the microscopic degrees of freedom present on the event horizon of a black hole? Given the limitations imposed by the Bekenstein bound\index{Bekenstein bound} (according to which the maximum amount of information that can be obtained concerning the microscopic state of the matter\index{microscopic state of matter!inside a surface} contained within any surface is proportional to the finite number of elementary units of area\index{elementary units of area!ordinary surface} on the surface), it would appear that this question actually applies to the microscopic configuration\index{microscopic configuration!matter} of matter under \textit{any} condition, regardless of the strength of the gravitational field (or that of any other field) on the surface through which information about this exact state must be obtained.

Therefore, it seems that the problem of identifying the fundamental degrees of freedom\index{fundamental degrees of freedom!matter} of matter which are associated with the binary measure of information\index{binary measure of information!two-dimensional boundary} encoded on a two-dimensional boundary is not one that concerns only those situations in which a black hole is present, even though its significance is made more obvious when we are actually dealing with an event horizon. I think that the fact that there exists a similar limit, regarding the measure of information\index{information limit!event horizons and ordinary surfaces} that can be encoded on both event horizons and ordinary surfaces means that we must admit the reality of what is occurring beyond the limits of black hole event horizons, despite the fact that the processes involved cannot be subject to direct observation (in the sense that an observer outside a black hole cannot obtain that information before the object evaporates through the emission of thermal radiation\index{thermal radiation!black hole}).

Regardless of the practical limitations which clearly exist for determining the exact nature of whatever microscopic degrees of freedom\index{microscopic degrees of freedom!matter} of matter are to be associated with the particular measure of missing information\index{missing information!black hole event horizon} encoded on the event horizon of a black hole, this problem should nevertheless be considered a tangible one, even if only because, in the case of an ordinary surface\index{ordinary surface!microscopic state information}, information about the state of those microscopic degrees of freedom could actually be obtained. In fact, we have no choice but to recognize the reality of the microscopic degrees of freedom\index{microscopic degrees of freedom!matter and interaction fields} of matter and its interaction fields, whose states are encoded in binary form on the event horizon of a black hole, even if only because their existence is necessary for the consistency of the semi-classical theory of black hole thermodynamics\index{black hole thermodynamics!semi-classical theory}.

What the semi-classical theory requires is that there does exist information about the exact microscopic state of the matter inside a black hole\index{black hole!microscopic state of matter} and that of the interaction fields\index{interaction field!microscopic state} produced by this matter, even though this information is missing from the classical description of a black hole\index{black hole!classical description} in terms of its observable, macroscopic physical parameters. We must, therefore, assume that the information could only be obtained through a determination of the exact configuration of some microscopic, quantum-gravitational degrees of freedom\index{quantum-gravitational degrees of freedom!black hole event horizon} on the surface delimited by the event horizon of the object. The fact that a consistent theory of black hole thermodynamics\index{black hole thermodynamics!consistent theory} actually exists means that this missing information must itself exist, as otherwise there would be no sense in attributing to those objects an entropy\index{entropy!black hole} and a temperature\index{temperature!black hole}.

In the context where it is understood that, from a physical viewpoint, any information\index{information!recognizable difference} must involve a recognizable difference, this assumption is supported by the existence of a relation of proportionality between the square of the mass of a black hole and its entropy, because any distinctive features must be carried by elementary particles and when the mass of a black hole grows the number of particles it contains grows and those particles can produce stronger interaction fields mediated by more interaction bosons\index{interaction bosons}. This observation would remain significant even if it was determined that the actual microscopic degrees of freedom\index{microscopic degrees of freedom!matter in black hole} which are allowed to vary (but for which information remains unavailable) for the matter that fell into a black hole do not consist of mere particle energies. If we recognize that the information\index{information!microscopic state of matter and interaction fields} which is necessary to describe the microscopic state of matter and its interaction fields can exist even without any observer knowing what this microscopic state actually is, then we are certainly allowed to assume that it is possible for a certain measure of information to exist, even when it becomes impossible to obtain this information due to the presence of a black hole event horizon.

The possibility for information\index{information!binary units} to be encoded in binary units in the detailed configuration of certain microscopic degrees of freedom\index{microscopic degrees of freedom!black hole event horizon} on the event horizon of a black hole is dependent on one basic assumption, which is that there is a finite, maximum level of accuracy applying to measurements of spatial distances, which must also apply to measurements of surface areas, such as those which are associated with event horizons. This is actually a prediction that is common to all current quantum theories of gravitation\index{quantum gravitation theories}. Thus, it is now understood that trying to describe the microscopic state of a system at a level of definition of spatial distances and during time intervals more precise than those provided by the Planck scale\index{Planck scale} would constitute a superfluous characterization of physical reality.

What was learned, more exactly, is that two events must be considered indiscernible, from the viewpoint of any measurement, when they would occur within intervals of space and time smaller than the natural scale of quantum-gravitational phenomena\index{quantum-gravitational phenomena!natural scale}. But, despite the fact that this constraint now appears clearly inescapable, it is still often ignored, as when someone is talking about what may have happened at a time shorter than the Planck time\index{Planck time} after the Big Bang. Here I will assume that the limitations imposed by quantum indeterminacy on the very structure of spacetime, which imply the existence of a minimum meaningful spatial distance\index{distance!minimum meaningful value}, constitute a fact which will gradually become as well established as the existence of elementary particles of matter and on which further insights can, therefore, be based.

What is remarkable is that it appears that the states of the microscopic quantum-gravitational degrees of freedom\index{quantum-gravitational degrees of freedom!black hole event horizon} on the event horizon of a black hole themselves also vary in a discrete way, which means that they actually provide a binary measure for the entropy\index{entropy!binary measure}, or missing information, which characterizes those objects. Even on an ordinary surface, the relevant microscopic degrees of freedom\index{microscopic degrees of freedom!ordinary surface} can only be this or that, or yes or no, rather than assume any value from a continuous spectrum of possibilities as we go from one discrete surface element\index{discrete elements of surface!binary values of physical attributes} to the next. It appears that, not only must we accept that space is divided in elementary units on the shortest scale, but we must also recognize that the values taken by the physical attributes associated with those discrete elements of surface can only be either one thing or another and nothing in between.

Now, despite the ambiguous character of the relationship between the microscopic degrees of freedom\index{microscopic degrees of freedom!black hole event horizon} which allow information to be encoded on the event horizon of a black hole and the exact microscopic state of the matter\index{microscopic state of matter!black hole} it contains (or that of the interaction fields present on the same boundary), it must be assumed that such a unique correspondence exists. What's more, given the size of the elementary units of surface\index{elementary unit of surface} on which the missing information concerning the microscopic state of matter and its interaction fields is encoded, it appears that we would be justified to assume that the microscopic degrees of freedom\index{microscopic degrees of freedom!matter and interaction fields} which we must identify are those which would apply to a description of matter on the Planck scale\index{Planck scale}.

I think that the existence of such a correspondence between the microscopic degrees of freedom on the event horizon of a black hole and the degrees of freedom of the particles it contains must be considered unavoidable, because, even in the case of an ordinary surface\index{ordinary surface!information limit}, we can never get more information concerning what is located beyond the surface than is obtainable by observing through it. But given that there exists a limit to the accuracy of measurements that can be effected on a surface (due to the existence of a minimum meaningful spatial distance\index{distance!minimum meaningful value}), then it follows that there must also be a limit to the amount of information that could be obtained through a detailed probing of the microscopic, binary degrees of freedom on that surface and this limit should naturally be expected to be proportional to the number of discrete surface elements\index{discrete elements of surface}, through which the information must flow.

It should not come as a surprise, therefore, that it is the area of a black hole event horizon\index{black hole!area of event horizon} that provides a measure of the amount of missing information concerning the exact microscopic state of the matter\index{microscopic state of matter!black hole} contained within such an object or that of the interaction fields which are produced by this matter. What is more difficult to explain is why this constraint does, in effect, appear to be relevant to what's actually taking place within a surface, such as the event horizon of a black hole, rather than merely to what we can \textit{tell} about what's going on in there. Despite the enduring uncertainty associated with this question, I believe that the following discussion will help clarify the nature of the relationship between the microscopic degrees of freedom\index{microscopic degrees of freedom!ordinary surface} on a surface and the microscopic state of the matter located within that surface.

\bigskip

\noindent Before I undertake the task of explaining why it is that the states of the elementary particles which have been absorbed by a black hole can become so constrained that they are allowed to match the required binary measure of missing information\index{information!binary measure} which is encoded on the event horizon of the object, it would be appropriate to first recall what the semi-classical analysis of black hole thermodynamics\index{black hole thermodynamics!semi-classical analysis} has revealed. What was learned, in effect, is that for a black hole of mass $M$ with an event horizon of area $A=4\pi R_S^2$, where $R_S$ is the Schwarzschild radius\index{Schwarzschild radius} of the object (which varies in proportion to its mass, but which also depends on its angular momentum and charge), the entropy is given by $S_{BH}=A/(4 A_P)$, where $A_P=l_P^2$ is the Planck unit of area\index{Planck unit of area} given in terms of the Planck length\index{Planck length}, which is defined as $l_P=(\hbar G/c^3)^{1/2}$ and the units are chosen so that Boltzmann's constant $k_B$ is equal to unity. 

A black hole\index{black hole!area of event horizon}, therefore, has an entropy that is determined by the area of its event horizon in elementary units of area\index{elementary units of area} equal to four Planck units of area. Given that entropy is simply a measure of the information that is missing from the description of a black hole\index{black hole!macroscopic parameters} in terms of its macroscopic parameters of mass, angular momentum, and charge, it seems that the amount of information encoded in the unobserved microscopic degrees of freedom\index{microscopic degrees of freedom!black hole event horizon} on the event horizon of the object is equal to its area\index{area!natural units} in natural units, which means that one binary unit of information\index{information!binary units} is encoded in every elementary unit of area on the surface of a black hole\footnote{
Despite the fact that the minimum unit of area\index{minimum unit of area!loop quantum gravity|nn} (called the area gap\index{area gap|nn}), which is derived from loop quantum gravity\index{loop quantum gravity|nn}, is not necessarily equal to a multiple of the Planck unit of surface\index{Planck unit of surface|nn} and is rather provided by a more complex equation that contains a free parameter that must be adjusted to match the results of the semi-classical theory, I believe that it is nevertheless appropriate to assume that the phenomenologically derived elementary unit of area\index{elementary units of area!phenomenological derivation|nn} provided by the semi-classical theory of black hole thermodynamics actually constitutes the true fundamental unit of surface\index{fundamental unit of surface|nn}, because the theoretically derived area gap is obtained on the basis of purely gravitational aspects of Planck-scale physics\index{Planck-scale physics!gravitational aspects|nn} and will need to be adjusted by taking into account the details of the currently incomplete unified theory of elementary particle interactions\index{unified theory of elementary particle interactions|nn} that must apply on this scale, unlike the area of the event horizon of a macroscopic black hole\index{macroscopic black hole!event horizon area|nn}, which also varies as a function of the electric charge and the angular momentum of the object.}.

The fact that the area and therefore, also, the entropy $S_{BH}$ of a black hole is not in general proportional to its mass $M$, but rather to its mass squared (so that entropy rises faster than the matter content), is merely a reflection of the fact that the gravitational field must contribute a larger portion of the entropy\index{black hole entropy!gravitational field contribution} of a black hole as its mass rises. However, if we recognize that the gravitational interaction is mediated by elementary particles, then we have no choice but to associate the entire amount of missing information encoded on the event horizon of a black hole with the states of the elementary particles contained inside the object, which would then include gravitons\index{graviton}.

It must be clear, though, that the information which is encoded in one elementary unit of area\index{elementary units of area!black hole event horizon} on the event horizon of a macroscopic black hole about the state of some particle it contains is not necessarily all the information that's necessary to determine the state of that particle, but merely one of the possibly many binary units of information necessary to completely determine this state. In any case, if an elementary unit of area on the event horizon of a black hole can encode the state of some physical attribute of one of the particles it contains, then it is necessary to recognize that the binary nature of the microscopic degrees of freedom\index{microscopic degrees of freedom!black hole event horizon} on the event horizon of a black hole is a reflection of the existence of states of matter which can only vary in a discrete way.

So, what are exactly those degrees of freedom prevailing for the particles which are trapped by the gravitational field of a black hole? If we were assuming that the information encoded in one elementary unit of area\index{elementary units of area!black hole event horizon} on the event horizon of a black hole actually provides a complete description of the state of a most-elementary particle inside the object, then it would appear necessary to assume that this state is completely definable by one single binary unit of information\index{information!binary units}. Under such conditions we should be seeking to identify a unique physical parameter, that reverses under application of a given discrete symmetry operation\index{discrete symmetry operations}, as being the binary degree of freedom of matter\index{binary matter degree of freedom} that is related to the single binary unit of information encoded in an elementary unit of area on the event horizon of a black hole.

But if we are to assume that the same fundamental space- and time-related parameters\index{space- and time-related parameters!fundamental} characterize the states of elementary particles under all conditions, then it rather seems that each of the previously defined $T$, $P$, $C$, and $M$ symmetry operations, should have their counterpart in the information necessary to characterize the state of a particle constrained by the gravitational field of a black hole. Indeed, all of those reversal operations allow to distinguish the sign or the direction of some physically significant attribute of elementary particles and there is no \textit{a priori} reason why only a subset of those attributes should need to be taken into account in characterizing the discrete degrees of freedom\index{discrete degrees of freedom} applying at a fundamental level inside an event horizon.

It must be clear that if all the discrete symmetry operations\index{discrete symmetry operations} which were defined in the previous sections of this chapter were considered to determine one distinct degree of freedom of a particle confined by the event horizon of a black hole, then we would need not one elementary unit of area\index{elementary units of area!black hole event horizon} on the event horizon of the object to encode the state of each elementary particle, but rather four. Indeed, with two yes or no questions we can determine the direction of propagation in time\index{direction of propagation in time} of elementary particles (reversed by $T$ or not reversed), as well as their direction of propagation in space \index{direction of propagation in space} (reversed by $P$ or not reversed), which already allows to distinguish four states of matter (identity being the state where neither space nor time is reversed). The distinctions which exist between each of those four states as they appear from the bidirectional-\index{bidirectional-time viewpoint} and the unidirectional-time viewpoints\index{unidirectional-time viewpoint} are illustrated in Figure \ref{fig:3.1} and Figure \ref{fig:3.2}.

With an additional yes or no question we can then determine the sign of action\index{sign of action} of particles (reversed by $M$ or not reversed), which doubles the number of states of matter that can be distinguished. But despite the fact that the $C$ symmetry operation, is merely a combination of the $P$ and $T$ operations, it does allow to reverse one additional, independent degree of freedom, as I will explain below. Thus, it seems that at least four binary units of information\index{binary unit of information} and an equal number of elementary units of area would be required to encode the exact state of an elementary particle confined by the event horizon of a black hole, if all of its physical attributes were to vary as discrete parameters.

What I will now explain is that the hypothesis that the state of the elementary particles inside the event horizon of a black hole can only vary as discrete parameters, thereby reflecting the structure underlying the previously defined discrete symmetry operations\index{discrete symmetry operations!underlying structure}, which together allow to operate a reversal of all space- and time-related attributes\index{space- and time-related attributes} of particles, is actually unavoidable. Thus, it will be made clear that the conditions which are imposed by the first law of black hole mechanics\index{black hole mechanics!first law} (from which is derived the measure of black hole entropy provided above) are not really incompatible with the fact that under ordinary circumstances (when no black hole event horizon is present) most physical attributes of matter particles appear to vary as continuous parameters, which would appear to rule out the possibility that the missing information concerning the microscopic degrees of freedom\index{microscopic degrees of freedom!matter in black hole} of matter inside a black hole (or those of the interaction fields on its surface) could all be encoded in a finite number of elementary units of area\index{elementary units of area!black hole event horizon} on the event horizon of such an object.

\bigskip

\noindent In order to clarify the situation regarding what variations are allowed for the various physical attributes of elementary particles when matter has become confined by the gravitational field of a black hole, we may first recall that the three macroscopic physical parameters characterizing a black hole\index{black hole!macroscopic parameters} are its total mass $M$ or energy $E$, its total charge $Q$, and its overall angular momentum $J$. To those three parameters, I would add the total momentum $p$, which is not usually considered to define the macroscopic state of a black hole (given that the object can always be described in the reference system relative to which it is at rest), but which I believe provides essential information required to identify the microscopic degrees of freedom\index{microscopic degrees of freedom!black hole entropy} which must be taken into account in determining the entropy of such an object.

It must be clear that each of those macroscopic parameters must be allowed to vary not just in magnitude, but also in sign or in direction. The total mass $M$, in particular, must be conceived as being either positive or negative, depending on the overall sign of energy of the black hole\index{black hole!sign of mass and energy}. This is also an aspect that is usually not taken into consideration in the conventional theory of black hole thermodynamics\index{black hole thermodynamics!conventional theory}, but which must be recognized as a necessary assumption in the context where the existence of negative-energy matter is theoretically unavoidable.

A different question would be to ask whether the sign of energy or action is a variable parameter for the particles forming a black hole. Given that I have already argued that negative-energy matter cannot be absorbed by a positive-energy black hole, it would seem that only positive energies need to be taken into account in describing the microscopic state of the matter\index{microscopic state of matter!black hole} that was captured by the gravitational field of a positive-mass black hole. One cannot assume, however, that all black holes with a given mass sign must, at all times, be formed only of particles with the same mass sign as that of the object itself, because even if no particle of energy sign opposite that of a given black hole can cross its event horizon from the outside, it is indisputable that a positive-energy black hole with a very large radius and a rather low density could form, despite the initial presence of some comparatively small amount of negative-energy matter inside the surface that is to become its event horizon.

Thus, it is not strictly forbidden, for a macroscopic positive-energy black hole, to contain negative-energy matter, even though this matter would only be allowed to be present inside the event horizon associated with such a black hole if it was already contained inside the surface that became this event horizon before the gravitational collapse\index{gravitational collapse} occurred. But, even if a positive-energy black hole was to contain negative-energy matter, this matter would not remain in this situation for very long, because it would rapidly be expelled by a repulsive gravitational force\index{repulsive gravitational force} equivalent in strength to that which is attracting the rest of the matter toward the central singularity, so that the black hole would actually end up containing exclusively particles having the same sign of energy as its own. Therefore, even if the sign of energy of the particles contained within an ordinary macroscopic surface\index{ordinary surface!information} was to constitute a relevant microscopic degree of freedom\index{microscopic degrees of freedom!sign of energy} (transformed by the $M$ symmetry operation) which, under particular circumstances, could contribute to the measure of information encoded on this surface, it is nevertheless appropriate to assume that positive energy black holes\index{black hole!stationary state} which have reached a stationary state are composed exclusively of positive-energy particles.

In fact, if I want to explain the results of the semi-classical theory of black hole thermodynamics\index{black hole thermodynamics!semi-classical theory}, I have no choice but to assume that the energy sign of every matter particle forming a black hole\index{black hole!energy sign of matter particles} is the same as the energy sign of the object itself, because the conventional theory is based on the hypothesis that positive-energy black holes\index{black hole!stationary state} exist in a stationary state (similar to thermal equilibrium\index{thermal equilibrium}) and are not in the process of releasing negative-energy particles, which means that they must be formed exclusively of positive-energy matter. I will, therefore, assume that a knowledge of the sign of mass of a macroscopic black hole\index{black hole!sign of mass}, derived from a measurement of the polarity of its gravitational field\index{gravitational field!polarity}, allows to determine the energy sign of \textit{all} the particles it contains, whose states are reflected in the detailed configuration of the microscopic degrees of freedom\index{microscopic degrees of freedom!black hole event horizon} on the event horizon of the object.

Under such conditions, it definitely cannot be assumed that the energy sign of matter particles, which is transformed by the action reversal symmetry operation $M$, is the only binary degree of freedom \index{binary degree of freedom!particle energy sign} of matter particles whose state is encoded in the elementary units of area\index{elementary units of area!black hole event horizon} on the event horizon of a black hole, which provide a measure of its entropy, because if that was the case, then, in the most common of situations, all the microscopic physical attributes of a black hole\index{black hole!microscopic physical parameters} would be fixed by a knowledge of the sign of mass of the object and no information would be missing from the macroscopic description. It would thus follow that the entropy of a black hole would always be null, which is certainly not a desirable conclusion, given that the semi-classical theory rather requires entropy\index{entropy!maximum} to be maximum when matter collapses into a black hole.

The constraint imposed by the sign of mass of a black hole\index{black hole!sign of mass} on the energy sign of the matter particles it contains may not be so significant, however, given that a determination of the sign of energy cannot alone be considered to exhaust the requirements of a complete description of the energy state of the matter particles forming a black hole\index{black hole!state of matter particles}, because, in principle, energy must also be allowed to vary in magnitude. For now, though, we may choose to leave aside that difficulty and examine what could be the contribution of the degrees of freedom which are transformed by the other three discrete symmetry operations\index{discrete symmetry operations} to the entropy of a black hole.

You may recall that in the context of the redefinition of the discrete symmetry operations which I proposed in a previous section of this chapter, the action-sign-preserving time reversal symmetry operation $T$ must be assumed to reverse the momentum, as well as the angular momentum of particles, along with all their non-gravitational charges, even if merely from the unidirectional-time viewpoint. The action-sign-preserving space reversal operation $P$, on the other hand, has absolutely no effect on the direction of angular momentum or the sign of charge, from any viewpoint, but must be considered to reverse the momentum and the handedness of particles (as indicated in Table \ref{tab:3.3}). Thus, taken together, the $T$ and $P$ symmetry operations would transform the directions or the polarities\index{polarity!elementary particle attributes} of all physical attributes of elementary particles present inside the event horizon of a black hole which add up to produce the total momentum $p$, angular momentum $J$, and charge $Q$ parameters that characterize the macroscopic state of a black hole\index{black hole!macroscopic parameters} with a given sign of mass.

Yet this does not necessarily mean that all that must be specified in order to determine the state of all matter particles inside a black hole are the signs of the microscopic parameters transformed by the $T$ and $P$ operations (reversed or not reversed for each of the two symmetry operations) for every elementary particle it contains. There is, in effect, no \textit{a priori} reason to assume that the momentum of elementary particles (like their energy) can vary only in sign and it would rather seem that, not only must the magnitude of this parameter be allowed to vary like that of energy, but its orientation must also be allowed to vary, not in a binary way like the sign of energy, but as a continuous two-dimensional angular variable, which would forbid its complete determination through a knowledge of the value that would be taken by one single binary degree of freedom\index{binary degree of freedom} (because there are three components to the momentum of every particle).

What's more, even if, under ordinary circumstances, an action-sign-preser\-ving reversal of time intervals generated by $T$ would affect the direction of angular momentum (because it would reverse momentum independently from position), it would not affect the handedness of particles, and in the context where we are trying to identify the microscopic state of elementary particles whose momentum direction would be fixed, it appears necessary to restrict our account of the spin state of matter particles to handedness. But while an action-sign-preserving reversal of space intervals obtained by applying $P$ would reverse the handedness\index{handedness of particles!reversal} of particles (because it would reverse their momentum without also reversing their spin), we have no reason to assume that the spin could not itself be reversed, independently from momentum. In fact just such a transformation would occur as a result of applying $C$, given that this symmetry operation would reverse both the spin and the handedness of particles from a unidirectional-time viewpoint\index{unidirectional-time viewpoint}.

Thus, it would appear that the handedness of particles\index{handedness of particles!independent degree of freedom} constitutes an additional microscopic degree of freedom that must be specified independently from the other degrees of freedom which are reversed by the $T$ and $P$ symmetry operations. As a consequence, only the momentum direction\index{momentum direction!space reversal dependence} of a given particle can be assumed to be entirely determined by its dependence on the space reversal symmetry operation $P$, while only the sign of charge\index{sign of charge!time reversal dependence} can be assumed to be entirely determined by its dependence on the redefined time reversal symmetry operation $T$, when the effects of such a transformation are considered from the unidirectional viewpoint.

Despite the fact that the $T$ operation reverses both momentum and charge, it certainly seems appropriate to assume, in effect, that, as far as those microscopic physical parameters\index{microscopic physical parameters!momentum and charge} are concerned, we are actually dealing with two distinct degrees of freedom, because momentum can also be independently reversed by the $P$ operation. But even though it may appear obvious that the sign of charge should be independent from the direction of momentum, it is reassuring to observe that, from a bidirectional-time viewpoint\index{bidirectional-time viewpoint}, this hypothesis is unavoidable given that the variation of the sign of charge only occurs from a unidirectional-time viewpoint\index{unidirectional-time viewpoint} and is actually the consequence of a reversal of time intervals obtained while leaving the sign of action invariant, which would reverse the sign of energy, but leave invariant the direction of momentum.

In any case, the outcome of this reflection is that we have to accommodate three independent, microscopic degrees of freedom\index{microscopic degrees of freedom} which are the sign of charge (or equivalently the sign of time intervals), which is reversed by $T$, the direction of momentum (or equivalently the sign of space intervals), which is reversed by $P$, and the handedness of elementary particles, which depends on both the direction of spin and that of momentum, and which can be reversed independently from space and time intervals by a $C$ symmetry operation. Thus, it seems that it should be possible to derive the state of all the relevant, \textit{macroscopic} physical parameters of a black hole\index{black hole!macroscopic physical parameters} from a knowledge of both the sign of energy or action of matter particles and the sign of those three physical attributes of elementary particles\index{elementary particles!sign of physical attributes}.

It is also important to mention that despite the fact that what I'm seeking to determine are the degrees of freedom which would apply on a very small scale, at which the fundamental interactions\index{fundamental interactions!unified} would presumably be unified, I'm nevertheless assuming that the sign of non-gravitational charges would remain a parameter distinct from the sign of action (the sign of gravitational charge), because the variation of the sign of charge would here occur merely as a secondary consequence of a reversal of the direction of propagation in time\index{direction of propagation in time!reversal} applied while reversing the sign of energy, which, even under such conditions, would need to be considered a significant change, clearly distinct from a reversal of action (which may also involve a reversal of time, but which would leave the sign of energy unchanged).

An additional assumption which appears reasonable and which is also necessary in order to maintain agreement with the results of the semi-classical theory of black hole thermodynamics\index{black hole thermodynamics!semi-classical theory}, is that it is only necessary to specify the sign of one unified charge\index{unified charge} in order to completely describe the state of the non-gravitational attributes of an elementary particle which are the source of an interaction field, on the quantum-gravitational scale\index{quantum-gravitational scale}. The validity of this hypothesis is made plausible by the current trend observed in particle physics\index{particle physics}, which clearly indicates that the distinctions we observe between the various charges characterizing matter particles in their lower energy states are not fundamental, thereby implying that the particles themselves are not distinguishable other than by the sign of their unified non-gravitational charge on the higher energy scale at which the known forces unify. Anyhow, given that, aside from gravitation, the only known interaction that could exert an effect on the surface of a black hole is the long-range electromagnetic interaction\index{electromagnetic interaction!long-range interaction}, then, it appears appropriate to assume that only the sign of electric charge may affect the thermodynamic properties of macroscopic black holes\index{macroscopic black hole!thermodynamic properties}.

If we recognize the appropriateness of the above remarks and we consider as relevant only those parameters which are transformed by the redefined discrete symmetry operations\index{discrete symmetry operations}, it seems that the degree of freedom associated with the handedness of particles\index{particle handedness} would provide an additional contribution (dependent on the direction of spin\index{spin!direction}) to the measure of missing information\index{missing information!particles inside black hole} concerning the state of matter and interaction field particles inside an ordinary Schwarzschild black hole\index{Schwarzschild black hole}. This contribution would add to those provided by the degrees of freedom associated with the sign of charge and the momentum direction of particles (which are dependent on the sign of time intervals and the sign of space intervals, respectively). It would, therefore, appear that we still need at least three binary units of information\index{binary unit of information} per elementary particle to completely determine even just the \textit{signs} of all the relevant physical parameters characterizing the microscopic state of matter and interaction field particles under the influence of a macroscopic, stationary black hole\index{black hole!stationary}.

\bigskip

\noindent It is while I was trying to visualize what would happen to a negative-mass object which would find itself inside some surface that was about to become the event horizon of a positive-mass black hole that I realized that both positive- and negative-mass particles would actually be submitted to very restrictive constraints when experiencing the effects of the gravitational field which exists inside the region delimited by the event horizon of a black hole.

A negative-energy particle which would happen to be located near the center of a positive-mass black hole at the time of its formation would, in effect, be repelled outward by a force as large as that it would experience inside the most powerful conceivable particle accelerator\index{particle accelerator!black hole}. While it is being ejected outside the event horizon, the negative-energy particle would reach an arbitrarily high (negative) energy and its negative momentum would also become arbitrarily large in the direction of the forming central singularity (considered as the point where the density of the dominant form of energy reaches its maximum theoretical limit), regardless of what its initial state of motion was. The nearer to the center of the object the particle would initially be, the larger its final negative energy would be when it would emerge from the event horizon of the positive-mass black hole.

But given that, in the case of a non-rotating black hole\index{black hole!non-rotating} at least, the force which accelerates the particle is always directed away from the forming singularity, it follows that the lateral components of its momentum\index{momentum!lateral components} would become completely negligible in comparison with the component of its negative momentum directed toward the singularity. Thus, if we were to consider only negative energy particles literally emerging from a positive-mass singularity, we would end up (for a non-rotating black hole, in the absence of collisions with collapsing positive-energy matter) obtaining particles reaching the event horizon with a maximum (negative) energy and a negative momentum invariably directed opposite the positive normal to the surface of the black hole. In other words, we would always obtain (in the absence of interferences) particles in a very specific state of energy and momentum.

The process would be even more constraining for a positive-mass object in the gravitational field of a positive-mass black hole, given the rising tidal effect\index{tidal effect!black hole} which, in this case, compresses the object laterally and stretches it vertically, as it is accelerated in the direction of the singularity. In such a case, we would necessarily end up with a very focused beam of particles whose lateral momenta would again be completely negligible (at least for a black hole\index{black hole!angular momentum} with a sufficiently low angular momentum).

The force attracting the particles toward the singularity of such a black hole would grow with time, from the moment when they cross the event horizon, eventually becoming so large that the energy of the particles would become as high as it can be, while the horizontal components of their momenta would become completely negligible in comparison with the vertical component of their positive momenta oriented toward the central singularity. It can be expected, in effect, that any residual, lateral motion would simply contribute to increase or decrease the total angular momentum of the black hole\index{black hole!total angular momentum}, whose rotation is shared by all the particles that fell into the singularity (as a consequence of collisions and relativistic frame dragging\index{relativistic frame dragging!black hole}), which means that it wouldn't contribute to the entropy of the object (as a measure of the amount of missing information concerning the state of its microscopic degrees of freedom\index{microscopic degrees of freedom!black hole}).

Thus, when a positive-energy particle reaches the singularity of a positive-mass black hole, its momentum\index{momentum!unidirectional variable} (in the reference system relative to which the object is not rotating) has basically become a unidirectional variable. In fact, it is already understood that space itself becomes analogous to unidirectional time\index{unidirectional time!spatial analog} for a positive-energy particle that crosses the event horizon of a positive-mass black hole, but what I came to realize is that this means that, under such conditions, momentum\index{momentum!fixed parameter} (along with the sign of space intervals) constitutes a fixed parameter, with a unique direction and a maximum magnitude. As a result, we once again obtain a unique final state of maximum energy and invariant momentum.

The crucial assumption in the present context is that a maximum magnitude of energy\index{energy magnitude!maximum value} actually exists for elementary particles. I believe that this conjecture is appropriate, given that, from a quantum-gravitational viewpoint\index{quantum-gravitational viewpoint}, the existence of a minimum meaningful time\index{time!minimum meaningful interval} interval or spatial distance\index{distance!minimum meaningful value} implies the existence of a maximum value for the magnitude of energy of individual particles.

When we add energy to an elementary particle it is possible to more accurately determine its position, which means that its energy becomes concentrated in a smaller region of space. Therefore, if the energy of a particle was to reach the point where it is so large that it would be concentrated in one single elementary unit of area\index{elementary units of area} (which occurs precisely when the energy of the particle equals the Planck energy\index{Planck energy}), it would follow that adding even more energy to it would not increase the mass contained within the same surface, but would produce a black hole with an event horizon containing more elementary units of area, which would contain more Planck-energy particles. Hence, there would be no sense in attributing one single elementary particle at the Planck scale\index{Planck scale} an energy larger than the Planck energy.

The situation we encounter here is somewhat similar to that which we have in quantum\index{quantum!chromodynamics} chromodynamics, where, beyond a certain threshold, the energy spent at trying to separate two oppositely charged quarks\index{oppositely charged quarks} in a meson no longer contributes to increase the distance-dependent attractive nuclear force between the two quarks, but merely ends up splitting the original particle into two new mesons, thereby neutralizing the force that existed between the original two quarks.

It appears necessary to assume, therefore, that the elementary particles\index{elementary particle!maximum energy} that reach a singularity, after having been accelerated by its gravitational field, must have a maximum energy, which is the Planck energy\index{Planck energy}. Given that it is not that difficult to visualize what would happen to a positive-energy particle which would cross the event horizon of a positive-mass black hole, it is surprising that it had never been fully realized what the outcome of such a process would mean for a description of the final states of matter particles submitted to a gravitational collapse\index{gravitational collapse}. But I believe that it is crucial to recognize, in order to clarify the whole question of black hole entropy, that what happens when positive-energy matter collapses into a black hole singularity is that every elementary particle invariably acquires a maximum positive energy and a correspondingly large momentum, characterized by a unique invariant direction (to which is associated a unique sign of space intervals) which is straight toward the singularity, regardless of the initial state of motion of the particles at the time they crossed the event horizon of the black hole.

Here it must be understood that, despite the fact that the wavelength of the light emitted by a positive-energy particle which is about to be absorbed by a positive-mass black hole would be submitted to a near infinite redshift\index{redshift!near infinite} (from the viewpoint of a remote observer not moving with respect to the event horizon of the object) and would show time\index{time!nearly standing still} as nearly standing still, we are nevertheless allowed to assume that the events occurring after the particle crosses the event horizon of the black hole can be described in a physically meaningful way.

It is certainly appropriate to consider, in particular, that a particle's momentum will keep increasing in the direction toward the singularity, as I'm suggesting would be the case, because time dilation\index{time dilation} does not mean that the particle itself becomes motionless, but merely that the signals it emits are submitted to a near infinite redshift by the gravitational field of the black hole (still from the viewpoint of a remotely located observer). Thus, despite the fact that signals would show the particle as apparently immobilized on the event horizon, we must still assume that, from its own perspective, this particle actually crosses the event horizon of the black hole and in a finite amount of time acquires an energy\index{energy!maximum} which, relative to the motionless central singularity, would be maximum.

Also, the idea put forward in many textbooks on relativity theory\index{relativity theory} that, from the viewpoint of an external observer, a positive-energy particle could, in fact, acquire a negative energy after it crosses the event horizon of a black hole, would only be appropriate if we were to consider that the negative gravitational potential energy\index{gravitational potential energy!particle energy} reduces the energy of the particle itself into negative territory. But in fact, this is not an appropriate approach to defining the energy of matter (especially in the context where the true properties of negative-energy matter are understood to make such a notion implausible), because as far as this potential energy is concerned we are actually dealing with a distinct contribution to the total measure of energy, which is that of the gravitational field. The truth is that the kinetic energy of the particle itself would keep increasing to arbitrarily large values, even if this energy is compensated by a growing negative contribution to the energy of the gravitational field associated with the interaction of this particle with the rest of the matter in the black hole.

However, one may perhaps question the conclusion that momentum\index{momentum!fixed magnitude} would have a fixed magnitude for any positive-energy particle that reaches the singularity of a positive-mass black hole, in the context where the rest mass\index{rest mass!variable parameter} itself may be a variable parameter. It is true, in effect, that the magnitude of this momentum would depend on the rest mass of the particle which is accelerated in the gravitational field of the black hole, given that all masses have the same acceleration and are therefore subjected to the same velocity increase.

But in the context where we are dealing with final kinetic energies which are so large, it appears appropriate to assume that the energy associated with the rest mass\index{rest mass!negligible energy} of the particles which are reaching a singularity, after having been absorbed by a black hole, would be negligible or null\footnote{
In fact, even if that wasn't the case, from the viewpoint of a unified theory of interactions\index{unified theory of interactions|nn} it should probably be the case that a unique magnitude of mass would exist for the most-elementary particles\index{elementary particles!unique mass magnitude|nn} which compose the currently known matter and radiation particles and this would have consequences similar to those discussed here.}.
 Therefore, if the energy of those particles is the Planck energy\index{Planck energy} (the maximum physically meaningful measure of energy), we are allowed to conclude that the final magnitude of their momentum would always be the maximum theoretically meaningful value of momentum\index{momentum!maximum value} which can be carried by a massless particle (and associated quantum mechanically with the smallest meaningful measure of spatial distance\index{distance!smallest measure}). Under such conditions, we would have no choice but to recognize that the final magnitude of the momentum\index{momentum magnitude!invariant attribute} of all particles reaching a black hole singularity actually constitutes an invariant attribute, just like the direction of this momentum.

Now, I initially thought that it would be appropriate to assume that if space actually comes to an end for matter that reaches a singularity, then momentum\index{momentum!conjugate attribute to position}, as the conjugate attribute to position, simply cannot continue to evolve after the final stages of a gravitational collapse\index{gravitational collapse!final stages} and would remain in the final state I have identified above for the whole lifetime of the black hole. But some relatively recent results from loop quantum gravity\index{loop quantum gravity} appear to show that the final state of a gravitational collapse is not a singularity, but merely a state of maximum matter density\index{maximum matter density state}, which would immediately be submitted to a `quantum bounce'\index{quantum bounce} that would turn the collapse into a process of outward expansion. It is sometimes argued that this might be problematic, given that, if a black hole was to expel matter, it seems that entropy could decrease in the process.

Yet, given that black hole evaporation\index{black hole!evaporation} does involve a \textit{local} decrease of entropy for the black hole itself (independently from its environment) over its entire lifetime, then the prediction that the singularity would decay may not be as paradoxical as one might assume. In fact, I think that if black holes do evaporate, then something like the quantum bounce must occur, so that there remains no singularity in the final state, when the mass of the black hole\index{black hole mass!minimal} itself has become minimal. The perceived problem merely arises when we fail to recognize that the near infinite time dilation\index{time dilation!near infinite} that is attributable to the enormous gravitational field of a black hole implies that the process of gravitational collapse and the following quantum bounce that would take place over a finite and relatively short time from the viewpoint of the matter that falls toward the singularity, actually appear to occur over the entire lifetime of the object from the viewpoint of an external observer\footnote{
After I wrote those lines I realized that Carlo Rovelli\index{Rovelli, Carlo|nn}, one of the original contributors to the theory of loop quantum gravity\index{loop quantum gravity|nn}, has more recently arrived at the same conclusion.}.

Thus, from the viewpoint of an external observer, the whole gravitational collapse\index{gravitational collapse} process, as well as the quantum bounce\index{quantum bounce} that follows it, would take place over the arbitrarily long period of time during which the black hole would exist (until it evaporates completely). The quantum bounce, if it could be observed from outside the event horizon, would, therefore, appear as a very slow process during which all the matter particles that ever fell toward the singularity would reverse their collapsing motion and begin to expand outward, eventually reaching the horizon at which point they would be released in the form of thermal radiation\index{thermal radiation!black hole}, as the black hole slowly evaporates. But given that the energy originally contained in the objects which were absorbed by the black hole can only be released in high-entropy form through the emission of such thermal radiation, it follows that no violation of the second law of thermodynamics\index{second law of thermodynamics!violation} would be observed.

Despite what is sometimes suggested, therefore, the process that takes place following a generic quantum bounce is different from a white hole\index{white hole} (conceived as the time-reverse of a black hole gravitational collapse), because the matter which is released following the bounce has high entropy and does not consist in the same macroscopic objects that originally fell through the event horizon. Yet it must also be the case that it is as a result of the quantum bounce that the energy of the particles which were captured by the gravitational field of a black hole can be released in higher entropy form as Hawking radiation\index{Hawking radiation}. It may well only be the widespread ignorance of the unavoidable character of this conclusion that prevents us from acknowledging the fact that the information\index{information!state of particles inside black hole} concerning the state of the particles which were absorbed by a black hole is not really lost as a result of the evaporation process, but is actually contained in the microscopic state of the emitted radiation.

It is simply the fact that the prediction that a black hole must emit radiation was originally derived from a semi-classical theory and is, therefore, dependent on the hypothesis that the gravitational field\index{gravitational field!microscopic uniformity hypothesis} is microscopically uniform and itself devoid of small-scale structure that explains that Hawking's\index{Hawking, Stephen} original approach cannot account for the presence of information in the flow of energy that emerges from a black hole as it decays\footnote{
What is happening according to the most knowledgeable experts \cite{Perez-1} is that quantum correlations with the discrete, high-energy degrees of freedom\index{discrete high-energy degrees of freedom|nn} present in the radiation remain unobservable on a macroscopic scale and could only be detected by performing observations on the Planck scale\index{Planck scale|nn} (at very high energy) and this is why the radiation appears to remain thermal from a semi-classical perspective, despite the fact that it does contain the information that was encoded in the microscopic degrees of freedom\index{microscopic degrees of freedom!black hole event horizon|nn} on the event horizon of the decaying black hole.}.
 But it must be clear that, for an observer outside a black hole, the information about the matter that was captured by its gravitational field does not vanish from reality, but actually remains encoded in the microscopic degrees of freedom\index{microscopic degrees of freedom!black hole event horizon} on the event horizon at all times, because, due to time dilation\index{time dilation!black hole} and space contraction\index{space contraction!black hole} (in the radial direction), a matter particle would appear to spend most of its time in a particular location on the event horizon\index{event horizon} itself, right until the moment when its energy is released as thermal radiation\index{thermal radiation!black hole}, after a period equivalent to the entire lifetime of the black hole\index{black hole!entire lifetime} (from the viewpoint of an observer outside a black hole, everything that happens to the particles as they reach the singularity would seem to take place on the event horizon itself).

As a result, any entanglement of a particle\index{particle entanglement!black hole} outside a black hole with a particle crossing its event horizon remains in effect right until the time when the energy of this very same particle is released, after an arbitrary long period corresponding to the lifetime of the object. Therefore, it is not appropriate to assume that the outside particle becomes entangled with the whole thermal radiation or that this entanglement is necessarily lost whenever a particle crosses the event horizon of a black hole. But all difficulties we encounter while trying to account for the conservation of information\index{conservation of information} in the presence of black holes arise from assuming one or another of those two incorrect assumptions, which means that the perceived problems are actually the result of a misunderstanding. In fact, there never really was any paradox associated with the loss of information\index{information!loss paradox} for matter that falls into a black hole, because, as I mentioned above, we do know that classical general relativity and the hypothesis that the spacetime structure is smooth and uniform down to the smallest scale is not valid in a quantum-mechanical context, while it is also clear that a certain measure of information must be associated with any microscopic structure that exists on such a scale.

In any case, if we are willing to recognize that the description provided by current quantum theories of gravitation\index{quantum gravitation theories} constitutes the most accurate account of the process of black hole gravitational collapse\index{gravitational collapse} that one can presently derive, then it follows from the preceding analysis that over the entire lifetime of an ordinary Schwarzschild black hole\index{Schwarzschild black hole!lifetime}, the particles with the same energy sign as that of the object would spend most of their time either falling, with maximum momenta directed toward the singularity, or rising, with maximum momenta directed in the exact opposite direction (as would occur after the quantum bounce\index{quantum bounce} takes place). This is because, the time dilation\index{time dilation!maximum} effect is maximum when the particles are near the singularity and are either falling onto it with maximum energy or moving away from it with maximum energy, so that from an external viewpoint they would, in effect, appear to spend most of their time in either one of those two states.

The discrete nature of the microscopic quantum-gravitational degrees of freedom\index{quantum-gravitational degrees of freedom!black hole event horizon} on the event horizon of such a black hole would, therefore, be a reflection of the fact that the momentum states\index{momentum state of particles!discrete variable} of the matter particles it contains themselves constitute discrete variables. The detailed configuration of those microscopic event-horizon degrees of freedom must, in effect, be considered to reflect the momentum state of every particle contained in the black hole at the time immediately before or immediately after it reaches the singularity of the object. But given that, from an external viewpoint, the reversal of a particle's momentum occurs during the most time-dilated portion of the gravitational collapse\index{gravitational collapse!most time-dilated portion} process, while the details of this process are unconstrained and unpredictable, then it follows that it is not possible for an observer outside the black hole to tell in which of those two possible momentum states any given particle is at any given time, if she cannot obtain detailed information about the configuration of the microscopic degrees of freedom on the event horizon of the object, because the precise time at which a particle reverses its motion constitutes a random variable and could be any particular moment during the whole lifetime of the black hole.

If we agree on the plausibility of the above conclusions concerning the state of any particle involved in a gravitational collapse\index{gravitational collapse!simplest kind} of the simplest kind, then we need to recognize that it is not just the sign of energy of the matter particles inside a macroscopic, stationary black hole\index{black hole!stationary} that would be constrained by the gravitational field of the object. It now appears that not only must the signs of energy of the matter particles which are present in the final stages of a gravitational collapse\index{gravitational collapse!final stages} be considered to be completely determined by a knowledge of the sign of mass of the object, but the magnitude of those energies\index{magnitude of energy!invariant parameter} is also to be considered an invariant parameter, which, therefore, cannot contribute to the entropy of the black hole.

What's more, it would appear that the momentum of the particles\index{particle momenta!binary degrees of freedom} which are constrained by the presence of an event horizon can only point in one of two opposite directions, which means that it can only vary as a binary degree of freedom (along with the sign of space intervals which is transformed by the space reversal operation $P$). It should be clear, therefore, that while the energy states of the matter particles which were captured by the gravitational field of a non-rotating, stationary black hole without charge cannot contribute to the entropy of such an object (once its own sign of energy is known), the momentum of each of those particles does contribute one binary unit to the measure of missing information encoded in the quantum-gravitational degrees of freedom\index{quantum-gravitational degrees of freedom!black hole event horizon} on the event horizon of such a black hole and also imparts microscopic structure on the gravitational field\index{gravitational field!microscopic structure}, thereby allowing it to contribute to the entropy of the object as well.

It seems appropriate to conclude, therefore, that three binary degrees of freedom can freely vary for every matter particle\index{matter particles!three binary degrees of freedom} in the final stages of a gravitational collapse\index{gravitational collapse!final stages} into a macroscopic, stationary black hole\index{black hole!stationary} with zero total charge and no angular momentum. The first is the vertical momentum direction\index{vertical momentum direction!particles inside black hole} (or the sign of space intervals) and the other two are the ones I previously identified as being the sign of charge (or the sign of time intervals) and the handedness\index{handedness!matter particles} of matter particles. All three attributes could potentially contribute to the measure of missing information concerning the microscopic state\index{microscopic state!matter inside black hole} of matter inside such a black hole.

Once the momentum direction of a given particle is fixed it appears, in effect, that the handedness of the particle\index{particle handedness} would still be allowed to vary and should, therefore, contribute one additional binary degree of freedom\index{binary degree of freedom} which would vary upon a reversal of the direction of spin relative to this momentum direction. But it also appears that there should be a contribution to the entropy of a black hole by the sign of charge of the particle, that varies upon a reversal of its direction of propagation in time\index{direction of propagation in time}. Even if we assume that there exists only one type of charge, with one given polarity, for the elementary particles present on the unification scale\index{unification scale!one type of charge}, certainly information should be needed to specify whether this charge is positive or negative from the viewpoint of unidirectional time\index{unidirectional-time viewpoint}.

I have explained why three binary units of information\index{binary unit of information} would appear to be necessary and sufficient to determine the state of all unconstrained physical attributes of any positive-energy particle under the influence of a non-rotating, positive-energy black hole\index{black hole!non-rotating} that reached a stationary state and therefore it may appear that three elementary units of area\index{elementary units of area} are required to encode the state of each matter and interaction field particle inside a black hole, because the semi-classical theory of black hole thermodynamics\index{black hole thermodynamics!semi-classical theory} allows each elementary unit of area on the event horizon of such a black hole to encode only one binary unit of information.

The truth, however, is that if we want to explain the results of the semi-classical theory in the case of extremal black holes\index{extremal black hole} with either maximum electric charge or maximum angular momentum and spin, then what we need is not three, but rather four binary degrees of freedom\index{binary degrees of freedom!particle inside black hole} to characterize the state of every elementary particle inside an ordinary black hole. To see what justifies this conclusion it is necessary to first recall how the above formula for the entropy of a black hole was derived. The most basic requirement that must be obeyed by any stationary black hole is the following equation, called the first law of black hole mechanics\index{black hole mechanics!first law}, that relates the observable macroscopic parameters of such an object:
\begin{equation}\label{eq:3.1}
\frac{\kappa}{2\pi}\frac{1}{4}dA=dE-\Omega dJ-\Phi dQ \, ,
\end{equation}
where $A$ is the area of the event horizon of the black hole in Planck units\index{Planck units}, $E$ its total energy, $\Omega$ its angular velocity\index{angular velocity!black hole}, $J$ its angular momentum\index{angular momentum!black hole}, $\Phi$ its electrostatic potential\index{electrostatic potential!black hole}, $Q$ its total electric charge\index{electric charge!black hole}, and finally $\kappa$ its surface gravity\index{surface gravity!black hole} (which is not determined only by the strength of the gravitational field on the surface of the object, but also by that of the electromagnetic field and by the magnitude of the angular momentum of the black hole).

Given the obvious similarity between this equation and the conventional first law of thermodynamics\index{first law of thermodynamics}, which applies to all thermodynamic systems in the absence of black hole event horizons, one is led to suggest that the following relation holds as a thermodynamic requirement for any black hole\index{black hole!stationary state} in a stationary state:
\begin{equation}\label{eq:3.2}
T_H dS_{BH}=dE-\Omega dJ-\Phi dQ \, ,
\end{equation}
where $T_H$ would be the temperature of the object and $S_{BH}$ its entropy. But it was found, based on independent theoretical evidence, that a black hole with a surface gravity $\kappa$ would actually have a temperature $T_H=\kappa/(2\pi)$ (called the Hawking temperature\index{Hawking temperature}), which means that the entropy of the object that enters the above equation (\ref{eq:3.2}) is $S_{BH}=A/4$, where the units are chosen so that Boltzmann's constant $k_B=1$.

One interesting aspect of the above equations is that they indicate that the area of the event horizon\index{event horizon area!black hole charge and angular momentum} of a black hole, which is a measure of its entropy, can also vary as a result of changes to the electric charge or the angular momentum of the object (effected while keeping the other parameters constant). In fact, equations can be derived \cite{Jacobson-1} from this fundamental thermodynamic relation\index{fundamental thermodynamic relation} which determine exactly how the area of a stationary black hole of mass $M$ varies as a function of its charge $Q$ and its angular momentum per unit of mass $J/M$:
\begin{equation}\label{eq:3.3}
A=4\pi(2M^2+2M\mu-Q^2) \, ,
\end{equation}
in which $\mu$ is given by
\begin{equation}\label{eq:3.4}
\mu=\sqrt{M^2-Q^2-(J/M)^2} \, .
\end{equation}
For a non-charged and non-rotating black hole\index{black hole!non-charged and non-rotating}, $Q$ and $J$ equal zero and $\mu$ simply equals $M$, so that the area of the event horizon of the object is determined by its mass alone. Here it appears to be the fact that $Q$ produces an interaction field distinct from the gravitational field, while $J$ merely reduces the strength of the gravitational field (by producing an oppositely-directed inertial gravitational field\index{inertial gravitational field!rotation} whose existence is attributable to the rotation of the black hole relative to the global inertial reference system\index{global inertial reference system}) that explains why $Q$ is present twice in the general equation for the area $A$, while $J/M$ is present only once (in the $\mu$ parameter).

Now, it must be clear that when $Q$ is maximum and all particles inside a black hole of mass $M$ have the same sign of charge, the strength of the electromagnetic field\index{electromagnetic field strength!black hole event horizon} on the event horizon of the object is maximum, but the amount of missing information\index{missing information!electric charge signs} concerning the electric charge sign of the matter particles it contains is null. Under more general conditions, however, it is not possible to obtain information about the sign of charge of a specific matter particle inside a black hole from observable properties of the electrostatic potential\index{electrostatic potential!black hole event horizon} $\Phi$ on the event horizon of the object, which must be uniform over its entire surface, just like the gravitational field.

This means that the hypothesis that black holes\index{black hole!no hair hypothesis} have no `hair' remains valid, from a macroscopic viewpoint, even for electrically charged black holes, as required by the zeroth law of black hole mechanics\index{black hole mechanics!zeroth law}, interpreted as a constraint of thermal equilibrium\index{thermal equilibrium}. Nevertheless, it must be recognized that, under ordinary circumstances, the microscopic degrees of freedom of the electromagnetic field\index{electromagnetic field!microscopic degrees of freedom} on the surface of a black hole do contain information, even though the amount of it that is missing must decrease when the charge of a black hole\index{black hole!charge} with a fixed value of mass is growing and a larger proportion of the particles it contains have the same sign of charge, because the microscopic structure that may be present in a static electromagnetic field\index{electromagnetic field!microscopic structure} is imparted by local variations in the sign of charge.

Thus, there can be entropy in the distribution of charge signs inside a black hole only as long as the total charge of the object is not maximum, like the electrostatic potential on its surface, because unlike the energy sign of matter particles, which doesn't vary for a stationary black hole\index{black hole!stationary}, and similarly to the vertical momentum direction\index{vertical momentum direction!particles inside black hole} of particles, the direction of propagation in time\index{direction of propagation in time} of particles (which determines the sign of their charge from a unidirectional viewpoint\index{unidirectional viewpoint}), may vary for matter particles inside a black hole\index{black hole!non-maximum charge}, but only when the total charge of the object and the electrostatic potential\index{electrostatic potential!black hole surface} on its surface are not maximum. One can, therefore, expect that it is only when the charge of a black hole is null, that one binary unit of information\index{binary unit of information} is missing concerning the microscopic state of each and every matter particle inside the object, that would allow to determine the sign of their charge, or the direction in which they are propagating in time and from which depend their polarities from a unidirectional-time viewpoint.

As far as non-gravitational interactions are concerned, the situation we usually encounter is one where the total charge is, in effect, null, even when large amounts of positive and negative charges are present inside a black hole. Such situations are analogous to that which is occurring when the entropy associated with the energy of matter particles is merely submitted to the constraint of the Bekenstein bound\index{Bekenstein bound} and both positive- and negative-energy matter can be present together inside an ordinary surface.

But contrarily to what happens with gravitational entropy\index{gravitational entropy} under conditions where the energy sign of matter particles is subject to more variation, the entropy associated with the electromagnetic field\index{electromagnetic field!entropy} actually increases when the amount of missing information\index{missing information!sign of charge} concerning the sign of charge of matter particles inside a black hole increases (due to a more balanced distribution of electric charge signs\index{electric charge signs!balanced distribution} inside the object), despite the fact that the field strength diminishes on its surface, because the field itself necessarily contains more microscopic structure\index{microscopic structure!electromagnetic field} when there is more variation in the sign of charges inside the black hole, just like it does, in general, when the density of the matter carrying those charges is growing (I will have more to say about this in section \ref{sec:4.7}).

A different question would be to ask whether the handedness of particles\index{particle handedness!black hole entropy} contributes to the entropy of a black hole\index{black hole!angular momentum} independently from its angular momentum. Given that the spin of matter particles is a particular instance of angular momentum, it is necessary to assume that the spin\index{spin!particles inside black hole} of particles inside a black hole may contribute to enhance the angular momentum of the object, when more particle spins inside the black hole are oriented in the same direction as its angular momentum. But whereas the angular momentum, as a macroscopic parameter of black holes, never contributes any microscopic degrees of freedom, the spin of particles may contribute to the entropy of such an object, and this is what differentiates the two attributes under conditions where the spins would be oriented more or less randomly. If this was not the case, then only the non-gravitational charge of matter particles and its related interaction field would contribute to the entropy of a black hole\index{black hole entropy!non-gravitational charges}, along with the momentum direction of matter particles and the gravitational field they produce, but under such conditions it would become impossible to obtain the right measure of entropy in the case of extremal black holes\index{extremal black hole}, as I will soon explain.

It seems, therefore, that the handedness of particles\index{handedness of particles!missing information} should, in effect, contribute one bit of missing information per elementary particle, that could be encoded in one elementary unit of area\index{elementary units of area!black hole event horizon} on the event horizon of a black hole, when the sum of the spins of individual particles inside the object is null, and this contribution would arise from an absence of knowledge about the spin direction of those particles.

The situation with handedness is somewhat different from that involving the momentum of matter particles inside an ordinary Schwarzschild black hole, however, because while the vertical momenta\index{vertical momenta!particles inside black hole} usually add up to zero, like the spins, the strength of the gravitational field\index{gravitational field!microscopic structure} on the surface of the object is not null under such conditions, but rather maximum, despite the fact that it does contain microscopic structure due to the variable momentum direction of particles, because the density of positive or negative matter energy is maximum. This is unlike the inertial gravitational field\index{inertial gravitational field!spin angular momentum} attributable to the angular momentum produced by the spins, which contains no microscopic structure\index{microscopic structure!inertial gravitational field} when all the spins are oriented in the same direction and provide a maximum contribution to the angular momentum of a black hole\index{black hole!angular momentum}. This inertial gravitational field, therefore, only adds to the entropy contributed by the handedness of matter particles when the portion of angular momentum attributable to the spin of particles is not maximum and the spins are not all aligned inside the black hole.

Now, in section \ref{sec:4.3} I will explain that it is necessary to assume that the sign of charge of a given type of elementary particles\index{elementary particle!sign of charge}, say positive-action electrons, can vary independently from the direction of propagation in time\index{direction of propagation in time} of the particles, contrarily to what is usually assumed based on what appears to be undeniable observational evidence in favor of the opposite conclusion. At this point into my discussion, the hypothesis that there may exist electrons\index{electrons!positive charge propagating forward in time} propagating a positive electric charge forward in time (which would be different from a positron propagating a negative charge backward in time) may appear to lack any experimental justification or theoretical motivation, but keep in mind that I will provide such a justification when I will further discuss the issue in chapter \ref{chap:4}.

Anyhow, once it is recognized that an electron propagating a positive charge and a negative energy backward in time would differ from an ordinary electron propagating a negative charge and a positive energy forward in time (despite the apparent equivalence of the signs of charge and energy of those two types of particle from a unidirectional-time viewpoint\index{unidirectional-time viewpoint}), then it follows that an additional microscopic degree of freedom of matter must be taken into account which has to do with the polarity of what we may call the \textit{bidirectional charge}\index{bidirectional charge!polarity} of particles (that which is independent from their direction of propagation in time). This physical attribute would then be transformed by a bidirectional-charge-reversal symmetry operation\index{bidirectional-charge-reversal symmetry operation}, which may be denoted $C_B$, and would allow to differentiate between ordinary baryonic matter and baryonic dark matter\index{baryonic dark matter}.

In such a context it would still be possible to determine the direction of propagation in time\index{direction of propagation in time} of a given particle, even if the sign of charge of this particle can also vary independently from its direction of propagation in time, precisely because matter with reversed bidirectional charges\index{reversed-bidirectional-charge matter!dark matter} would need to be dark from the viewpoint of an observer made of ordinary matter (for reasons that will be clarified in section \ref{sec:4.3}). Thus, the interaction field whose polarity varies as a function of the sign of charge, or the direction of propagation in time of ordinary matter particles, is not the exact same field as that whose polarity varies as a function of the sign of charge, or the direction of propagation in time of particles with reversed bidirectional charges\index{reversed-bidirectional-charge particles!interactions} (which do interact among themselves like ordinary matter particles), as if two different types of charge were involved, even at the most fundamental level of description.

An additional contribution to black hole entropy\index{black hole entropy!bidirectional charge signs} must, therefore, arise when information is missing about the bidirectional charge sign of elementary particles that would add to that which concerns their direction of propagation in time. Under such conditions the bidirectional charge sign\index{bidirectional charge sign!missing information} of elementary particles must contribute one additional bit of missing information per elementary particle, that must be encoded in one elementary unit of area\index{elementary units of area!black hole event horizon} on the event horizon of a black hole containing those particles.

Thus, while the distinction between elementary particles and antiparticles with non-reversed bidirectional charges contributes one bit per elementary particle to the amount of missing information encoded on the event horizon of an electrically neutral black hole, which decreases as the charge of the object increases and the strength of the component of the electromagnetic field\index{electromagnetic field!non-reversed bidirectional charges} produced by those particles grows, the distinction between such particles and reversed-bidirectional-charge particles and anti-particles contributes one additional bit per elementary particle to the entropy of an electrically neutral black hole\index{black hole!electrical neutrality}. This contribution can only decrease when the difference between the magnitudes of the total positive and negative bidirectional charges of the object grows. Therefore, even if the positive and negative bidirectional charges of a black hole are maximum, if their magnitudes are equal, the entropy contained in the bidirectional charge of particles is still maximum, because there are as many particles with a positive bidirectional charge as there are with a negative bidirectional charge and under such conditions only the entropy associated with the direction of propagation in time\index{direction of propagation in time!black hole entropy} of particles is null.

It is important to understand that even when more positive- than negative-bidirectional-charge particles are present inside an event horizon, it is not easier to tell what the bidirectional charge sign\index{bidirectional charge sign!direction of propagation in time} of elementary particles actually is, unless the difference between the number of particles propagating a positive bidirectional charge in a given direction of time and the number of particles propagating the same bidirectional charge in the opposite direction of time, is larger than the difference between the number of particles propagating a negative bidirectional charge in a given direction of time and the number of particles propagating the same bidirectional charge in the opposite direction of time, because opposite-bidirectional-charge particles\index{opposite-bidirectional-charge particles} remain indistinguishable from the viewpoint of gravitation.

It is only when all particles with a given bidirectional charge sign propagate in the same direction of time and only particles with that bidirectional charge sign are present inside a black hole, that the strength of a single one of the two non-gravitational fields\index{non-gravitational field!maximum strength} can be maximum and the entropy associated with both the bidirectional charge sign\index{bidirectional charge sign!black hole entropy} of elementary particles and their direction of propagation in time can be minimum, along with that of the interaction fields produced by reversed and non-reversed bidirectional charges.

If those non-gravitational fields\index{non-gravitational interaction field!absence of microscopic structure} cannot contribute to the entropy of a black hole, it is because they contain no microscopic structure, given that they are produced by a perfectly homogeneous distribution of charges which all have the same polarity. Thus, it is not just the measure of missing information contained in the sign of bidirectional charge\index{bidirectional charge sign!missing information} and the direction of propagation in time\index{direction of propagation in time!missing information} of matter particles that must be null for an extremal black hole\index{extremal black hole!maximum charge} with a maximum charge and a maximally strong electrostatic field, but also that which could have been contained in the interaction fields produced by those charges. It is only under conditions where the total positive and negative bidirectional charges of a black hole\index{black hole!matter degrees of freedom} are null that the matter degrees of freedom associated with both the direction of propagation in time and the bidirectional charge sign of particles provide a maximum contribution to the entropy of the object, along with their associated interaction fields.

It is merely the fact that the bidirectional charge of particles inside a black hole\index{black hole!particle bidirectional charge} cannot be determined from observable physical properties of the object, when the magnitudes of its total positive and negative bidirectional charges are equal, that requires its entropy to reflect this absence of knowledge, because in principle one would be allowed to tell the sign of bidirectional charge\index{bidirectional charge sign!particles inside black hole} of all the particles inside an electrically neutral black hole\index{black hole!electrical neutrality} if one knew that the object formed out of positive bidirectional charges only, even if there are as many of those charges propagating in the future as there are propagating in the past direction of time. Knowledge about the sign of bidirectional charge of the matter out of which an electrically neutral black hole formed wouldn't allow one to tell what the direction of propagation in time\index{direction of propagation in time!particles inside black hole} of the particles it contains is, but only knowledge about the direction of propagation in time of particles inside a black hole\index{black hole!total charges} allows to determine its total positive and negative bidirectional charges, so that if the entropy of a black hole only varies as a function of the magnitude of its charges, then knowledge of the sign of bidirectional charge of the matter particles it contains wouldn't affect its entropy.

What requires the entropy of a black hole associated with the bidirectional charge sign\index{bidirectional charge sign!black hole entropy} of particles to be maximum, regardless of the actual ratio of positive- to negative-bidirectional-charge particles, except when the magnitude of the positive bidirectional charge of the object is actually larger or smaller than the magnitude of its negative bidirectional charge, is the fact that a reversal of the bidirectional charge\index{bidirectional charge!reversal} of all particles inside a black hole would have no observational consequences, even if all those particles had the same sign of bidirectional charge, unless those conditions are met. An increase in the proportion of particles propagating in the same direction of time always contributes to reduce the entropy of a black hole, but it is only when this increase is not the same for positive- and negative-bidirectional-charge particles that the portion of entropy attributable to the bidirectional charge sign of particles is reduced from its maximum value.

Unlike the gravitational field\index{gravitational field!entropy}, which may contain entropy when the density of positive or negative matter energy is maximum (despite the fact that the sign of energy of matter particles does not vary under such conditions), the non-gravitational interaction field\index{non-gravitational interaction field!maximum charge density} produced by a maximum charge density (which corresponds to the case where the bidirectional charges of all matter particles have the same polarity and the same direction of propagation in time) doesn't contain entropy, because its strength grows when the entropy contained in the sign of bidirectional charge and the direction of propagation in time of matter particles decreases. When the charge becomes maximum, the microscopic degrees of freedom of the non-gravitational field attributable to the direction of propagation in time and the sign of bidirectional charge of matter particles, which could have contributed to the entropy of the object, become completely specified, as all of the field sources have the same polarity and propagate in the same direction of time, and this is what explains that the field itself does not contribute to the entropy of such a black hole.

What emerges from the above considerations, basically, is that the contribution by the sign of non-gravitational charges\index{non-gravitational charge sign!black hole entropy} and the associated interaction fields to the entropy of a black hole may be as much as two times larger than what one would expect if only particles with non-reversed bidirectional charges\index{bidirectional charge!non-reversed} existed, because information may be missing about both the bidirectional charge sign\index{bidirectional charge sign!missing information} and the direction of propagation in time\index{direction of propagation in time!missing information} of particles. This conclusion is actually much more unavoidable than one might expect, not only because the existence of particles with reversed bidirectional charges\index{reversed-bidirectional-charge particles} is itself unavoidable, but because it is only under such conditions that one can obtain the right measure of missing information or entropy from the first law of black hole mechanics\index{black hole mechanics!first law} when extremal black holes\index{extremal black hole} are involved, as I will soon explain.

I have already made clear that once the sign of energy of a black hole is fixed, the sign of action of the particles it contains (which is transformed by the $M$ symmetry operation) is also determined, which would normally mean that three discrete variables\index{discrete variables!elementary particle states} remain undetermined concerning the exact state of each of the particles under its influence, which are the direction of a particle's momentum\index{momentum direction} (upward or downward), or equivalently the sign of space intervals\index{sign of space intervals} associated with its motion (which is transformed by the $P$ symmetry operation), the particle's handedness\index{particle handedness} (spin-up or spin-down), which can be independently transformed by a $C$ symmetry operation, and the non-gravitational charge sign\index{non-gravitational charge sign} of the particle (positive or negative), or equivalently its direction of propagation in time\index{direction of propagation in time} (which is transformed by the $T$ symmetry operation and which determines its particle or antiparticle nature from a unidirectional-time perspective\index{unidirectional-time perspective}).

But if a quantum theory of gravitation\index{quantum gravitation theories} is to eventually constitute a unified theory of all interactions\index{unified theory of interactions}, it can be expected that additional information would need to be provided to specify the sign of the bidirectional charge\index{bidirectional charge sign!most-elementary particle} of a most-elementary particle (which is transformed by the bidirectional-charge-reversal symmetry operation\index{bidirectional-charge-reversal symmetry operation} $C_B$ and which determines what type of non-gravitational field the particle interacts with). That is to say, information would need to exist which would be encoded in the microscopic, quantum-gravitational degrees of freedom\index{quantum-gravitational degrees of freedom!black hole event horizon} on the event horizon of a black hole and which would allow to determine whether a particle has a reversed or non-reversed bidirectional charge sign as it propagates either forward or backward in time.

What I'm suggesting we need to recognize is that, not only are those the only fundamental physical attributes\index{fundamental physical attributes} which are allowed to vary under such conditions (as a result of applying the discrete $P$, $T$, $C$, and $C_B$ symmetry operations) and which, therefore, allow to completely characterize the state of any particle that could be present inside the event horizon of a black hole, and not only is it possible for the information that is required to determine the value of each of those parameters to be encoded on the event horizon of a black hole, but in fact, this is the only information that \textit{could} be encoded on the surface of such an object.

Before I can explain how I arrived at such a conclusion, however, it is necessary to examine how the first law of black hole mechanics\index{black hole mechanics!first law} (equation (\ref{eq:3.1})) could be made compatible with the outcome of the preceding analysis. It turns out that the simplest way to accommodate the concept of bidirectional charge is to rewrite the first law in a slightly different way, by introducing an additional term that would account for the contribution of reversed bidirectional charges\index{reversed bidirectional charge!black hole entropy} to the entropy of a black hole
\begin{equation}\label{eq:3.5}
\frac{T_H}{4}dA=dE-\Omega dJ-\Phi_+dQ_+ -\Phi_-dQ_- \, ,
\end{equation}
where $T_H=\kappa/(2\pi)$ is still the Hawking temperature\index{Hawking temperature} determined by the surface gravity\index{surface gravity} $\kappa=4\pi\mu /A$ of a black hole, while $\Phi_+$ is the electrostatic potential\index{electrostatic potential!non-reversed bidirectional charge} on the surface of a black hole that is produced by its non-reversed bidirectional electric charge $Q_+$, and $\Phi_-$ is the electrostatic potential\index{electrostatic potential!reversed bidirectional charge} on the surface of the same black hole that is produced by its reversed bidirectional electric charge $Q_-$. Here, the previously encountered parameter $\mu$ from equation (\ref{eq:3.4}) is redefined as
\begin{equation}\label{eq:3.6}
\mu=\sqrt{M^2-Q_+^2-Q_-^2-(J/M)^2} \, ,
\end{equation}
while the equation for the area of a black hole event horizon as a function of its mass, angular momentum, and charges becomes
\begin{equation}\label{eq:3.7}
A=4\pi(2M^2+2M\mu-Q_+^2-Q_-^2) \, .
\end{equation}

A notable aspect of the above equation (\ref{eq:3.5}), which would also apply to the original equation (\ref{eq:3.1}), is that it indicates that, while the area of the event horizon of a black hole\index{black hole event horizon!area variation} must grow when its energy or its mass rises and all its other macroscopic parameters are kept constant, it must also be the case that an increase of either angular momentum or any of the two bidirectional charges, occurring while the other macroscopic parameters are kept constant, would contribute to decrease the area of the event horizon of the object, which constitutes a measure of its entropy.

But it must be clear that the reduction of the area of the event horizon of a black hole that comes with a larger positive or negative charge or a larger angular momentum merely reflects the fact that matter entropy is smaller for a black hole\index{black hole!matter entropy} with the same mass under such conditions, even though this is actually a consequence of the fact that the non-gravitational field\index{non-gravitational field!black hole charge} produced by the charge of the object and the inertial gravitational field\index{inertial gravitational field!black hole angular momentum} attributable to its angular momentum (that exists only in the reference system relative to which the object is not rotating) contribute to reduce the force produced by the gravitational field of the object that determines the area of its event horizon (for particles with the same signs of energy and charge). The truth is that if the entropy of a charged or rotating black hole is smaller, this is due to the fact that the compensation of forces that results from a stronger non-gravitational field or a larger angular momentum contributes to reduce the number of microscopic degrees of freedom\index{microscopic degrees of freedom!momentum of matter particles} which would otherwise exist in the momentum of matter particles.

Thus, if the angular momentum\index{angular momentum!gravitational field entropy} associated with the rotation of a black hole contributes to reduce the entropy contained in the gravitational field (which is attributable to the variable momentum direction of matter particles), this is not simply because the inertial gravitational field\index{inertial gravitational field!rotation} produced by rotation relative to the global inertial reference system\index{global inertial reference system} reduces the strength of the gravitational field produced by the black hole, but because it reduces the randomness of the distribution of momentum directions\index{momentum directions!randomness reduction} for matter particles, due to the fact that a larger angular momentum requires more particle momenta to be oriented in the direction of the rotation. But the same conclusion applies for the total spin of matter particles, whose growth is occurring at the expense of a diminution of the contribution by the handedness of matter particles\index{handedness of particles!diminished entropy contribution} to the entropy of the object, which becomes minimal when the spins are all pointing in the same direction and contribute maximally to the angular momentum of the object.

What happens when the angular momentum\index{black hole angular momentum!maximum value} of a black hole reaches its maximum value (at which the strength of the inertial gravitational field produced by its rotation and spin equals that of the gravitational field produced by its mass) is that all the energy is in the rotational motion\index{rotational motion!energy} and no energy can be associated with the vertical momentum of matter particles\index{vertical momentum of matter particles!absence of energy}, so that no entropy can be contained in this matter degree of freedom. For a rapidly rotating black hole the inertial gravitational field reduces the gravitational attraction of the matter inside the object, which means that matter particles are no longer submitted to a gravitational field of maximum strength and their momenta are no longer constrained to their maximum theoretical value in the vertical direction (unlike is the case for an ordinary Schwarzschild black hole\index{Schwarzschild black hole}), but are rather all aligned in the direction of the rotational motion. Under such conditions their contribution to the angular momentum of the object is maximal, but they contribute nothing to its entropy and only the sign of bidirectional charge\index{sign of bidirectional charge!black hole entropy} or the direction of propagation in time\index{direction of propagation in time!black hole entropy} of matter particles can contribute to the entropy of the black hole, along with their associated non-gravitational fields\index{non-gravitational field!black hole entropy}.

It is important to understand that, even though the degree of freedom of matter particles inside a black hole which is associated with their handedness\index{particle handedness!discrete parameter} varies as a discrete parameter under measurement and may be null, like their momentum, for the whole object, the total spin of a black hole\index{black hole!non-vanishing total spin} does not necessarily vanish, and must be allowed to contribute to the strength of the angular momentum of a black hole\index{black hole angular momentum!spin contribution}, which means that its growth may also contribute to reduce the entropy of the matter it contains. But when the spins of particles are all oriented so as to produce an angular momentum of maximum magnitude the handedness of matter particles\index{handedness of particles!non-variable attribute} is no longer a variable attribute.

The growth of electromagnetic field strength\index{electromagnetic field strength!black hole surface} on the surface of a black hole is also correlated with a reduction in the entropy contained in the momentum direction of particles\index{momentum direction of particles!entropy} inside the object, given that when the strength of the electromagnetic field equals that of the gravitational field, the electric force may oppose the gravitational force to the point where the momentum of matter particles\index{momentum of matter particles!null value inside black hole} is no longer constrained to its maximum value, but is rather null, because the repulsive non-gravitational forces\index{repulsive non-gravitational forces!particles inside black hole} between the particles inside the black hole are then as large as the gravitational forces that bind them together in the inner singularity. This can be expected to arise because the gravitational attraction that would normally give rise to a quantum bounce\index{quantum bounce} is then neutralized by the electrical repulsion of the matter particles in the singularity, while the electrical repulsion\index{electrical repulsion!neutralization} that would normally require matter to disperse is itself neutralized by the mutual gravitational attraction of those same particles, thereby giving rise to a static state\index{static state!black hole singularity}.

But given that, when the charge of a black hole is null, it is the randomness of the vertical momentum state of matter particles\index{vertical momentum of matter particles!black hole entropy} that allows them to contribute to the entropy of the object and to impart microscopic structure on the gravitational field\index{gravitational field!imparted microscopic structure}, thereby allowing it to contribute to this measure of entropy as well, then it follows that when the charge of a black hole is maximum, momentum direction no longer contributes to the entropy of the object, even through the gravitational field. This conclusion becomes unavoidable once one recognizes that it is also impossible for a homogeneous charge distribution with maximum positive or negative density to impart microscopic structure on the electromagnetic field\index{electromagnetic field!imparted microscopic structure} that could in turn be imparted to the gravitational field it produces. As a result, only the handedness of particles\index{handedness of particles!black hole entropy} contributes to the entropy of a black hole with maximum charge, because under such conditions there is no entropy in the sign of bidirectional charge\index{bidirectional charge sign!entropy contribution} or the direction of propagation in time\index{direction of propagation in time!entropy contribution} of matter particles either. The handedness of particles is then determined relative to an arbitrary direction associated with the \textit{macroscopic} momentum state of the black hole, however, as there is no longer any particle momentum in the vertical direction towards the singularity for particles inside such an object.

What must be clear is that, while the energy of both matter and interaction field particles\index{interaction field particles!attractive gravitational force} contribute to the attractive gravitational force of a charged black hole, it cannot be the case that the interaction field produced by the charge of particles merely provides an additional (positive) contribution to the entropy of such an object, that would \textit{always} add to that which is normally contained in the momentum direction of matter particles and the microscopic structure of the gravitational field\index{gravitational field!microscopic structure} they produce, because if that was the case, then it would be impossible for the entropy of such a black hole to match the measure of missing information obtained from the first law of black hole mechanics\index{black hole mechanics!first law}, which is reduced when the charge of a black hole increases and the strength of the electromagnetic field approaches that of the gravitational field. But this doesn't mean that the interaction field produced by the charge of particles contributes negatively to the entropy of the object, merely that the strength of the non-gravitational field\index{non-gravitational field strength!growth conditions} can only grow when the entropy contained in the sign of charge of matter particles is decreasing to zero, while when this matter entropy is null the interaction field it produces, itself contains no entropy.

It seems that the effects of rotation\index{rotation!reduction of gravitational attraction} or those which are produced by a non-zero electric charge\index{electric charge!reduction of gravitational attraction} on the gravitational attraction of positive-energy matter are, in fact, similar to those which we would attribute to the presence of negative-energy matter\index{negative-energy matter!reduction of gravitational attraction}, given that the presence of negative-energy matter inside a surface containing mostly positive-energy matter merely reduces the gravitational attraction of the positive-energy matter. Thus, when more gravitationally repulsive matter is present inside a black hole, the area of the event horizon\index{event horizon area!reduction} that contains this matter is smaller than if didn't contain gravitationally repulsive matter. The only difference with the case where it is the charge or the angular momentum that balances the gravitational attraction is that a positive-energy black hole containing negative-energy matter would necessarily have a smaller mass, as well as a larger temperature (as a result of the reduction of the area of its event horizon).

To summarize what I have explained so far, it seems that there exist four contributions of equal magnitude to the entropy of an ordinary Schwarzschild black hole\index{Schwarzschild black hole!entropy}, one by the vertical momentum of matter particles (which also imparts structure to the gravitational field and allows it to contribute to this measure of entropy), one by the handedness\index{handedness of particles} of matter particles, one by the sign of bidirectional charge\index{bidirectional charge sign} of matter particles, and one by their direction of propagation in time\index{direction of propagation in time}.

But in the case of an electrically neutral, extremal black hole\index{extremal black hole!maximum angular momentum} with maximum angular momentum and spin, only the entropy associated with the sign of bidirectional charge and the direction of propagation in time of matter particles, as well as that which is contained in the associated non-gravitational fields\index{non-gravitational field!black hole entropy}, is not null and contributes to the entropy of the object, because the momentum of matter is all contained in the rotational motion of the object and contributes nothing to its entropy, while the spins are all aligned so as to contribute maximally to the angular momentum of the black hole and therefore do not contribute to its entropy either. In the case of a non-spinning, extremal black hole\index{extremal black hole!maximum charge} with maximum charge, only the handedness of matter particles contributes to the entropy of the object, because the momentum is null for all matter particles in the singularity and contributes nothing to entropy, which means that it cannot impart microscopic structure on the gravitational field\index{gravitational field!imparted microscopic structure} either.

More specifically, a black hole is extremal when it exerts perfectly balanced opposite forces on a particle with its own energy and charge signs, which from a conventional viewpoint would happen when either $Q^2=M^2$ and $J=0$, or $(J/M)^2=M^2$ and $Q=0$, so that $\mu=0$ in equation (\ref{eq:3.4}). From the alternative viewpoint proposed here, however, a black hole would be extremal when $Q_+^2=M^2$, $Q_-=0$, and $J=0$, or when $Q_-^2=M^2$, $Q_+=0$, and $J=0$, but not when $Q_+^2=(M/2)^2$ and $Q_-^2=(M/2)^2$ and neither $Q_+$ nor $Q_-$ is maximal, even if $J=0$. Thus all bidirectional charge must be either reversed or non-reversed for a black hole to be extremal due to having a maximum charge, because when this maximum charge is evenly distributed between reversed and non-reversed bidirectional charges $\mu=\sqrt{M^2/2}\neq 0$ in equation (\ref{eq:3.6}), when $J=0$.

Now, for an ordinary Schwarzschild black hole\index{Schwarzschild black hole} with $Q_+^2=Q_-^2=0$ and $J=0$ we have
\begin{equation}\label{eq:3.8}
\mu=\sqrt{M^2}=M
\end{equation}
and
\begin{equation}\label{eq:3.9}
A=4\pi(2M^2+2M^2)=16\pi M^2 \, ,
\end{equation}
so that the entropy has the maximum value it can have for a black hole\index{black hole!maximum entropy} with mass $M$, which is certainly appropriate, given that all four binary degrees of freedom\index{binary degrees of freedom!particle inside black hole} of particles contribute to the entropy of such an object.

For an extremal black hole\index{extremal black hole!maximum angular momentum} with a maximum angular momentum per unit of mass $(J/M)^2=M^2$ and $Q_+^2=Q_-^2=0$, so that we have
\begin{equation}\label{eq:3.10}
\mu=\sqrt{M^2-M^2}=0 \, ,
\end{equation}
which means that
\begin{equation}\label{eq:3.11}
A=8\pi M^2 \, ,
\end{equation}
and the entropy is exactly one half that of an ordinary Schwarzschild black hole with the same mass, which is also appropriate, given that under such conditions there are no contribution to the entropy of a black hole by the binary degree of freedom associated with the vertical momentum direction\index{vertical momentum direction!particles inside black hole} of particles and that which is associated with their handedness\index{handedness!particles inside black hole}, which leaves only the two contributions by the sign of bidirectional charge\index{bidirectional charge sign!particles inside black hole} and the direction of propagation in time\index{direction of propagation in time!particles inside black hole} of matter particles to contribute to the entropy of the object.

For an extremal black hole\index{extremal black hole!maximum charge} with either $Q_+^2=M^2$ or $Q_-^2=M^2$ and $J=0$ the magnitude of the positive bidirectional charge or that of the negative bidirectional charge is maximal and equals the magnitude of the mass of the object, so that again we have
\begin{equation}\label{eq:3.12}
\mu=\sqrt{M^2-M^2}=0 \, ,
\end{equation}
which means that
\begin{equation}\label{eq:3.13}
A=4\pi(2M^2-M^2)=4\pi M^2 \, ,
\end{equation}
and the entropy is exactly one fourth that of an ordinary Schwarzschild black hole\index{Schwarzschild black hole} with the same mass and is the smallest possible entropy for an extremal black hole\index{extremal black hole!smallest entropy} with mass $M$, which is in line with what I have argued above, given that under such conditions there are no contribution to the entropy of a black hole by the binary degree of freedom associated with the vertical momentum direction of particles, nor by the two binary degrees of freedom which are associated with the sign of bidirectional charge and the direction of propagation in time of particles, which leaves only the degree of freedom provided by the handedness of matter particles to contribute to the entropy of the object.

In the present context it would be incorrect to argue that the vertical momentum\index{vertical momentum!particles inside black hole} of matter particles is singled out as the only attribute of particles inside a black hole whose entropy cannot be reduced to zero, because, in fact, for an extremal black hole with maximum charge, the entropy of matter contained in the momentum direction of particles is actually null, simply because the momentum of all matter particles is itself null, while for an extremal black hole\index{extremal black hole!maximum angular momentum} with maximum angular momentum, the entropy of matter contained in the momentum direction of particles is null as well, because all momenta are oriented so as to contribute maximally to the angular momentum of the object and are entirely determined by the value of this macroscopic parameter. But while such configurations are extremely unlikely, they are not completely impossible, because it is not impossible, but merely unlikely for a black hole to form with only positive-charge particles (given the attractive electrostatic force\index{electrostatic force} that exists between opposite-charge particles and the repulsive force that exists between particles with the same charge sign) or with particles whose spins would all contribute in the same way to the angular momentum of the object, as would be required for an extremal instance of black hole to form.

It must also be clear that it is only the fact that the microscopic state of the gravitational field\index{gravitational field!microscopic state} depends on the vertical momentum\index{vertical momentum!particles inside black hole} of matter particles, that allows the gravitational field to contribute to the entropy of a black hole, so that this contribution can be reduced when the entropy contained in the momentum of matter particles is reduced due to the object having a large angular momentum or charge (which reduces the randomness of particle momenta). As I previously explained, variations in the energy sign of matter particles does not allow the gravitational field to contribute to the entropy of a black hole as there is \textit{always} only one sign of energy for the matter particles in the maximum density state of a stationary black hole\index{black hole!stationary}.

In the context of the above developments it also becomes possible to explain why it is that the entropy of an extremal black hole\index{extremal black hole!non-zero entropy} is not reduced to zero when the electromagnetic field or the inertial gravitational field\index{inertial gravitational field!rotation} produced by rotation reach their maximum strength and become as strong as the gravitational field of the object, under which conditions a little more charge, or a little more angular momentum would be expected to eliminate the event horizon and to produce a naked singularity\index{naked singularity}. It is the fact that, even when the non-gravitational charge\index{non-gravitational charge!maximum} is maximum, there still remains information in the handedness\index{particle handedness} of particles, while when it is the angular momentum\index{angular momentum!maximum} which is maximum there still remains information in the sign of bidirectional charge\index{bidirectional charge sign!information} and the direction of propagation in time\index{direction of propagation in time!information} of the particles which produce the gravitational field of the black hole that explains that, as long as there is an event horizon, it is not possible to reduce the entropy of the object to zero, even though the temperature $T_H=\kappa/(2\pi)$ is null for an extremal black hole\index{extremal black hole!null temperature} (given that when $\mu=0$ then $\kappa=0$).

It is already well-known, however, that what explains that the temperature of an extremal black hole (with maximum charge, or angular momentum and spin) is null, even though its entropy is not zero, is the fact that for such a black hole the force which is attributable to the gravitational field is entirely compensated by that which is attributable to the non-gravitational field or that which is attributable to the angular momentum, which means that a particle with the same sign of charge and energy as the object would not accelerate on the surface of the object and would therefore experience no thermal (Unruh) radiation\index{Unruh radiation} in the vacuum (that would otherwise be observed as Hawking radiation\index{Hawking radiation} by a remote observer). Thus, if extremal black holes\index{extremal black hole!zero surface gravity} have zero surface gravity, then it means that interaction fields other than the gravitational field do contribute to determine the surface gravity $\kappa$ of a black hole\index{black hole!temperature} and therefore also its temperature (even though they do not necessarily reduce it, because a smaller event horizon area\index{event horizon area} may contribute to increase the surface gravity\index{surface gravity!black hole} and the temperature of the object, while charge and angular momentum contribute to decrease this area).

But while the strength of non-gravitational fields does contribute to determine the area of the event horizon of a black hole and the temperature of the object, only the attractive gravitational force exerted on positive-energy particles by a positive-energy black hole is responsible for the fact that the information which is encoded in the microscopic degrees of freedom\index{microscopic degrees of freedom!black hole event horizon} on the event horizon of such an object cannot be obtained.

It is the fact that the forces which are attributable to the angular momentum\index{angular momentum!black hole} and the non-gravitational charge\index{non-gravitational charge!black hole} of the object oppose its gravitational force, for particles with the same sign of energy and the same signs of charges, that explains that those forces contribute to reduce the area of the event horizon of a black hole\index{black hole!reduction of event horizon area}, which is a measure of the entropy of the object, because it is only when an interaction field is attractive among particles with the same sign of `charge' (in the most general sense of the word that would include gravitational charges) that the process of charge accumulation\index{charge accumulation!irreversibility} is irreversible, as a larger charge increases the attractive force of the object which allows it to attract even more charge. It is, therefore, the attractive nature of the gravitational field\index{gravitational field!attractive nature} between particles with the same sign of energy that allows mass to contribute positively to the entropy of a black hole, unlike electric charge and angular momentum.

Another advantage of the alternative formulation of the first law of black hole mechanics\index{black hole mechanics!alternative formulation of first law} proposed here is that it allows the amount of missing information encoded in every elementary unit of area\index{elementary units of area!black hole event horizon} on the surface of a black hole to remain constant when the area of an event horizon is reduced, as a result of the presence of more electric charge or a larger angular momentum, while the mass remains unchanged, like the number of matter particles. If the area of the event horizon of a black hole\index{black hole!reduction of event horizon area} with a given mass is reduced to half the value it would otherwise have as the object is made extremal through rotation, while the number of matter particles remains unchanged, then the amount of missing information per elementary particle of matter that is contained in such an extremal black hole\index{extremal black hole!missing information per particle} must definitely be half as large as it would be for an ordinary Schwarzschild black hole\index{Schwarzschild black hole}. What I have explained is why this must be the case.

It must be clear, therefore, that it is really the area of the event horizon alone that decreases, while the mass of a black hole is not reduced when more charge is present and angular momentum is larger, because, in fact, a portion of this mass is attributable to the energy of the electromagnetic field\index{electromagnetic field!energy} produced by the charge and another is attributable to the momentum of particles\index{momentum of particles!rotation around axis} in rotation around the axis of the object. Thus, extremal black holes\index{extremal black hole!maximum bidirectional charge} with maximum positive or negative bidirectional charges must be recognized as the true states of maximum energy density\index{maximum energy density state} (with which is associated the smallest possible event horizon\index{event horizon!smallest area for given mass} area for a given value of mass), because it is not possible for a black hole to have both the value of angular momentum and spin of an extremal black hole\index{extremal black hole!absence of charge} with no charge and the value of charge of an extremal black hole\index{extremal black hole!absence of angular momentum} with no angular momentum, because such an object would have no event horizon and would no longer actually be a black hole.

If we are willing to recognize the validity of the arguments on which those deductions are based, it would then follow that we now have an explanation, not only for the fact that the states of the matter particles which are trapped by the gravitational field of a black hole\index{black hole!state of matter particles} vary as discrete variables\index{discrete variables!matter particles inside black hole}, but also for why it is that four such variables actually contribute to the measure of missing information which must be encoded on the event horizon of an ordinary black hole, while only one or two of those variables contribute to the measure of missing information in the case of extremal black holes\index{extremal black hole!missing information} with a maximum charge or angular momentum and spin.

It is quite remarkable that such an exact quantitative result can be derived from logical arguments made in the context of a semi-classical approximation\index{semi-classical approximation}. I believe that this conclusion, more than any other, illustrates the effectiveness of an unconventional approach, such as the one I came to adopt, for solving certain kinds of problems of particular importance in fundamental theoretical physics. Thus, it is now actually possible to at least confirm the existence of a definite relationship between the microscopic quantum-gravitational degrees of freedom\index{quantum-gravitational degrees of freedom!black hole event horizon} on the event horizon of a black hole and actual states of the matter it contains. What held the key to a better understanding of the exact nature of the degrees of freedom which characterize the states of matter particles submitted to a gravitational collapse\index{gravitational collapse!matter particle degrees of freedom} was the recognition that, for matter particles reaching a black hole\index{black hole!singularity} singularity, the only relevant variables are the signs of all those physical parameters which are transformed by the redefined discrete symmetry operations\index{discrete symmetry operations}.

It is also remarkable that vertical momentum\index{vertical momentum!gravitational collapse} should be one of the only attributes of elementary particles (along with their handedness and their charge signs) that is not constrained to any specific value by the conditions prevailing in the final stages of a gravitational collapse\index{gravitational collapse!final stages} (that is to say, in the moments immediately preceding or following a generic quantum bounce\index{quantum bounce!generic}) and that it must contribute as a discrete variable to the measure of information that is missing about the microscopic state of matter confined by the event horizon of an ordinary, Schwarzschild black hole\index{Schwarzschild black hole!microscopic state of matter}. This is certainly one of the most significant outcomes which has emerged from my re-examination of the question of discrete symmetries as it arises in a semi-classical context.

\bigskip

\noindent If we now return to the more general case, in which the density of matter is not large enough to produce an event horizon and the possibility for positive- and negative-action matter to be present together inside a surface cannot be ignored, it transpires that this is a situation in which more information would be required to describe the energy and momentum states of matter particles, because more states of motion are allowed for the particles in the period before such a configuration reaches a stationary state. In fact, even when an event horizon associated with a positive-mass black hole is present, it is clear that while a positive-energy particle would be drawn toward the center of mass of the object during the collapsing phase, a negative-energy particle which would be present in the same place at the same moment, would be repelled outward by a force of similar magnitude (to the extent that the average cosmic density of positive-energy matter can be neglected). Thus, in such a case, we would need to take into account at least one additional binary degree of freedom\index{binary degree of freedom!particle energy sign}, associated with the sign of energy or action of the matter particles present inside the surface, from which would depend the directions of the space intervals associated with their momentum states.

But this would actually be the simplest case, as more complex states of motion would be allowed if the matter was not contained within a surface that constitutes a black hole event horizon, because under such conditions not only would the orientations of the momentum of particles be allowed to vary, but it seems that their magnitudes could also vary significantly. It is important to understand, however, that the validity of the Bekenstein bound\index{Bekenstein bound} would be preserved even if more information was required to determine the energy and momentum states of matter particles under those less constraining conditions.

This is, again, because while more information may be required to determine the microscopic state of matter\index{microscopic state of matter!ordinary surface} inside a surface when the energy magnitudes\index{energy magnitude} and the momentum orientations\index{momentum orientation} of elementary particles inside the surface are not fixed, this information growth is offset by a decrease in the amount of information required to determine the microscopic state of the gravitational field\index{microscopic state of gravitational field!missing information}, when it is made weaker on that surface due to the lower (nearer to zero) positive or negative density of matter energy associated with such configurations, which may arise as a consequence of the presence of both positive- and negative-energy matter inside the surface. Indeed, while the amount of microscopic structure contained in the gravitational field\index{gravitational field!microscopic structure} is growing when the distribution of matter energy becomes more inhomogeneous, it is also growing along with density of matter (as I will explain in section \ref{sec:4.7}), even if the matter distribution becomes more homogeneous, because a portion of the structure it contains is attributable to the variable momentum direction\index{momentum direction of particles} of matter particles.

Now, it may appear contradictory that under ordinary circumstances, when no macroscopic event horizon\index{macroscopic event horizon} is present and the distribution of matter energy is smoother, it is more difficult to tell the exact energy and momentum states of the matter particles present within a surface\index{ordinary surface!particle energy and momentum states}. How could it be more difficult, in effect, to determine the microscopic state of the matter inside an ordinary surface, when it seems that you can more easily observe what is going on inside such a surface? But, in fact, all I have argued is that there must be information encoded on the event horizon of an ordinary Schwarzschild black hole\index{Schwarzschild black hole!missing information} about the momentum states of the matter particles which were captured by the gravitational field of the object, not that this information can actually be obtained by an external observer.

First of all, it must be clear that from the viewpoint of an observer standing outside a black hole\index{black hole!macroscopic parameters}, away from the event horizon, the only information that is readily available about the object is contained in the value of its macroscopic parameters of total energy, momentum, angular momentum, and electric charges. Due to the microscopic nature of the event-horizon\index{event horizon!degrees of freedom} degrees of freedom which encode the information about the momentum, the handedness, the bidirectional charge\index{bidirectional charge} sign, or the direction of propagation in time of the individual matter particles which are trapped inside a black hole, the only way this information could be obtained is by performing very precise measurements right on the surface of the object. Indeed, as I explained above, from the viewpoint of an observer outside a black hole, what happens to the particles that reach its singularity would appear to take place right on the surface of the object, due to time dilation\index{time dilation!black hole} and the contraction of distances\index{contraction of distances!black hole} in the direction of the singularity. Thus, in principle, information about the momentum state of any given elementary particle inside a black hole\index{black hole!availability of information}, could be obtained by directly measuring the state of the relevant quantum-gravitational degrees of freedom\index{quantum-gravitational degrees of freedom!black hole event horizon} on the surface of the object.

The problem, however, is that, in order to obtain detailed knowledge about the exact state of those microscopic degrees of freedom, you would need to approach the surface of the object with the appropriate measuring device to the point where your distance from it would be no larger than the scale of quantum-gravitational\index{quantum-gravitational scale} phenomena. You may then be able to perform the required measurements, but given the enormous difference between the gravitational potential\index{gravitational potential!black hole event horizon} on the event horizon of the object and that just above it, if you tried to send back a signal encoding the information you have been able to obtain, this signal could not be received by a remote observer (or even by the higher portions of the measuring device itself) before the black hole\index{black hole!evaporation} evaporates through the emission of Hawking radiation\index{Hawking radiation}, at which point the information would actually be contained in the microscopic state of the radiation\index{microscopic state of radiation} itself and could be obtained by examining the quantum-gravitational degrees of freedom on the ordinary surface\index{ordinary surface!quantum-gravitational degrees of freedom} enclosing it.

What happens, therefore, is that due to time dilation\index{time dilation!black hole} and space contraction\index{space contraction!black hole} there is an \textit{absolute} limitation that prevents any observer outside a black hole\index{black hole!unavailability of information} from obtaining the information that does exist, right on the event horizon of the object, about the momentum direction, the sign of charge, and the handedness\index{particle handedness} of the matter particles it contains and it is the insurmountable character of this practical limitation that allows one to assume that the information missing from a description of the state of a black hole in terms of its four macroscopic physical parameters\index{black hole!four macroscopic physical parameters} must remain unknown, as a matter of principle, despite the fact that it does exist. As a result, we are justified to conclude that those macroscopic parameters provide a natural definition of coarse-graining\index{coarse-graining!natural definition} that does not exist in the case of an ordinary surface\index{ordinary surface!coarse-graining}, whose information content is limited merely by the Bekenstein bound\index{Bekenstein bound}, under which conditions it is possible to obtain detailed information\index{information!position or momentum of particles} about the state of any physical attribute of elementary particles.

Even though the information about the microscopic state of matter inside a black hole\index{black hole!microscopic state of matter} is encoded in the microscopic degrees of freedom\index{microscopic degrees of freedom!black hole event horizon} on the event horizon of such an object, from a practical viewpoint this information cannot be obtained before it is made irrelevant. With the appropriate experimental means, the exact state of the quantum-gravitational degrees of freedom\index{quantum-gravitational degrees of freedom!ordinary surface} on a surface can be determined down to the most intricate details, but when this surface is the event horizon of a black hole, this information remains inaccessible to the outside world and the related measure of entropy\index{entropy!objective measure} becomes objectively defined, as there is no alternative choice of coarse-graining\index{coarse-graining} that could provide a more complete description of the states of the matter particles which were captured by the gravitational field of the object. It is only in the case of an ordinary surface\index{ordinary surface!availability of information}, that it is possible to obtain detailed information about the momentum direction, the sign of charge, and the handedness of the elementary particles it contains (even when those physical attributes are allowed to freely vary), by examining the quantum-gravitational degrees of freedom\index{quantum-gravitational degrees of freedom!ordinary surface} on that surface.

This is a very significant limitation, because it is ultimately the non-subjective character of that portion of entropy variation\index{entropy!non-subjective variation} which is attributable to the growing strength of gravitational fields that enables one to conceive of the temporal irreversibility\index{temporal irreversibility!objective property} that characterizes the evolution of certain macroscopic physical systems as being an objective property, even under conditions where gravitation does not appear to be involved, given that, as I will emphasize in section \ref{sec:4.8}, all the entropy growth that is taking place in our universe must ultimately be attributed to the initial conditions of low gravitational entropy\index{gravitational entropy!initial conditions} that existed in the remote past. The real difficulty here consists in recognizing that it is possible for information to remain absolutely unknown, concerning what takes place on the event horizon of a black hole, despite the fact that this information does not vanish from reality. It is only once one recognizes that it is not self-contradictory to assume that a definite reality\index{reality!existence without experimental knowledge} of some kind always exists, even under conditions where we cannot have direct experimental knowledge of what this reality actually is, that one can appreciate the validity of this statement (I will have more to say about this in section \ref{sec:5.7}).

Of course, this conclusion would also apply, from the viewpoint of a negative-energy observer, for the microscopic state of the negative-energy matter\index{microscopic state of matter!black hole} which is under the influence of a negative-energy black hole\index{negative-energy black hole!unavailability of information}. It is not possible for the information that is missing from the macroscopic description of such an object about the states of the negative-energy matter particles it contains to be communicated by a negative-energy observer located just outside its event horizon (using negative-energy photons) to an observer located farther away, so that this information must be considered to be missing from the viewpoint of negative-energy observers.

One may wonder, though, whether it would be possible for a \textit{positive-energy} observer to obtain more information about the microscopic state of negative-energy matter under the same circumstances? But that does not appear possible, because, due to gravitational repulsion\index{gravitational repulsion!negative-energy black hole}, a positive-energy observer would never be able to reach the surface of a negative-energy black hole, on which the missing information about the state of negative-energy matter particles inside the object would be located.

An observer must approach the event horizon of a black hole to within a quantum-gravitational unit of distance\index{quantum-gravitational unit of distance} in order to determine the state of the microscopic degrees of freedom on its event horizon\index{microscopic degrees of freedom!black hole event horizon}, which means that, even if it would be possible for a positive-energy observer outside a negative-energy black hole to communicate information to observers farther from the surface of the object (using positive-energy photons), this information must remain unknown, as a matter of principle, for positive-energy observers as well, because it cannot even be obtained, due to the fact that such an observer cannot get to within a quantum-gravitational unit of distance of the event horizon of the black hole, because of the gravitational repulsion. An objective measure of entropy\index{entropy!negative-energy black hole} must, therefore, be associated with negative-energy black holes and indeed with the gravitational field of any overdensity in the negative-energy matter distribution, even from the viewpoint of positive-energy observers.

Thus, what really differentiates black hole entropy\index{black hole entropy} from the measure of information associated with an ordinary surface\index{ordinary surface!information}, enclosing a smooth matter distribution with a lower energy density, is that, in the case of an ordinary, non-rotating and electrically neutral black hole\index{black hole!unavailability of information}, even if one wanted to obtain more information about the momentum state, or the handedness, or the sign of charge of a given particle inside the event horizon of the object, this would be absolutely impossible. Thus, while it seems that, if one could obtain information about the state of a microscopic quantum-gravitational degree of freedom\index{quantum-gravitational degrees of freedom!black hole event horizon} in a maximally precise spatial location on the event horizon of a black hole, this would allow one to determine the state of some physical attribute of an elementary particle which is encoded in this elementary unit of area\index{elementary units of area}, from a practical viewpoint this is impossible.

On the other hand, what's most significant regarding those situations where no macroscopic black hole is present and the entropy\index{entropy!gravitational field} of the gravitational field is not maximum is that we are necessarily dealing with transitional states which will, in general, continue to evolve until the configuration described above is reached. Thus, the negative-energy matter which may be present inside a surface containing mostly positive-energy matter will eventually be expelled from that surface, while the positive-energy matter will keep collapsing on itself until it forms a black hole. When all the negative-energy matter is released from a surface containing a larger proportion of positive-energy matter, the total mass contained within the surface actually \textit{increases} and this means that its gravitational entropy grows larger in the process.

We are therefore in a situation where a surface containing less matter (but not less mass) can have a larger entropy. This counter-intuitive outcome is allowed, because when negative-energy matter is released outside such a surface, the total amount of information required to describe the microscopic state of the matter\index{microscopic state of matter!inside a surface} still contained within the surface diminishes (because less matter is present), but at the same time the strength of the gravitational field\index{gravitational field!on a surface} on the surface, which is attributable to its energy content, grows larger (because the mass itself is growing), while, as I will explain in section \ref{sec:4.7}, this has for consequence to increase the amount of missing information required to describe the microscopic state of the gravitational field\index{microscopic state of gravitational field!missing information} itself and it just happens that the magnitude of this increase is larger than the decrease in matter entropy which can be attributed to the release of negative-energy matter outside the surface.

A negative-energy particle inside a surface containing more positive-energy matter does contribute (positively) to the amount of missing information concerning the microscopic state of the matter\index{microscopic state of matter!inside a surface} within the surface, but at the same time it reduces the amount of missing information concerning the microscopic state of the gravitational field, so that, overall, the amount of missing information is smaller than it would be without the presence of the negative-energy particle. The more general situation, in which only the Bekenstein bound\index{Bekenstein bound} is required to apply, is therefore not incompatible with the results I have derived from a study of stationary black holes\index{black hole!stationary}, from which all matter with an energy sign opposite that of the object has been expelled.

In fact, it seems that, from a fundamental viewpoint, there is no difference between the situation we observe in general, when opposite-energy particles are allowed to be present within a surface, and that which arises when we are considering the surface delimited by the event horizon of a stationary black hole, because all that really changes when a black hole event horizon forms is that a maximum amount of entropy\index{entropy!concentration in gravitational field} becomes concentrated in the gravitational field on its surface, while the entropy associated with the energy of matter particles becomes minimal. The same conclusion applies to the entropy associated with the sign of charge and the handedness of matter particles\index{particle handedness}, which decreases when more charges share the same polarity and more spins share the same direction. But while the electromagnetic field produced by a charged black hole grows in strength when more charges with the same sign are present inside the object, the field doesn't provide a larger contribution to the amount of information encoded on the surface, as does a stronger gravitational field, because the charge distribution becomes more homogeneous and no degree of freedom exists in the matter that could impart microscopic structure to this non-gravitational field.

The fact that the presence of negative-energy matter within a positive-energy black hole\index{positive-energy black hole!negative-energy matter} would only be temporary (even from the viewpoint of an external positive-energy observer) and would always give way to a stationary state in which only positive-energy matter would remain inside the surface delimited by the event horizon of the object, suggests that such end states play a role in gravitational physics which is analogous to that which is played by thermal equilibrium states\index{thermal equilibrium state!gravitational physics analogy} in statistical mechanics\index{statistical mechanics}. But the real question, regarding the Bekenstein bound\index{Bekenstein bound}, is how it can be that, under the more general conditions in which it applies, the energy and the momentum states of matter particles\index{particle energy and momentum states!ordinary surface} located within a surface are allowed to vary in a continuous way, not just in magnitude, but (in the case of momentum) also in spatial orientation, while the measure of information encoded on the surface must still be provided in binary form.

What my investigations have led me to understand is that if the states of the microscopic, quantum-gravitational degrees of freedom\index{quantum-gravitational degrees of freedom!ordinary surface} on an ordinary surface are restricted to vary as discrete parameters (which requires the microscopic state of the matter\index{microscopic state of matter!inside a surface} located inside the surface to be describable using binary units of information), it is because event horizons\index{event horizon!quantum-gravitational scale} are actually always present on the smallest scale, where quantum fluctuations in the energy of the gravitational field\index{gravitational field energy!quantum fluctuations} continuously give rise to the formation of \textit{elementary black holes}\index{elementary black hole} with masses of the order of the Planck mass\index{Planck mass}.

It is clear that the presence of energy fluctuations on the Planck scale\index{Planck scale!energy fluctuations} does not imply that the energy of particles\index{particle energies!maximum value} whose position inside an ordinary, macroscopic surface is not determined to within a Planck unit of surface\index{Planck unit of surface} must always be fixed to some maximum value, but the fact that such fluctuations are omnipresent when we reach this scale means that elementary black holes are actually the substance of physical space and time at this level of precision of measurement. Therefore, when we perform maximally precise measurements on a macroscopic surface\index{macroscopic surface!maximally precise measurement}, in order to obtain information about the state of a specific matter particle inside of it, what we determine are the states of some particles, present in the vacuum, which are under the influence of elementary black holes located on this surface. Under such conditions it is necessary to conclude that any matter or interaction field particle present \textit{on} the surface, that may carry the quantum-gravitational degrees of freedom which encode the information about the state of a matter particle \textit{inside} the surface, would be submitted to constraints similar to those which apply to matter inside a macroscopic black hole\index{macroscopic black hole}.

Thus, if we were to perform a measurement of the microscopic degrees of freedom\index{microscopic degrees of freedom!ordinary surface} in a most precise location on an ordinary, macroscopic surface, what we would actually determine is the state of some discrete quantum-gravitational degrees of freedom on the event horizon of an elementary black hole\index{elementary black hole!quantum-gravitational degrees of freedom} that would encode the state of the elementary particles\index{elementary particle!inside elementary black hole} it contains. What's crucial to understand is that there exist four different types of elementary black holes\index{elementary black hole!four types} containing any possible combination of observed and unobserved physical attributes. The elementary black hole with the smallest possible event horizon area is one that contains a single particle with a maximum energy, confined inside an elementary unit of area\index{elementary units of area} that encodes one binary unit of information\index{binary unit of information!elementary black hole} concerning the state of this particle.

Given that the \textit{macroscopic} black hole of mass $M$ with the smallest event horizon area would be an extremal black hole\index{extremal black hole!maximum charge} with a maximum charge, whose area is given by $A=4\pi M^2$ (where the mass is in Planck units\index{Planck units}), then it would appear necessary to assume that the elementary black hole\index{elementary black hole!smallest} with the smallest event horizon area would be a charged black hole, whose mass would be the Planck mass\index{Planck mass!black hole} $m_P$, and which would have an event horizon area $A=4\pi m_P^2$. But as the conventional formula for the area of a sphere with a radius equal to the Planck length\index{Planck length} $l_P$ would be $A_{min}=4\pi l_P^2$, this suggest that, on the Planck scale\index{Planck scale}, the area of such a black hole may actually be the elementary unit of area, because in Planck units those two areas are precisely the same, while if we were instead using the formula for the event horizon of an ordinary Schwarzschild black hole\index{Schwarzschild black hole} to determine the size of an elementary unit of area\index{elementary units of area!Planck scale} on the Planck scale, we would obtain a value four times larger, which therefore cannot really be \textit{the} elementary unit of area on this scale.

Another essential hypothesis is that the elementary particles which are under the influence of an elementary black hole\index{elementary black hole} necessarily carry one elementary, positive or negative unit of charge\index{elementary unit of charge} and one of spin\index{elementary unit of spin}. This hypothesis is justified by the fact that it is required in order that every elementary particle inside a \textit{macroscopic} black hole be allowed to always carry four binary units of information\index{binary unit of information!particle inside black hole} (three of which relate to their charge and spin) under the most general circumstances, so that even when two or three of those physical attributes are determined by the macroscopic parameters of an extremal black hole\index{extremal black hole!missing information}, the appropriate amount of missing information remains that must be encoded on their event horizons.

What may not be obvious, here, is that while it would be impossible for a macroscopic black hole\index{macroscopic black hole!maximum charge and angular momentum} to have, all at once, a maximum charge and a maximum angular momentum, in the case of the smallest elementary black holes\index{elementary black hole!smallest}, with a mass equal to the Planck mass\index{Planck mass} and an event horizon equal to one elementary unit of area\index{elementary units of area}, this is possible, because only one matter or radiation particle can be present inside such an object, while a charged particle cannot contribute to electrically or non-gravitationally repel itself inside an elementary black hole, which means that the object cannot disintegrate as a result of electrostatic repulsion. But the absence of internal structure on a scale smaller than an elementary unit of area would also make it impossible for the inertial gravitational force\index{inertial gravitational force!elementary black hole spin} produced by the spin of such an elementary black hole (in the reference system in which it is not rotating) to repel the one particle that is present inside the object, which means that the object cannot disintegrate as a result of its rotation either, as would a macroscopic black hole with maximum charge and spin.

Therefore, the force produced by the non-gravitational charge\index{non-gravitational charge!elementary black hole} of the smallest elementary black holes merely contributes to balance the gravitational force exerted by such an object on a particle with the same physical attributes located \textit{outside} of it, just like it would in the case of a macroscopic, extremal black hole\index{extremal black hole!maximum charge} with a maximum charge and no angular momentum.

Now, if we recognize that the most-elementary particles, with the highest energy magnitudes, cannot be localized within a surface smaller than an elementary unit of area\index{elementary units of area!most-elementary particles}, then it follows that while the elementary particles present inside the smallest elementary black holes\index{elementary black hole!smallest} can have a charge and a spin, they can have no \textit{vertical} momentum\index{vertical momentum!particle inside elementary black hole}, as there is no space within which an elementary particle\index{elementary particles} could accelerate inside a surface whose area is itself minimum. This does not simply mean that the vertical momentum\index{vertical momentum!elementary unit of area} of a particle located within an elementary unit of area is null, as would be the case for matter particles inside an extremal, macroscopic black hole with a maximum value of charge, but really that there is no internal momentum degree of freedom\index{momentum degree of freedom!particle inside elementary black hole} for a particle inside the smallest of elementary black holes. This conclusion is unavoidable, because, from a quantum-mechanical viewpoint, it is not possible to assume that the momentum of a particle whose position is determined to such a high degree of precision could be null.

The smallest elementary black holes\index{elementary black hole!observable degrees of freedom} would, therefore, be, all at once, elementary particles and black holes, as they would carry four observable degrees of freedom, like every truly elementary particle with a maximum magnitude of energy. Those degrees of freedom are the sign of action, the momentum direction, the sign of bidirectional charge\index{bidirectional charge sign!elementary black hole} and the direction of propagation in time\index{direction of propagation in time!elementary black hole}. The state of each of those physical attributes can be determined by measuring either the state of motion of the object or the interaction fields it produces, as only one particle is the source of the fields, while this particle can only interact with either reversed or non-reversed bidirectional charge\index{reversed and non-reversed bidirectional charges} particles with the same sign of action. But the smallest elementary black holes\index{elementary black hole!event horizon degrees of freedom} would also have one event horizon degree of freedom encoding the state of their handedness, given that the spin of the one particle they contain is not the source of an interaction field in the reference system in which this intrinsic angular momentum\index{angular momentum!intrinsic} is not null. Obtaining the information that is missing about this binary degree of freedom, therefore, requires coming into direct contact with the surface of the object, just as is the case for macroscopic black holes.

An elementary black hole\index{elementary black hole!internal structure} composed of two elementary particles, however, does have an internal structure, which means that its two component particles can each have a maximum momentum directed toward or away from their center of mass, as long as the sum of those two momenta is null for the whole object, which may require twice as many degrees of freedom to exist for each \textit{matter} particle inside such a black hole, that would be encoded in twice as many elementary units of area\index{elementary units of area!elementary black hole} on the event horizon of the object. Elementary black holes\index{elementary black hole!internal momentum degrees of freedom} which have internal momentum degrees of freedom must, therefore, contain at least two elementary particles, which can move along an axis. But the same is true for the elementary black holes\index{elementary black hole!angular momentum} which have an angular momentum distinct from the sum of the spins carried by their component particles, which must contain at least two elementary particles in rotation around an axis.

When a particle is contained within an elementary black hole inside of which at least two matter or radiation particles are present, its vertical momentum\index{vertical momentum!particle inside elementary black hole} still varies in a discrete way, because its energy is still maximum and the sign of space intervals\index{sign of space intervals} still needs to be either positive or negative, but due to quantum indeterminacy\index{quantum indeterminacy!momentum orientation}, it is impossible to specify the orientation of the momentum of the particle any more precisely than there are elementary units of area encoding the information about this degree of freedom on the surface of the elementary black hole itself, as if this momentum could point along any direction normal to the surface of that fraction of elementary units of area within which the particle is confined. Under such conditions, there always exist only two possibilities for the momentum state of a particle\index{momentum state of particles!two possibilities}, which can therefore be specified using one single binary unit of information\index{binary unit of information}.

What's crucial to understand is that only the elementary black holes which contain a single elementary particle, confined to an elementary unit of area\index{elementary units of area!elementary black hole}, can have both a non-zero charge and a non-zero spin, while this is impossible for those with a mass equal to twice the Planck mass\index{Planck mass} (which contain two elementary particles), for the same reason that macroscopic, extremal black holes\index{extremal black hole!maximum charge and angular momentum} cannot have, all at once, a maximum charge and a maximum angular momentum, which is that their component particles would then be submitted to repulsive forces larger than the attractive gravitational force that binds them together inside the objects, which would therefore no longer be black holes.

The smallest elementary black holes\index{elementary black hole!smallest} containing two matter or radiation particles must, therefore, carry a non-zero charge, but they must also have a null total spin and no angular momentum (see Table \ref{tab:3.5}). Those elementary black holes\index{elementary black hole!two Planck masses} would have a mass equal to two Planck masses and an event horizon area $A=4\pi (2m_P)^2=4A_{min}$ that amounts to four times the elementary unit of area defined above. The two elementary particles they contain would have equal bidirectional charges propagating in the same direction of time and would each necessarily have a spin, but those two spins would need to have opposite directions, which could both vary, but which would remain undetermined from an observational viewpoint (as the total spin of the elementary black hole\index{elementary black hole!spin} would remain null regardless of their actual directions). The signs of bidirectional charge\index{bidirectional charge sign} and the directions of propagation in time\index{direction of propagation in time} would not be variable degrees of freedom for those elementary black holes\index{elementary black hole!charge}, because both attributes would be fixed by the total charge of the object.

\begin{table}
\begin{center}
\begin{tabular}{c|c|c|c|c}
mass & charge & spin & area & information \\ \hline\hline
$m_P$ & $q_{\pm}$ & $s_{\pm}$ & $4\pi l_P^2$ & 1 bit \\ \hline
$2m_P$ & $2\times q_{\pm}$ & $0$ & $4\times 4\pi l_P^2$ & 4 bits \\ \hline
$2m_P$ & $0$ & $2\times s_{\pm}$ & $8\times 4\pi l_P^2$ & 8 bits \\ \hline
$2m_P$ & $0$ & $0$ & $16\times 4\pi l_P^2$ & 16 bits
\end{tabular}
\end{center}
\caption[The four different types of elementary black holes and their physical parameters of mass, charge, spin, event horizon area, and missing information]{The four different types of elementary black holes\index{elementary black hole!four types} and their physical parameters of mass, charge, spin, event horizon area, and missing information. Here $m_P$ is the Planck mass\index{Planck mass}, $q_{\pm}$ is an elementary unit of reversed or non-reversed bidirectional charge\index{bidirectional charge} that can be positive or negative, $s_{\pm}$ is an elementary unit of spin that can be either up or down, and $l_P$ is the Planck length\index{Planck length}.}\label{tab:3.5}
\end{table}

Elementary black holes\index{elementary black hole!internal momentum degrees of freedom} with such attributes would have no internal momentum degrees of freedom either, as the two particles composing the objects would have the same signs of charge and would be submitted to mutual, repulsive, non-gravitational forces whose magnitude would be equal that of the mutual, attractive, gravitational forces between them (given that all interactions\index{interaction!maximum strength} have the same maximum strength on such a scale). The area of the event horizons of the smallest elementary black holes containing two matter particles would, therefore, encode the state of only one binary degree of freedom per matter particle, but given that the variable spin of each of those particles would impart structure on the gravitational field (through the local variations of inertial gravitational field\index{inertial gravitational field!particle spin} they would produce), then two more binary degrees of freedom\index{binary degrees of freedom!particle inside elementary black hole} would need to be encoded on the event horizon of such an object, which is what explains that it must have an area equal to four times an elementary unit of area\index{elementary units of area!elementary black hole}.

The elementary black holes\index{elementary black hole!null charge and non-zero spin} which contain two matter or radiation particles and which have a null total charge, but a non-zero total spin and a non-zero angular momentum, would have event horizon areas twice as large as those of the smallest elementary black holes\index{elementary black hole!smallest} containing two matter particles (despite the fact that the two kinds of objects would have the same total mass). Indeed, the spins of the two particles inside such a black hole would no longer be variable degrees of freedom, as they would both point in the same direction, while both the sign of bidirectional charge\index{bidirectional charge sign!particles inside elementary black hole} and the direction of propagation in time\index{direction of propagation in time!particles inside elementary black hole} of the particles could freely vary, as long as the bidirectional charge signs of the two particles are the same and their directions of propagation in time opposite one another. But such elementary black holes\index{elementary black hole!internal momentum degrees of freedom} would also have no internal momentum degrees of freedom, as the inertial gravitational field\index{inertial gravitational field!rotation} associated with the rotation of the elementary black hole and the total spin of its component particles would compensate the gravitational attraction that the two particles exert on each other on such a scale.

The second smallest elementary black hole containing two matter particles would, therefore, have twice as many binary degrees of freedom per matter particle as the smallest such elementary black hole\index{elementary black hole!particle degrees of freedom}, but those degrees of freedom would each impart as much additional microscopic structure on the non-gravitational field\index{non-gravitational field!microscopic structure} as the variable spin would impart on the gravitational field of the smallest elementary black hole with two matter particles, which requires the event horizon of the object to have twice the area of the smaller one.

Now, the smallest elementary black holes\index{elementary black hole!zero charge and spin} that would have neither charge, nor angular momentum would necessarily also contain two oppositely charged elementary particles, which would now have opposite spins, as the only elementary black holes\index{elementary black hole!non-zero charge and spin} containing fewer particles would have both a non-zero charge and a non-zero spin. A non-spinning and non-charged elementary black hole would, therefore, have four variable, binary degrees of freedom per particle, which are the sign of bidirectional charge\index{bidirectional charge sign}, the direction of propagation in time\index{direction of propagation in time}, the vertical momentum direction\index{vertical momentum direction!particles inside elementary black hole}, and the handedness\index{handedness!particles inside elementary black hole}, because the spin of each of its two component particles could have any of two directions without affecting the total spin of the object.

Twice as many binary degrees of freedom would need to be encoded on the event horizon of such an elementary black hole than on that of a non-charged elementary black hole with angular momentum and spin, due to the fact that the degree of freedom associated with the internal momentum of the two matter particles would impart the same amount of microscopic structure on the gravitational field\index{gravitational field!imparted microscopic structure} as would the degree of freedom associated with their variable spin. As a result, the event horizon area of the object would be twice as large as that of the third largest elementary black hole, while it would be four times larger than that of the second largest elementary black hole, which is that with a charge and no spin.

Various measurements performed on the four different types of elementary black holes\index{elementary black hole!four types} that can be present on a macroscopic surface would allow to determine the microscopic state of the matter and interaction fields contained within that surface. But even though the smallest of the elementary black holes\index{elementary black hole!smallest} described above can be expected to carry all the physical attributes of a most-elementary particle\index{most-elementary particle!physical attributes} with the highest possible energy, it would not be appropriate to argue that the elementary particles\index{elementary particles!elementary black holes} which are present inside an ordinary, macroscopic surface may themselves be composed of elementary black holes, because the energy of those particles is not, in general, a multiple of the Planck energy\index{Planck energy}, which means that elementary particles with much lower energies must exist which are not contained within an event horizon.

It must be clear, therefore, that the energy of matter particles\index{particle energies!ordinary surface} inside an ordinary, macroscopic surface would not necessarily be the Planck energy, which means that the position of a particle need not always be determined to within a quantum-gravitational unit of distance\index{quantum-gravitational unit of distance}, as would be the case when the state of the particle is constrained by the gravitational field of a macroscopic Schwarzschild black hole\index{Schwarzschild black hole}. This also means that a low-energy particle inside an ordinary surface wouldn't be under the influence of one single elementary black hole\index{elementary black hole}. Under such conditions, more elementary units of area\index{elementary units of area} may be needed to determine the state of a particle inside the surface. Yet, given that to each of those elementary units of area corresponds the state of one discrete quantum-gravitational degree of freedom on the event horizon of an elementary black hole\index{elementary black hole!quantum-gravitational degrees of freedom}, then it follows that the measure of information\index{information!binary measure} required to determine the state of a particle inside the surface must still be provided in binary form.

This is all due to the fact that the states of the physical attributes of the matter or interaction field particles which are present in the elementary black holes\index{elementary black holes!around ordinary surface} around an ordinary, macroscopic surface can only vary in a discrete way, as is the case for particles inside a macroscopic black hole, while it is those elementary units of area\index{elementary units of area!elementary black hole} which encode the information\index{information!matter particle inside surface} about the state of a matter particle present in a low energy state \textit{inside} the surface. On the quantum-gravitational scale\index{quantum-gravitational scale}, there are always only two possibilities for the state of the degrees of freedom of elementary particles encoded on the surface of the elementary black holes containing them, which can therefore be specified using one single binary unit of information\index{binary unit of information}. As a result, every elementary unit of area\index{elementary units of area!ordinary surface} on an ordinary surface still contains the same amount of information as would an elementary unit of area\index{elementary units of area!macroscopic black hole} on the surface of a macroscopic black hole.

Even in the absence of a macroscopic event horizon\index{macroscopic event horizon}, it would, therefore, appear appropriate to assume that the information\index{information!microscopic state of matter inside surface} that can be obtained through a surface, about the microscopic state of the matter inside that surface, is encoded in binary units corresponding to specific elements of area on the surface. But this actually occurs only when we assume that elementary black hole event horizons must always be present, locally, when measurements are performed on a quantum-gravitational scale of precision.

Under such conditions it would appear that more binary units of information\index{binary unit of information} would be needed in order to determine the state of each matter particle present inside an ordinary, macroscopic surface\index{ordinary surface!state of matter particles}, because in addition to specifying the sign of the space interval associated with the motion of a particle, we would also need to specify the orientation of its momentum, as well as the sign and the magnitude of its energy, because the momentum of particles would not be restricted to vary as a discrete parameter, which means that more than four elementary units of area\index{elementary units of area!ordinary surface} on the surface may be required to encode the state of each of the individual matter particle it contains.

In fact, the amount of information required to determine the states of the matter particles inside a surface would be the largest when the gravitational and other interaction fields\index{interaction fields on a surface} on the surface would be as weak as they can get and the energy, the momentum, the charge, and the spin of matter particles, would be as uniformly and as randomly distributed as they can be in the available space. But this means that there would be a compensation between the larger amount of information\index{information!microscopic state of matter} required to specify the microscopic state of matter and the smaller quantity of information required to describe the exact microscopic state of its gravitational field\index{microscopic state of gravitational field!information} and this is what would allow the Bekenstein bound\index{Bekenstein bound} to continue to apply.

\section{Negative temperatures\label{sec:3.11}}

It is not a widely known fact that while temperatures are usually confined to positive values, it is nevertheless unavoidable that some physical systems be attributed negative temperatures\index{negative temperature} under certain conditions. Those who have considered the issue have recognized, in effect, that negative measures of temperature must necessarily occur when we are dealing with certain macroscopic systems with a finite number of energy levels\index{energy levels!finite number}. What happens is that, as temperature rises it must in general be assumed that more energy states become available for the constituent particles, so that the amount of missing information or entropy\index{entropy!dependence on temperature} is itself rising. Therefore, entropy must be assumed to be minimum when a system is at zero temperature. But for systems with a finite number of energy levels, it turns out that, as temperature increases, we may reach the point where entropy is maximum\index{entropy!maximum}, so that temperature\index{temperature!infinite} must be considered infinite.

This may occur, for example, in the case of a spin system in a magnetic field\index{spin system in magnetic field}, where the number of orientation states of each nuclei is finite. For such a system, the lowest energy configuration is that where all the spins are in the direction of the magnetic field, while the highest energy configuration is that which occurs when all the spins are oriented in the direction opposite that of the magnetic field. At infinite temperature all spins would be oriented in the most random way, with as many spins oriented in the direction of the magnetic field as there would be in the opposite direction. If we were to add more energy to a system in such a state, we would witness a decrease of its entropy\index{entropy!decrease}, as more spins would become oriented in the direction opposite the magnetic field and less information would be required to describe the unknown microscopic state of the system.

Given that, according to the first law of thermodynamics\index{first law of thermodynamics}, when the volume of a system and the number of particles it contains are kept invariant, its temperature merely defines the relationship which exists between its thermal energy\index{thermal energy} and its entropy, through the formula $dS=dE/T$ (where $dE$ is the amount of heat\index{heat} absorbed by a system with a temperature $T$ in the small time interval during which it evolves between two thermal equilibrium states\index{thermal equilibrium state}), then  it follows that if an increase of thermal energy produces a decrease of entropy, it must necessarily be the case that the temperature\index{temperature!negative} of the system has become negative.

But if adding more energy decreases the entropy only slightly when it reaches its maximum point, at which the temperature\index{temperature!infinite} is infinite, then it means that the temperature is not `minus zero', but actually `minus infinity'. Thus, as even more energy is added to the system, the entropy\index{entropy!minimum} would gradually decrease back to a minimum, at which point the negative temperature would actually reach the zero value again. In the case of the spin system, this point would be reached when \textit{all} the spins would be oriented in the direction opposite that of the magnetic field and no further change could occur. I may also mention that it was found that when we combine two such systems which happen to have opposite temperatures of equal magnitude, the outcome must be a system with infinite temperature.

It must be understood that, despite common expectations to the effect that temperature\index{temperature!positive-definite quantity} is a positive-definite quantity, the conclusion that negative temperatures may occur in nature is not just a consequence of adopting some particular definition for what temperature\index{temperature!reference scale} is, or of choosing a particular reference scale for this quantity, because the negative values of temperature discussed here are significant even when we are using an `absolute' measure of temperature\index{temperature!absolute measure}, like that provided by the Kelvin scale\index{Kelvin scale}, for which the zero level is not fixed to some arbitrary, not universally significant quantity, as is the case with the Celsius or Fahrenheit scales\index{Celsius or Fahrenheit scales}. Specialists are unequivocal concerning the fact that negative temperatures cannot be avoided in a general context, because they are associated with actual states of certain macroscopic systems with a finite number of energy levels\index{energy levels!finite number}.

What I would like to point out is that black holes are somewhat similar, from a thermodynamic viewpoint, to those more conventional systems for which negative temperatures\index{negative temperature} are allowed. Indeed, I have already explained in section \ref{sec:2.12} that it appears necessary to attribute to ordinary negative-energy systems a measure of temperature that is itself negative. This negative temperature becomes necessary in order to allow the same relationship between entropy variation and heat absorption\index{heat absorption!negative} to remain valid in the context where negative-energy systems absorb thermal energy\index{thermal energy!negative-definite quantity} or heat as a negative-definite quantity, even though the entropy of matter can be expected to rise when such a change takes place, as is normally the case when positive-energy systems absorb positive heat.

Temperature\index{temperature!intensive measure of thermal energy}, in effect, is merely an intensive measure of thermal energy and therefore, if we have no choice but to recognize that thermal energy itself is negative for a negative-energy system, then it only makes sense to conclude that the temperature of such a system is negative. Thus, it is only the magnitude of temperature that becomes more evenly distributed when two opposite-energy systems exchange heat, in agreement with the previously noted observation, from section \ref{sec:2.10}, that kinetic energy\index{kinetic energy!negative-definite quantity} is exchanged as a negative-definite quantity by negative-energy systems, even when they interact with positive-energy systems (while it must be absorbed and released as a positive-definite quantity by positive-energy systems).

But given that negative-energy black holes attract negative-energy matter and repel positive-energy matter gravitationally, then it seems necessary to assume that such black holes would have a negative temperature\index{negative temperature!negative-energy black holes} as well and that they would radiate negative-energy particles, even if merely in order to satisfy the requirement of symmetry under exchange\index{exchange symmetry!positive- and negative-energy matter} of positive- and negative-energy matter. In fact, what requires a negative-energy black hole to have a negative temperature and to radiate negative-energy particles, or negative heat (which is a positive change for the energy of such an object), is the fact that this is the only way the area of its event horizon and its entropy\index{entropy decrease!black hole} can diminish in the process, just as they do when a positive-energy black hole radiates positive-energy particles.

If negative-energy black holes do have a negative temperature, however, then it should not come as unexpected that there exists a certain correspondence between the thermodynamics of black holes\index{black hole thermodynamics} and the thermodynamic phenomenon of spin systems in a magnetic field\index{spin system in magnetic field}. Of particular significance is the fact that, as a positive-energy black hole\index{black hole!evaporation} evaporates through the emission of thermal radiation and its mass decreases toward zero (in positive territory), its positive temperature\index{temperature!infinite} must rise until it becomes infinite (which occurs when the object reaches a mass of the order of the Planck mass\index{Planck mass!black hole}), at which point, if we were to continue to remove energy from it (by actually adding negative energy) its mass would start to increase into \textit{negative} territory with an initial temperature that would be infinite, but negative, and which would decrease in magnitude (toward zero) as the negative mass of the object increases. Of course, the dependence of temperature on total energy is not exactly the same here as in the case of spin systems, given that a larger-mass black hole would have a lower temperature. But if we consider only the relationships between thermodynamic properties, then the analogy is valid.

Also, if we were able to combine a positive-energy black hole\index{positive-energy black hole!positive temperature} (to which is associated a positive temperature) with a negative-energy black hole\index{negative-energy black hole!negative temperature} (to which is associated a negative temperature), then what we would obtain is not a zero temperature\index{zero temperature} object, but an object with a larger and possibly infinite temperature\index{temperature!infinite} (just like when we combine two opposite-temperature systems in the conventional theory). This could happen, for example, if a negative-energy black hole was present inside a surface that was about to become the event horizon of a large enough positive-energy black hole. In such a case, the temperature of the positive-energy black hole would actually be larger, in the time before the negative-energy object is expelled from its event horizon, because the mass of the larger black hole would be reduced by that of the smaller object while a smaller-mass black hole must have a higher temperature.

The fact that such a beautiful correspondence between the semi-classical theory of black hole thermodynamics\index{black hole thermodynamics!semi-classical theory} and the classical thermodynamics of macroscopic systems with a finite number of microscopic energy levels\index{energy levels!finite number} is allowed to occur, under the hypothesis that two signs of energy are relevant for a description of the thermodynamics of black holes, constitutes an additional argument for recognizing the legitimacy of this theoretically motivated insight. In fact, I'm surprised that the conclusion drawn by specialists, concerning the unavoidable character of the concept of negative temperature\index{negative temperature}, was never considered to imply that energy itself should be allowed to vary in sign rather than only in magnitude.

But as I have always believed that the true motivation behind the wide\-spread idea that energy can only be positive originates from the thermodynamic conception of energy\index{thermodynamic conception of energy} as a measure of heat\index{heat!positive-definite quantity} (which is itself a positive-definite quantity from a conventional viewpoint), then I was quite satisfied when I learned that this most thermodynamic concept of all, the temperature, must itself vary in sign. If there is no reason to assume that negative temperatures cannot have a clear significance in physical theory, and if it turns out that they must ultimately be associated with the state of objects whose energy is negative, then we have one less argument for assuming that the concept of negative energy itself cannot be given clear meaning.

\chapter{Cosmology and Irreversibility\label{chap:4}}

\section{The outstanding problems of cosmology\label{sec:4.1}}

The situation we face today in the field of theoretical cosmology\index{theoretical cosmology} can be resumed by mentioning two broad categories of problems. The first issue has to do with dark energies\index{dark energies!problem} in general and the consequences of the existence of invisible forms of matter and energy on the gravitational dynamics of visible matter. One of the main difficulties regarding dark energies has to do with explaining how it is possible for the density of vacuum energy to be as low as one observes it to be, while not being exactly null.

With the discovery that the rate of expansion of space\index{expansion rate!acceleration} is accelerating, it has become necessary to recognize that some invisible form of positive energy with negative pressure\index{negative pressure} is present in empty space and in the present theoretical context the only plausible explanation we have for this phenomenon is that it is a consequence of zero-point vacuum fluctuations\index{zero-point vacuum fluctuations}. But such a small value for the cosmological constant\index{cosmological constant!unexpectedly small value} is unexpected and therefore one is encouraged in seeking alternative and more exotic interpretations for this dark energy\index{dark energy}. In the first portion of the present chapter I will explain that it is, in fact, still possible to assume that dark energy is attributable to the existence of a non-vanishing average value for the density of vacuum energy\index{vacuum energy density!average value} and I will show that this hypothesis is not invalidated by the otherwise inexplicably small, but non-zero value of the cosmological constant.

The progress I will achieve while solving the problem of the nature of dark energy will also allow me to identify the probable origin of the discrepancy between the present value of the Hubble constant\index{Hubble constant!present value} $H_0$ associated the current rate of expansion of space\index{expansion rate!present value} derived from observations of the cosmic microwave background\index{cosmic microwave background} and the value of the same parameter associated with the current rate of expansion derived from observations of the luminosity of type Ia supernovae\index{type Ia supernovae}, which was recently identified as an additional difficulty for the standard model of cosmology\index{standard model of cosmology}.

Another aspect of the problem of dark energies\index{dark energies!problem} has to do with the phenomenon of missing mass\index{missing mass!phenomenon} which arises because it appears that the visible material that is present in galaxies and clusters of galaxies does not provide enough gravitational force to explain the motion of the astronomical objects that compose those large-scale structures\index{large-scale structures}. Here, one of the main objectives usually consists in trying to determine the exact nature of the dark-matter particles\index{dark-matter particles} which are assumed to contribute additional gravitational attraction around visible structures in the positive-energy matter distribution. Despite all the efforts which were devoted to this task, this is a problem which has remained unsolved.

But as I will soon demonstrate, it is possible, in the context of the developments which were introduced in the second chapter of this report, to explain most of the missing-mass effects\index{missing mass!effect} observed around galaxies and clusters as being another, perhaps more unexpected, consequence of the existence of zero-point vacuum fluctuations\index{zero-point vacuum fluctuations}. However, the presence of underdensities in a uniform distribution of negative-energy matter can also be expected to contribute to the missing-mass effect experienced by positive-energy objects under particular circumstances and therefore I will examine the consequences of such a phenomenon on the formation of large-scale structures.

The other broad category of issues we are currently dealing with in cosmology could be called the inflation problem\index{inflation!problem}. This may sound paradoxical, as inflation presently constitutes a dominant paradigm for theoretical cosmology\index{theoretical cosmology} and is still believed to provide solutions to many serious problems in the field. If I'm allowed to speak about a problem concerning inflation, it is because there does exist a series of issues which were most accurately described by the originators of inflation theory\index{inflation theory} and which have long been considered to be appropriately solved by one or another instance of such a model, until it became clear that the theory actually offers so much predictive freedom that it is nearly unfalsifiable.

As the following discussion progresses, it will become clear that what made the inflation paradigm\index{inflation!paradigm} so successful is mainly an absence of alternative solution to the various problems it was originally proposed to address. Given that I believe that the most important contribution of the originators of inflation theory was to show that there does remain decisive, unresolved issues in cosmology\index{cosmology!unresolved issues}, which could perhaps be solved using this theory, then I will not refrain from discussing those issues as a genuine category of problem to which new solutions can be proposed, even in the context where we do not reject the basic idea that there may have occurred a short period of exponentially accelerated expansion\index{exponentially accelerated expansion} in the first instants of the Big Bang.

Two different aspects of the inflation problem\index{inflation!problem} will be discussed in this chapter. The first aspect has to do mainly with the problem of flatness\index{flatness problem}, or the fact that the present rate of expansion of space\index{expansion rate!unnatural present value} on the cosmological scale appears to be set to some unnatural value, which requires an extremely precise adjustment of parameters in the initial Big Bang state\index{initial Big Bang state}. I will explain that in the context of the progress I have achieved while solving the cosmological constant problem\index{cosmological constant!problem}, this difficulty arises merely because we fail to appropriately recognize that the constraint of relational definition\index{constraint of relational definition!physical attributes} of physical attributes must also apply to the energy of the universe\index{universe!energy}.

The other aspect of the inflation problem which I will address is the horizon problem\index{horizon problem}, which has to do with the fact that it is not possible to explain the uniformity of the very-large-scale distribution of matter energy\index{very-large-scale matter energy distribution} as being a consequence of smoothing processes\index{smoothing processes} that would obey the principle of local causality\index{principle of local causality}. Two further issues actually constitute particular aspects of the horizon problem. They are the smoothness problem\index{smoothness problem} and the problem of topological defects\index{topological defects!problem}. Actually, the smoothness problem would not exist if it was not for the fact that it is usually assumed that a solution to the horizon problem would have for consequence to leave the universe perfectly homogeneous, therefore requiring an independent explanation for the fact that some inhomogeneities nevertheless remained in the primordial matter distribution\index{primordial matter distribution}, which gave rise to present-day structures. It will be shown that inflation is not required to solve this problem and perhaps also that which is associated with the rarity of topological defects, given that those difficulties arise merely as a consequence of the inappropriateness of inflation theory\index{inflation theory} as a solution to the horizon problem.

The one truly amazing consequence of the particular approach I followed in dealing with the horizon problem, however, is that it allowed me to gain a new perspective on another decisive issue which is not always recognized as a problem for cosmology, despite the fact that it can be traced back to the particular boundary conditions\index{boundary conditions!Big Bang} which were in effect at the Big Bang. This is the problem of the origin of the thermodynamic arrow of time\index{thermodynamic arrow of time!problem of origin}, which is probably the most serious difficulty currently facing cosmology. It is merely the fact that the problem is so old, and has remained unsolved for so long, that explains that it is often not recognized as a problem for cosmology, as if we had long ago given up trying to resolve it. But the developments which have been introduced in the preceding two chapters and those which will be discussed in the second portion of the current one will allow to confirm the cosmological nature of the issue and will culminate in the elaboration of a plausible explanation for how it can be that a fully time-symmetric fundamental theory\index{time-symmetric theory} can conspire to enforce boundary conditions which give rise to irreversible evolution\index{irreversible evolution} and the second law of thermodynamics\index{second law of thermodynamics}.

We therefore have two broad categories of problem in cosmology, which are the problem of dark energies\index{dark energies!problem} and the inflation problem\index{inflation!problem} and which each involve several different aspects. I will first discuss the cosmological constant problem\index{cosmological constant!problem}, along with the problem of missing mass\index{missing mass!problem}, as particular aspects of the problem of dark energies, which will then allow me to approach the problem of structure formation\index{structure formation!problem} from a new perspective. The progress achieved while solving the cosmological constant problem will then enable me to provide a satisfactory solution to the flatness problem\index{flatness problem} as one particular aspect of the inflation problem. Then I will discuss the problem of the origin of time asymmetry\index{time asymmetry!problem of origin} and in the process contribute significant insights into the nature of gravitational entropy\index{gravitational entropy}. This will allow me to propose an alternative solution to the horizon problem\index{horizon problem} and the related issue of the origin of primordial inhomogeneities\index{primordial inhomogeneities} as another aspect of the inflation problem.

\section{The cosmological constant problem\label{sec:4.2}}

One of the key parameters of the standard model of cosmology\index{standard model of cosmology} that remains unexplained is certainly that which we call the cosmological constant. If there is often reticence to assume that the cosmological constant is a manifestation of the energy contained in zero-point vacuum fluctuations\index{zero-point vacuum fluctuations}, it is certainly because it is normally expected that the density of energy contained in the vacuum at the present epoch should be either precisely null (due to some unknown symmetry principle) or much larger in magnitude than the energy density we may associate with the observed, current value of the cosmological constant\index{cosmological constant!current value}. It appears much more natural, therefore, to assume that we are rather dealing with some dark energy\index{dark energy} of unknown nature whose density could vary with the expansion of space, like that of matter. If dark energy is merely a material substance with negative pressure\index{negative pressure}, then it would appear natural to assume that it should now have a density similar to that of matter, while it seems rather unlikely that vacuum energy would happen to have a density comparable to that of matter (visible and dark), given that the density of vacuum energy is usually assumed to be unaffected by expansion.

Thus, either dark energy is not vacuum energy, in which case we have no idea what its material nature is, or we restrict ourselves to known phenomena and we recognize that it must be vacuum energy, in which case we need an explanation for the observed similarity between the current average value of the energy density of matter and that of vacuum fluctuations, that is to say, we need to explain how it can be that the vacuum contains so little energy, and yet does not provide a null contribution to the universe's energy budget, as we usually assume should have been the case if some symmetry principle was responsible for the fact that this energy is much smaller than the natural value associated with the quantum-gravitational scale\index{quantum-gravitational scale!natural energy value}, which is more than 120 orders of magnitude larger than the observed value.

I find it significant that the problem associated with the small value of the cosmological constant\index{cosmological constant!small value} is usually recognized to be a disagreement between the viewpoint of experimentalists and that of theoreticians, because, from that perspective, it becomes apparent that resolving the issue will necessarily require reconsidering the validity of certain hypotheses we take for granted in the current theoretical context. First of all, it must be acknowledged that, despite the fact that the empirical determination of a positive value for the cosmological constant contributed to reinforce the traditional belief that any energy density that could be associated with this parameter should probably be positive, in the context of the progress achieved in the second chapter of this report this restriction would be totally unjustified.

The truth is that the average value of vacuum energy could certainly have been negative, as the only thing we can be certain about is that it is the observer-independent sum of all positive and negative contributions to vacuum energy density\index{vacuum energy density!positive and negative contributions} which would have an effect on the rates of expansion\index{expansion rate!opposite-energy observers} of space experienced by positive- and negative-energy observers, unlike would be the case with a material substance like quintessence\index{quintessence} with pressure opposite its energy sign, which would only influence the expansion rate measured by a positive-energy observer through its positive energy component, as any uniform matter distribution with both a positive- and a negative-energy component.

Therefore, in the context of the developments discussed in section \ref{sec:2.7}, it may perhaps look like quintessence has an advantage over vacuum energy as a candidate for dark energy\index{dark energy}, in that it could produce the desired effect even when the material contains just as much positive energy as it contains negative energy. But I will show that this is not really the case and that the advantage rather goes to vacuum energy for at least originating from known physical principles applying to known forms of matter, or forms of matter whose existence can be deduced from know principles.

There is a certain similarity between the prediction of an arbitrarily large magnitude of energy in zero-point vacuum fluctuations\index{zero-point vacuum fluctuations!arbitrarily large energy} and the old problem of the ultraviolet divergence of black body radiation\index{black body radiation problem} which was solved by the creation of quantum theory\index{quantum theory}. I believe that the commonly met suggestion that a cut-off may come about in the calculation of the density of vacuum energy, which would be associated with the quantized nature of space\index{quantized nature of space} at the most-elementary level is certainly appropriate, but it is also insufficient to solve the cosmological constant problem\index{cosmological constant!problem}, because such a cut-off would simply decrease the energy contributions from their potentially infinite values to very large values associated with the characteristic scale of quantum-gravitational phenomena\index{quantum-gravitational phenomena!characteristic scale} and those various energy contributions would still need to cancel out in order to produce the much smaller observed value.

This is precisely the problem we face right now: the required cancellation must occur by chance out of a myriad of potentially enormous, independent contributions to the energy of the vacuum. The validity of the hypothesis that space itself must be submitted to quantization (so that there must exist a maximum theoretical value of vacuum energy density\index{vacuum energy density!maximum theoretical value}) is certainly quite inevitable, especially in the context of the developments introduced in section \ref{sec:3.10} concerning black hole entropy\index{black hole entropy} and the relationship between discrete symmetry operations\index{discrete symmetry operations} and the microscopic state of the matter that crossed the event horizon of such an object. But even if this assumption is well-founded, it is simply inadequate all by itself to reconcile the theoretically derived and observationally inferred values of vacuum energy density.

In fact, I believe that we have no choice but to assume that some symmetry principle must be responsible for the almost perfect cancellation that gives rise to the observed small average value of vacuum energy density\index{average vacuum energy density}, because under current assumptions there would be virtually no limit to the expected value of this parameter, which would then be more likely to have a relatively high positive or negative value. However, I also share Feynman's\index{Feynman, Richard} opinion that it may not be quantum field theory\index{quantum field theory}, or the preferred Grand Unified Theory\index{Grand Unified Theories}, which needs to be modified in order to accommodate such a requirement, but rather our current classical theory of gravitation\index{classical gravitation theory}. Indeed, the generalized gravitation theory\index{generalized gravitation theory} I have introduced in chapter \ref{chap:2} has allowed me to identify a new category of matter particles with negative action sign, with which we may naturally expect to be associated a contribution to the energy of zero-point vacuum fluctuations\index{zero-point vacuum fluctuations!energy contribution} which would be opposite that associated with positive-action matter particles.

It is true that there are already both positive and negative contributions to the density of vacuum energy\index{vacuum energy density!positive and negative contributions} in the context of conventional theories, but it is simply too unlikely that the required outcome could arise by chance from an extremely precise cancellation of the countless, independently varying, positive and negative contributions which are normally taken into account. What I'm suggesting is that there exists a whole new class of contributions whose total energy must necessarily compensate the sum of all currently considered contributions to the energy of the vacuum.

In the context where there must be a symmetry under exchange\index{exchange symmetry!positive- and negative-energy matter} of positive- and negative-energy matter, we are allowed to expect that the energy of the vacuum should actually be null, because negative-energy observers would necessarily experience vacuum fluctuation processes\index{vacuum fluctuation processes} which contribute energies that are the exact opposite of those contributed by the vacuum fluctuation processes which are experienced by positive-energy observers and which are the only type of vacuum fluctuations currently taken into account from the viewpoint of conventional quantum field theory\index{quantum field theory!conventional}. This is a consequence of the fact that, while only one category of positive and negative energy fluctuations directly interacts with positive-energy matter, both categories of zero-point fluctuations\index{zero-point fluctuations!positive and negative energy} exert a \textit{gravitational} force on positive-energy matter and must be taken into account in determining the current value of the cosmological constant\index{cosmological constant!positive-energy observers} measured by a positive-energy observer.

From my viewpoint, the presently considered negative contributions to the density of vacuum energy\index{vacuum energy density!negative contributions} provided by certain virtual processes\index{virtual processes} present as zero-point vacuum fluctuations\index{zero-point vacuum fluctuations} would become the positive contributions of those same processes in the negative-action portion of quantum field theory\index{quantum field theory!negative-action portion} (that which describes the processes which directly affect negative-energy matter other than through their gravitational influence) and the currently considered positive contributions to vacuum energy density\index{vacuum energy density!positive contributions} provided by other virtual processes, also present as zero-point vacuum fluctuations, would become the negative contributions of the same processes in the negative-action portion of quantum field theory. This would be true despite the fact that, as I explained in section \ref{sec:3.9}, there are actually four distinct action reversal symmetry operations\index{action reversal symmetry operations}, which can be violated in different proportions, because, when we are considering all possible processes occurring in the vacuum, we are actually dealing with the outcome of all those operations combined and as I explained in the same section, there must be invariance under such a combination of all action reversal symmetry operations that relate positive-energy matter to negative-energy matter.

Thus, all currently considered contributions to the energy density of the vacuum, whether they are positive or negative, must have a counterpart of equal magnitude and opposite sign, which guarantees a cancellation of all contributions, regardless of the details of the Grand Unified Theory\index{Grand Unified Theories} chosen to describe elementary particles and their interactions. It is not the conclusion that there are no unexpected cancellations among the multiple independent terms which add up to produce the total energy density of that portion of vacuum fluctuations experienced by positive-energy observers which is wrong, but the ignorance of the fact that there is a corresponding set of contributions, experienced only by negative-energy observers, whose distinguishing feature is that all of its terms contribute energies which are naturally the opposite of those which are already taken into account, as a consequence of the requirement of symmetry under exchange\index{exchange symmetry!positive- and negative-energy matter} of positive- and negative-energy matter.

It is merely the fact that no fully consistent theory, incorporating the concept of negative-energy matter, had ever been formulated that justified the implicit assumption that no contributions of the kind proposed here needed to be taken into account, because from that perspective the whole idea that virtual processes could take place in the vacuum that would interact merely with negative-energy matter appeared meaningless, as no such matter was believed to exist in our universe.

The usual remark to the effect that it is highly unlikely that all contributions to the energy of the vacuum could conspire to produce a vanishing density is justified, but only in the context where the sole class of contributions which is recognized to exist is that which is associated with those zero-point vacuum fluctuations\index{zero-point vacuum fluctuations} and virtual particles which exert a direct influence on positive-energy matter. However, if we recognize the unavoidable character of the assumption that negative action states are not forbidden, then it would seem that we can now predict a vanishing value for the density of vacuum energy\index{vacuum energy density!vanishing value}. It is no longer necessary to assume that there occurs a miraculous conspiracy, that results in the numerous, currently envisaged, independent contributions to the density of vacuum energy adding up to produce a number several orders of magnitude smaller than those individual terms.

Such a precisely adjusted set of independent contributions is no longer required to exist in order for the right outcome to be derived. We are not really looking for compensations among multiple unconstrained parameters, but for an overall cancellation among two identical sets of parameters, whose corresponding elements have equal magnitudes and opposite signs, even on the low-energy scale at which the symmetries associated with the Grand Unified Theory\index{Grand Unified Theories} are spontaneously broken. This does not mean that there must be a cancellation of energy fluctuations locally on the Planck scale\index{Planck scale}, however, because, as I mentioned in section \ref{sec:3.10}, even the sign of energy must be considered a variable parameter when measured on such a scale (in the absence of a macroscopic event horizon\index{macroscopic event horizon} to constrain the states of matter particles) and it is merely on the scale at which classical gravitation theory\index{classical gravitation theory} applies that a cancellation of positive and negative contributions to the density of vacuum energy\index{vacuum energy density!cancellation of opposite contributions} is allowed to occur.

What's surprising, therefore, is not that the cosmological constant\index{cosmological constant!small present value} is presently so small, but rather that it is, in effect, not perfectly null. But even if this may not be as serious a problem as that of the discrepancy between current estimates of vacuum energy density and the actual value of this parameter provided by astronomical observations (given that in the present case the amplitude of the required adjustment is much smaller than that which would have to occur in the context of a conventional model), it would not be appropriate to assume that the progress achieved so far in this section constitutes a complete solution to the cosmological constant problem\index{cosmological constant!problem}.

What I will now explain is that, despite the fact that it is natural to expect that there should be a perfect compensation between the currently considered contributions to vacuum energy density and the additional contributions arising from the presence of those virtual particles\index{virtual particles} which directly interact only with negative-energy matter, it is nevertheless possible, in principle, for the cosmological constant\index{cosmological constant!arbitrarily large value} to be arbitrarily large, even though it does appear that, for some reason, the magnitude of vacuum energy density was negligible compared to the magnitude of positive and negative matter energy densities in the very first instants of the Big Bang.

Faced with the dilemma presented here, I must acknowledge that I initially tried to explain how it can be that we appear to measure a small but non-vanishing value for the cosmological constant\index{cosmological constant!small non-vanishing value} by assuming that, in fact, the cosmological constant\index{cosmological constant!null value} is actually null, while the effects we attribute to it, instead of being the consequence of a non-zero density of vacuum energy, are rather a consequence of the presence of a very-large-scale inhomogeneity in the invisible negative-energy matter distribution. Indeed, as I explained in section \ref{sec:2.7}, an overdensity of negative-energy matter should produce an outward-directed (repulsive) gravitational force\index{repulsive gravitational force!negative-energy matter overdensity} on positive-energy matter. Thus, if we happen to be located near the center of such a very-large-scale overdensity of negative-energy matter\index{very-large-scale overdensity!negative-energy matter} we should expect to observe a `local' acceleration of the rate of expansion that would merely be a consequence of the presence of this inhomogeneity in the invisible distribution of negative-energy matter.

In fact, it was also suggested by others that just the opposite might be occurring and that we may be located inside a very-large-scale \textit{underdensity} in the distribution of \textit{positive-energy} dark matter\index{very-large-scale underdensity!positive-energy dark matter}, which would produce a `local' acceleration of the rate of expansion\index{expansion rate!local acceleration} of space experienced by positive-energy observers. But it is precisely here that a problem occurs with my own original hypothesis, because it was later shown that the acceleration of the rate of expansion of space which was revealed by observations of high-redshift, type Ia supernovae\index{type Ia supernovae!high-redshift} is incompatible with any such an explanation. In fact, in the context where there is a constraint on the amplitude of density fluctuations arising from the near uniformity of the temperature of the cosmic microwave background\index{cosmic microwave background!amplitude of density fluctuations}, it appears that there simply could not have existed inhomogeneities of sufficiently large magnitude to provide an alternative explanation of the acceleration of expansion.

What's more, if we recognize the observational and theoretical necessity for the universe to have a critical density\index{critical density!positive energy} of positive energy, then we have an additional argument to justify the rejection of such an explanation for the acceleration of the rate of expansion, because we actually need the additional positive energy that would be contained in the vacuum in order to reach the critical density\index{critical density!positive vacuum energy contribution}, which cannot be provided by positive-energy matter alone\footnote{
The same argument can also be used to rule out the possibility that dark energy\index{dark energy!conventional negative-energy matter|nn} could actually consist of gravitationally repulsive negative-energy matter of the conventional kind, which would repel both positive-energy matter and negative-energy matter itself, because such material would contribute negatively to the energy budget and while it would not form local structures, it would interfere with current estimates concerning the initial rate of expansion\index{initial expansion rate|nn} of space at the Big Bang (which allow to successfully predict the observed, relative abundances of light chemical elements\index{relative abundances of light chemical elements|nn}), when its density would be much larger.}.
 It is necessary to acknowledge, therefore, that despite the fact that, in the context of the developments proposed in the preceding chapters, we may expect the natural value of average vacuum energy density\index{average vacuum energy density!natural zero value} to be zero, there must nevertheless exist an imbalance between the positive and negative contributions to vacuum energy density\index{vacuum energy density!imbalance between opposite contributions}. What must be understood is that this imbalance cannot be attributed to a violation of the symmetry under exchange\index{exchange symmetry!positive- and negative-energy matter} of positive- and negative-energy matter, which is a necessary requirement of the constraint of relational definition of physical attributes\index{constraint of relational definition!physical attributes}.

At this point it is necessary to recall the definition of the vacuum-energy term\index{vacuum-energy term} that emerged from the generalized gravitational field equations\index{generalized gravitational field equations} developed in section \ref{sec:2.14}. There, I proposed that the value of vacuum energy density which exists in the absence of any matter and whose uniform portion would be associated with the cosmological constant\index{cosmological constant!uniform portion of vacuum energy density} measured by a positive-energy observer be defined as the sum of the natural vacuum-stress-energy tensors\index{natural vacuum-stress-energy tensors} $\gamma^{-+}\bm{V}_P^{++}$ and $-\bm{V}_P^{-+}$, which provide the maximum positive and negative values of energy density contributed by those portions of zero-point vacuum fluctuations\index{zero-point vacuum fluctuations} that directly interact only with negative- and positive-energy matter (respectively), but which both exert a gravitational influence on positive-energy matter:
\begin{equation}\label{eq:4.1}
\bm{T}_{V}^+=\gamma^{-+}\bm{V}_P^{++}-\bm{V}_P^{-+}
\end{equation}

From that particular viewpoint it would appear clearly inappropriate to consider the existence of a `bare' cosmological constant\index{cosmological constant!bare}, distinct from that which would be associated with the energy contained in zero-point vacuum fluctuations, because the cosmological term\index{cosmological term} $\bm{T}^{++}_{\Lambda}=\Lambda\bm{g}^{++}$ that enters the original form of the gravitational field equations\index{gravitational field equations!original form} associated with a positive-energy observer (with $\Lambda$ as the cosmological constant) must now be understood to consist of the uniform portion of the locally variable vacuum-energy term\index{vacuum-energy term!locally variable} $\bm{T}_{V}^+$ measured at a given epoch of cosmic time\index{cosmic time} and this means that it must be considered a particular form of vacuum energy, even in a purely classical context.

What is significant in the above equation is the appearance of the metric conversion factor\index{metric conversion factors} $\gamma^{-+}$ in front of the positive contribution to vacuum energy density, which becomes necessary once we recognize that the portion of vacuum fluctuations that cannot be directly experienced by a positive-energy observer (other than through its gravitational influence) is, in effect, the one that provides a maximum positive contribution $\gamma^{-+}\bm{V}_P^{++}$ to the density of vacuum energy\index{vacuum energy density!maximum opposite contributions}, while the portion that can be directly experienced only by a positive-energy observer would be the one that provides a maximum negative contribution $-\bm{V}_P^{-+}$. This is what justifies submitting the maximum positive contribution to the same metric conversion factor as applies to the measures of negative-energy matter density effected by positive-energy observers, because, in the absence of direct interactions, it cannot be assumed that the metric properties of space and time\index{metric properties of space and time} which govern this portion of vacuum fluctuations are necessarily the same as those experienced by a positive-energy observer.

In section \ref{sec:2.14} I mentioned, in effect, that the $\gamma^{-+}$ conversion factor is the mathematical object that allows to map the metric properties of spacetime experienced by negative-energy observers onto those experienced by positive-energy observers. But if the portion of zero-point vacuum fluctuations\index{zero-point vacuum fluctuations!maximum contribution} that provides a maximum positive contribution to the density of vacuum energy, is directly experienced only by negative-energy observers, then from the viewpoint of positive-energy observers the measure of energy density involved must be submitted to the same metric conversion factor as applies to measures of negative-energy matter density.

It must be understood, therefore, that the maximum positive contribution to the density of vacuum energy\index{vacuum energy density!maximum positive contribution} considered here is not the sum of all positive contributions directly experienced by both positive- and negative-energy observers, but really the sum of all contributions, positive and negative, which are directly experienced only by a negative-energy observer and which must necessarily produce a maximum positive outcome (given that negative-energy matter must by definition consist of voids in positive vacuum energy\index{void in positive vacuum energy} and such voids cannot be considered not to interact with that very portion of vacuum energy in which they develop and propagate).

Thus, while the hypothesis that the sum of all contributions to the density of vacuum energy\index{vacuum energy density!sum of all contributions} experienced by a positive-energy observer produces a negative number (while the sum of all such contributions which are directly experienced by a negative-energy observer produces a positive number) may at first, perhaps, appear arbitrary, it is actually unavoidable in the context where it cannot be assumed that the density of negative matter energy itself could be directly determined (other than through its gravitational influence) by a positive-energy observer, while that would be allowed if such an observer could directly measure the actual value of the maximum positive contribution to the density of vacuum energy that would be reduced by the presence of negative-energy matter, as I explained in section \ref{sec:2.14}. If one considers the measure of vacuum energy density that is contributed by the maximum positive-energy term as it is perceived by a positive-energy observer, it must, therefore, be submitted to metric conversion\index{metric conversion}. But even though the necessity of such a mapping is justified by the absence of direct interaction between positive- and negative-energy matter, its legitimacy can only be understood based on considerations of a cosmological nature.

First of all, it must be noted that the magnitude of positive vacuum energy density which would be associated with the natural vacuum-stress-energy tensor\index{natural vacuum-stress-energy tensors} $\bm{V}_P^{+-}$ that is directly experienced by an observer made of negative-energy matter is an invariant quantity, which according to the requirement of symmetry under exchange\index{exchange symmetry!positive- and negative-energy matter} of positive- and negative-energy matter should be the same as that which is provided by the magnitude of negative vacuum energy density associated with the natural vacuum-stress-energy tensor $\bm{V}_P^{-+}$ that is directly experienced by a positive-energy observer.

Therefore, in the context where the vacuum-energy term\index{vacuum-energy term} does not vanish to zero, it follows that if there must be a difference between the maximum positive and negative contributions to vacuum energy density\index{vacuum energy density!maximum opposite contributions}, it can only arise because the metric properties of spacetime\index{metric properties of spacetime} that determine the magnitude of the positive contribution, as it is perceived by a positive-energy observer (through its gravitational influence), are not the same as those that determine the magnitude of the same positive contribution, as it is perceived by a negative-energy observer (and similarly for the maximum negative energy contribution). What I'm suggesting is that this means that the appearance of the metric conversion factors\index{metric conversion factors} in the definition of the net values of vacuum energy density is a consequence of the fact that the volume of space contained within a given boundary may vary depending on whether this volume is measured by a positive- or a negative-energy observer, so that the same invariant magnitude of vacuum energy density can provide contributions of different magnitudes for observers of opposite energy signs.

Now, when I introduced the notion of observer-dependent gravitational fields\index{gravitational field!observer dependence}, which gives rise to observer-dependent metric properties\index{metric properties!observer dependence}, I emphasized that it must be recognized that there is still a correspondence between the local topology of space\index{local topology of space} associated with positive-energy observers and that which is associated with observers of opposite energy sign. Thus, the set of events occurring in spacetime must be the same regardless of the way the metric properties of spacetime are perceived, which also means that every particle that is present inside a surface parameterized using the metric properties of space associated with a negative-energy observer must also be present within a corresponding surface parameterized using the metric properties of space associated with a negative-energy observer, even when the volume contained inside the surface varies as a function of the sign of energy of the observer.

In such a context, even when the ratio of the average cosmic densities\index{average cosmic matter densities!ratio} of positive- and negative-energy matter is fixed from the viewpoint of any given observer (as must be the case before the early annihilation of matter with antimatter\index{early matter-antimatter annihilation}, for reasons I will explain later), the average densities of both positive- and negative-energy matter could be different from the viewpoints of observers with opposite energy signs, which do not share the same metric properties of space, whenever a discrepancy emerges between the scale factors\index{scale factor} experienced by opposite-energy observers.

To visualize the nature of the relationships between the measures of energy density perceived by positive- and negative-energy observers on a cosmological scale, it may help to consider the analogy provided by the case of a universe\index{universe!two-dimensional space and closed geometry} with two-dimensional space and closed geometry. More specifically, we may imagine two spherical surfaces centered on the same point (in three-dimensional space) which would represent the entire volumes of space experienced by opposite-energy observers\footnote{
It must be clear that the situation described here is only valid as an analogy, because, as I will explain in section \ref{sec:4.5}, in a more realistic context it is not even possible for space to be closed from both the viewpoint of positive-energy observers and that of negative-energy observers.}.
 It would then be appropriate to assume (for reasons that will be discussed later) that initially, in the very first instants of the Big Bang, the two surfaces both have minimum areas which correspond to a state of maximum positive and negative energy densities\index{maximum positive and negative energy densities}. Under such conditions, the average densities of positive- and negative-energy matter particles determined using the metric properties of spacetime\index{metric properties of spacetime} associated with one of the surfaces would initially be exactly the same as those which are determined using the metric properties of spacetime associated with the other surface.

But even if we assume that the average cosmic matter densities\index{average cosmic matter densities} only vary as a result of expansion (additional variations can be expected to arise as a result of the early annihilation of matter with antimatter\index{early matter-antimatter annihilation}), as space expands and the two closed surfaces grow in size, any difference in their expansion rates would make their respective areas to differ. Yet, even if such a divergence was to develop, to each position of a particle on the smaller surface would still correspond a unique position on the larger surface associated with observers of opposite energy sign and to each boundary on the smaller surface would correspond one larger boundary on the other surface. In the absence of any \textit{local} variations in the metric properties of spacetime experienced by opposite-energy observers, the only difference which would characterize the matter distributions observed on the two surfaces would, therefore, be the difference between the magnitudes of their average densities, which would follow from the fact that the same particles occupy spherical surfaces with different total areas.

Even in the absence of local space curvature, therefore, it seems that the metric properties of space and time\index{metric properties of space and time} could differ for observers of opposite energy signs, because, on the cosmic scale, regions of space delimited by corresponding boundaries (associated with observers of opposite energy signs) could have different volumes depending on the sign of energy of the observer that determines this volume, if the scale factor\index{scale factor} determined by positive-energy observers is different from that which is determined by negative-energy observers. This is due to the fact that the present average densities of positive- and negative-energy matter measured by a positive-energy observer are allowed to differ from those measured by a negative-energy observer, even when there was no difference, initially, between the scale factors experienced by observers with opposite energy signs, given that it is possible for the rate of expansion\index{expansion rate!positive- and negative-energy observers} measured by positive-energy observers to differ, or to have differed at some point, from that which is measured by negative-energy observers.

I believe that what is implied by the appearance of the metric conversion factors\index{metric conversion factors} in the proposed definitions of the density of vacuum energy, therefore, is that the magnitudes of the maximum, positive and negative contributions $\gamma^{-+}\bm{V}_P^{++}$ and $-\bm{V}_P^{-+}$ to the energy density of the vacuum\index{vacuum energy density!maximum opposite contributions} determined by a positive-energy observer can be made to differ as a consequence of the fact that opposite-energy observers do not necessarily share the same metric properties of spacetime, even on the global scale, where it can be expected that matter is homogeneously distributed. The rule would be that when the scale factor is measured as being proportionately smaller by a positive-energy observer, the density of the maximum positive contributions to the energy of the vacuum (which cannot be directly measured by such an observer) would be increased from the viewpoint of this observer, in comparison with the density of the maximum negative contributions to the energy of the vacuum which can be directly measured by the same observer, so that according to equation (\ref{eq:4.1}) above, the average density of vacuum energy\index{average vacuum energy density!positive} would be positive and our positive-energy observer would measure a positive cosmological constant\index{cosmological constant!positive-energy observers} $\Lambda$.

This outcome would be attributable to the fact that, from the viewpoint of an observer that measures a smaller volume of space on the cosmological scale, those vacuum fluctuations whose invariant maximum energy density can only be directly measured by an observer of opposite energy sign would actually take place within a comparatively larger volume and would therefore appear to have a higher positive or negative energy density (when projected on the smaller volume of space perceived by the first observer) which means that they would provide a larger contribution than the vacuum fluctuations whose invariant maximum energy density our observer can directly measure.

A definite relationship would therefore exist between the net value of uniform vacuum energy density\index{uniform vacuum energy density} or the cosmological constant\index{cosmological constant} and the difference between the scale factors\index{scale factor} determined by observers with opposite energy signs, which is made even more significant by the fact that the cosmological constant must itself modify the rates of expansion\index{expansion rate!positive- and negative-energy observers} experienced by positive- and negative-energy observers which determine those scale factors. Thus, if the current value of the cosmological constant is positive, it means that any volume of space, enclosed by a sufficiently large boundary, that would be determined using the metric properties of spacetime\index{metric properties of spacetime!positive-energy observers} experienced by positive-energy observers must presently be smaller than the corresponding volume which would be determined based on the metric properties of spacetime\index{metric properties of spacetime!negative-energy observers} experienced by negative-energy observers. I'm therefore allowed to predict that if those volumes were exactly the same in the initial Big Bang state, as I will propose in section \ref{sec:4.5}, then space must have expanded at a smaller rate, from the viewpoint of positive-energy observers, during a certain portion of the universe's history, in comparison with the rate at which it expanded from the viewpoint of negative-energy observers.

The problem that may seem to arise, under such conditions, is that the smaller scale factor\index{scale factor!positive-energy observers} presently experienced by positive-energy observers can be expected to produce a positive cosmological constant\index{cosmological constant!positive value} that would actually contribute to accelerate the rate of expansion\index{expansion rate!acceleration} of space determined by positive-energy observers and to reduce any difference between this measure of the scale factor and that which is determined by negative-energy observers. Indeed, while a positive cosmological constant would contribute to accelerate the expansion of space from the viewpoint of a conventional, positive-energy observer (due to the larger contribution of its negative pressure\index{negative pressure}), it would contribute to decelerate the expansion rate\index{expansion rate!deceleration} for a negative-energy observer (again as a result of its negative pressure), which would have for consequence to reduce the divergence between the scale factors\index{scale factors!reduced divergence} associated with observers of opposite energy signs.

It may, therefore, appear that the current conditions could only be realized if the observed positive value of the cosmological constant did not arise as a result of positive-energy observers experiencing a smaller scale factor, but rather as a consequence of those same observers experiencing a \textit{larger} scale factor\index{scale factor!positive-energy observers}, that would then contribute to further accelerate the rate of expansion of space experienced by those the same observers. This is the reason why I originally thought (as I mentioned in section \ref{sec:2.14}) that the empirical evidence appears to support the hypothesis that, contrarily to what I have proposed above, positive-energy observers should directly experience a maximum contribution to the density of vacuum energy\index{vacuum energy density!maximum positive contribution} that happens to be positive (while negative-energy observers should directly experience a maximum contribution that is negative).

One must recognize, in effect, that if one was to assume that the sum of all contributions to the density of vacuum energy\index{vacuum energy density!sum of experienced contributions} which are directly experienced by a positive-energy observer actually produces a maximum positive number, then a different form of the generalized gravitational field equations\index{generalized gravitational field equations!alternative form} would have to be adopted, such that, from the viewpoint of a positive-energy observer, the metric conversion factor\index{metric conversion factors} would rather apply to the negative portion of the maximum contribution to vacuum energy density\index{vacuum energy density!maximum negative contribution} (while, from the viewpoint of a negative-energy observer, it would apply to the positive portion of the maximum contribution to vacuum energy density), thereby giving rise to a modified version of the vacuum-energy term\index{vacuum-energy term!modified version}:
\begin{equation}\label{eq:4.2}
\bm{T}_{V}^+=\bm{V}_P^{++}-\gamma^{-+}\bm{V}_P^{-+}
\end{equation}
(this equation is to be compared with equation (\ref{eq:4.1}) above). If I had originally believed that this alternative form of the vacuum-energy term\index{vacuum-energy term!alternative form} was a more appropriate choice to model the evolution of the average densities of matter energy, it is because I had difficulty seeing how the universe could otherwise have evolved in such a way that the scale factor\index{scale factor!negative-energy observers} experienced by negative-energy observers could have become so much larger, in comparison with the scale factor experienced by positive-energy observers\index{scale factor!positive-energy observers}, that the cosmological constant\index{cosmological constant!divergence of scale factors} which results from this divergence could have grown into a positive value that is much larger than the average density of positive-energy matter experienced by positive-energy observers (which would already be larger than the magnitude of the average density of negative-energy matter experienced by negative-energy observers).

This would seem to constitute a serious problem, because according to the definition of the generalized gravitational field equations that gives rise to the final version of the vacuum-energy term\index{vacuum-energy term!final version} provided by equation (\ref{eq:4.1}), one would expect that any divergence that develops between the scale factors\index{scale factors!divergence reduction} experienced by opposite-energy observers would rather tend to be reduced by the gravitational force attributable to the pressure of the vacuum. Thus, it appeared desirable to assume that the alternative form of the vacuum-energy term provided by equation (\ref{eq:4.2}) applies, because that would allow vacuum energy to produce the very conditions which allow it to grow even larger. But in fact, this would not be an appropriate conclusion.

First of all, it must be clear that what we measure, through astronomical observations, at the present epoch, is an acceleration of the rate of expansion\index{expansion rate!acceleration} that is experienced only by observers with our own sign of energy, while observers with an opposite sign of energy could measure a different variation of the rate of expansion, not just because the same vacuum energy would exert an opposite gravitational force on negative-energy matter, but because only the average density of positive-energy matter influences the rate of expansion\index{expansion rate!positive-energy observers} determined by positive-energy observers, while only the average density of negative-energy matter influences the rate of expansion\index{expansion rate!negative-energy observers} determined by negative-energy observers.

Now, given that we do observe the present, average density of positive matter energy (both visible and dark) to be somewhat smaller than the current, average, positive density of vacuum energy, then what we can expect, based on the final form of the vacuum energy term provided by equation (\ref{eq:4.1}), is that the current, average density of vacuum energy and the present magnitude of the cosmological constant\index{cosmological constant!decreasing value} will be reduced to ever smaller values. However, it is not possible to conclude, from those considerations alone, that the magnitude of the positive cosmological constant must have been decreasing at all times in the past, as a result of the opposite effects it exerts on the expansion rates\index{expansion rate!positive- and negative-energy observers} experienced by positive- and negative-energy observers. This is not due merely to the fact that the average value of vacuum energy density\index{average vacuum energy density!null initial value} must have already been null initially if the scale factors experienced by positive- and negative-energy observers were themselves precisely equal in the first instants of the Big Bang (as I will suggest one needs to assume in section \ref{sec:4.5}), it is also due to the fact that the average density of matter may not only change as a result of expansion, but also as a consequence of the early annihilation of matter with antimatter\index{early matter-antimatter annihilation}.

Even if the magnitudes of the average, initial densities of positive- and negative-energy matter\index{average matter densities!equal initial magnitudes} (and antimatter) were exactly the same, from the viewpoint of all observers, in the very first instants of the Big Bang, they could come to differ in a relatively large proportion later on, due to the potentially distinct violations of time reversal symmetry\index{violation of time reversal symmetry} which can be experienced by positive- and negative-energy matter. It is, in effect, under conditions where more matter than antimatter particles with a given sign of action (or vice versa) are produced, as a result of a violation of time reversal symmetry, during the very first instants of the Big Bang (by processes I will describe in section \ref{sec:4.3}) that a certain portion of baryonic matter\index{baryonic matter} is allowed to survive the later processes of pair annihilation\index{pair annihilation!processes}. But if the violation of time reversal symmetry that gives rise to those violations of matter-antimatter symmetry\index{matter-antimatter symmetry!violations} is not as substantial for negative-action matter as it is for positive-action matter (which is always possible, as I have explained in section \ref{sec:3.9}), then the magnitude of the average density\index{average matter densities!differing magnitudes} of negative-energy matter could become significantly smaller than that of positive-energy matter following the early annihilation of matter with antimatter, even if those two magnitudes were identical initially.

What's significant here is that the energy that is released in the course of those matter-antimatter annihilation\index{matter-antimatter annihilation!processes} processes is radiation energy\index{radiation energy}, while the density of this energy decreases more rapidly than that of matter energy as a result of expansion, which means that it does not contribute to decelerate the rate of expansion\index{expansion rate!deceleration} experienced by an observer with the same sign of energy as much as matter itself over time. Therefore, the rates of expansion\index{expansion rate!positive- and negative-energy observers} experienced by positive- and negative-energy observers may be influenced differently by the presence of matter, during the matter-dominated era\index{matter-dominated era}, even if the magnitudes of the densities of matter with opposite energy signs where exactly the same initially, from the viewpoint of any observer (which is unavoidable, as I will explain in section \ref{sec:4.5}).

As a consequence, the deceleration of the rate of expansion experienced by positive-energy observers could have been almost exactly the same as that experienced by negative-energy observers in the radiation-dominated era\index{radiation-dominated era} (while the cosmological constant\index{cosmological constant!null value} would have had a null value) and the rate of expansion\index{expansion rate!positive- and negative-energy observers} determined later on (during the matter-dominated era) by the same positive-energy observers could have become smaller than that which is determined by negative-energy observers, if the contribution of positive matter energy to the deceleration of the rate of expansion experienced by positive-energy observers had become larger than the contribution of negative matter energy to the deceleration of the rate of expansion experienced by negative-energy observers. This is what one can actually expect to happen whenever significantly more positive- than negative-action matter is allowed to survive the early annihilation of matter with antimatter\index{early matter-antimatter annihilation}. But under such conditions it can be expected that the cosmological constant\index{cosmological constant!growth} would grow, from its initial zero value toward larger positive values, during the matter-dominated era\footnote{
It is important to note that the early annihilation of baryons with antibaryons\index{early baryon-antibaryon annihilation|nn} would not make the rates of expansion\index{expansion rate!positive- and negative-energy observers|nn} experienced by positive- and negative-energy observers to vary much during the radiation-dominated era\index{radiation-dominated era|nn}, because the rate of expansion was then determined by the density of radiation and all particles are relativistic during the radiation-dominated era, so that if there is no overall change to the energy density of matter plus radiation, then the rates of expansion\index{expansion rate!deceleration|nn} should continue to decelerate as if all the matter was still present.}.

Now, astronomical observations do appear to show that the average density of baryonic negative-energy matter\index{baryonic matter!average densities} measured by a positive-energy observer is currently much smaller than the average density of baryonic positive-energy matter measured by the same kind of observer (for reasons I will discuss in section \ref{sec:4.3}), and this means that the expansion of the universe must have been slowed down to a greater extent during the matter-dominated era\index{matter-dominated era}, from the viewpoint of a positive-energy observer, compared with what a negative-energy observer experienced. The only conclusion we can draw, therefore, is that the cosmological constant did grew to a larger positive value during the early portion of the matter-dominated era (before the epoch of decoupling\index{decoupling!epoch}) despite the fact that a positive, average value of vacuum energy density\index{average vacuum energy density!positive} would contribute to accelerate the rate of expansion\index{expansion rate!acceleration and deceleration} determined by positive-energy observers and to decelerate that which is determined by negative-energy observers, thereby moderating its own growth rate.

Thus, even if the positive, average density of vacuum energy\index{average vacuum energy density!dominance over matter energy} did not become dominant over that of positive matter energy until recent times, it must have had a significant influence on the early rates of expansion\index{early expansion rates!opposite-energy observers} of positive- and negative-energy matter determined by positive- and negative-energy observers, until it was reduced to a smaller value as a result of its dependence on the difference between the scale factors\index{scale factors!opposite-energy observers} experienced by opposite-energy observers, which was itself reduced as a result of the very influence exerted by the average density of vacuum energy on those expansion rates (which was maximum very early on, during the matter dominated era, when the magnitude of vacuum energy density was highest).

Indeed, while a positive cosmological constant\index{cosmological constant!positive value} would only be allowed to actually produce an acceleration of the rate of expansion of space\index{expansion rate!acceleration} experienced by a positive-energy observer under conditions where the average density of positive \textit{matter} energy determined by a positive-energy observer would be sufficiently small, compared to the average density of positive \textit{vacuum} energy\index{average vacuum energy density!positive}, this does not mean that the difference between the scale factors\index{scale factors!opposite-energy observers} experienced by opposite-energy observers, that gives rise to a positive cosmological constant, can only be reduced when the average density of vacuum energy\index{average vacuum energy density!dominance over matter energy} becomes dominant over that of positive matter energy, because the same average value of vacuum energy density may then already be dominant over the average density of \textit{negative} matter energy, whose magnitude would be smaller than that of positive matter energy and its negative pressure\index{negative pressure} may, therefore, already contribute to decelerate the rate of expansion of space\index{expansion rate!deceleration} experienced by negative-energy observers and in such a way reduce the divergence between the scale factors\index{scale factors!divergence} and diminish the positive value of average vacuum energy density\index{average vacuum energy!diminution of positive value}.

I believe that this is what allows the Hubble tension\index{Hubble tension} (concerning the disagreement between low-redshift estimates of the current value of the Hubble constant\index{Hubble constant!present value} $H_0$ \cite{Burns-1} \cite{Riess-2} and estimates of the same parameter deduced from observations of the cosmic microwave background\index{cosmic microwave background} \cite{Ade-1}) to be eased, ultimately, because a substantially larger early value for the positive, average density of vacuum energy\index{average vacuum energy density!larger early value}, similar in magnitude to the early density of baryonic positive-energy matter\index{baryonic positive-energy matter!early density}, would allow the largest perturbations in the temperature of cosmic microwave background\index{cosmic microwave background!largest temperature perturbations} radiation to occur on a smaller scale, from the viewpoint of a positive-energy observer (because it would take less time for the average matter density to become small enough that the decoupling of matter from radiation\index{decoupling of matter and radiation} is allowed to occur, as required for the cosmic microwave background to be released), while this appears necessary in order to raise the current value of the Hubble constant derived from the standard model of cosmology (based on measurements of those CMB temperature fluctuations) up to the larger value that is determined by the direct method.

But it must be clear that even though the average positive density of vacuum energy\index{average vacuum energy density!self-reduction} was smaller than the average density of positive matter energy until relatively recently and even though it exerts an influence that tends to diminish its own magnitude, the fact that this positive vacuum energy\index{positive vacuum energy!dominance over positive matter energy} became dominant over positive matter energy on the global scale is not unexpected in the context where the average density of positive matter energy is allowed to decrease faster than the difference between the average matter densities determined by opposite-energy observers, as a result of the expansion of space.

Therefore, despite the fact that the average density of vacuum energy\index{average vacuum energy density!null initial value} must have been null in the very first instants of the Big Bang, it is possible for it to grow into a positive value larger than that of positive matter energy (from the viewpoint of a positive-energy observer), at which point it would begin to accelerate the rate of expansion\index{expansion rate!acceleration} experienced by positive-energy observers (rather than merely cause it to decelerate less rapidly). Those deductions would appear to agree with astronomical observations, which indicate that the dominance of the average density of vacuum energy\index{average vacuum energy density!dominance over matter energy} over that of matter energy occurred only recently on the cosmic time scale, given that the rate of expansion\index{expansion rate!deceleration} of space is observed to be decelerating immediately before it began accelerating, in the most recent period of the universe's history.

The fact that it wouldn't be appropriate to assume that the portion of zero-point vacuum fluctuations\index{zero-point vacuum fluctuations!maximum contribution} which produces a maximum negative contribution to the energy of the vacuum cannot directly interact (other than gravitationally) with matter of positive energy sign, even though the presence of such matter must be equivalent to an absence of energy in this very portion of vacuum fluctuations would, therefore, appear to constitute sufficiently strong a motive to conclude that the final form of the generalized gravitational field equations\index{generalized gravitational field equations!final form}, from which is derived the vacuum-energy term\index{vacuum-energy term} provided by equation (\ref{eq:4.1}) above, is the one that must be retained. Thus, contrarily to what I had initially envisaged, the constraint that allows to determine whether the maximum value of the density of energy associated with those vacuum fluctuations\index{vacuum fluctuations!interaction with positive-energy matter} that interact with positive-energy matter is positive or negative is not \textit{purely} empirical, but also constitutes an unavoidable consistency requirement.

To avoid confusion, however, it must be understood that a positive cosmological constant\index{cosmological constant!positive value} contributes to accelerate the rate of expansion\index{expansion rate!acceleration} of space measured by a positive-energy observer and not merely to accelerate the rate of expansion of positive-energy matter, because the same metric conversion factor\index{metric conversion factors} that is involved in determining the average value of vacuum energy density\index{average vacuum energy density} also affects the measure of negative-energy matter density determined by a positive-energy observer, as is made perfectly clear in the formulation of the generalized gravitational field equations introduced in section \ref{sec:2.14}.

As a result, even in a universe in which the average densities of positive- and negative-energy matter\index{average matter densities!equal initial magnitudes} are of equal magnitudes initially, from the viewpoint of any observer, if what we may call the \textit{specific density} of negative-energy matter\index{specific density of negative-energy matter} (that which is measured by a negative-energy observer) had grown comparatively smaller than the specific density of positive-energy matter\index{specific density of positive-energy matter} (that which is measured by a positive-energy observer) \textit{before} the early annihilation of most baryons with their antibaryon\index{early baryon-antibaryon annihilation} counterparts, the average density of negative-energy matter which enters the gravitational field equations\index{gravitational field equations!positive-energy observer} associated with a positive-energy observer would have remained as similar as it originally was to the average, specific density of positive-energy matter\index{specific density of positive-energy matter!average value} (that which is measured by a positive-energy observer) right until the annihilation process began.

This would be due to the fact that the presence of the metric conversion factor in the second term of the decomposed generalized gravitational field equations\index{generalized gravitational field equations} (\ref{eq:2.25}) associated with a positive-energy observer produces the same increasing effect on measures of negative-energy matter density as applies to the positive instance of natural vacuum-stress-energy tensor\index{natural vacuum-stress-energy tensors!positive instance}, while this later increase is what gives rise to a net positive value for the average density of vacuum energy.

Of course, a similar effect would then occur for the measures of average positive-energy matter density entering the gravitational field equations\index{gravitational field equations!negative-energy observer} associated with a negative-energy observer, because, despite the fact that the average, \textit{specific} density of positive-energy matter would have become comparatively larger than that of negative-energy matter during that very short period, the average density of positive-energy matter that is physically significant for a negative-energy observer would have actually decreased in comparison with that measured by a positive-energy observer, along with the average, specific density of negative-energy matter, as a consequence of the presence, in the gravitational field equations, of the metric conversion factor\index{metric conversion factors} $\gamma^{+-}$ associated with a negative-energy observer, which must give rise to the same unique cosmological constant\index{cosmological constant!unique} (so that it must have an effect opposite that which arises from the metric conversion factor $\gamma^{-+}$ associated with a positive-energy observer).

To return to the analogy of the two embedded two-dimensional spherical surfaces\index{embedded two-dimensional spherical surfaces analogy} representing the spatial volumes of a closed universe which are experienced by opposite-energy observers, we may determine (through cosmological observations) the average density of negative-energy matter on the smaller surface associated with positive-energy observers in a universe with a positive cosmological constant\index{cosmological constant!positive value}, in order to predict the future evolution of the distribution of negative-energy matter. But, in doing so, we would have to take into account the fact that the surface over which the negative-energy particles actually evolve has a larger area, so that the distribution of negative-energy matter would appear to be deflated as it is projected on the surface over which positive-energy particles evolve. The average density of negative-energy matter which would be `observed' on that surface would therefore be higher than the `real' density which would be determined based on measures of distances associated with the larger surface on which negative-energy particles evolve.

As a consequence, the ratio of the average density of negative-energy matter to that of positive-energy matter\index{ratio of average matter densities!invariance} obtained while using the measures of area associated with the smaller surface would remain identical to what it was initially, when the two surfaces had exactly the same minimum area, but again, only as long as the magnitudes of the densities of positive- and negative-energy matter are not made to differ as a result of the annihilation of matter with antimatter\index{early matter-antimatter annihilation} that takes place before the end of the radiation-dominated era\index{radiation-dominated era}. This, I believe, is the true significance of the transformation that is accomplished when one considers the stress-energy tensor\index{stress-energy tensors!negative-energy matter} of negative-energy matter in the form under which it is combined with the appropriate metric conversion factor\index{metric conversion factors} in the generalized gravitational field equations\index{generalized gravitational field equations} from section \ref{sec:2.14}.

What this means is that the average density of negative-energy matter over which are measured any density perturbations which may potentially affect the gravitational dynamics of positive-energy matter is not the specific density of negative-energy matter\index{specific density of negative-energy matter} which is measured by negative-energy observers, but a measure of matter density dependent on the metric properties of spacetime\index{metric properties of spacetime} specific to positive-energy observers and which varies as a function of the rate of expansion measured by such observers. Thus, the variation of the average density of negative-energy matter which takes place either before or after the early phase of matter-antimatter annihilation is always assessed by a positive-energy observer based on the rate of expansion of space related to his own measures of distance and duration, which on a cosmological scale are influenced only by the average densities of positive-energy matter and vacuum energy and the same is true for the density of positive-energy matter measured by a negative-energy observer. This is why the ratio of the average cosmic densities of positive- and negative-energy matter must be considered an invariant quantity that is only allowed to vary as a consequence of the early annihilation of matter with antimatter.

There is no \textit{a priori} motive, therefore, to assume that if space is expanding at a certain rate from the viewpoint of a positive-energy observer, then it should expand at the same rate from the viewpoint of a negative-energy observer, even during the radiation-dominated era\index{radiation-dominated era} (before the early annihilation of matter with antimatter\index{early matter-antimatter annihilation} had a sizable effect on the rates of expansion). It remains, however, that during the first instants of the Big Bang the average densities of positive- and negative-energy matter\index{average matter densities!equal initial magnitudes} can be expected to have had exactly the same magnitude, so that the early rates of expansion\index{early expansion rates!opposite-energy observers} measured by positive- and negative-energy observers should themselves correspond, to an arbitrarily high degree of precision, as I will explain in section \ref{sec:4.5}.

Anyhow, it transpires that even the positive cosmological constant\index{cosmological constant!positive value} must affect positive- and negative-energy matter in the same way from the viewpoint of a positive-energy observer, because any acceleration or deceleration of the rate of expansion of space\index{expansion rate!acceleration or deceleration} would depend merely on the metric properties of spacetime\index{metric properties of spacetime} associated with the gravitational field that this positive-energy observer experiences, despite the fact that the same average density of energy of zero-point vacuum fluctuations\index{zero-point vacuum fluctuations!average energy density} would influence the rate of expansion of matter in a different way from the viewpoint of a negative-energy observer. On the cosmological scale, the rate of expansion of space does not differ depending on the sign of energy of the expanding matter, but depending on the sign of energy of the observer who measures the expansion.

It must be emphasized again that the rule invoked above for justifying that the maximum positive contribution to the average density of vacuum energy\index{average density of vacuum energy!maximum positive contribution} is predominant when the scale factor\index{scale factor!positive- and negative-energy observers} determined by negative-energy observers is larger than that which is determined by positive-energy observers, simply follows from the fact that in such a case the metric conversion factor\index{metric conversion factors} associated with the measurements of negative-energy matter densities effected by a positive-energy observer transforms the magnitude of the specific density of negative-energy matter\index{specific density of negative-energy matter} (measured by a negative-energy observer) to a larger value, while the magnitude of the density of the maximum positive contribution to the energy of vacuum\index{vacuum energy density!maximum positive contribution} fluctuations that can be directly measured by a negative energy observer is an invariant quantity, so that when it is submitted to the same metric conversion as applies to negative matter energy, it would appear to be increased in comparison with the magnitude of the density of the maximum negative contribution to the energy of vacuum\index{vacuum energy density!maximum negative contribution} fluctuations that can be directly measured by a positive energy observer, thereby giving rise to a positive cosmological constant.

It should be clear, however, that it is really the \textit{specific} value of negative-energy matter density\index{specific density of negative-energy matter} measured by a negative-energy observer that is transformed by the metric conversion factor which enters the gravitational field equations\index{gravitational field equations!positive-energy observer} associated with a positive-energy observer and not the measure of negative stress-energy\index{negative stress-energy} that is observationally determined (through its gravitational influence) by a positive-energy observer. If such a transformation is necessary, it is merely as a consequence of the impossibility to directly compare the average density of matter energy determined by a negative-energy observer, or the average energy density of those vacuum fluctuations\index{vacuum fluctuations!experienced by negative-energy observers} which are directly experienced by such an observer, with the average density of matter energy experienced by a positive-energy observer, or with the average density of energy of those vacuum fluctuations\index{vacuum fluctuations!experienced by positive-energy observers} which are directly experienced only by a positive-energy observer, due to the fact that it is not possible to directly compare the measures of spatial volume\index{spatial volumes!opposite-energy observers} effected by such opposite-energy observers on a cosmological scale.

Those limitations are made unavoidable not just by the absence of direct interactions between positive and negative energy matter, but also as a consequence of the fact that the presence of a globally uniform distribution of negative-energy matter\index{negative-energy matter!globally uniform distribution} exerts no influence on the gravitational field experienced by a positive-energy observer. But there is no reason to expect that the density of vacuum energy\index{density of vacuum energy!position dependent} itself cannot vary with position, because it remains that the metric conversion factors\index{metric conversion factors!locally-variable parameters} were defined as locally-variable parameters and if that is allowed, then there is no \textit{a priori} motive to assume that variations of vacuum energy density cannot occur, above those associated with the presence of ordinary matter itself (as voids in the homogeneous distributions of positive or negative vacuum energy\index{void in positive or negative vacuum energy!ordinary matter}). In the following section I will explain what the freedom for the vacuum-energy term\index{vacuum-energy term!function of position} to vary as a function of position really means and how this property actually becomes an advantage of the particular interpretation of dark energy\index{dark energy} developed above.

It is also important to mention that when it is recognized that all positive contributions to vacuum energy must have a negative counterpart of equal magnitude, the whole notion of false vacuum\index{false vacuum} with a larger than usual energy density becomes somewhat irrelevant, at least from a gravitational viewpoint, given that, under such circumstances, a non-zero cosmological constant\index{cosmological constant!non-zero value} can only arise when there exists a difference between the metric properties of spacetime\index{metric properties of spacetime!opposite-energy observers} perceived by observers with opposite energy signs and not as a consequence of the actual nature of the processes taking place in the vacuum. Thus, when we say that a symmetry is broken in a low-energy vacuum state\index{low-energy vacuum state!broken symmetry}, what we should really mean is that the matter particles in this vacuum interact in a manner that is different from that by which the same particles interact when they are cooled in a different way in the same vacuum, or by which they interact at higher energies. But that does not mean that the vacuum itself is physically different, in particular with regards to its energy content.

Of course, given that I have described matter as being equivalent to missing vacuum energy, I must recognize that the fact that matter can behave in different ways depending on how a symmetry is broken may nevertheless justify that we refer to the products of such symmetry breakings as consisting of different vacuums. In any case, I think that it would no longer be appropriate to argue that, as ordinary baryonic matter\index{baryonic matter!ordinary} contributes less than 5 percent to the average positive density of energy, then 95 percent of all matter must be considered of unknown nature, because if dark energy\index{dark energy!vacuum energy}, which comprises more than 70 percent of the density of positive energy, really is vacuum energy, then a significant portion of it would consist in the exact same matter particles continuously fluctuating in and out of existence in their virtual form.

Now, if one demands an explanation for the smallness of the cosmological constant\index{cosmological constant!explanation of smallness} in the context of the above description of its origin, one would have to explain why it is that the scale factors\index{scale factors!opposite-energy observers} and the rates of expansion\index{expansion rate!opposite-energy observers} experienced by positive- and negative-energy observers (which we may call the \textit{specific expansion rates}\index{specific expansion rates} of positive- and negative-energy matter) were so similar in the very first instants of the Big Bang that they only began to differ significantly during the matter-dominated era\index{matter-dominated era}, as a result of the early annihilation of matter and antimatter\index{early matter-antimatter annihilation} (which appears to have been more complete for negative-action matter).

Despite the fact that a smaller specific rate of expansion\index{specific expansion rate!positive-energy matter} of positive-energy matter would produce a larger positive cosmological constant\index{cosmological constant!larger positive value}, which would contribute to accelerate the rate of expansion\index{expansion rate!positive-energy observers} of space measured by a positive-energy observer, thereby reducing the difference between this expansion rate and the specific expansion rate of negative-energy matter\index{specific expansion rate!negative-energy matter}, which would allow to reduce the magnitude of the cosmological constant\index{cosmological constant!magnitude reduction}, there is no doubt that the average value of vacuum energy density\index{average vacuum energy density} could have been much larger (into positive or negative territory) initially, even before the early annihilation of baryons and antibaryons\index{early baryon-antibaryon annihilation} had an effect on the specific rates of expansion\index{specific expansion rates} of positive- and negative-energy matter, in which case its present magnitude would still be much larger than the measured value.

Here it would appear that one may have no choice but to invoke the weak anthropic principle\index{weak anthropic principle}, because it is not sufficient to recognize that the magnitude of the cosmological constant\index{cosmological constant!self-reduction} must have been reduced to a certain extent as a result of the negative feedback exerted by the average density of vacuum energy on itself. Indeed, according to Steven Weinberg\index{Weinberg, Steven} \cite{Weinberg-1}, the current value of the cosmological constant\index{cosmological constant!current value} is so close to the maximum limit imposed by the anthropic principle that it would appear that, if it is not much larger, this may simply be a consequence of the fact that a larger value would be incompatible with the existence of an observer\index{observer!existence constraint}. What I will explain in section \ref{sec:4.5} is that, in the context where we impose a requirement of null energy on the universe\index{universe!requirement of null energy} as a whole, it becomes possible to assume that it is really anthropic selection\index{anthropic selection} which, alone, requires that the average density of vacuum energy\index{average density of vacuum energy!small current value} be as small as it is currently observed to be.

In any case, the validity of the approach advocated here is not compromised by the fact that it once seemed that empirical data was perhaps favorable to the hypothesis that the cosmological constant\index{cosmological constant!invariance hypothesis} has not changed much during the recent history of the universe, because, given the current smallness of the observed average value of vacuum energy density (compared to the natural scale of quantum-gravitational phenomena\index{quantum-gravitational phenomena!natural scale}), it is natural to expect that the current rate at which the cosmological constant\index{cosmological constant!current reduction rate} is being reduced, which is determined by the very magnitude of this average density of vacuum energy, would have remained too small to be detected. But more precise measurements may allow to reveal this variation, which is one clear prediction of the approach proposed here, that may allow to confirm its validity\footnote{
After I wrote the latest of the earlier versions of this document it was found that some puzzling astronomical observations of baryon acoustic oscillations\index{baryon acoustic oscillations|nn} \cite{AbdulKarim-1} could be resolved if it is assumed that the (average) density of dark energy\index{dark energy!average density|nn} has, in effect, been slowly decreasing since the epoch when the cosmic microwave background\index{cosmic microwave background|nn} was released.}.

Even under such conditions, however, it is fortunate that it would appear that the effect of a non-zero cosmological constant\index{cosmological constant!effect of non-zero value} is not to produce an even larger average (positive or negative) density of vacuum energy, as would have been the case if the vacuum-energy term\index{vacuum-energy term!alternative equation} that enters the generalized gravitational field equations\index{generalized gravitational field equations} had been that which is provided by the alternative equation (\ref{eq:4.2}) above, because, in such a case, one would have to conclude that despite the fact that the cosmological constant is still relatively small, it should eventually become much larger, while the specific rate of expansion of positive-energy matter\index{specific expansion rate!acceleration} should accelerate ever more rapidly, as a result of its own growth. But I cannot be considered guilty of having chosen the right form of the vacuum-energy term based on a desire to avoid the prospect of predicting such a precocious end of the world\index{precocious end of world}, given that, as I have explained above, I had originally assumed that the alternative form of the equations was actually the right one, on the basis of what appeared to be an unavoidable empirical constraint and despite the discomforting outcome of such a choice. It is only in order to achieve greater consistency that I was later forced to recognize that this position is untenable.

To resume the situation, it transpires that the problem of the cosmological constant\index{cosmological constant!problem} was complicated by the fact that it no longer appeared possible to explain the value of this parameter as being the outcome of a symmetry principle, when astronomical observations began to show that it is not exactly zero. This is because any violation of symmetry would likely produce a value of average vacuum energy density\index{average vacuum energy density!outcome of symmetry violation} much closer to the natural scale of energy associated with quantum gravitation\index{quantum gravitation!natural energy scale}. What I have explained is that it is the necessary invariance under exchange of positive- and negative-energy matter\index{positive- and negative-energy matter!invariance under exchange} (which is justified by the requirement of relational definition of physical attributes\index{requirement of relational definition!physical attributes} discussed in chapters \ref{chap:2} and \ref{chap:3}) that allows one to expect a perfect cancellation of all contributions to the average density of vacuum energy\index{average vacuum energy density!cancellation of all contributions}, in the absence of a divergence between the scale factors\index{scale factors!divergence} experienced by opposite-energy observers, while it is possible to assume (as I will explain in section \ref{sec:4.5}) that it is the weak anthropic principle\index{weak anthropic principle} which, alone, explains that this divergence of the scale factors was not as large as it could have been initially, thereby allowing the current value of the cosmological constant\index{cosmological constant!small present value} to be as small as it is observed to be.

I believe that the fact that such a relatively simple but highly efficient solution to what has been called `the mother of all physics problem\index{mother of all physics problem}' had never been seriously considered before is simply a consequence of the preconceived opinion that negative-energy matter cannot exist, which is a consequence of both irrational prejudice and what always appeared to be the insurmountable difficulties preventing a consistent description of gravitationally repulsive matter\index{gravitationally repulsive matter}.

\section{Missing mass and dark matter\label{sec:4.3}}

In this section I would like to discuss the impact of the developments introduced in the earlier portions of this report on our understanding of the phenomenon of missing mass\index{missing mass!phenomenon}\footnote{
It must be clear that what I'm referring to here is the general phenomenon that is usually attributed to the presence of dark matter\index{dark matter|nn} and not that of voids in a matter distribution (even though I will suggest that those two phenomena may sometimes be related). If I chose this slightly ambiguous and rarely used denomination it is because the problem I'm referring to is more general than the dark-matter problem\index{dark matter!problem|nn} itself, which merely consists in identifying a potential candidate for the weakly-interacting massive particles\index{weakly-interacting massive particles|nn} whose presence is usually assumed to explain this missing-mass effect\index{missing mass!effect|nn}.},
 which is currently believed to always arise solely from the presence of additional, unseen, but normally-gravitating positive-energy matter. What will emerge from those considerations is that additional effects, similar to those we normally attribute to ordinary dark matter\index{dark matter}, must be taken into account in the context of a cosmological model based on the generalized gravitation theory\index{generalized gravitation theory} introduced in chapter \ref{chap:2}. As a result, it is no longer necessary to assume that conventional dark matter is responsible for most of the missing-mass effect observed at the present epoch around visible positive-energy galaxies and clusters.

Thus, while I will suggest that it is necessary to recognize the existence of an unexpected component of dark, but normally-gravitating baryonic matter\index{baryonic dark matter}, which could be responsible for a small portion of those missing-mass effects, I will also explain that, for the main part, the phenomenon of missing mass appears to merely be a secondary effect of the presence of energy attributable to zero-point vacuum fluctuations\index{zero-point vacuum fluctuations!energy}. Before delving into this important issue, however, I will explore another dimension of the dark-matter phenomenon which has been altogether ignored until now and which has to do with the gravitational attraction attributable to the presence of voids in an otherwise uniform negative-energy matter distribution\index{void in negative-energy matter distribution!gravitational attraction}.

I have already mentioned in section \ref{sec:2.7} that certain forces which could not be distinguished from those conventionally attributed to positive-energy dark matter\index{dark matter!positive energy} would arise in the presence of an underdensity in an otherwise uniform distribution of invisible negative-energy matter. It must be assumed, in effect, that a uniform distribution of negative-energy matter exerts no gravitational force on positive-energy particles, due to the fact that the void of universal proportion in the positive portion of vacuum energy\index{void in positive vacuum energy!universal proportion} that is equivalent to such a uniform matter distribution leaves no surrounding positive vacuum energy to exert an uncompensated gravitational attraction\index{uncompensated gravitational attraction} that would be equivalent to a gravitational repulsion\index{gravitational repulsion} exerted by the negative-energy matter itself. But this means that a local absence of negative-energy matter from such a uniform matter distribution would be equivalent to the presence of a local overdensity of positive vacuum energy\index{positive vacuum energy!overdensity}, which must produce a gravitational field identical to that which would arise from the presence of a corresponding, additional amount of positive-energy matter, from the viewpoint of positive-energy particles.

If the interaction between positive- and negative-energy matter is governed by the principles enunciated in section \ref{sec:2.13}, it would appear that such a phenomenon could in principle occur around positive-energy matter overdensities, given that such structures would repel negative-energy matter and thus create underdensities in this negative-energy matter distribution that could potentially enhance the gravitational attraction of the positive-energy objects, locally, if the average density of negative-energy matter and the magnitude of the underdensities are large enough. In fact, the same phenomenon should arise from the presence of voids in the positive-energy matter distribution, which can be expected to exert a gravitational attraction on any negative-energy matter (either visible or dark) that would be present around those voids.

What we can expect to occur, as we approach the center of mass of a sufficiently large overdensity in the positive-energy matter distribution, is that an increasingly smaller density of negative-energy matter would be present, because a larger fraction of it would not be able to overcome the repulsive gravitational force\index{repulsive gravitational force} exerted by the overdensity. The local reduction in the density of negative-energy matter that is attributable to the gravitational repulsion exerted by the positive-energy matter overdensity should therefore grow, along with the magnitude of the missing mass effect\index{missing-mass effect!magnitude}, as we approach the center of the structure. But, clearly, this cannot continue indefinitely, because the average density of negative-energy matter over which the underdensity is measured has a finite magnitude, which, at the present epoch at least, is much smaller than the average density of positive-energy matter in a typical galaxy\index{galaxies!average matter density}.

When the point is reached at which the magnitude of the underdensity of negative-energy matter attributable to the gravitational repulsion of the positive-energy matter overdensity equals the magnitude of the average density of negative-energy matter itself, it becomes impossible to further reduce this matter density. This marks the limit beyond which the gravitational attraction attributable to the presence of such an underdensity can no longer grow and actually becomes insignificant in comparison with the gravitational attraction that is attributable to the density of positive-energy matter within the structure.

In such a context, it should be clear that it is not possible to conclude that a potential contribution, by negative-energy matter underdensities, to the missing-mass effect\index{missing mass!effect} around visible positive-energy structures could make contributions of a distinct nature unnecessary, because, for this to happen, the current magnitude of the average density of negative-energy matter (visible and dark) would need to be much larger than the currently inferred average density of positive-energy matter (both visible and dark), so that underdensities of sufficiently high magnitude could exist that would explain all of the missing-mass effects presently attributed to positive-energy dark matter\index{dark matter!positive energy}. If the current average density of negative-energy matter determined by a positive-energy observer is merely as large as that of positive-energy matter, or if it is actually even smaller, as would appear to be required by observations (I will return to this question later in this section), then it simply isn't large enough to allow a replication of all the missing-mass effects around visible structures, which are known to involve equivalent matter densities hundreds of times larger than the average density of ordinary baryonic matter\index{baryonic matter!ordinary}.

Therefore, the portion of missing mass effect that can be attributed to the presence of underdensities in the negative-energy matter distribution is limited, due to the fact that the average cosmic density of negative-energy matter has a finite value which is relatively small at the present moment, compared to the density of matter inside most visible structures. Negative-energy matter underdensities can therefore be expected to have accelerated the process of structure formation\index{structure formation!process} in the positive-energy matter distribution only at the epoch, in the remote past, when the average matter density was still relatively large and the matter distribution homogeneous enough on the scale of the structures considered, because any hypothetical missing-mass effect that could be attributed to the presence of underdensities in the negative-energy matter distribution can only be concentrated around positive-energy structures if negative-energy matter is otherwise smoothly distributed around those structures.

The problem we face when we try to attribute to the presence of negative-energy matter underdensities most of the missing-mass effect around positive-energy galaxies\index{galaxies!positive energy} is that, even though negative-energy matter may be more homogeneously distributed than positive-energy matter, at the present epoch, on the scale of galaxies and clusters, the average density of negative-energy matter is presently much smaller than the density of visible matter inside those structures, as I mentioned above. There is a real possibility, however, that negative-energy matter underdensities could have exerted an influence on the gravitational dynamics\index{gravitational dynamics!positive-energy matter} of positive-energy matter in the early universe, even on the scale of individual stars and galaxies, and this is certainly not an undesirable prospect, given that, despite all the progress which was achieved in the last decades to model the formation of large-scale structures\index{large-scale structures!formation}, the currently favored theory of structure formation, involving only positive-energy cold dark matter\index{cold dark matter}, is still inadequate in certain respects.

It is well-known, in effect, that the most recent observations (see in particular Ref. \cite{Labbe-1}) have kept revealing the presence of well-developed galaxies and clusters\index{galaxies and clusters!unexpectedly large masses} of galaxies with masses much larger than expected, at increasingly larger redshifts, corresponding to an epoch when there shouldn't be any such structures according to current models. It is my belief that those difficulties will be alleviated once we recognize that, in the remote past, there existed significant contributions to the missing-mass effect\index{missing mass!effect} which arose from the presence of underdensities of considerably high magnitudes in a relatively uniform distribution of negative-energy matter, which were produced as a result of the gravitational repulsion\index{gravitational repulsion} exerted on negative-energy matter by positive-energy matter overdensities.

Even though we may have reasons to expect that no significant amount of baryonic negative-energy matter\index{baryonic negative-energy matter} survived the early period of matter-antimatter annihilation\index{early matter-antimatter annihilation}, there are also good reasons to believe that negative-energy dark matter\index{dark matter!negative energy} (dark from the viewpoint of observers made of baryonic negative-energy matter) was present in the primordial universe\index{primordial universe}, with an average density comparable to that of non-baryonic, positive-energy dark matter\index{dark matter!non-baryonic}, because most of this matter wasn't submitted to matter-antimatter annihilation like baryonic matter\index{baryonic matter} (for reasons I will soon discuss) and must, therefore, still be present in the universe today. As a result, the gravitational attraction attributable to the presence of negative-energy matter underdensities must have played an important role (which need not be attributed only to hypothetical, positive-energy, cold dark matter\index{cold dark matter!particles} particles) in the formation of the primordial inhomogeneities\index{primordial inhomogeneities} that gave rise to visible, present day structures.

It is necessary to assume, in effect, that when the distribution of negative-energy matter was still sufficiently dense and uniform, as must have been the case in the primordial universe (for reasons I will explain later in this chapter), the gravitational repulsion of the overdense structures which developed in the positive-energy matter distribution should have triggered the formation of negative-energy matter underdensities concentrated mostly around those developing structures, thereby allowing them to develop more rapidly, by enhancing their gravitational attraction\index{gravitational attraction!enhanced}.

What cannot be assumed, however, is that overdense structures were present in the negative-energy matter distribution as well, at the same epoch, on a similar scale, which could have produced stellar- or galactic-size underdensities in the positive-energy matter distribution, that would have similarly accelerated the growth of those negative-energy structures\index{negative-energy structures}. This is a hypothesis that is both theoretically unnecessary and observationally doubtful, because the presence of overdense negative-energy objects on a galactic scale should have exerted recognizable effects that would have been revealed already, by weak gravitational lensing\index{weak gravitational lensing!experiments} experiments. But given that I will argue, later in this section, that dark matter overdensities\index{dark matter overdensities} have a tendency to form and to grow where baryonic matter overdensities\index{baryonic matter overdensities} with the same sign of energy are located, while baryonic negative-energy matter\index{baryonic negative-energy matter!present absence} appears to be virtually absent at the present epoch in our universe, then those observations are quite understandable.

It must be clear, though, that even if a small portion of the missing-mass effect\index{missing mass!effect} observed around present-day structures could still be attributed to the presence of negative-energy matter underdensities (particularly on larger scales), those contributions would not allow the average density of positive energy in our universe to reach its critical value, because any contributions to the energy budget from inhomogeneities in the negative-energy matter distribution would cancel out on a global scale, if those inhomogeneities developed in an originally smooth distribution of negative-energy matter\index{negative-energy matter!smooth original distribution} (which I will argue to be a necessary assumption in section \ref{sec:4.9}), given that there would then be as much spread out overdensities as there are localized underdensities in this matter distribution.

Thus, even independently from any other considerations, it is necessary to recognize that the presence of negative-energy matter underdensities cannot contribute significantly to the observed missing-mass effect around positive-energy objects at the present epoch, given that the effect is already known to require the contribution of a density of gravitationally-attractive matter energy about as large as that which would bring the total density of positive energy to its theoretically and empirically required critical value.

\bigskip

\noindent Now, if one recognizes that the presence of negative-energy matter underdensities would never allow to explain a significant portion of the missing-mass effects\index{missing mass!effect} which are observed around visible positive-energy structures at the present epoch, then one must admit that there definitely exist additional contributions of unknown origin to positive matter energy in our universe. Faced with the undeniable evidence that a certain form of dark matter\index{dark matter} must exist, the normal reaction is to seek to identify a weakly-interacting particle\index{weakly-interacting particle}, different by necessity from all known particles, that might constitute a viable candidate for this dark matter. But for various reasons, despite the fact that all attempts at detecting and identifying such a particle have failed, it is still believed that dark matter should actually consist of particles that do not interact with ordinary matter \textit{only} through the gravitational interaction.

I believe that what really motivates this view is the fact that, if dark matter interacts with the rest of matter only through the very weak gravitational interaction, then it may, in effect, become impossible to determine the nature of those dark-matter particles by experimental means, which justifies that we concentrate, instead, on trying to identify a particle that does interact with ordinary matter through one of the other known forces. But what if we could deduce from certain observable properties of ordinary matter that there \textit{must} exist positive-energy matter particles which can only interact with ordinary matter through gravitational forces?

At this point you may recall the discussion from section \ref{sec:2.2} concerning the experimental requirement for positrons to be electrons carrying a negative electric charge backward in time, instead of being a similar particle actually carrying a positive charge forward in time. I explained that such a constraint must exist in the context where it is recognized that a certain condition of continuity of the flow of time\index{condition of continuity of flow of time!particle world-line} along the world-lines of elementary particles must apply to all particle-antiparticle annihilation processes. The restriction that there be continuity in the true direction of propagation in time\index{direction of propagation in time!true direction} along an elementary particle world-line in spacetime becomes relevant in the context where it is recognized that there does exist a fundamental time-direction degree of freedom\index{fundamental time-direction degree of freedom} distinct from the observed direction of motion of elementary particles. Compliance with such a continuity requirement would imply that particle-antiparticle creation and annihilation\index{particle-antiparticle!creation and annihilation} processes can only occur as the kind of events during which a particle bifurcates in spacetime to start propagating in the opposite direction of time and not as a chance encounter of two opposite-charge particles propagating in the same direction of time.

What requires one to assume that \textit{ordinary} antiparticles\index{antiparticles!backward-in-time-propagating particles} (those that routinely take part in interactions involving ordinary matter) are indeed backward-in-time-propagating particles (and not particles propagating opposite charges forward in time) is the fact that when we view any transformation along a particle world-line as a continuous process, then ordinary antiparticles must always be considered to propagate in the direction of time opposite that in which the corresponding particles are propagating, given that the annihilation of an ordinary particle with an ordinary antiparticle must be allowed to occur with the same probability for all such pairs and cannot only take place for those pairs where the two particles would happen to be propagating in opposite directions of time.

But this requirement would also impose that events cannot occur which would involve a positive-energy particle that propagated a negative charge forward in time turning into an identical negative-energy particle that would continue to exist in the \textit{future}, but that would actually be a particle arriving from that future while propagating a positive charge backward in time and which, from the unidirectional-time viewpoint\index{unidirectional-time viewpoint}, would appear as a particle carrying the same positive energy and the same negative charge forward in time. What makes such a transformation not just implausible, but actually impossible is the fact it would imply that the continuous path of a particle in spacetime\index{particle path in spacetime!reversal} (the arrow along a particle world-line in a Feynman diagram\index{Feynman diagrams!particle arrow}) could abruptly reverse when, by chance, a particle would meet an oppositely charged version of the exact same particle arriving from the future, that would be propagating backward in time.

When a condition of continuity of the flow of time\index{condition of continuity of flow of time} applies along the world-lines of elementary particles\index{elementary particle world-line}, if certain \textit{ordinary} electrons are propagating a negative electric charge forward in time, while other \textit{ordinary} electrons were allowed to propagate a positive electric charge backward in time, then certain electrons could not annihilate with certain positrons (those that would propagate an opposite electric charge in the same direction of time) with which they would nevertheless be allowed to interact, while it is known experimentally that no such a restriction to electron-positron annihilation exists (all known electrons can annihilate with all known positrons).

Thus, even if some of the electrons that propagate in a particular direction of time could have a negative charge, while others would have a positive charge, we must consider as empirically forbidden for particles with such opposite \textit{bidirectional charges}\index{opposite bidirectional charges} (the invariant measures of charge which are not affected by a reversal of the direction of propagation in time\index{direction of propagation in time!reversal} of elementary particles which we have met in section \ref{sec:3.10}) to transform into one another or to interact with one another, at least under ordinary circumstances. No particle of any given kind that propagates some non-gravitational charge forward or backward in time can decay into, or interact with a particle of the same kind that would propagate an opposite charge either forward or backward in time and which would otherwise appear to consist of the exact same particle (two such particles propagating in opposite directions of time, would not merely consist of a particle and its antiparticle, they would actually have the same sign of energy and the same sign of charge, from the viewpoint of unidirectional time\index{unidirectional-time viewpoint}).

What I'm suggesting, therefore, is that the fact that particle-antiparticle annihilation\index{particle-antiparticle!annihilation} processes are allowed to occur for any particle-antiparticle pair does not mean that there can be discontinuities in the direction of the flow of time\index{direction of flow of time!particle world-line} along a particle world-line in spacetime, but really that particles with opposite bidirectional charges\index{opposite-bidirectional-charge particles!absence of interactions} cannot interact with one another or transform into one another, under ordinary circumstances. Thus, I propose that we recognize the existence of a fundamental rule which can be formally expressed using the following definition\footnote{
It will be made clear in the latter portion of chapter \ref{chap:5} that what justifies this rule, from a fundamental viewpoint, is really the principle of local causality\index{principle of local causality|nn} as it applies to particle propagation processes from the viewpoint of a time-symmetric, quantum-mechanical description of reality\index{quantum-mechanical description of reality!time-symmetric|nn}, in the context where we require all causes to originate from within the universe in which the processes involved take place.}:
\begin{quote}
\textbf{Condition of continuity of the flow of time\index{condition of continuity of flow of time}}: There must always be a continuous flow in the true direction of propagation in time\index{direction of propagation in time!particle world-line} along a particle world-line in spacetime for elementary fermions, even when particle-antiparticle creation and annihilation processes are involved from a unidirectional-time viewpoint.
\end{quote}
I will soon explain what justifies (from a theoretical viewpoint) the validity of the empirical rule that particles propagating a given charge forward or backward in time cannot interact (other than gravitationally) with similar particles propagating an opposite bidirectional charge forward or backward in time (even when it may appear that the condition of continuity of the flow of time would not be violated) and therefore cannot transform (under ordinary circumstances) into such particles either, even in the course of particle-antiparticle creation and annihilation\index{particle-antiparticle!creation and annihilation} processes.

It should already be clear, however, that even when the proposed constraint applies, the charge of a particle (not necessarily the electric charge, but any non-gravitational charge) can still vary on a continuous world-line (as when a blue quark turns into a red quark, or a neutrino turns into an electron), because all that is required is that a particle with a given charge does not change into an identical particle with an opposite \textit{bidirectional} charge (the measure of charge which is independent from the direction of propagation in time) along such a continuous path, particularly in the case of fermions, so that if a particle was initially propagating a given charge forward in time, it can still be assumed to propagate the same charge in the same direction of time, unless a particle-antiparticle annihilation process takes place and the same charge begins propagating backward in time. Of course, a similar conclusion would apply for a particle propagating a given charge backward in time, which must continue to propagate the same charge in the same direction of time unless a particle-antiparticle \textit{creation} process takes place and the same charge begins propagating forward in time.

The difficulty, here, consists in recognizing that there are actually very good reasons to assume that a condition of continuity of the flow of time\index{condition of continuity of flow of time} must be imposed in the context where an antiparticle must be considered to be an ordinary particle propagating the same charge backward in time with reversed energy. What we do, from a conventional viewpoint, is that we simply ignore the possibility that an electron\index{electrons!propagation of positive charge backward in time}, for example, may exist that would propagate a positive charge and a negative energy backward in time, by assuming, as a matter of coordinative definition\index{coordinative definition!particle bidirectional charge sign}, that a positive-action electron always propagates a negative charge forward in time, while a positive-action positron always propagates the same negative charge backward in time, as if there were no other possibilities for the bidirectional charge sign (which makes the issue of continuity irrelevant).

This is similar to what we do when we exclude negative energy states propagating forward in time\index{negative energy states!propagating forward in time}, or positive energy states propagating backward in time\index{positive energy states propagating backward in time}, by assuming that they are nonphysical states\index{nonphysical states}. In the present case, however, it is not even understood that in doing so we are deliberately choosing to exclude certain states of matter from our description of reality, because it looks like all that is involved is a definition. But that is not the case, and if the choice of which positive-action electrons propagate a negative charge forward in time and which propagate a positive charge (along with a negative energy) backward in time is, in effect, a simple matter of definition, the decision to exclude as nonphysical those electrons which, according to this definition, would propagate a positive charge forward or backward in time can only be justified on the basis of observational evidence.

One may argue that this distinction is insignificant, because the validity of the conventional approach is in fact empirically confirmed, given that it does provide a theoretical framework whose predictions agree perfectly well with observational constraints. Or does it? We still have a serious problem in theoretical cosmology\index{theoretical cosmology!serious problem}, because we do not know what most of the matter in our universe is made of. Could it be that there is in fact something wrong with some of the implicit choices which were made a long time ago, while we were trying to make sense of the newly developed mathematical framework of relativistic quantum theory\index{relativistic quantum theory!mathematical framework}, before everybody even knew about the existence of dark matter\index{dark matter}? Is it possible that there does exist in our universe positive-action electrons\index{electrons!positive bidirectional charge} with positive bidirectional charges and positive-action protons\index{protons!negative bidirectional charge} with negative bidirectional charges and that those particles actually constitute a non-negligible portion of the normally-gravitating dark matter, along with the positive-action neutrons composed of negatively-charged up quarks and positively-charged down quarks propagating forward or backward in time?

I do recognize that serious difficulties may be associated with this idea, because, even if one acknowledges the fact that, from an empirical viewpoint, positively-charged electrons propagating either forward or backward in time should not be allowed to interact with, or to transform into ordinary electrons propagating negative electric charges in any direction of time, or to interact with ordinary protons propagating positive electric charges forward or backward in time, one still needs to explain what justifies this limitation from a theoretical perspective. What's more, even if we could justify the absence of interactions between ordinary matter particles and their dark-matter\index{dark-matter particles!reversed bidirectional charges} counterparts with reversed bidirectional charges, then it would remain to explain why it is that those particles do interact through the gravitational interaction.

I believe, however, that those difficulties do not decisively rule out the existence of such baryonic dark-matter\index{baryonic dark-matter!reversed-bidirectional-charge particles} particles and that it is possible to understand, by making use of the developments already introduced in this report, why reversed-bidirectional-charge particles should, in effect, be dark, despite the fact that they can also be expected to interact gravitationally with the rest of matter, thereby allowing them to contribute to the missing-mass effect\index{missing mass!effect} around visible positive-energy structures.

What I have come to understand is that the difficulty one may face while trying to explain the absence of electromagnetic interactions between electrons\index{electrons!negative and positive bidirectional charges} with negative bidirectional charge and electrons with positive bidirectional charge arises merely because one ignores the fact that the previously defined constraint regarding the continuity of the flow of time along a particle world-line\index{constraint of continuity of flow of time!interaction mediating particles} must also apply in the case of the particles that mediate the interactions between elementary particles of matter.

The problem is that, according to the conventional interpretation, the world-lines of interaction bosons\index{interaction boson world-line!abrupt termination} would appear to abruptly come to an end when they are absorbed by a matter particle, just like they would seem to come into existence discontinuously when they are emitted, either by a fermion or another interaction boson. While this may not appear to violate any principle, a certain tension clearly exists between the conventional description of those particle absorption and emission\index{particle absorption and emission!conventional description} processes and the previously discussed constraint regarding the continuity of the flow of time along a particle world-line.

But, instead of arguing indefinitely as to why such discontinuities are allowed to occur, despite the fact that they may be at odds with certain rules that seem to apply in the case of fermions, I would suggest that we simply assume that in fact the flow of time along the world-lines of elementary particles\index{particle world-line!absence of interruption of flow of time} is never really interrupted, given that the bosons mediating the non-gravitational interactions\index{non-gravitational interaction!mediating bosons} between elementary particles of matter, somehow, allow charges to propagate along two opposite directions of time all at once, \textit{as if} spin-one interaction bosons\index{spin-one interaction bosons!composite particles} were composite particles made of a fermion and an anti-fermion which need not carry the same charges.

One important characteristic of such an alternative description is that, if there must, in effect, be a continuity of the flow of time along the world-lines of all elementary particles, then, in the context where the interaction bosons would allow a propagation of charges along two opposite directions of time all at once, it follows that the direction of propagation in time\index{direction of propagation in time!interacting particles} of the interacting particles would actually be allowed to remain unchanged during any such interaction process, because time flows in and out of the boson at each interaction vertex\index{interaction vertex!in and out time flow}. This is, of course, in accordance with the previously stated conclusion to the effect that the condition of continuity of the flow of time\index{condition of continuity of flow of time!particle world-line} along the world-lines of elementary particles forbids the transformation of a particle propagating a given charge forward in time into an apparently identical particle propagating an opposite charge backward in time and therefore it seems that it is really the necessary continuity of the flow of time that imposes that the interaction bosons be described as always propagating charges in two opposite directions of time all at once, so that the flow of charge\index{flow of charge!absence of interruption} itself is never interrupted.

From the viewpoint of this equivalent description of elementary particle interaction\index{elementary particle interactions!equivalent description} processes, it would follow that, for any interaction vertex, time would flow from the incoming fermion into the interaction boson and from the interaction boson into the outgoing fermion (or from the outgoing fermion into the interaction boson and from the boson into the incoming fermion if this particle is propagating backward in time) and the same must be happening at the other vertex of an interaction diagram.

An examination of the diagrams describing the interactions between elementary particles, such as those represented in Figure \ref{fig:4.1} and Figure \ref{fig:4.2}, clearly shows that this hypothesis agrees with the description of all known interaction processes involving a variation of the charges\index{charges variation!interacting matter particles} of the interacting matter particles that must be carried by the interaction bosons\index{interaction bosons!charge carrying}, as well as with that of all neutral interaction\index{neutral interactions} processes involving particles with integer charges, but only when we assume that the bidirectional charge signs\index{bidirectional charge sign!interacting particles} of the interacting particles (those which are attributed to matter particles which are propagating forward in time, when they are observed from the unidirectional-time viewpoint\index{unidirectional-time viewpoint}) must be either both non-reversed or both reversed, regardless of whether the interacting particles are propagating forward or backward in time.

\begin{figure}
\begin{center}
\begin{picture}(380,290)
\newsavebox{\diagff}
\savebox{\diagff}{
\put(10,0){\vector(1,3){20}}
\put(30,62){\vector(-1,3){20}}
\put(33,59){\vector(3,1){60}}
\put(92,82){\vector(-3,-1){60}}
\put(115,20){\vector(-1,3){20}}
\put(94,82){\vector(1,3){20}}}
\newsavebox{\diagfp}
\savebox{\diagfp}{
\put(10,0){\vector(1,3){20}}
\put(30,62){\vector(-1,3){20}}
\put(33,59){\vector(3,1){60}}
\put(92,82){\vector(-3,-1){60}}
\put(114,142){\vector(-1,-3){20}}
\put(95,80){\vector(1,-3){20}}}
\put(40,140){\usebox{\diagff}}
\put(32,260){$E^+$}\put(157,280){$E^+$}
\put(30,225){$q^{-1/3}$}\put(150,255){$q^{+2/3}$}
\put(80,190){$q^{+2/3}$}\put(100,220){$q^{-1/3}$}
\put(30,165){$q^{+2/3}$}\put(150,185){$q^{-1/3}$}
\put(97,175){$W^{+}$}
\put(210,140){\usebox{\diagfp}}
\put(202,260){$E^+$}\put(327,280){$E^-$}
\put(200,225){$q^{-1/3}$}\put(320,255){$q^{-2/3}$}
\put(252,190){$q^{+2/3}$}\put(272,221){$q^{-2/3}$}
\put(200,165){$q^{+2/3}$}\put(320,185){$q^{+1/3}$}
\put(270,175){$?$}
\put(40,0){\usebox{\diagfp}}
\put(32,120){$E^+$}\put(157,140){$E^-$}
\put(44,85){$b^{+}$}\put(150,115){$b^{+}$}
\put(81,50){$r^{+}$}\put(115,82){$b^{+}$}
\put(44,25){$r^{+}$}\put(150,45){$r^{+}$}
\put(97,35){$G_{r\overline{b}}$}
\put(210,0){\usebox{\diagff}}
\put(202,120){$E^+$}\put(327,140){$E^+$}
\put(214,85){$r^{-}$}\put(320,115){$r^{+}$}
\put(250,50){$r^{+}$}\put(284,82){$r^{-}$}
\put(214,25){$r^{+}$}\put(320,45){$r^{-}$}
\put(270,35){$?$}
\end{picture}
\end{center}
\caption[Alternative Feynman diagrams for flavor- and color-changing interactions between quarks]{Alternative Feynman diagrams\index{Feynman diagrams!flavor- and color-changing interactions} for flavor- and color-changing interactions between quarks. Here $q^{+2/3}$ and $q^{-1/3}$ represent the magnitudes and the signs of two bidirectional electric charges\index{bidirectional electric charges}, while $b^{+}$ and $b^{-}$ or $r^{+}$ and $r^{-}$ represent a quark bidirectional color and anti-color, and $E^+$ and $E^-$ are the energy signs relative to the direction of propagation in time\index{direction of propagation in time}, which corresponds to the direction of the arrows relative to the vertical axis. The diagrams on the left represent processes which are allowed to occur, while the diagrams on the right represent processes which are not allowed to occur based merely on the requirement that the bidirectional charge signs\index{bidirectional charge sign!invariance} be left invariant along the direction of the flow of time.}\label{fig:4.1}
\end{figure}

\begin{figure}
\begin{center}
\begin{picture}(380,150)
\savebox{\diagff}{
\put(10,0){\vector(1,3){20}}
\put(30,62){\vector(-1,3){20}}
\put(33,59){\vector(3,1){60}}
\put(92,82){\vector(-3,-1){60}}
\put(115,20){\vector(-1,3){20}}
\put(94,82){\vector(1,3){20}}}
\savebox{\diagfp}{
\put(10,0){\vector(1,3){20}}
\put(30,62){\vector(-1,3){20}}
\put(33,59){\vector(3,1){60}}
\put(92,82){\vector(-3,-1){60}}
\put(114,142){\vector(-1,-3){20}}
\put(95,80){\vector(1,-3){20}}}
\put(40,0){\usebox{\diagfp}}
\put(32,120){$E^+$}\put(157,140){$E^-$}
\put(40,85){$q^{-1}$}\put(150,115){$q^{-1}$}
\put(80,50){$q^{-1}$}\put(112,82){$q^{-1}$}
\put(40,25){$q^{-1}$}\put(150,45){$q^{-1}$}
\put(92,35){$\gamma/Z^0$}
\put(210,0){\usebox{\diagff}}
\put(202,120){$E^+$}\put(327,140){$E^+$}
\put(210,85){$q^{-1}$}\put(320,115){$q^{+1}$}
\put(250,50){$q^{-1}$}\put(282,82){$q^{+1}$}
\put(210,25){$q^{-1}$}\put(320,45){$q^{+1}$}
\put(270,35){$?$}
\end{picture}
\end{center}
\caption[Alternative Feynman diagrams for flavor-conserving electroweak interactions between fermions with integer charges]{Alternative Feynman diagrams\index{Feynman diagrams!flavor-conserving interactions} for flavor-conserving electroweak interactions between fermions with integer charges. Here again $q^{-1}$ and $q^{+1}$ represent the magnitudes and the signs of the electric charges, as determined from a bidirectional-time viewpoint\index{bidirectional-time viewpoint}, while $E^+$ and $E^-$ are the energy signs relative to the direction of propagation in time\index{direction of propagation in time}. It is only for processes of the kind described in the diagram on the left that the sign of the bidirectional charge\index{bidirectional charge sign!discontinuous variations} carried by the interacting matter particles does not vary discontinuously along the direction in which time is flowing and what is observed is that only processes of this kind actually occur in nature, even when there is no actual change in the charges of the interacting particles.}\label{fig:4.2}
\end{figure}

Indeed, while the above defined condition is satisfied for those processes where both of the interacting particles are propagating a certain bidirectional charge with a unique given sign either forward or backward in time, it cannot occur for the same processes where only one of the interacting particles is propagating a reversed bidirectional charge (in any direction of time). In the latter case, either a bidirectional charge\index{bidirectional charge!reversal} would have to transform into an opposite bidirectional charge in the direction in which time is flowing as a charge is propagating from one of the two particles into the interaction boson\index{interaction bosons} and then along the ingoing or outgoing trajectory of the other particle, or else a similar transformation would need to take place along the spacetime trajectory of one or the other of two interacting particles as a result of the interaction (which would require the interaction boson to propagate opposite bidirectional charges in opposite directions of time, while no interaction particle with such properties is known to exist).

It is the fact that the bidirectional charge signs\index{bidirectional charge sign!discontinuous variation} of elementary particles would have to vary discontinuously along the direction in which time is flowing in the diagram describing an interaction among elementary particles (that which is indicated by the direction of the arrows) that explains that, from the viewpoint of the above description of interaction processes, the only interactions which are allowed to take place between two particles are those involving particles which both have either a reversed or a non-reversed sign of bidirectional charge\index{reversed and non-reversed bidirectional charges!interacting particles}, although this is only explicitly apparent in the case of an interaction during which there is an exchange of charge that is carried by the interaction boson\index{interaction bosons!charge carrying}.

But even when a neutral interaction boson\index{neutral interaction boson!continuous bidirectional time flow} is involved, bidirectional time must flow continuously from the original interacting matter particle into the forward-propagating component of the boson and then into the ingoing or outgoing world-line\index{world-line!ingoing or outgoing} of the other interacting matter particle (depending on its direction of propagation in time\index{direction of propagation in time}), despite the fact that it may appear like no specific charge is being propagated by the interaction boson. In such a case one must still assume, in effect, that a charge is being propagated forward in time, while another is being propagated backward in time by the interaction boson, but given that those two charges would actually be the exact same bidirectional charge, then from a unidirectional-time viewpoint\index{unidirectional-time viewpoint} they would have opposite polarities, so that the interaction boson itself would appear to have a null charge.

In fact, even in the case of neutral interactions\index{neutral interactions} involving particles with differing fractional charges\index{fractional charges} it seems that this continuity would be observed to apply, but only if we consider that the current elementary fermions\index{fermions!composite particles} themselves are composite particles made of three more elementary fermions carrying either no electric charge or an elementary electric charge\index{elementary electric charges} $q=\pm 1/3$ (whose polarity is the same for all three particles) which they can propagate in any direction of time, independently from one another, as illustrated in Figure \ref{fig:4.3}. It would then be those component particles which would interact when certain neutral interaction bosons\index{neutral interaction boson} are exchanged between fermions, so that even when neutral interactions between particles with differing fractional charges are involved, two identical, fractional charges would be propagated in opposite directions of time by the interaction boson.

Under such conditions, even fermions which are not usually considered to be anti-particles could actually be composed of both particles and anti-particles always carrying either no electric charge at all, or the same bidirectional charge\index{bidirectional charge!component particles} (with the same magnitude and polarity) either forward or backward in time. I believe that the fact that this hypothesis allows to extend the domain of applicability of the condition of continuity of the flow of time\index{condition of continuity of the flow of time!domain of applicability} to all known elementary particle processes\index{elementary particle processes}, as required from a phenomenological viewpoint, constitute a strong indication to the effect that it is generally valid and cannot be ignored. This may well be the decisive argument in favor of a composite model\index{composite model!quarks and leptons} for quarks and leptons, if the difficulties such a model would involve can themselves be overcome.

\begin{figure}
\begin{center}
\begin{picture}(370,350)
\savebox{\diagff}{
\put(10,0){\vector(1,3){20}}
\put(30,62){\vector(-1,3){20}}
\put(33,59){\vector(3,1){60}}
\put(92,82){\vector(-3,-1){60}}
\put(115,20){\vector(-1,3){20}}
\put(94,82){\vector(1,3){20}}}
\savebox{\diagfp}{
\put(10,0){\vector(1,3){20}}
\put(30,62){\vector(-1,3){20}}
\put(33,59){\vector(3,1){60}}
\put(92,82){\vector(-3,-1){60}}
\put(114,142){\vector(-1,-3){20}}
\put(95,80){\vector(1,-3){20}}}
\newsavebox{\diagquarks}
\savebox{\diagquarks}{
\put(0,0){\vector(1,3){20}}
\put(20,62){\vector(-1,3){20}}
\put(7,0){\vector(1,3){20}}
\put(27,62){\vector(-1,3){20}}
\put(14,0){\vector(1,3){20}}
\put(34,62){\vector(-1,3){20}}
\put(35,59){\vector(3,1){60}}
\put(94,82){\vector(-3,-1){60}}
\put(116,142){\vector(-1,-3){20}}
\put(96,80){\vector(1,-3){20}}
\put(123,142){\vector(-1,-3){20}}
\put(103,80){\vector(1,-3){20}}
\put(130,20){\vector(-1,3){20}}
\put(110,82){\vector(1,3){20}}}
\put(20,200){\usebox{\diagff}}
\put(12,320){$E^+$}\put(137,340){$E^+$}
\put(10,285){$q^{+2/3}$}\put(130,315){$q^{-1/3}$}
\put(60,250){$?$}\put(98,282){$?$}
\put(10,225){$q^{+2/3}$}\put(130,245){$q^{-1/3}$}
\put(77,235){$\gamma/Z^0$}
\put(200,200){\usebox{\diagff}}
\put(192,320){$E^+$}\put(317,340){$E^+$}
\put(190,285){$q^{+2/3}$}\put(310,315){$q^{+1/3}$}
\put(240,250){$?$}\put(278,282){$?$}
\put(190,225){$q^{+2/3}$}\put(310,245){$q^{+1/3}$}
\put(260,235){$?$}
\put(20,40){\usebox{\diagquarks}}
\put(7,165){$E^+$}\put(22,165){$E^+$}\put(37,165){$E^+$}
\put(124,185){$E^-$}\put(139,185){$E^-$}\put(154,185){$E^+$}
\put(7,43){$q^0$}\put(17,25){$q^{+1/3}$}\put(42,43){$q^{+1/3}$}
\put(104,63){$q^{+1/3}$}\put(139,45){$q^{+1/3}$}\put(154,63){$q^{+1/3}$}
\put(17,6){$Q_{future}^{+2/3}$}\put(139,27){$Q_{future}^{-1/3}$}
\put(59,89){$q^{+1/3}$}\put(84,120){$q^{+1/3}$}
\put(75,75){$\gamma/Z^0$}
\put(200,40){\usebox{\diagquarks}}
\put(187,165){$E^+$}\put(202,165){$E^+$}\put(217,165){$E^+$}
\put(304,185){$E^-$}\put(319,185){$E^-$}\put(334,185){$E^+$}
\put(187,43){$q^0$}\put(197,25){$q^{+1/3}$}\put(222,43){$q^{+1/3}$}
\put(284,63){$q^{-1/3}$}\put(319,45){$q^{-1/3}$}\put(334,63){$q^{-1/3}$}
\put(197,6){$Q_{future}^{+2/3}$}\put(319,27){$Q_{future}^{+1/3}$}
\put(239,89){$q^{+1/3}$}\put(264,120){$q^{-1/3}$}
\put(265,75){$?$}
\end{picture}
\end{center}
\caption[Alternative Feynman diagrams for flavor-conserving electroweak interactions between fermions with fractional charges]{Alternative Feynman diagrams\index{Feynman diagrams!flavor-conserving interactions} for flavor-conserving electroweak interactions between fermions with fractional charges. Here $q^{+2/3}$ and $q^{-1/3}$ represent the bidirectional electric charges\index{bidirectional electric charges} of quarks\index{quarks!component particles} or their hypothesized component particles and $E^+$ and $E^-$ are the energy signs relative to the direction of propagation in time\index{direction of propagation in time}, while $Q_{future}^{+2/3}$ and $Q_{future}^{-1/3}$ are the charges of the composite quarks\index{quarks!composite particles}, as observed from the forward-in-time viewpoint. It is only for processes of the kind described in the diagrams on the left that the sign of the bidirectional charge\index{bidirectional charge sign!discontinuous variations} carried by the interacting component particles does not vary discontinuously along the direction in which time is flowing, while only processes of this kind are observed to occur.}\label{fig:4.3}
\end{figure}

The important point, however, is that the identical bidirectional charges which are emitted and absorbed at an interaction vertex\index{interaction vertex!emitted and absorbed charges} by a particle with a non-reversed bidirectional charge\index{bidirectional charge!non-reversed} must themselves both consist of non-reversed bidirectional charges, if the condition of continuity of the flow of time\index{condition of continuity of flow of time} is to be satisfied at this interaction vertex. But given that this must also be the case at the other interaction vertex of an interaction diagram, it means that if the other particle involved in the interaction has a reversed bidirectional charge, then those conditions cannot be fulfilled, because in such a case the neutral interaction boson would need to propagate two identical, \textit{reversed} bidirectional charges in opposite directions of time, which would be emitted and absorbed by this other particle, while this would not agree with the condition that exist at the first interaction vertex. Thus, regardless of whether the charges of two interacting particles vary or not, it is only in those cases where the charges of both particles are either reversed or non-reversed that the interaction is actually allowed to occur.

For this conclusion to apply, all that is required is that there always exists a clear (even though relationally defined) distinction between what constitutes an ordinary matter particle and what constitutes a particle with reversed bidirectional charge\index{reversed bidirectional charge}, even when we are dealing with particles which do not carry the exact same charges, and this means that, even observable particles with different charges must share a certain physical attribute which would only reverse if the sign of the bidirectional charge carried by those particles (that which is independent of the direction in which a particle is propagating in time) was allowed to reverse, while it remains unaffected by a mere reversal of the direction of propagation in time\index{direction of propagation in time!reversal} that actually leaves the sign of charge invariant from the viewpoint of bidirectional time\index{bidirectional-time viewpoint}. But as I suggested above, there are reasons to believe that we will only be able to fully understand what justifies the rule described here when we obtain a more complete theory of elementary particles\index{elementary particles!more complete theory} which would allow a description of quarks and leptons\index{quarks and leptons!composite particles} (and perhaps also of interaction bosons\index{interaction bosons!composite particles}) as composite particles\index{composite particles}.

In any case, if the constraint of continuity of the flow of time\index{constraint of continuity of flow of time!elementary particle world-lines} along an elementary particle world-line extends to interaction bosons in the way suggested here, then it would appear that no interaction can occur that would involve two identical matter particles with opposite bidirectional charges (those observed while following the direction of propagation in time of the particles) propagating in any direction of time, or even merely two different particles, when only one of them has a non-reversed sign of bidirectional charge which is propagating in either the past or the future direction of time. I believe that this would be a simple consequence of the fact that no such an interaction could ever be described as a process during which the polarities of all the \textit{non-gravitational} attributes of the elementary particles involved (and this excludes spin and the sign of energy) remain unchanged (are not subject to discontinuous reversal) as we follow the direction of the flow of time\index{direction of flow of time!particle world-line} along their respective world-lines, from one of the two interacting matter particles into the interaction boson and then back into the other matter particle, either forward or backward in time.

What must be clear is that there is a difference between the description of an interaction process according to which none of two interacting particles has a reversed sign of bidirectional charge\index{reversed bidirectional charge sign} and the alternative description according to which one of the particles would have a reversed bidirectional charge sign. Thus, even if it may appear that a quantum-mechanically equivalent description of a certain interaction process could exist that would be obtained by simply reversing both the sign of charge and the direction of propagation in time\index{direction of propagation in time!reversal} for one of the interacting particles, we would have to conclude that the description for which the sign of charge is reversed, independently from the direction of its propagation in time, is actually distinct from that which does not involve such a reversal. This distinction would simply be a consequence of the fact that, while the sign of charge that is reversed in the apparently equivalent description of the process would not appear to be reversed from the viewpoint of unidirectional time\index{unidirectional-time viewpoint}, along which the particle is observed, from a bidirectional viewpoint\index{bidirectional-time viewpoint} this charge would nevertheless be reversed and this constitutes a physically significant change.

Therefore, I'm allowed to conclude that the rule which is implicitly assumed to apply, from a conventional viewpoint, to the effect that no positive-action particle that would be propagating charges opposite those of ordinary particles in the opposite direction of time need be considered to exist, is only appropriate in the sense that it is not possible for any such particle to interact with ordinary particles, at least through the exchange of interaction bosons\index{interaction bosons!non-gravitational forces} associated with non-gravitational forces. But it is also very clear that this does not mean that particles with reversed bidirectional charges\index{reversed-bidirectional-charge particles!existence} cannot exist, because, from a theoretical viewpoint, this conclusion would be as unjustified as that which would amount to argue that ordinary particles themselves cannot exist.

The distinction between electrons\index{electrons!propagation of positive charge backward in time} propagating positive charges backward in time and ordinary electrons\index{electrons!propagation of negative charge forward in time} propagating negative charges forward in time is only a relational distinction, in the sense that a positively-charged electron propagating backward in time can only be distinguished from an ordinary electron through the fact that it actually has a charge that is opposite that of the ordinary electron, even while it propagates in a direction of time opposite that in which an ordinary electron propagates, but those are not absolutely characterized properties and an electron with positive bidirectional charge is only different from an electron with negative bidirectional charge in the exact same way an electron with negative bidirectional charge is different from an electron with positive bidirectional charge and it is not possible to distinguish one from the other, except through those mutual relationships. If there is no intrinsic or absolute distinction between particles in those two different states, however, then it means that none of them can be considered more real than the other. In other words, both kinds of particles must be assumed to exist, even though matter composed of reversed-bidirectional-charge particles\index{reversed-bidirectional-charge particles!dark matter} must by necessity be dark from the viewpoint of ordinary matter.

Now, obviously, the only way that such a conclusion could come out as not totally meaningless is if the gravitational interaction is not affected by the condition of continuity of the flow of time\index{condition of continuity of flow of time!particle world-line} along the world-lines of elementary particles, because otherwise there should be no interaction at all between ordinary positive-energy particles and positive-energy particles with reversed bidirectional charges. But I believe that this is actually unavoidable, because it is clear from the above discussion that it is merely the non-gravitational charges of elementary particles that must not be subjected to any discontinuous reversal along their respective world-lines. The gravitational interaction\index{gravitational interaction!fundamental distinction} is fundamentally distinct from all other interactions in this respect, given that it is neutral with respect to all non-gravitational charges, which is not really the case with other neutral interactions\index{neutral interactions} that couple to charge (even though they are mediated by interaction bosons\index{interaction bosons!neutral} that do not appear to carry any charge).

This essential distinction, which is unique to the gravitational interaction, appears to be what allows opposite-bidirectional-charge particles\index{opposite-bidirectional-charge particles!gravitational interaction} with the same sign of action to interact gravitationally (and attractively) with one another. The fact that gravitons couple only to energy, while the sign of energy or action is not affected by a reversal of bidirectional charge\index{bidirectional charge reversal!invariance of action sign} means that gravitation\index{gravitation!truly neutral interaction} is the only truly neutral interaction, which therefore remains unaffected by the condition of continuity of the flow of time that prevents other interactions between opposite-bidirectional-charge particles\index{opposite-bidirectional-charge particles!non-gravitational interactions}. Given that gravitons\index{graviton!spin-two interaction boson} have a spin that is twice as large as that of other interaction-mediating bosons, it is tempting to suggest that this neutrality is preserved as a result of the fact that gravitons allow charges to propagate along two opposite directions of time \textit{in two different ways} all at once, for each individual interaction, thereby allowing the changes produced by one of those exchange process to be neutralized by those generated by the other process, in observance of the condition of continuity of the flow of time, \textit{as if} spin-two gravitons\index{graviton!composite particle} where composite particles made of two spin-one bosons.

Anyhow, it seems that the only conclusion that can be drawn is that, despite the fact that positive-energy matter with reversed bidirectional charges\index{reversed-bidirectional-charge matter!dark matter} is dark, it would actually exert attractive gravitational forces on ordinary positive-energy matter particles, and in the presence of inhomogeneities, also, indirect, repulsive gravitational forces\index{indirect repulsive gravitational force} on all negative-energy matter particles, regardless of their bidirectional charge signs.

It must be assumed, in effect, that there also exist negative-action particles\index{negative-action particles!reversed bidirectional charges} with reversed bidirectional charges, which would be dark from the viewpoint of `ordinary' negative-energy observers as well. But due to the requirement of symmetry under exchange\index{exchange symmetry!positive- and negative-energy matter} of positive- and negative-energy matter, any such negative-energy particle would be gravitationally attracted to other negative-energy matter particles and gravitationally repelled by positive-energy matter overdensities, regardless of whether this matter carries positive or negative bidirectional charges. What allows negative-energy matter carrying reversed bidirectional charges to exert a gravitational force on positive-action particles (in the presence of inhomogeneities in this dark, negative-energy matter distribution) is the fact that \textit{all} negative-action particles are equivalent to the presence of voids in the positive-energy portion of the vacuum\index{void in positive vacuum energy}, while, under appropriate conditions, such voids \textit{necessarily} exert indirect gravitational forces on positive-action particles.

\bigskip

\noindent To summarize what I have discussed so far, it seems that additional, attractive gravitational forces could, in principle, be exerted on positive-energy matter, as a consequence of the presence in the negative-energy matter distribution of underdensities which develop as a consequence of the gravitational repulsion\index{gravitational repulsion} exerted by positive-energy matter overdensities. We have no choice, however, but to assume that positive-energy dark matter\index{dark matter!positive energy} is present in our universe in one form or another, because the magnitude of the underdensities that could exist today in the negative-energy matter distribution is much too small to explain a significant portion of the missing-mass\index{missing mass!effect} effects observed around visible, present day structures, due to the fact that the average density of negative-energy matter in which such underdensities could develop is itself much too small.

I must acknowledge, however, that it is not possible to conclude that the missing-mass effect is attributable, for the most part, to the presence of reversed-bidirectional-charge particles\index{reversed-bidirectional-charge particles}. Indeed, if most of the dark matter that is assumed to be responsible for the missing-mass effect was composed of baryonic dark matter\index{baryonic dark matter!self-interaction} particles which interact with themselves through the same forces by which ordinary baryonic matter\index{baryonic matter!ordinary} particles interact, then it would be more difficult to explain the near spherical shape of dark-matter halos\index{dark-matter halos!near spherical shape}, or certain observations of colliding clusters of galaxies\index{colliding clusters of galaxies} which show that, while the detectable high-energy gas\index{high-energy gas} originally present in the clusters is stripped of the galaxies as a result of such a collision, most of the dark matter is unaffected by the process.

I initially thought that this difficulty may simply be a consequence of the fact that we ignore the possibility that baryonic dark matter could be more susceptible to collapse into stars and other high-density objects at a very early stage if its average density happens to be larger than that of visible baryonic matter\index{baryonic matter!visible} (I will soon explain how I later came to realize that this is not possible and that the average density of normally gravitating baryonic dark matter must equal that of ordinary baryonic matter). Under such conditions baryonic dark matter would no longer interact with itself, other than gravitationally, on a larger scale, when galaxies begin to form later on, which could perhaps allow to explain the near spherical shape of dark-matter halos. For the same reason, it would have appeared appropriate to assume that the dark matter present inside colliding clusters is mostly unaffected by the collisions, just like the visible stars which are present within the galaxies, despite the fact that the dark matter particles that form those astronomical objects are allowed to interact electromagnetically among themselves.

For this to be a valid hypothesis, however, one would need to assume that a density of baryonic dark matter\index{baryonic dark matter} much larger than that of visible baryonic matter exists in the form of massive, compact, halo objects\index{massive compact halo objects} or MACHOs. But even though early studies seemed to indicate that the existence of a large amount of matter in the form of invisible MACHOs was not completely ruled out, because what really motivated the commonly held opinion that there cannot exist enough MACHOs to provide a sizable portion of the dark matter\index{dark matter} was merely the impossibility for those objects to be formed of ordinary baryonic matter\index{baryonic matter!ordinary} (whose presence would affect the predictions of Big Bang nucleosynthesis\index{Big Bang nucleosynthesis}), more recent astronomical observations \cite{Zumalacarregui-1} do confirm that there cannot be such a large portion of normally-gravitating matter in the form of MACHOs (regardless of the nature of their constituent particles). Thus, it is no longer possible to assume that a sufficiently large number of such objects could exist that would be composed of baryonic matter\index{baryonic matter!reversed-bidirectional-charge particles} with reversed bidirectional charges\footnote{
This is not to say that there cannot exist any large astronomical objects composed of reversed-bidirectional-charge matter\index{reversed-bidirectional-charge matter!astronomical objects|nn}, as it is quite possible, in fact, that invisible stars and planets made of such matter are present in our own region of the universe.}
 (which would not have affected the predictions of Big Bang nucleosynthesis).

As a consequence, it is necessary to recognize that the above discussed difficulties associated with the hypothesis that dark-matter particles\index{dark-matter particles!self-interaction} may interact electromagnetically with themselves (like ordinary baryonic matter) can only be surmounted if most of the observed missing-mass effect\index{missing mass!effect} is attributable to a phenomenon distinct from those I have discussed so far. This doesn't mean that none of the dark matter can consist of baryonic matter with reversed bidirectional charges (this is not ruled out by the new observations), but merely that it is not possible to conclude that the necessary existence of such matter provides a valid explanation for most of the missing-mass effect observed around visible galaxies and clusters.

Now, even aside from the question of the origin of the missing-mass effect, it would certainly be interesting to know what the average density of positive-energy matter\index{positive-energy matter!reversed bidirectional charges} with reversed bidirectional charges actually is in our universe. Does a theoretical constraint exist that would allow one to tell what this density actually is? In fact, it turns out that this value can be determined with great accuracy. To see how this is possible it will be necessary to re-examine the problem of matter-antimatter asymmetry\index{matter-antimatter asymmetry!problem} in the light of the progress already achieved in this report. First of all, is it appropriate to conclude that the absence of antimatter in our universe reflects an absolute lopsidedness\index{absolute lopsidedness!direction of time} with respect to the direction of time?

This is an important question, because, while it is usually recognized that thermodynamic time asymmetry\index{thermodynamic time asymmetry} is probably not the cause of the violations of $T$ symmetry which have been observed in certain high-energy experiments, the direction of time singled out by $T$ violations can be related to the thermodynamic arrow of time\index{thermodynamic arrow of time} and this might allow one to conclude that, from a phenomenological viewpoint, our universe\index{universe!absolute lopsidedness} is characterized by an absolute lopsidedness. Given that the time reversal symmetry operation can now be understood to involve a transformation of matter into antimatter, the question of whether there actually exists such a preferred direction in time\index{preferred time direction!matter-antimatter asymmetry} would, in effect, be equivalent to ask if there really is an absolutely definable asymmetry associated with the predominance of matter over antimatter in our universe?

I once thought that what would allow symmetry with respect to the direction of time to be regained, despite the observed asymmetry between matter and antimatter\index{matter-antimatter asymmetry}, would be the existence of a certain constraint that would require that whenever there is an overabundance of forward-in-time propagating positive-action particles\index{forward-in-time propagating positive-action particles} over positive-action particles propagating backward in time\index{backward-in-time propagating positive-action particles}, there should be an overabundance of unobservable, negative-action particles propagating backward in time\index{backward-in-time propagating negative-action particles} over forward-in-time propagating negative-action particles\index{forward-in-time propagating negative-action particles}. It would then merely be the fact that the backward-propagating particles which do exist in large numbers cannot be observed, due to the fact that they would have an opposite sign of action and would therefore be dark, that would make it seem like there is a smaller number of such particles, compared to forward-propagating particles.

But, as I realized later, this solution would not be valid, because, as I explained in section \ref{sec:3.9}, the violation of $T$ or $C$ symmetry that must be involved in giving rise to the observed positive-action, matter-antimatter asymmetry cannot be assumed to directly compensate that which would give rise to the similar, relationally-defined $T$ or $C$ asymmetry that \textit{may} be affecting negative-action matter and antimatter, because $T$ and $C$ can be violated to a different degree for negative-action matter, as long as invariance under a combined $PTC$ symmetry operation\index{combined symmetry operation} is independently preserved by processes involving this type of matter.

Thus, it cannot simply be assumed that what happens is that the matter-antimatter asymmetry\index{matter-antimatter asymmetry!negative-energy matter} is reversed for negative-energy matter and that there is actually the same number of otherwise identical particles propagating the same sign of energy and the same sign of bidirectional charge\index{bidirectional charge} in opposite directions of time, when we appropriately take into account the contribution of the unseen negative-energy matter. What I will explain in section \ref{sec:4.9} is that once it is recognized that there is a possibility for time to extend past the initial Big Bang singularity\index{initial Big Bang singularity}, following a hypothetical quantum bounce\index{quantum bounce}, then it is no longer necessary to conclude that there is an absolute lopsidedness\index{absolute lopsidedness!matter-antimatter asymmetry}, that one could attribute to the apparent asymmetry between matter and antimatter. The question that remains, however, is whether there exist more ordinary electrons and protons than there exist electrons and protons with reversed bidirectional charge signs?

Here, I must first mention that it is well understood already that the difference between the density of matter and that of antimatter in the universe today is a consequence of the fact that, due to a certain violation of $CP$ symmetry (which would imply a violation of time reversal symmetry\index{time reversal symmetry!violation} $T$), there was a little more matter than antimatter in the primordial Big Bang state\index{primordial Big Bang state}, so that some matter was allowed to survive the annihilation of matter with antimatter\index{matter-antimatter annihilation} that took place when the temperature became too low for ordinary pair creation\index{pair creation and annihilation!ordinary} to compensate the related process of pair annihilation. The problem is that it is not known what the exact origin of the asymmetry is that gave rise to this overabundance of matter over antimatter\index{overabundance of matter over antimatter!origin}.

Now, what I realized is that, given that particles with opposite bidirectional charges\index{opposite-bidirectional-charge particles!gravitational interaction} do interact gravitationally with one another, then under conditions where gravitation is strong enough, as was the case in the first instants of the Big Bang\index{Big Bang!first instants} (despite the very high homogeneity of the initial matter distribution\index{initial matter distribution!very high homogeneity}), it should be possible for pairs of particles with opposite bidirectional charges\index{opposite-bidirectional-charge particles!pair creation} propagating in the \textit{same} direction of time to be created out of gravitational radiation\index{gravitational radiation} energy.

While all sorts of particle-antiparticle pairs\index{particle-antiparticle pairs!creation and annihilation} can be created in the first instants of the Big Bang out of all sorts of radiation, most such pairs annihilate back to radiation just to be recreated in the same way soon after, and those creation processes may last until relatively late times, as long as the interactions involved are strong enough for the processes to persist, despite the reduction of temperature associated with expansion. But once a pair of particles was created out of gravitational radiation, at the epoch in the far past when space was expanding at an arbitrarily large rate, the particles so produced were allowed to move away from one another rapidly enough that they were no longer able to annihilate back to gravitational radiation (given that, on the characteristic scale of quantum-gravitational phenomena\index{quantum-gravitational phenomena!characteristic scale}, the distance between the particles had become too large and the magnitude of their energies too low), which means that the creation process had become permanent.

If those particles had opposite bidirectional charges\index{opposite bidirectional charges!particle-antiparticle pair creation} and if there are more matter particles than antiparticles created in such a way, however, then this imbalance would itself become permanent, because ordinary matter particles and particles with reversed bidirectional charges cannot interact other than gravitationally. For such an imbalance to be initiated, a violation of $T$ or $C$ symmetry would need to arise which would affect the production of pairs of opposite-bidirectional-charge particles\index{opposite-bidirectional-charge particles!forward-in-time propagation} propagating forward in time in the exact opposite way it would affect the production of pairs of opposite-bidirectional-charge particles\index{opposite-bidirectional-charge particles!backward-in-time propagation} propagating backward in time, even though it would not affect the production of pairs\index{pair production!same-bidirectional-charge particles} of particles with the same bidirectional charge signs propagating in opposite directions of time (precisely because in such a case both directions of time are involved in every process).

Thus, the first of the following two graviton decay processes\index{graviton!decay process} (where $+2/3$ and $-2/3$ are the \textit{bidirectional} electric charges\index{bidirectional electric charges} of an up quark and the overline indicates that the quark is propagating backward in time) must have occurred a little more often than the second and the same must be true for all types of particles:
\begin{displaymath}
g\rightleftharpoons u^{+2/3}+u^{-2/3}
\end{displaymath}
\begin{displaymath}
g\rightleftharpoons \bar{u}^{+2/3}+\bar{u}^{-2/3}
\end{displaymath}
I believe that this is what explains that a little more ordinary matter than ordinary antimatter\index{matter and antimatter!ordinary} was produced in the first instants of the Big Bang\index{Big Bang!first instants}. But if more positively charged up quarks\index{positively charged up quarks!forward-in-time propagation} propagating forward in time than positively charged up quarks\index{positively charged up quarks!backward-in-time propagation} propagating backward in time are produced by gravitational radiation, then only positively charged up quarks propagating forward in time would remain once conventional particle-antiparticle\index{particle-antiparticle!annihilation} annihilation processes are over.

While it is not required that the exact same numbers of ordinary up quarks and anti-up quarks be produced by the processes described above, it is still necessary, however, that the exact same numbers of particles propagating opposite bidirectional charges\index{opposite-bidirectional-charge particles!direction of propagation in time} in the same direction of time be produced by those pair creation events. As a result, if there is an overabundance of protons propagating a positive bidirectional charge\index{positive-bidirectional-charge protons and antiprotons} forward in time over antiprotons propagating the same charge backward in time, from the viewpoint of an observer made of such protons, then there should exist an overabundance of protons propagating a negative bidirectional charge\index{negative-bidirectional-charge protons and antiprotons} forward in time over antiprotons propagating the same bidirectional charge backward in time, from the viewpoint of an observer made of such negative-bidirectional-charge protons, which means that no asymmetry actually exists that would have to do with an overabundance of ordinary particles and antiparticles\index{ordinary particles and antiparticles} over particles and antiparticles with reversed bidirectional charges\index{reversed bidirectional charge!particles and antiparticles}.

Given that it is possible for the symmetry under a reversal of the direction of time\index{time reversal symmetry!violation} to be violated, even when time direction\index{time direction!relational parameter} remains a relationally defined parameter, and given that the gravitational interaction allows opposite-bidirectional-charge particles\index{opposite-bidirectional-charge particles!gravitational interaction} to interact without violating the constraint imposed by the requirement of continuity of the flow of time\index{requirement of continuity of flow of time}, I must then conclude that nothing would forbid a violation of matter-antimatter symmetry\index{matter-antimatter symmetry!violations} from developing as a result of more forward- than backward-in-time-propagating pairs being created (or vice versa).

Thus, contrarily to what I had envisaged at a certain point, it is not the number of negatively charged, negative-action electrons\index{negatively charged negative-action electrons} propagating backward in time, that must be the same as the number of negatively charged, positive-action electrons\index{negatively charged positive-action electrons} propagating forward in time at the present moment in our universe, but really the number of positively charged, positive-action electrons\index{positively charged positive-action electrons} propagating forward in time. As a result, the density of matter with reversed bidirectional charge signs\index{reversed-bidirectional-charge matter!density} (arising from the presence of those particles that survived the early phase of matter-antimatter annihilation\index{early matter-antimatter annihilation}) must be the same as that of matter with non-reversed bidirectional charge\index{non-reversed-bidirectional-charge matter!density} signs in our universe, while it is the density of negative-energy matter\index{negative-energy matter!density} that is allowed to differ substantially from that of positive-energy matter, at least once the primordial annihilation of matter particles with their antimatter counterparts has taken place.

Therefore, it is possible to assume that the average density of baryonic negative-energy matter\index{baryonic negative-energy matter!average density} is currently much smaller than that of baryonic positive-energy matter\index{baryonic positive-energy matter!average density}, even if those two densities must have been equal initially, because the ratio of matter to antimatter\index{ratio of matter to antimatter!primordial universe} in the primordial universe may have been closer to unity for negative-energy matter, so that a larger portion of negative-energy matter\index{negative-energy matter!particle-antiparticle annihilation} would have been subjected to particle-antiparticle annihilation. But if that is the case, then even baryonic negative-energy matter\index{baryonic negative-energy matter!reversed bidirectional charges} with reversed bidirectional charges would be nearly absent at the present time, as its density must equal that of baryonic negative-energy matter\index{baryonic negative-energy matter!non-reversed bidirectional charges} with non-reversed bidirectional charges, which means that following the early annihilation of baryons with antibaryons\index{early baryon-antibaryon annihilation}, the magnitude of the total, average density of negative-energy matter could differ from that of positive-energy matter by as much a measure as the whole average density of baryonic positive-energy matter (both visible and dark).

Of course this does not mean that no negative-energy matter would remain in the universe, because there was initially a large portion of negative-energy matter\index{negative-energy matter!non-baryonic dark matter} in the form of non-baryonic dark matter (as I will explain below). But, as I came to realize, the possibility that there may no longer exist a significant proportion of negative-energy matter\index{negative-energy matter!baryonic matter} in baryonic form actually constitutes a requirement, from an observational viewpoint, as it implies that there can be no significant, localized overdensities in the negative-energy matter\index{negative-energy matter!stellar and galactic scale overdensities} distribution on stellar and galactic scales (for reasons that will be discussed in the following section), in agreement with the limits imposed by astronomical observations (concerning weak gravitational lensing\index{weak gravitational lensing!experiments} experiments in particular).

The prediction that more baryonic positive-energy matter\index{baryonic positive-energy matter!reversed bidirectional charges} with reversed bidirectional charges than baryonic negative-energy matter of the same kind can survive the early annihilation of matter with antimatter\index{early matter-antimatter annihilation} is also an essential element of the solution to the Hubble tension\index{Hubble tension} which I proposed in the preceding section, given that it allows the cosmological constant\index{cosmological constant!growing positive value} to grow to a larger positive value more rapidly during the early portion of the matter-dominated era\index{matter-dominated era}, because, under such conditions, the difference between the contribution of positive-energy matter and that of negative-energy matter to the deceleration of their specific rates of expansion\index{specific expansion rates!deceleration} is even larger than expected, due to the fact that the difference between the densities of positive and negative energy matter is twice as large as would have been the case if only ordinary, baryonic positive-energy matter (with non-reversed bidirectional charges) had survived the annihilation, while this allowed the specific rates of expansion\index{specific expansion rates!faster than expected divergence} to diverge more rapidly and the cosmological constant\index{cosmological constant!larger than expected growth} to grow larger than would otherwise be possible, before it decreased back to a lower value as a result of the moderating influence it was itself exerting on the divergence of the specific expansion rates of positive- and negative-energy matter.

It must be clear that the above discussed equality between the number of particles with non-reversed bidirectional charge\index{non-reversed-bidirectional-charge particles} signs and that of particles with reversed bidirectional charges\index{reversed-bidirectional-charge particles} is not just a mere possibility, but that the nature of the phenomenon responsible for the predominance of ordinary matter over ordinary antimatter implies that there must necessarily be as many particles of a given kind with a positive bidirectional charge as there are particles of the same kind with a negative bidirectional charge propagating in the same direction of time. Thus, one can avoid having to conclude that a preferred, absolutely defined sign of bidirectional charge\index{bidirectional charge!preferred sign} exists, as would be the case from a conventional viewpoint. But, given that I have argued (based on independent motives) that it is not possible for absolutely (non-relationally) defined physical attributes\index{absolutely defined physical attributes} to exist at a fundamental level, then we may consider that the prediction that there should exist dark matter\index{dark matter!reversed bidirectional charges} with reversed bidirectional charges in our universe and that this matter should be as abundant as ordinary matter itself at the present epoch is both observationally and theoretically adequate.

One obvious, but nonetheless significant consequence which would emerge, if the above proposed solution to the problem of the origin of the asymmetry between matter and antimatter\index{matter-antimatter asymmetry!origin problem} is valid, is that despite the fact that a condition of continuity of the flow of time\index{condition of continuity of flow of time} must apply that forbids all non-gravitational interactions between particles with opposite bidirectional charges\index{opposite-bidirectional-charge particles!non-gravitational interactions}, it should nevertheless be possible for such particles to interact through the gravitational interaction, because this assumption is necessary to derive that solution.

If we do not allow for opposite-bidirectional-charge particles\index{opposite-bidirectional-charge particles!decay of graviton} propagating in the same direction of time to be produced by the decay of a graviton, then no imbalance between the number of baryons and that of antibaryons can develop, because only particle-antiparticle pairs involving particles with the same sign of bidirectional charge propagating in opposite directions of time could be created and in such a case no violation of $T$ or $C$ symmetry could give rise to the required imbalances. Thus, the previously discussed argument, to the effect that it should be expected, based on independent motives, that particles with opposite bidirectional charges\index{opposite-bidirectional-charge particles!gravitational interaction} can interact gravitationally, without violating the condition of continuity of the flow of time along a particle world-line (as a result of the absolute neutrality of the gravitational interaction\index{gravitational interaction!absolute neutrality}), appears to be well-founded.

\bigskip

\noindent What I would like to explain, now, is that it is actually another phenomenon, made unavoidable by the existence of negative-energy matter, but not associated with the presence of voids in this matter distribution, that is ultimately responsible for most of the missing-mass effects\index{missing mass!effect}. You may recall that I mentioned in section \ref{sec:4.2} that from the viewpoint of the particular interpretation of the metric conversion factors\index{metric conversion factors} I have proposed and which allows the emergence of a non-zero value for the cosmological constant\index{cosmological constant!non-zero value}, it should be possible for vacuum energy density\index{vacuum energy density!position dependence} to vary with position, in addition to have a non-zero value on the global scale. But, in the context of this particular interpretation, it appears that if local variations of vacuum energy density\index{vacuum energy density!local variations} do arise, then they could only be attributable to the fact that local differences may develop between the metric properties of space\index{metric properties of space!opposite-energy observers} experienced by positive-energy observers and those experienced by negative-energy observers.

What I have come to understand is that, in fact, such variations are unavoidable, given that the presence of an inhomogeneity in the positive- or negative-energy matter distribution produces a local variation of the metric properties of spacetime\index{metric properties of spacetime!local variations} which, for a positive-energy observer, is different from that which is experienced by a negative-energy observer. The possibility, for opposite-energy observers, to experience differing metric properties of space as a result of the presence of matter inhomogeneities (which is allowed when it is not possible to directly compare such observer-dependent measures of distance) implies that vacuum energy can vary locally, depending on the strength of the gravitational fields\index{gravitational field strength!local source} produced by local sources, as long as there is no compensation between the local gravitational field attributable to positive-energy matter and that attributable to negative-energy matter.

A positive-energy observer, for example, would measure a positive increase in vacuum energy density\index{vacuum energy density!positive local increase} and a stronger attractive gravitational field, locally, in the presence of a positive-energy matter overdensity, because the spatial volume\index{volume of space!positive-energy observer} around that overdensity would appear smaller for such an observer given that from her viewpoint space is contracted and time dilated by the presence of positive-energy matter, while space is dilated and time contracted by the presence of a positive-energy matter overdensity, from the viewpoint of a negative-energy observer, which means that the maximum positive contribution to the density of vacuum energy\index{vacuum energy density!maximum positive contribution} which is directly experienced only by a negative-energy observer would be larger than the maximum negative contribution\index{vacuum energy density!maximum negative contribution} which is directly experienced only by our positive-energy observer, given that the volume of space in which the zero-point vacuum fluctuations\index{zero-point vacuum fluctuations} are taking place which give rise to this maximum negative contribution would appear proportionately smaller to this positive-energy observer.

However, the same positive-energy observer would measure a \textit{negative} increase in vacuum energy density\index{vacuum energy density!negative local increase} and a stronger \textit{repulsive} gravitational field, locally, in the presence of an overdensity in the negative-energy matter distribution, because the measures of spatial volume around that overdensity would appear larger to her, given that the curvature of spacetime\index{curvature of spacetime!positive-energy observer} attributable to such an overdensity, would give rise to space dilation and time contraction\index{space dilation and time contraction} for such an observer, while it would give rise to space contraction\index{space contraction} from the viewpoint of an observer of opposite energy sign, which means that the maximum negative contribution to the density of vacuum energy which is directly experienced by the positive-energy observer would be larger than the maximum positive contribution which is directly experienced by a negative-energy observer, so that more negative vacuum energy would appear to be present in the volume of space around such an overdensity\footnote{
I must warn the reader to be careful in asserting the validity of those conclusions, as it is very easy to make a mistake when evaluating the effects of the curvature of spacetime\index{curvature of spacetime!effect on spatial volume|nn} produced by various matter configurations on the volume of space present within the boundary surrounding a matter inhomogeneity\index{matter inhomogeneity!surrounding boundary|nn}. I am myself guilty of having once arrived at the exact opposite conclusion as that stated above and not having immediately realized that it was an error, because it produced the desired outcome in the context where the alternative form of the vacuum-energy term\index{vacuum-energy term!alternative form|nn} (entering the generalized gravitational field equations\index{generalized gravitational field equations|nn} introduced in section \ref{sec:2.14}), which I later abandoned, appeared to require the validity of that conclusion.}.

Those results are a simple consequence of the fact that the different metric properties of spacetime\index{metric properties of spacetime!opposite-energy observers} experienced by opposite-energy observers imply different volumes of space\index{volume of space!observer dependence}, even locally, and therefore also different measures for the maximum positive and negative contributions to the density of vacuum energy\index{vacuum energy density!maximum opposite contributions} provided by the natural vacuum-stress-energy tensors\index{natural vacuum-stress-energy tensors} (either $\gamma^{-+}\bm{V}_P^{++}$ and $-\bm{V}_P^{-+}$ for positive-energy observers, or $\gamma^{+-}\bm{V}_P^{--}$ and $-\bm{V}_P^{+-}$ for negative-energy observers) which add up to those arising from the observer-dependent measures of cosmological scale factor\index{scale factor!observer dependence}, as if we were dealing with an additional stress-energy tensor\index{stress-energy tensors!additional instance}, similar to that of ordinary matter and independent from the usually considered cosmological term\index{cosmological term} $\bm{T}^{++}_{\Lambda}=\Lambda\bm{g}^{++}$ or $-\bm{T}^{+-}_{\Lambda}=-\Lambda\bm{g}^{--}$ associated with the positive  cosmological constant\index{cosmological constant!positive} $\Lambda$.

It would, therefore, appear that if the maximum contribution to the density of vacuum energy which is directly experienced by a positive-energy observer was positive (while that which is directly experienced by a negative-energy observer was negative), so that, for such an observer, the metric conversion factor\index{metric conversion factors} $\gamma^{-+}$ would rather apply to the natural vacuum-stress-energy tensor that contributes negatively to the density of vacuum energy, thereby requiring the alternative form of the vacuum energy term\index{vacuum-energy term!alternative equation} provided by equation (\ref{eq:4.2}) to apply, then the curvature of spacetime\index{curvature of spacetime!positive-energy object} around a positive-energy object would not increase the mass of the object, but rather decrease it, by providing a negative contribution to its total energy.

This means that even from an observational viewpoint, it is preferable to assume that the final form of the generalized gravitational field equations\index{generalized gravitational field equations!final form} I have proposed in section \ref{sec:2.14} (on the basis of consistency requirements) and which implies that the positive cosmological constant\index{cosmological constant!contribution to own growth} does not contribute to its own growth over time, is the correct one, given that it allows one to predict that dark matter\index{dark matter!enhancement of gravitational field strength} actually contributes to enhance the strength of the gravitational field produced by an astronomical object. It is then merely the fact that the local variations of vacuum energy density\index{vacuum energy density!local variations} involved are correlated, under most circumstances, with the presence of local inhomogeneities in the distribution of baryonic matter\index{baryonic matter!inhomogeneities}, due to the fact that such inhomogeneities are usually required to trigger the development of local variations in the density of vacuum energy, that allows them to provide the long sought explanation of the missing-mass effect\index{missing-mass effect!particular aspect of dark energy} as being a particular aspect of the phenomenon of dark energy.

What must be understood is that, even if \textit{local} fluctuations in the density of negative-energy matter can be measured by a positive-energy observer, given that they do exert an influence on the gravitational field experienced by such an observer, this does not mean that the gravitational fields associated with the presence of negative-energy matter inhomogeneities cannot give rise to additional effects of a gravitational nature, arising from the response of vacuum energy fluctuations to the presence of those gravitational fields. In fact, even in the absence of any inhomogeneity in the negative-energy matter distribution, there may arise local variations of vacuum energy density as a result of the presence of positive-energy matter inhomogeneities, and the gravitational fields attributable to those local variations of vacuum energy density would actually affect the motion of both positive- and negative-energy objects.

Therefore, I believe that what explains most of the missing-mass effect\index{missing-mass effect} around visible positive-energy structures is the fact that the gravitational fields produced by those inhomogeneities in the matter distribution give rise to such local variations of the density of vacuum energy\index{vacuum energy density!local variations} which must necessarily be concentrated around the visible structures and which must themselves give rise to further variations of vacuum energy density, arising from the gravitational fields produced by those very same concentrations of vacuum energy\index{vacuum energy!concentration}. The crucial point here, therefore, is that the positive vacuum energy which is produced by the curvature of space attributable to the presence of positive-energy matter must itself contribute to produce additional space curvature\index{space curvature}, similar to that which would be produced by ordinary positive-energy matter, which in turn produces additional positive vacuum energy.

The problem we would normally face, in such a context, is that it would seem that the mass of an astronomical object would be allowed to increase without limit, as the growth of mass arising from the concentration of vacuum energy would trigger the formation of an even larger concentration of vacuum energy that would further increase the mass of the object. But, in fact, this situation is no more problematic than that which arises as a result of ordinary gravitational instability\index{gravitational instability}, because, as I will explain below, the portion of dark matter\index{dark matter!local variations of vacuum energy density} attributable to local variations of vacuum energy density already existed in a macroscopically homogeneous form on the cosmic scale, before it accumulated around ordinary matter overdensities, yet we already know that gravitational instability cannot produce catastrophic outcomes when the matter distribution is relatively homogeneous, given that the energy of the gravitational field\index{gravitational field energy!positive-energy object} generated by a positive-energy object is opposite the energy of the source, while the field\index{gravitational field!self-interaction} also interacts with itself.

Therefore, the growth of mass attributable to local variations of vacuum energy should be limited, especially since the gravitational interaction itself is very weak. This does not mean that no such an effect would exist, however. In fact, it appears that once we recognize that there must exist a portion of zero-point vacuum fluctuations\index{zero-point vacuum fluctuations} that directly interacts (other than gravitationally) only with baryonic negative-energy matter\index{baryonic negative-energy matter}, then it is not possible to avoid the conclusion that such local variations of vacuum energy density\index{vacuum energy density!local variations} and a phenomenon which we may call \textit{vacuum dark matter}\index{vacuum dark matter} would arise which would have consequences similar to those we normally attribute to the presence of conventional dark matter\index{dark matter!conventional} (such as weakly-interacting massive particles\index{weakly-interacting massive particles}), as it would actually contribute to significantly increase the mass of any astronomical object present on a sufficiently large scale, without raising the density of baryonic matter\index{baryonic matter}.

Now, I must admit that for a long time I, myself, believed that local variations of vacuum energy density could not constitute a solution to the missing-mass problem\index{missing mass!problem}, because I thought that the equivalent mass attributable to vacuum dark matter would not be allowed to contribute to the total energy of matter\index{total energy of matter} that is required to bring the average density of positive energy to its critical value, given that there would also exist negative contributions to the energy of matter, that would arise from those local variations of vacuum energy density\index{vacuum energy density!local variations} which are attributable to the presence of negative-energy matter overdensities, which I thought would cancel out the additional positive contributions and prevent the density of positive matter and vacuum energy from reaching its critical value, while such negative contributions also appeared unavoidable.

In other words, I had forgotten about the idea, because, when I first considered this possibility, I thought that, given that the energies involved were particular instances of vacuum energy, then the positive and the negative contributions should add up to produce a null average density of vacuum-dark-matter energy\index{vacuum-dark-matter energy!null average density} that would not allow to increase the average density of energy attributable to positive-energy matter and the uniform portion of vacuum energy associated with the positive cosmological constant\index{cosmological constant!positive} to its critical value, while this appeared to be required from a theoretical viewpoint, particularly in the context of inflationary cosmology\index{inflationary cosmology}.

Also, when I began seriously considering the possibility that some local variations of vacuum energy density\index{vacuum energy density!local variations} attributable to the gravitational field of large astronomical objects could be responsible for the phenomenon of missing mass\index{missing mass!phenomenon}, I had actually (but inappropriately) come to believe that voids in the uniform negative-energy matter distribution\index{void in uniform matter distribution!missing-mass effect} could provide an alternative explanation to most of the missing-mass effects around visible structures and therefore I didn't see the need that there was to explain the missing-mass effect as being the outcome of an inhomogeneous distribution of vacuum energy attributable to the presence of matter, even if the existence of such a phenomenon appeared unavoidable.

It is only much later that I came to understand that the fact that we are dealing here with \textit{local} variations in the distribution of vacuum energy means that, from a gravitational viewpoint, vacuum dark matter\index{vacuum dark matter!gravitational equivalence with ordinary matter} would be equivalent to the presence of ordinary matter, which allows one to expect that the gravitational attraction of positive vacuum-dark-matter energy\index{vacuum-dark-matter energy!gravitational attraction} would not be compensated by a gravitational repulsion attributable to the presence of negative vacuum-dark-matter energy\index{vacuum-dark-matter energy!gravitational repulsion}, on the cosmological scale (on which negative vacuum-dark matter energy\index{vacuum-dark matter energy!homogeneous distribution} is homogeneously distributed). But for the exact same reason, we should also expect that a local measure of positive vacuum-dark-matter energy\index{positive vacuum-dark-matter energy!non-negative pressure}, unlike a positive cosmological constant\index{cosmological constant!positive value}, should not exert a negative pressure that would repel positive-energy matter, either locally or globally (because the average density of positive vacuum-dark-matter energy\index{positive vacuum-dark-matter energy!decreasing average density} does not remain constant in an expanding universe).

What is unique, in effect, about the interpretation of the inhomogeneous character of the distribution of vacuum energy\index{vacuum energy!inhomogeneous distribution} discussed here (which is derived from the generalized gravitational field equations\index{generalized gravitational field equations} introduced in section \ref{sec:2.14}) is that, despite the fact that vacuum dark matter\index{vacuum dark matter!both vacuum and matter energy} is a form of vacuum energy, that must consequently be dark, it nevertheless contributes to the gravitational dynamics of the universe on a global scale in much the same way ordinary matter does. Indeed, if most of the missing-mass effect\index{missing-mass effect!local variations of vacuum energy density} is attributable to local variations of vacuum energy density, then the gravitational forces exerted by dark matter must be similar to those which are attributable to the presence of voids in the otherwise uniform distribution\index{void in uniform distribution of vacuum energy} of positive and negative vacuum energies, while, as I explained in section \ref{sec:2.7}, the void of cosmic proportions\index{void of cosmic proportions!positive vacuum energy} in the positive portion of the distribution of vacuum energy, which arises from the presence of a globally homogeneous distribution of negative matter energy\index{negative matter energy!homogeneous distribution}, exerts no gravitational force on positive-energy matter and has no effect on the specific rate of expansion\index{specific expansion rate!positive-energy matter} of positive-energy matter (the rate of expansion of space which is determined by positive-energy observers).

The one crucial aspect that differentiates vacuum dark matter from ordinary matter, therefore, is the fact that, as I explained in section \ref{sec:2.7}, if we are to conceive of negative-energy matter as missing positive vacuum energy\index{missing positive vacuum energy!negative-energy matter}, then we have no choice but to assume that what is missing from zero-point vacuum fluctuations\index{zero-point vacuum fluctuations} under such conditions is not just positive energy, but also some positive or negative non-gravitational charges, as the missing virtual particles\index{missing virtual particles!real matter} which are equivalent to the presence of real matter also carry charges in addition to energy. The charges which are missing\index{missing non-gravitational charge!opposite-sign charge} in the electrically (or non-gravitationally) neutral vacuum\index{vacuum!electrical neutrality} are equivalent to the presence of charges of opposite signs, which appear to be carried by the matter particles.

The above proposed concept of vacuum dark matter\index{vacuum dark matter!concept}, however, makes it clear that the presence of dark matter does not result from a local absence of virtual particles from zero-point vacuum fluctuations, that would need to be correlated with an absence of charge, but arises from local differences in the metric properties of spacetime\index{metric properties of spacetime!local differences} experienced by opposite-energy observers, which alter the relative values of the maximum positive and negative contributions to the density of vacuum energy\index{vacuum energy density!maximum opposite contributions} without affecting the electrical (or non-gravitational) neutrality of the vacuum. In other words, while ordinary positive-energy matter\index{positive-energy matter!missing negative energy and missing charges} consists of both missing negative energy and missing positive \textit{or} negative charge, positive-energy vacuum dark matter\index{positive-energy vacuum dark matter!excess of positive vacuum energy} is only equivalent to an excess of positive vacuum \textit{energy} and must remain electrically or non-gravitationally neutral\index{vacuum dark matter!non-gravitational neutrality} (it does not carry any non-gravitational charges), which explains that it is, in effect, dark. This is actually the only aspect that differentiate vacuum dark matter from ordinary matter, from an observational viewpoint.

Dark matter\index{dark matter!hybrid form of matter}, therefore, appears to be a hybrid form of matter that shares some physical properties with the uniformly distributed portion of vacuum energy\index{vacuum energy!uniformly distributed portion} that gives rise to the cosmological constant\index{cosmological constant}, even though it produces gravitational forces which are equivalent to those produced by ordinary matter, due precisely to the fact that its presence is attributable to \textit{local} variations of vacuum energy density\index{vacuum energy density!local variations} and because its average density must vary like that of ordinary matter, as a result of expansion.

Thus, again, we can expect that as long as negative vacuum-dark-matter energy\index{negative vacuum-dark-matter energy} is uniformly distributed on a global scale, like ordinary negative matter energy, its average density $-\bm{\bar{T}}_{VM}^{-+}$ does not really affect the specific rate of expansion of positive-energy matter\index{specific expansion rate!positive-energy matter} and does not contribute to the critical energy density\index{critical energy density!positive-energy observers} that is relevant to positive-energy observers, unlike the truly uniform distribution of negative vacuum energy\index{vacuum energy!uniform distribution} which would be associated with a negative cosmological constant\index{cosmological constant!negative}, and this what justifies replacing the vacuum-energy term\index{vacuum-energy term} $\bm{T}_{V}^+$ from equation (\ref{eq:2.23}) with the irregular vacuum-energy term\index{irregular vacuum-energy term} $\bm{\widetilde{T}}_{V}^+=\bm{T}_{V}^{+}-(-\bm{\bar{T}}_{VM}^{-+})$ from equation (\ref{eq:2.24}) in the most explicit form of the generalized gravitational field equations\index{generalized gravitational field equations} (\ref{eq:2.25}) associated with positive-energy observers. It is only when negative vacuum-dark-matter\index{negative vacuum-dark-matter energy!accumulation} energy accumulates around massive astronomical objects, that its presence becomes apparent to both positive and negative energy observers.

Now, it must be clear that the total quantity of energy contained in positive-energy dark matter\index{positive-energy dark matter!invariant total energy}, like that contained in negative-energy dark matter, does not change with time on the largest scale (even when the amount of positive or negative energy contained in ordinary matter itself varies, as a result of the annihilation of matter with antimatter\index{matter-antimatter annihilation}), despite the fact that the portion of missing-mass effects\index{missing-mass effect!local variations of vacuum energy density} attributable to local variations in the density of vacuum energy only becomes apparent when macroscopic inhomogeneities develop in the matter distribution and those energies become more concentrated around large astronomical objects.

Thus, the additional amount of positive energy that is present around positive-energy galaxies, but that cannot be accounted for by the presence of baryonic matter\index{baryonic matter}, was already present, in more diffuse form, before the formation of those structures, even though it was then exerting a significant gravitational force only on the global scale. What makes this possible is the fact that, even though the distribution of matter and radiation energy in the early universe\index{early universe!matter and radiation energy distribution} was very homogeneous from a macroscopic viewpoint (a hypothesis that is theoretically and observationally unavoidable, as I will explain in section \ref{sec:4.9}), positive- and negative-energy matter particles couldn't occupy the exact same positions in the initial state\index{initial state!maximum matter energy densities} of maximum positive and negative matter energy densities and this means that, on the quantum-gravitational scale\index{quantum-gravitational scale}, there existed variations of considerable magnitude in the density of both positive and negative matter energies, to which were associated equally large amounts of vacuum-dark-matter energy\index{vacuum-dark-matter energy}.

As the universe expanded and the average density of matter decreased, along with the average kinetic energy\index{average kinetic energy!matter and radiation particles} of matter and radiation particles, the macroscopically homogeneous distribution of vacuum dark matter\index{vacuum dark matter!macroscopically homogeneous distribution} (which existed in diffuse form as a consequence of the presence of microscopic inhomogeneities in the primordial distribution of matter and radiation energy) was allowed to spread into the available space, along with the rest of matter, and it is only when the small-amplitude inhomogeneities which were present on larger scales in the early matter distribution\index{early matter distribution!small-amplitude inhomogeneities} (including that of vacuum dark matter\index{vacuum dark matter!early distribution}) began to grow, later on, as a result of gravitational instability\index{gravitational instability}, that vacuum dark matter began to enhance the gravitational attraction of visible, overdense structures.

But this vacuum-dark-matter energy\index{vacuum-dark-matter energy} was not created as a result of the development of those macroscopic inhomogeneities, despite the fact that the gravitational forces it exerts are apparent only in those places where matter is inhomogeneously distributed on a macroscopic scale and the curvature of space\index{curvature of space} is more developed. In section \ref{sec:4.5} I will explain that if that was not the case, and the amount of positive-energy dark matter\index{positive-energy dark matter!growing amount} attributable to local variations of vacuum energy density\index{vacuum energy density!local variations} was actually growing in the universe (along with that of negative-energy dark matter), difficulties would arise, even if the total energy of matter (comprising the contributions of both positive- and negative-energy dark matter) was conserved in the process.

The fact is that what we usually describe as a homogeneous matter distribution\index{homogeneous matter distribution!smaller-scale inhomogeneities} actually contains inhomogeneities on a smaller scale and this additional amount of structure produces gravitational fields or spacetime curvature which affect local measures of vacuum energy density\index{vacuum energy density!local measures} (how this is possible will become clearer once the reader learns about certain unexpected microscopic properties of gravitational fields\index{gravitational field!microscopic properties} in section \ref{sec:4.7}). As a result, even in places where no macroscopic inhomogeneities are present in the matter distribution\index{matter distribution!absence of macroscopic inhomogeneities}, there usually exist local variations in the metric properties of spacetime\index{metric properties of spacetime!local variations}, attributable to the presence of microscopic inhomogeneities in such a macroscopically homogeneous matter distribution and those inhomogeneities result in the presence of both positive and negative vacuum-dark-matter energy\index{vacuum-dark-matter energy}.

It is only when the density of positive-energy matter grows larger than its average value on a macroscopic scale, at the expense of the creation of an underdensity of equal magnitude in the surrounding, homogeneous, positive-energy matter distribution, that the uniformly distributed positive-energy portion of vacuum dark matter\index{vacuum dark matter!positive-energy portion} becomes rarefied in the underdensity and more concentrated around the developing positive-energy structure, where it can begin to exert a gravitational pull on nearby positive-energy matter. At the same time, the uniformly distributed negative vacuum-dark-matter\index{vacuum-dark-matter!negative-energy portion} energy becomes rarefied in the positive-energy matter overdensity and more concentrated around the void that formed (as a consequence of the formation of this positive-energy structure) in the surrounding positive-energy matter distribution, where it can now exert a measurable gravitational attraction on nearby \textit{negative-energy} matter that adds to that which is produced by the void itself.

It must be clear, again, however, that vacuum dark matter is not created by the structure formation\index{structure formation!process} process. It was already present in the initial Big Bang state, before macroscopic inhomogeneities\index{macroscopic inhomogeneities!matter distribution} began to grow in the matter distribution, only, this vacuum dark matter was then more homogeneously distributed in space, from a macroscopic viewpoint and only exerted its full influence on the gravitational dynamics of the universe\index{gravitational dynamics of universe} as a whole. This means that there is no growth in the total amount of positive dark matter energy\index{positive dark matter energy!constant total amount} when a local overdensity develops in the positive-energy matter distribution, even if it is indeed the curvature of space\index{curvature of space!positive-energy matter overdensity} attributable to this overdensity that is responsible for the presence of vacuum dark matter\index{vacuum dark matter}, because this overdensity formed through the accumulation of both baryonic positive-energy matter\index{baryonic positive-energy matter} and positive vacuum-dark-matter energy\index{positive vacuum-dark-matter energy} and it is only because the initial vacuum-dark-matter\index{vacuum dark matter!homogeneous initial distribution} distribution was macroscopically very homogeneous that its influence is allowed to grow with time as the curvature of space itself grows (locally) on a macroscopic scale.

From an observational perspective, it would appear possible to confirm that dark matter\index{dark matter!local variations of vacuum energy density} is, for the most part, an outcome of local variations in the density of vacuum energy, because currently available data indicates \cite{McGaugh-1} that there is a strong correlation, in general, between the gravitational acceleration\index{gravitational acceleration!all matter inside an orbit} attributable to the total amount of matter inside an orbit (say around the center of a galaxy) and the gravitational acceleration\index{gravitational acceleration!baryonic matter inside an orbit} attributable to the baryonic matter inside that orbit (and this correlation would probably be even stronger if we were taking into account the presence of baryonic dark matter\index{baryonic dark matter}). Indeed, if the presence of dark matter\index{dark matter!effect of the curvature of space} must be considered to be an effect of the curvature of space (attributable to the matter that is present in a region of space) on the local measures of vacuum energy density, then the more gravitational acceleration that there is as a consequence of the presence of baryonic matter, the more distinct the metric properties of spacetime\index{metric properties of spacetime!opposite-energy observers} experienced by opposite-energy observers must be that gave rise to the accumulation of vacuum dark matter\index{vacuum dark matter!accumulation} around that particular location.

Even though the importance of the empirically derived relationship that allows to confirm the validity of those conclusions is often overlooked, it would certainly be a significant problem if it was to remain unexplained, as would be the case from the viewpoint of a more conventional interpretation of the missing-mass effect\index{missing-mass effect!conventional interpretation} (given that in such a context dark matter\index{dark matter!conventional} is simply an additional component of matter whose existence does not depend directly on the presence of ordinary matter).

But the conclusion that there must exist a relationship between the amplitude of the gravitational acceleration attributable to visible positive-energy matter overdensities and the amplitude of the missing-mass effect\index{missing-mass effect!amplitude} would also imply that, even within galaxies and clusters, the dark matter\index{dark matter!concentration around visible structure elements} should be more concentrated around the visible elements of the structure. While this result is certainly unexpected, it does, in fact, agree with some recent observations \cite{Meneghetti-1}, which indicate that there is a greater than expected concentration of gravitational lensing\index{gravitational lensing!concentration around galaxies in cluster} around individual galaxies within clusters. There is, thus, a strong motive to prefer an interpretation of the missing-mass effect\index{missing-mass effect!local variations of vacuum energy density} as being a consequence of local variations in the density of vacuum energy, which must exert gravitational forces whose strengths are correlated to those of the forces produced by the baryonic matter inhomogeneities\index{baryonic matter!inhomogeneities} which are present inside the same orbits.

Given that vacuum dark matter\index{vacuum dark matter!clumping} exerts its own gravitational field, it must then be subjected to clumping just like conventional dark matter, despite the fact that it really is vacuum energy. This means that the observations which indicate that large overdensities of visible matter can sometimes become separated from their dark matter component (as a result of collisions between galaxy\index{galaxies!cluster collisions} clusters or in the course of galaxy\index{galaxies!mergers} mergers) can be easily explained, unlike would be the case if the currently unexplained correlations discussed above (between the gravitational acceleration attributable to the total amount of matter inside an orbit and the gravitational acceleration attributable to baryonic matter alone) were the result of a more profound modification of the laws that govern the gravitational dynamics\index{gravitational dynamics!modified} of astronomical objects (such as envisaged in the context of the theory known under the MOND\index{MOND theory} acronym).

Indeed, once separated from the structure that allowed its mass to grow, such a dark-matter object\index{dark-matter object!self-sustaining} could continue to exist all by itself, sustained merely by its own gravitational field, while only a minimum measure of vacuum dark matter\index{vacuum dark matter!minimum measure} would be left in the visible structure around which it was originally located\footnote{
From that viewpoint, it would seem that the galaxies\index{galaxies!dark-matter free|nn} which do not appear to contain much dark matter are not galaxies which produce no local variations of vacuum energy density\index{vacuum energy density!local variations|nn}, or which did not grow out of the accumulation of vacuum dark matter\index{vacuum dark matter!accumulation|nn}, but merely galaxies which were stripped off of that portion of their dark matter attributable to the presence of inhomogeneities in the distribution of vacuum dark matter\index{vacuum dark matter!inhomogeneities|nn} itself, as a result of encounters with more massive structures (a conclusion which is supported by the latest computer simulations\index{computer simulations!dwarf galaxies|nn} \cite{Moreno-1} involving dwarf galaxies, once dark matter is recognized to originate from local variations of vacuum energy density).}.
 This is a considerable advantage of the original approach proposed here which, once again, appears to confirm the validity of the generalized gravitation theory\index{generalized gravitation theory} developed in the earlier portions of this report.

\section{Large-scale structure\label{sec:4.4}}

I remember, as a teenager, before I even learned about the existence of dark matter\index{dark matter}, having been deeply amazed and puzzled after reading in the newspaper that astrophysicists had determined that most of the visible matter in the universe, including our own galaxy, was located on the surface of giant voids\index{giant voids!matter distribution} of truly enormous proportions forming a bubble-like pattern\index{bubble-like pattern!matter distribution} in the matter distribution. I cannot say that I already expected, back then, that I would eventually be involved in developing a model that would help explain this troubling observation, but I did feel very strongly that this was something I needed to better understand. Anyhow, this stunning discovery and the mystery that initially surrounded it helped shape my early approach to the problem of gravitation in a way that turned out to be highly fruitful. What is truly remarkable is that the problem of voids\index{problem of voids} has endured to this day, as we kept discovering empty regions of increasingly larger sizes that still defy conventional explanations, despite all the progress which was achieved in developing cosmological models that can more accurately reproduce those features.

I believe that the introduction of negative-energy matter will have a significant impact on theories of structure formation\index{structure formation!theories}. What emanates from the results discussed in the preceding section is that the formation of overdensities in the primordial distribution of positive-energy matter must have been accelerated by the presence of negative-energy matter underdensities, which developed as a result of the gravitational repulsion exerted by those positive-energy matter overdensities and whose effects on developing positive-energy structures would be similar to those which are usually attributed to ordinary, positive-energy dark matter. What needs to be emphasized, in this regard, is that, given that, in the early universe, the average matter density (not just that of negative-energy matter) was much larger than it currently is, then it follows that the underdensities which were present in the distribution of negative vacuum-dark-matter energy\index{negative vacuum-dark-matter energy} had a significant influence on positive-energy matter.

As I previously mentioned, certain observations \cite{Labbe-1}, performed after the first versions of the present document were published, appear to confirm that the formation of galaxies\index{galaxy formation!acceleration} must have been accelerated, in the primordial universe, by effects of an unknown origin, because the observations in question have revealed the existence of galaxies which are too large at such an early epoch for their presence to be explained using more conventional models (which do not involve negative-energy matter). In fact, it had been known for some time that the first large elliptical galaxies\index{large elliptical galaxies} appear to have formed too early after the Big Bang for their creation to be easily explainable using conventional models. But if we recognize that the presence of negative-energy matter underdensities must have played an important role on such a scale in the remote past, when the average density of negative vacuum-dark-matter energy\index{negative vacuum-dark-matter energy!average density} was much larger, and its distribution much more homogeneous, then those mysteries can be explained quite straightforwardly.

It is important to understand that the observationally confirmed absence of localized overdensities in the distribution of negative vacuum-dark-matter energy\index{negative vacuum-dark-matter energy!overdensities}, on the scale of individual stars and galaxies means that there was no similar influence on negative-energy matter arising from the presence of underdensities in the early distribution positive-energy matter on smaller scales, because only the presence of negative-energy matter overdensities would allow the depth of those underdensities to grow large enough, on this particular scale, that they could significantly accelerate the formation of structures\index{structure formation} in the negative-energy matter distribution.

But then, one may ask why it is that there were no significant overdensities in the negative vacuum-dark-matter energy distribution on smaller scales during the matter-dominated era\index{matter-dominated era}, even though (as I will suggest in section \ref{sec:4.5}) the average density of negative vacuum-dark-matter energy\index{negative vacuum-dark-matter energy!average density} was always very similar to that of positive vacuum-dark-matter energy\index{positive vacuum-dark-matter energy!density}? I wrestled with that question for a long time before I realized that the absence of negative-energy matter overdensities is due to the fact that only a negligible amount of baryonic negative-energy matter\index{baryonic negative-energy matter} has survived the early annihilation of matter with antimatter\index{early matter-antimatter annihilation}.

What happens is that the presence of baryonic matter\index{baryonic matter} is necessary for initiating the formation of structures in a distribution of vacuum dark matter\index{vacuum dark matter} on smaller scales, because only such matter can reduce its average kinetic energy\index{average kinetic energy!reduction} through the emission of radiation, given that vacuum dark matter has no charge and does not interact with itself or with ordinary matter other than gravitationally. It is well-known, in effect, that collapsing astronomical structures\index{collapsing astronomical structures} can only begin to form through gravitational instability\index{gravitational instability} when they are allowed to release thermal energy\index{thermal energy!release} through radiation, as otherwise their internal pressure\index{internal pressure!astronomical structures} remains too large to allow them to stabilize. But if one may assume that vacuum dark matter shares some of the physical properties of baryonic matter, due to the fact that zero-point vacuum fluctuations\index{zero-point vacuum fluctuations} involve the same particles fluctuating in and out of existence in their virtual form, then one would be justified to assume that, under the same conditions, vacuum dark matter\index{vacuum dark matter!cold} is merely as cold as it would be if it was composed of baryonic matter (both quarks and leptons)\footnote{
It would be problematic to assume that it is not possible to say whether vacuum dark matter\index{vacuum dark matter!cold or hot|nn} is either cold or hot or anything in between, because that would mean that its physical properties are undefined, despite the fact that it must behave unambiguously in any particular circumstances. But the fact that it appears that vacuum dark matter must be as cold as baryonic matter is certainly not undesirable given that it was shown that only cold dark matter\index{cold dark matter|nn} can give rise to a bottom-up process of structure formation\index{bottom-up process of structure formation|nn} of the kind that is favored by astronomical observations.}.

Therefore, on a scale where ordinary baryonic matter is allowed to collapse into stable structures only through the emission of radiation, the density of vacuum dark matter itself can only begin to grow if baryonic matter overdensities\index{baryonic matter overdensities} with the same sign of energy are already present, because without the additional gravitational attraction produced by such an overdensity, vacuum dark matter would rather tend to disperse, if it is merely as cold as ordinary baryonic matter.

Thus, it would appear that it is the possibility for baryonic negative-energy matter and antimatter\index{baryonic matter and antimatter!early annihilation} to have annihilated more completely than baryonic positive-energy matter and antimatter in the early universe, that explains that no significant, gravitationally repulsive overdensities\index{gravitationally repulsive matter overdensities!early universe} of negative-energy matter were present in the early universe (and therefore also at later times) on the scale of stars and galaxies. Such an outcome can be expected to occur whenever a more limited violation of time reversal symmetry\index{time reversal symmetry!violation} (or indeed an absence of such violation) affects the creation of opposite-bidirectional-charge particles and antiparticles\index{opposite-bidirectional-charge particles and antiparticles!creation} with negative action than affects the creation of opposite-bidirectional-charge particles and antiparticles with positive-action.

What must be understood is that the presence of negative-energy matter overdensities on the scale of stars and galaxies is not unavoidable, because, even if baryonic negative-energy matter\index{baryonic negative-energy matter} must have been present in the universe initially, it can be almost completely absent during the matter-dominated era\index{matter-dominated era}, unlike negative-energy vacuum dark matter\index{negative-energy vacuum dark matter}, which is not submitted to matter-antimatter annihilation\index{matter-antimatter annihilation}, given that it has no charge, which allows its average density to remain unchanged following the early annihilation of baryons with antibaryons\index{early baryon-antibaryon annihilation}. Anyhow, one must conclude that immediately after most of the baryonic negative-energy matter and antimatter annihilated, during the radiation-dominated era\index{radiation-dominated era}, the average density of baryonic negative-energy matter\index{baryonic negative-energy matter!negligible average density} did became completely negligible, because this is the only way one can explain that no overdensities later developed in the negative-energy matter distribution, on smaller scales, whose presence could have been revealed by weak gravitational lensing\index{weak gravitational lensing!experiments} experiments.

That is not to say, however, that there can be no overdensities in the distribution of negative vacuum-dark-matter energy\index{negative vacuum-dark-matter energy!overdensities}. In fact, it appears necessary to assume that vacuum-dark-matter overdensities with very large negative masses would be present inside the largest voids\index{void in positive-energy galaxy distribution!largest voids} in the distribution of positive-energy galaxies, which do exert a localized gravitational pull on this vacuum dark matter. Given that the gravitational attraction exerted on negative vacuum-dark-matter energy by voids in the positive-energy matter distribution grows along with their size, it is possible to conclude, in effect, that most of the overdensities which are present today in the distribution of negative vacuum-dark-matter energy should be concentrated in the largest voids in the positive-energy matter distribution. In the absence of baryonic negative-energy matter overdensities\index{baryonic negative-energy matter!overdensities}, it would appear that only such large voids can exert a gravitational attraction large enough to allow negative vacuum-dark-matter energy overdensities to form and to grow to such proportions that they may actually influence the process of structure formation\index{structure formation!process} in the positive-energy matter distribution.

In fact, the presence of such very-large-scale overdensities in the negative-energy matter distribution is not only allowed by current observational data, it is actually required in order to explain the size of those voids. In the absence of large negative-energy matter overdensities located within the largest voids in the positive-energy matter distribution, the unexpectedly large size of those voids would need to be explained through biasing\index{biasing}, an approach which amounts to assume, without justification, that galaxies have a tendency to form preferably in those regions where the density of the cold dark matter\index{cold dark matter} is already larger, at the epoch of recombination\index{recombination!epoch}.

If I do not agree that biasing is the appropriate solution to the presence of unexpectedly large voids in the galaxy distribution\index{galaxy distribution!unexpectedly large voids}, it is because it seems to me that biasing merely amounts to impose the required matter distribution without explaining it. It cannot even be assumed that the solution offered by the biasing hypothesis is merely incomplete and that what one must do is explain why the largest overdensities in the primordial distribution of positive vacuum-dark-matter energy\index{positive vacuum-dark-matter energy!primordial distribution} formed predominantly in those regions where baryonic positive-energy matter\index{baryonic positive-energy matter!overdensities} overdensities were located, as I have suggested would occur at later times on smaller scales, because any correlation between those two types of inhomogeneities that may have existed in the initial Big Bang state\index{initial Big Bang state} would have been lost, early on, due to the fact that any preexisting baryonic matter inhomogeneity would be wiped out on smaller scales, before new ones would be allowed to form during the matter-dominated era\index{matter-dominated era}, influenced by the presence of the original inhomogeneities which had continued to develop in the distribution of vacuum dark matter\index{vacuum dark matter distribution}, on a larger scale\footnote{
The fact that vacuum dark matter\index{vacuum dark matter|nn} is only submitted to the gravitational interaction is what explains that dark matter inhomogeneities\index{dark matter!inhomogeneities|nn} were allowed to grow larger, sooner than the inhomogeneities which were present in the baryonic matter\index{baryonic matter!inhomogeneities|nn} distribution, despite the fact that those two types of fluctuations had the same scale-invariant magnitude\index{scale-invariant magnitude!initial matter density fluctuations|nn} in the initial Big Bang state. For observational reasons, it is not possible to assume that vacuum dark matter\index{vacuum dark matter!overdensities|nn} overdensities began to grow at the same time as those in the baryonic matter distribution, but that the presence of vacuum dark matter merely accelerated the rate of structure formation\index{structure formation!acceleration|nn} that took place subsequently.}.

It should be clear though, that given the relatively small average value of the current density of positive-energy matter, only the largest of the voids in the positive-energy matter distribution\index{void in positive-energy matter distribution!largest voids} should produce a localized gravitational attraction on negative vacuum-dark-matter energy\index{negative vacuum-dark-matter energy!measurable overdensity} that would be strong enough to allow a measurable overdensity of such matter to exist inside those structures (the effect cannot have been more substantial in the early universe, because, despite the larger average density of positive vacuum-dark-matter energy\index{positive vacuum-dark-matter energy!early average density}, the amplitude of fluctuations in the density of matter energy\index{matter energy density!amplitude of early fluctuations} was then smaller on such a large scale). Thus, despite the absence of small-scale overdensities in the early distribution of negative vacuum-dark-matter energy, the presence of sufficiently large voids in the distribution of positive-energy galaxies\index{void in positive-energy galaxy distribution} should allow gravitational instability\index{gravitational instability!negative vacuum-dark-matter energy} to arise in this distribution of negative vacuum-dark-matter energy, which, once triggered, would give rise to a self-amplifying process that could produce more overdensity on such a scale.

It is the fact that baryonic positive-energy matter overdensities can only gravitationally repel negative vacuum-dark-matter energy and make it spread, that explains that, in the absence of baryonic negative-energy matter overdensities\index{baryonic negative-energy matter overdensities!absence}, the only place where the density of this vacuum dark matter was allowed to grow and to begin exerting an influence on positive-energy matter is in the large-scale underdensities that formed in the positive-energy matter distribution. But it appears that the presence of large and growing overdensities of negative vacuum-dark-matter energy\index{negative vacuum-dark-matter energy!growing overdensities} inside those voids in the positive-energy matter distribution would allow the voids themselves to grow larger than expected at an earlier time, due to the gravitational repulsion\index{gravitational repulsion} exerted by those negative vacuum-dark-matter energy overdensities on the surrounding positive-energy galaxies.

There are still reasons to expect, therefore, that negative-energy matter must have contributed to the process of structure development\index{structure development!very large scale} on a very large scale, not just because this is not ruled out from an observational viewpoint, but also because it would actually help explain certain observations mentioned above, like the size of the largest voids in the galaxy distribution\index{largest voids in galaxy distribution!size} or the mass of the earliest galaxies\index{earliest galaxies!mass}. But given that the gravitational repulsion attributable to a large negative-energy matter overdensity would produce a local acceleration of the rate of expansion\index{expansion rate!local acceleration} of positive-energy matter similar to that which is attributable to the void in the expanding positive-energy matter distribution in which it would normally be located, then it follows that the presence of the overdensity would only enhance the magnitude of this phenomenon. I think that this is what explains that it is not immediately apparent that this local acceleration of the rate of expansion of visible positive-energy matter is sometimes attributable in part to the presence of gravitationally repulsive matter\index{gravitationally repulsive matter}.

However, the presence of a negative-energy matter overdensity within such a large void in the expanding positive-energy matter distribution would produce some distinctive effects, as it implies that the structure can actually exert a gravitational repulsion\index{gravitational repulsion} on the surrounding positive-energy matter. This is certainly a positive development, given that it has been known for some time that certain voids, apparent on a very large scale in the positive-energy matter distribution, do produce a larger than expected local acceleration of the rate of expansion\index{expansion rate!local acceleration} for galaxies located on their periphery, a phenomenon which had remained unexplained until now.

When large negative-energy matter overdensities are present inside the largest voids in the visible positive-energy matter distribution, it is also easier to reconcile our theory of structure formation\index{structure formation!theories} with those observations which show that there is a much smaller number of galaxies in the Local Void\index{Local Void} than is predicted by computer simulations\index{computer simulations!galaxy formation} of galaxy formation, because any galaxy that would form in the void would rapidly be expelled to the periphery by the repulsive gravitational forces\index{repulsive gravitational force} attributable to the presence of this negative-energy matter. Also, given that the density of negative vacuum-dark-matter energy in the Local Void would not be as low as it would in our galactic neighborhood\index{galactic neighborhood}, it follows that the portion of missing-mass effect\index{missing-mass effect!more localized} attributable to negative-energy matter underdensities would be more localized around those galaxies located closer to the center of the void and this must have accelerated their formation. That may explain why a larger than expected number of very large galaxies in the Local Sheet\index{Local Sheet} are located on the periphery of the Local Void instead of in the more crowded areas of the Local Sheet, where most of the visible matter is concentrated.

The fact that the only observed departures from what is currently predicted to occur on the scale of individual galaxies do not involve gravitational repulsion\index{gravitational repulsion}, but merely require the existence of additional gravitational attraction localized over the visible structures, would appear to confirm that those deviations are attributable to the presence of early underdensities in the negative vacuum-dark-matter energy\index{negative vacuum-dark-matter energy!underdensities} distribution. It remains, though, that negative-energy matter is the source of additional gravitational instability\index{gravitational instability!negative-energy matter} which does not arise only from stronger gravitational attraction, but also from the gravitational repulsion exerted on positive-energy matter by the overdensities of negative-energy matter which are present on a very large scale.

As a result, starting from the same relatively smooth initial distribution of positive-energy matter\index{positive-energy matter!smooth initial distribution} that is revealed by the low amplitude of cosmic microwave background\index{cosmic microwave background!temperature fluctuations} temperature fluctuations, we can actually expect inhomogeneities to develop more rapidly in this matter distribution, due to the presence of smaller-scale underdensities\index{smaller-scale underdensities!negative vacuum-dark-matter energy} and larger-scale overdensities\index{larger-scale overdensities!negative vacuum-dark-matter energy} in the distribution of negative vacuum-dark-matter energy. But given that a certain portion of the fluctuations which are observed in the temperature of CMB radiation can be attributed to the presence of inhomogeneities in the primordial distribution of negative vacuum-dark-matter energy\index{negative vacuum-dark-matter energy!primordial inhomogeneities}, then one must conclude that the magnitude of the inhomogeneities which were present, at the epoch of last scattering\index{last scattering!epoch}, in the positive-energy matter distribution is smaller than what seems to be indicated by observations of CMB temperature fluctuations, which means that their magnitude has not necessarily developed to become larger than one would otherwise expect, particularly on smaller scales (I will return to this question in section \ref{sec:4.9}).

Now, even though it appears that most of the baryonic negative-energy matter\index{baryonic negative-energy matter} has vanished at a very early time, shortly after the quark-hadron transition\index{quark-hadron transition}, it cannot be the case that the only negative-energy matter that remains in the universe today is vacuum dark matter, because negative-energy neutrinos\index{negative-energy neutrinos}, at least, must have survived the early period of matter-antimatter annihilation\index{early matter-antimatter annihilation} and be present with an average density as large as that of positive-energy neutrinos, along with negative-energy radiation\index{negative-energy radiation}. Indeed, even if neutrinos and antineutrinos existed in the exact same numbers initially, it cannot be expected that they annihilated completely with one another, given their low mass and the weakness of their interactions. But this negative-energy matter\index{negative-energy matter!hot} would have been hot when it decoupled from the rest of matter and radiation and therefore it must have remained homogeneously distributed and cannot have contributed much to the process of structure formation\index{structure formation!process}, particularly on smaller scales.

In fact, given that it is known that positive-energy neutrinos themselves cannot exert much influence on the formation of structures in the positive-energy matter distribution, due to their low mass, then it is certainly not necessary to take into account any potential effect that could be attributed to the presence of underdensities in the distribution of negative-energy neutrinos\index{negative-energy neutrinos!homogeneous distribution}, even if it seems that the fact that this matter is still homogeneously distributed on the scale of galaxies and clusters could allow the missing-mass effect\index{missing-mass effect!negative-energy neutrino underdensities} attributed to those underdensities to be localized around gravitationally repelling positive-energy structures.

What must be retained from all this is that the additional influence which may be exerted, under certain conditions, by negative vacuum-dark-matter energy\index{negative vacuum-dark-matter energy!inhomogeneities} inhomogeneities on the process of structure formation\index{structure formation!process} in the positive-energy matter distribution may be significant enough to have given rise to structures which are more developed than those which are predicted to arise by the conventional cold dark matter model\index{cold dark matter model} at various epochs, either at early times, due to the presence of smaller-scale underdensities, or at later times, due to the presence of larger-scale overdensities in the distribution of negative-energy dark matter.

This is not a problem, but rather an advantage of the proposed approach, because it is no secret that the most recent observations have revealed the existence, on both a large and a very large scale, of structures whose existence has become increasingly more difficult to reconcile with conventional models of structure formation\index{structure formation!conventional models}. It is obvious to me that such observations, and the bubble-like pattern of the very-large-scale matter distribution\index{bubble-like pattern!very-large-scale matter distribution} in general, can be much more easily explained if we allow for the existence of a parallel distribution of dark, gravitationally repulsive, negative-energy matter which is gravitationally attracted to itself like positive-energy matter.

\bigskip

\noindent Before concluding this section, I would like to mention the existence of another astronomical phenomenon which can be expected to occur as a consequence of the presence of very-large-scale overdensities in the distribution of negative vacuum-dark-matter energy\index{negative vacuum-dark-matter energy!overdensities}. It involves an effect which might be called \textit{repulsive gravitational lensing}\index{repulsive gravitational lensing} and which is merely the counterpart to ordinary gravitational lensing that would be produced when the visible light from a distant source would be gravitationally repelled while it travels through a negative-energy matter overdensity, on its way to our telescopes. While ordinary gravitational lensing\index{gravitational lensing!arcs of light} produces arcs of light, repulsive gravitational lensing\index{repulsive gravitational lensing!blobs of light} would produce blobs of light, because they would distort the image of the background structures in such a way that the objects observed would appear to be more densely packed in space, behind the invisible negative-energy matter overdensity located in the foreground.

But there is also a possibility that ordinary gravitational lensing could be enhanced by the presence of underdensities in the negative-energy matter distribution, superposed on positive-energy objects located in the foreground. Such an effect would not be easily distinguishable from that which is produced by the presence of positive-energy matter overdensities, however, while that which we can expect to arise from the presence of negative vacuum-dark-matter energy overdensities can be expected to exist only inside the largest voids in the positive-energy matter distribution and to be very diffuse, given that negative vacuum-dark-matter energy\index{negative vacuum-dark-matter energy!very smooth distribution} is very smoothly distributed, due to the absence of baryonic negative-energy matter overdensities\index{baryonic negative-energy matter overdensities!absence}. This may explain why weak gravitational lensing\index{weak gravitational lensing!experiments} experiments have not yet revealed the existence of a phenomenon of this kind that could definitely be attributed only to the presence of gravitationally repulsive matter\index{gravitationally repulsive matter}.

\bigskip

\noindent In face of the mounting difficulties we have encountered in recent years in trying to make sense of a growing amount of unexpected empirical results, I think that the time has come to recognize that simply adjusting the free parameters of the cold dark matter model\index{cold dark matter model!free parameters} is no longer an adequate approach for addressing the challenges raised by the observed large-scale features of our universe. But even if the words `dark matter' are contained in the name of the currently favored cosmological model, it does not mean that rejecting this model requires completely abandoning the idea that invisible forms of energy may play a role in the development of large-scale structures\index{large-scale structures!development}, because it remains that a certain phenomenon attributable to local variations of vacuum energy density\index{vacuum energy density!local variations} can be expected to have consequences similar to those which were once attributed to conventional cold dark matter\index{cold dark matter!conventional}. Thus, I believe that what is required to make the current models more acceptable is merely an additional ingredient that would strengthen the gravitational forces responsible for sculpting the large-scale matter distribution in ways which would allow to appropriately describe certain phenomena that would otherwise remain unexplained.

It is merely the fact that a void in the positive-energy matter distribution is expected to produce a local acceleration of the rate of expansion\index{expansion rate!local acceleration} of space from the viewpoint of a positive-energy observer that prevents us from drawing the obvious conclusion that unseen matter must be present in the largest voids in the galaxy distribution\index{galaxy distribution!largest voids} that produce a similar effect through gravitational repulsion\index{gravitational repulsion}, thereby allowing those empty spherical structures to grow to unexpectedly large sizes. The early proposals that the largest voids might have formed as a consequence of explosive processes that would have taken place in the early universe were thus based on the right intuition, but they failed because they did not involve gravitation as the repulsive force. It would therefore be the traditional reluctance to consider the possibility that gravitationally repulsive matter\index{gravitationally repulsive matter} may exist, as well as the ignorance of the fact that such matter must necessarily be dark and be gravitationally attracted to itself, that would explain the difficulties we experience in trying to make sense of the most recent data regarding the processes that take place in our universe on a very large scale.

\section{Flatness and the zero-energy condition\label{sec:4.5}}

In the introductory section of this chapter I mentioned that there are two broad aspects to what I call the inflation problem\index{inflation!problem}, which are the flatness problem\index{flatness problem} and the horizon problem\index{horizon problem}. Here I would like to discuss the first category of issues. Despite the commonly held belief that this problem has been solved by inflation theory\index{inflation theory}, I think that it is still important to understand the difficulties it raises for cosmology, given that the validity of inflation has not yet been definitely confirmed and even if there occurred an initial phase of accelerated expansion\index{initial phase of accelerated expansion}, it may not necessarily produce the desired outcome.

As I previously mentioned, the flatness problem arises from the fact that the present, average density of positive matter and vacuum energy\index{positive matter and vacuum energy!present average density} appears to be fixed to its critical value, while we have no idea what the constraint is that would require such an extremely precise adjustment of parameters as would have to occur in the early stages of the Big Bang in order to produce the observed outcome. The problem is that if the faintest of deviation away from a critical rate of expansion\index{critical expansion rate} had taken place at such an epoch, it would have given rise to a much larger deviation away from flatness at later times, while what we observe is a universe\index{universe!flat space} with a flat space and a positive density of energy that is still critical to a very good degree of precision. The truth, therefore, is that according to current knowledge, the Big Bang model\index{Big Bang model!incompleteness}, while mathematically consistent, is nevertheless incomplete, given that the initial conditions of our universe\index{universe!initial conditions}, it would seem, cannot be determined or explained by the theory.

Of course, this does not mean that we can't determine a unique rate of expansion\index{expansion rate!evolved backward in time} of space at any time in the past by evolving the current state backward in time, which would actually allow us to predict that the density of positive matter, radiation, and vacuum energy\index{positive matter and vacuum energy!critical density} $\rho(t)$ was closer to its critical value $\rho_c(t)$ (associated with a density parameter\index{density parameter} $\Omega=\rho/\rho_c$ equal to 1) in the past. Only, we cannot explain why the current density of positive energy\index{positive energy!current density} $\rho_0$ itself is fixed to its critical value $\rho_{c,0}$ to such a high degree of precision. Thus, while conventional relativity theory\index{relativity theory!conventional} enables a positive-energy observer to predict what the rate of expansion of the universe was at different times in the past, given the current densities of positive matter, radiation, and vacuum energy\index{positive matter and vacuum energy!current densities}, according to the conventional approach this is only true in as much as the rate of expansion of space\index{expansion rate!present value} at the present time is empirically determined through a measurement of the Hubble constant\index{Hubble constant} $H_0$, but the model remains well-defined for any value of $\rho_0$ and $H_0$.

Yet, I believe that there is much less freedom than is usually assumed in fixing the initial values of the specific rates of expansion\index{specific expansion rates!initial variation} that give rise to the present specific densities\index{specific densities!present values} of positive- and negative-energy matter. What I will now explain is that despite the conventional assumption to the effect that this initial condition is left unconstrained in the standard Big Bang model\index{standard Big Bang model} (without inflation\index{inflation}), there does exist an unavoidable requirement for the current density of matter and vacuum energy\index{matter and vacuum energy!current density} to be very precisely equal to the critical density\index{critical density} associated with a flat space\index{flat space}, from the viewpoint of both positive- and negative-energy observers.

One thing must be clear before we attempt to explain the current flatness of space\index{flatness of space!cosmic scale} on the cosmic scale and this is that there is an upper limit to the positive and negative contributions to the density of matter energy\index{matter energy density!upper limits}. This means that space cannot continue to contract (in the past direction of time) beyond the point at which a maximum amount of matter and radiation energy of positive or negative energy sign is contained in every elementary unit of area\index{elementary units of area}. It would be incorrect to assume that the initial value of the density parameter\index{density parameter!initial value} $\Omega$ cannot be determined, due to the `fact' that the density of matter\index{matter density!infinite initial value} is infinite in the very first instant of the Big Bang.

From a quantum-gravitational viewpoint\index{quantum-gravitational viewpoint}, there is no time zero\index{time zero} at which the density of matter is infinite, only a minimum significant time\index{minimum significant time} at which positive and negative energy densities\index{positive and negative energy densities!maximum magnitude} have a finite maximum magnitude. Given that, in the context of my interpretation of matter as being equivalent to missing vacuum energy\index{missing vacuum energy}, a maximum value of matter energy density is determined by the natural vacuum-stress-energy tensors\index{natural vacuum-stress-energy tensors} which enter the generalized gravitational field equations\index{generalized gravitational field equations} introduced in section \ref{sec:2.14}, then this must be assumed to be the maximum magnitude of the positive and negative contributions to the density of matter energy in the state that emerges from the initial singularity\index{initial singularity}.

What needs to be explained, therefore, is merely why it is that the rate of expansion of space\index{expansion rate!critical value} was set to its critical value to an almost prefect level of precision when the universe emerged from this state of maximum positive and negative matter energy densities\index{maximum matter energy densities!state} that is uniquely determined by the natural scale of quantum-gravitational phenomena\index{quantum-gravitational phenomena!natural scale}. The initial positive and negative densities of matter\index{initial matter energy densities!non-arbitrary values} and radiation energy are not arbitrary, but the problem is that there is too much freedom in fixing the initial rate of expansion\index{expansion rate!initial value} which determines how rapidly the average density of matter\index{average matter density!rate of decrease} decreases at later times. From a conventional viewpoint, it would appear that the initial rate of expansion that gives rise to a universe with flat space\index{universe!flat space} at the present time is merely one alternative among an enormous range of possibilities.

But while the current value of gravitational potential energy\index{gravitational potential energy!universe} for the universe as a whole (which is fixed by the present, average density of positive matter and vacuum energy\index{positive matter and vacuum energy!present average density} $\rho_0$) and the current value of kinetic energy of expansion\index{kinetic energy of expansion} (which is determined by the present value of the Hubble constant $H_0$) appear to constitute free parameters of the standard model of cosmology\index{standard model of cosmology!free parameters}, I will explain that they are not really independent variables in the context where energy must be null for the universe\index{universe!null energy} as a whole. In fact, under such conditions, the initial value of the rate of expansion measured by a positive-energy observer must be adjusted not merely to a level of precision that would allow space to keep expanding at a near critical rate until the present epoch, but to such an extent that space can be expected to remain perfectly flat for all time on the largest scale. I will show that this constraint can only be fulfilled when a maximum density of negative-energy matter\index{negative-energy matter!maximum initial density} is assumed to have been originally present throughout the universe, alongside that of positive-energy matter.

In the context of the model I have proposed to integrate negative-energy matter to gravitation theory, it may seem like the presence of such negative-energy matter would change nothing to the conclusion that flat space is an unlikely possibility for the present state of the universe\index{universe!unlikeliness of current flat space}, because a uniform distribution of negative-energy matter\index{negative-energy matter!uniform distribution} exerts no influence on the gravitational dynamics\index{gravitational dynamics!positive-energy matter} of positive-energy matter on the largest scale, for reasons I have explained in section \ref{sec:2.7}. The present specific density of negative-energy matter\index{specific density of negative-energy matter!present value} would in fact be independently subjected to the same excess of freedom as affects that of positive-energy matter, given that the variation of the specific rate of expansion of negative-energy matter\index{specific expansion rate of negative-energy matter!variation} is determined only by the density of matter with the same energy sign, while it would appear that there is as much freedom in setting the initial value of this expansion rate\index{expansion rate!initial value} as there is in fixing the initial value of the specific rate of expansion of positive-energy matter\index{specific expansion rate!positive-energy matter}.

Anyhow, if the universe had a negative space curvature\index{universe!negative space curvature} from the viewpoint of a negative-energy observer, this would not merely be a consequence of the fact that the energy of matter that determines the expansion rate measured by such an observer is negative, as if negative-energy matter could accelerate its specific rate of expansion\index{specific expansion rate!acceleration by gravitational repulsion} through the gravitational repulsion it would exert on itself, like we would expect from a conventional viewpoint, because, as I explained in section \ref{sec:2.4}, negative-energy or negative-mass matter does not exert a gravitational repulsion\index{gravitational repulsion!negative-energy matter} on matter with the same energy sign. Thus, in principle, the universe\index{universe!closed positively curved space} could just as well have a closed, positively curved space from the viewpoint of a negative-energy observer, because gravitational repulsion\index{gravitational repulsion!non-absolute attribute} is not an absolute attribute of matter with a given energy sign. Yet, despite this state of affairs, it turns out that the presence of negative-energy matter is, in fact, required (as I mentioned above) to explain why it is that we are allowed to expect that the universe\index{universe!perfectly flat space} should have a perfectly flat space from the viewpoint of a positive-energy observer.

Although the alternative solution I will propose to the flatness problem\index{flatness problem} is quite simple, it was actually one of the results which I had the most difficulty deriving among those that figure in this report. Part of the difficulty arose from the fact that there are conflicting accounts of what constitute the many contributions to the energy budget of the universe\index{universe!energy budget} and how their magnitudes may vary as a function of the values assumed by various cosmological parameters\index{cosmological parameters}. Thus, while I always had the intuition that, in the context where the presence of negative-energy matter cannot be ignored, a natural solution to the flatness problem might become possible once we recognize the necessity to appropriately apply the principle of conservation of energy\index{principle of conservation of energy!Big Bang} to the Big Bang, it was not clear which contributions could balance one another out exactly in order to produce a universe out of nothing\index{universe out of nothing}.

But when I finally figured out what the various contributions to the energy budget of the universe are in the presence of negative-energy matter, and which must be considered independent from which others, and which would need to have the same magnitudes in the initial Big Bang state\index{initial Big Bang state}, then it became clear that, under such conditions, one must actually observe space to expand at precisely the critical rate\index{critical expansion rate} when we require the energy to be null for the universe\index{universe!requirement of null energy} as a whole. Before I explain why it is exactly that applying this theoretically motivated constraint may have such far-reaching implications, however, I would like to describe what the motives are that justify assuming that the energy of the universe must, in effect, be null, even when creation out of nothing is not itself an actual requirement for the universe\index{universe!creation out of nothing}.

I already discussed the importance and the unavoidable character of the constraint imposed by the requirement of relational definition\index{requirement of relational definition!physical attributes} of physical attributes in the preceding two chapters of this report. Basically, what must be understood, concerning the problem at hand, is that the total energy of the universe\index{universe!total energy} constitutes one such physical property which definitely cannot violate the rule that it be defined in a purely relational way. What is implied by such a requirement is that, even if the Big Bang was not considered to constitute a creation event at which any conserved physical quantity must be created out of nothing, from the viewpoint of an observer of any energy sign the universe\index{universe!vanishing total energy} would still need to have a vanishing total energy and therefore a \index{universe!null average energy density} null average energy density.

One might argue, in effect, that the requirement of conservation of energy\index{conservation of energy!Big Bang} does not apply to such a singular event as the Big Bang at which time\index{time!coming into existence} itself may come into existence, or alternatively that the initial Big Bang state\index{initial Big Bang state!beginning of time} does not even constitute an absolute beginning to time, given that evolution could perhaps be continued to times before the initial singularity\index{initial singularity} if a quantum bounce\index{quantum bounce} occurs, as implied by the most promising quantum gravitation theories\index{quantum gravitation theories!most promising}. But when we recognize the unavoidable nature of the constraint of relational definition\index{constraint of relational definition!energy} of the physical attribute of energy, it emerges that it is impossible for the universe\index{universe!impossibility of non-zero energy} as a whole to have a non-zero energy, even if the Big Bang does not constitute a creation event at which any conserved quantity must be created in equal positive and negative amounts.

This conclusion simply follows from the fact that if it was possible to measure a non-zero value of energy for the universe\index{universe!positive or negative energy} as a whole, then this value would have to be either positive or negative and this would allow the particular direction of time\index{direction of time!propagation of positive or negative energy} relative to which this positive or negative energy would propagate to be singled out as an absolutely defined time direction\index{time direction!absolute}, as would necessarily happen in the absence of negative-energy matter if there was no compensation of the positive energy of matter\index{energy of matter!universe} by a negative energy of the gravitational field\index{energy of gravitational field!universe}. A similar condition would apply to the momentum of the universe\index{universe!momentum}, given that any non-zero momentum of matter for the universe\index{universe!non-zero momentum of matter} as a whole would allow to single out an absolutely defined direction of space\index{space direction!absolute} as being that along which this positive or negative momentum is directed, as would happen if positive- and negative-energy matter were moving all at once in opposite directions of space and there was no compensation by a momentum of the gravitational field\index{gravitational field momentum!universe} or, more specifically, by that of the equivalent component of the curvature of space\index{curvature of space}.

This means that the total energy of the universe\index{universe!total energy and momentum}, just like its total momentum, must remain null under all circumstances, if one is to avoid giving preferred status to one particular direction of space or time that would be significant on the scale of the universe as a whole, as if this direction could be related to some reference point outside the universe, in violation of the requirement of relational definition of physical attributes\index{requirement of relational definition!physical attributes}. The validity of this argument is reinforced by the fact that, even in the context where positive and negative energy observers are allowed to experience different rates of expansion\index{expansion rate!observer dependence} of space on the global scale, any variation of the average, specific density of negative-energy matter\index{specific density of negative-energy matter!average value} relative to that of positive-energy matter\index{specific density of positive-energy matter!average value} must remain unobservable from the viewpoint of a positive-energy observer (even though negative matter energy can be transformed into negative radiation energy as a result of matter-antimatter annihilation\index{matter-antimatter annihilation}).

Based on the developments introduced in section \ref{sec:4.2}, it would appear, in effect, that when the average, specific density of negative-energy matter\index{specific density of negative-energy matter!relative variation} varies relative to that of positive-energy matter, as a consequence of a divergence of their specific rates of expansion\index{specific expansion rates!divergence} (the rates of expansion experienced by negative- and positive-energy observers, respectively), the ratio of the average densities of positive- and negative-energy matter\index{ratio of average matter densities!invariance} determined by a positive-energy observer must remain invariant, because the density of negative-energy matter measured by such an observer is modified by the same metric conversion factor\index{metric conversion factors} which fixes the average density of vacuum energy\index{average density of vacuum energy}, while the magnitude of this vacuum energy density depends only on the magnitude of the divergence between the scale factors\index{scale factors!divergence} experienced by opposite-energy observers. As a result, if the density of matter energy is null in the initial Big Bang state\index{initial Big Bang state!null density of matter energy}, it always remains null, from the viewpoint of any observer, as expansion takes place.

For those reasons, I believe that it would be incorrect to assume that it may not be absolutely necessary to require the universe\index{universe!requirement of null energy} to have a null value of energy when time comes into existence at the Big Bang, or when matter\index{matter!creation out of nothing} is not created out of nothing due to the fact that it already existed prior to the Big Bang, which may seem to imply that the principle of conservation of energy\index{principle of conservation of energy} would not be required to apply. The fact that, by taking a different stance, I will achieve significant progress in determining the variation of the rate of the expansion of space\index{expansion rate!early variation} in the early universe will serve, I hope, to vindicate the legitimacy of my viewpoint.

Now, what most people already recognize concerning the energy content of the universe\index{universe!energy} is that, for a universe\index{universe!flat geometry} with a flat geometry and a zero cosmological constant\index{cosmological constant!null value}, the negative gravitational potential energy\index{negative gravitational potential energy!positive-energy matter} of positive-energy matter and radiation is balanced by the positive kinetic energy of expansion\index{kinetic energy of expansion} of this matter. When that is not the case, then an additional amount of energy is present that is attributable to the gravitational field\index{gravitational field!energy} itself (or the curvature of space\index{curvature of space!energy}) and this energy tends to become dominant very rapidly as space expands (regardless of whether it is positive or negative), because, while the gravitational potential energy of matter decreases in inverse proportion to the volume, the energy associated with the curvature of space decreases as the inverse of the surface enclosing that volume.

What may be difficult to understand is the fact that the kinetic energy of expansion is actually a property of the expanding space (because, in principle, it could exist even in a universe\index{universe!empty} entirely devoid of matter), which means that it must be considered an energy of the gravitational field itself and not really an energy of matter, despite the fact that the sign of this energy varies as a function of the sign of energy of the matter which is the source of the gravitational potential energy that may balance this kinetic energy of expansion.

Anyhow, the initial value equation\index{initial value equation}, which is derived from the general-relativistic gravitational field equations\index{gravitational field equations!general-relativistic} for a positive-energy observer and which expresses the conservation of gravitational energy\index{conservation of gravitational energy!expanding universe} for an expanding, homogeneous and isotropic universe, is usually written as
\begin{equation}\label{eq:4.3}
E=K+V(a)=\left(\frac{1}{a}\frac{da}{dt}\right)^2+\left(\frac{-8\pi\rho}{3}-\frac{\Lambda}{3}+\frac{k}{a^2}\right)=0
\end{equation}
where $E$ is the gravitational energy of the universe\index{universe!gravitational energy}, $K$ is the kinetic energy of expansion, and $V(a)$ is the Friedmann potential\index{Friedmann potential} as a function of the scale factor\index{scale factor} $a(t)$ in the presence of a cosmological constant\index{cosmological constant} $\Lambda$ for a universe with an average matter density $\rho$. Here the spatial curvature parameter\index{spatial curvature parameter}, which I redefine as $-k/a^2$ and which is always precisely equal to zero for a universe\index{universe!flat space} with flat space, appears as just one particular (reversed) contribution to the Friedmann potential, but when it is possible to assume that the magnitude of the cosmological constant was negligible initially, this equation can be rewritten as
\begin{equation}\label{eq:4.4}
E_g=K+U(a)=\left(\frac{1}{a}\frac{da}{dt}\right)^2-\left(\frac{8\pi\rho}{3}\right)=\frac{-k}{a^2}
\end{equation}
which clearly shows that the spatial curvature parameter is an outcome of the imperfect cancellation of the gravitational potential energy of matter\index{gravitational potential energy of matter} by the kinetic energy of expansion.

Thus, whenever the gravitational potential energy of matter $U(a)$ is not matched by a kinetic energy of expansion\index{kinetic energy of expansion} $K$ that's exactly its opposite, the energy $E_g$ associated with the gravitational field\index{gravitational field!energy} or the curvature of space\index{curvature of space!energy} itself, which is given by $-k/a^2$, is not zero and contributes to alter the expansion rate. If $k$ is positive this excess of gravitational energy is negative, which means that the negative gravitational potential energy of matter contributes predominantly to determine the gravitational field, as must be the case when the source of this gravitational field has positive energy, while when $k$ is negative there is a positive excess of gravitational energy, which means that the positive kinetic energy of expansion (which is also an energy of the gravitational field) contributes predominantly to determine the gravitational field of the universe\index{universe!gravitational field}. The gravitational energy\index{gravitational energy} $E_g$ associated with the present value of the spatial curvature parameter $-k/a^2$ must therefore be considered to consist of a residual measure of energy, which could in principle assume any positive, negative, or null value depending on the current value of the scale factor and on whether $k$ is equal to $-1$, $+1$, or $0$.

There is no \textit{a priori} reason, however, to assume that the measure of energy associated with the curvature of space on the cosmological scale should be the same for positive- and negative-energy observers at the same epoch, because the kinetic energy of expansion varies as a function of the rate of expansion\index{expansion rate!observer dependence}, which is an observer-dependent quantity in the context where, as I explained in section \ref{sec:2.7}, only the average density of positive-energy matter contributes to determine the gravitational field that influences the expansion rate measured by a positive-energy observer, while, in principle, a negative-energy observer could measure different magnitudes for both the average density of negative-energy matter and the rate of expansion it contributes to determine (even before the early annihilation of matter with antimatter\index{early matter-antimatter annihilation}), for reasons I discussed in section \ref{sec:4.2}.

It must be clear that, even though it is usually assumed that, in a general-relativistic context, the initial value equation\index{initial value equation} expresses the requirement of gravitational energy conservation\index{gravitational energy conservation!universe} for the universe as a whole, what the original form of the equation really means is that when an additional term, which is provided by the negative of the spatial curvature parameter\index{spatial curvature parameter} $-k/a^2$, is added to the equation that would otherwise express the nullity of the gravitational energy of the universe\index{universe!null gravitational energy}, then this energy can be conserved even in those cases where it would not really be null initially, but it does not really amount to require that the universe comes into existence with zero gravitational energy. What equation (\ref{eq:4.4}) means is that, once it is assumed that the cosmological constant\index{cosmological constant!negligible initial value} $\Lambda$ is negligible initially, then it is only when the free parameter $-k/a^2$ associated with the energy contained in curvature of space\index{curvature of space!energy} is zero that the positive kinetic energy of expansion\index{kinetic energy of expansion} $K$ can balance the negative gravitational potential energy\index{negative gravitational potential energy!positive-energy matter} $U(a)$ attributable to the presence of positive-energy matter.

The true measure of gravitational energy\index{gravitational energy!universe} for the universe as whole, therefore, is really the energy $E_g$ which is associated with the curvature of space (which would justify that we refer to this energy as the actual gravitational energy of the universe) and it is only when this energy is null that the gravitational field does not contribute energy on the cosmological scale. But it is usually assumed that this spatial curvature parameter can also be positive or negative and the universe\index{universe!positive or negative curvature} be positively or negatively curved, so that the degree of curvature at any given time would depend on the initial value of the kinetic energy of expansion\index{kinetic energy of expansion!initial value} when the density of matter and radiation was maximum. It must be acknowledged, however, that from the viewpoint of positive-energy observers at least, space does have a flat geometry, to a relatively good degree of precision, and this means that there must be a reason why the spatial curvature parameter has a null value.

I believe that what allows the value of gravitational energy $E_g$ associated with the spatial curvature parameter to be null for an expanding zero-energy universe\index{expanding zero-energy universe} is the fact that the gravitational potential energy of matter\index{gravitational potential energy of matter!positive-energy observer} experienced by a positive-energy observer can be arbitrarily large, even when negative-energy matter is present and the total energy of matter\index{energy of matter!null total value} itself is null. Indeed, when a large density of negative-energy matter\index{negative-energy matter!large initial density} is present initially, a universe\index{universe!flat space} with flat space can actually have zero energy\index{universe!zero energy}, despite the fact that from a conventional viewpoint it would appear that if the energy contained in the gravitational field\index{gravitational field!null energy} we experience was null (as would occur if the negative gravitational potential energy\index{negative gravitational potential energy} of matter was compensated by the positive kinetic energy of expansion\index{kinetic energy of expansion!positive}) the energy of the universe\index{universe!positive energy} would still be positive (because the energy of matter would not cancel out). It is only from a conventional perspective that it would appear impossible to require our spatially flat universe to have zero energy.

In the absence of a large density of negative-energy matter, the primordial universe\index{universe!positive space curvature} would actually need to have a positive space curvature in order to have zero energy, because only then could the negative energy contained in the gravitational field\index{gravitational field!negative energy} compensate the large positive energy of matter (while the gravitational field of a universe with negative space curvature\index{universe!negative space curvature} would contribute even more positive energy, as the positive kinetic energy of expansion would overcompensate the negative gravitational potential energy of positive-energy matter to provide a positive gravitational energy of curvature\index{gravitational energy!curvature} $E_g>0$). In fact, it seems that it is only for a closed universe\index{universe!closed and non-expanding} that does not expand at all, that the positive energy of matter could be entirely compensated by the residual energy of the gravitational field\index{gravitational field!residual energy} in the initial state of maximum positive-energy matter density\index{initial state!maximum positive-energy matter density}, because according to certain accounts, when the energy density is that high, the gravitational potential energy is actually equal, in magnitude, to the energy of matter. Again, however, the problem is that the universe\index{universe!highly curved space} doesn't have a highly curved space, but in all likeliness, an almost perfectly flat one.

At this point, it is important to mention that the idea that the energy of the universe\index{universe!null energy} should perhaps be required to be null is not new. Thus, it was once suggested \cite{Tryon-1} that the universe\index{universe!fluctuation into existence} could fluctuate into existence if the positive energy of matter could be compensated by its negative gravitational potential energy\index{negative gravitational potential energy}, at least in a primordial state of very high matter density. The problem was that it appeared that a universe\index{universe!creation out of nothing} with such a highly curved space could never be produced as a fluctuation out of nothing, because, if it actually has zero energy, it would only be allowed to expand for a very short period of time before immediately recollapsing back to the vacuum. Creation out of nothing was eventually salvaged from this severe failure by assuming that once in a while inflationary expansion\index{inflationary expansion} may occur when a universe is fluctuating out of the vacuum, which would enable its expansion rate to start growing exponentially, thereby giving rise to a flat space\index{flat space} which would keep expanding indefinitely.

I will not immediately discuss any motives we may have to resist appealing to inflation theory\index{inflation theory} in order to obtain an expanding, zero-energy universe or indeed to solve any other problem in cosmology, but given that, very early on, I chose to explain known facts with principles which are themselves known to be valid with absolute certainty (even if certain consequences of applying those fundamental principles may not yet be recognized as unavoidable), then I was led to explore the possibility that there may exist a more natural way by which the requirement of null energy could be satisfied in the presence of negative-energy matter.

In order to proceed in this direction, however, one must first acknowledge that if the negative gravitational potential energy\index{negative gravitational potential energy!positive-energy matter} of positive-energy matter exactly balances the positive kinetic energy of expansion\index{kinetic energy of expansion!positive} for a universe\index{universe!flat space} with flat space, then this gravitational potential energy cannot also balance the positive energy of matter itself, as earlier proposals required assuming. This doesn't mean that the magnitude of the gravitational potential energy experienced by a positive-energy observer cannot be equal to the magnitude of positive matter energy initially, only that this is not an appropriate and sufficient condition for obtaining a zero-energy universe\index{universe!zero energy}. In fact, as I mentioned above, it does appear, according to certain accounts, that in the initial Big Bang singularity\index{initial Big Bang singularity} (or indeed any other singularity) the positive energy of matter is equal in magnitude to its negative gravitational potential energy and this is precisely the reason why it was so difficult for me to realize that it is not appropriate to merely require the gravitational potential energy to compensate the energy of matter in order to obtain a universe with zero energy.

What I have realized is that, in a zero-energy universe, any residual measure of gravitational field energy\index{gravitational field energy!residual measure} associated with the initial value of the spatial curvature parameter\index{spatial curvature parameter!initial value} $-k/a^2$ determined using the metric properties of spacetime\index{metric properties of spacetime!positive-energy observers} experienced by positive-energy observers must necessarily balance the residual energy of matter\index{matter energy!residual value} obtained by adding the opposite contributions of positive- and negative-energy matter. Now, if the spatial curvature parameter\index{spatial curvature parameter!null value} is null, like gravitational energy itself, in the case of a universe with flat space (for which the kinetic energy of expansion\index{kinetic energy of expansion} experienced by an observer with a given energy sign precisely balances the gravitational potential energy of matter\index{gravitational potential energy of matter} with the same sign of energy), then it can only mean that, in such a case, the total energy of matter\index{total energy of matter!zero} must itself add up to zero. Normally that would not be possible, because only an empty universe\index{empty universe} would have a null, average density of matter energy. But in the presence of negative-energy matter, a universe\index{universe!arbitrarily high initial matter energy densities} with arbitrarily high initial matter energy densities can actually have a null total matter energy, as long as the average densities of positive- and negative-energy matter\index{average matter densities!equal initial magnitudes} have exactly the same magnitude initially.

Now, the idea that negative matter energy could compensate positive matter energy, from the viewpoint of a positive-energy observer, may perhaps appear problematic in the context where positive-action particles do not directly interact with negative-action particles. One must keep in mind, however, that negative-energy matter inhomogeneities, at least, do have an effect on positive-energy matter and if the presence of a homogeneous distribution of negative-energy matter is without consequences for positive-energy observers, given that it exerts no influence on the expansion rate they experience, it is merely because a uniform distribution of negative-energy matter\index{negative-energy matter!uniform distribution} is equivalent to a void of universal proportion in the positive portion of vacuum energy\index{void in positive vacuum energy!universal proportion}. But this void does exist, from the viewpoint of a positive-energy observer, and therefore its presence constitutes an objective fact\footnote{
In section \ref{sec:4.7} I will explain that there is actually a measure of gravitational information, attributable to the presence of a uniform distribution of negative-energy matter\index{uniform matter distribution!gravitational information|nn}, whose value is independent from the sign of energy or action of an observer and which, therefore, allows to confirm the validity of the hypothesis that the presence of this matter constitutes an objective fact, even for positive-energy observers.}.
 We may, in fact, consider that the way by which negative-energy matter contributes to determine the gravitational field experienced by positive-energy observers on a global scale is precisely by reducing the total energy of matter, which allows the magnitude of the energy of the gravitational field\index{gravitational field energy} to itself be reduced when the sum of all energies is required to be null for the universe\index{universe!null energy} as a whole.

Anyhow, either the negative energy of matter remains totally uncompensated by the energy of the gravitational field associated with a positive-energy observer, in which case this gravitational field energy would alone need to compensate the positive energy of matter, which would imply that there can be no contribution by a positive kinetic energy of expansion\index{kinetic energy of expansion!positive} (so that the universe should not expand at all), or both the positive and the negative portions of the energy of matter must be compensated by the same gravitational field energy, in which case expansion is allowed to occur, but we must explain why the total, average density of matter energy\index{average matter energy density!near-zero initial value} was initially so close to zero that the energy of the gravitational field of the universe\index{universe!null gravitational field energy} (associated with the global curvature of space\index{global space curvature}) was itself required to be perfectly null. Clearly, the second option is the only one that \textit{could} be viable, and therefore I will concentrate on explaining why the total, average density of matter energy which balances the energy of the gravitational field for the universe as a whole cannot be as arbitrarily large as one might expect.

It must be clear, first of all, that even in a universe with zero energy\index{universe!zero energy}, it would be possible for the magnitude of the average density of positive matter energy to be larger or smaller than that of the average density of negative matter energy, either in the initial state\index{initial state!maximum matter energy densities} of maximum matter energy densities or at later times. In the absence of an appropriate constraint this would, in effect, be allowed as long as the difference between the magnitudes of the positive and negative energies of matter is compensated, from the viewpoint of a given observer, by the energy of the gravitational field of the universe\index{universe!gravitational field energy}, which varies as a function of the kinetic energy of expansion\index{kinetic energy of expansion} determined by that observer. It must be clear, however, that only the gravitational field experienced by a positive-energy observer can contribute to cancel out any non-zero, average energy density of matter determined by such an observer.

Under such conditions, the magnitudes of the positive and negative contributions to the energy of the universe could be equal initially, even if the average densities of positive and negative matter energy were not themselves equal, and therefore the total energy could in principle be null regardless of the amount of energy contained in the gravitational field. It may, therefore, seem like a condition of null energy\index{null energy condition!universe} for the universe as a whole does not provide sufficiently strong a constraint to necessarily give rise to a universe\index{universe!flat space} with flat space. But, in fact, I came to realize that this condition is much more constraining than one may expect for gravitational energy and the rate of expansion and that it actually allows to predict that the geometry of our universe can only be observed to be flat on the largest scale.

It is important to point out, however, that the null value of the total energy of matter\index{total energy of matter!null value} cannot be fixed as an independent consistency requirement, because that would require assuming that there cannot even be local fluctuations away from this zero energy for the matter that is present in the primordial state\index{primordial state!local fluctuations in matter energy}, while this is required in order to explain the observed inhomogeneities present in the initial distribution of matter energy on a scale larger than the cosmological horizon\index{cosmological horizon}. But in the absence of such a constraint, local fluctuations above or below the average zero value of matter energy density could, in effect, be present in the initial Big Bang state\index{initial Big Bang state}, even if the average densities of positive and negative matter energy\index{average positive and negative matter energy densities} were required to cancel out on the global scale, so as to allow the zero-energy universe\index{universe!zero energy} to have a flat geometry, as long as there is, in effect, as much overdensity as there is underdensity in the positive- and negative-energy matter distributions on a sufficiently large scale. Such fluctuations in matter energy would simply need to be compensated by \textit{local} variations in the kinetic energy of expansion\index{kinetic energy of expansion!local variations}, above or below the value associated with a critical expansion rate\index{critical expansion rate}.

Even when the average positive and negative matter energy densities\index{average matter energy densities!maximum values} are at their maximum, in the first instants of the Big Bang\index{Big Bang!first instants}, local fluctuations in the total density of matter energy\index{total density of matter energy!fluctuations} are possible for a zero-energy universe, given that local variations of gravitational field energy\index{gravitational field energy!local variations} can compensate local variations in the energy of matter and maintain the sum of matter and gravitational field energies\index{matter and gravitational field energies!zero sum} to a null value. If there was less positive than negative matter energy in a certain location initially, then there would simply need to be more positive gravitational energy and therefore more positive kinetic energy\index{kinetic energy of expansion!positive} of expansion from the viewpoint of positive-energy observers and less negative kinetic energy of expansion\index{kinetic energy of expansion!negative} from the viewpoint of negative-energy observers.

Thus, \textit{locally} at least, the null value of energy can arise from a compensation between the energy of matter and the energy of the gravitational field, because a local variation in the energy of the gravitational field (attributable to a local variation of the kinetic energy of expansion\index{kinetic energy of expansion!local variations} above or below the value associated with a critical expansion rate\index{critical expansion rate}) can be made to compensate any local difference between the magnitude of the density of positive matter energy and that of negative matter energy, just like the global measure of gravitational field energy\index{gravitational field energy!global measure} which is attributable to the difference between the observer-dependent gravitational potential energy of matter\index{gravitational potential energy of matter!observer dependence} and the observer-dependent kinetic energy of expansion\index{kinetic energy of expansion!observer dependence} could in principle compensate any difference between the magnitudes of the \textit{average} cosmic densities of positive and negative matter energy. However, in section \ref{sec:4.9} I will explain that a certain unavoidable constraint actually limits the amplitude of those fluctuations in the initial Big Bang state and therefore it cannot be expected that there would occur large deviations from zero gravitational energy\index{gravitational energy!local deviations from zero value} locally if this condition is also obeyed globally.

But even if local fluctuations in the density of matter energy are clearly unavoidable, it remains to explain why it is that such a compensation of matter energy by gravitational energy is not allowed to take place on a global scale, as required if space is to be flat for the universe\index{universe!flat space} as a whole. Indeed, as I mentioned above, if a residual gravitational energy associated with the spatial curvature parameter\index{spatial curvature parameter!residual gravitational energy} $-k/a^2$ could also compensate a difference in the magnitude of the initial, average densities of positive and negative matter energy on a global scale, then it should be possible for the magnitude of the kinetic energies of expansion\index{kinetic energy of expansion} experienced by positive- and negative-energy observers to be larger or smaller than the magnitude of the gravitational potential energies\index{gravitational potential energy} of their associated matter. Under such conditions, the rates of expansion of space\index{expansion rate!critical value} would no longer need to be critical, even in a zero-energy universe\index{universe!zero energy}.

It is certainly true that a homogeneous distribution of negative-energy matter exerts no influence on the specific expansion rate\index{specific expansion rate!positive-energy matter} of positive-energy matter which determines the kinetic energy of expansion measured by a positive-energy observer, but this is significant merely in the sense that only the energy of the gravitational field\index{gravitational field energy!positive-energy observer} perceived by a positive-energy observer can contribute to the energy\index{energy!global scale} that must add up to zero on a global scale, from the viewpoint of such an observer. For reasons I previously mentioned, it is still necessary to assume that both the positive and the negative energy of matter contribute to the total energy of the universe\index{universe!total energy} measured by a positive-energy observer.

What must be clear also is that, in the context where the energy of the universe is required to be null, if space was positively curved and closed\index{positively curved and closed space} from the viewpoint of a positive-energy observer, it would need to be negatively curved and open\index{negatively curved and open space} from the viewpoint of a negative-energy observer. Indeed, the gravitational field of a universe\index{universe!positive space curvature} whose space would be positively curved, from the viewpoint of a positive-energy observer, would have a negative energy and could, therefore, only compensate an excess of positive matter energy (through a reduction of the positive kinetic energy of expansion\index{kinetic energy of expansion!positive}). But while it is true that, even from the viewpoint of a negative-energy observer, an excess of positive matter energy would require the contribution of a gravitational field with negative energy\index{gravitational field energy!negative}, such a gravitational field would be associated not with a smaller positive kinetic energy of expansion, but with a larger negative kinetic energy of expansion\index{kinetic energy of expansion!negative} and a higher than critical expansion rate\index{expansion rate!higher than critical}, which would actually give rise to a universe\index{universe!open space} with an open space.

If the total energy of matter was instead negative initially (before the early annihilation of matter and antimatter\index{early matter-antimatter annihilation} took place), as would occur if negative-energy matter particles contributed more energy than positive-energy matter particles on the average, then the opposite would be true and the universe\index{universe!closed space} would need to have a closed space from the viewpoint of a negative-energy observer and an open space from that of a positive-energy observer. Now, while those two mutually exclusive configurations may appear to merely consist of two additional possibilities, no different from the case where the total average density of matter energy\index{total average matter energy density!null initial value} happens to be null initially, just like the energy of the gravitational field of the universe\index{universe!null gravitational field energy}, there is actually a very important distinction between the case of a universe\index{universe!flat geometry} with flat geometry and that of the curved space configurations. This essential difference has to do with the fact that, in the case of a flat space\index{flat space!observer independent openness}, the geometry of the universe would be open from both the viewpoint of a positive-energy observer and that of a negative-energy observer, while in all the other possible cases it seems that the universe\index{universe!open or closed space} would need to have an open space for an observer with a given energy sign and a closed space for an observer with the opposite energy sign.

I believe that if the average density of positive matter energy must exactly compensate the average density of negative matter energy in the initial Big Bang state\index{initial Big Bang state!average matter energy densities}, even though local deviations away from the zero energy of matter\index{zero energy of matter!local deviations} are allowed to be present to a certain extent (as long as they are compensated by opposite local deviations away from zero gravitational energy\index{zero gravitational energy!local deviations}), it is precisely because, in the absence of any other contribution to the energy budget, if matter energy was not null, then the universe\index{universe!observer-independent openness of space} could not have an open space from the viewpoint of all observers. If an excess of positive or negative gravitational energy\index{gravitational energy!compensation of matter energy} was allowed to compensate a deficit of positive or negative matter energy (respectively) on a global scale, then this excess gravitational energy would give rise to a universe\index{universe!open or closed space} whose space would be open for one observer and closed for an observer with opposite energy sign.

But given that the difference between the volume of a universe\index{universe!volume} with a closed space and that of a universe with an open space would in principle be infinite, it follows that such a configuration would be characterized by an arbitrarily large, positive or negative, average density of vacuum energy\index{average vacuum energy density!arbitrarily large value}. Indeed, from the viewpoint of the developments discussed in section \ref{sec:4.2}, it would follow that if gravitational energy\index{gravitational energy!negative global value} was negative globally and the universe had a closed space from the viewpoint of a positive-energy observer and an open space from the viewpoint of a negative-energy observer, as would appear to be required in order to compensate a positive total energy of matter\index{total energy of matter!positive}, the average density of vacuum energy\index{average vacuum energy density!maximum values} should be set to its maximum positive value, while if the opposite was true and gravitational energy was instead positive, as would appear to be required if it is to compensate a negative total energy of matter\index{total energy of matter!negative}, then the average density of vacuum energy should be set to its maximum negative value right at the Big Bang\footnote{
It should be clear that it cannot be assumed that the total energy of matter\index{total energy of matter|nn} in a universe\index{universe!non-zero space curvature|nn} with a non-zero curvature of space is compensated by an opposite energy that would be contained in the vacuum as a result of this curvature, because in a universe\index{universe!gravitational energy and space curvature|nn} with negative gravitational energy and positive space curvature the average density of vacuum energy\index{average vacuum energy density!positive|nn} would actually be positive and would add to the positive energy of matter, thereby requiring an even larger negative energy for the gravitational field, that would make the positive density of vacuum energy even larger.}.

The problem is that the maximum positive or negative value of the cosmological constant\index{cosmological constant!maximum positive or negative value} which would be associated with \textit{any} such configuration would appear to forbid the emergence of an observer\index{observer emergence!forbidding conditions}, because even if the gravitational force exerted by the cosmological constant on the specific expansion rates\index{specific expansion rates} of positive- and negative-energy matter generally contributes to reduce the magnitude of the average density of vacuum energy\index{average vacuum energy density!self-reduction}, if this magnitude had been maximum right at the beginning of the expansion process, then it could never have been reduced to a level favorable to the emergence of an observer\index{observer emergence!favorable conditions}, at least not before the average matter density itself would have become too low to allow for the development of structures\index{structure development} (which we may assume to be essential for the emergence of an observer). Only a universe\index{universe!precisely balanced initial energies} with precisely balanced initial contributions to the energy of matter and therefore, also, to the energy of the gravitational field, is allowed to be experienced as a long lasting process by a physical observer that is part of that universe, when it is appropriately required that the universe\index{universe!requirement of null energy} itself has null energy.

It is only when space is flat\index{flat space!global scale} on a global scale and there is no energy in the gravitational field\index{gravitational field!absence of energy} that the magnitude of the average density of vacuum energy\index{average vacuum energy density!maximum theoretical value} can be different from its maximum theoretical value initially. But given that this is required if an observer is to be present at some point in the universe to measure any value of gravitational energy, then one must conclude that the kinetic energy of expansion\index{kinetic energy of expansion!positive-energy observer} determined by a positive-energy observer would always precisely compensate the gravitational potential energy\index{gravitational potential energy!positive-energy matter} attributable to positive-energy matter and the same would be true (independently) for negative energy matter and its gravitational field, from the viewpoint of a negative energy observer. What I'm suggesting is that this is allowed to occur, in the case of a zero-energy universe\index{universe!zero energy}, when the total energy of matter\index{energy of matter!null total initial value} itself is null in the very first instants of the Big Bang\index{Big Bang!first instants} (before the early annihilation of most baryons with their antibaryon\index{early baryon-antibaryon annihilation} counterparts), as becomes possible in the presence of negative energy matter.

When this is properly understood, it becomes clear that the `extra' principle which would allow to fix the rate of expansion of space\index{expansion rate!critical value} on a global scale to its critical value is nothing else but the requirement of relational definition\index{requirement of relational definition!physical attributes} of physical attributes, which requires the sum of all energies to be null, for the universe\index{universe!null energy} as a whole, from the viewpoint of both positive- and negative-energy observers. In the context of the generalized gravitation theory\index{generalized gravitation theory} introduced in the second chapter of this report, and given the interpretation that was proposed in section \ref{sec:4.2} for the vacuum-energy term\index{vacuum-energy term}, this constraint actually allows to determine which solution of the gravitational field equations\index{gravitational field equations!appropriate solution} is the appropriate one for a description of the expanding universe.

It is, therefore, by applying this very basic principle, in the context where it is recognized that negative-energy matter must also contribute to the universe's\index{universe!initial energy budget} initial energy budget, that it becomes possible to explain not only why there is an expansion of space, but why it is that the rate of this expansion\index{expansion rate!critical value} is still critical, even long after the Big Bang. Space is flat and the rate of expansion remains critical, because the universe\index{universe!open space} must have an open space from both the viewpoint of positive-energy observers and that of negative-energy observers and the precision with which the initial rate of expansion\index{initial expansion rate!precision of adjustment} was adjusted to its critical value is merely a reflection of the exactness of this requirement\footnote{
It must be noted that the same constraint allows one to expect that there is no difference between the average states of motion of positive- and negative-energy matter\index{average states of motion!global scale|nn} on the global scale that could have given rise to a non-zero momentum for the universe\index{universe!non-zero momentum|nn} as a whole, because such a momentum for matter would need to be compensated by an opposite momentum of the gravitational field\index{gravitational field momentum!compensation of momentum of matter|nn}, and if the gravitational field\index{gravitational field!non-zero energy|nn} (or the appropriate component of the curvature of space) had a non-zero momentum on a global scale, it would also need to have a non-zero energy and this is not possible for a universe\index{universe!flat geometry|nn} with flat geometry.}.

It must be clear, however, that in the context where the initial, average density of negative-energy matter can be reduced to a greater extent than that of positive-energy matter following the annihilation of baryonic matter and antimatter\index{baryonic matter and antimatter annihilation} that takes place early on, during the Big Bang, it is possible for the average value of vacuum energy density\index{average vacuum energy density!growing value}, or the cosmological constant\index{cosmological constant!growth}, to grow from its initial zero value toward a larger, positive value during the matter-dominated era\index{matter-dominated era}, because under such circumstances the specific rate of expansion of positive-energy matter\index{specific expansion rate!positive-energy matter} is reduced more rapidly than that of negative-energy matter, due to the larger gravitational pull exerted by positive-energy matter, which allows the scale factors\index{scale factors!divergence} experienced by opposite-energy observers to diverge. But it is not to be expected that this divergence could develop to an arbitrarily large magnitude, because the weak anthropic principle\index{weak anthropic principle!cosmological constant growth} also forbids the cosmological constant from becoming so large, as a result of this divergence, that it would no longer be compatible with the presence of a (positive-energy) observer at the present time.

What must be retained from all this is that, if it was not for the fact that the presence of a homogeneous distribution of negative-energy matter exerts no influence on the specific expansion rate of positive-energy matter (as explicitly stated in the formulation of principle 6 from section \ref{sec:2.13} and for reasons I have explained in section \ref{sec:2.7}), then, even if the total energy of matter\index{energy of matter!null total initial value} was null initially, it would not be possible to conclude that the actual expansion rate\index{actual expansion rate!critical rate} must have been the critical rate which is determined by the density of positive matter energy alone, because under such conditions the gravitational potential energy of matter\index{gravitational potential energy of matter!zero} that would need to be balanced by the kinetic energy of expansion\index{kinetic energy of expansion} would actually be zero (because the average density of matter energy\index{average density of matter energy!null} that would determine the strength of the gravitational field would itself be null initially), which means that the kinetic energy of expansion would also need to be zero and the universe should not expand at all.

But if the requirement of null energy did not apply to the gravitational field of the universe\index{gravitational field of universe!requirement of null energy} and space did expand, as we would normally assume, then the expansion rate\index{expansion rate!absence of deceleration} would not be submitted to any deceleration and the universe would explode like a negatively-curved universe\index{universe!negatively curved} with a null matter density. The independence of the specific expansion rates\index{specific expansion rates!independence} of positive- and negative-energy matter from the presence of matter with an opposite energy sign, which follows from my description of negative-energy matter as consisting of voids in the positive-energy portion of the vacuum\index{void in positive vacuum energy}, is therefore an essential ingredient of the alternative solution to the problem of flatness\index{flatness problem!alternative solution} that is proposed here. This condition is especially constraining in the context where the initial matter distribution\index{initial matter distribution!macroscopic homogeneity} must be highly homogeneous on a macroscopic scale (for reasons I will explain in section \ref{sec:4.9}), so that there cannot even exist significant local perturbations of the specific rate of expansion\index{specific expansion rate!local perturbations} of matter with a given energy sign by matter with an opposite energy sign on a large scale.

It is only after I realized that, from the viewpoint of a positive-energy observer, the presence of negative-energy matter does not contribute to determine the gravitational potential energy\index{gravitational potential energy!universe} of the universe (which in the case of a universe\index{universe!flat geometry} with an overall flat geometry is compensated by the kinetic energy of expansion\index{kinetic energy of expansion}), that I was able to understand that, despite what is usually assumed, it is, in effect, not only the current \textit{variation} of the specific rate of expansion of positive-energy matter\index{specific expansion rate of positive-energy matter!variation} which is determined in part by its energy density, but actually also the current specific rate of expansion\index{specific expansion rate!current value} itself. It took me a certain time to recognize that the variation of the specific rate of expansion of positive-energy matter must depend on the density of (positive) matter energy, as most people may consider obvious (but unlike one would perhaps expect for a universe\index{universe!null matter energy} with null matter energy), yet my questioning has allowed me to realize that the relation which exists between the rate of expansion\index{expansion rate!dependence on matter density} of space and the density of matter energy is actually much more constraining than is usually assumed.

As a result, I'm allowed to conclude that, even in the absence of an early phase of inflationary expansion\index{inflationary expansion}, it is not necessary to assume that the present, average density of positive matter and vacuum energy\index{average density of positive matter and vacuum energy!critical value} is critical purely for aesthetic reasons, because, in fact, it is possible to explain why the universe is so perfectly balanced, when one recognizes the necessity to properly apply the requirement of relational definition\index{requirement of relational definition!physical attributes} of physical attributes to the energy of the universe\index{universe!energy} as a whole, which requires it to remain null even when it is hypothesized that bidirectional time\index{bidirectional time!continuation past initial Big Bang singularity} may be continued past the initial Big Bang singularity, following a quantum bounce\index{quantum bounce}, so that matter\index{matter!creation out of nothing} doesn't need to be created out of nothing.

Indeed, under such conditions, the fact that an observer can only measure a value of vacuum energy density that is compatible with the conditions of her own existence, implies that it is not merely the total energy content of the universe\index{universe!null total energy} that is observed to be precisely null, but also the total energy of matter\index{energy of matter!null total initial value} in the initial Big Bang state, as well as the gravitational energy\index{gravitational energy!universe} of the universe. Therefore, it is now possible to understand that the flatness of space\index{flatness of space!basic consistency requirement} is not a mere possibility that emerged as a byproduct of an uncertain process of inflationary expansion, but rather constitutes a basic consistency requirement that must be satisfied by any viable cosmological model\index{cosmological models!viable}.

\bigskip

\noindent When I will discuss the horizon problem\index{horizon problem}, in section \ref{sec:4.9}, I will explain what justifies assuming that the distribution of positive and negative matter energies in the initial Big Bang state was sufficiently homogeneous that no macroscopic event horizons\index{macroscopic event horizon} (understood as any event horizon larger than that which is associated with an elementary black hole\index{elementary black hole} with a mass equal to one Planck mass\index{Planck mass}) would be present on any scale. But it can already be appreciated that, in the context where the initial distribution of matter energy is uniform to a very high degree and the \textit{local} rates of expansion\index{local expansion rates!variations} experienced by positive- and negative-energy observers only vary in such a way as to allow the kinetic energy of expansion\index{kinetic energy of expansion} to compensate any difference between the amplitudes of their opposite energy densities, as I'm here assuming, then the expansion of space\index{expansion of space!isotropy} must remain almost perfectly isotropic on the largest scale; which certainly constitutes an appropriate conclusion from an observational viewpoint.

The fact that, from a conventional perspective, such an outcome would only be allowed to happen as the consequence of an early phase of inflationary expansion\index{inflationary expansion}, therefore, no longer constitutes a decisive argument in favor of inflation theory\index{inflation theory}, because from my perspective an initial phase of accelerated expansion\index{accelerated expansion} is no longer necessary to produce such an outcome.

In the context where the sum of all energies which can be measured by a given observer must be null, for the universe\index{universe!null energy} as a whole, it also emerges that the often met remark, to the effect that the observed equilibrium between open and closed space\index{equilibrium between open and closed space} is improbable, as it requires a delicate balance between the kinetic energy of expansion\index{kinetic energy of expansion} and the gravitational potential energy of matter\index{gravitational potential energy of matter}, is irrelevant, because, on the basis of the hypothesis that an observer must be allowed to exist in the universe to determine the rate of expansion\index{expansion rate} of space, such an observation, far from being improbable, is actually unavoidable.

The tentative solution to the problem of flatness\index{flatness problem} provided by inflation theory\index{inflation theory}, therefore, appears to simply be unnecessary, because even when the initial density of positive-energy matter is very high, the energy of the gravitational field\index{gravitational field energy!universe} is required to be null in a zero-energy universe\index{universe!zero energy}, which means that the universe must necessarily have a critical density\index{critical density!positive matter energy} of positive matter energy and enough kinetic energy to keep expanding forever (at an ever slower rate), even if a non-zero cosmological constant\index{cosmological constant!non-zero value} develops later on, as a result of a variation of the ratio of positive to negative matter energy densities\index{ratio of positive to negative matter energy densities} attributable to the annihilation of matter with antimatter\index{matter-antimatter annihilation}, because even a negative average value of vacuum energy density\index{average vacuum energy density!self-reduction} would have a tendency to be reduced under its own influence, instead of allowing space to recollapse from the viewpoint of a positive-energy observer.

It would therefore appear that the idea that the initial push of inflation\index{inflation!initial push} is necessary to explain that there is any expansion at all is incorrect, because, if a state of maximum positive and negative matter energy densities\index{maximum matter energy densities!state} must exist in the universe at some point (if such a condition needs to be satisfied independently from whether there is expansion or not, as I will suggest in section \ref{sec:4.9}), then expansion at a proportionately high rate does actually become an absolute necessity, if gravitational energy is to be null at all times for the universe\index{universe!null gravitational energy} as a whole, independently.

But even if it is the presence of an observer that requires this latter condition to be satisfied, this does not mean that it is necessary to appeal to the weak anthropic principle\index{weak anthropic principle} in order to explain the fact that the universe has not yet recollapsed, because what is required by the presence of an observer is not merely that the universe is still expanding at the appropriate rate for life to exist, but really that it has a perfectly null cosmological constant\index{cosmological constant!perfectly null initial value} initially (which can only happen when the rate of expansion of space\index{expansion rate!critical value} is critical). It is true, though, that if the universe did not expand at a rate that would have been too large or too small to allow for the emergence of an observer\index{observer emergence}, it is not only because the cosmological constant was null initially and the rate of expansion critical, but also because the average density of vacuum energy\index{average density of vacuum energy!growth} did not later grow to a much larger value that would have accelerated or decelerated the critical rate of expansion\index{expansion rate!acceleration or deceleration} to such an extent that it would have become incompatible with the presence of an observer.

In the context where bidirectional time\index{bidirectional time!continuation past initial Big Bang singularity} may extend past the initial singularity, one would also need to recognize that it is not necessary for matter\index{matter!creation out of nothing} to be created out of nothing, because matter and radiation could have been present before the Big Bang that would have been submitted to a quantum bounce\index{quantum bounce} as a result of the contraction of space that would have taken place in the future direction of time. In such a case, there would be no meaning to ask how it is that matter was created, because matter simply exists and is not produced by the Big Bang. In fact, in chapter \ref{chap:5} I will explain that there may be good reasons to believe that this persistence in time is actually an essential requirement and that it may need to be extended in a certain, more unexpected way, in order to avoid the hypothesis that matter must have been created out of truly nothing.

From this viewpoint, it would appear that it is only when we ignore the limitations imposed on the magnitude of positive and negative matter energy densities in a quantum-gravitational context, that a problem may arise with the fact that matter appears to be present in the very first instants of the Big Bang, even if it cannot be created out of nothing. Ironically, it is precisely because we assume the existence of an early phase of inflationary expansion\index{inflationary expansion} that must leave the universe totally empty, that we need to justify the presence of matter in our universe, by assuming that it was created at a later time by a process of reheating attributable to inflation\index{inflation!reheating process}. But given that inflation theory\index{inflation theory} may no longer be required to explain flatness itself, then it is certainly not inadequate to conclude that there may, after all, be no substance to the problem of matter creation\index{matter-creation problem}. The idea that only inflation theory allows to explain the relatively large `initial' density of positive-energy matter would then be incorrect, because, in fact, the hypothesis that there occurred an early phase of inflationary expansion is precisely what makes it more difficult to explain the existence of a hot Big Bang\index{hot Big Bang}.

Now, if there actually exists a history\index{history!before initial state of maximum matter density} unfolding past the initial state of maximum matter density, then it is necessary to assume that the same density of matter energy must have existed in the moments immediately \textit{preceding} the initial singularity\index{initial singularity!preceding moments}, as existed in the moments immediately following it. This means that the expansion rate, following the quantum bounce\index{quantum bounce} in the \textit{past} direction of time, must be as large as it was in the moments immediately following the singularity in the future direction of time (from the viewpoint of both positive- and negative-energy observers), given that this expansion rate\index{expansion rate!critical value} must be critical if energy is to be null on the other side in time of the state of maximum matter density\index{maximum matter density state} as well. Therefore, it seems that the conditions necessary for the emergence of an observer\index{observer emergence!necessary conditions} may also exist in that portion of history.

In fact, the initial expansion rates\index{initial expansion rate!past and future of initial singularity} can only be equal on both sides in time of the singularity under the condition that the distribution of matter energy in the `final' state which would be reached while space collapses in the \textit{future} direction of time, in that unknown portion of history taking place before the Big Bang, is as homogeneous as the distribution of matter that provides the `initial' boundary conditions for the current one, because otherwise macroscopic inhomogeneities could potentially survive the quantum bounce that would influence the local rates of expansion\index{local expansion rates}. In section \ref{sec:4.9} I will explain why this hypothesis is appropriate and it will then be clear that it does not even need to be confirmed by making use of the detailed mathematical framework of a quantum theory of gravitation\index{quantum gravitation theories!mathematical framework}.

It is remarkable that despite our ignorance of the exact nature of the laws which apply at the Planck time\index{Planck time}, it is nevertheless possible to predict with arbitrarily high precision what the rate of expansion of space\index{expansion rate!initial value} was when the average densities of positive and negative matter energy were maximum. But it is also possible to predict that, regardless of what happens to the ratio of the average densities of positive- and negative-energy matter\index{ratio of average positive- and negative-energy matter densities} (as a result of matter-antimatter annihilation\index{matter-antimatter annihilation} in particular), both the average, specific density of negative matter energy plus vacuum energy and the average, specific density of positive matter energy plus vacuum energy must remain critical if they originally were, given that a flat geometry\index{flat geometry!constant radius of curvature} is the one configuration whose radius of curvature does not change with time.

When the cosmological constant\index{cosmological constant!growth from initial zero value} grows from its initial zero value into a larger positive value, there is more positive vacuum energy to accelerate the specific rate of expansion\index{specific expansion rate!positive-energy matter} of positive-energy matter at later times, but this additional positive energy also contribute to the total density of energy that determines the curvature of space\index{curvature of space!positive-energy observer} experienced by a positive-energy observer, which means that this density remains critical if it initially was and the same is true for a negative-energy observer. This is allowed as a consequence of the fact that the uniform portion of vacuum energy\index{uniform portion of vacuum energy!independent conservation} is conserved independently from the energy of matter and can actually be created, even when it does not exist initially, because it is compensated by its own contribution to gravitational potential energy\index{gravitational potential energy!vacuum energy contribution} which, under such conditions, can actually grow (reach larger negative or positive values) while space is expanding, exactly as would occur during a hypothetical phase of inflationary expansion\index{inflationary expansion}.

At this point, one may recall the conclusion I arrived at in section \ref{sec:4.3} to the effect that positive vacuum-dark-matter energy\index{vacuum-dark-matter energies!global invariance} (like negative vacuum-dark-matter energy) cannot be assumed to rise on the global scale, despite the fact that the density of vacuum-dark-matter energy\index{vacuum-dark-matter energy density!local growth} should grow in various locations, along with the inhomogeneity of the visible matter distribution. In light of the developments introduced in this section, it would appear that this conclusion is fully justified, because if the amount of positive-energy dark matter was allowed to grow in such a way, then, in the context where the universe\index{universe!flat space} had a flat space and the matter distribution was highly homogeneous in the initial Big Bang state\index{initial Big Bang state!homogeneous matter distribution} (or in the state immediately following inflation), if the growth of inhomogeneity that follows was to contribute to increase the total amount of positive vacuum-dark-matter energy, this would slow down the rate of expansion determined by a positive-energy observer, because only the positive density of vacuum-dark-matter energy would exert an influence on this expansion rate. As a consequence, the universe\index{universe!positive space curvature} would acquire a positive space curvature on the cosmic scale (while a similar phenomenon would be experienced by negative-energy observers).

The problem is that this means that the energy of the gravitational field of the universe\index{universe!negative gravitational field energy} would become negative, while the energy of matter and radiation could still be null in principle (as there may occur a similar growth of negative vacuum-dark-matter energy density). As a result the condition of null energy would be violated for the universe\index{universe!requirement of null energy} as a whole. We may, therefore, conclude that it is, in effect, an absolute requirement for vacuum dark matter to already be present in homogeneously distributed form (still distinct from the uniform portion of vacuum energy\index{uniform portion of vacuum energy!cosmological constant} associated with the cosmological constant), before its density begins to grow locally, as a result of the formation of inhomogeneities in the matter distribution.

\bigskip

\noindent To summarize, we are in a situation where the average densities of positive and negative matter energy must reach their maximum values, which are determined by the natural scale of quantum-gravitational phenomena\index{quantum-gravitational phenomena!natural scale}, at the exact same time in the past, because in a zero-energy universe\index{universe!zero energy} this allows the energy of the gravitational field\index{gravitational field energy!null} to also be null and space to be flat from the viewpoint of both positive- and negative-energy observers. If this was not the case, then opposite-energy observers would necessarily measure opposite values for the non-vanishing gravitational energy of the universe\index{universe!non-vanishing gravitational energy} and this would result in opposite space curvatures\index{opposite space curvatures!opposite-energy observers}, which would give rise to an average density of vacuum energy\index{average vacuum energy density!maximum magnitude} of maximum magnitude that would be incompatible with the emergence of an observer\index{observer emergence} at later times. But this condition of null gravitational energy\index{null gravitational energy condition} can only be satisfied when the kinetic energy of expansion\index{kinetic energy of expansion} measured by a positive-energy observer precisely balances the opposite gravitational potential energy of positive matter and vacuum energy\index{gravitational potential energy!positive matter and vacuum energy} and this is what explains that space still expands at a critical rate long after the Big Bang.

The problem that there was with the conventional approach is that, if we required energy to be null for the universe as whole, we could not balance the very large positive density of matter energy (characteristic of quantum-gravitational phenomena) that existed initially in our flat space universe\index{universe!flat space}, so that it always appeared inappropriate to try to justify the flatness of space as being the consequence of a condition of null energy that would apply to the universe as a whole, despite the fact that gravitational energy\index{gravitational energy!flat universe} itself really is null for a flat universe (given that the kinetic energy of expansion is the exact opposite of the gravitational potential energy of matter and vacuum fluctuations). This is the reason why we failed to understand that applying a condition of zero energy to the universe\index{universe!requirement of null energy} could actually provide the basis for an explanation of the flatness of space that does not require assuming that the null energy of the gravitational field determined by a positive-energy observer is a mere coincidence or an outcome of inflationary expansion\index{inflationary expansion}.

\section{The problem of time asymmetry\label{sec:4.6}}

It is remarkable that, at this point into my discussion, I have already been able to provide independent solutions to two of the worst fine-tuning problems\index{fine-tuning problems!cosmology} of cosmology, guided merely by an unwavering confidence in the validity of well-known physical principles. It is significant, also, that both the solution to the cosmological constant problem\index{cosmological constant!problem} and that which was proposed to the flatness problem\index{flatness problem} involved considering the balancing effects of negative-energy matter in order to provide additional constraints on the values of physical parameters. But before I can address other aspects of the inflation problem\index{inflation!problem}, it will be necessary to delve a little deeper into what really constitute the many facets of the problem of time asymmetry\index{problem of time asymmetry} from a classical viewpoint. This will allow me to properly identify the nature of the deep contradiction that still dwells at the heart of theoretical physics\index{theoretical physics!deep contradiction}, as a result of the apparent incompatibility between the time-symmetric laws\index{time-symmetric laws!classical mechanics and particle physics} of classical mechanics and particle physics and the unidirectional laws\index{unidirectional laws!thermodynamics and statistical mechanics} of thermodynamics and statistical mechanics.

Before engaging in a discussion of the problem of time asymmetry what one must first decide is whether temporal irreversibility\index{temporal irreversibility} is real, or whether it is a mere consequence of the way we describe the state of a system. It has been argued, in effect, that it is only as a consequence of adopting a particular coarse-graining\index{coarse-graining} and due to the choice that is made regarding what details of the microscopic state\index{microscopic state!ignored details} of a system are to be ignored, that temporal irreversibility occurs. If that was the case, then the continuous increase of entropy\index{entropy increase!subjective notion}, which under certain conditions appears to characterize the evolution of physical systems with a large number of microscopic degrees of freedom\index{microscopic degrees of freedom}, would be a purely subjective notion, significant merely in the context where there are practical limitations on our ability to perceive the evolution of a physical system down to its most intricate details.

Under such conditions, even if entropy (as a measure of the number of possible, distinct, microscopic states of a system that are compatible with an appropriate choice of observable macroscopic parameters\index{observable macroscopic parameters}) was to vary, the changes which are taking place would have no fundamental significance and the observation of certain regularities regarding entropy growth would not require explanation, given that the quantity involved would merely be a subjective aspect of reality. But despite the fact that this idea is still quite popular among those who have not seriously studied the question of the origin of thermodynamic time asymmetry\index{thermodynamic time asymmetry!origin}, it is no longer viewed by most specialists as an appropriate solution to the problem of the origin of time irreversibility\index{time irreversibility!origin}, but rather as an attempt at easily disposing of the problem without really explaining anything.

It was pointed out by Roger Penrose\index{Penrose, Roger} that the growth of entropy involved in most irreversible thermodynamic processes\index{irreversible thermodynamic processes} is so large that it is only marginally dependent on the choice of coarse-graining\index{coarse-graining!choice}. Thus, it appears that the degree of \textit{appropriateness} of any particular coarse-graining itself varies dramatically in the course of certain processes which are occurring all the time in our universe. But the truth is that, even if we were to follow the evolution of all the microscopic physical parameters\index{microscopic physical parameters} of a large system in a non-equilibrium state\index{non-equilibrium state} with as much precision as quantum theory allows, certain aspects of this evolution could still be characterized as unidirectional, despite the fact that a maximum amount of information\index{information!maximum amount} would be available about the system.

What this means is that we are not just shuffling an initially well-ordered deck of cards\index{deck of cards!shuffling analogy} (to use a simple analogy) which would merely be losing a subjective amount of structure. When we are considering an ordinary deck of cards, all configurations are equivalent, despite the particular significance we attach to the `ordered' configuration. But in our universe the changes which are taking place when entropy is observed to be growing can be characterized in a more objective way, due to the nature of that portion of entropy\index{entropy!gravitational interaction} that is attributable to the gravitational interaction. Indeed, as I have explained in section \ref{sec:3.10}, the measure of entropy associated with black hole event horizons\index{black hole event horizon!entropy} does not grow merely as a consequence of our adoption a certain \textit{arbitrary} definition regarding what parameters should characterize the macroscopic state of such a system and what information remains unavailable, and therefore it gives rise to a less subjective notion of time irreversibility\index{time irreversibility!less subjective notion}. Another distinction of the evolution which is actually taking place on a macroscopic scale in our universe is that the probability to return to a former state of lower entropy never stops diminishing, because the entropy\index{entropy growth!absence of limit} is in principle allowed to grow without limit.

It must be clear, though, that time irreversibility\index{time irreversibility!expansion of space} is not just a consequence of the expansion of space. It was once suggested, in effect, that the growth of entropy associated with all irreversible processes could be a consequence of universal expansion, given that the thermodynamic arrow of time\index{thermodynamic arrow of time} is oriented in the same direction as what is sometimes called the cosmological arrow of time\index{cosmological arrow of time}, which is merely the direction of time in which space is expanding. But it was later pointed out that this assumption is inappropriate, because in such a context one would need to assume that the thermodynamic arrow of time\index{thermodynamic arrow of time!reversal} should immediately reverse under conditions where space would begin contracting, while there is no independent motive to justify this hypothesis.

The expansion of space is a global phenomenon, while an expanding gas in a container is a local phenomenon which we have no reason to expect would be so drastically affected by what happens to the relative motion of distant galaxies as to start behaving anti-thermodynamically and retract into a smaller volume the moment space would begin contracting on a global scale. This conclusion is certainly appropriate, given that if we were to assume that space contraction alone is sufficient to give rise to a reversal of the thermodynamic arrow of time then we would probably also have to assume that the arrow of time reverses locally in the presence of a strong enough, attractive gravitational field, while of course there is no experimental evidence at all that this is happening.

It is usually understood, however, that while we are allowed to consider entropy as missing information, an objective characterization of temporal irreversibility\index{temporal irreversibility!objective characterization} does not require assuming that information\index{information!irreversible loss} is irreversibly lost when entropy is rising. It is certainly true that, when the exact evolution of a system that is not in a state of thermal equilibrium\index{thermal equilibrium state} cannot be followed down to its most intricate details, we may lose sight of information concerning its exact microscopic state\index{exact microscopic state} and therefore more information than is available afterward may be needed to describe it. But if ignorance is growing, it is only because the macroscopic parameters we use to describe the state of such a system are leaving aside an increasingly larger portion of the information that would be required to accurately describe its exact microscopic state.

Thus, even if certain macroscopic physical parameters\index{macroscopic physical parameters!irreversible evolution} which allow to objectively assess the growth of entropy evolve irreversibly, the amount of structure present on a microscopic scale remains unchanged as those transformations are taking place. It is simply the fact that, regardless of how well chosen they are, \textit{macroscopic} parameters are leaving aside an increasing amount of the information that would be necessary to provide a full description of the structure contained in the exact microscopic state of a system, that makes it look like information is being irreversibly lost when the number of undetermined microscopic states which can potentially be occupied is growing with time.

In other words, it is merely because it is impossible to keep track of all the changes which are taking place in the most detailed description of the state of a system that the amount of missing information\index{missing information!irreversible growth} is growing with time in an irreversible way, but no information\index{information!loss} is really lost and no microscopic structure is vanishing from reality in the course of any process that may be characterized as irreversible. When one recognizes that there does exist a minimally coarse-grained\index{minimally coarse-grained state} definition of the state of a macroscopic system associated with what would be a maximum level of knowledge of its microscopic configuration\index{microscopic configuration!maximum knowledge} (regardless of whether or not information about this microscopic state can actually be obtained by a given observer at a specific moment), then one has no choice but to also recognize that it provides a measure of information\index{information!conservation} that is rigorously conserved.

In the next section, I will show how certain usually unrecognized variations in the amount of information required to describe the exact microscopic state of the gravitational field\index{gravitational field!exact microscopic state} are crucially involved in allowing information to be conserved, even when black holes are present and the growth of entropy constitutes a more objective change. But it is already possible to acknowledge that the conclusion that entropy growth does not require the minimally coarse-grained measure of information\index{minimally coarse-grained measure of information} to vary is appropriate from a theoretical viewpoint, because the conservation of information is a requirement of quantum unitarity\index{quantum unitarity} (or of Liouville's theorem\index{Liouville's theorem} in a classical context).

Now, if entropy is indeed increasing in the future, from the viewpoint of an objectively determined choice of coarse-graining\index{coarse-graining}, then it means that entropy was definitely smaller in the past. What is deduced from observations, in fact, is that entropy continuously decreases in the past, in every place we look and as far back in time as we can probe. This is a condition that is far more constraining than simply assuming that the universe is not in a state of thermal equilibrium\index{thermal equilibrium state!universe} at the present time, which would certainly also allow entropy to grow larger in the future.

What we might be justified to expect is that entropy should rise in the past, just as it does in the future, given that it is not already maximum at the present moment. This would appear to be implied by the fact that there is a higher probability that such states be reached as evolution takes place randomly, because there is a much, much larger number of allowed microscopic states\index{microscopic states} compatible with a condition of higher entropy than there are microscopic states compatible with a condition of lower entropy. Only for an isolated system\index{isolated systems}, with a finite number of microscopic degrees of freedom\index{microscopic degrees of freedom!finite number}, would there be a chance that evolution could momentarily take place toward a lower entropy state as a mere statistical possibility. Such entropy\index{entropy!fluctuation} fluctuations would not constitute violations of the second law of thermodynamics\index{second law of thermodynamics!probabilistic nature}, given that this law is probabilistic in nature. Thus, we may consider that the evolution we observe to be taking place in general in the future direction of time is in line with expectations arising from statistical mechanics\index{statistical mechanics}.

The real problem is with the past. Due to the time-symmetric nature of fundamental physical laws\index{fundamental physical laws!time symmetry} it would appear, in effect, that when a macroscopic physical system with many independent microscopic degrees of freedom evolves in the past direction of time, starting from a present non-equilibrium state\index{non-equilibrium state} of relatively low entropy, its entropy should grow (regardless of the details of its microscopic configuration) for the exact same reason that we expect its entropy to grow in the future, when evolution occurs in a random way. But in our universe, entropy was clearly not larger in the past than it now is and the truth is that there is no evidence from astronomical observations that any large-scale, entropy-decreasing phenomenon\index{large-scale entropy-decreasing phenomenon} has ever taken place and no written account of any person having ever observed any significant departure from constant, or continuously increasing entropy at any occasion in our entire history.

Thus, while we can determine the probability of the statistically significant properties of the future state of a system from a knowledge of its current state, the probability of past states\index{probability of past states} cannot in general be appropriately estimated based on that same knowledge. In fact, even if entropy was continuously increasing in the past, from its present non-maximum value, we may still have a problem, because from the forward-in-time\index{forward-in-time!viewpoint} viewpoint the evolution that would have taken place in the past would have occurred with diminishing entropy in the future and this aspect would also be unexplained, unless we are dealing with a \textit{momentary} fluctuation. Thus, it seems that what must be explained is not merely why it is that entropy does not increase in the past, but why it is not already maximal and unchanging in both the past and the future.

It was suggested that the conclusion that entropy should increase in the past may not be valid, because even a macroscopic system with a very large number of independent microscopic degrees of freedom\index{microscopic degrees of freedom!very large number} could perhaps be so carefully prepared that it would be allowed to retrace an unnatural entropy decreasing evolution\index{entropy decreasing evolution!initial state preparation} as it evolves backward in time. Thus, it was argued that it is the details of the present microscopic state of the universe\index{microscopic state of universe!present state} that explains that it evolves toward apparently less probable states in the past. But unsurprisingly, this argument dates back to a time when quantum\index{quantum!chance} chance and classical instability\index{classical instability} had not yet been discovered. In the present theoretical context, however, such an argument simply no longer makes sense, despite the fact that it is often still used to try to justify the kind of evolution that is taking place in the past direction of time. The hypothesis that a reversal of the motion of every particle\index{reversal of particle motions!lower entropy evolution} in an irreversibly evolving system would bring it back to its preceding lower entropy state would actually be true only for a very limited period of time, as short in fact as the system is large and its entropy growth in the future significant\footnote{
The experiments which are sometimes mentioned as having confirmed that a reversal of the motion of all particles in the final state of a macroscopic system are observed to induce anti-thermodynamic evolution\index{anti-thermodynamic evolution!reversal of particle motions|nn} are misleading, because the processes involved take place under carefully controlled conditions, where random perturbations\index{random perturbations!absence|nn} are absent over the totality of the short period during which the phenomena occur and therefore they merely confuse us into believing that the mystery of the continuous diminution of entropy that is taking place in the past direction of time is explainable as being the mere consequence of an improbable configuration of the present microscopic state\index{microscopic state!improbable present configuration|nn}, while this is clearly impossible under more general conditions.}.

It is certainly right that a true reversal of time\index{time reversal!reversal of particle motions and rotations} would actually have to involve more than a simple reversal of the motion and rotation of all components of a system, as I explained in chapter \ref{chap:3}, but even if such a time reversal operation was applied to the whole universe, there is absolutely no reason to believe that, in the absence of any constraint, the past evolution would be likely to evolve toward lower entropy states, because the only violation of symmetry\index{symmetry violation!time reversal} that might occur as a result of such a time reversal would not be such as to allow anti-thermodynamic evolution\index{anti-thermodynamic evolution}.

In any case, even if we were to assume that a system could be so carefully prepared that despite the known sensibility to initial conditions\index{sensibility to initial conditions} which exists even in a classical deterministic context and despite the inherently random nature of quantum processes\index{quantum processes!randomness}, the system would nevertheless follow an evolution so unlikely that its entropy would be continuously decreasing all the way back to the first instants of the Big Bang with absolute precision, we would still be left with having to explain why it is that the present state of the universe\index{universe!unlikely present state} happens to be of such an unlikely nature that it allows this kind of awkward evolution to take place. Clearly, this attempt at explaining the occurrence of the lower entropy states into which the whole universe evolves in this direction of time we call the past cannot be considered satisfactory.

What is problematic, also, with the assumption that the entropy-reducing evolution which we observe to take place in the past direction of time could be the mere outcome of a precise adjustment of the present microscopic state of the universe\index{microscopic state of universe!adjustment of present state} is that, even if we take this as an explanation for the diminishing entropy, we still cannot explain why such an adjustment does not occur for the future instead of the past, because even if that was the case it would simply seem like the past is replaced with the future and the future with the past and we would still not be able to explain why there is, in effect, an asymmetry.

What we should actually expect to observe, if it was a precise adjustment of initial conditions\index{adjustment of initial conditions!time-asymmetric processes} that explained the occurrence of time-asymmetric processes, is a situation where entropy would be continuously decreasing in various regions of the universe whose initial microscopic states\index{initial microscopic states!careful preparation} would have been carefully prepared so as to produce anti-thermodynamic evolution\index{anti-thermodynamic evolution}, but not all in the same direction of time, that is, not all in the past direction for all locations. There is absolutely no reason to expect that such carefully prepared systems would all be set so as to evolve with diminishing entropy in only one particular direction of time, because time itself does not impose such a requirement. But we do not observe multiple, oppositely-directed arrows of time\index{arrows of time!multiple oppositely-directed instances} in our universe and this is precisely what would have to be explained for such an approach to be made valid.

We cannot assume that the reason why entropy-decreasing evolution\index{entropy-decreasing evolution!unlikeliness of final conditions} is not occurring toward the future, from time to time, in certain locations, is that the precise initial conditions required to produce it are too unlikely, while we would also be assuming that the precise `final' conditions required to produce a decrease of entropy in the past are, for their part, allowed to occur, even if they are no less improbable. The rules of probability\index{rules of probability!initial conditions} applied to initial conditions would lead us to predict that entropy should increase in the past, just as it increases in the future, and therefore they cannot alone explain the existence of a thermodynamic arrow of time\index{thermodynamic arrow of time}, even if they do at least explain why it is that entropy does not decrease in the future.

Now, even if we were to recognize that the situation in which multiple coexisting systems\index{coexisting systems!opposite thermodynamic arrows of time} would be set to evolve with decreasing entropy in opposite directions of time would probably be highly unstable, as the precise configuration required to produce a decrease of entropy in a given region would be subject to interference by what happens in another region where entropy would be decreasing in the opposite direction of time, there is no reason to believe that such a mixture of oppositely evolving systems should, through some kind of interference, give rise to a universe with a single well-defined direction of its thermodynamic arrow of time, as required by observations, that is, by our memory of past events. What must be clear is that, if we do not expect to frequently observe such carefully prepared systems evolving with diminishing entropy in the future, then we should not expect to observe the entire universe itself to evolve in such an unnatural way in the past, but this is precisely what is happening all the time, and if that is indeed the case then there must be another explanation to it.

It is only as a consequence of the fact that, for practical reasons, our thought processes are always functioning in the direction of time in which entropy is rising (thereby giving rise to a psychological arrow of time\index{psychological arrow of time}) that we usually fail to recognize that the kind of evolution that takes place in the past direction of time is amazingly abnormal from a purely probabilistic viewpoint. Thus, while it is certainly true that the present state of the universe is relatively unlikely configured, for example in the sense that, if time was reversed, a local tendency for particle trajectories\index{particle trajectories!divergence} to diverge would momentarily turn into one for particle trajectories to converge\index{particle trajectories!convergence}, while a tendency for wave fronts\index{wave fronts!spreading} to spread would turn into one for wave fronts\index{wave fronts!convergence} to converge, this is explainable as merely being a consequence of the fact that the original state in the past that gave rise to the present state was, itself, in a highly unlikely configuration, even from a purely macroscopic viewpoint. It's not the present states which are inexplicably configured, but really the initial state (in the distant past) that gave rise to them.

One of the oldest attempts at solving the problem of the origin of the thermodynamic arrow of time\index{thermodynamic arrow of time!problem of origin}, which is usually recognized as inadequate, was originally proposed by Ludwig Boltzmann\index{Boltzmann, Ludwig}, the originator of the kinetic theory of gases\index{kinetic theory of gases}. It was based on the recognition that there always occur fluctuations to lower entropy states for randomly-evolving, isolated systems\index{isolated systems!entropy decreasing fluctuations} which are in a state of thermal equilibrium\index{thermal equilibrium state}. Over a very long time-scale, it should sometimes happen that those fluctuations would be so significant as to bring even a system in thermal equilibrium into a state with an entropy so low that any subsequent evolution would likely be characterized by a continuous increase of entropy.

Thus, it was proposed that the universe\index{universe!isolated system}, as the ultimate isolated system, really starts in a maximum entropy state, which would presumably be a likely state to be randomly chosen as our initial conditions\index{initial conditions!universe}, and then remains in such a state during most of its existence, but that once in a while, as it evolves in either the past or the future, it simply fluctuates to a much lower entropy state from which it would naturally be expected to evolve with continuously increasing entropy back to its more likely, maximum entropy state in the same arbitrarily determined direction of time, which we would then call the future, regardless of its actual (relative) orientation. The fact that such an evolution would perhaps appear to be similar to that which we presently observe to occur at the level of the universe as a whole then suggests that this is what explains the continuous growth of entropy in one single direction of time that characterizes the evolution of all systems which have not yet reached a state of thermal equilibrium.

It should be clear, however, that in such a context, the only reason we would have to expect to observe the universe in a phase of continuously growing entropy, instead of finding it in one of the much, much more common phases of unchanging maximum entropy would be that this entropy growth is necessary for the presence of an observer which can witness such an evolution. Indeed, the fact that we are allowed to experience a memory of past events and to have a persistent conscious existence is dependent on the condition that there exists a well-defined thermodynamic arrow of time\index{thermodynamic arrow of time!observer existence condition}.

The problem, however, is that if such a requirement was to be satisfied merely as a consequence of the occurrence of an entropy diminishing fluctuation\index{entropy diminishing fluctuation!maximum entropy state} in an otherwise unchanging maximum entropy state, then we should not expect to observe entropy to be so low in all parts of the universe and as far back in time as the epoch of the Big Bang. A much more localized and ephemeral entropy diminishing fluctuation\index{entropy diminishing fluctuation!localized and ephemeral}, that would provide the observer with no records of a pervasive and long-lasting, time-asymmetric history\index{time-asymmetric history!pervasive and long-lasting}, would do just as well for allowing this kind of phenomenon to occur at the present time and given that such a fluctuation would be more likely to occur than a long-lived fluctuation involving the entire universe, then based on this kind of argument what we should experience is a short-lived fluctuation.

The question, therefore, remains: Why is the universe\index{universe!irreversible evolution} evolving irreversibly in one single direction of time in all locations and throughout its entire lifetime? One cannot hope to satisfy the requirement imposed by the time-symmetric nature of fundamental physical laws\index{fundamental physical laws!time symmetry} by simply postulating that the universe actually evolves without any constraint, either in the past or the future, because that would leave the very property of irreversibility unexplained. As Boltzmann\index{Boltzmann, Ludwig} himself appears to have realized, the entropy-fluctuating universe scenario\index{entropy-fluctuating universe scenario} is ineffective for explaining this very constraining aspect of reality and therefore cannot count as a valid solution to the problem of the origin of time asymmetry\index{time asymmetry!problem of origin}.

Now, the fact that I'm suggesting that the random nature of elementary physical processes\index{elementary physical processes!randomness} and the sensibility to initial conditions\index{sensibility to initial conditions} is what allows one to reject the possibility that it could be a precise adjustment of the present microscopic conditions\index{present microscopic conditions!precise adjustment} that would completely explain the diminution of entropy that is observed to take place in the past direction of time does not mean that I'm agreeing with the opinion that time irreversibility\index{time irreversibility!fundamental and irreducible} is occurring at a fundamental and irreducible level in our description of physical processes, as was once proposed by some of those who pioneered the study of chaotic systems\index{chaotic systems}. I do not believe that we must equate unpredictability and randomness with irreducible time asymmetry, even if, in its most general form, statistical mechanics\index{statistical mechanics}, as a probabilistic theory, is dealing with systems in non-equilibrium states\index{non-equilibrium state!irreversible evolution} whose evolution is inherently irreversible.

The fact that quantum field theory\index{quantum field theory!instance of statistical mechanics} can be considered to be a more fundamental instance of statistical mechanics, while it is definitely a time-symmetric theory\index{time-symmetric theory}, clearly indicates that my position is justified\footnote{
I will explain in the latter portion of chapter \ref{chap:5} under which conditions irreversibility can be expected to enter the quantum-mechanical description of elementary particle processes\index{elementary particle processes|nn} and what consequences this must have for observations.}.
 It would certainly not be appropriate to abdicate the requirement of time reversal symmetry\index{time reversal symmetry!requirement} simply to provide an explanation for the observed unidirectionality of thermodynamic processes\index{thermodynamic processes!unidirectionality}, in the context where our most valuable physical theories are all time-symmetric at the most-elementary level of description.

The difficulties that we are experiencing in trying to identify the constraint that would allow to derive irreversible evolution from time-symmetric physical laws\index{time-symmetric physical laws} should not become a justification for abandoning some of the requirements we have very good reasons to believe must constitute part of a fully satisfactory solution. We would not be wise to reject a theoretical framework that works so well, even if it may seem that it cannot explain every aspect of reality, simply to follow an alternative approach which also cannot be made to describe all significant aspects of reality. The challenge consists in actually explaining time irreversibility\index{time irreversibility}, not in decreeing that it is the foundation of reality, when this would require abandoning most of everything else we have learned. I believe that the fact that we have not yet been able to achieve this objective is not an indication that our most fundamental theories are wrong, but merely a proof that we still do not fully understand all the consequences of the physical principles upon which they were built.

It is important to note, in this regard, that it has also been proposed that it is perhaps a fundamental irreversibility of the quantum measurement process\index{quantum measurement process!fundamental irreversibility} that allows to explain the asymmetry of the evolution in time of observable physical phenomena that does not appear to characterize the evolution that takes place in between measurements. But while I do not want to immediately enter into a discussion of how irreversibility intertwine with quantum theory, I must point out that it would be circular reasoning to assume that it is the measurement process that gives rise to thermodynamic irreversibility\index{thermodynamic irreversibility}, while it is already recognized that it is the irreversibility of the processes taking place in the environment with which a quantum system interacts that is involved in giving rise to the decoherence effect\index{decoherence effect} that characterizes all quantum measurements.

But even if we were to follow such a route, it is not clear what would explain that this same unidirectionality does not operate toward the past instead of the future. After all, there is no sign of an intrinsic asymmetry regarding the direction of time in the equations of quantum theory\index{quantum theory!absence of intrinsic asymmetry}. Why would quantum evolution always pick the same one particular direction of time instead of another during those processes that can be qualified as measurements? Once again, even if for pure convenience it was assumed to be the case that quantum theory, or a hypothetical process of actualization of potentialities\index{actualization of potentialities}, was to show preference for one direction of time instead of another, we would still be left with as great a mystery to explain, because time itself does not provide the means for such a distinctive aspect to arise, given that a satisfactory explanation of its own unidirectional character is currently missing.

I do agree that time irreversibility\index{time irreversibility} (just like time itself) is real and constitutes an objective aspect of physical reality and is not just a consequence of some arbitrary choice regarding the level of coarse-graining\index{coarse-graining!level}, but what I will try to demonstrate is that the suggestion that it is no longer appropriate to conceive of reality in terms of elementary particles obeying time-reversible physical laws\index{time-reversible physical laws!elementary particles} is not justified, even when we are dealing with complex systems\index{complex systems!strong non-linearity} which exhibit strong non-linearity or highly irreversible evolution. As we will progress, it will become clear that the idea that there should be no laws more fundamental than those which currently apply only under those particular conditions is excessive in proportion to the very specific nature of that most extraordinary property of physical reality we are trying to explain.

\section{Gravitational entropy\label{sec:4.7}}

Now that I have properly defined and circumscribed the problem of time asymmetry\index{problem of time asymmetry}, I would like to discuss one further problem that arises when one acknowledges the objective nature of the growth of entropy\index{entropy!objective growth} that follows the formation of a black hole or indeed of any overdensity in the matter distribution and that is attributable, in part, to a local increase in the amount of missing information required to describe the microscopic state of the gravitational field\index{microscopic state of gravitational field!missing information}. As a consequence of the progress I have achieved in better understanding the origin of the growth of gravitational entropy\index{gravitational entropy!origin of growth} that takes place under such circumstances, I will be able to provide a definitive solution, not only to the problem of the unaccounted growth of missing information\index{missing information growth!event horizons} associated with the formation of event horizons, but also to the problem of the violation of the conservation of information\index{conservation of information!violation} which appears to take place in the context where the expansion of space is continuously creating new, elementary, quantum-gravitational units of area\index{elementary units of area!expansion of space} in the vacuum.

What is already known, concerning gravitational entropy, is that it grows when the mass of an astronomical object and the strength of its gravitational field are rising. Thus, when gravitational attraction is involved, the natural tendency for matter to spontaneously disperse into a larger volume of space is overcome and the decreasing entropy of matter\index{entropy of matter!decrease} that follows is overcompensated by the growth of missing gravitational field information\index{missing gravitational field information!growth}. In fact, we currently have no exact definition for the entropy attributable to the gravitational field in a general context and it is merely a knowledge of the exact formula for black hole entropy that allows one to estimate the magnitude of this entropy in the absence of event horizons.

In any case, the prevailing character of gravitational entropy\index{gravitational entropy} means that, when a large enough amount of matter is present in a given volume of space, particles with the same sign of energy are allowed to become more densely packed, because such an evolution is favored from a thermodynamic viewpoint, in the context where there are more possible microscopic configurations of the gravitational field\index{microscopic configuration of gravitational field} compatible with a state of higher matter energy density. Only the expansion of space could perhaps allow this natural tendency to be surmounted, if it was to become rapid enough that it would allow the growth of matter entropy\index{matter entropy growth!cosmic expansion} to overcompensate the growth of gravitational entropy\index{gravitational entropy growth!formation of inhomogeneities} that occurs as a result of the formation of inhomogeneities.

In section \ref{sec:3.10} I have explained that black holes\index{black hole!macroscopic physical parameters} provide us with a unique set of macroscopic physical parameters, which allow a natural definition of coarse-graining\index{coarse-graining!natural definition} and therefore an objective measure of entropy growth\index{entropy growth!objective measure}. This is due to the fact that the only information that can be obtained by an observer outside a black hole about the state of matter inside the object is provided by its four macroscopic parameters of total energy, angular momentum, and positive and negative bidirectional charges\index{bidirectional charge}. I have also explained that the first law of black hole mechanics\index{black hole mechanics!first law} allows one to expect that the amount of information which is missing about the microscopic state of matter\index{microscopic state of matter!missing information} as a result of those limitations is equal to the area of the event horizon of the black hole\index{black hole event horizon!area} in elementary units of area\index{elementary units of area} equal to four Planck areas\index{Planck unit of area}.

Thus, an exact quantitative measure of missing information is associated with the area of a black hole event horizon and given that the choice of this area as a macroscopic coarse-grained parameter\index{macroscopic coarse-grained parameter!unique choice} is unavoidable, as it is physically impossible for an observer outside a black hole to obtain a more detailed description of the microscopic state\index{microscopic state of matter and interaction fields} of matter and its interaction fields than is provided by a knowledge of its four macroscopic physical parameters, then it becomes possible to conclude that an objective definition of entropy\index{entropy!objective definition} exists under such circumstances, which is provided by the value of this area, which itself depends only on the value of those macroscopic physical attributes.

Therefore, any change to entropy which is reflected in a variation of the surface area of a black hole\index{black hole!variation of surface area} constitutes a non-subjective change which cannot be attributed merely to a particular choice of macroscopic physical parameters\index{macroscopic physical parameters!particular choice}, as under such conditions it is simply impossible to provide a more detailed description of the microscopic state of matter and its gravitational field\index{gravitational field!microscopic state}, that would allow to reduce entropy to a smaller value. Something essential, therefore, differentiates the measure of information associated with the area of the event horizon of a black hole\index{black hole event horizon!area} from that which is associated with an ordinary surface that is not an event horizon, for which only the Bekenstein bound\index{Bekenstein bound} must apply, and this distinction has to do with the availability of information.

It remains, though, that the information loss\index{information!loss} which appears to be taking place when a black hole absorbs low-entropy matter cannot be considered real, because, as I explained in section \ref{sec:3.10}, it seems that the information about the microscopic state of the matter that was submitted to gravitational collapse is encoded in binary form in the microscopic, quantum-gravitational degrees of freedom\index{quantum-gravitational degrees of freedom!black hole event horizon} on the event horizon of the object and is released when the black hole\index{black hole!thermal radiation} decays through the emission of thermal radiation.

Once it is recognized that no information needs to vanish from reality, even when a unique definition of coarse-graining\index{coarse-graining!unique definition} exists that gives rise to a non-subjective measure of entropy, what must be understood is that, even though no information is lost as a result of the absorption of matter by a black hole, the existence of a measure of missing information\index{missing information!area of black hole event horizon} proportional to the area of the event horizon of such an object implies that there is a real growth in the amount of information\index{information!growth} that would be required to completely specify the state of all the microscopic, binary degrees of freedom on the surface of a black hole of increasing mass, which reflects the existence of a growing amount of microscopic structure in the gravitational field\index{gravitational field!microscopic structure}.

It should be clear, in effect, that the amount of missing information which would be required to specify the exact state of all the \textit{matter} particles which were captured by the gravitational field of a macroscopic black hole\index{macroscopic black hole!entropy growth} is not large enough to account for its entropy growth, because the entropy of a black hole is proportional to the area of its event horizon in elementary units of area\index{elementary units of area} (equal to four Planck units of area\index{Planck unit of area}) and for an ordinary Schwarzschild black hole\index{Schwarzschild black hole} (with zero charge and no angular momentum) this variable grows in proportion to the square of the mass of the matter absorbed by the object (in Planck units of energy\index{Planck unit of energy}), while what I have shown in section \ref{sec:3.10} is that the information which is necessary to completely describe the state of one Planck-energy\index{Planck-energy particle} particle inside the event horizon of such a black hole can always be encoded in four elementary units of area, regardless of the mass of the object.

Thus, it appears that the amount of information which would be required to completely specify, with the highest possible precision, the microscopic state of matter\index{microscopic state of matter!black hole} is actually decreasing when matter is captured by the gravitational field of a black hole. But if this measure of information is not growing when the mass of a black hole is rising, while the total amount of missing information (the entropy) is growing faster than the mass of the object (which rises in proportion to its matter content), then one has no choice but to recognize that the amount of missing information which would be required to describe the microscopic state of the gravitational field\index{microscopic state of gravitational field!missing information} is indeed rising when local gravitational fields grow stronger, at least in the presence of a macroscopic event horizon\index{macroscopic event horizon}.

It should be clear, therefore, that when matter particles assemble into a black hole, the number of microscopic degrees of freedom\index{microscopic degrees of freedom!gravitational field} of the gravitational field grows larger locally, even while the number of microscopic degrees of freedom\index{microscopic degrees of freedom!matter} of matter is being reduced as a result of the constraints exerted on the states of motion of elementary particles by the gravitational field of the object. Thus, while some information about the exact microscopic state of the matter\index{microscopic state of matter!missing information} that fell into a black hole is missing from a description of the object in terms of its observable macroscopic physical parameters\index{macroscopic physical parameters!black hole}, an even larger amount of information, concerning the exact microscopic state of the gravitational field\index{microscopic state of gravitational field!missing information}, goes missing as well, while the mass of the black hole is growing, which also contributes to increase its entropy. As a result, the amount of missing information appears to be growing faster than would be allowed if information\index{information!conservation} was conserved, that is to say, it grows faster than the progression of our ignorance concerning the intricate details of the microscopic state of matter.

Now, given that I will later argue that the growth of inhomogeneities in the matter distribution, which is the source of stronger gravitational fields, provides the dominant contribution to entropy increase\index{entropy increase!dominant contribution} in our universe (because the entropy of matter\index{entropy of matter!expansion of space} itself does not change much as a consequence of the expansion of space), then it would appear that temporal irreversibility\index{temporal irreversibility!gravitational entropy growth} arises mostly as a consequence of the growth of gravitational entropy. What is crucial to understand, under such conditions, is that the irreversible character of this evolution as well as the growth in the amount of missing information which is giving rise to it cannot be considered subjective features of reality, precisely because, under appropriate conditions, they can be associated with the presence of black hole event horizons\index{black hole event horizon!natural boundaries} which constitute natural boundaries, enabling a unique definition of coarse-graining\index{coarse-graining!unique definition} that is entirely determined by the strength of local gravitational fields.

It must, in effect, be recognized that a growing amount of information is required to describe in complete detail the microscopic structure that emerges in the gravitational field\index{gravitational field!microscopic structure} when overdensities develop, either in the baryonic matter\index{baryonic matter!overdensities} distribution, or in the distribution of vacuum dark matter\index{vacuum dark matter!overdensities} (that which is attributable to local variations of vacuum energy density\index{vacuum energy density!local variations}). I believe that it is merely because we do not benefit from the guidance of a fully developed quantum theory of gravitation\index{quantum gravitation theories} that we haven't yet realized that the amount of missing information\index{missing information!growth} is actually growing faster than would appear to be allowed, when a gravitational field gains in strength as a consequence of a local increase in the magnitude of the energy density of matter (we often hear about people claiming that information\index{information!loss} may be lost when matter falls into a black hole, but to my knowledge no one has ever suggested that there may be something wrong with the growth of missing gravitational field information\index{missing gravitational field information!growth}).

The problem that emerges under such conditions is that the growth in the amount of information necessary to describe the microscopic state of the gravitational field\index{microscopic state of gravitational field!information growth}, which can be expected to occur when stronger gravitational fields develop as a consequence of a local growth in the density of matter, would appear to violate the constraint of conservation of information\index{conservation of information!violation} that is imposed by quantum theory. It must be the case, therefore, that the additional measure of gravitational field information, which appears to be produced when a black hole forms, already existed before it contributed to the gravitational entropy\index{gravitational entropy!black hole} associated with such an object, just like the measure of information contributed by matter itself, so that the total amount of information required to describe the exact microscopic state of our universe\index{microscopic state of universe!information invariance} does not really change as a result of those transformations.

Given that what allows the additional growth in the amount of missing gravitational field information\index{missing gravitational field information!growth} that is taking place as a consequence of a local increase in the density of matter to be objectively characterized is merely the fact that it occurs as a result of adopting the natural definition of coarse-graining\index{coarse-graining!natural definition} that is associated with the presence of macroscopic event horizons\index{macroscopic event horizon}, then it follows that this growth of missing information can only be compensated by a change in the amount of information which is itself independent from any arbitrary choices regarding the coarse-graining and therefore we can already expect that this compensation arises from additional changes in the strength of gravitational fields attributable to local variations in the density of matter.

In fact, what allows me to conclude that the amount of missing information is growing faster than would appear possible, when a stronger gravitational field develops as a result of the formation of a matter overdensity, are not merely my findings concerning the nature of the microscopic degrees of freedom\index{microscopic degrees of freedom!matter in black hole} of matter trapped by the gravitational field of a black hole, but the very fact that it also appears necessary to assume that a diminution of information\index{information!decreasing gravitational field strength} occurs when the strength of a gravitational field decreases, as a result of the formation of an \textit{underdensity} in the uniform, large-scale matter distribution; which suggests that it is only as a consequence of the fact that there arises a compensation between those two variations that the measure of gravitational field information\index{gravitational field information!global invariance} can be left invariant on a global scale, regardless of how much it varies locally.

What I'm suggesting, more exactly, is that, given that, from the viewpoint of a positive-energy observer, a larger amount of information is required to completely describe the detailed microscopic state of the gravitational field\index{microscopic state of gravitational field!information growth} when it grows stronger as a result of a local increase in the density of positive-energy matter, then it should necessarily be the case that a correspondingly smaller amount of information would be required in order to completely describe, with the same level of precision, the microscopic state of the gravitational field\index{microscopic state of gravitational field!information decrease} associated with a lower than average density of positive-energy matter, at least from the viewpoint of a positive-energy observer who measures a weaker attractive gravitational field as a result of the presence of such a void in the uniform positive-energy matter distribution\index{void in positive-energy matter distribution!weaker gravitational field}.

You may recall that, in section \ref{sec:2.7}, I mentioned that, from the viewpoint of a positive-energy observer, the presence of a void in the cosmic distribution of positive-energy matter results in a local \textit{absence} of gravitational attraction which can give rise to a local acceleration of the rate of expansion\index{expansion rate!local acceleration} of space. But if this local decrease in the density of matter actually produces a weaker attractive gravitational field, then it would appear appropriate to assume that, from the viewpoint of positive-energy observers at least, a lesser amount of information would be required to describe the microscopic state of the gravitational field as a result of the presence of such a void in the matter distribution.

In the context where it would be necessary to assume that the matter distribution was very uniform in the first instants of the Big Bang (as I will suggest in section \ref{sec:4.9}), the validity of this hypothesis would imply that, from the viewpoint of positive-energy observers at least, any additional increase in the amount of information necessary to describe the unobserved microscopic state of the gravitational field\index{microscopic state of gravitational field!information growth} attributable to the formation of a local positive-energy matter overdensity would be compensated by an exactly corresponding decrease of gravitational field information\index{gravitational field information!compensating decrease} attributable to the formation of the positive-energy matter underdensity that must develop in the surroundings of this overdensity in order to allow it to form.

Actually, even from the viewpoint of a negative-energy observer who measures a stronger attractive gravitational field\index{gravitational field!void in matter distribution} in the presence of such a void in the uniform positive-energy matter distribution, a similar conclusion would follow, because as I will explain below, despite the growing strength of the gravitational field which is produced as a result of the development of such an astronomical structure, the amount of information contained in the \textit{microscopic} state of its gravitational field\index{microscopic state of gravitational field!information decrease} must still decrease in proportion with the amount of positive-energy matter that is missing in it and the same conclusion would apply in the case of a void in the uniform negative-energy matter distribution\index{void in matter distribution!information decrease}, from the viewpoint of a positive-energy observer, despite the fact that such a structure would produce an attractive gravitational field similar to that of a positive-energy object, from this particular viewpoint.

As a result, it becomes possible to deduce that, despite the fact that real changes take place locally in the measure of gravitational field information\index{gravitational field information}, when the matter distribution is growing more inhomogeneous, information\index{information!conservation} is always rigorously conserved, even when this evolution involves an alteration of the macroscopic parameters associated with black hole event horizons\index{black hole event horizon!macroscopic parameters}.

To summarize, it appears necessary to assume that when the density of matter grows larger than its average value locally, there is an increase in the amount of missing information that is not attributable merely to our growing ignorance concerning the exact microscopic state of matter\index{microscopic state of matter!growing ignorance}, but which is due in part to an actual increase in the amount of information necessary to describe the microscopic state of the gravitational field\index{microscopic state of gravitational field!information growth}\footnote{
In fact, it appears that even when a non-gravitational force field\index{non-gravitational force field!microscopic state information|nn} grows in strength, additional information may be required to describe the exact microscopic state of this field of interaction and under such conditions a compensating change would need to take place in the environment. This is perhaps a desirable outcome given that even the fields associated with non-gravitational interactions\index{non-gravitational interaction field!entropy|nn} can be expected to carry their own specific measures of entropy, as I explained in section \ref{sec:3.10}.}.
 However, when the density of positive-energy matter becomes smaller than its average value locally, there occurs a corresponding \textit{decrease} in the amount of information necessary to describe the microscopic state of the gravitational field\index{microscopic state of gravitational field!information decrease}, at least from the viewpoint of a positive-energy observer. For such an observer, it is the decrease of gravitational field information, attributable to the formation of this underdensity in the macroscopically uniform distribution of positive-energy matter, that compensates the additional unaccounted increase in the amount of missing gravitational field information\index{missing gravitational field information!growth} which is attributable to the formation of the corresponding positive-energy matter overdensity and which would otherwise give rise to a violation of the condition of conservation of information\index{conservation of information!violation}.

It is crucial to understand, however, that the decrease of gravitational field information that occurs, from the viewpoint of a positive-energy observer, when the density of positive-energy matter is being reduced below its average value locally does not translate in an overall reduction of gravitational entropy\index{gravitational entropy!overall reduction}. If the density of positive-energy matter is only allowed to decrease in a given region of space when a compensating increase takes place in its vicinity, it follows that the diminution of information\index{information diminution!formation of matter underdensity} that occurs as a result of the formation of such an underdensity in the matter distribution only serves to increase the amount of missing information which would be necessary to describe the unobserved microscopic state of the gravitational field\index{microscopic state of gravitational field!missing information} associated with the corresponding matter overdensity.

Given that, when the density of matter is growing locally, the amount of information that goes missing about the microscopic state of the gravitational field must itself grow, then it is necessary to conclude that a certain amount of information which was available prior to the formation of such an inhomogeneity, would now be missing and would merely contribute to increase the objective measure of gravitational entropy\index{gravitational entropy growth!formation of matter overdensity} that is attributable to the formation of the overdensity whose mass grew at the expense of the formation of the underdense structure. Therefore, even if the amount of information necessary to describe the exact microscopic state of the gravitational field\index{microscopic state of gravitational field!information decrease} is allowed to diminish in a forming matter underdensity, it can be expected that gravitational entropy will nevertheless be growing overall, as any measure of entropy does under ordinary conditions when information becomes unavailable, despite the fact that it does not vanish from reality.

What allows gravitational entropy\index{gravitational entropy!global rise} to rise globally at the expense of a local decrease in the measure of information contained in the same force field is the fact that those changes in the density of matter which produce a growth of gravitational entropy are always the outcome of gravitational attraction, even when the gravitational fields attributable to the forming overdensities are experienced as repulsive, as must be the case for observers with an energy sign opposite that of the matter whose density is rising locally. As a consequence, the changes involved have the character of irreversibility and are always likely to take place when gravitation\index{gravitation!predominance} is predominant, which is certainly appropriate, given that they are actually favored from a thermodynamic viewpoint.

Now, there is no question that if a void in the negative-energy matter distribution was deep enough over a sufficiently large region of space, then, from the viewpoint of a positive-energy observer, it could give rise to an attractive gravitational field so strong that it would produce an event horizon\index{event horizon!void in negative-energy matter distribution} similar to that of a positive-energy black hole. But while one may be tempted to assume that the thermodynamic properties of such a void would be identical to those of a positive-energy black hole\index{positive-energy black hole!positive temperature} and that the structure would have a positive temperature and would radiate positive-energy particles, that would be an incorrect conclusion.

Once it is understood that the thermal radiation\index{thermal radiation!black hole} emitted by a conventional black hole arises as a consequence of the thermodynamic requirement that local energy differences be smoothed out, even in the presence of event horizons, then it becomes clear that a positive-energy black hole must lose positive energy if its mass is to decrease in the process. Given that negative-energy matter cannot cross the event horizon of a positive-energy black hole\index{positive-energy black hole!loss of positive energy} and remain inside such an object, it follows that this loss of positive energy can only occur through the emission of positive-energy particles outside the event horizon. A positive-energy black hole, therefore, \textit{releases} positive heat\index{positive heat!emission} in its environment and in the process necessarily reduce its surface area and its entropy, which therefore requires the temperature of the object to be positive.

But it is precisely here that the distinction between a positive-energy black hole and a sufficiently large void in the negative-energy matter distribution would arise, from the viewpoint of a positive-energy observer, because the tendency to reach thermal equilibrium\index{thermal equilibrium} would not produce the same outcome in the case of the void in a negative-energy matter distribution, despite the similarity of the gravitational fields associated with both kinds of configuration. Indeed, while the gravitational field produced by a sufficiently large void in the negative-energy matter distribution must be equivalent to that of a positive-energy black hole from the viewpoint of a positive-energy observer outside the structure, in the case of the void, the uniformity of the distribution of energy cannot be re-established through the emission of positive-energy particles by the void, as there is no way such a radiative process could allow the void to regain the negative energy it lost as it was growing, even if positive-energy particles were present inside the structure and could surmount the growing gravitational attraction exerted on them as they would stray from the center of mass of the void.

What would happen, therefore, is that thermodynamic equilibrium would be reached through the \textit{absorption} of negative-energy particles from the surrounding higher-density matter distribution, a phenomenon which is not forbidden, as it would for a positive-energy black hole, because a void in the negative-energy matter distribution does not exert any gravitational force on negative-energy particles and only the gravitational attraction of the surrounding negative-energy matter overdensities can be expected to prevent negative-energy particles from `escaping' into the inner, low-density region, when the gravitational forces attributable to those overdensities do not cancel out locally, despite the spherical symmetry of the void itself.

Thus, even if the surface gravitational field\index{surface gravitational field!void in matter distribution} of a sufficiently large void in the negative-energy matter distribution must be identical to that of a positive-energy black hole from the viewpoint of a positive-energy observer outside the void, as long as the contribution of the positive vacuum-dark-matter energy\index{positive vacuum-dark-matter energy}, which is present inside the structure, can be ignored, one can expect that the void would not reach equilibrium through the emission of positive heat\index{positive heat!emission}, but rather through the absorption of negative heat\index{negative heat!absorption}. But the outcome of this process would not be a decrease in the gravitational entropy\index{gravitational entropy!decrease} of the structure itself, but rather a decrease in the gravitational entropy of the surrounding negative-energy matter distribution which actually requires the temperature\index{temperature!negative} of the surrounding negative-energy matter overdensities to be negative, in accordance with the fact that the thermal radiation\index{thermal radiation!negative energy} which is released \textit{inside} the structure has negative energy.

This conclusion is inescapable, because from the viewpoint of negative-energy observers, such a void in the uniform distribution of negative-energy matter would not produce a repulsive gravitational field\index{repulsive gravitational field}, but merely an absence of attractive gravitational field and while this means that the void itself cannot have a temperature, nothing would prevent the surrounding negative-energy matter overdensities themselves from having a \textit{negative} temperature. Thus, even from the viewpoint of negative-energy observers, it can be expected that the surrounding negative-energy matter distribution would release negative heat\index{negative heat!emission} into the void, which would allow the gravitational entropy of this surrounding matter distribution to be reduced in the process, as matter would become more homogeneously distributed and the void would be replenished with negative-energy matter, thereby allowing the rise in matter entropy\index{matter entropy!increase} to overcompensate the loss of gravitational entropy.

Therefore, while a positive-energy black hole\index{positive-energy black hole!positive temperature} would have a positive temperature, due to the fact that a reduction of its positive energy, through the emission of positive-energy radiation\index{positive-energy radiation!emission}, would give rise to a reduction of its surface area and its gravitational entropy\index{gravitational entropy!decrease} (at the expense of a larger growth of matter entropy in its environment), from the viewpoint of a positive-energy observer the event horizon of a sufficiently large void in the negative-energy matter distribution\index{void in negative-energy matter distribution!event horizon} would have a negative temperature\index{negative temperature!void in negative-energy matter distribution} that would only be apparent inside the structure. This is a consequence of the fact that it is actually the gravitational entropy of the surrounding negative-energy matter distribution that would be reduced as the void decays through the absorption of negative-energy particles and not really that of the void itself, which is already minimal.

From the viewpoint of positive-energy observers, it is the time dilation\index{time dilation!void in matter distribution} produced by the gravitational field of such a void in the negative-energy matter distribution that would require negative-energy particles to be released inward, through the event horizon of the structure, as thermal radiation\index{thermal radiation!void in matter distribution} (just as is the case for the negative-energy particles which are released outward through the event horizon of a negative-energy black hole). Again, however, it must be clear that the formation of the void itself does not give rise to a growth of missing gravitational field information\index{missing gravitational field information!growth} or entropy, despite the fact that the structure does produce an attractive gravitational field.

Even if the surface gravitational field\index{surface gravitational field!void in matter distribution} of a sufficiently large void in the negative-energy matter distribution may actually be identical to that of a positive-energy black hole\index{positive-energy black hole}, one must conclude that the thermodynamic properties of those two kinds of matter inhomogeneities are not the same, because, even from the viewpoint of a positive-energy observer, such a void would not have a positive temperature and would not radiate any positive-energy particles outward and this is merely a reflection of the fact that the amount of information required to describe the microscopic state of the gravitational field\index{microscopic state of gravitational field!information decrease} must diminish, locally, from the viewpoint of a positive-energy observer, when a void forms in the uniform negative-energy matter distribution, just like it does from the viewpoint of a negative-energy observer.

Yet, it would appear that the requirement that, even from the viewpoint of a negative-energy observer, the measure of \textit{missing} gravitational information, or entropy\index{gravitational entropy!global rise}, should be rising, globally, when such a void forms in the negative-energy matter distribution, is not incompatible with this conclusion, because when the density of matter decreases in a given region of space it must increase in its surroundings and a local increase in matter density always produces an increase of gravitational entropy\index{gravitational entropy!local growth} that compensates the decrease of gravitational information\index{gravitational information!local decrease} that occurs as a result the diminution of matter density in the surrounding environment, which means that more information goes missing about the microscopic state of the gravitational field as a result of those changes.

It is quite unavoidable that, even from the viewpoint of positive-energy observers, a void in the uniform negative-energy matter distribution\index{void in negative-energy matter distribution!event horizon} that is sufficiently large to produce an event horizon, would reach thermal equilibrium\index{thermal equilibrium} through the absorption of negative-energy radiation, instead of through the emission of positive-energy radiation, because the emission and the absorption of thermal radiation\index{thermal radiation emission and absorption!objective fact} is an objective fact that cannot depend on the energy sign of an observer (unlike the strength or the polarity of a gravitational field or the curvature of spacetime\index{curvature of spacetime!observer dependence}). From the viewpoint of the interpretation proposed here this objectivity is preserved, because even a negative-energy observer would observe an inward flow of negative-energy radiation in the presence of such a void in the negative-energy matter distribution\index{void in matter distribution!inward flow of radiation}. Only, from the viewpoint of such an observer, this thermal radiation would not appear to be produced by the gravitational field of the void itself, but rather by those of the negative-energy matter overdensities present in its environment.

But if the assumption that the amount of information required to describe the microscopic state of the gravitational field\index{microscopic state of gravitational field!information decrease} decreases locally, from the viewpoint of a positive-energy observer, upon the formation of an underdensity in a homogeneous negative-energy matter distribution is to be considered valid, it must be further justified from a more elementary perspective. I will now explain what justifies this conclusion. What must be clear, once again, is that, despite the apparent similarity between the gravitational field experienced by a positive-energy observer in the presence of a void in the \textit{negative-energy} matter distribution and that which is experienced by the same observer in the presence of positive-energy matter, there nevertheless exists a fundamental difference between those two categories of objects, which arises from the fact that positive-energy matter does not consist of voids in a negative-energy matter distribution, but is rather equivalent to voids in the negative-energy portion of the \textit{vacuum}, as was emphasized in section \ref{sec:2.7}.

I believe that what explains that the formation of a void in the uniform negative-energy matter distribution would give rise to a negative change in the amount of information required to describe the microscopic state of the gravitational field, from the viewpoint of a positive-energy observer, while more information would be required to describe the microscopic state of the gravitational field as a consequence of the presence of a void of similar magnitude in the negative-energy portion of the vacuum\index{void in negative vacuum energy!gravitational information} (which can be liken to the presence of positive-energy matter), is the fact that in the absence of matter (including any vacuum dark matter\index{vacuum dark matter} attributable to local variations of vacuum energy density\index{vacuum energy density!local variations}), the distribution of vacuum energy\index{vacuum energy distribution!real uniformity} is really uniform on all scales, while the macroscopically homogeneous distribution of matter\index{homogeneous matter distribution!microscopic inhomogeneity} (baryonic or dark) in which a void may be produced is not really uniform on a microscopic scale.

In the absence of any matter there are no persistent density variations in the distribution of vacuum energy\index{vacuum energy distribution!persistent density variations}, such as those which would be associated with the presence of real (by opposition to virtual) particles\index{real particles}, and removing energy from such a perfectly uniform distribution cannot be assumed to reduce the amount of structure that would have been present initially, on a microscopic scale, in the gravitational field which is attributable to the presence of this energy. This is unlike the situation we have when we are dealing with what would normally be considered a homogeneous matter distribution\index{homogeneous matter distribution!small-scale density variations}, in which there actually exist persistent, smaller scale variations in the density of matter energy, which generate microscopic structure in the gravitational field itself, which may not be apparent from a macroscopic viewpoint, but which can be as strong as the average density of matter energy is high.

Thus, when the density of matter\index{matter density!local reduction} is reduced, locally, in a \textit{macroscopically} uniform, negative-energy matter distribution, the amount of microscopic structure contained in the gravitational fields\index{gravitational field!microscopic structure} which are present in this matter distribution, as a result of its own small-scale inhomogeneity, must decrease and this can only mean that, in such a case, even from the viewpoint of a positive-energy observer who measures a stronger, attractive gravitational field as a result of those changes, less information is required to describe the exact microscopic configuration of the gravitational field\index{microscopic configuration of gravitational field!information decrease} or the curvature of spacetime\index{curvature of spacetime!microscopic configuration}.

By contrast, when we increase the density of positive-energy matter\index{positive-energy matter density!local increase} in a local region of space (which amounts to add more voids in the homogeneous, negative portion of vacuum energy\index{void in homogeneous negative vacuum energy}), we produce microscopic gravitational fields\index{microscopic gravitational fields} which were not present beforehand in this region and it is only appropriate that, in such a case, the amount of information required to determine the microscopic structure of the gravitational field\index{microscopic structure of gravitational field!information growth} should be growing locally, from the viewpoint of any observer. This is all a consequence of the fact that the presence of more matter energy allows particles to exert stronger attractive gravitational forces on each other, so that removing negative energy from a local portion of the \textit{vacuum} normally has for consequence to generate additional microscopic variations of the gravitational field\index{microscopic variations of gravitational field}, given that it is equivalent to increasing, either the number of positive-energy matter particles, or the magnitude of their energies.

What must be understood is that, even though a uniform distribution of negative-energy matter\index{negative-energy matter!uniform distribution} contributes nothing to the gravitational field which is experienced by positive-energy observers on a large scale, it cannot be assumed that no contribution to the amount of information necessary to describe the microscopic state of the gravitational field\index{microscopic state of gravitational field!information} arises from the macroscopically uniform portion of the negative-energy matter distribution, because smaller-scale inhomogeneities in this matter distribution do exert gravitational forces on positive-energy matter, despite the fact that those forces cancel out on a large scale.

Thus, even though, for a positive-energy observer, a local absence of negative-energy matter is gravitationally equivalent to the presence of positive-energy matter, as I explained in section \ref{sec:2.7}, it is nevertheless only in the presence of more negative-energy matter that the amount of information contained in the gravitational field\index{gravitational field!information increase} increases, because what a local absence of negative-energy matter means is that the positive portion of vacuum energy\index{vacuum energy!more homogeneous distribution} is more homogeneously distributed locally, while more uniformity in the distribution of vacuum energy means less microscopic structure and therefore less information in the gravitational field\index{gravitational field!information decrease}, even if this gravitational field would appear to be stronger, on a macroscopic scale, from the viewpoint of positive-energy observers, when the density of negative-energy matter decreases below its average value in a given region of space, because under such conditions fewer matter particles with variable momentum states are present to impart microscopic structure on the gravitational field\index{gravitational field!microscopic structure}.

Unlike a local reduction in the density of negative vacuum energy\index{negative vacuum energy!local reduction} that could be liken to the presence of positive-energy matter, a local reduction in the average density of negative-energy matter would, therefore, give rise to a diminution in the amount of gravitational field information, even from the viewpoint of a positive-energy observer, and this is reflected in the fact that a void in the negative-energy matter distribution\index{void in negative-energy matter distribution!event horizon} that is sufficiently large to produce an event horizon does not have the same thermodynamic properties as a positive-energy black hole, despite the similarity of the gravitational fields produced by the presence of both kinds of astronomical structures from the viewpoint of an external, positive-energy observer. What this really means, is that, from a \textit{microscopic} viewpoint, the gravitational field experienced by a positive-energy observer, in the presence of a void in the uniform negative-energy matter distribution, is equivalent to that which is produced by the uniform portion of vacuum energy\index{uniform portion of vacuum energy!absence of microscopic structure} associated with the cosmological constant\index{cosmological constant}, which, as I will explain below, must be assumed to lack any microscopic structure.

In such a context, it becomes possible to actually explain, not only why it is that the amount of information which would be necessary to describe the microscopic state of the gravitational field\index{microscopic state of gravitational field!information decrease} must diminish, from the viewpoint of any observer, when a void forms in the uniform, large-scale positive- or negative-energy matter distribution, but also why it is that the amount of missing information about the microscopic state of the gravitational field\index{microscopic state of gravitational field!missing information} is actually growing, from the viewpoint of any observer, when the density of positive or negative matter energy is increasing locally, as one would normally expect, based on the results of the semi-classical theory of black hole thermodynamics\index{black hole thermodynamics!semi-classical theory}.

This argument, concerning the distinction between local voids in the negative portion of vacuum energy\index{void in negative vacuum energy} and local voids in the macroscopically uniform distribution of negative-energy matter\index{void in uniform negative-energy matter distribution}, would also justify assuming that, even the gravitational field attributable to an apparently uniform distribution of positive-energy matter\index{uniform matter distribution!gravitational field information} would contribute a certain measure of information, from the viewpoint of a positive-energy observer, despite the fact that it is usually assumed that only the gravitational fields associated with the presence of macroscopic inhomogeneities in the distribution of positive-energy matter\index{inhomogeneities in matter distribution!gravitational field information} contain information.

It would seem, in effect, that if locally reducing the density of positive-energy matter produces a decrease of information in the gravitational field, from the viewpoint of both positive- and negative-energy observers, then, even on a cosmic scale, a certain amount of information should be contained in the gravitational field\index{gravitational field!information} produced by this uniformly distributed matter, which would be reduced as a result of the expansion of space\index{expansion of space!reduction of gravitational information}. This reduction would occur because the global decrease in the average density of positive-energy matter\index{average density of positive-energy matter!global decrease} particles and the consequent diminution of their average kinetic energy\index{average kinetic energy!positive-energy matter particles}, that must take place as space expands, would reduce the magnitude of microscopic inhomogeneities in the distribution of positive matter energy (independently from any local density variation that would take place as a result of gravitational instability\index{gravitational instability}), which means that the strength of the gravitational fields\index{gravitational field strength!microscopic scale} which are present on a microscopic scale would also be reduced and would, therefore, contain a smaller amount of information.

The situation, here, is similar to that which arises in the presence of macroscopic gravitational fields\index{macroscopic gravitational field}, only, in the present case we are dealing with additional degrees of freedom which are normally left out of a classical description of the gravitational field\index{gravitational field!classical description} attributable to a uniform matter distribution. In fact, the same condition of conservation of information\index{conservation of information} which imposes a compensation between the various local variations in the measure of gravitational field information\index{gravitational field information!compensation of local variations} attributable to the formation of macroscopic matter inhomogeneities, actually requires that a certain variation in the measure of information\index{information!variation with average matter density} associated with the microscopic structure of gravitational fields\index{gravitational field!microscopic structure} present in a macroscopically homogeneous matter distribution be allowed to take place as the average density of matter energy decreases following expansion.

It had already been proposed, in effect, that the expansion of space\index{expansion of space!information increase} should perhaps be considered to produce an increase of information on the quantum-gravitational scale\index{quantum-gravitational scale}, given that it would appear to continuously create additional elementary units of area\index{elementary units of area!creation} in the vacuum, in apparent violation of the theoretical requirement regarding the conservation of information\index{conservation of information!violation}. I believe that this suggestion is valid, because, according to the developments introduced in section \ref{sec:3.10}, it appears that a larger volume of space would imply the existence of a larger number of elementary black holes\index{elementary black hole} in the vacuum and a determination of the state of the microscopic (quantum-gravitational) degree of freedom\index{quantum-gravitational degrees of freedom!elementary black hole} which is encoded on the surface of each of those ephemeral objects (whose existence is attributable to quantum fluctuations) would require additional binary units of information.

Thus, if information must always be conserved\index{conservation of information}, it would appear necessary to assume that, as expansion takes place and the average density of matter decreases, a reduction in gravitational field information\index{gravitational field information!reduction with expansion} should take place that would be attributable to this reduction in the average matter density\footnote{
In fact, the expansion of space would also produce an increase in the number of microscopic degrees of freedom\index{microscopic degrees of freedom!handedness of particles|nn} associated with the handedness of elementary particles and the direction of propagation in time of positive and negative bidirectional charges\index{bidirectional charge!direction of propagation in time|nn}, but those changes would be compensated, independently, by a diminution in the number of microscopic degrees of freedom contained in the inertial gravitational field\index{inertial gravitational field!angular momentum|nn} associated with angular momentum and in the two components of the interaction field associated with the unified non-gravitational charge\index{unified non-gravitational charge|nn}, that would result from the decreasing strength of those interaction fields.}.
 For this to be a valid proposal, however, it must be recognized that a variation of the average density of vacuum energy\index{vacuum energy!variation of average density} that is normally associated with the cosmological constant\index{cosmological constant!variation} does not contribute to alter the total amount of information contained in the microscopic state of the gravitational field\index{microscopic state of gravitational field!information}, despite the fact that, like the varying average densities of positive- and negative-energy matter, the varying average density of vacuum energy provides a variable contribution to the gravitational fields which influence the rates of expansion\index{expansion rate!opposite-energy observers} of space determined by positive- and negative-energy observers.

The conclusion that a variation in the average value of vacuum energy density\index{average vacuum energy density!variation} associated with the cosmological `constant' cannot result in a variation of gravitational field information\index{gravitational field information!variation} actually constitutes an essential requirement if information is to be conserved on a cosmological scale, because, if such a variation was taking place, gravitational field information\index{gravitational field information!density of matter and vacuum energy} would vary as a result of changes occurring in the average density of both matter energy and vacuum energy and this would be problematic, because it would be possible for the variation of gravitational field information which would occur over the entire lifetime of the universe, as a result of those changes, to not be compensated by the variation of information\index{information variation!growth of spatial volume} attributable to the growing volume of space.

This would be a consequence of the fact that only the average density of matter\index{average density of matter!scale factor} necessarily diminishes as the scale factor grows and could be reduced to a minimum value if the volume of space was to become arbitrarily large, or be raised to a larger value if the volume of space was to be reduced to a smaller value (in either the past or the future direction of time). The magnitude of the average density of vacuum energy\index{average density of vacuum energy!volume independence} associated with the cosmological constant, on the other hand, could in principle become larger (at least temporarily) in a universe of growing volume and diminishing matter density, or else become smaller in a collapsing universe with growing matter density and a diminishing volume of space, with no possible compensation of the changes taking place in the total amount of gravitational field information\index{gravitational field information!absence of compensation}, thereby precluding information\index{information!conservation} from being conserved as it must be, that is, independently for positive- and negative-energy observers who may experience different rates of expansion and different average matter densities.

However, once it is recognized that global changes in the strength of gravitational fields\index{gravitational field strength!global changes} attributable to a variation of the average density of vacuum energy\index{average vacuum energy density!variation} associated with the cosmological constant\index{cosmological constant!variation} do not contribute any changes to the amount of information necessary to describe the microscopic state of the gravitational field\index{microscopic state of gravitational field!information} and therefore need not be taken into account in balancing the growth of information\index{information growth!rising volume of space} associated with the rising volume of space produced by expansion (given that all persistent inhomogeneities in the distribution of vacuum energy\index{vacuum energy distribution!persistent inhomogeneities} are actually equivalent to the presence of matter), then those difficulties no longer exist.

As space expands, the average density of matter and radiation (including the density of dark matter\index{dark matter!local variations of vacuum energy density} attributable to local variations of vacuum energy density) would, in effect, be continuously decreasing, along with the measure of information required to specify the microscopic state of the gravitational field attributable to the presence of this matter, which was originally maximum (this is allowed, given that a reduction of negative-energy matter density contributes like a reduction of positive-energy matter density to lower the measure of gravitational field information, despite the opposite sign of the variation of energy density involved). But at the same time, the amount of information\index{information!growth} associated with the number of elementary units of area\index{elementary units of area!co-moving volume} present within a co-moving volume (or, more accurately, the number of elementary units of area\index{elementary units of area!two-dimensional boundary} on the two-dimensional boundary of the same volume) would grow to some arbitrarily large value, thereby compensating the decrease of gravitational field information\index{gravitational field information!decrease} that is associated with the diminishing, average matter density, while a similar compensation would occur in the case of a contracting space\index{contracting space}\footnote{
If those conclusions are appropriate, it would mean that the idea proposed by certain authors that the size of the elementary units of area\index{elementary units of area!growing size|nn} determined by the natural scale of quantum-gravitational phenomena\index{quantum-gravitational phenomena!natural scale|nn} is perhaps growing with time, so that the amount of information associated with the total volume of space\index{total volume of space!information|nn} would be constant despite expansion, which should eventually give rise to a `Big Snap'\index{Big Snap|nn} that would rip everything apart, can be considered unnecessary and this is certainly appropriate, given that no such an event seems to be occurring.}.

I believe that this is the strongest argument which can be formulated to the effect that it is appropriate to consider that, even though a void in a macroscopically uniform distribution of negative-energy matter\index{void in uniform matter distribution!gravitational information} produces an attractive gravitational field from the viewpoint of a positive-energy observer, it should contribute nothing to the measure of gravitational field information, because this conclusion is what allows me to assume that a variation in the average density of the uniform portion of vacuum energy\index{uniform portion of vacuum energy!gravitational information} associated with the cosmological constant\index{cosmological constant} itself doesn't give rise to a variation of gravitational field information, due to the fact that the microscopic state of the gravitational field\index{microscopic state of gravitational field!information} produced by this uniform portion of the distribution of vacuum energy actually contains no information, while this assumption is a necessary element of the solution proposed here to the problem of the conservation of information\index{conservation of information!expansion of space}, in the context of a quantum-gravitational description of the expansion of space.

If I have properly conveyed the nature of the insights which have allowed me to arrive at such a conclusion, then it should be clear that there is no longer a problem with the fact that the expansion of space\index{expansion of space!production of information} appears to produce information which, in principle, could be obtained, even though it remains missing in the absence of measurements performed on a quantum-gravitational scale of precision\index{quantum-gravitational scale of precision}, just like most of the information contained in the gravitational field associated with a macroscopically homogeneous matter distribution\index{homogeneous matter distribution!gravitational field information}. From my viewpoint, even if this growth in the amount of information concerning the microscopic state of the gravitational field\index{microscopic state of gravitational field!empty space} associated with empty space must be considered real, it cannot give rise to a net increase in the total amount of information required to completely specify the microscopic state of the gravitational field\index{microscopic state of gravitational field!matter distribution} associated with both empty space and the matter distribution, for the universe as a whole.

\bigskip

\noindent Before concluding this section I would like to add a few important remarks concerning the role played by voids in balancing the gravitational field information\index{gravitational field information} budget. First of all, it must be emphasized, again, that if one assumes that there is a measure of gravitational information associated with a macroscopically uniform distribution of negative-energy matter\index{uniform matter distribution!gravitational information}, even from the viewpoint of a positive-energy observer, then it becomes possible for the presence of a uniform distribution of negative-energy matter to have an objective significance, from the viewpoint of a positive-energy observer, even when it is recognized that the average density of negative matter energy\index{average density of negative matter energy} cannot exert a gravitational force on positive-energy matter. There are good reasons to believe, therefore, that it is only in the limit where no matter of any energy sign is present and only vacuum energy remains, that the energy distribution\index{energy distribution!absence of microscopic structure} becomes totally devoid of microscopic structure that could be imparted to the gravitational field and that it is only under such conditions that the amount of information contained in the microscopic state of the gravitational field\index{microscopic state of gravitational field!null information} must be null.

If one was to assume, instead, that a null measure of gravitational information is associated with a macroscopically uniform distribution of negative-energy matter\index{negative-energy matter distribution!gravitational information}, from the viewpoint of positive-energy observers, then the amount of missing gravitational field information\index{missing gravitational field information!growth} would need to rise to a larger positive value, from the null value it would have in the absence of inhomogeneities in the negative-energy matter distribution, when the magnitude of the density of negative-energy matter would be either increasing above its average value (along with the strength of repulsive gravitational fields\index{repulsive gravitational field}) or decreasing below this same average value (along with the strength of attractive gravitational fields), and this would mean that, from the viewpoint of positive-energy observers, information\index{information!creation out of nothing} could be created out of nothing and would not be conserved, unlike energy itself, when inhomogeneities form in the negative-energy matter distribution. But under such conditions there would also be a requirement for the temperature of a sufficiently large void in the negative-energy matter distribution\index{void in negative-energy matter distribution!temperature} to be positive, because entropy would rise as a result of its creation, despite the fact that, once formed, no positive-energy thermal radiation would be released outside its event horizon, as I explained above.

What may perhaps make this latter hypothesis appear appropriate is that it would allow the measure of information contained in the microscopic state of the gravitational field\index{microscopic state of gravitational field!information} to always be proportional to the strength of macroscopic gravitational fields\index{macroscopic gravitational field!strength}. But in fact, there is no reason to assume that it should always be the strength of macroscopic gravitational fields that determines the measure of information that is contained in their microscopic states, just like it would be incorrect to assume that the presence of a maximally strong electromagnetic field on the surface of a black hole\index{black hole!maximally strong electromagnetic field} would require the amount of missing information\index{missing information!microscopic state of electromagnetic field} contained in the microscopic state of this interaction field to be maximum instead of null (as I explained in section \ref{sec:3.10}). Therefore, I believe that it is necessary to assume that the amount of information contained in the gravitational field\index{gravitational field information!decrease} actually decreases, even from the viewpoint of a positive-energy observer, when a void forms in the uniform negative-energy matter distribution, as one would expect in the context where it is understood that this measure of information must be null in the total absence of negative-energy matter.

If it is not appropriate to define gravitational information\index{gravitational information!zero value} as being zero, from the viewpoint of positive-energy observers, in the absence of any macroscopic inhomogeneities in the negative-energy matter distribution, it is because, even in the absence of any macroscopic inhomogeneities, there is information in the gravitational field associated with such a matter distribution, due the fact that matter is not a perfectly uniform fluid\index{perfectly uniform fluid}. What happens is that the gravitational fields attributable to the presence of microscopic matter inhomogeneities\index{microscopic matter inhomogeneities!gravitational field strength} are stronger and more variable in space, in the presence of a larger \textit{average} density of matter energy, and this is why more information is required to describe their microscopic states under such conditions, despite the apparent uniformity of the matter distribution. It should be clear, therefore, that macroscopic underdensities in a uniform distribution of negative-energy matter are not associated with a negative value of gravitational field information\index{gravitational field information!negative value} (whatever that would mean), but merely with a lower than average amount of information, even from the viewpoint of positive-energy observers.

What one must understand is that the contribution to gravitational field information\index{gravitational field information!average matter energy density} by the average density of matter energy does not arise merely from the uniform portion of the matter distribution, but from the presence of inhomogeneities in this macroscopically uniform matter distribution\index{macroscopically uniform matter distribution!inhomogeneities} and in the absence of microscopic inhomogeneities, this contribution would be null, even though it varies only as a function of the magnitude of the average density of positive or negative matter energy. If this was not the case, a similar variation in the density of the uniform portion of vacuum energy\index{uniform portion of vacuum energy!density variation} would produce a similar change in gravitational field information\index{gravitational field information!variation} and this information\index{information!conservation} would no longer be conserved in the course of expansion, as I explained above.

To say the truth, even a homogeneous matter distribution with a maximum positive or negative energy density may be allowed to contribute to gravitational information\index{gravitational information!homogeneous matter distribution} (unlike a homogeneous distribution of vacuum energy with non-zero average density), because under such conditions it is still possible for the variable momentum direction of particles to impart microscopic structure on the gravitational field\index{gravitational field!microscopic structure}. But under ordinary conditions, it is the fact that, even when positive-energy matter particles are as homogeneously distributed as they can be, a certain amount of structure is still present in the gravitational field on a microscopic scale (because the particles only occupy a small portion of the space which is available to them), that explains that it is possible for gravitational information to exist, even in a macroscopically homogeneous matter distribution.

It is this measure of information that decreases when the average density of matter decreases and the strength of microscopic gravitational fields\index{microscopic gravitational field strength!reduction} is reduced, as a result of the decreasing magnitude of microscopic matter inhomogeneities\index{microscopic matter inhomogeneities!magnitude decrease} that is attributable to the expansion of space\index{expansion of space} and the diminution of the average kinetic energy of matter\index{average kinetic energy of matter particles!diminution} and radiation particles. One must, therefore, distinguish the hypothesis that matter and radiation particles are homogeneously distributed (which can be satisfied when no macroscopic inhomogeneities are present in the matter distribution) and the hypothesis that the energy of matter\index{energy of matter!homogeneous distribution} itself is homogeneously distributed (which may not be satisfied, even when no inhomogeneities are present in the matter distribution).

Thus, even in what one would consider to be a perfectly homogeneous matter distribution\index{homogeneous matter distribution!energy inhomogeneities}, where no local variations would exist in the density of matter particles, above or below the average density, there would normally remain enough energy inhomogeneities to produce microscopic gravitational fields\index{microscopic gravitational fields!information} which would contain a certain amount of information and this is also true for a positive-energy observer regarding the distribution of negative-energy matter, despite the fact that the homogeneous portion of the negative-energy matter distribution exerts no gravitational force on positive-energy matter. It is the inhomogeneities which are present in a distribution of matter energy that contribute to the measure of gravitational field information\index{gravitational field information!reduction} that can be reduced by the expansion of space and inhomogeneities in the distribution of negative matter energy do exert a gravitational force on positive-energy matter (even though the force itself cancels out on the largest scale).

In case this is not yet completely clear, I must also mention that from a practical viewpoint, even for a local void in the positive-energy matter distribution\index{void in matter distribution!gravitational field information}, the amount of gravitational field information would not necessarily be null, because negative vacuum-dark-matter energy\index{vacuum-dark-matter energy!void in matter distribution} would normally be present in the void (as a result of the curvature of space that is experienced under such conditions by negative-energy matter) and this matter would contribute to raise the amount of gravitational information\index{gravitational information!vacuum-dark-matter energy} contained in the void. It is only the void itself which gives rise to a local reduction of gravitational field information, despite the fact that, from the viewpoint of a negative-energy observer, it gives rise to a local gravitational field similar to that which is produced by the vacuum dark matter it contains.

It would be incorrect to assume that the change of gravitational information which is associated with the formation of a void in the positive-energy matter distribution cannot be opposite that which is associated with the local growth of negative vacuum-dark-matter energy density that takes place as a consequence of the formation of this void. Anyhow, when the contribution of vacuum dark matter is taken into account, it follows that the gravitational entropy associated with the presence of a void in the positive-energy matter distribution\index{void in matter distribution!non-zero gravitational entropy} would always remain larger than zero and would be continuously rising as more negative-energy matter concentrates in the structure.

\section{The initial singularity\label{sec:4.8}}

What emerges from the preceding reflection concerning the character of gravitational entropy\index{gravitational entropy} is that, while the amount of information required to describe the microscopic state of the gravitational field\index{microscopic state of gravitational field!information growth} is growing in those places where matter is becoming more densely packed, an equal amount of information is being lost at the same time in the gravitational field\index{gravitational field!information loss} as a consequence of the resulting diminution of density which is taking place in the surrounding matter distribution. Yet, given that, ultimately (when a black hole forms), the information that is gained becomes missing information\index{missing information} which is no longer accessible from an observational viewpoint, while the information that is lost was in principle available (as the original matter distribution was not constrained by the presence of an event horizon), then gravitational entropy must be assumed to rise whenever the matter distribution is becoming more inhomogeneous.

What I will now explain is how significant this conclusion actually is, in the context where the initial matter distribution\index{initial matter distribution!inexplicably high uniformity} at the Big Bang appears to have been of inexplicably high uniformity. Thus, I will argue that, for what regards temporal irreversibility\index{temporal irreversibility}, it is the measure of gravitational entropy that distinguishes the state that emerged from the past Big Bang singularity\index{past Big Bang singularity} and the state into which our universe will evolve in the far future (independently from whether it continues to expand or collapses back on itself). This discussion will set the stage for the more significant developments which will be introduced in the next section and which will provide the actual explanation for the existence of the thermodynamic arrow of time\index{thermodynamic arrow of time}, as a cosmological phenomenon.

It is important to note, first of all, that there is no paradox associated with the fact that the universe still evolves irreversibly, while the initial state at the Big Bang\index{initial Big Bang state!thermal equilibrium} was already one of near perfect thermal equilibrium, because, as Roger Penrose\index{Penrose, Roger} first pointed out \cite{Penrose-1}, under such conditions it is only the portion of entropy which excludes the contribution of gravitational fields that is maximum. In fact, what transpires from the developments introduced in the previous section, concerning gravitational entropy\index{gravitational entropy}, is that it is precisely the smoothness of the initial matter distribution\index{initial matter distribution!smoothness} (which is reflected in the uniformity of the temperature of the cosmic microwave background\index{cosmic microwave background!temperature uniformity}) that is responsible for having allowed the universe to evolve irreversibly at later times, because under such conditions it is the growth of matter inhomogeneities which must have provided the dominant contribution to irreversible entropy growth in our universe, since the epoch of matter-antimatter annihilation\index{matter-antimatter annihilation!epoch}.

What really needs to be explained, therefore, is not why the universe evolves irreversibly, despite the initial state of thermal equilibrium, but why matter was actually so homogeneously distributed initially that gravitational entropy was almost perfectly null, even if that would appear to be a highly unlikely configuration to begin with, in the context where a much larger number of possibilities exist for the microscopic state of matter and its gravitational field\index{microscopic state of matter and gravitational field}, which would not be characterized by such a uniform matter distribution and an absence of primordial black holes\index{primordial black holes!absence}.

In order to explain those facts, it is necessary to identify the nature of the constraint imposed by the fundamental, time-symmetric physical laws\index{time-symmetric physical laws} on the boundary conditions\index{boundary conditions!Big Bang} at the Big Bang that is responsible for the very high level of homogeneity and the very low gravitational entropy\index{gravitational entropy!very low initial value} that characterizes this initial state. We must, therefore, once again transcend our natural reluctance to apply the known principles of physics to the Big Bang if we are to avoid having to modify the laws themselves in order to achieve greater overall consistency. It would be incorrect to assume that proposing a solution to the problem of the origin of time asymmetry\index{time asymmetry!problem of origin} that relies on the application of certain constraints to the initial conditions at the Big Bang would be akin to requiring divine intervention. The most fundamental principles must be assumed to be valid under absolutely all conditions, including those that existed during the Big Bang. I believe that it is our failure to acknowledge the importance of this requirement that explains most of the difficulties we currently face in theoretical cosmology\index{theoretical cosmology}.

But before we can achieve some real progress in understanding why time irreversibility\index{time irreversibility} occurs, we must first recognize that the source of most changes to entropy that take place after the early annihilation of baryons with antibaryons\index{early baryon-antibaryon annihilation} is actually to be found in the growing strength of local gravitational fields. It is as a consequence of gravitational attraction that the stars, in particular, can form and are allowed to release their radiation and it is due to gravitational collapse\index{gravitational collapse} that black holes, as the objects with the highest entropy density, can form and grow more massive at the expense of a local reduction of matter density in their environment. But one need not assume that this is due to the `fact' that gravitation is always attractive, as all that is required is that it be attractive among particles with the same sign of energy, which allows gravitational energy\index{gravitational energy!mass squared proportionality} and therefore also gravitational entropy\index{gravitational entropy!mass squared proportionality} to be proportional to the square of the mass of an object, instead of being merely proportional to its mass, as does matter entropy.

In such a context, it may appear that a much larger number of possible, initial microscopic states\index{initial microscopic states} would be characterized by the presence of an abundance of primordial black holes\index{primordial black holes} and other matter density fluctuations, while those initial conditions would not have had as much potential for allowing subsequent evolution to take place irreversibly. However, given that the presence of primordial black holes would have disturbed the process of structure formation in the initial matter distribution in ways which would have had observable consequences at the present epoch, then it seems necessary to assume that the initial Big Bang state was virtually free of black holes and therefore it remains to explain why the universe was in such an unlikely configuration at the Big Bang.

One thing that should be clear is that the weakness of the gravitational interaction\index{gravitational interaction!relative weakness} in comparison with the other fundamental forces and the fact that it became predominant over those other interactions only during the matter-dominated era\index{matter-dominated era}, does not mean that no constraint that would be imposed on the magnitude of gravitational fields (or the curvature of space\index{curvature of space!matter inhomogeneities} attributable to the presence of matter inhomogeneities), could be involved in determining the early conditions which are responsible for the existence of the thermodynamic arrow of time\index{thermodynamic arrow of time}. If gravitational instability\index{gravitational instability} allowed structures to begin developing only at a relatively late time in the matter distribution, this is due precisely to the fact that the initial matter distribution\index{initial matter distribution!uniformity} was so uniform to begin with and this is a condition that applies on the magnitude of gravitational fields. But it would certainly be inappropriate to assume that a constraint which would be imposed on the initial magnitude of local gravitational fields\index{local gravitational fields!initial magnitude} would not have much impact as a consequence of the very fact that the magnitude of those gravitational fields was, in effect, so small initially.

In section \ref{sec:4.5} I have explained why we can actually expect the universe to be expanding. But the fact that we are not instead observing it to be contracting at the present moment can only be explained as being the consequence of another fact, which is that the magnitude of inhomogeneities in the matter distribution is decreasing continuously in this direction of time relative to which space is contracting. If we perceive the universe to be expanding, it is simply because our memory formation process\index{memory formation!thermodynamic process}, as a thermodynamic process, only occurs in the direction of time in which the inhomogeneity of the matter distribution\index{inhomogeneity of matter distribution!growth} and the associated gravitational entropy\index{gravitational entropy!growth} are growing, which means that if the inhomogeneity of the matter distribution was instead growing in the direction of time relative to which the universe is contracting, we would then necessarily perceive space to be contracting.

This is actually all that can be meant when we say that we experience the universe to be expanding, because in fact we do also `observe' space to be contracting, but merely in the sense that we also have knowledge of the contraction of space\index{contraction of space!past direction of time} that occurs in the past direction of time, as we could witness by watching a backward-running movie of the same events. Thus, what explains that the universe is observed to be expanding (what explains that the cosmological arrow of time\index{cosmological arrow of time} is oriented in the same direction as the thermodynamic arrow of time\index{thermodynamic arrow of time}) is the fact that gravitational entropy is practically null in the initial Big Bang state\index{initial Big Bang state!null gravitational entropy}, while it is allowed to grow to arbitrarily large values at later times and this means that if we want to explain why it is that we observe an expanding universe, then we must first explain why it is that its initial state was characterized by such a low gravitational entropy.

But, in the context where one must acknowledge the presence of negative-energy matter in our universe, the fact that the density of matter was much larger in the past does not make the initial smoothness of the matter distribution\index{initial smoothness of matter distribution} more unexpected. Even when the volume of the universe\index{volume of universe!arbitrarily large value} is arbitrarily large, a uniform matter distribution is not more likely to be observed, because a diluted matter distribution could still contain inhomogeneities and produce arbitrarily strong gravitational fields on a very large scale, as a result of the fact that negative-energy matter can be concentrated in regions of space distinct from those occupied by positive-energy matter, even if the magnitude of the average densities of both types of matter is arbitrarily small. Thus, in the absence of additional constraints, the most likely configuration for the initial distribution of positive- and negative-energy matter\index{initial matter distribution!most likely configuration} would appear to always be one of higher inhomogeneity, because there exist a lot more microscopic configurations of matter and its gravitational field for which the portions of positive- and negative-energy matter that existed before the early annihilation of matter with antimatter\index{early matter-antimatter annihilation} were not mixed up in a perfectly uniform manner.

In any case, the distributions of positive- and negative-energy matter\index{matter distribution!unexpectedly high homogeneity} were, in effect, homogeneous to an unexpectedly high degree in the initial Big Bang state. If this hadn't been the case, then macroscopic event horizons\index{macroscopic event horizon} would abound in the primordial universe\index{primordial universe} and even if the magnitude of density fluctuations was not large enough to prevent space from expanding in most locations, what we would observe (either around us at the present moment or in the cosmic microwave background\index{cosmic microwave background} at the epoch of last scattering\index{last scattering!epoch}) would be a much different world. The problem, therefore, is that it seems that the observable universe should have begun its evolution in a state where positive- and negative-energy matter would already be highly inhomogeneously distributed and strong local gravitational fields\index{local gravitational fields!high initial strength} would be present, with which would be associated an arbitrarily large measure of gravitational entropy. But if the initial state was not of such a nature, then it means that something must have constrained the universe to have a much lower gravitational entropy initially, because this does not appear to be a natural configuration to begin with when all possibilities are allowed.

It should be clear, however, that the simple fact that space must not have collapsed into a black hole\index{black hole!singularity} singularity, locally, if an observer is to be present to witness an absence of inhomogeneities, does not provide strong enough a constraint to explain that the matter distribution was as smooth as one can deduce it was, initially. Matter could be much more inhomogeneously distributed than it currently is and the conditions necessary for the emergence of an observer\index{observer emergence!necessary conditions} would still be satisfied in most locations, even if a large number of primordial black holes\index{primordial black holes} had been present initially. It is merely the fact that the inhomogeneity is not as pronounced as it \textit{could} have been that is unexplained.

What must be understood is that the homogeneity of the initial distribution of matter energy is not apparent only in the low magnitude of local variations in the density of positive-energy matter, but must also be apparent in the near absence of very-large-scale disparities between the positive- and negative-energy matter distributions. This particularity is especially significant in the context where one of the only differences which would exist during a recollapsing phase\index{recollapsing phase} of the universe's history (if such an evolution was actually allowed to happen), would have to do with the fact that, in the recollapsing phase, the dissociation of the positive- and negative-energy matter distributions\index{opposite-energy matter distributions!dissociation} would actually be much more pronounced, as a result of the gradual polarization of the matter distribution\index{polarization of matter distribution} along energy sign which can be expected to occur in the context where particles with the same sign of energy are submitted to mutual gravitational attraction, while concentrations of matter with opposite energy signs gravitationally repel one another.

For the present discussion to be meaningful, however, one must also understand that there are strong motives for believing that, even in the presence of negative-energy matter, if the initial matter distribution\index{initial matter distribution!homogeneity} is sufficiently homogeneous, it is still appropriate to consider that there should arise a state in the past which, from a classical viewpoint, would consist in a spacetime singularity\index{spacetime singularity!past}. What should be clear, based on the developments introduced in section \ref{sec:2.7}, is that a globally homogeneous distribution of negative-energy matter (regardless of what its current average density might be) would exert no influence on the rate of expansion of space\index{expansion rate} experienced by positive-energy observers and would not diminish the strength of the gravitational field attributable to the presence of this matter (despite the fact that negative-energy matter inhomogeneities do exert gravitational forces on positive-energy particles).

This conclusion follows from the proposed description of negative-energy matter as being equivalent to voids in the positive-energy portion of the vacuum\index{void in positive vacuum energy} and the acknowledgment that the void of cosmic proportions in positive vacuum energy\index{void of cosmic proportions!positive vacuum energy} that must be associated with a homogeneous distribution of negative-energy matter cannot give rise to uncompensated gravitational attraction\index{uncompensated gravitational attraction} from a surrounding distribution of positive vacuum energy, which would otherwise be the source of the gravitational repulsion\index{gravitational repulsion} that would arise from the presence of such a void.

Once this is recognized, it becomes possible to predict that if the initial matter distribution\index{initial matter distribution!sufficient homogeneity} is sufficiently homogeneous on the largest scale, then nothing can prevent the formation of the trapped surface\index{trapped surface} which according to classical theorems would give rise to a past spacetime singularity\index{spacetime singularity!past}, even if one of the axioms of the theorems is that matter must always have positive energy. It is only the inappropriateness of the conventional description of negative-energy matter\index{negative-energy matter!conventional description} as being the source of absolutely repulsive gravitational fields\index{absolutely repulsive gravitational fields} that makes it seem like the presence of such matter could prevent the formation of a past singularity (or the occurrence of a state of maximum positive- and negative-energy matter densities\index{state of maximum matter densities}, as one would rather need to assume from a quantum-gravitational viewpoint\index{quantum-gravitational viewpoint}).

It would, therefore, appear that the very uniformity of the initial matter distribution\index{initial matter distribution!uniformity} which is responsible for giving rise to the existence of a thermodynamic arrow of time\index{thermodynamic arrow of time} is actually required in order that the existence of a past spacetime singularity be considered unavoidable. This is a decisive observation whose significance will be made more explicit in the following section. I must emphasize, however, that what I have in mind when I'm referring to an initial singularity\index{initial singularity!maximum matter energy densities} is not a state where the laws of physics would actually break down, but simply a state where the average, positive and negative densities of matter energy would reach their maximum theoretical values, determined by the natural vacuum-stress-energy tensors\index{natural vacuum-stress-energy tensors} which enter the generalized gravitational field equations\index{generalized gravitational field equations} introduced in section \ref{sec:2.14}.

In any case, if we are to assume that there must, in effect, be a singularity, or a state of maximum positive and negative matter energy densities at the beginning of (unidirectional) time, then it seems necessary to assume that this singularity is also different, in certain respects, from an ordinary black hole\index{black hole!singularity} singularity. First of all, even if the initial state that emerged from the past singularity at the Big Bang had been highly inhomogeneous, it would not be expected to have given rise to the same evolution as that through which a future Big Crunch singularity\index{future Big Crunch singularity} would go from a backward-in-time viewpoint, because, whereas the state that would emerge from a final Big Crunch singularity, in a universe like ours, would evolve back to a more homogeneous state, a state that would emerge from an initial Big Bang singularity\index{initial Big Bang singularity} with the same level of matter-energy inhomogeneity would not evolve toward a more homogeneous state, because in our universe future evolution is unconstrained.

Therefore, a highly inhomogeneous distribution of matter energy emerging from a past singularity could only evolve toward an even more inhomogeneous state (if there does not exist any limit to gravitational entropy\index{gravitational entropy!growth} growth), otherwise it would not evolve at all, from a thermodynamic viewpoint, as it would already be in one of its most likely maximum gravitational entropy\index{gravitational entropy!maximum} states. As a result, no time-reversed gravitational collapse\index{time-reversed gravitational collapse} or white hole\index{white hole} phenomenon would occur that would release objects of lower entropy, as we would expect to happen in the course of a time-reversed Big Crunch\index{time-reversed Big Crunch}.

But while the Big Bang is not the time-reverse of a Big Crunch, or black hole gravitational collapse, it also appears that the initial singularity\index{initial singularity} is different from a future singularity\index{future singularity} owing to the fact that the state that emerges from it is not characterized by large fluctuations in matter energy density, with which would be associated a very large gravitational entropy, such as would be the case for the final state of a generic future singularity. What is occurring in the past direction of time, in our universe, therefore, is not what one would expect to happen as a mere consequence of the contraction of space\index{contraction of space}. The initial singularity was of such a nature that it could not constitute the outcome of a generic gravitational collapse\index{generic gravitational collapse} of the kind that would occur in a universe in which the matter distribution is growing ever more inhomogeneous due to gravitational instability\index{gravitational instability}. The universe is changing as it collapses in the past direction of time, but not in the way one would expect in the absence of a constraint that operates a continuous decrease in the magnitude of the inhomogeneities present in the distribution of matter energy.

What's significant, as well, is that the presence of past singularities\index{past singularity} appears to be restricted to the one known initial singularity from which the Big Bang emerged, even if there does exist solutions of the gravitational field equations\index{gravitational field equations} that would appear to describe processes which would be the time-reverses of a black hole gravitational collapse\index{black hole gravitational collapse!time-reverse}. All the evidence indicates that the hypothetical white hole\index{white hole} processes which could be described using those solutions never occur in our universe. I believe that if those solutions do not represent processes that can be observed in the forward direction of time in our universe, it is because they would allow gravitational entropy\index{gravitational entropy!decrease} to decrease in this direction of time, while such an evolution is thermodynamically unlikely in the absence of a specific constraint.

If they were allowed to arise, white holes would expel low-entropy matter at an arbitrarily high rate, which would reduce their masses and the area of their event horizons faster than would be allowed as a consequence of the emission of macroscopically thermal radiation\index{macroscopically thermal radiation} (this has nothing to do with negative-energy black holes expelling positive-energy matter), so that the processes would involve a decrease of gravitational entropy\index{gravitational entropy!uncompensated local decrease} in the future that wouldn't be compensated by a larger increase of matter entropy. It should be clear, therefore, that the Big Bang does not constitute a white hole, even though it originates from a past singularity.

The only motive one might have to assume that white holes could exist would be that, in all likeliness, the gravitational entropy\index{gravitational entropy!black hole} of a black hole should rise in the past, just like it does in the future, so that, from the forward-in-time viewpoint, the evolution that would take place during the same period of time would consist in a fluctuation involving a decrease of gravitational entropy\index{gravitational entropy!decrease} that would persist until the present moment is reached. But the problem is that, even if our present state would seem to require the occurrence of such a phenomenon, there appears to be something that constrain evolution in the past direction of time to take place with continuously decreasing gravitational entropy\index{gravitational entropy!continuous decrease in past}, despite the apparent improbability of this evolution, and this is precisely what remains unexplained.

If white holes\index{white hole} are never observed, therefore, it is simply because such processes would require a decrease of gravitational entropy in the future (which is unlikely), or equivalently, a continuous increase of gravitational entropy\index{gravitational entropy!continuous increase} in the past (which for some reason appears to be forbidden). Therefore, if we can understand why the state that emerged from the initial Big Bang singularity\index{initial Big Bang singularity!gravitational entropy} had minimum gravitational entropy, then we may also be allowed to understand why there is only one such past singularity.

Given that black holes are the objects associated with the highest possible density of gravitational entropy\index{density of gravitational entropy}, it would appear that in the absence of any constraint the most likely configuration for the initial state of the universe would be one for which the initial distribution of positive- and negative-energy matter\index{opposite-energy matter distributions!polarization} would be completely polarized, in such a way that all the matter would be contained in opposite-energy black holes\index{opposite-energy black holes!arbitrarily large masses} with arbitrarily large masses, whose magnitude would be limited solely by the amount of matter in the universe, because there exist a much larger number of microscopic states compatible with such a macroscopic configuration.

But even though it may appear that our universe\index{universe!closed geometry} could actually be contained in one of two such opposite-energy black holes if it had a closed geometry and a relatively low matter density, this hypothesis would not agree with observations, because there would be no low-gravitational-entropy Big Bang\index{Big Bang!low gravitational entropy} in such a universe, not only from the viewpoint of an observer outside the opposite-energy black holes, but even from the viewpoint of an observer inside one of the two objects. In fact the most likely state in which we would expect such a universe to be, not just initially, but at all times, would still be that of a spacetime singularity\index{spacetime singularity!maximum gravitational entropy} with maximum gravitational entropy, which means that it would still remain to explain why this is not what we observe to be the past and present state of our actual universe.

What must be understood is that there is no \textit{a priori} motive for assuming that a high level of polarization of the positive- and negative-energy matter distributions\index{opposite-energy matter distributions!polarization} could not apply to the initial Big Bang state\index{initial Big Bang state!statistically favored configuration} (regardless of the fact that the matter density is then maximum) if such a configuration is, in effect, favored from a statistical viewpoint, because a universe that would evolve without constraint, as space is contracting in the past direction of time, would be much more likely to reach such a configuration than one of minimum gravitational entropy\index{gravitational entropy!minimum}.

Now, it was once suggested that the smoothness of the initial matter distribution\index{initial matter distribution!apparent smoothness} might only be apparent and that a state of higher inhomogeneity might have existed initially, that was later made uniform through various smoothing processes\index{smoothing processes}. But given that such processes would have released a large amount of heat that would have modified the temperature of the cosmic microwave background\index{cosmic microwave background!temperature modifications} to an extent that appears to be incompatible with measurements, then it appears that, even if the smoothing could occur at the appropriate time and on the appropriate scale, its outcome would not agree with observational constraints.

Furthermore, if the distributions of positive- and negative-energy matter\index{matter distribution!amplification of inhomogeneities} had been highly inhomogeneous before any such process could smooth it out, the magnitude of those inhomogeneities would have been rapidly amplified under the effect of the gravitational interaction and it would have become even more difficult to give rise to the homogeneous distribution of matter energy that is revealed by measurements of the temperature of cosmic microwave background\index{cosmic microwave background} radiation. Indeed, the same argument implies that the initial Big Bang state\index{initial Big Bang state!imperfect uniformity} cannot have been \textit{perfectly} uniform, otherwise the process of structure formation\index{structure formation!process} could not have occurred rapidly enough to allow for the timely development of stars, galaxies and other large-scale structures\index{large-scale structures!development} which means that the constraint responsible for the high level of homogeneity of the initial state must not be so restrictive that it would imply a complete absence of energy fluctuations.

Of course, in the presence of negative energy matter, an additional contribution to gravitational instability\index{gravitational instability} exists that more readily triggers the formation of inhomogeneities. But given that most of the baryonic negative-energy matter\index{baryonic negative-energy matter} vanished following the early annihilation of matter with antimatter\index{early matter-antimatter annihilation}, then smaller-scale fluctuations in the negative-energy matter distribution\index{negative-energy matter distribution!smaller-scale fluctuations} were not allowed to grow as much as they otherwise would by the time the cosmic microwave background was released, while larger-scale fluctuations in the distribution of negative vacuum-dark-matter energy\index{negative vacuum-dark-matter energy!larger-scale fluctuations} only began to grow when they were encompassed by the cosmological horizon\index{cosmological horizon}. Therefore, there weren't much additional perturbations to the temperature of CMB radiation as a result of the presence of negative-energy matter. But it remains that energy fluctuations, even if only those attributable to local variations of the sign of energy of matter particles\index{particle energy sign!local variations}, would need to be present initially if inhomogeneities are to develop at a later time, on a larger scale.

What constitutes the most significant difficulty for the smoothing hypothesis\index{smoothing hypothesis}, however, is the fact that the existence of cosmological horizons\index{cosmological horizon} would have forbidden any such process from ironing out inhomogeneities above the scale determined by the size of the horizon, at the time when the CMB was released, and therefore we should not observe uniformity on the largest scale if the homogeneity of the distribution of matter energy is attributable to smoothing processes obeying the principle of local causality\index{principle of local causality}. An intrinsic limit is actually imposed on such processes, that would prevent them from producing the kind of homogeneous state which emerged from the Big Bang and therefore it appears appropriate to conclude that, regardless of any other difficulty, conventional smoothing processes\index{conventional smoothing processes} should probably not be considered a viable explanation for the homogeneity of the initial distribution of matter energy\index{initial distribution of matter energy!homogeneity}.

As a consequence of the clear inadequacy of conventional smoothing processes and in the absence of a better alternative, it is still widely believed that inflationary expansion\index{inflationary expansion} may be the cause of the very high homogeneity of the universe's initial matter distribution which is reflected in the small amplitude of cosmic microwave background\index{cosmic microwave background!temperature fluctuations} temperature fluctuations. However, I think that the occurrence of this hypothetical process of exponentially accelerated expansion would not be of much help in explaining the observed time asymmetry\index{time asymmetry!cosmic evolution} that characterizes cosmic evolution, because there is no reason to expect that a contracting universe\index{contracting universe!evolution toward more homogeneity} would evolve toward a more homogeneous configuration during the epoch that would precede a hypothetical phase of exponentially accelerated \textit{contraction}\index{exponentially accelerated contraction}, which would then take the universe back to a more likely state of maximum inhomogeneity.

If inflation theory\index{inflation theory} could perhaps explain why the universe evolves in an otherwise unnatural way (from the viewpoint of the growth of gravitational entropy\index{gravitational entropy!growth}), between the moment when matter emerges from the initial singularity\index{initial singularity} and the instant at which inflation ceases, it could not explain why it evolves toward greater homogeneity from far in the future and back toward the time at which the universe would presumably begin to contract at an exponentially accelerated rate into the initial singularity, now with naturally growing inhomogeneity.

Even if inflation may give rise to a homogeneous universe forward in time, a Big Crunch\index{Big Crunch!decreasing inhomogeneity} would not be expected to occur with decreasing inhomogeneity forward in time, unless the state immediately preceding the exponentially accelerated contraction\index{exponentially accelerated contraction!final singularity} into the final singularity would be required to be as smooth as the state which was produced in the past following ordinary inflation. But assuming that this would occur would amount to require that causality\index{causality!backward-in-time operation} operates backward in time from the final singularity, instead of forward in time from the initial singularity, because, from the viewpoint where causality operates from the past toward the future, a Big Crunch\index{Big Crunch!increasing inhomogeneity} would be more likely to occur with increasing inhomogeneity in the future, right up to the moment when a process of inflationary expansion\index{inflationary expansion!reverse process} would perhaps take place in reverse and merely increase the inhomogeneity that would already exist even further and produce an inhomogeneous final state.

Assuming that this is not what occurs would amount to postulate without motive that classical (unidirectional) causality\index{unidirectional causality!backward-in-time operation} must rather operate backward, from the instant at which matter emerges from the future Big Crunch singularity\index{future Big Crunch singularity} and until the moment when the universe would begin recollapsing, after having reached its maximum volume, so that the period of inflationary expansion\index{inflationary expansion} that would occur backward in time from the instant at which matter emerges from the final singularity would give rise to a homogeneous state \textit{after} inflation, in the past direction of time. But there is no \textit{a priori} reason not to assume, instead, that it is a highly inhomogeneous `final' state existing \textit{before} the phase of exponentially accelerated contraction that gives rise to the inhomogeneous state that would develop in the future direction of time as a result of this exponentially accelerated contraction, as we may expect based on the hypothesis that causality\index{causality!forward-in-time operation} still operates forward in time.

The problem is that the hypothesis that classical causality operates forward in time from the past singularity is necessary for the conclusion that inflationary expansion\index{inflationary expansion} would necessarily produce a homogeneous state, because if it was assumed that it is the events in the future which can \textit{irreversibly} influence what occurs backward in time until the moment when matter would start contracting at an exponentially accelerated rate back into the initial singularity\index{initial singularity}, then the state we would expect to obtain following the initial phase of inflationary expansion, from the forward-in-time viewpoint, would still be a state of maximum inhomogeneity\index{maximum inhomogeneity state!outcome of inflation}, while this does not correspond to reality.

What must be understood is that, even if we simply interchange future and past, we are still facing a mystery, because, if it is the future that influences the past in an irreversible way and if inflationary expansion takes place backward in time, so as to smooth out the state emerging from a future Big Crunch singularity\index{future Big Crunch singularity}, instead of giving rise to a homogeneous state forward in time, beginning from the inhomogeneous state that emerged from the past singularity, then we simply reverse the direction in which irreversible evolution\index{irreversible evolution!reversed direction} would take place and we still have no explanation for why unidirectional causality\index{unidirectional causality} does, in effect, operate in this particular direction of time (which we would then call the future) and not in the opposite one, while this is precisely what we are trying to explain.

Classical causality\index{classical causality!thermodynamic time asymmetry}, or the rule that past events always have an influence on future events and not the opposite, is simply a consequence of thermodynamic time asymmetry or time irreversibility\index{time irreversibility}, and if this property is assumed to characterize our universe without question, then it cannot be used to explain time irreversibility itself. Therefore, assuming that inflationary expansion\index{inflationary expansion} necessarily produces a highly homogeneous state, from a more likely inhomogeneous state, amounts to assume without justification the very outcome we want to derive, which means that inflation is not valid as an explanation of the origin of time asymmetry\index{time asymmetry!origin} that would arise from the necessity of a homogeneous initial state (following inflation).

The fact that inflation\index{inflation!finely-tuned initial conditions} itself requires finely-tuned initial conditions to occur does not even need to be taken into account in order to conclude that this hypothetical phenomenon is insufficient to explain the observed asymmetry of the evolution of gravitational entropy\index{gravitational entropy!asymmetry of evolution}. To actually explain the unlikely homogeneous state that emerged from inflation during the Big Bang, using the hypothesis of inflationary expansion itself, we would have to predict that this process operates in both the future and the past directions of time to produce a homogeneous state out of the generic inhomogeneous initial states\index{inhomogeneous initial states!generic} that would emerge from both the initial and (hypothetical) final singularities\index{initial and final singularities} and this would require that the direction of time relative to which inhomogeneities are growing mysteriously reverses when the universe starts contracting, when its volume would be maximum (or at any, arbitrarily-chosen, intermediary time, indeed) and as I previously explained there is absolutely no reason to expect that a reversal of the thermodynamic arrow of time\index{thermodynamic arrow of time!reversal} associated with the growth of gravitational entropy\index{gravitational entropy!growth} would occur when space would begin contracting on a global scale.

At this point it is necessary to mention that a variation of the more conventional attempt at explaining cosmological time asymmetry\index{cosmological time asymmetry} by making use of inflation theory\index{inflation theory} which was proposed more recently \cite{Carroll-1} postulates that it is through the process of creation out of `nothing'\index{creation out of nothing} that symmetry with respect to the direction of time can be reintroduced in our description of cosmic history\index{cosmic history!time symmetry}. What is proposed is that the initial state of our universe\index{initial state of universe!extended vacuum state} is actually an extended vacuum state, which we may perhaps consider to be a likely state from a thermodynamic viewpoint, given that under such conditions and when only positive-energy matter is allowed to exist, the entropy of matter\index{entropy of matter!maximum} itself (if there is any) would seem to be maximum.

Of course this is not the state in which our universe began, according to observations, but it might be possible to assume that what happened is that the universe\index{universe!local fluctuation in extended vacuum} emerged out of a local fluctuation in this extended vacuum and that it is inflationary expansion\index{inflationary expansion} that is responsible for having allowed the high-density state so produced to start expanding at a critical rate and if this is indeed the case then the history of the universe could perhaps be considered to necessarily begin in a state (preceding inflation) that is not so unlikely from the viewpoint of its entropy, even if this would otherwise be unexpected. The idea is that this kind of process could take place in both the past and the future, starting from this initial, extended vacuum state, and in such a way produce a globally time-symmetric history for the universe.

The problem I have with this description, however, is not merely that the proposed solution is dependent on the validity of the hypothesis that there occurred an early phase of inflationary expansion, which can only produce the desired outcome under highly unlikely initial conditions of a distinct nature. The more unavoidable difficulty has to do with the fact that, as a tentative explanation of time asymmetry\index{time asymmetry!tentative explanation}, it would suffer from the same reliance shared by more conventional approaches on the implicit assumption that there is already a favored direction of time\index{favored direction of time!implicit assumption}. Indeed, despite what is usually assumed, an extended vacuum state\index{extended vacuum state!prior phase of expansion} can only arise out of a prior phase of expansion that would occur in the future direction of time. If a large volume of space is to remain nearly empty for a sufficiently long time that a localized fluctuation of vacuum energy \index{vacuum energy fluctuation!universe creation} is perhaps allowed to give rise to the creation of an entire universe, then this space must have been expanding prior to the creation event and this expansion can only take place in one direction of time at once.

In the following section I will explain, in effect, that it is not possible to assume the existence of an expanding low-density universe\index{expanding low-density universe} that would not have emerged out of a state of maximum matter density\index{maximum matter density state} at some point in the past and if the universe did emerge from a past singularity\index{past singularity}, then the most likely possibility is still that its initial state\index{initial state!high gravitational entropy} was an inhomogeneous state of high gravitational entropy. But it will also become clear later on that, in the presence of negative-energy matter, it is not possible for inflation, alone, to produce a homogeneous state out of a heterogeneous distribution of positive- and negative-energy matter inhomogeneities. In fact, even if one assumed that our universe\index{universe!local fluctuation in extended vacuum} emerged out of a fluctuation in the extended vacuum state that would follow this initial expansion phase, there would still be no reason to assume that the state that would be produced by this second phase of inflationary expansion\index{inflationary expansion!second phase} would itself have a low entropy, unless we assume that classical (unidirectional) causality\index{unidirectional causality!forward-in-time operation} operates forward in time (as I explained above), which, again, amounts to simply assume the validity of the result we are seeking to derive.

If we believe that the initial conditions at the Big Bang must be subjected to the same constraint of likeliness\index{constraint of likeliness!initial conditions} as applies to the configurations of matter which are reached through random evolution under more general circumstances, then the fact that it does not appear that this initial state could have been produced by chance alone, means that there must be an explanation for this anomaly, but this explanation cannot be found in the usually favored cosmological models\index{favored cosmological models!inflation theory} based on inflation theory.

It must be noted, again, that the anthropic principle\index{anthropic principle} would be of no use in trying to achieve such a goal, because, if the initial conditions\index{initial conditions!free determination} are freely determined, they would not be required to be as highly constrained as they appear to have been when matter emerged from its initial state of maximum positive and negative matter energy densities. A state as homogeneous as that which appears to have existed in the remote past is so unlikely to have arisen randomly that, even the chance occurrence of an observer\index{observer!chance occurrence} in a universe with a thermodynamically less favorable initial state\index{initial state!thermodynamically less favorable} would be a more likely phenomenon, in comparison. If the universe initially had such a low gravitational entropy\index{gravitational entropy!low initial value}, it is because it necessarily had to go through such a constrained state at least once in its lifetime. What I will now explain is why this conclusion should have been expected all along.

\section{The horizon problem and irreversibility\label{sec:4.9}}

So, here we are, having actually ruled out the possibility that the high degree of homogeneity of the matter distribution in the primordial universe could be due to any conventional or inflationary smoothing processes\index{smoothing processes!conventional or inflationary}, but with apparently no option left to explain this remarkable fact. Although this outcome may be quite perplexing, the attentive reader may have already perceived a glimmer of light on the cosmological horizon. Indeed, when one carefully examine the causes of the failure of inflationary scenarios at explaining thermodynamic time asymmetry\index{thermodynamic time asymmetry}, I think that one cannot avoid getting the feeling that it is the very fact that there exists a state of maximum positive and negative matter energy densities\index{maximum matter energy densities} in the remote past that must constitute the basis of a consistent explanation of the origin of the anti-thermodynamic evolution\index{anti-thermodynamic evolution!past direction of time} that is taking place in the past direction of time in our universe, because this is the only aspect of our universe which is correlated with the state of minimum gravitational entropy\index{gravitational entropy!minimum}.

What I will now explain is that there is actually a requirement for the magnitude of the densities of positive and negative matter energy to be maximum at a certain point in the history of the universe that does not just follow from the fact that space must expand or contract and this actually allows to explain why it is that the universe did not came into existence in an extended vacuum state\index{extended vacuum state} with negligible positive and negative matter energy densities. But, quite remarkably, this same requirement is also responsible for having produced a maximum-density state so exceptionally configured that it guarantees that all future evolution will take place irreversibly.

Before embarking on an explanation of how it can be that an arrow of time\index{arrow of time} was allowed to emerge as a result of the existence of a past singularity\index{past singularity}, however, I would like to first recall my earlier discussion from section \ref{sec:4.5} concerning the requirement for the universe\index{universe!requirement of null energy} to have a null energy and the fact that it is possible in principle for matter energy\index{matter energy} to be compensated by the energy of the gravitational field\index{gravitational field energy} through a variation of the kinetic energy of expansion\index{kinetic energy of expansion}.

There, I mentioned that the zero-energy condition\index{zero-energy condition}, alone, does not require that the energy of matter be uniformly distributed, because, even if this constraint may require gravitational energy itself to be null globally, it would not prevent the energy of matter and that of the gravitational field from varying in opposite ways \textit{locally}, as long as there is an overall compensation between all positive and negative contributions to the energy of the gravitational field. But if the energy of the matter that was present in the very first instants of the Big Bang had been as inhomogeneously distributed as it can be, macroscopic black hole event horizons\index{macroscopic black hole!event horizon} (any event horizon produced by a black hole more massive than an elementary black hole\index{elementary black hole} with a mass equal to one Planck mass\index{Planck mass}) would abound in the early universe. So, why is it that observations rather seem to show that the initial distribution of matter energy\index{initial distribution of matter energy!homogeneity} was highly uniform, with very few black holes?

One can only begin to understand the cause of the homogeneity of the matter distribution that emerged out of the past singularity\index{past singularity} when one acknowledges that what is significant with our current description of the physics of the early universe is the explicit assumption that the cosmological horizon\index{cosmological horizon} (sometimes called the particle horizon\index{particle horizon}) begins to grow at the exact moment when the average density of matter is maximum. But why should causality\index{causality} have anything to do with the magnitude of the average densities of positive and negative matter energy?

I must admit that I always had difficulty accepting the very validity of the notion that the universe\index{universe!set of causally independent entities} could have come into existence as a set of independent entities, not causally related to one another, due to the presence of non-overlapping cosmological horizons in the primordial state. The conclusion that the limited velocity of causal signals\index{causal signals!limited velocity} would forbid interactions between sufficiently distant regions of the universe, however, appears unavoidable. The conventional idea is that this is not a problem as long as causal relationships\index{causal relationships} can be established at later times in the future through the propagation of effects\index{propagation of effects!relativistic velocities} at relativistic velocities. This is what would allow the causally unrelated parts\index{causally unrelated parts!Big Bang} which are assumed to exist at the Big Bang to form one single universe.

What also seemed problematic to me was the assumption that the cosmological horizon\index{cosmological horizon} must begin to grow at the precise moment when the density of matter and radiation is maximum (or infinite, as one would assume from a classical viewpoint). This idea appears to be motivated by the hypothesis that time begins with the past singularity\index{past singularity!beginning of time}. Here, again, I just couldn't understand the appropriateness of a picture that most people accepted as valid without a second thought. But this led me to develop a better understanding of the conditions imposed by the principle of local causality\index{principle of local causality!initial Big Bang state} on the initial Big Bang state that turned out to be crucial for explaining the high degree of homogeneity of the primordial matter distribution\index{primordial matter distribution!homogeneity} that is responsible for the existence of the thermodynamic arrow of time\index{thermodynamic arrow of time} and unidirectional causality itself.

First of all, I think that it is important to mention that the notion that the size of the cosmological horizon increases with time, as the universe itself expands, contains an implicit assumption that is not always recognized for what it is. Indeed, when one considers that the horizon encompasses an increasingly larger portion of space in the future, one is actually presuming the validity of the classical principle of causality\index{classical principle of causality}, that is, of the idea that causes always precede their effects. But it is actually always \textit{past} causes that produce \textit{future} effects. It is never assumed that a future cause could produce an effect in the past. This is usually appropriate, as we experience time in a unidirectional way as a consequence of the fact that the thermodynamic arrow of time always operates from past to future and never in the opposite direction.

But when we are considering that no signal was allowed to propagate farther than the distance reached by the cosmological horizon\index{cosmological horizon} at any given time after the Big Bang, we are implicitly assuming that it is only the past that can influence the future and that effects propagate in the future direction of time, from causes which originate in the initial singularity\index{initial singularity}. In other words, we are assuming the existence of a preferred direction in time (the future) and a preferred instant (that of the past singularity) at which effects begin to propagate. It must be clear, however, that this is an assumption and that there is no \textit{a priori} reason not to assume that classical causality instead operates toward the past from the instant at which a hypothetical future Big Crunch singularity\index{future Big Crunch singularity} would be formed, in which case the size of the cosmological horizon would already encompass all of space, or at least a very large portion of it, at the Big Bang.

The truth is that even classical causality\index{classical causality} could begin to operate at any given instant of time, in both the past or the future, if entropy is not already maximum at this particular moment, and under such conditions it would appear more natural to assume that the cosmological horizon begins to spread at the present moment, both forward and backward in time. Therefore, if what we are seeking to explain is the existence of a preferred direction in time\index{preferred time direction}, then we cannot simply assume the validity of the classical concept of a cosmological horizon expanding from the Big Bang in the future direction of time. We cannot claim that there is a problem with the homogeneity of the large-scale matter distribution\index{large-scale matter distribution!homogeneity problem}, if this problem arises as a consequence of assumptions concerning the size of the cosmological horizon which are only meaningful in the context where there is a preferred direction to causal signals\index{causal signals!preferred direction} which originates from this very same homogeneity. What we must provide is a consistent justification for the very validity of this particular choice of a horizon concept. We must explain why this particular state in the past was configured in such a way that it allowed classical (unidirectional) causality to be a meaningful concept that came into effect at the exact moment when space itself began to expand out of the past singularity.

Despite those difficulties, I came to recognize the validity of the limitations imposed by the existence of cosmological horizons. I believe that what allows this concept to be acceptably formulated is simply the fact that, ultimately, as we consider increasingly earlier times, the size of the horizon would actually reach the limit imposed by quantum theory on the classical definiteness of any measure of spatial distance\index{distance!limit to classical definiteness}. When the size of the cosmological horizon passes below the limit at which the uncertainty that is intrinsic to quantum phenomena would apply to spacetime relationships\index{spacetime relationships!quantum uncertainty} themselves, it is certainly no longer appropriate to assume that the limited velocity of signal propagation\index{signal propagation!limited velocity} forbids the existence of causal relationships\index{causal relationships} between regions of space separated by distances larger than the size of the cosmological horizon, but smaller than this characteristic scale of quantum-gravitational phenomena\index{quantum-gravitational phenomena!characteristic scale}, as there are no classically well-defined relationships of distance and duration below that scale.

In such a context, it would be suitable to assume that there may, after all, exist causal relationships\index{causal relationships} between all components of the universe which were in contact with one another to within an elementary quantum-gravitational unit of area\index{quantum-gravitational unit of area} at the Planck time\index{Planck time}, if we also have good reasons to expect that the area delimited by the cosmological horizon\index{cosmological horizon!elementary unit of area} was then equal to this elementary unit of area, within which the gravitational field and the metric properties of spacetime\index{metric properties of spacetime!quantum indefiniteness} where submitted to quantum indefiniteness. In fact, from a quantum-gravitational viewpoint\index{quantum-gravitational viewpoint}, it may be preferable to simply recognize that there is nothing smaller than the elementary unit of surface\index{elementary unit of surface} associated with this particular scale. But given that causality is a feature of the classical (relativistic) spacetime structure\index{classical spacetime structure!causality}, this means that there would be no sense in imposing limitations on signal propagation\index{signal propagation!limitations} below that scale.

Therefore, when the size of the cosmological horizon reaches the natural limit imposed by quantum gravitation\index{quantum gravitation}, as it contracts in the past, if the most-elementary particles\index{particles!most-elementary} of matter and radiation are allowed to be in contact with one another to within such an elementary unit of area (equal to a small multiple of the Planck unit of area\index{Planck unit of area}), then no smaller components would remain causally unrelated in the initial Big Bang state, which is probably sufficient a condition to impose, regarding the necessity for the universe\index{universe!ensemble of causally interrelated elements} to form a global ensemble whose elements (the elementary particles) are allowed to remain causally interrelated at all times, as a result of having been in direct contact with one another when the size of the cosmological horizon\index{cosmological horizon!minimum size} was minimal.

Now, this simple formulation of the requirement which I believe allows the universe to exist as the ensemble of all those things which are physically related to one another and to nothing else may appear benign, even if adequate, but in fact it can be attributed the most amazing consequences, in the context where it is recognized that a density of negative-energy matter as large as that of positive-energy matter must have existed in the initial Big Bang state\index{initial Big Bang state!negative-energy matter} (for reasons I have explained in section \ref{sec:4.5}). Thus, I would suggest that all the elementary particles originally present in our universe at the Big Bang be required to have been in contact with at least one other particle to within an elementary quantum-gravitational unit of area\index{quantum-gravitational unit of area} at the Planck time\index{Planck time}. More specifically, I propose that the following condition must apply.
\begin{quote}
\textbf{Global entanglement constraint}\index{global entanglement constraint}: There must exist a entire space-like hypersurface\index{space-like hypersurfaces}, at one particular moment of cosmic time\index{cosmic time}, over which all elementary particles, regardless of their energy sign, are in contact with at least one neighboring elementary particle of either positive or negative energy sign to within an elementary quantum-gravitational unit of area\index{quantum-gravitational unit of area}, in a state of maximum positive and negative matter energy densities\index{maximum matter energy densities}.
\end{quote}
If this condition is fulfilled, then any elementary particle that is present in the universe today would have been in direct contact with another elementary particle that was in contact with another such particle, and so on, when the size of the cosmological horizon\index{cosmological horizon!size of most-elementary particle} was equal that of a most-elementary particle, which means that no component of the universe\index{universe!component} could then exist that would be causally unrelated to the other components of the same universe.

Of course, the existence of such a smallest, physically significant causal horizon does not mean that the limits usually imposed by the size of cosmological horizons on the propagation of causal signals\index{causal signals!propagation} no longer apply, but merely that they need not apply at times earlier than the Planck time\index{Planck time} after the initial singularity\index{initial singularity}. It must be clear that there would be no sense in speaking about the ultimate horizon as being that which would be associated with the epoch at which the whole visible universe would be contained within a single, elementary, quantum-gravitational unit of surface\index{quantum-gravitational unit of surface}, because, once the magnitude of the average densities of positive and negative matter energy is maximum and there is a matter particle with a Planck energy\index{Planck energy} in every elementary unit of area\index{elementary units of area}, then no further contraction is possible, as all tentative quantum theories of gravitation\index{quantum gravitation theories} appear to confirm.

What this means is that it wouldn't even make sense to impose a condition of causal contact\index{causal contact condition} on a state that would be reached at an even earlier time, to which would be associated a measure of area smaller than the fundamental unit provided by quantum-gravitational theories. But even if the constraint of global entanglement\index{global entanglement constraint} concerns the state of the universe at the Planck time, it would be incorrect to assume that only the detailed knowledge of a fully developed quantum theory of gravitation would allow us to say anything meaningful regarding the state of the universe at such an early time. Anyhow, we still need to explain why it is that the positive- and negative-energy matter distributions\index{matter distribution!almost perfect homogeneity} were almost perfectly homogeneous, on a scale larger than the size of the cosmological horizon\index{cosmological horizon!maximum matter energy densities} at the time when the magnitude of the densities of positive and negative matter energy was maximum, as required if the growth of this cosmological horizon\index{cosmological horizon growth!unidirectional phenomenon}, as a unidirectional phenomenon, is to actually begin at that particular instant of time. This is a particularly difficult question given that large-scale homogeneity is precisely what would appear to be forbidden by the existence of such a horizon.

The implications of the global entanglement constraint only emerge in the context where it is recognized that event horizons (such as those associated with black holes) can constitute potential barriers\index{potential barrier} which are impossible to overcome. It must be clear, first of all, that even though certain positive-energy particles could be prevented from coming into contact with other positive-energy particles, as a consequence of being contained within the macroscopic event horizons of positive-energy black holes, if only positive-energy matter existed this would not justify imposing a limit on the amplitude of primordial density fluctuations\index{primordial density fluctuations!amplitude limit}, because in such a case, regardless of the presence of macroscopic event horizons\index{macroscopic event horizon}, all matter particles would eventually end up being in contact with their neighbors in the initial state of maximum matter density, because the contraction of space\index{contraction of space!past direction of time} that takes place backward in time, toward the initial Big Bang state, would lead to the merger of all the event horizons which were originally present and their spacetime singularities, as in a generic Big Crunch\index{Big Crunch!generic} process.

Under such conditions, all the particles which may now be isolated by the presence of event horizons would nevertheless merge into one initial state\index{initial state!maximum positive matter energy density} of maximum positive matter energy density, where every matter particle with a Planck energy\index{Planck energy} would occupy an elementary unit of area\index{elementary units of area} and be in contact with the surrounding particles present in this initial singularity\index{initial singularity}. Thus, if only positive-energy matter was present in our universe, it would seem that the global entanglement constraint\index{global entanglement constraint} could be satisfied in the initial state without gravitational entropy\index{gravitational entropy!minimum} being minimal, because even if strong local gravitational fields and macroscopic event horizons existed in the instants immediately preceding the formation of the initial singularity (in the past direction of time), all elementary particles would nevertheless be allowed to come into contact with their neighbors in the maximum-density state, because those are attractive gravitational fields.

When negative-energy matter is present, however, things become more complicated, because gravitational repulsion\index{gravitational repulsion}, unlike gravitational attraction, may forbid local contacts, by giving rise to insurmountable potential barriers\index{potential barrier!insurmountable} for those particles which are confined within the event horizons of \textit{macroscopic} black holes\index{macroscopic black hole!opposite energy signs} with opposite energy signs. Therefore, if the constraint of global entanglement requires contact between neighboring elementary particles at the Planck time\index{Planck time}, regardless of their energy sign, it follows that event horizons can be expected to be absent initially on all but the smallest scale, even though macroscopic black holes are allowed to form at later times. If this was not the case, then certain particles could exist in our universe that would not be causally related to the rest of it, which I believe would involve a contradiction.

In the absence of a condition of global entanglement\index{global entanglement condition}, the most likely initial state, from a purely statistical viewpoint, would be one for which all the matter in the universe would be concentrated in the smallest possible number of opposite-energy black holes\index{opposite-energy black holes!arbitrarily large masses} with arbitrarily large masses which would already be in a state of maximum gravitational entropy\index{gravitational entropy!maximum}. But this was not allowed to constitute our boundary conditions\index{boundary conditions!Big Bang} at the Big Bang simply because, under such conditions, the spacetime singularities\index{spacetime singularity} at the center of the objects could never come into contact with one another in the state of maximum matter density\index{maximum matter density state} or at any other time (because gravitational entropy\index{gravitational entropy} can only grow as time passes), while this is required by the global entanglement constraint.

In the presence of negative-energy matter, global entanglement actually constitutes a very constraining requirement, because any sufficiently large fluctuation in the initial density of positive or negative matter energy would give rise to the presence of a macroscopic event horizon\index{macroscopic event horizon} that would forbid the condition from applying. Therefore, such large fluctuations in the density of positive- and negative-energy matter must be completely absent in the first instants of the Big Bang and can only develop gradually at later times, in an initially smooth and homogeneous matter distribution. The mass of any black hole that is now present in the universe must, therefore, diminish continuously in the past direction of time, as we approach the initial singularity\index{initial singularity}, so as to allow the condition of homogeneity imposed on the initial matter distribution\index{initial matter distribution!condition of homogeneity} to be satisfied, despite the fact that it is actually the past condition that gives rise to the future configuration, in the context where the condition that applies on the initial Big Bang state\index{initial Big Bang state!minimum gravitational entropy} is, in effect, one of minimum gravitational entropy, from which the classical (unidirectional) principle of causality\index{classical principle of causality} itself can be expected to emerge.

What was so puzzling about the previously unexplained fact that an ever smaller number of microscopic configurations seems to be available for matter evolving in the past direction of time, under the influence of the gravitational interaction, was that no such a decrease in the number of allowed microscopic states of matter\index{microscopic state of matter!allowed states} is observed in the future direction of time. As a consequence of this limitation, predictions of a statistical nature, such as those made using quantum theory\index{quantum theory!statistical predictions}, are always valid only for evolution toward the future, while evolution toward the past cannot, in general, be accurately predicted (the probability of prior events cannot be determined from that of subsequent events, while the probability of future events can usually be determined from that of past events), which is annoying, given that the equations of the theory are symmetric under a reversal of the direction of time.

But this is not a consequence of the fact that information concerning the state in which a system will evolve is only available for the past and not the future, because it is possible to recognize, retrospectively, the absence of statistically significant constraints\index{statistically significant constraint!future evolution} that would apply to future evolution by considering the future of an initial state at a time when this future is now itself in the past. This is in contrast with the evolution that can be observed to take place at the same time toward the past and which reveals that systems can only come to occupy a subset of their theoretically allowed microscopic states whose only distinctive property is its lower entropy.

What remained unexplained, therefore, is the fact that an ensemble of systems started in the same macroscopic state evolves to occupy all available microscopic states in the future, while a similar ensemble, started in the same macroscopic state, usually evolves only to past states characterized by a lower entropy and more particularly, a lower gravitational entropy\index{gravitational entropy!lower past values}. But I have now explained that this diminution in the number of allowed microscopic states\index{microscopic state of matter!allowed states} toward the past originates from the necessity that all the elementary particles present in the initial state\index{initial state!maximum matter energy densities} of maximum positive and negative matter energy densities at the Big Bang come into contact with their neighbors of any energy sign, in order that there exist causal relationships\index{causal relationships!components of universe} between all independently evolving components of the universe.

The unnatural evolution that takes place in the past direction of time is a direct consequence of the limitation imposed on the initial state by the constraint of global entanglement\index{global entanglement constraint} in the presence of negative-energy matter and it would not merely characterize a small portion of all possible universes, but really all universes governed by the known, fundamental principles of physics\index{fundamental principles of physics}, in which negative-energy matter\index{negative-energy matter!initial density} is present with an initial density whose average magnitude equals that of positive-energy matter. Remarkably enough, this unrecognized, but necessary condition allows to explain why it is that only the gravitational component of entropy\index{gravitational entropy!non-maximum initial value} was not maximum at the Big Bang, while the entropy of matter\index{matter entropy!arbitrarily large initial value} was allowed to be arbitrarily large, which is certainly appropriate given that the universe was then already in a state of thermal equilibrium\index{thermal equilibrium state!universe}. The constraint of global entanglement only limits the magnitude of entropy attributable to the gravitational field and this is exactly what we need from an observational viewpoint.

It must be clear that the fact that a perfectly uniform distribution of negative-energy matter exerts no gravitational influence on positive-energy matter does not allow one to assume that it is not necessary to take into account the presence of negative-energy matter in trying to identify the nature of the constraint that gives rise to a homogeneous, initial distribution of positive matter energy, because it is precisely the magnitude of local inhomogeneities in the distribution of positive and negative matter energy which needs to be constrained and negative-energy matter inhomogeneities do have an effect on positive-energy matter. In fact, negative-energy matter always exerts an influence on positive-energy matter under conditions of maximum matter density and null average energy density, because, locally, elementary black holes\index{elementary black hole} are necessarily present (as I have explained while discussing the problem of black hole entropy\index{black hole entropy!problem} in section \ref{sec:3.10}) and the energy distribution is never perfectly smooth and homogeneous, given that two elementary black holes with opposite energies cannot be located in the same elementary units of area\index{elementary units of area}.

The constraint of global entanglement\index{global entanglement constraint}, therefore, merely imposes that the positive- and negative-energy matter particles which were present in the initial state\index{initial state!maximum matter energy densities} of maximum matter energy densities be as homogeneously distributed as necessary for an absence of \textit{macroscopic} event horizons\index{macroscopic event horizon} with an area larger than one elementary unit of area, that would contain more than one matter or radiation particle, because it is only under such conditions that the most-elementary particles\index{particles!most-elementary} (with the highest possible positive and negative energies), submitted to the gravitational fields of the smallest elementary black holes (with an event horizon area equal to one elementary unit of area), can all be in direct contact with one another, regardless of their energy signs, when the area delimited by the cosmological horizon\index{cosmological horizon!smallest unit of area} is no larger than the smallest physically significant unit of area, as required if they are to be part of the same universe.

It is appropriate to impose such a condition, even in the context where one recognizes that there are no direct interactions between positive- and negative-energy particles\index{opposite-action particles!absence of direct interactions}, while there nevertheless exists an indirect repulsive gravitational force\index{indirect repulsive gravitational force} between opposite-energy objects which is strong enough to prevent macroscopic opposite-energy black holes\index{macroscopic opposite-energy black holes!absence of contact} from coming into contact with one another, because what is involved here is not an interaction propagated across space and time, but a minimum unit of area\index{minimum unit of area!causal relationships} within which there need not even be an exchange of energy (mediated by an interaction boson\index{interaction bosons}) for causal relationships to exist. The very meaningfulness of the constraint of global entanglement is, in fact, dependent on the hypothesis that there exists a minimum, physically significant unit of area\index{minimum physically significant unit of area}, within which no causal signal\index{causal signals!propagation} needs to propagate for entanglement to be established and which corresponds precisely to the area of the event horizon of the smallest elementary black holes, containing only one elementary particle\index{elementary particle!maximum energy} with a maximum positive or negative energy.

The particles which are under the influence of two such elementary black holes\index{elementary black holes!particle contact} with opposite energy signs are not forbidden from coming into contact, despite the fact that they cannot interact directly with one another, because there is no distinction between the size of an elementary particle present on the quantum-gravitational scale\index{quantum-gravitational scale!size of elementary particle} and that of the event horizon of the smallest elementary black hole\index{elementary black hole!smallest} that may contain it. This particularity is precisely what allows to explain that the information which is encoded on the event horizon\index{event horizon!availability of information} of such an elementary black hole can be obtained by an observer of any energy sign, while that which is encoded on the event horizon of macroscopic black holes\index{macroscopic black hole} always remains inaccessible to observation, as I explained in section \ref{sec:3.10}.

This is an important point, because if the possibility for two elementary black holes\index{elementary black holes!entanglement} with minimum surface areas and opposite energies to become entangled was assumed to mean that the matter particles inside two \textit{macroscopic} opposite-energy black holes\index{macroscopic black holes!opposite energies} can themselves come into contact with one another despite gravitational repulsion\index{gravitational repulsion}, then we would have to conclude that the particles inside a macroscopic positive-energy black hole could become entangled with those inside a similar negative-energy black hole, even though the opposite-energy singularities\index{opposite-energy singularities} (or the surfaces enclosing the states of maximum energy density\index{maximum energy density state} at the center of the black holes) cannot themselves come into contact with one another.

Therefore, it is merely the fact that the elementary particles inside a macroscopic, positive-energy black hole cannot come into contact with those inside a macroscopic, negative-energy black hole, that constitutes a limitation to the existence of causal relationships\index{causal relationships!particles inside opposite-energy black holes} between those particles. If all the matter in the universe was initially concentrated in two macroscopic black holes with opposite energy signs, the elementary particles inside one of the objects would necessarily remain causally independent from those inside the other black hole, due to the insurmountable gravitational repulsion\index{gravitational repulsion!opposite-energy black holes} that exists between the two objects and this is what explains that it is not possible for gravitational entropy to be maximum in the initial Big Bang state\index{initial Big Bang state!gravitational entropy}. But the presence of extremal, macroscopic black holes\index{extremal black hole} is also forbidden, given that opposite-energy black holes do not interact electromagnetically with one another, so that the gravitational repulsion\index{gravitational repulsion!extremal black holes} they exert on one another cannot be neutralized by an attractive electromagnetic force.

What's interesting is that, contrarily to the situation we would have if inflationary expansion\index{inflationary expansion} was assumed to be responsible for the smoothness of the initial matter distribution\index{initial matter distribution!smoothness}, it is now possible to explain why it is that the constraint that gives rise to a homogeneous initial state is necessarily effective in only one direction of time. Thus, gravitational entropy\index{gravitational entropy!continuous decrease} can be expected to decrease continuously in the past direction of time from its current intermediary value, even if this would appear to be a very unlikely evolution for the universe to go through from a statistical viewpoint, because if the present inhomogeneity is not reduced, then the smooth merger of the positive- and negative-energy matter distributions\index{opposite-energy matter distributions!smooth merger} that is required for the global entanglement of all particles to take place would not happen. This reduction of gravitational entropy can now be understood to occur regardless of whether space is expanding or contracting, as long as we are, in effect, approaching the instant at which is formed the unique past singularity\index{past singularity} on which the condition of global entanglement\index{global entanglement condition} is imposed.

It is, therefore, simply the fact that the condition that applies to the initial singularity is precisely one of minimum gravitational entropy\index{gravitational entropy!minimum initial value}, from which can emerge a phenomenon of unidirectional causality\index{unidirectional causality} that operates toward the future, from that particular instant of time, that requires the evolution that takes place at all later times to be such that it allows an initial state obeying this condition to be reached in the past direction of time, because under such circumstances causality must, in effect, operate in one unique direction of time, thereby allowing past causes to produce irreversible effects in the future.

If quantum theory only works for predicting future events it is because all possibilities are allowed for evolution toward the future, while only a limited subset of potentialities can be actualized in the past, as a consequence of the constraint that is continuously being exerted on past evolution by the condition of global entanglement\index{global entanglement condition}, which imposes a minimum gravitational entropy on the state which existed in the past, when the density of matter was maximum. It is quite remarkable that this apparent backward teleology\index{backward teleology!apparent} can be shown to arise from the existence of an inescapable constraint that applies on one particular state only, but even more surprising is the fact that this can be achieved despite the time symmetry\index{time symmetry!physical laws} of the physical laws involved, which gave no hint of having the potential to produce such a manifestly irreversible evolution\index{irreversible evolution}.

It is important to emphasize that, in the context of this explanation of temporal irreversibility\index{temporal irreversibility!explanation}, all physical systems, regardless of how isolated they may have become at the present time, must evolve with continuously decreasing gravitational entropy\index{gravitational entropy!continuous decrease in past} in the same past direction of time, because they are all submitted to the same unavoidable constraint applying to the same unique past state of maximum positive and negative matter energy densities\index{maximum matter energy densities!unique state}. This constraint, therefore, is stronger than any condition that would be imposed independently on the present state of one or another system in order to favor an evolution to lower entropy states in a given arbitrarily-chosen direction of time.

Given that all microscopic processes are fundamentally unpredictable, a constraint that would apply merely to the present microscopic state of a system that is not in a thermal equilibrium state\index{thermal equilibrium state} could not, alone, impose on this system that it evolves with decreasing gravitational entropy\index{gravitational entropy!decrease} over a very long period of time in either the past or the future, regardless of how carefully the system is prepared. The global entanglement constraint\index{global entanglement constraint}, however, necessarily applies to all physical systems which are part of the same universe and exerts its influence incessantly in the same unique direction of time (toward the initial singularity\index{initial singularity}) and in such a way gives rise to a temporal asymmetry\index{temporal asymmetry!shared property} which is actually shared by all systems, including any branch systems\index{branch systems} which may no longer be in contact with their environment. In the present context, this temporal parallelism\index{temporal parallelism} is a simple consequence of the fact that all physical systems in the universe are led by a common condition which applies to the state they occupied when the cosmological horizon\index{cosmological horizon} began to spread and which originates from the requirement that they actually be part of the same universe.

If the initial or final conditions\index{initial or final conditions!current microscopic states} imposed on current microscopic states cannot alone explain the temporal parallelism of branch systems, it is because, even if this would be possible for the evolution that takes place in the future direction of time, the fact that, for all practical purpose, such isolated systems\index{isolated systems} never evolve toward a state of higher gravitational entropy\index{gravitational entropy!increase} in the past direction of time, like they do in the future, but rather nearly \textit{always} evolve to even lower gravitational entropy\index{gravitational entropy!decrease} states in this direction of time, means that it is not the conditions imposed on their current microscopic states which alone determine their past evolution.

It is precisely the fact that the condition of global entanglement\index{global entanglement condition} must, as a matter of consistency, apply to all particles in the universe that guarantees that every branch system\index{branch systems}, without any exception, must obey the same constraint of decreasing gravitational entropy in the same direction of time toward the initial singularity\index{initial singularity}. The parallel thermodynamic behavior\index{parallel thermodynamic behavior!isolated branch systems} of isolated branch systems can be expected to occur as a result of the fact that any system that is part of a given universe, regardless of how isolated it might have become, must have been entangled with the rest of the matter in this universe at the Big Bang in order that causal relationships\index{causal relationships!components of universe} be established between all components of the universe and this implies that even those portions of the universe which are now isolated must follow the same kind of gravitational entropy decreasing evolution that is necessary for achieving this global entanglement at some point in the past.

The parallelism of the asymmetry of thermodynamic evolution\index{asymmetry of thermodynamic evolution!parallelism} can only be explained if there exists a constraint that requires the diminution of the gravitational entropy\index{gravitational entropy!decrease} of all systems in the past direction of time, regardless of how isolated they have become and independently from what their initial or final microscopic states\index{microscopic states!initial or final} are at the present time and the fact that such parallelism is actually observed under all circumstances clearly shows the validity of the arguments that allowed me to determine the nature of this constraint. Thus, even though the boundary conditions\index{boundary conditions!Big Bang} at the Big Bang are fixed arbitrarily, it is not impossible for a constraint to apply on those conditions that gives rise to irreversible evolution\index{irreversible evolution}, even in the context where the fundamental physical laws\index{fundamental physical laws!time reversal invariance} which apply under such conditions are all invariant under a reversal of the direction of time.

What is important to understand is that a state of maximum positive and negative matter energy densities\index{maximum matter energy densities!necessary occurrence} must necessarily occur at one moment or another of unidirectional time for the global entanglement\index{global entanglement} of all elementary particles to be achieved in the presence of negative-energy matter and given that such a state would not likely be characterized by an absence of macroscopic event horizons\index{macroscopic event horizon!absence} unless it constitutes the mandatory unique event at which global entanglement is enforced on the universe, then one must conclude that our Big Bang really is this unique event. In such a context, the presence of an initial singularity\index{initial singularity!universe existence requirement} would no longer be a mere fortuitous consequence of the fact that space is expanding, but would be an essential requirement for the existence of any universe obeying the known fundamental principles of physics.

I believe that it is the widespread ignorance of this fact that explains that it took so much time for all the consequences of the presence of a Big Bang to be properly understood and appreciated. To the usual three elements of observational evidence in favor of the existence of a Big Bang\index{Big Bang!three elements of observational evidence} which are the observation that space is expanding, the accuracy of the prediction of light element abundances\index{light element abundances!prediction}, and the detection of the cosmic microwave background\index{cosmic microwave background!detection}, I would, therefore, suggest that one adds the theoretical argument concerning the very necessity of a state of maximum matter density, which is made conspicuous by the undeniable character of our shared experience of a thermodynamic arrow of time\index{thermodynamic arrow of time!shared experience}.

Once we recognize that there actually exists an independent requirement for the presence of an initial state of maximum positive and negative matter energy densities\index{maximum matter energy densities!necessary occurrence} in the history of our universe, then any attempt at explaining the apparent unlikeliness of the homogeneous distribution of matter particles in this initial state\index{initial state!homogeneous matter distribution}, by assuming that it is the consequence of an early phase of inflationary expansion\index{inflationary expansion} that originated from a local fluctuation in a high-entropy, extended vacuum state\index{extended vacuum state!local fluctuation}, can no longer be considered satisfactory. Indeed, if it is governed by the principle of local causality\index{principle of local causality} (as an unavoidable feature of relativity theory\index{relativity theory}), such an extended space would need to have emerged out of a prior state of maximum positive and negative matter energy densities at which global entanglement\index{global entanglement} would have been allowed to take place, which means that this state would necessarily have minimum gravitational entropy\index{gravitational entropy!minimum}, regardless of whether inflation happened or not\footnote{
This conclusion is made even more unavoidable in the context where it is understood that for the universe to expand and the kinetic energy of expansion\index{kinetic energy of expansion!positive|nn} not to be null, in a universe\index{universe!zero energy|nn} with zero energy, matter must be present initially that contributes to the negative gravitational potential energy\index{gravitational potential energy!negative|nn} that must be balanced by this positive kinetic energy of expansion, so that it is not possible to conceive of a space that would have always existed in an extended vacuum state\index{extended vacuum state|nn}.}.

Thus, comments to the effect that it would become impossible to explain the existence of a thermodynamic arrow of time\index{thermodynamic arrow of time} if there only existed one single universe in all of space and one single Big Bang in all of time appear to be misguided, because, in fact, it seems that the truth is to be found in the exact opposite statement. It is as a result of having tried very hard to understand why it is that there should, in effect, be a unique initial state of maximum matter density\index{initial state!maximum matter density} for the universe, by first acknowledging that this is a perfectly legitimate hypothesis, that I was allowed to achieve progress in identifying the cause of the homogeneity of the initial distribution of matter and energy that gave rise to temporal irreversibility\index{temporal irreversibility}.

If gravitational entropy\index{gravitational entropy!increase} does indeed rise in only one particular direction of time, it is because only evolution away from the initial singularity\index{initial singularity}, either in the future or in the past, can be expected to be left unaffected by the constraint of global entanglement\index{global entanglement constraint}, which actually gives rise to a well-defined thermodynamic arrow of time, independent of whether space is expanding or contracting. It is, therefore, possible to understand why it is that classical causality\index{classical causality} operates from past to future in the portion of history that follows the initial singularity and also why it is that the cosmological horizon\index{cosmological horizon} only begins to spread outward at the Big Bang. It is the fact that the constraint of global entanglement would only be required to apply once, even if the universe was to return to a state of maximum matter density at some point in the future, that explains that the evolution that takes place from the moment at which this condition is enforced is not symmetric in time.

Thus, it is incorrect to argue, as certain authors do, that in order not to assume the very outcome we are seeking to derive (the temporal irreversibility\index{temporal irreversibility}), it is required that any condition that applies to some initial state should also apply to a final state of the universe's history. Once it is understood that there need only be one state of maximum matter density and minimum gravitational entropy\index{gravitational entropy!minimum} in any given universe, then the kind of evolution which can be expected to take place in the direction of time toward that unique state, either in the past or in the future direction of time, would necessarily be different from that occurring in the opposite direction and this allows to explain time asymmetry\index{time asymmetry!explanation} without assuming it in the first place. What I'm assuming here is not the asymmetry itself, but merely the uniqueness of the state which allows it to arise. I'm not picking up a unique direction of time, I'm merely identifying the necessary conditions that must apply to the distribution of matter energy at one single moment of time and it just happens that those conditions are so unlikely to ever be satisfied again by chance alone that any later or earlier evolution can be expected to take place irreversibly.

Now, if \textit{bidirectional} time\index{bidirectional time!extension past initial singularity} does extend past the `initial' singularity, following a quantum bounce\index{quantum bounce}, we can expect space to be expanding at a critical rate and the density of matter to be decreasing from its maximum value immediately after the event (in the past direction of time), while the inhomogeneity of the matter distribution\index{inhomogeneity of matter distribution!minimum} would still need to be minimum if there is to be any continuity in the evolution of the microscopic state of matter\index{microscopic state of matter!continuity past initial singularity} and its gravitational field as we pass the point of maximum positive and negative energy densities. But this means that, even for the portion of history the precedes the initial singularity, the thermodynamic arrow of time\index{thermodynamic arrow of time} would (initially at least) have the same direction as the cosmological arrow of time\index{cosmological arrow of time} associated with expansion and would actually be opposite that we observe on our side in time of the initial singularity\index{initial singularity}. As a result, the area of black hole event horizons\index{black hole event horizon!growing area} and the associated gravitational entropy\index{gravitational entropy!increase} would be growing toward the past (which any observer then present would consider to be her future), which means that, in the future direction of time, the same objects would evolve as white holes\index{white hole} emerging from generic (high entropy) past singularities\index{past singularity!generic}.

It would, therefore, be inappropriate to propose that it is simply because a condition of low gravitational entropy applies to \textit{all} past singularities\index{past singularities!condition of low gravitational entropy} that the matter that emerged from the initial Big Bang singularity\index{initial Big Bang singularity} was so uniformly distributed. Anyhow, it must be clear that, if the thermodynamic arrow of time is indeed reversed as soon as the instant of the initial singularity is reached, then whatever occurred during the portion of history that preceded the Big Bang would remain unknowable to observers in the current portion of history. This would be true for the exact same reason that events located in our future cannot be known in advance, which is attributable to the fact that classical (unidirectional) causality\index{classical causality!mutually consistent records} and the formation of mutually consistent records of events only take place in the direction of time relative to which entropy is rising.

Still regarding the possibility for bidirectional time\index{bidirectional time!extension past initial singularity} to extend past the initial singularity, I believe that it would be inappropriate to assume that, if this hypothesis is valid, it would become impossible to explain the low gravitational entropy of the initial Big Bang state\index{initial Big Bang state!low gravitational entropy} by imposing a condition on the initial singularity. It was argued, in effect, that if there was a history prior to the Big Bang\index{history prior to Big Bang!final singularity}, the final singularity\index{final singularity!high gravitational entropy} which would be produced in the future direction of time (which would constitute our initial singularity) would likely be in a high gravitational entropy state (as any future state reached after a long period of random evolution), which would require the state following it (our initial state) to have a similar configuration.

But in fact, it is exactly the opposite which is true and the state preceding the initial singularity\index{initial singularity!homogeneous preceding state} must actually be very homogeneous, because the constraint of global entanglement\index{global entanglement constraint} applies to the singularity itself, while it is the evolution away from it, in \textit{any} direction of time, which is unconstrained. Continuity merely imposes that the configuration be similar on both sides of the initial state, but it does not allow one to determine what this configuration actually is. It is, in effect, only in the absence of an appropriate constraint to be imposed on the initial singularity that gravitational entropy\index{gravitational entropy!maximum} would have to be maximum in both the immediate past and the immediate future of the initial state and indeed at all times. Not recognizing this would, again, amount to favor one particular direction of time (that relative to which entropy would be assumed to grow before the initial singularity) without justification, instead of explaining why such a preferred time direction\index{preferred time direction!natural emergence} naturally emerges, as I have done.

When time is actually unfolding toward states of higher gravitational entropy\index{gravitational entropy!growth past initial singularity} past the `initial' Big Bang singularity, then the history of the universe\index{history of universe!global time symmetry} is allowed to be globally symmetric with respect to past and future, not just because the thermodynamic arrow of time\index{thermodynamic arrow of time!reversal past initial singularity} is reversed in that portion of history which is unfolding past the initial singularity, but also because it can be expected that the unequal violations of time reversal symmetry\index{time reversal symmetry!unequal violations} which explain the matter-antimatter asymmetry\index{matter-antimatter asymmetry} that is observed on our side in time of the initial singularity, will have an opposite effect in the portion of history that precedes the initial singularity\index{initial singularity!preceding history}, because the direction of propagation in time\index{direction of propagation in time!favored direction} that is favored by the violation of time reversal symmetry $T$ remains the same for that portion of history taking place before the initial singularity, while that relative to which entropy is growing and matter-antimatter annihilation\index{matter-antimatter annihilation} is happening is reversed.

But this means that, in the portion of history that may be unfolding before the Big Bang, there will be more positive-action antimatter\index{positive-action antimatter} propagating negative energy toward the past than positive-action matter propagating positive energy in the future, following the early annihilation of matter with antimatter\index{early matter-antimatter annihilation}, while the opposite is true on our side in time of the initial singularity (for reasons which were discussed in section \ref{sec:4.3}), so that, overall, symmetry with respect to the direction of time will be preserved.

Thus, it is perhaps not just possible, but actually compulsory, to assume that there is, in effect, a history, not so unlike our own, that unfolds in the past direction of time before the instant of the initial singularity. It must be clear, however, that when I say that the thermodynamic arrow of time\index{thermodynamic arrow of time!reversal} reverses when the state of maximum matter density is reached in the past, I do not mean that an observer living in this portion of history would be allowed to witness white hole\index{white hole} phenomena and other violations of the second law of thermodynamics\index{second law of thermodynamics}. In the present context, the thermodynamic arrow of time associated with the variation of gravitational entropy\index{gravitational entropy} would reverse for all processes, without any exception, at the exact moment when the universe would begin expanding, following the quantum bounce\index{quantum bounce}, and therefore no entropy-decreasing processes\index{entropy-decreasing processes} could actually be observed.

The picture that develops, therefore, is that of a universe for which gravitational entropy is growing continuously in both the future and the past of a state of maximum positive and negative matter energy densities\index{state of maximum energy densities!past and future entropy growth}, characterized by a highly homogeneous distribution of positive- and negative-energy matter particles. This irreversible evolution\index{irreversible evolution!independent from expansion and contraction} can be expected to continue regardless of whether space keeps expanding or eventually begins to recontract. But in the context where gravitational entropy\index{gravitational entropy!continuous increase} is continuously growing, as a consequence of the polarization of the positive- and negative-energy matter distributions\index{polarization of matter distribution}, it follows that, if there is an infinite amount of matter in the universe, there may never arise a state of maximum stability, equivalent to thermal equilibrium\index{thermal equilibrium}, where gravitational entropy\index{gravitational entropy!maximum} would become maximum and would no longer rise.

Under such conditions, it cannot be expected that the universe will ever evolve back to a state similar in every respects to the state in which it was at the Big Bang, because the probability that such a universal Poincar\'{e} return\index{universal Poincar\'{e} return} would occur is not merely low, it is decreasing all the time (in section \ref{sec:5.12} I will provide a decisive argument to the effect that there is not enough time for such a return to occur). We may, therefore, be justified in describing the evolution that takes place on a cosmic scale, in both the future and the very far past, as truly irreversible. The ancient view of a universe reaching its heat death\index{heat death} and remaining in this sterile randomly fluctuating state forever may well be incompatible with the most basic theoretical constraints governing its birth process and later evolution, which rather bespeak of its potential for eternal vitality.

\bigskip

\noindent At this point it is necessary to mention that even though the positive- and negative-energy matter particles which were present in the first instants of the Big Bang must be as homogeneously distributed as required to avoid the presence of macroscopic event horizons\index{macroscopic event horizon}, the constraint of global entanglement\index{global entanglement constraint} does not impose on matter energy that it be perfectly homogeneously distributed (which would be impossible anyhow, given that, even in a perfectly smooth state of maximum positive- and negative-energy matter densities\index{maximum matter density state!perfect smoothness}, the sign of energy would need to vary from one elementary black hole\index{elementary black hole!energy sign} to another). Thus, there can still be fluctuations in the densities of positive and negative matter energy on all scales in the initial state\index{initial Big Bang state!matter energy density fluctuations} that emerged from the Big Bang singularity and it is those fluctuations that would give rise to present-day structures.

But in the context where both positive and negative matter energies were, in effect, very uniformly distributed in the first instants of the known Big Bang, as a consequence of the requirement of global entanglement, if the energy of the gravitational field\index{gravitational field energy!compensation of matter energy} must everywhere compensate any non-zero energy of matter arising from unequal variations in the magnitude of positive and negative matter energy densities, in order that the universe\index{universe!null energy} have a null energy, as I proposed in section \ref{sec:4.5}, then it becomes possible to conclude that the universe\index{universe!isotropic expansion} must expand isotropically to a very good degree of precision, even in the absence of an early phase of inflationary expansion\index{inflationary expansion}, because under such conditions the kinetic energies of expansion\index{kinetic energy of expansion} experienced by positive- and negative-energy observers, which determine the energy of the gravitational field, must not vary much with position, as the amplitude of fluctuations in matter energy density\index{matter energy density!low amplitude of fluctuations} is itself very low. What's more, if it is actually the case that the expansion is nearly isotropic around every point, due to the requirement that this expansion rate be fixed by the matter density, then it follows that the matter distribution\index{matter distribution!persistence of large scale homogeneity} must remain highly homogeneous on the largest scale, as expansion proceeds.

The uniformity of the expansion rate\index{expansion rate!uniformity} also allows one to deduce that the temperature of the cosmic microwave background\index{cosmic microwave background!homogeneous temperature} should be homogeneous even on a scale larger than the size of the cosmological horizon\index{cosmological horizon} at the epoch of last scattering\index{last scattering!epoch}, because the absence of macroscopic event horizons\index{macroscopic event horizon!requirement of absence} is required on all scales and this imposes very stringent conditions on the fluctuations of matter energy density that could be observed, even on the largest scale. This means that no smoothing process\index{smoothing processes} is required to make the temperature of the cosmic microwave background uniform, because the distribution of matter energy and the expansion rate were mostly uniform on all scales right from the beginning, even if the size of the cosmological horizon decreases more rapidly than the scale factor\index{scale factor} as we approach the initial Big Bang singularity\index{initial Big Bang singularity} in the past direction of time, so that regions which are now in contact must have been causally disjoint at the epoch of decoupling\index{decoupling!epoch} (despite the existence of causal relationships\index{causal relationships} between all elementary particles\index{elementary particles} which were present in the universe at this epoch).

When one properly recognizes the limitations imposed by the global entanglement constraint\index{global entanglement constraint} on the initial state at the Big Bang, the horizon problem\index{horizon problem} simply no longer exists and no independent assumption is required to confirm the relevance of the cosmological principle\index{cosmological principle} for a description of the early universe. There is no longer any mystery associated with the fact that only one parameter (the scale factor) is required to describe the state of the universe at all but the most recent epoch. In fact, it would now appear that the cosmological principle must be obeyed as accurately as we are considering increasingly larger regions of space, corresponding to times increasingly closer to the initial singularity\index{initial singularity}.

Despite the enormous densities and the extreme temperature that characterize the Big Bang, it would, therefore, be possible to determine the general properties of the initial state with much greater precision than is usually assumed, even if we do not known what the exact unified theory of interactions\index{unified theory of interactions} is that must apply under such conditions. To be sure, the usual assumption that, in order to obtain a homogeneous matter distribution\index{homogeneous matter distribution!cosmic scale} on the cosmic scale it is necessary for the entire observable universe to have been contained within the cosmological horizon\index{cosmological horizon} at some point in the past (which would be impossible without an early phase of inflationary expansion\index{inflationary expansion}), can now be recognized as inappropriate and unnecessary.

In the context where it is indeed the magnitude of local fluctuations in the primordial distribution of matter energy\index{primordial matter energy distribution!magnitude of fluctuations} which is restricted by the global entanglement constraint\index{global entanglement constraint}, it would also follow that arguments to the effect that topological defects\index{topological defects} should have been abundantly produced during the Big Bang may no longer be as significant as they used to be. Of course, even from a conventional viewpoint, one must be careful when considering the prediction that there should have formed in the universe a large number of magnetic monopoles\index{magnetic monopoles} or cosmic strings\index{cosmic strings}, because the validity of the Grand Unified Theories\index{Grand Unified Theories} on which those deductions are based hasn't yet been experimentally confirmed. However, some of those predictions appear to be largely independent from the details of the theories from which they are derived and therefore cannot be ignored.

What I have realized is that the relatively low abundance of topological defects\index{topological defects!low abundance} may simply be a consequence of the fact that they are very-high-energy objects, similar, in certain respects, to naked singularities\index{naked singularity} of the future kind. The presence, in the early universe, of compact objects that would concentrate such large amounts of positive or negative energy in such small volumes of space may simply be incompatible with the requirement of smoothness of the initial distribution of matter energy\index{initial matter energy distribution!requirement of smoothness} which arises from the requirement of absence of macroscopic event horizons\index{macroscopic event horizon!requirement of absence} that is imposed by the constraint of global entanglement. Indeed, magnetic monopoles are sometimes described as magnetically charged black holes\index{black hole!magnetically charged} and if this characterization is appropriate, then we should certainly not expect objects of this kind to have developed out of the highly homogeneous distribution of matter and radiation energy that is required to exist in the very first instants of the Big Bang\index{Big Bang!first instants} by the condition of global entanglement.

The rarity of topological defects at the end of the GUT era\index{GUT era} may, therefore, simply be a consequence of the fact that the amplitude of fluctuations in the density of positive and negative matter energies is too small initially to allow the production of highly dense topological defects at such an early epoch. I believe that this constraint, which simply does not exist from a conventional viewpoint, imposes strong limits on the presence of topological defects. In any case, the fact that the vacuum itself is a much different phenomenon, in the context where its natural, average energy density is actually null, certainly contributes to explain why it is that conventional expectations regarding the cosmological consequences of symmetry-breaking phase transitions\index{symmetry-breaking phase transitions!cosmological consequences} are not reflected in what we already have of experimental evidence.

Those explanations may not be as satisfactory as the solutions I have provided to other aspects of the inflation problem\index{inflation!problem}, but given that, according to the most knowledgeable experts, there is only a very small chance that, in the absence of additional constraints on the initial, pre-inflationary state\index{initial pre-inflationary state!additional constraints}, the conditions could be met that would allow inflationary expansion\index{inflationary expansion} to occur and to last for a sufficiently long time that it could actually reduce the density of topological defects\index{topological defects} to acceptable levels, then we may have no choice but to recognize that the constraint discussed above provides a more solid foundation for explaining the rarity of those theoretical objects. In any case, the fact that the physics of topological defects is still relatively uncertain means that it is not possible to rule out the validity of the tentative explanation provided here, even if at this point it may not be entirely conclusive.

Now, it is sometimes argued that the distribution of matter energy was so uniform at the time when the cosmic microwave background\index{cosmic microwave background} was released that what remains unexplained is really that the temperature of the CMB was not perfectly smooth and free of any fluctuations initially. But I believe that this smoothness problem\index{smoothness problem} is a mere consequence of the fact that we do not properly understand what gives rise to the high level of uniformity of the initial distribution of matter energy\index{initial distribution of matter energy!uniformity}. It is only when we assume that perfect smoothness is produced by default that we must invoke a cause in order to explain the fact that there actually existed fluctuations in the density of matter energy as far back in time as the epoch of last scattering\index{last scattering!epoch}.

Given that, in the context of the above discussed solution to the horizon problem\index{horizon problem}, it is merely the upper bound of fluctuations in matter energy density which is constrained, then it is to be expected that certain local variations in matter energy density would necessarily be present, as the absence of macroscopic event horizons\index{macroscopic event horizon!absence} can be satisfied even when some fluctuations are present. Therefore, if the initial state\index{initial state!random choice} is still chosen randomly, as it should, it would likely not be perfectly smooth. But we do not need local perturbations that propagate across vast distances in order to give rise to the correlations observed on the largest scales in the temperature of the cosmic microwave background\index{cosmic microwave background!large scale correlations}. The cause of the existence of correlations on scales larger than the cosmological horizon\index{cosmological horizon} is the constraint of global entanglement\index{global entanglement constraint}, which requires some level of smoothness even in the presence of inhomogeneities, thereby giving rise to a certain uniformity of the distribution of matter energy\index{matter energy distribution!initial uniformity} which need not have involved the propagation of effects, as it must have been already present initially.

The usually favored approach to the problem of the origin of primordial inhomogeneities\index{primordial inhomogeneities!problem of origin} in the distribution of matter energy, which involves assuming that they arise as a consequence of the presence of irreducible quantum fluctuations\index{quantum fluctuations!initial energy distribution} in the initial energy distribution of whatever fields where then present in the vacuum, only makes sense in the context where inflationary expansion\index{inflationary expansion} is assumed to generate an otherwise perfectly homogeneous `initial' state\index{initial state!perfect homogeneity} out of a much smaller volume of space. From my viewpoint, what must be explained is not the presence of inhomogeneities, but the overall uniformity of the initial distribution of matter energy\index{initial matter energy distribution!overall uniformity}, which, in the absence of a specific constraint, should not be observed. The natural configuration for the initial Big Bang state\index{initial Big Bang state!natural configuration} is not one of perfect smoothness and there is no need to invoke a particular effect to generate the observed fluctuations, which are allowed to be present on any scale, as long as they do not violate the condition of global entanglement.

What is truly remarkable is that the spectrum of fluctuations in the initial density of matter energy which is deduced from observations of cosmic microwave background\index{cosmic microwave background!temperature fluctuations} temperature fluctuations is a (nearly) scale-independent spectrum\index{scale-independent spectrum of fluctuations} of the Harrison-Zel'dovich\index{Harrison-Zel'dovich spectrum of fluctuations} type (for fluctuations larger than the scale of the cosmological horizon\index{cosmological horizon} at the epoch of the decoupling\index{decoupling!epoch} of matter from radiation) while this is the only spectrum which, according to specialists, does not allow the magnitude of early fluctuations in matter energy\index{matter energy fluctuations!absence of divergence on all scales} density to diverge on either large or small scales and which, therefore, does not give rise to the creation of a large number of primordial black holes\index{primordial black holes}, while those are precisely the conditions which are required by the theoretical constraint of global entanglement\index{global entanglement constraint}.

In this context, it must be clear that the requirement that the spectrum of temperature fluctuations in the cosmic microwave background\index{cosmic microwave background!scale-invariant spectrum} be scale-invariant is not a unique property of inflationary cosmology\index{inflationary cosmology}. Also, the idea that only inflation allows the initial perturbations in the distribution of matter energy\index{initial matter energy perturbations!beginning of oscillation} to begin oscillating (between compressions and rarefactions) at the same epoch of cosmic time\index{cosmic time}, as required to explain the existence of harmonics in the spectrum of cosmic microwave background temperature fluctuations\index{spectrum of CMB temperature fluctuations!harmonics}, may not be justified, because any scale-invariant spectrum of fluctuations\index{scale-invariant spectrum of fluctuations!cosmological horizon} in the distribution of matter and radiation energy which would be present initially on scales larger than the cosmological horizon, would have the same consequences. But this is precisely what we can expect to occur as a consequence of the constraint of global entanglement, in the presence of negative-energy matter. Thus, even in the absence of inflationary expansion\index{inflationary expansion}, the required fluctuations would already exist in the `initial' matter distribution and those perturbations would all begin oscillating as soon as they are encompassed by the cosmological horizon, which means that on a given angular scale all maximum compressions and rarefactions\index{maximum compressions and rarefactions!primordial fluctuations} would be reached at the same time.

The fact that the power spectrum of fluctuations in the temperature of the cosmic microwave background\index{cosmic microwave background!nearly scale invariant spectrum} appears to only be \textit{nearly} scale invariant, on the other hand, may not be a consequence of inflation either. It is usually assumed, in effect, that it is because the gravitational repulsion driving inflationary expansion\index{inflationary expansion!weakening gravitational repulsion} becomes weaker as the process is occurring, that the smaller-scale fluctuations, which are produced later by the inflation process, have a smaller amplitude. But what may be happening, instead, is that, given that there remained only a negligible portion of baryonic negative-energy matter\index{baryonic negative-energy matter} following the early annihilation of matter with antimatter\index{early matter-antimatter annihilation}, then one can expect that inhomogeneities in the negative-energy matter distribution were less developed on smaller scales when the CMB was released, because, even though there existed as much negative vacuum-dark-matter energy\index{negative vacuum-dark-matter energy!small-scale fluctuations} before and after the annihilation process, no significant small-scale fluctuations could develop in this matter distribution in the absence of baryonic negative-energy matter\index{baryonic negative-energy matter} (as I have explained in section \ref{sec:4.4}).

Thus, while the distribution of negative vacuum-dark-matter energy would be inhomogeneous on larger scales at the epoch of last scattering\index{last scattering!epoch}, it would be more homogeneous at the same epoch on smaller scales. But given that negative-energy matter inhomogeneities\index{negative-energy matter inhomogeneities!CMB fluctuations} do contribute to the presence of temperature fluctuations in the cosmic microwave background at the epoch of last scattering (as a result of the gravitational forces they exert on positive-energy matter and radiation), then there naturally arises a deficit of fluctuations on smaller scales, even when inflation is not assumed to be responsible for producing the inhomogeneities which were present on both smaller and larger scales at the epoch of last scattering. I believe that this is the true reason why the spectrum of CMB temperature fluctuations\index{spectrum of CMB fluctuations!smaller-than-one power-law index} is slightly tilted (with a power-law index smaller than one) compared to a pure Harrison-Zel'dovich spectrum\index{Harrison-Zel'dovich spectrum of fluctuations!pure}.

It should be clear, therefore, that negative-energy matter does have an effect on the observed properties of cosmic microwave background temperature fluctuations. Of all the measurements concerning the spectrum of CMB temperature fluctuations\index{spectrum of CMB fluctuations!angular scale of fluctuations}, the only ones which would remain mostly unaffected by the presence of negative-energy matter are those which regard a determination of the angular scale of fluctuations from which is derived the average density of positive matter and vacuum energy, because the trajectories of positive-energy photons are not affected by the presence of negative-energy matter on the largest scale (particularly when its density is as uniform as it must be on smaller scales, following the early annihilation of matter with antimatter\index{early matter-antimatter annihilation}). The appropriateness of this conclusion appears to be confirmed by the fact that estimates of the average density\index{average density estimates!positive-energy matter} of positive-energy matter (both visible and dark) based on measurements of the spectrum of CMB temperature fluctuations produce a value largely equivalent to that which is derived using more direct methods.

I must acknowledge, however, that discrepancies have emerged more recently (see in particular Ref. \cite{Troster-1}) between the average density of matter derived from CMB measurements and that which is inferred from weak gravitational lensing\index{weak gravitational lensing} of nearer structures and it is possible that we will only be able to resolve those difficulties once the various effects of the presence of negative-energy matter on the process of structure formation\index{structure formation!process}, in both the early universe (before decoupling) and at a more recent epoch, are properly taken into account and the consequences of a variation of the magnitude of the cosmological constant\index{cosmological constant!magnitude variation} are fully worked out.

As I explained above, we cannot expect that the inhomogeneities which were present on smaller scales in the early distribution of negative-energy matter were as developed as those which were present at the same epoch in the positive-energy matter distribution. But given that in the presence of negative vacuum-dark-matter energy\index{negative vacuum-dark-matter energy!additional inhomogeneities} there should nevertheless be more inhomogeneities in the matter distribution than we could attribute to positive-energy matter, at least on a very large scale, and therefore more fluctuations in the temperature of the cosmic microwave background\index{cosmic microwave background!additional fluctuations}, then it would appear necessary to revise the magnitude of density fluctuations attributable to positive-energy matter downward, as I mentioned in section \ref{sec:4.4}. The presence of negative-energy matter inhomogeneities would, therefore, allow to ease the tension that emerged from even more recent observations \cite{Asgari-1} which indicate that the current magnitude of positive-energy matter density fluctuations\index{positive-energy matter density fluctuations!current magnitude} is smaller than the value deduced from CMB temperature fluctuations extrapolated to the present epoch (a problem known technically as the $S_8$ tension\index{S8 tension@$S_8$ tension}).

Indeed, if the actual magnitude of positive-energy matter density fluctuations\index{positive-energy matter density fluctuations!actual magnitude} at the epoch of last scattering\index{last scattering!epoch} can be assumed to be smaller than the value currently deduced from observations of CMB temperature fluctuations, due to the fact that some of the observed fluctuations are attributable to the presence of negative-energy matter, then it is natural to expect that the magnitude of fluctuations in the positive-energy matter distribution obtained by evolving the observed fluctuations which existed at the epoch of last scattering forward in time to their present state, should be larger than the actual value which is obtained from local observations, even if is true that the presence of negative-energy matter must have accelerated the process of structure formation\index{structure formation!acceleration} in the early universe, on smaller scales, and at later times, on larger scales.

It is certainly possible that certain characteristic features of the spectrum of cosmic microwave background\index{cosmic microwave background!spectrum of temperature fluctuations} temperature fluctuations, distinct from those mentioned above, could also be explained by taking into account the effects that would be attributable to the presence of very-large-scale inhomogeneities in the distribution of negative vacuum-dark-matter energy\index{negative vacuum-dark-matter energy!very-large-scale inhomogeneities}, either in the initial state, or in the space through which radiation propagates before reaching our detectors. One very clear implication of a cosmological model based on the ideas developed in this report, however, is that it is no longer necessary to assume that gravitational waves\index{gravitational waves!stretching of space by inflation} must have been produced as a result of the stretching of space produced by inflation, starting from a highly inhomogeneous initial state\index{initial state!highly inhomogeneous}, and therefore it may be futile to search for an unmistakable sign of the existence of those gravitational waves in the polarization of cosmic microwave background radiation\index{polarization of CMB radiation!gravitational waves}.

\section{A criticism of inflation theory\label{sec:4.10}}

Now that I have provided alternative solutions to all aspects of the inflation problem\index{inflation!problem}, I would like to offer a constructive criticism of inflation theory\index{inflation theory!criticism} itself and explain why it may no longer constitute an appropriate response to the most enduring difficulties facing theoretical cosmology\index{theoretical cosmology}. It must be clear, however, that I do not claim to have proven that inflation theory is wrong or that the phenomenon it describes did not occur. Indeed, what I have shown is simply that inflation is no longer \textit{necessary} to solve the flatness problem\index{flatness problem} and that there may no substance to the related problem of matter creation\index{matter creation problem}, outside of inflation theory itself. I then proceeded to explain that an alternative solution to the horizon problem\index{horizon problem} and the related problem of smoothness\index{smoothness problem} can be formulated that may also go some way in solving the related problem of topological defects\index{topological defects!problem}.

But this does not mean that the hypothesis that there occurred an initial phase of accelerated expansion\index{initial phase of accelerated expansion} is not valid and that we should no longer expect something like inflation to have happened, only that the existence of such a phenomenon may not be required for explaining the puzzling features of the universe which are giving rise to the inflation problem. I find it significant, however, that of all the major difficulties facing cosmology, the cosmological constant problem\index{cosmological constant!problem} is the one for which inflation theory was never allowed to provide an appropriate answer, as this is definitely an issue that can only be addressed in the context of the generalized gravitation theory\index{generalized gravitation theory} proposed in chapter \ref{chap:2}. There may have been truth, then, to the long forgotten suggestion that the same insights which would turn out to be required in order to solve the cosmological constant problem may allow to do away with the other outstanding difficulties of cosmology, which would otherwise need to be addressed by resorting to inflation theory.

Thus, while inflation may not be invalidated as a theory, it appears that there are more natural solutions, based on more unavoidable theoretical and observational constraints, not only to the inflation problem\index{inflation!problem} itself, but to certain other important issues as well. In fact, I'm now in position to provide satisfactory answers to practically all the remaining outstanding problems of theoretical cosmology\index{theoretical cosmology!outstanding problems}, including the problem of the origin of the thermodynamic arrow of time\index{thermodynamic arrow of time!problem of origin} and that of the nature of dark matter\index{dark matter!problem}.

But what should motivate one to recognize the necessity for an alternative approach to cosmology such as the one I have proposed in the preceding sections, is the fact that even some of the originators of inflation theory\index{inflation theory!originators} have, more recently, expressed doubts concerning the usefulness of the theory for solving any of the problems to which it was originally believed to provide a satisfactory answer, because inflationary expansion\index{inflationary expansion!finely-tuned initial conditions} itself requires finely-tuned initial conditions to be initiated and to give rise to the desired outcome. Those criticisms, however, are usually overlooked, because of what appears to be the overwhelming evidence in favor of inflation theory\index{inflation theory!overwhelming favorable evidence} that was provided by the discovery that the universe\index{universe!flat space} has a flat space and that the primordial distribution of matter\index{primordial matter distribution!homogeneity above horizon scale} and radiation energy is homogeneous above the scale of the horizon (as revealed by observations of the cosmic microwave background\index{cosmic microwave background}).

Indeed, it is definitely the fact that a universe with a density parameter\index{density parameter} $\Omega_0=1$ was always favored by inflation theory, even at a time when it appeared that lower values of $\Omega_0$ were favored by observations, that is responsible for having transformed this theory into the paradigm it is today, when it was later found that this parameter is, in effect, equal to unity and space actually is perfectly flat on the largest scale. But given that I have shown that, in the presence of negative-energy matter, when we appropriately require the universe\index{universe!zero energy} to have zero energy, the specific expansion rates\index{specific expansion rates!critical value} of positive- and negative-energy matter are required to be critical to an arbitrarily high degree of precision, based merely on the assumption that an observer must be present in the universe to measure those parameters, then it would appear that the flatness of space\index{flatness of space!generic property of observable universe} is not valid as a confirmation of the validity of inflation theory, but is actually a generic property of any observable universe obeying the known principles of physics.

In this context, what was wrong was really the early expectation that, by default (in the absence of inflationary expansion\index{inflationary expansion}), space should be observed to be highly curved at the present time, given that perfect flatness requires very unlikely initial conditions on which there appeared to be no constraint. In fact space must be perfectly flat at all times in an observable zero-energy universe\index{observable zero-energy universe!perfect flatness of space}, but what I have tried to explain is that inflation has nothing to do with that and therefore flatness does not provide an unmistakable confirmation of the validity of inflation theory\index{inflation theory!unmistakable confirmation of validity}.

Of course, if all I had done was to show that the flatness problem\index{flatness problem} does not occur, even in the absence of inflation, when the universe\index{universe!null energy} is required to have a null energy and observer selection effects\index{observer!selection effect} are taken into account, then it would not be possible to conclude that inflation is unnecessary, because there would still be a problem associated with the observed large-scale homogeneity of the early matter distribution\index{early matter distribution!homogeneity}. But given that, when negative-energy matter is present in the primordial state and one recognizes the necessity for all elementary particles\index{elementary particles!causal relationships} in the universe to be causally related to one another, it actually becomes necessary for the initial matter distribution that existed in the very first instants of the Big Bang to be uniform enough that no macroscopic event horizon\index{macroscopic event horizon} is present, then it follows that the overall homogeneity of the temperature of CMB radiation is no longer a fact in need of some explanation.

The presence of a high initial density of positive-energy matter, on the other hand, no longer requires a hypothetical, post-inflation reheating\index{reheating!post-inflation process} process, dependent on very specific conditions, to have occurred, once one recognizes that there is no need for matter\index{matter!creation out of nothing} to be created out of nothing during the Big Bang, given that it must have already been present prior to the initial maximum density state, before being submitted to a quantum bounce\index{quantum bounce}. It is actually only when we assume that there occurred an initial phase of accelerated expansion\index{initial phase of accelerated expansion} that we need matter to be created at some point during the Big Bang and therefore it would certainly be wrong to assume that inflation constitutes an absolute requirement for the existence of matter in our universe. In fact, when all the dust has settled, it appears that not much evidence remains to possibly confirm that inflation really occurred. But again, that does not mean that inflation theory\index{inflation theory!consistency and observational accuracy} is inconsistent or observationally inaccurate, merely that the phenomenon of inflation is not required to produce the apparently unlikely `initial' conditions which were previously thought to remain unexplained outside the realm of this theory.

Concerning the flatness problem\index{flatness problem}, however, it transpires that if the \textit{specific} expansion rates\index{specific expansion rates!critical value} of positive- and negative-energy matter were fixed to their critical value by inflation alone, while only the magnitudes of the initial, average densities of positive and negative matter energy were required to be equal by the condition of null energy\index{null energy condition} (so that the gravitational potential energies\index{gravitational potential energy} and the kinetic energies of expansion\index{kinetic energy of expansion} would be left unconstrained by the same condition), then it would be difficult to explain how the average, specific densities of the two opposite-energy matter distributions could remain mostly equal following inflation (but before the early annihilation of baryons with antibaryons\index{early baryon-antibaryon annihilation}), as required if the cosmological constant\index{cosmological constant!larger value} is to not be much larger than it appears to have been at this remote epoch (which would also imply a larger present value).

There is no reason, in effect, to assume that the outcome of inflation would be exactly the same from both the viewpoint of positive-energy observers and that of negative-energy observers, while this is required if the specific densities\index{specific densities!similar magnitudes} are to be of similar magnitude following inflation. Thus, it would appear that the explanation of the flatness of space\index{flatness of space!explanation} provided in section \ref{sec:4.5} is actually an absolute requirement, under such conditions, and cannot merely be considered an alternative possibility.

In fact, the difficulty discussed here would be even worse if one was to assume that the early phase of inflationary expansion\index{inflationary expansion} was not merely produced by the presence of some hypothetical inflaton field\index{inflaton field}, but was instead the outcome of the existence of a large average density of vacuum energy\index{average density of vacuum energy!large initial value} in the initial Big Bang state, because in such a case the process would necessarily have opposite effects of considerable magnitudes on the expansion rates measured by observers with opposite energy signs.

Thus, while the space experienced by a positive-energy observer could be driven to inflate exponentially, the space experienced by a negative-energy observer may actually be made to collapse back into a singularity, which means that the `initial' densities of positive- and negative-energy matter measured by such an observer, following inflation, would remain maximum, while those measured by a positive-energy observer (before reheating\index{reheating}) would become negligible, which once again is not quite compatible with observations, which indicate that the expansion rates and the spatial volumes experienced by opposite-energy observers are still similar at the present epoch, due precisely to the small value of the cosmological constant\index{cosmological constant!small value}. If inflation\index{inflation!additional fine-tuning} is a product of the energy of zero-point vacuum fluctuations\index{zero-point vacuum fluctuations}, it would therefore appear that additional fine-tuning would be required to make the theory viable.

What must be clear is that there is only one (positive or negative) value for the average density of vacuum energy\index{average density of vacuum energy!one single value} at any given time and if the magnitude of this value is too large for too long a period of time initially, then there may be conflict with observations, even independently from whether matter is present initially (as I'm assuming) or is created through reheating after inflation (as must be assumed from a conventional viewpoint).

In the context of the approach I favor, this problem does not exist, because the average value of vacuum energy density is not affected by the changes which are taking place in the vacuum and depends only on the difference between the scale factors\index{scale factors!opposite-energy observers} experienced by opposite-energy observers. What's more, for a zero-energy universe\index{universe!zero energy}, when it is recognized that negative-energy matter must be present in the initial state, there is a requirement for space to be perfectly flat and the average density of vacuum energy\index{average density of vacuum energy!null initial value} to be null initially, from both the viewpoint of positive-energy observers and that of negative-energy observers. Under such conditions, even if a non-zero average density of vacuum energy\index{average density of vacuum energy!non-zero value} can arise when a difference develops, at later times, between the average density of positive matter energy and that of negative matter energy, the weak anthropic principle\index{weak anthropic principle} provides sufficiently strong a constraint to allow one to expect that the current value of the cosmological constant\index{cosmological constant!current value} should nevertheless be as small as it is observed to be.

Concerning the solution potentially offered by inflation theory\index{inflation theory} to the horizon problem\index{horizon problem} and more specifically to the unexplained uniformity of the initial matter distribution, it was already pointed out by Roger Penrose\index{Penrose, Roger} that the usual assumption, to the effect that inflation would take the universe from a highly inhomogeneous initial state\index{initial state!high inhomogeneity} to a perfectly smooth one, appears doubtful in the context where the initial state\index{initial state!maximum gravitational entropy} would, in effect, be characterized by a maximum gravitational entropy. But those remarks were made before we even had a theory of gravitation that allowed for the presence of negative-energy matter in the initial state.

From the viewpoint of the developments introduced in chapter \ref{chap:2}, it would seem that it is definitely impossible to assume that a universe with an arbitrarily large initial gravitational entropy\index{gravitational entropy!arbitrarily large initial value} could be rendered homogeneous through accelerated expansion, as there is no limit to how polarized the initial matter distribution\index{initial matter distribution!polarization along energy sign} could be along energy sign, just like there is no limit to the amount of matter that can be present in the universe. The opposite-energy black holes\index{opposite-energy black holes!initial Big Bang state} that could be present in the initial Big Bang state, if it was not for the limitation exerted by the constraint of global entanglement\index{global entanglement constraint} on density fluctuations, could be as massive as the radius of curvature of the universe\index{universe!radius of curvature} is large and would concentrate all the matter in the universe in their gravitationally repelling singularities\index{gravitationally repelling singularities}, which means that no amount of expansion could ever result in a homogeneous matter distribution. Thus, if negative-energy matter does exist, it seems that inflation alone could not prevent the `initial' distribution of matter energy from being highly inhomogeneous and this provides additional motive to believe that the process is not necessary, even if it still cannot be ruled out that it might have occurred\footnote{
In fact, the only way inflation could perhaps give rise to a homogeneous state from an initial state of arbitrarily large gravitational entropy, in the presence of negative-energy matter, is by inflating an elementary unit of area\index{elementary units of area|nn} to the current size of the observable universe\index{observable universe!current size|nn}. But if one assumes that there does, in effect, exist such an elementary unit of area, then one must also recognize that there is a maximum theoretical value of positive and negative matter energy densities\index{matter energy density!maximum theoretical value|nn} and under such conditions it is not possible to assume that the whole observable universe could find itself inside just one elementary unit of area, which means that it is still impossible for inflation to give rise to a homogeneous distribution of matter energy.}.

One particular aspect of the horizon problem\index{horizon problem} which is currently believed to have been solved by inflation theory has to do with predicting the spectrum of density fluctuations\index{spectrum of density fluctuations!initial matter distribution} in the initial positive-energy matter distribution. Indeed, the fact that observations of the cosmic microwave background\index{cosmic microwave background} have revealed a nearly scale-invariant spectrum of density fluctuations\index{nearly scale-invariant spectrum of density fluctuations}, of the kind that is predicted by inflation theory\index{inflation theory!strongest empirical element of evidence}, is often considered to provide the strongest element of empirical evidence in favor of this theory.

But given that a scale-invariant spectrum of density fluctuations\index{scale-invariant spectrum of density fluctuations} of the Harrison-Zel'dovich\index{Harrison-Zel'dovich spectrum of fluctuations} type would be typical of any theory according to which the presence of macroscopic black hole\index{macroscopic black hole!event horizon} event horizons is forbidden on all scales and space itself does not have a characteristic scale, as is the case for a universe with a critical energy density\index{critical energy density} and an infinitely large radius of curvature\index{radius of curvature!infinite}, then it seems that a well-behaved cosmological model, that would describe a universe\index{universe!null energy} with null energy, would also predict a scale-invariant spectrum of density fluctuations, given that it would require macroscopic black holes to be absent in the initial Big Bang state\index{initial Big Bang state!absence of macroscopic black holes} and space to be flat on a global scale. Therefore, once again, the observations cannot be assumed to provide a definitive confirmation of inflation theory and this conclusion is reinforced in the context where, even the fact that the spectrum of CMB temperature fluctuations\index{spectrum of CMB fluctuations!imperfect scale invariance} is not perfectly scale-invariant can be explained as being a consequence of the near absence of baryonic negative-energy matter\index{baryonic negative-energy matter!absence}, as I explained in the preceding section.

Now, it has been hailed that the fact that certain versions of inflation theory may allow the `true' universe to be comprised of many regions like the known universe, separated by arbitrarily large portions of inflating space\index{inflating space} in which new `universes' like our own are born all the time, could be a positive development, given that it seems increasingly more likely that some properties of our universe are constrained by the weak anthropic principle\index{weak anthropic principle}. Indeed, one of the implications of the existence of such otherwise unexplained properties is that it makes plausible the idea that there must be more than one possible instance of physical reality\index{physical reality!multiple instances}, so that the anthropically constrained universe we observe can exist as a mere possibility whose improbable nature need not be explained by appealing to divine intervention.

Some of us, however, appear to favor, for some mysterious reason, that all of those realities, instead of just existing on their own, be somehow tied (however loosely) to the universe we do experience, as if this was a requirement of the multiverse concept\index{multiverse!concept}. This conception of the multiverse has been appropriately renamed the `megaverse'\index{megaverse} by Leonard Susskind\index{Susskind, Leonard} and now enjoys respectable status as if it had been proved right by the `successes' of inflation theory\index{inflation theory}. But given that we may now have to reconsider the degree of inevitability of the phenomenon of inflation, it would appear that all this extraneous amount of inflating vacuum\index{inflating vacuum} may no longer be as appealing as it once was.

In any case, one thing should be clear and it is that eternal inflation\index{eternal inflation} is not necessary for making the multiverse concept a viable notion and in fact the emergence of a megaverse concept in inflation theory may actually constitute a problem for this approach to cosmology, given that it may indefinitely postpone the moment at which the global entanglement\index{global entanglement!enlarged universe} of this whole enlarged universe would occur in the past, while this must be considered a necessity, as I explained at length in the preceding section (if global entanglement is required only within the bubble universes\index{bubble universes}, then the megaverse\index{megaverse!ensemble of causally related parts} itself could not be assumed to exist as an ensemble of causally related parts in the first place). This remark is all the more relevant in the context where, unlike the megaverse, the existence of a thermodynamic arrow of time\index{thermodynamic arrow of time} (understood as being a consequence of the constraint of global entanglement\index{global entanglement constraint}) is an observable fact with undeniably real consequences.

The most enduring problem facing inflation theory\index{inflation theory!most enduring problem}, however, remains the fact that it is still as difficult today as it was back when the model was introduced decades ago to identify what is the deep principle from which it would emerge as an unavoidable aspect of physical reality. If such a foundation cannot be developed, we will perhaps eventually need to recognize that what was provided by inflation theory was a solution that was useful merely because of the absence of a better alternative. It is not appropriate, in the context where a more natural explanation of facts is available, to just keep adjusting the free parameters\index{free parameters!inflation theory} of a theory which is supposed to determine the very boundary conditions\index{boundary conditions!universe} applying to the universe as a whole.

At this point it is not just questionable whether inflation can actually solve any of the outstanding problems of Big Bang cosmology\index{Big Bang cosmology!outstanding problems}, it is even uncertain whether it is still possible to assume that the phenomenon occurred at all. Under such circumstances, only our inherent resistance to paradigm change\index{paradigm change} may prevent us from acknowledging the eventual failure of the theory. But if there is any reason to believe that inflation, in effect, did not occur, it would have to be the fact that it is not merely a single one of the difficulties originally assumed to be solved by the theory that can be explained away in the context of negative-energy matter cosmology, but nearly all aspects of what was once the inflation problem\index{inflation!problem}.

\chapter{Quantum Theory and Causality\label{chap:5}}

\section{The problem of interpretation\label{sec:5.1}}

In the preceding chapters of this report, I offered original solutions to several outstanding problems in the fields of gravitational physics and cosmology, which were all based on an alternative interpretation of the concepts of time reversal\index{time reversal!alternative concept} and negative energy. First of all, I introduced a generalized, classical theory of gravitation\index{classical gravitation theory!generalization} that is consistent with the possibility that elementary particles could exist that would propagate negative energy forward in time. Based on the understanding that it is necessary to distinguish between a fundamental, bidirectional concept of time\index{bidirectional time concept} associated with the propagation of elementary particles and the classical, unidirectional concept of time\index{unidirectional time!concept} associated with thermodynamic irreversibility\index{thermodynamic irreversibility}, I was then led to introduce a more consistent formulation of the time reversal symmetry operation\index{time reversal symmetry!operation} that was shown to be relevant to a description of the fundamental states of matter particles on the quantum-gravitational scale\index{quantum-gravitational scale}.

I also showed that the hypothesis that negative-energy matter was present alongside positive-energy matter in the first instants of the Big Bang allows the formulation of a satisfactory solution to the problem of the origin of thermodynamic time asymmetry\index{thermodynamic time asymmetry!origin} as being the outcome of a certain condition that must be imposed on this initial state in order that all the elementary particles present in the universe be causally related to one another. But given that the bidirectional concept of time that underlies this approach constitutes a challenge to our conventional concept of causality\index{causality!conventional concept} and the idea that causes always precede their effects, then it becomes necessary to introduce a revised concept of causality\index{causality!revised concept} that would allow to take into account the time-symmetric nature of elementary particle processes\index{elementary particle processes!time-symmetric nature}. This is one of the objective I'm seeking to achieve in this chapter.

Part of the progress I have accomplished concerning this issue emerged from an investigation of the significance of certain puzzling aspects of the currently favored interpretations of quantum theory\index{quantum theory!currently favored interpretations} which did not appear to be connected with the issue of time directionality\index{time directionality}, but which were of interest in their own right. Yet, some of the results I have obtained regarding the issue of time directionality in gravitational physics turned out to be necessary for developing a solution to the remaining problems that still stand in the way of a truly consistent interpretation of quantum theory\index{quantum theory!consistent interpretation}.

Thus, in the present chapter, I would like to address not only the question of time directionality as it arises in a quantum mechanical context, but also, and more specifically, the important problem of the interpretation of quantum theory\index{quantum theory!problem of interpretation} itself. I will show, in particular, that it is now possible to provide a realistic picture\index{realistic picture!quantum phenomena} of quantum phenomena that does not violate the principle of local causality\index{principle of local causality}, even though it is not incompatible with the consequences of quantum entanglement\index{quantum entanglement} and the existence of non-local correlations\index{non-local correlations}. This improved understanding will then be used to provide a complete and definitive solution to the quantum measurement problem\index{quantum measurement problem} that allows to explain the emergence and the persistence of a quasiclassical world\index{quasiclassical world!emergence and persistence}.

But some of the most significant contributions I will offer in this chapter consist in showing that there really is a problem with some aspects of our current understanding regarding quantum physics. Two categories of questions I will try to address more specifically are not always distinguished from one another and together constitute the problem of the interpretation of quantum theory. I will explain, however, that they must be considered as independent questions in need of separate answers. There is, thus, a problem of interpretation concerning the mathematical framework of quantum theory\index{quantum theory!mathematical framework} in general and the distinctive features of quantum physics, which are mainly the use of probability amplitudes\index{probability amplitudes} instead of classical probabilities\index{classical probabilities} and the existence of entangled states\index{entangled state}, which are both unavoidable features of physical reality.

It is commonly believed that the problem here does not have to do with the inappropriateness of our interpretation, but with the inappropriateness of our conventional approach to understanding reality. Yet, it is not clear, from the perspective of currently available interpretations, how quantum entanglement, in particular, and the kind of non-locality\index{non-locality} it makes possible, is to be understood in a manner that is consistent with the classical principle of local causality\index{principle of local causality!classical}. Thus, if it may be appropriate to argue that there is no real difficulty, in the context of the conventional interpretation of quantum theory\index{quantum theory!distinctive aspects}, with the distinctive aspects of its formalism for what concerns probability amplitudes and the existence of quantum interferences\index{quantum interferences} in general, I will show that this posture is not justified for what concerns the problem of quantum non-locality\index{quantum non-locality}, which is actually a question in need of an answer. Thus, answers will be offered to a problem which I call the \textit{quantum reality problem}\index{quantum reality!problem} and which includes the problem of quantum non-locality\index{quantum non-locality!problem} as a particular aspect.

This problem must be distinguished from the associated problem that is usually referred to as the \textit{quantum measurement problem}\index{quantum measurement problem}. Those who are not actively working on this particular issue often believe that it too may not be real, or, alternatively, that it was entirely solved by more recent developments that showed how the evolution of a quantum system is affected as soon as it becomes entangled with certain irreversible processes taking place in its environment. But, as a handful of researchers have already understood, this opinion is not warranted and even though real progress has indeed been achieved in trying to solve the quantum measurement problem and more generally the problem of the emergence of `quasiclassicality'\index{emergence of quasiclassicality!problem}, some related questions remain unanswered and it is precisely those I will address.

However, if you happen to be among those who are convinced that there is no longer a problem with quantum measurement, then I would ask you a very simple question: what is the cause of the irreversibility that characterizes the evolution of the environment degrees of freedom\index{environment degrees of freedom!irreversible evolution} with which a quantum system becomes entangled during measurement and which is necessary for explaining decoherence\index{decoherence}? Clearly, an appropriate answer to that question must be provided before the problem may be considered to have been solved and this is what I have tried to achieve in the previous chapter. But, as I will explain, that is not the only difficulty. In order to clarify this complex situation I will, therefore, need to draw on the insights I have gained while solving the problem of the origin of time asymmetry\index{time asymmetry!problem of origin}, but I will also have to build on the insights I have gained while solving the quantum reality problem\index{quantum reality!problem}, which illustrates how important it was, in effect, to first solve that perhaps more intangible problem.

It is quite remarkable that, in order to answer those two categories of questions, it is possible to rely on the most appropriate of the already existing mathematical frameworks within which quantum theory\index{quantum theory!existing mathematical frameworks} is currently formulated and it is not necessary to alter the foundations of the theory. I must immediately point out, however, that there is something terribly wrong with the often met remark to the effect that, choosing which of the existing interpretations of quantum theory\index{quantum theory!existing interpretations} is the correct one is a mere matter of taste, given that they are all mathematically equivalent and therefore all constitute equally valid proposals, which all agree with observations. Before any progress can be achieved, what needs to be understood is that most interpretations are not equally appropriate, but rather all equally incorrect or incomplete.

It would be misleading, therefore, to argue that the problem is that there are too many viable candidates for an interpretation of quantum theory\index{quantum theory!interpretation}, because in fact \textit{none} of the currently available proposals is fully consistent, either from a logical viewpoint, or regarding the requirement that the obtained theory be compatible with all observable aspects of physical reality. This state of affairs can only mean one thing and it is that further progress is required to formulate the one interpretation that will meet both of those requirements. I believe that the original results I have unveiled in the preceding chapters of this report provide some of the missing elements which are required to achieve just such a leap forward in our understanding of quantum mechanics\index{quantum mechanics!fully coherent theory}, which will, at last, allow it to become a fully coherent theory.

\section{A simple analogy\label{sec:5.2}}

One particular event from my early years in elementary school contributed more than any other in developing my awareness of the deep structure of physical reality. I do not remember much about the many events that happened during the period of my life when I was acquiring many of the skills which I'm still using today (like writing and calculating), but I still remember perfectly well that when I was about eight years old my teacher once gave me and each of the other children in my class a few copper wires, an electrical battery, and a tiny light bulb, with as a mission to figure out how to produce light using only those components. This may seem like an easy task and most of the kids did, in effect, manage to achieve the assigned objective quite rapidly. Yet, even though I was usually considered a fast learner in most traditional academic disciplines, I really had trouble finding out how to obtain the desired result.

I believe that this is because, even as a kid, I always preferred to actually understand things, rather than simply be satisfied with learning about the finished answers I was proposed. So, rather than just trying to combine the elements in every possible way and be satisfied once I had accomplished my homework, like the other children, I tried very hard to understand what the rule could be that would justify that a certain arrangement does, in effect, produce light. I don't know why I had such an inclination, but it has remained with me all my life as I began to develop an interest in the sciences and learned about the unsolved mysteries of modern physics. What I have since realized, with retrospect, is that somewhere in this simple laboratory experiment was hidden the answer to some of the most enduring problems facing fundamental theoretical physics\index{fundamental theoretical physics!most enduring problems}.

The first lesson I have learned from this experiment with the light bulb and the battery\index{light bulb and battery experiment} is that there is a polarity to most physical attributes. The battery has a positive and a negative pole, and so does any light bulb, and it is only by taking this aspect into consideration that one is allowed to understand what constitutes a successful configuration for producing light. In the first two chapters of this report I have discussed at length how this aspect is relevant even in a gravitational context, where the sign of action is the decisive physical property that is involved in determining the attractive or repulsive nature of the gravitational interaction\index{gravitational interaction!attractive or repulsive nature} between two particles. I also emphasized the purely relational nature of any polarity\index{polarity!relational nature}, whether it regards the sign of electric charge\index{sign of electric charge}, or the direction of propagation in time\index{direction of propagation in time} of a particle. Only the difference or the identity between any such attribute of a system and that of another system has a physical significance.

But the most difficult part in devising an electrical setup that works consisted in understanding the role of the wires. What is essential to learn, in effect, is that the experiment can only work if the wires are arranged so as to form a circuit that goes from one pole of the battery into one pole of the light bulb and then back from the opposite pole of same light bulb into the unconnected pole of the same battery. I only realized that this must be so when I carefully examined the light bulb and saw that there is a special kind of wire inside of it that connects the two poles from within, thereby suggesting that, for some reason, the setup must form a closed circuit\index{closed circuit}.

After I came to understand this requirement, however, I was not only faced with the problem of understanding why only such a configuration would produce the desired outcome, but also with the difficulty of understanding what was the role of the battery in allowing light to be produced by the light bulb. It is only much later in my life that I learned that what the light bulb does, from a fundamental viewpoint, is simply dissipate the energy that is stored in a useful form in the battery, as a result of the friction that is exerted on the electrical current that flows through the part of the circuit located inside this light bulb.

This is an expression of the second law of thermodynamics\index{second law of thermodynamics} in its purest form. The very objective of producing the circuit is to allow the current to dissipate the energy that is contained in the battery, by producing an enormous number of light particles that expand out of the system irreversibly. The whole mystery associated with the irreversibility of time is contained in this little experiment and with it the solution to the quantum measurement problem\index{quantum measurement problem}. Any circuit that produces a useful outcome, i.e. one that is observable and which has an effect on its surroundings (the light turns on), must dissipate energy that was originally present in the universe in a well-ordered configuration.

I will eventually explain what is the essential role of irreversibility in allowing the emergence of the quasiclassical character of reality that is revealed by any process of quantum measurement, but first, I would like to point out the profound significance of the property of closure that is imposed on any operational electric circuit\index{electric circuit!closure property}. For anyone who works as an electrician, the notion of a closed circuit is omnipresent, but it is also forgotten somehow, in a practical context, where one always works with pairs of polarized wires\index{pair of polarized wires} in which the two branches of a circuit are always contained in a single cable that invariably goes from energy source to appliance, over large distances, as if what was involved was one single flow from source to sink, similar to the flow of water inside a pipe.

Thus, it is easy to forget that one is always dealing with a closed circuit\index{closed circuit}, however complicated it might be. I believe that what explains some of the difficulties we encounter in quantum physics is that we have always learned to work only with pairs of `polarized wires' and this is why we fail to understand that what we are dealing with, in general, is not a single process that unfolds from initial conditions to final measurement, from `source' to `sink', but really a \textit{closed causal chain}\index{closed causal chain}, similar to the closed electrical circuit of my childhood experiment.

It is the fact that, for a certain reason that will be discussed later, we are always working with portions of a `closed circuit' which are highly stretched and extended in the direction of configuration space along which unidirectional time\index{unidirectional time} unfolds and whose two polarized components are constrained to evolve along very similar trajectories in this configuration space, that explains that we have been allowed to ignore the fact that we are actually always dealing with two distinct processes which are the oppositely polarized portions of a causal chain that closes up on itself, like a functioning electrical circuit\index{electric circuit}. We always model very long causal chains\index{causal chains}, similar to electric cables, that extend not along a distance in space, but along the direction in quantum mechanical configuration space\index{quantum mechanical configuration space} relative to which unidirectional time\index{unidirectional time} unfolds and for this reason we have never realized that what the polarized character of this causal chain really means is that we are dealing with a `closed circuit'.

I will explain how the simple analogy discussed here can be developed into a rigorous interpretation of experimental facts that allows to provide, not only a consistent explanation of the persistence of quasiclassicality\index{persistence of quasiclassicality}, but also a realistic and fully intuitive picture of quantum phenomena that reproduces their non-local character without violating the principle of local causality\index{principle of local causality}. When we will reach that point, it will become possible to actually understand why it is, in effect, that the causal chain associated with the history of the universe as a whole closes up on itself like any electrical circuit that produces light.

\section{Time-symmetric causality\label{sec:5.3}}

It is somewhat strange that it is Richard Feynman\index{Feynman, Richard} himself who once remarked that one question he believes to be unanswerable or unscientific is the one that asks why it is that we are allowed to guess from one part of the universe what the rest of it will do? Indeed, it has become pretty clear to me that, if this is possible, it is simply because things propagate, not just in space, but also in time and as Feynman himself was one of the first to understand, not just forward in time, but also backward, from the future toward the past. In fact, this is the essence of causality\index{causality} in the context where one recognizes that bidirectional time\index{bidirectional time} itself is an unavoidable aspect of reality. The events that form the universe are all related to one another and to nothing else by the network of causal relationships\index{causal relationships!network} that is established by the propagation of elementary particles\index{elementary particle!propagation} between those events, across spatial distances and through time, both forward and backward.

The results I have discussed in section \ref{sec:4.9}, regarding the role of the constraint of global entanglement\index{global entanglement constraint} in giving rise to thermodynamic time asymmetry\index{thermodynamic time asymmetry} appear to confirm that the existence of causal relationships\index{causal relationships!local contact and propagation}, established through local contact and propagation, is, in effect, essential for a consistent description of physical reality. Another significant conclusion from the preceding chapters of this report, that concerns time directionality\index{time directionality} more specifically, is that a distinction must be made between the traditional concept of time direction associated with the thermodynamic arrow of time\index{thermodynamic arrow of time} and a more fundamental concept of time direction, which has to do with the direction of propagation in time\index{direction of propagation in time!elementary particles} of elementary particles and which merely distinguishes two opposite directions without favoring one of them in any particular way (under most circumstances).

Thus, even at a semi-classical level of description, there already emerges a notion of bidirectional causality\index{bidirectional causality} more fundamental than the classical, unidirectional concept of causality\index{causality!classical or unidirectional concept}, according to which causes always precede their clearly distinguishable effects in the same unique and invariant direction of time\index{direction of time!unique and invariant}. From this more fundamental viewpoint, there is no longer an \textit{absolute} distinction between causes and effects\index{causes and effects!absence of absolute distinction} and all that one can meaningfully ask is whether a certain event does exert an influence on another event taking place at a different time, either earlier in the past or later in the future (which would affect the probability that one of the events is observed when the other is).

Those who have seriously examined the question usually recognize that the idea that the present can influence the future, but not the past, is not entirely correct. When calculating correlation probabilities\index{correlation probabilities}, we must take into account the effect of the future on the present whenever there are antiparticles in the final state, because antiparticles\index{antiparticles!propagation toward past} are most appropriately described as particles propagating from the future toward the past. In fact, there appears to be no real distinction, at the \textit{elementary} level of description, between the past and the future, and despite the fact that the future\index{future!unknown but unique} remains unknown to present observers, it is as unique as the past and we are merely discovering what that future is as we progress irreversibly towards it. What we consider to be our control over the future is not a complete illusion, of course, because there are correlations between what we do now and what happens later, but those correlations are no more real than the more subtle correlations which are predicted to occur by quantum theory and which arise from the effects exerted by certain present events on certain events in their past.

As I will explain below, it is merely the fact that entropy is always growing only in the future (from a global viewpoint) that makes it look like we only have control over the future, while the future itself appears to exert no influence on the present and the past, because this is what explains that the present can exert multiple recognizable effects\index{multiple recognizable effects!future} on the future, while the opposite is so unlikely as to be virtually impossible. But this is also why we can have no information about the future\index{information!about future}, while every future outcome appears to be possible, which makes it seem like we have a certain freedom over what future outcomes\index{future outcomes!freedom to cause} we choose to cause that does not exist for the past, as if causality\index{causality!present toward future} could only operate from the present toward the future.

What I'm suggesting, therefore, is that, on the level of elementary particles, where thermodynamic time asymmetry\index{thermodynamic time asymmetry} is not a meaningful concept, causality is not constrained to always operate from past to future, which means that causes and effects\index{causes and effects!sign of time intervals} cannot be distinguished based merely on the sign of the time interval between the events they relate. Thus, while it may still be necessary to assume that causes precede their effects, this can actually be achieved in any of the two directions of time in which the particles conveying the effects are propagating. At a fundamental level of description there simply is no restriction regarding the direction of time in which causality operates and this means that, from the unidirectional-time viewpoint\index{unidirectional-time viewpoint!effects preceding causes}, effects can actually precede their causes. But instead of saying that, under certain circumstances causes may actually follow from their effects, while effects would give rise to their causes, it is more appropriate to define causes and effects\index{causes and effects!bidirectional definition} in a more fundamental bidirectional way, so that effects\index{effects!forward- or backward-in-time propagation} can propagate either forward or backward in time, but always in the direction in which the particle that produces the effect is propagating in time.

The absence of an \textit{absolute} distinction between causes and effects\index{causes and effects!absence of absolute distinction} does not mean that the relativistic concept of a future light cone\index{future light cone!relativistic concept}, clearly distinct from its past equivalent, is wrong. But it does mean that there is no \textit{a priori} reason to differentiate the causal structure of spacetime\index{causal structure of spacetime} that arises as a consequence of the limits imposed on the propagation of causal signals\index{causal signals!limits on propagation toward future} in the future from that which arises as a consequence of the constraints imposed on the possible propagation of causal signals\index{causal signals!limits on propagation toward past} in the past.

Yet, it would be incorrect to argue that only correlations exist between various past and future events at a fundamental level of description and that causes cannot be distinguished from their effects in any way, because what the bidirectional, or time-symmetric nature of causality\index{causality!time-symmetric or bidirectional} implies is merely that there is no absolute (non-relationally defined) distinction between the past and the future for what concerns the propagation of effects\index{propagation of effects!distinction between past and future} and that only a relatively defined notion of time directionality\index{time directionality!relative definition} is involved, from a semi-classical perspective, which still allows to distinguish a direction as being that relative to which a given effect\index{effects!direction of propagation in time} propagates in time. Actually, it seems that the possibility for effects\index{effects!forward- or backward-in-time propagation} to propagate either forward or backward in time is precisely what gives its meaning to the degree of freedom which is associated with the direction of propagation in time\index{direction of propagation in time} of elementary particles.

What must be understood is that the only invariably true notion of causality\index{causality!time-symmetric or bidirectional} is the time-symmetric, or bidirectional one, while classical, thermodynamic causality\index{classical thermodynamic causality}, or unidirectional causality\index{unidirectional causality}, is valid only as a consequence of the existence of the constraint of low gravitational entropy\index{constraint of low gravitational entropy!initial state} that applies on the initial state at the Big Bang and is not a fundamental property of nature. In fact, what I have shown in section \ref{sec:4.9} is that a certain condition of local causality\index{condition of local causality!no time asymmetry} that is not \textit{a priori} asymmetric in time can be used to explain the observed thermodynamic time asymmetry\index{thermodynamic time asymmetry!explanation} from which unidirectional time\index{unidirectional time!emergence} and classical causality\index{classical causality!emergence} emerge.

It is now possible, therefore, to appreciate the fact that the direction of time relative to which entropy grows and information flows is independent from the direction of propagation in time\index{direction of propagation in time!elementary particles} of the elementary particles involved in producing such a change. The direction of propagation in time of an elementary particle, which determines its particle or antiparticle nature, merely allows to assess whether the particle propagates an effect in the past or in the future, while the flow of information\index{information flow!independence from direction of propagation in time} is a higher level property that is fixed merely by the macroscopic boundary conditions\index{macroscopic boundary conditions} imposed on a process, regardless of whether it involves particles or antiparticles.

Thus, a classical, unidirectional causal chain\index{unidirectional causal chain} can be differentiated by the fact that it invariably involves a unique event in the past exerting a recognizable effect on multiple space-like separated events in the future\footnote{
In fact, as emphasized by Lawrence Sklar\index{Sklar, Lawrence|nn} \cite{Sklar-1}, it is not always possible to associate the asymmetry that distinguishes causes from effects with thermodynamic time asymmetry\index{thermodynamic time asymmetry|nn}, other than by relying on the property of parallelism of the direction of time\index{parallelism of direction of time|nn} that allows to `project' the thermodynamic asymmetry characterizing certain causally related pairs of events\index{causally related pairs of events|nn} onto other pairs of causally related events where the cause is not so easily differentiated from the effect (as in the case of certain mechanical or astronomical processes where friction and dissipation\index{friction and dissipation!apparent absence|nn} are not manifestly apparent). But this only strengthens the validity of the position I'm defending, to the effect that causality is not, by necessity, an asymmetric property.}.
 In this particular sense, it transpires that unidirectional causality\index{unidirectional causality} does, in effect, always operate from past to future in our universe, as no single event in the future has ever been observed to exert a unique recognizable influence on multiple separate events in the past that would actually involve a flow of information\index{information!flow from future to past} from that future toward its past. But, again, this does not mean that a future event cannot influence a past event, merely that this cannot occur in a way that would allow the formation of mutually consistent records\index{mutually consistent records!of future event} of this future event. It is not causality, in the fundamental sense, that is asymmetric in time, but the making of records\index{making of record!asymmetry in time} by which it is usually made manifest. This asymmetry has already been recognized to arise from the existence of a thermodynamic arrow of time\index{thermodynamic arrow of time} associated with the continuous decrease of entropy that takes place in the past direction of time.

It is not true, therefore, that observations show that causal chains\index{causal chains!future toward past} never run from the future toward the past, because all that we know is that future events\index{future events!no recognizable consequences on past events} are never observed to exert the same kind of recognizable consequences on the past, as past events exert on the future. The absence of recognizable effects, at an earlier time, from an event that would have taken place at a later time, can always be attributed, not to the absence of backward-in-time causality\index{backward-in-time causality} and to a \textit{fundamental} character of unidirectionality\index{fundamental unidirectionality}, but to the fact that entropy always increases only in the future direction of time, while records can only be formed in the direction of time relative to which entropy actually increases.

Given that, in chapter \ref{chap:4}, I have explained that the thermodynamic arrow of time\index{thermodynamic arrow of time!non-fundamental property} is not a fundamental property of nature, but arises instead from a condition regarding the entanglement of every elementary particle in the initial state\index{initial state!maximum matter density} of maximum matter density at the Big Bang, which, in the presence of anomalously gravitating negative-energy matter, imposes on all matter that it be as homogeneously distributed as necessary for an absence of macroscopic event horizon\index{macroscopic event horizon!absence}, then it clearly follows that there is absolutely no rational motive to argue that backward-in-time causation\index{backward-in-time causation!rational justification} is forbidden in our universe. Indeed, all the observable properties of naturally occurring processes can be explained without relying on this assumption, while the requirement of a relational definition\index{requirement of relational definition!direction of propagation in time} of the direction of propagation in time implies that backward causation\index{backward causation} must exist at a fundamental level, given that it cannot be distinguished in any absolute way from forward-in-time\index{forward-in-time!causation} causation.

I believe that it is, again, our failure to recognize the full significance of Feynman's\index{Feynman, Richard} (or maybe St\"{u}ckelberg's\index{St\"{u}ckelberg, Ernst}) description of antiparticles\index{antiparticles!backward-in-time-propagating particles} as particles propagating backward in time that is responsible for our ignorance of the necessity (and not just the possibility) of adopting a time-symmetric concept of causality\index{causality!time-symmetric concept} in the quantum realm. Once it is understood that there is a requirement for causality to be described in a time-symmetric way (due to the existence of backward-in-time propagation), then what we are facing is no longer merely the problem of understanding how the future can influence the known past, but really how it can be that a unique future\index{future!uniqueness} may itself be causally related to this unique past\index{past!uniqueness}, despite the fact that there is obviously more than one possibility for the outcome of future measurement results, even when a complete knowledge of the present state of a system is available.

The relative nature of the order in time of space-like separated events\index{space-like separated events!relative time order}, which is implied by the special theory of relativity\index{special relativity!theory}, means that an event that appears to be in the future of another event, for a certain observer, is actually in its past, for a different observer in a different state of motion, and therefore if a unique past is causally related to the experienced present, then the future can only be similarly unique. In a time-symmetric context\index{time-symmetric context}, it is not just the teleological character of backward causation\index{backward causation!teleological character} that would need to be justified, but the equivalent teleological character of ordinary forward-in-time causality\index{forward-in-time causality!teleological character}. If one insists that there is a problem with the possibility of causally influencing the known past, then one must at least also admit that this problem could not be distinguished from that which would arise from the fact that the past also influences the future, while a unique (even though unpredictable) future is associated with the known unique present\index{present!uniqueness}.

What makes it look like the present state of the universe is not causally related to one particular future is simply the fact that all future states compatible with current boundary conditions\index{current boundary conditions} are, in principle, allowed to be the outcome of random evolution from a given present state (more than one outcome is usually possible for future measurements), while not all past states compatible with current conditions are allowed as `final' states for backward-in-time evolution\index{backward-in-time evolution}, in the context where the constraint of global entanglement\index{global entanglement constraint} discussed in section \ref{sec:4.9} exerts a limit on entropy growth in the past. Thus, from a practical viewpoint, regardless of what happens, certain constraints are always present for evolution toward the past and it is the fact that those constraints are precisely such as to prevent one present\index{present!no recognizable effects on past} event from having recognizable effects on multiple past events that makes it seem like current conditions exert no influence on past evolution.

In fact, this is precisely the nature of the difference between what we usually call causality\index{causality!thermodynamic asymmetry between causes and effects} and which relates to the thermodynamic asymmetry between causes and effects and the kind of causality\index{causality!time-symmetric} that is involved in a time-symmetric context. But once it is realized that the past and the future are not distinct, from the more general perspective of time-symmetric causality, then it clearly follows that if we are willing to accept that the future can be influenced by what happens in the present, as confirmed by our direct experience of reality, then it is also necessary to recognize that the present can itself affect that past, in a certain way, so that imposing final conditions\index{final conditions} is no less appropriate than imposing initial conditions, as long as these conditions are not those responsible for the observed thermodynamic time asymmetry\index{thermodynamic time asymmetry} itself.

One commonly encountered misconception regarding backward causation\index{backward causation} in general, as well as in a quantum context, is that if the future is allowed to causally affect the past, then we can no longer be confident that the past is what it seems to be, because it can be altered by future events. This is usually provided as an argument against backward-in-time causation because, as everyone knows, the past is unique and unalterable and therefore any approach that would allow the future to `change' the past is certainly based on incorrect assumptions. But what is incorrect with this apparently logically unassailable conclusion is the idea that exerting an effect on the past would involve altering an observable fact\index{observable fact!alteration} from the past which we already know has occurred, just like we are allowed by our apparent free-will\index{free-will!apparent} to alter the course of future history.

In fact, that is just an inappropriate understanding of the meaning of backward causation, because if an event in the future affects the outcome of an observation in the past, this `outcome' has already taken place at the moment in the past at which it was observed and the fact is not changed, from an alternative counterfactual, at the moment when the effect of this future cause\index{future cause!past effect} reaches the event in the past which it contributes to determine. In other words, it is not possible to change history using the kind of time-reversed causal chain\index{time-reversed causal chain} that is allowed by fundamental theories. History\index{history!globally consistent whole} is the outcome of all effects, from both the past and the future, and is experienced only once as such a globally consistent whole. No known fact is altered or changed by effects propagating from the future, as any change that would be produced by future causes would need to have already taken effect at the time at which the fact first occurred. The `effects' that the future may exert on the past would always be made conspicuous merely through the influence they would have on correlation probabilities\index{correlation probabilities} established after the fact (when that future itself becomes a known past).

Now, it must be clear that the condition imposed on special-relativistic transformations that they should preserve the direction of all time-like intervals\index{time-like intervals!preservation of direction} (the causal ordering postulate\index{causal ordering postulate}) is not incompatible with the conclusion that backward-in-time causation\index{backward-in-time causation} must be allowed at a certain level, because all that is required by relativity theory\index{relativity theory} is that, if a causal chain\index{causal chain!past to future or future to past} operates from the past toward the future (as we usually assume to always be the case), then the same causal chain cannot be found to operate from the future toward the past as a result of such a transformation (a change of reference system\index{reference system!change}). But that does not mean that a distinct causal chain cannot operate from the future toward the past, merely that, if such a backward-propagating causal chain\index{backward-propagating causal chain} exists, it too cannot be turned into a causal chain propagating in the opposite direction of time, which in this case would be toward the future.

In any case, the fact that the causal time\index{causal time!relativity theory} of relativity theory is not unidirectional is certainly not a problem, because, even in such a context, unidirectionality is allowed to emerge from the global entanglement constraint\index{global entanglement constraint} which imposes a condition of low gravitational entropy\index{condition of low gravitational entropy} at the Big Bang, as I have explained in section \ref{sec:4.9}. What explains the time asymmetry of causal processes\index{causal processes!time asymmetry} is not a fundamental property of unidirectionality applying to causal chains\index{causal chains!fundamental unidirectionality}, but the particular boundary conditions\index{boundary conditions} which apply to the initial state\index{initial state!maximum matter density} of maximum matter density which must have existed in the very first instant of the Big Bang. What makes a flow of information\index{information!flow from future to past} from the future toward the past impossible is not a limitation that would be arbitrarily imposed on the direction in which causality operates, but a distinct constraint that limits the growth of entropy in the direction of time toward the initial Big Bang state. Thus, as long as an effect does not propagate faster than the relativistic speed limit\index{relativistic speed limit!propagation of effects}, it cannot give rise to a violation of the classical, unidirectional principle of causality\index{unidirectional principle of causality!violation}, whether the effect propagates forward or backward in time.

It is not true that the scientific method\index{scientific method} excludes the possibility that `final causes'\index{final causes} of any kind might exist, despite the fact that time-symmetric causality\index{time-symmetric causality!relativity theory} appears to be allowed by relativity theory, because what can be scientifically demonstrated is merely that entropy does not increase in the past, not that there is no backward-in-time propagation of effects\index{backward-in-time propagation of effects}. What explains that we have become naturally suspicious, regarding the possibility that effects could propagate backward in time, is only the fact that from a classical perspective it never appeared necessary, or even possible to describe an object or a component of an object as propagating backward in time, while the time asymmetry that characterizes most causal processes was always observed to operate from past to future, which made it look like a fundamental requirement.

This prejudice remained in effect, even when it became clear that backward causation\index{backward causation} is a necessary assumption, in the context where one must recognize that certain particles do propagate backward in time (even though, from a unidirectional-time viewpoint\index{unidirectional-time viewpoint}, they are observed as oppositely charged particles propagating forward in time, which are involved in the same entropy increasing processes as their ordinary matter counterparts). The teleological problem of time\index{teleological problem of time}, which is often believed to arise in the context where present conditions are allowed to influence the known past, or even when one recognizes that a unique future\index{future!uniqueness} is associated with the known present, is not a true problem, but merely follows from the psychological expectation of unidirectional causality\index{unidirectional causality!psychological expectation} that we inherited from our thermodynamically constrained experience of reality\index{reality!thermodynamically constrained experience} and which does not reflect any fundamental limitation on the propagation of effects. There is no other explanation for the widespread belief that causality must always operate forward in time.

What must be understood, then, is that it is not merely the order in which time flows that varies for an antiparticle\index{antiparticles!order of time flow}, but really the fundamental direction in which effects propagate in time. When a particle propagates backward in time, the direction of time in which it may come to influence other particles is actually reversed and it is merely the thermodynamic arrow of time\index{thermodynamic arrow of time} and the direction in which classical causality\index{classical causality!invariant direction} operates that remain unchanged. If there is any meaning to be associated with a concept of causality, from a fundamental viewpoint, then causality must be allowed to operate backward in time and future events must be allowed to exert an influence on past events. There cannot be a distinction between causal order\index{causal order!direction of propagation in time} and the direction of time in which an elementary particle propagates an effect, even though this direction is a relatively defined physical property and is only significant in relation to the direction of time in which another such particle is propagating a similar effect.

This requirement may perhaps appear doubtful, given that it seems that the concept of particle trajectory\index{particle trajectories!quantum mechanical concept} may no longer be relevant in a quantum mechanical context, where multiple distinct histories\index{multiple distinct histories!simultaneous occurrence} would be required to take place all at once in order to account for the statistics of quantum processes\index{quantum processes!statistics}. But I will explain, later in this chapter, why this apparent lack of uniqueness of particle trajectories is not an obstacle to a proper understanding of causality as actually depending on the direction of propagation in time of elementary particles. Causal order\index{causal order!locally-variable property} may be a locally-variable physical property, but it is not arbitrary, even when backward-in-time causation\index{backward-in-time causation} is allowed to take place.

\section{Closed causal chains and time travel\label{sec:5.4}}

As is already apparent from a semi-classical viewpoint\index{semi-classical viewpoint}, the time-symmetric nature of causality\index{causality!time-symmetric} does not merely imply that there is no absolute distinction between causes and effects\index{causes and effects!absence of absolute distinction}, it also means that a certain event can, all at once, influence another event and be influenced by that very same event. In other words, not only is there no absolute difference between causes and effects, but the cause of a certain event can also be an effect of the same event; although this circularity can only be appropriately described in the context of a purely quantum-mechanical model of reality such as the one I will propose in a latter portion of this chapter. It must be clear, however, that the possibility that such closed causal chains\index{closed causal chain} may occur does not constitute a valid motive to reject the whole concept of time-symmetric causality and backward-in-time causation\index{backward-in-time causation}, because, as I will explain, it is possible to provide a consistent description of such phenomena without encountering logical contradictions.

Reichenbach's\index{Reichenbach, Hans} insistence \cite{Reichenbach-1} (ch. 2, sec. 21) that one must be able to differentiate causes and effects\index{causes and effects!temporal order} independently from their temporal order, if we are to avoid the occurrence of closed causal chains, is not totally inappropriate, however, because, as I mentioned in the previous section, the direction of time in which causal chains\index{causal chains!direction of propagation in time} operate is determined locally by the direction of propagation in time of the particles involved and therefore it is not fixed merely by the global time order of causally related events. But it is precisely the fact that one can distinguish causes from effects in such a way that allows backward-in-time causation to occur, while this is what makes closed causal chains unavoidable, thereby requiring such phenomena to be properly described and interpreted at the most fundamental level.

Yet, even at the level where closed causal chains\index{closed causal chain} may occur, it is certainly necessary to require that no inconsistencies can arise, which would involve an incompatibility between some known present and some known past. What I will eventually explain is that there is actually a requirement for histories\index{histories!requirement of absence of self-contradiction} not to be self-contradictory and this condition can be satisfied, not merely despite the fact that causality also operates backward in time, but as a very consequence of the reality of backward causation\index{backward causation}.

In any case, it is certainly incorrect to argue that there is empirical evidence to the effect that closed causal chains are forbidden, because, in the course of elementary particle interactions\index{elementary particle interactions}, particle-antiparticle loops\index{particle-antiparticle!loop} are often encountered that constitute just such a phenomenon, which can be adequately described, even from a semi-classical perspective\index{semi-classical perspective}. Once again, I believe that the problem here does not have to do with the possibility that closed causal chains themselves may occur, but rather concerns the hypothesis that classical, unidirectional causality\index{unidirectional causality} could sometimes operate in the past direction of time along a closed causal chain. I will soon return to this question, but what should be clear already is that it is, in effect, only at the level where unidirectional causality operates that the order of events in time should be absolutely distinguishable and that no closed causal chains would be likely to arise. But, as I have mentioned already, this is a distinct issue, because at the fundamental level, where time-symmetric causality\index{causality!time-symmetric} operates, thermodynamic time asymmetry\index{thermodynamic time asymmetry} is ineffective and any restriction that would be imposed by the existence of the thermodynamic arrow of time\index{thermodynamic arrow of time} would be irrelevant.

One significant outcome of the existence of closed causal chains\index{closed causal chain} is that it is not always possible to establish the time order of events\index{time order of events} in an absolute way, because one event occurring along such a causal chain can be considered to occur both before and after another event occurring along the same causal chain, even if the events are uniquely ordered from the macroscopic viewpoint of thermodynamic time\index{thermodynamic time}. The topological order of time\index{topological order of time} is always clear locally, along a particle world-line\index{particle world-line}, but globally (even on a small scale) it must be determined in a purely relational way (as dependent on an arbitrarily-chosen reference point along a given circular trajectory), like any physically significant property.

It is important to realize that the existence of closed causal chains does not introduce additional unpredictability above that which is already assumed to characterize quantum evolution, because the current framework already involves some backward-in-time propagation\index{backward-in-time propagation} (I will further explain, in section \ref{sec:5.8}, what motivates the idea that backward causation\index{backward causation} is involved in determining correlation probabilities\index{correlation probabilities} in quantum mechanics). But even in a deterministic context, the fact that the present cause of a certain future event could itself be affected by this very same event would imply that it is not possible for the future to be determined by the past alone, because the future itself would be involved in determining the past that determines this very future.

Thus, it seems that even in the context of a hidden-variable model\index{hidden-variable models!fundamental randomness}, backward-in-time causation would imply that reality must remain fundamentally random and not merely unpredictable (as when we have insufficient knowledge concerning the exact present state of a system, from a conventional viewpoint). Of course, the simple fact that the cause of a future event can be located not in its past, but in its own future, also implies that even when a complete knowledge of the present quantum state of a system and its environment is available, it is not possible to identify all the causes which exert an influence on its future evolution, which means that from the viewpoint of time-symmetric causality\index{time-symmetric causality!unavoidable unpredictability} a certain measure of unpredictability is unavoidable that would not be present from a conventional viewpoint.

\bigskip

\noindent It is usually recognized that the problem which would be raised by what might be called a time travel\index{time travel!experience} experience has to do with the fact that such a phenomenon may allow the kind of closed causal chain\index{closed causal chain} in which the classical, \textit{unidirectional} principle of causality\index{classical principle of causality!violation} would be violated. More specifically, given the assumption that we are free to decide how we influence the future, in the context where our evolution is taking place irreversibly toward what would normally be the unknown future, it may appear that a time-traveler, arriving from the future, would be able to alter the course of a known history in the same way a normally evolving person is allowed to influence the unknown future.

The problem here does not only have to do with the fact that we don't know why such unidirectional-causality-violating evolution is never observed, it also concerns the fact that if we are, in effect, free to influence the course of events taking place along the direction in which our thought processes are functioning, then it would appear that by `traveling' back in time we might be able to alter a known future and to modify the course of events in a way that would be incompatible with the very possibility that the experience itself might have occurred, thereby giving rise to a time travel\index{time travel!paradox} paradox.

Although time travel has never been observed to occur and therefore remains a purely hypothetical problem for physics, the standard answer to the questions it raises is often believed to be David Deutsch's\index{Deutsch, David} proposal \cite{Deutsch-1}, based on the many-worlds interpretation of quantum mechanics\index{quantum mechanics!many-worlds interpretation}. What Deutsch suggested, basically, is that every time a paradox would be expected to occur that would involve an observer arriving from the future and altering the past conditions that gave rise to the future state that allowed the process to happen, history\index{history!splitting branches} would `split' into alternative branches, in the sense that both the initial history (in which the backward-evolving observer did not change the future) and its modified version (were the observer does effect a change that would prevent her from having effected this very same change) would occur, but would cease to interfere quantum mechanically with one another. Thus, it is proposed that there is an alternative future\index{alternative future} for every possibility that might be produced as a result of the influence exerted by a future event on a past event through backward causation\index{backward causation}.

I'm not sure what most people make of this description, but the problem I have with it is that I just can't figure out how it actually makes things any better. If we say that an observer arrives from the future and changes the past, then this past must be assumed to have already taken the `effect' into account when it first occurred and therefore it must be such that it allows the said future to occur, as I previously explained. So, how could this future be made different by such an altered past?

Clearly the problem with the hypothetical problem of a future `cause'\index{future cause} influencing a past `effect'\index{past effect} only occurs when we assume that there can actually arise inconsistencies or contradictions in the observed historical description of events\index{historical description of events!contradictions}. But when it is assumed that an observer can arrive from the future and change the past\index{past!alteration} to which she was causally related, it is not possible to say that the future\index{future!alteration} is merely altered from what it `originally' was by the presence of the observer, because the observer herself could not even have arrived from the future in such an altered version of history\index{history!altered version}. How could one possibly argue that a new future is written in an alternative branch of the universe's history\index{history!alternative branch}, as a result of the arrival of an observer from the future, if the backward-propagating influence of that observer did not even occur, in this alternative branch of history?

What the many-worlds approach\index{many-worlds approach} purports to show is that inconsistencies and contradictions can actually arise in our historical description of facts, but that this is acceptable, because the future always adapts to the inconsistencies it itself generates. But this is just non-sense, because if a future is such that it influences a given past, then this past must be such that it necessarily gives rise to this unique future\index{future!uniqueness} and this is not made to `happen' by some hypothetical splitting process\index{splitting process} taking place at a given arbitrarily-chosen moment, it is just how things actually are all along, in both the past and the future. Also, if we are to allow for the existence of other universes, then by definition those multiple universes\index{multiple universes!causal independence} should be causally independent from one another and things happening in one universe should not be allowed to influence what is taking place in another universe.

I believe that what is missing from our current understanding is an acknowledgment of the fact that a universe\index{universe!ensemble of causally related events}, by necessity, actually consists of a unique ensemble of events causally related to one another and to nothing else (as a consequence of the requirement of relational definition\index{requirement of relational definition!physical attributes} of physical attributes which was discussed in the preceding chapters of this report). From such a viewpoint, if an event in the past is influenced by the presence of an event in the future, then this past event cannot be causally related to a different future, but only to the future that actually influenced it. Thus, it becomes a fundamental requirement for the history\index{history!absence of self-contradictions} of a universe to form a consistent whole, free of self-contradictions.

Of course, we never experience time travel\index{time travel}, so this issue only has to do with elementary particles propagating backward in time and in this realm quantum field theory\index{quantum field theory} already does a very good job of consistently describing physical reality and predicting facts. In this particular sense, Deutsch's\index{Deutsch, David} proposal is a solution to a problem that does not exist and this becomes especially obvious in the context where, as I will later explain, the many-worlds interpretation of quantum theory\index{quantum theory!many-worlds interpretation} is not required to make sense of the quantum measurement process\index{quantum measurement process} and can even be understood to have consistency problems of its own (which does not mean that the multiverse concept\index{multiverse!concept}, as a distinct hypothesis, cannot be considered valid and fully justified). I believe that the strange and convoluted reality that emerges from such a description merely illustrates the kind of complications we would run into if we adopted an interpretation of quantum theory involving multiple splitting versions of history\index{history!multiple splitting versions}, not occurring within causally independent universes. In any case, if the many-worlds approach cannot even be made to work in a quantum-mechanical context, what motive do we have to invoke it in order to explain problems occurring at a classical level?

Historical consistency\index{historical consistency!requirement} requires that if a process is allowed to exert an influence backward in time, then this process must, in effect, evolve toward the very same past that gave rise to the future which is causally influencing this past, because the process must at all times remain causally related to the same external reality, otherwise nothing at all could be assumed to be causally related to anything else. But, then why is it, in effect, that we never experience time travel, if backward-in-time causation\index{backward-in-time causation} must be allowed to happen? Does this prohibition have to do with the fact that if it did not apply, then real contradictions might be made to occur whenever such closed causal chains\index{closed causal chain} would form?

To answer those questions, I must first point out that what would really differentiate time travel\index{time travel!experience} experiences from the backward-in-time propagation\index{backward-in-time propagation!elementary particles} of elementary particles that is routinely observed in laboratories is the fact that, with time travel, a macroscopic and thermodynamically constrained system, such as a living human being, would need to evolve with its thermodynamic arrow of time\index{thermodynamic arrow of time!reversal} reversed and pointing toward the past instead of the future. From the viewpoint of an observer not part of the process, this evolution would be seen as a local violation of the second law of thermodynamics\index{second law of thermodynamics!local violation}, or the principle that entropy never decreases in the future.

If the time traveler really travels back in time, then, as he does, he would not just remember what happened at his past \textit{destination}\index{past destination}, but also what happened in the future (which to him would appear to also be a past), thereby allowing information\index{information!flow toward past} to flow from the future toward the actual past. But this means that, from the viewpoint of a normal observer, the processes of memory\index{memory!formation process} formation and all the other irreversible processes\index{irreversible processes} usually involved in allowing a person to experience time\index{time!unidirectional phenomenon} as a unidirectional phenomenon would all appear to function backward for the time traveler, as the process is taking place, even if the time traveler is not composed of particles (like ordinary antiparticles\index{antiparticles!backward-in-time-propagating particles}) usually considered to be propagating backward in time.

At this point, it is necessary to recall the discussion from section \ref{sec:4.9} concerning the origin of thermodynamic time asymmetry\index{thermodynamic time asymmetry!origin} in a universe like ours. There, I explained that it is the inescapable nature of the constraint of global entanglement\index{global entanglement constraint} (which must be imposed in order to allow causal relationships\index{causal relationships!ensemble of elementary particles} to be established between all elementary particles in the initial Big Bang state) that explains the parallel nature of thermodynamic time asymmetry\index{thermodynamic time asymmetry!parallel nature}, or the fact that there doesn't coexist opposite thermodynamic arrows of time\index{thermodynamic arrow of time!unique direction} in different regions of the same universe, even for temporarily isolated branch systems\index{branch systems}.

Thus, in a universe in which negative-energy matter must have been present initially, with a density similar to that of positive-energy matter (for reasons I explained in section \ref{sec:4.5}), gravitational entropy\index{gravitational entropy!continuous decrease} (and therefore all entropy) must be continuously decreasing as we approach the instant in the past (which corresponds with the very first instant of the Big Bang\index{Big Bang!very first instant}) where the entanglement\index{entanglement!all elementary particles} of all elementary particles must take place, because if macroscopic black holes\index{macroscopic black hole} had been present at the Planck time\index{Planck time}, the elementary particles which would have been contained inside opposite-energy black holes\index{opposite-energy black holes} would not have been allowed to come into contact with one another in this initial state or at any time in the future, due to the insurmountable gravitational repulsion\index{gravitational repulsion!insurmountable} that would exist between such macroscopic black holes. As a result, all systems, regardless of how isolated they have become, must evolve with decreasing entropy in the past direction of time, because all matter particles, without any possible exception, must become entangled with the rest of the matter in the universe if they are to actually be components of that universe.

Of course, the point here is that if time travel\index{time travel!unobserved phenomenon} is never experienced or observed, it is not because backward-in-time causation\index{backward-in-time causation} is impossible at a fundamental level, but merely because entropy must be continuously decreasing in the past and only in the past (because no global entanglement constraint\index{global entanglement constraint} applies to future singularities), which means that the conditions necessary for the thermodynamic arrow of time\index{thermodynamic arrow of time!past direction of time} to be experienced in the past direction of time are not merely unlikely, they are actually forbidden, for all practical purpose. Unidirectional causality\index{unidirectional causality!past toward future} only operates from the past toward the future, because it would take a very significant fluctuation for entropy to temporarily decrease in the future, from a present state of non-maximum entropy, but given that this would be required for time travel to occur, then it is possible to understand why we never experience such phenomena.

Classical, unidirectional causality is reflected in the fact that it would take only a little change in the past to allow a present event not to have occurred, while, in general, enormous changes would be required to take place in the future for some present event not to have occurred, which is causally influenced by this future. This asymmetry is precisely what is enforced by the global entanglement constraint, when it is assumed that causal relationships\index{causal relationships!ensemble of elementary particles} must be established between all elementary particles in the initial singularity. For time travel to be possible, this thermodynamic time asymmetry\index{thermodynamic time asymmetry!local reversal} would need to be reversed locally for the whole duration of the process and the unlikeliness of such an evolution is responsible for the fact that time travel\index{time travel!controlled phenomenon} is virtually impossible, at least as a controlled phenomenon.

Therefore, it is not possible, in practice, to be involved in a closed causal chain\index{closed causal chain} while remembering what occurred at a later time (no information can be transferred from the future toward the past), even if this restriction does not affect the possibility for microscopic systems to be involved in such causal chains, as long as historical consistency\index{historical consistency} is preserved (I will explain in section \ref{sec:5.8} how this condition is enforced at the fundamental level). This means that so-called `knowledge' paradoxes\index{knowledge paradox} are also unlikely to occur.

It was suggested, in effect, that, if time travel\index{time travel} was possible, there could arise situations where some valuable piece of information (say a beautiful treatise about the physics of time directionality) would be brought back from the future that would not have existed before it arrived from that future, but which would nevertheless become available as a result of the process, so that it can later be brought back to the present, thereby raising questions as to its origin. But given that what would be required for such a paradox to occur is a sustained local increase of entropy toward the past\index{sustained local increase of entropy toward past} (or equivalently a sustained local decrease of entropy toward the future\index{sustained local decrease of entropy toward future}), then it follows that, while the communication of information\index{information!communication from future} from the future is not fundamentally impossible, it is nevertheless as unlikely as the \textit{instantaneous} production of a large amount of complex structure out of a state of thermal equilibrium\index{thermal equilibrium state!production of complex structure}, which, again, is not fundamentally impossible, but merely ridiculously unlikely\footnote{
It must be clear, that even if a system that is interacting thermodynamically with its environment could potentially experience a sustained decrease of entropy in the future (at the expense of a larger increase of entropy in the environment), such changes would not allow information\index{information!communication from future|nn} to be communicated from the future toward the past, because they are always produced by causes that originate in the past.}.

Thus, from my viewpoint, despite the fact that backward causation\index{backward causation} is allowed to occur, it is not possible for information to be communicated backward in time. In fact, if it was not for the constraint that is responsible for the diminution of entropy in the past, all evolution would be symmetric with respect to the direction of time, at all levels, and under such conditions there would be no way for information\index{information!absence of flow to past and future} to flow from either the past or the future, as all systems would remain in a state of thermal equilibrium\index{thermal equilibrium state} (if that was possible) and no record making process\index{record making process} would ever be allowed to take place. Only under such conditions would it be possible to directly appreciate the fact that the future is not fundamentally different from the past.

It is, therefore, possible to understand that what prevents the classical, unidirectional principle of causality\index{classical principle of causality!violation} from being violated is not merely the fact that if such violations could arise, then closed causal chains\index{closed causal chain} could be produced that would give rise to time travel\index{time travel!paradox} paradoxes. Instead, this limitation is an outcome of the global entanglement constraint\index{global entanglement constraint!past entropy growth} and the restriction it imposes on the growth of entropy in the past direction of time, which is clearly not a constraint of irreversibility that is imposed at a fundamental level, but rather one that emerges from the particular boundary conditions\index{boundary conditions!initial Big Bang state} which apply to the initial Big Bang state. The frequently encountered remark, to the effect that objects can move in any direction of space, but not in any direction of time (at least when they are restricted to not move faster than light in a vacuum), is only true in the sense that it is not possible to produce a local reversal of the thermodynamic arrow of time\index{thermodynamic arrow of time!impossibility of reversal} that would last a significant period of time; it does not mean that a particle cannot propagate backward in time under appropriate conditions.

What is important to understand is that, not only would entropy be observed to decrease in the future during a hypothetical time travel phenomenon\index{time travel phenomenon!observable process} (which would require it to increase in the past), but under normal conditions the process would remain observable all along as an entropy diminishing process taking place forward in time, even after a hypothetical time travel\index{time travel!paradox} paradox would have been produced. Indeed, an observer\index{observer!backward-in-time evolution} which would be evolving backward in time, from a thermodynamic viewpoint, would still be causally influenced by the events taking place at the moment of unidirectional time\index{unidirectional time} which would appear to her as the present, so that she would observe the same sequence of events about which she already had knowledge, only now those events would take place in the reverse order. But she would also be allowed to causally influence those events at all times and those events would influence her future, even before the process terminates in the remote past.

Therefore, if a time traveler\index{time traveler!memory of future} does remember the future as she evolves backward in time, then this future can only be the one which she already witnessed, even though it may seem that she would be free to alter this particular future, because if the process is allowed to occur it is because the future is such that it allows the process to actually arise and this means that \textit{at all times} the present itself is necessarily such that it allows that particular future (about which information would be available) to happen. If that was not the case then the time traveler\index{time traveler!absence of knowledge of original future} would not have advanced knowledge of the original future she wants to alter, but would rather remember the alternate future\index{alternate future!time traveler} she intends to create, which means that she would not be better off than an ordinary observer at telling what that future is that she would try to alter and therefore we have no reason to believe that she would be allowed to \textit{voluntarily} effect such a change.

What makes the time travel\index{time travel!paradox impossibility} paradoxes themselves impossible, however, is not the fact that they would require entropy to grow in the past, but the very same constraints that would forbid the occurrence of a factual contradiction\index{factual contradiction!fundamental viewpoint} from a fundamental viewpoint, as when elementary particles\index{elementary particle!backward-in-time propagation} are propagating backward in time, without being involved in anti-thermodynamic evolution\index{anti-thermodynamic evolution}. In this particular sense, it is true that the problem of time travel\index{time travel problem!quantum-mechanical context} can be fully resolved only in a quantum-mechanical context, but as I previously indicated (and for reasons that will be discussed later) this does not mean that one must invoke a hypothetical splitting process\index{splitting process!multiple branches of history}, involving multiple branches of history, as is proposed in the context of the many-worlds interpretation of quantum theory\index{many-worlds interpretation!quantum theory}. In any case, it is now possible to appreciate that what makes time travel\index{time travel!impossibility}, itself, impossible is not the fact that it may allow forbidden contradictions to occur, but really the improbability for processes to occur with decreasing entropy in the future.

But even if one was allowed to travel back in time, as a result of a phenomenal anti-thermodynamic fluctuation\index{anti-thermodynamic fluctuation}, one would not be allowed to alter one's own future, despite the fact that this is unexpected from our everyday viewpoint. Even under such conditions, there would necessarily occur events that would enforce historical consistency\index{historical consistency!enforcement} and this would happen despite the fact that, under normal conditions, we seem to be free to modify the future at will. It is simply the fact that we are used to experience the future as unknowable in advance that explains that it appears doubtful that we would \textit{not} be able to alter the course of reality\footnote{
In order to understand how historical consistency\index{historical consistency|nn} can be obeyed, even under such circumstances, it may help to notice that if knowledge about some future was to become available to a given observer, a prediction of her actions would have to take into account the fact that the prediction\index{prediction!influence on outcome|nn} itself can influence the outcome.}.
 We usually have no factual knowledge about the future\index{future!absence of factual knowledge} and this is why we never run into the possibility of being prevented from making a decision that we may expect would alter a known fact about the future. The requirement that history\index{history!global consistency} be globally consistent appears to have unexpected consequences merely because we are not used to experience a reality in which information would be available about what has not yet occurred.

We are accustomed to observe that present actions exert an influence on the probability that such or such a future occurs, but this is only a reflection of the fact that there exist correlations between the past and the future, which are the result of both forward- and backward-in-time propagated effects and it does not mean that it is impossible for a unique future\index{future!uniqueness} to be causally related to the unique past\index{past!uniqueness}. It is merely the fact that all possibilities are usually allowed for the unknown future, while only a subset of them is allowed for the known past (due to the constraint responsible for the diminution of gravitational entropy\index{gravitational entropy}) that justifies the impression we all share of being able to exert a certain control over the future which does not apply for the past, because it is precisely those limitations which imply that we can obtain knowledge about the past (which, therefore, appears unalterable), but not about the future (which, therefore, appears modifiable).

From a fundamental viewpoint, the future is not different from the past (even if it cannot be determined in advance) and we do know that the past cannot be changed from what it already is. If we never remember the future and if we are never confronted with the limitations to free-will\index{free-will!limitations} which exist as a result of the requirement of historical consistency\index{historical consistency!requirement}, it is simply due to the fact that information\index{information!absence of flow from future} does not usually flow from the future toward the past. This is probably the most important lesson that can be learned from the study of hypothetical time travel\index{time travel!experience} experiences, in the context of time-symmetric causality\index{causality!time-symmetric}: we are causally related to one unique past.

But this is also true for the future. We live in an unpredictable universe and while it is certainly true that what we choose to do now has an effect on what will happen tomorrow, based on the most rational explanation of both classical and quantum-mechanical phenomena, it is necessary to recognize that we are causally related to only one such future and even if we were to obtain, in advance, knowledge about what this unique future\index{future!uniqueness} actually is, events would have to unfold in such a way that the consistency of history would remain inviolable.

\section{Advanced waves and time asymmetry\label{sec:5.5}}

Since Maxwell\index{Maxwell, James Clerk} introduced his electromagnetic field equations\index{electromagnetic field equations!retarded and advanced solutions}, more than a hundred fifty years ago, it has been known that there exist both retarded and advanced solutions to those equations (this is equivalent to say that Maxwell's equations\index{Maxwell's equations} do not distinguish the future from the past). The retarded solutions\index{retarded solutions!Maxwell's equations} describe the propagation of positive-energy electromagnetic waves\index{electromagnetic waves} leaving a point source and spreading into a growing volume of space as time passes. The usually rejected advanced solutions\index{advanced solutions!Maxwell's equations}, on the other hand, would describe the propagation of electromagnetic waves of opposite energy sign leaving a point source and spreading into a growing volume of space in the past direction of time. This is usually described as the hypothetical phenomenon of a spherical and concentric positive-energy electromagnetic wave converging on a point source in the future direction of time\footnote{
The positive value of the energy of this converging wave\index{converging wave|nn}, which allows the `source' to gain energy as a result of the absorption process, arises from the fact that, as I explained in chapter \ref{chap:2}, a negative-energy photon\index{negative-energy photon!backward-in-time propagation|nn} propagating backward in time is always observed as a positive-energy photon propagating forward in time, while a negative-energy photon\index{negative-energy photon!forward-in-time propagation|nn} propagating forward in time would not even be allowed to interact with ordinary matter.}.
 From this equivalent viewpoint, it is obvious that the advanced solutions represent a kind of process that cannot occur, because, from the unidirectional-time viewpoint\index{unidirectional-time viewpoint}, one never observes light waves\index{light waves!convergence on source}, or any other kind of electromagnetic waves, converging on a `source' just to be absorbed by this source.

But while this observation reassures our commonsense expectations, the fact that the phenomenon described here never occurs, while there is no \textit{a priori} reason why it couldn't happen, still constitutes a profound mystery from a theoretical perspective. It is usually recognized, in effect, that if a valid theory describes a certain phenomenon and there is no good motive to assume that this phenomenon should be forbidden, then its occurrence is compulsory.

It is not enough to argue that what prevents the hypothetical phenomenon of a radio wave\index{radio wave!convergence on transmitter} produced by multiple sources in the environment converging in perfect spherical symmetry and with perfectly correlated phases onto a transmitter, where it would be absorbed, is the unlikeliness of the phenomenon, because, as I emphasized in chapter \ref{chap:4}, this is precisely what we `observe' to occur in the past direction of time and this evolution is clearly not the outcome of the singular nature of present conditions\index{present conditions!singular nature}. Given arbitrary initial conditions, what we should expect to observe are waves that would be diverging in the past, just like they do in the future, because this is, in fact, the most likely evolution when only the present conditions are fixed, even if, from the unidirectional-time viewpoint\index{unidirectional-time viewpoint}, such a process would appear unlikely.

If it is considered natural for electromagnetic waves\index{electromagnetic waves!spreading outward} to spread outward in the future, despite the fact that this means that they converge on their source in the past, then it should also be expected that electromagnetic waves would spread outward in the past, even if that means they would converge on their source in the future. Therefore, what remains unexplained is the asymmetry of the situation in which waves do not spread outward in the past, while they do so in the future of some arbitrarily-chosen initial state. The problem discussed here is all the more significant, given that it is not restricted to Maxwell's theory\index{Maxwell's theory}. Indeed, there exist advanced solutions to all relativistically invariant wave equations\index{relativistically invariant wave equations!advanced solutions}, including the equations that describes the propagation of electrons in quantum field theory\index{quantum field theory}.

Once again, this is a problem that Feynman\index{Feynman, Richard} visited, although it appears that he failed to resolve the issue. What he and John Wheeler\index{Wheeler, John} proposed was a theory \cite{Feynman-2} that would have allowed advanced electromagnetic waves\index{advanced electromagnetic waves} to be produced on an equal basis with retarded electromagnetic waves\index{retarded electromagnetic waves}, just to be canceled out through destructive interference\index{destructive interference}, as a consequence of the difference in opacity that seems to characterize the far past and the far future of our universe.

According to this model, retarded and advanced electromagnetic waves are always produced together in equal proportions and propagate in the future and the past respectively. But when the retarded wave is absorbed in the future, the absorbing process itself triggers the emission of an additional retarded wave of identical amplitude and given that this additional retarded wave is assumed to be completely out of phase with the original retarded wave, then all traces of its presence are erased. At the same time, the absorber also produces an advanced wave and if certain conditions are met, this advanced wave only serves to strengthen the retarded wave produced by the source through constructive interference\index{constructive interference}, while it also conspires to cancel out the advanced wave originally emitted by the same source through destructive interference, which may allow to explain the fact that it is not observed. The problem is that this \textit{absorber theory}\index{absorber theory} requires that there is more absorption in the future than in the past, while this possibility would appear unlikely in the context where space is expanding in the future direction of time and the matter density\index{matter density!maximum} is maximum at the Big Bang.

Other theories, based on similar assumptions (see for example Refs. \cite{Csonka-1} \cite{Partridge-1} \cite{Cramer-1}) and which tried to overcome the problems encountered by Feynman\index{Feynman, Richard} and Wheeler\index{Wheeler, John} through various alternative hypotheses (for example by assuming that the Big Bang\index{Big Bang!reflector of advanced radiation} acts as a reflector of all advanced radiation), have apparently also failed to produce a satisfactory solution to the problem of advanced waves\index{advanced waves!problem}. It seems that, whenever it is not independently assumed that, for some unknown reason, a fundamental asymmetry\index{fundamental asymmetry!interaction of matter with radiation} exists in the interaction of matter with radiation that would differentiate the far past from the far future, the desired outcome is never obtained.

The only way to reproduce the observed time asymmetry that characterizes wavelike processes\index{wavelike processes!time asymmetry} in our universe, using such a model, is by postulating that some asymmetry exists, which is responsible for reducing or increasing the amount of interference that takes place either in the past or in the future. But given that no convincing explanation exists that would justify this assumption, then it is apparent that it merely amounts to assume the very outcome we would like to explain. From the difficulties encountered with this kind of approach, it has become pretty clear that it is not possible to explain the absence of advanced waves\index{advanced waves!absence} as being a mere consequence of hypothetical interference effects.

I was only able to understand what explains the absence of advanced waves when I began considering the quantum aspect of this hypothetical phenomenon. Indeed, I already knew that backward-in-time propagation\index{backward-in-time propagation!elementary particles} was possible for elementary particles and therefore it seemed to me that what was not allowed was not really backward propagation itself, but merely the spreading of a backward-propagating wave\index{backward-propagating wave!spreading} into an increasingly larger region of space. I also knew that there was a requirement, imposed by the constraint of global entanglement\index{global entanglement constraint} which I had recently uncovered, that the evolution that takes place in the past direction of time be such that it gives rise to a continuous decrease of gravitational entropy\index{gravitational entropy!continuous decrease}. But, as elegantly explained by Olivier Costa de Beauregard\index{Costa de Beauregard, Olivier} \cite{Costa-1}, there is a certain equivalence between entropy increase toward the future and wave retardation\index{wave retardation!entropy increase toward future}, which is implied by Planck's definition of entropy\index{entropy!Planck's definition} and which arises from the quantized nature of electromagnetic radiation\index{electromagnetic radiation!quantized nature}.

In a quantum-mechanical context, entropy necessarily rises when an electromagnetic wave\index{electromagnetic waves!spreading} spreads into a larger volume of space, because, at any given time, the photons associated with an expanding wave front\index{expanding wave front!photons} can be detected anywhere on its growing surface. In fact, given that, from the viewpoint of relativistic quantum field theory\index{quantum field theory}, any wave equation\index{wave equation!propagation of elementary particles} is associated with the propagation of some elementary particle, it follows that entropy increase in the future is always associated with wave retardation, while the observation of advanced waves\index{advanced waves!entropy decrease toward future} would always imply that a decrease of entropy has taken place in the future direction of time. But this is equivalent to say that entropy would need to increase in the past for an advanced wave\index{advanced waves!entropy increase in past} to spread as it propagates backward, while this is precisely what is forbidden by the second law of thermodynamics\index{second law of thermodynamics}.

What is also unexpected, from a thermodynamic viewpoint, is the fact that, from the unidirectional-time viewpoint\index{unidirectional-time viewpoint}, the existence of advanced waves\index{advanced waves!work out of nothing} would seem to allow work to be generated out of nothing, when radiative energy\index{radiative energy!convergence on source} would converge on a `source'. But the existence of advanced waves\index{advanced waves!transmission of information from future} would also make possible the transmission of information from the present or the future toward the past. It is natural to expect, therefore, that this kind of process should be prevented from occurring by the same condition that explains thermodynamic time asymmetry\index{thermodynamic time asymmetry}.

It must be clear, however, that simply invoking the classical (unidirectional) principle of causality\index{unidirectional principle of causality} does not allow to solve the problem of the absence of advanced waves\index{advanced waves!problem of absence}, because, in the above discussed context, saying that there always exists a unique preferred direction in time for the propagation of effects\index{propagation of effects!preferred direction in time} merely amounts to restate the problem of advanced waves (which is also known as the problem of the electromagnetic arrow of time\index{electromagnetic arrow of time!problem}) without explaining why such a restriction is indeed observed to apply. In fact, the hypothetical phenomenon of time travel\index{time travel phenomenon} I have described in the preceding section would be one particular instance of backward-in-time communication\index{backward-in-time communication}, of the kind that would be allowed by the existence of advanced electromagnetic waves\index{advanced electromagnetic waves}, and therefore a solution to the problem of advanced waves would definitely rule out time travel.

Now, I mentioned in section \ref{sec:5.3} that the causal structure of spacetime\index{causal structure of spacetime} is not incompatible with the concept of backward-in-time causation\index{backward-in-time causation}, given that with every event is associated both a future and a past light cone\index{future and past light cones}, which reflect the existence of a speed limit imposed on the propagation of causal signals\index{causal signals!propagation in future or past} in either the future or the past. But it should also be clear by now that there is a difference between the kind of backward-in-time causation that may occur as a consequence of the propagation of an elementary particle\index{elementary particle!backward-in-time propagation} backward in time and the kind of causality we experience in a purely classical context and which is known to operate only forward in time.

Thus, while it is not observationally forbidden for an electron to propagate backward in time, an explanation of cosmological time asymmetry\index{cosmological time asymmetry} based on the global entanglement constraint\index{global entanglement constraint} would not allow this propagation to occur in such a way that the surface over which the presence of the electron could be detected at an \textit{earlier} time would be growing continuously along with the two-dimensional boundary of the past light cone\index{past light cone}. But this is precisely the kind of evolution that an advanced wave\index{advanced waves!quantum-mechanical viewpoint} would describe from a quantum-mechanical viewpoint and therefore what explains that advanced waves are absent is the constraint of global entanglement I have identified in section \ref{sec:4.9}, which enforces a continuous decrease of entropy in the past, as a consequence of the requirement that causal relationships\index{causal relationships!ensemble of elementary particles} be established between all the elementary particles which were present in the initial Big Bang singularity.

Our failure to observe advanced waves\index{advanced waves!absence of observation} must not, therefore, be interpreted as an indication that backward-in-time propagation\index{backward-in-time propagation}, or backward-in-time causation\index{backward-in-time causation} are forbidden, but rather as evidence that only a small subset of potentially available states is available as `final' conditions for backward-propagating particles\index{backward-propagating particles!final conditions}. Such particles are not only prevented from propagating faster than the speed of light in the past direction of time by the causal structure of spacetime\index{causal structure of spacetime} and the existence of a past light cone\index{past light cone}, they are also prevented from propagating in all possible directions of space in ways that would allow entropy to grow in the past. This means that the statistical predictions\index{statistical predictions!identically prepared systems}, obtained using quantum theory, for the evolution of a large number of identically prepared physical systems are not valid in the past direction of time and this is what explains that electromagnetic waves\index{electromagnetic waves!wave functions}, as particular instances of wave functions, are never observed in their advanced form.

In such a context, it becomes apparent that the only true virtue of the Feynman-Wheeler absorber theory\index{Feynman-Wheeler absorber theory} (aside from the fact that it was one of the first models which actually took the problem of advanced waves\index{advanced waves!problem} seriously) is that it sought to deduce the absence of advanced waves\index{advanced waves!boundary conditions} from boundary conditions imposed on the universe at large, instead of requiring that time asymmetry\index{time asymmetry!fundamental} be imposed at a fundamental level, which could only be satisfied by assuming that backward-in-time propagation is impossible. In any case, even if absorber theory had conveniently solved the problem of advanced waves, this solution would have remained problematic, because it would not have allowed to explain the origin of thermodynamic time asymmetry\index{thermodynamic time asymmetry!problem of origin} in a more general context (when quantum interferences\index{quantum interferences!absence} are absent).

From my viewpoint, the fact that there also exist advanced solutions to Dirac's relativistic equation\index{Dirac's relativistic equation!advanced solutions} for the electron allows to confirm the validity of the conclusion that the absence of advanced waves\index{advanced waves!absence} does not preclude backward-in-time propagation\index{backward-in-time propagation!photon}, because, while it is not possible to assess whether a given photon propagates forward or backward in time, in the case of electrons it is possible to differentiate a forward-in-time-propagating particle from a backward-in-time-propagating particle, given that, from a unidirectional-time perspective\index{unidirectional-time perspective!backward-in-time propagation}, such an electron is observed as a positron with its positive electric charge. Therefore, if we do observe positrons, it means that the irrelevance of advanced solutions cannot arise from the nonphysical nature of backward-in-time-propagating particles and must, in effect, be the outcome of the global entanglement constraint\index{global entanglement constraint}.

\section{Early interpretations\label{sec:5.6}}

To begin the portion of this chapter that deals with quantum aspects of reality more specifically, I would like to first describe what constitutes the distinctive characteristic of the revised interpretation of quantum theory\index{quantum theory!revised interpretation} I will propose. What I had already understood, even before I was able to solve the problem of advanced waves\index{advanced waves!problem}, is that the processes that constitute the essence of our experience of reality are all mirrored by similar processes which obey the same observable macroscopic conditions\index{observable macroscopic conditions}, but which take place in the opposite chronological order\index{chronological order!opposite}, in a parallel portion of history\index{parallel portions of history!causal independence} that must be assumed \textit{independent} from the viewpoint of local causality.

The hypothesis that history does not occur only once, but must happen a second time in the reverse order may appear arbitrary and unnecessary, given that we know of only one history, but, as I will explain, this possibility is actually made unavoidable by some of the most fundamental principles of physics and also reflects the basic mathematical structure of quantum theory\index{quantum theory!mathematical structure}. Even though I was not motivated only by the desire to produce a time-symmetric theory\index{time-symmetric theory} when I began developing this original approach, the final outcome does share a certain property of time symmetry with some early interpretations of quantum theory\index{quantum theory!early time-symmetric interpretations} which are based on the hypothesis that there must be an equivalence between initial and final conditions.

Given that most of those early time-symmetric interpretations constitute more or less elaborate (and more or less inappropriate) quantum-mechanical versions of the original absorber theory\index{absorber theory!quantum-mechanical version} discussed in the preceding section, then one may say that this theory is their common ancestor. In this respect, it is apparent that those time-symmetric quantum theories also share some of the above discussed weaknesses of the original, classical theory.

I believe that if this kind of approach is usually considered to have failed to produce a satisfactory interpretation of quantum theory\index{quantum theory!satisfactory interpretation}, despite the many advantages it offers (which will be discussed below), this is due in part to the fact that absorber theory\index{absorber theory!failure}, itself, is considered a failure. As a result, many generations of physicists were inoculated against time-symmetric approaches in general, even though a few well-informed specialists have recognized that the requirement of time symmetry\index{time symmetry!requirement} is essential to a consistent interpretation of quantum theory\index{quantum theory!consistent interpretation}. But it is also clear that this is not the only reason why the early attempts at formulating a time-symmetric version of quantum theory\index{quantum theory!time-symmetric version} did not arouse more interest, because, as I came to understand, they also contain hypotheses and constructs that make them inconsistent and inadequate as a representation of quantum reality\index{quantum reality!inconsistent representation}.

One of the first interpretation of quantum theory that sought to accommodate the requirement of time symmetry was that proposed by John Cramer\index{Cramer, John} \cite{Cramer-2} as an outcome of his work on the problem of advanced waves\index{advanced waves}. As such, it contains hypotheses which are very similar to those of the original absorber theory which I have identified as problematic. But the most important defect of this transactional interpretation\index{transactional interpretation!quantum theory}, in my opinion, is that, despite the fact that it is proposed as an alternative time-symmetric model, it actually involves a fundamental time asymmetry\index{time asymmetry!fundamental} that is incompatible with this basic requirement.

What Cramer proposed, basically, was that a kind of `handshake' process\index{handshake process} takes place whenever a quantum particle is emitted by a source and then propagates a certain distance before being absorbed by a detector. We may consider, for example, the traditional double slit experiment\index{double slit experiment} in which a particle must go from source to detector by passing through the slits. It is known that an accurate estimation of the probability for such a process to occur must take into account the existence of interferences between the individual probability amplitudes\index{probability amplitudes!interferences} associated with each of the paths through which the particle is allowed to go, whenever both slits are open.

What Cramer's handshake process involves is the emission of a classical wave\index{classical waves!offer} acting as an `offer', which is assumed to be sent by the source forward in time and which is allowed to propagate without constraint (it is assumed to go through both slits all at once), followed by the production of another such classical wave\index{classical waves!confirmation} that would constitute its `confirmation' and which would be sent by the detector backward in time (toward the initial emission event), upon absorption of the offer wave. The most problematic aspect of this description, from my viewpoint, is the fact that the confirmation wave must follow an evolution that is restricted to be compatible with the macroscopic constraints\index{macroscopic constraints} which would have existed if the particle (not the offer wave) had been restricted to follow the unique classical path\index{classical path!uniqueness} it is assumed to actually have taken as it propagated forward in time (the confirmation wave only comes back through one of the two open slits)\footnote{
In fact, Cramer\index{Cramer, John|nn} assumes that this handshake process\index{handshake process|nn} is actually repeated several times, for any single quantum process, and is responsible for the transfer of energy and other conserved physical attributes which take place during the process, but we may ignore this unnecessary aspect of the handshake process if it simplifies the discussion.}.

It is difficult to see how the advanced wave\index{advanced waves!distinct macroscopic constraints} could be submitted to macroscopic constraints which differ from those that apply to the retarded wave\index{retarded waves}, in the context where the observable macroscopic conditions\index{observable macroscopic conditions} of the experiment are fixed once and for all. But what is even more incomprehensible with this interpretation is that the evolution of the `confirmation' wave\index{confirmation wave} is actually required to reflect the fact that the particle took a certain path (say the upper slit), while the evolution of the `offer' wave\index{offer wave} would not be allowed to reflect the same fact (passage through both slits would initially be allowed).

This is how time asymmetry\index{time asymmetry!transactional model} is reintroduced in this transactional model, as a means to allow a unique, \textit{classically} well-defined history\index{classically well-defined history} to correspond with the process, despite the fact that the statistics of this quantum mechanical process\index{quantum mechanical process!statistics} can only be explained by assuming that the particle is not restricted to follow a unique path. Of course, even if those problems did not exist, there would still be a difficulty associated with the fact that this approach requires the existence of both classical waves\index{classical waves} and classical particles\index{classical particles!unique trajectories} (constrained to follow unique trajectories by those classical waves), while it is known that both concepts (which are shared by certain classical hidden-variable theories\index{classical hidden-variable theories}) are problematic from a quantum-mechanical viewpoint.

I believe that the source of the problems affecting Cramer's\index{Cramer, John} transactional interpretation of quantum theory\index{quantum theory!transactional interpretation} is to be found in the fact that it assumes that the retarded and advanced waves\index{retarded and advanced waves!same portion of history} are actually propagating in the same portion of history, because this is why it needs to be required that the quantum particle submitted to the constraint of those classical waves goes through only one slit, corresponding to this unique history\index{history!uniqueness}, which, in turn, requires a certain fundamental time asymmetry\index{time asymmetry!fundamental} to be introduced into the theory, in violation of the time-symmetric nature of its equations. Also, the fact that, as a particular instance of (quantum-mechanical) absorber theory\index{absorber theory!quantum-mechanical version}, Cramer's framework appears to require genuine wave emission and absorption to take place in the course of all quantum processes, may be problematic, because there are situations where quantum measurements\index{quantum measurement!interaction free} are performed without interaction.

Those difficulties are more significant than the additional problem that would arise in the context where it is not clear, from the viewpoint of Cramer's theory\index{Cramer's theory}, when it is exactly that the offer portion of the handshake process\index{handshake process!offer portion} would be completed, while the particle is propagating along its classical path\index{classical path}. If the offer portion of the handshake was to be completed when the particle reaches one of the two open slits, then the process would always be that which we expect to occur when one is allowed to observe through which slit the particle goes and under such conditions, the particle would follow a quasiclassical trajectory\index{quasiclassical trajectory!absence of interferences} (interferences would be absent), which is contrary to observation. Thus, there may be a difficulty associated with the apparent arbitrariness of the choice of which macroscopic conditions are necessary to trigger a handshake process\index{handshake process!triggering conditions} (do we have to wait for an observer to become aware of the outcome as John Von Neumann\index{Von Neumann, John} once proposed?).

But this is, in fact, the same quantum measurement problem\index{quantum measurement problem} as may affect a more conventional interpretation and therefore we are allowed to assume that any solution to this problem that would be proposed in a more conventional context would also apply to the transactional interpretation\index{transactional interpretation!quantum theory}. This is an important point, because this difficulty is sometimes proposed as an argument against all time-symmetric approaches to quantum theory\index{time-symmetric quantum theories!common counter-argument}, while, when it is properly understood, it no longer stands out as a problem that is specific to time-symmetric models. Of course, it would not be appropriate, either, to assume that Cramer's theory is equivalent to standard quantum theory\index{standard quantum theory}, as its author suggested, because the validity of the predictions derived from ordinary quantum mechanics\index{quantum mechanics!ordinary} does not depend on the existence of advanced waves\index{advanced waves}, while this hypothesis is essential in the context of the transactional interpretation. In fact, when the inadequacy of the boundary conditions\index{boundary conditions!destructive interference} that give rise to the destructive interference effects that would allow advanced waves to go unnoticed is recognized, the theory no longer even agrees with observation, which certainly makes it different from standard quantum mechanics.

What I'm suggesting that we retain from those alternative, semi-classical interpretations\index{semi-classical interpretations!quantum theory} of quantum theory is the notion that the squaring of the wave function\index{squaring of wave function!probability of occurrence}, which allows one to estimate the probability of occurrence of a process, is made necessary as a consequence of the fact that, somehow, two histories are involved in every quantum process\index{quantum process!two histories}. I believe that this is what explains that it is merely by multiplying the probability amplitudes\index{probability amplitudes!two histories} associated with those two histories that we can obtain (under appropriate conditions) the probability for the entire process to occur. Indeed, the squaring of the wave function (which is necessary to obtain the probability of a process in quantum mechanics) involves taking the complex conjugate of the probability amplitude\index{probability amplitudes!complex conjugate and time reversal} associated with one history before multiplying it with the probability amplitude associated with another history, while taking the complex conjugate of the wave function is equivalent to reversing the direction of time in which it varies.

Therefore, the core mathematical framework of quantum theory\index{quantum theory!core mathematical framework} already contains, in embryonic form, the requirement that each process be described as a history\index{history!unfolding forward and backward in time} that unfolds both forward and backward in time, for some mysterious reason. This otherwise puzzling requirement has been transformed by modern interpretations (such as the consistent-histories interpretation of quantum theory\index{consistent-histories interpretation!quantum theory}) into a condition, imposed (without any real justification) on certain pairs of minimally coarse-grained histories\index{minimally coarse-grained histories!pairs}, that they provide the probability of occurrence of a `consistent' history\index{consistent histories!probability of occurrence}. But in the process, it seems that the most important aspect of this requirement, which is the fact that the two histories forming such pairs actually take place in opposite directions of time, was lost and with it, the important insight we should have learned from early time-symmetric interpretations of quantum theory\index{quantum theory!early time-symmetric interpretations}.

At this point, it is important to mention that a more pragmatic approach to achieve symmetry with respect to the direction of time in quantum mechanics\index{quantum mechanics!time symmetry} had already been proposed by Aharonov\index{Aharonov, Yakir}, Bergmann\index{Bergmann, Peter} and Lebowitz\index{Lebowitz, Joel} \cite{Aharonov-1} (see also Ref. \cite{Aharonov-2} for a more recent review) long before Cramer\index{Cramer, John} introduced his transactional interpretation\index{transactional interpretation!quantum mechanics}. Unlike the transactional interpretation, this two-state-vector formulation of quantum mechanics\index{two-state-vector formulation!quantum mechanics} really is mathematically equivalent to the standard theory, but it does not seek to explain the time asymmetry of boundary conditions\index{boundary conditions!time asymmetry} and merely suggests that two state vectors are required to describe the state of a quantum system\index{state of quantum system!two state vectors}. One state vector\index{state vector!information from past measurements} contains all the information obtained from past measurements (as in the standard interpretation) and the other contains all the information that will be obtained, concerning the same system, in the future. Between measurements, those two state vectors follow a `unitary' evolution\index{unitary evolution!toward future or past} toward the future and toward the past respectively\footnote{
I use the term `unitary' to denote the \textit{deterministic} evolution of the wave function\index{wave function!deterministic evolution|nn} or state vector\index{state vector!deterministic evolution|nn} that takes place in the absence of a change (usually performed through an act of measurement\index{measurement!act|nn}) in the observational constraints\index{observational constraints!absence of change|nn} applied on a quantum system, because using the term `deterministic' would be misleading in the context where I will argue that the evolution of the system itself always takes place randomly.}.

What this means is that there is no longer a preference for the past over the future in determining the current state of a quantum system (a system can be submitted to both pre- and post selection\index{pre- and post selection!quantum system}, although the post selection\index{post selection!future measurement} is only apparent after a future measurement has actually been performed). Of course, there is a natural reluctance to recognize that it might be possible for a state vector\index{state vector!determined by future} to be determined by what `happened' in the future, instead of what happened in the past, but this is merely a consequence of the previously discussed prejudice toward a unidirectional conception of causality\index{causality!unidirectional conception}, which we inherited from our thermodynamically constrained experience of reality\index{experience of reality!thermodynamic constraint} and it does not rest on any rationally formulated argument.

It must be clear that, despite the equivalence between the two-state-vector formalism\index{two-state-vector formalism} and standard quantum theory\index{standard quantum theory}, it has been shown that post selection\index{post selection!effect of future measurement on past state}, or the effect of a future measurement on the past state of a system, is not an optional feature of quantum theory, but arises even in the simplest and most conventional of quantum-mechanical experiments. Indeed, in certain interferometer experiments\index{interferometer experiment} (called delayed-choice experiments\index{delayed-choice experiments}) which bear enormous resemblance to the classical double slit experiment\index{double slit experiment} and which will be discussed in section \ref{sec:5.9}, the choice of performing either a measurement that determines through which path a photon went on its way to the detector, or a measurement that reveals the existence of quantum interferences\index{quantum interferences!two possible paths} attributable to the presence of two possible paths, can be delayed to long after the particle has actually traveled most of the distance to the detector and it does, in effect, influence what the particle did back when it was just leaving the source.

The reality of such post selection effects has, therefore, been experimentally confirmed and contrarily to what is sometimes suggested, it is not possible to assume that no post selection occurs as long as we reject the requirement of a realistic interpretation of quantum phenomena\index{quantum phenomena!realistic interpretation} (because it is not possible to ignore this requirement, as I will explain later). Thus, somehow, the path taken by a photon can be influenced by a measurement that takes place long after the actual process is over\footnote{
It should be clear that I'm not suggesting that post selection\index{post selection|nn} would allow information\index{information!flow from future|nn} to flow from the future, or that it would allow one to change an observable fact\index{observable fact!alteration|nn} from the past which has already been established. For reasons I have already mentioned, backward causation\index{backward causation|nn}, as would occur in the context of a consistent, time-symmetric interpretation of quantum theory\index{quantum theory!time-symmetric interpretation|nn}, is incompatible with both of those conclusions.}.
 Only a time-symmetric approach to quantum theory\index{quantum theory!time-symmetric approach} that recognizes the existence of a backward-evolving state\index{backward-evolving state!quantum theory} allows to explain those facts while remaining within the confines of the principle of local causality\index{principle of local causality}.

Now, even though some of the originators of the two-state-vector formulation\index{two-state-vector formulation!quantum theory} of quantum theory are hesitant to assume the reality of the backward-evolving state\index{backward-evolving state!two-state-vector formalism} that enters the formalism, it is clearly possible to assume that we are indeed dealing with a distinct state that evolves somewhat independently from the forward-propagating state\index{forward-propagating state!two-state-vector formalism}, but which is subjected to the same macroscopic experimental conditions\index{macroscopic experimental conditions}. What I'm proposing is that in order to go beyond early time-symmetric models\index{early time-symmetric models} one must, in effect, recognize that a whole causally independent history\index{causally independent history!opposite time direction} exists, whose particles are propagating in a direction of time opposite that in which the corresponding particles of the present history are propagating.

I believe that, in order to accommodate the requirement of time symmetry\index{time symmetry!requirement}, it is not enough to simply assume that semi-classical waves\index{semi-classical waves!backward-in-time propagation} are propagating backward in time, in the same portion of history, because, as I have already explained, advanced waves\index{advanced waves} are forbidden to exist by the constraint of global entanglement\index{global entanglement constraint!time asymmetry} that gives rise to time asymmetry in our universe. The problem, here, usually is that, even though two instances of Schr\"{o}dinger equation\index{Schr\"{o}dinger equation!two instances} appear to exist, which would allow to describe the propagation of the wave function\index{wave function propagation!future or past} in either the future or the past, only the equation that describes the evolution of the retarded portion of the wave function\index{wave function!retarded portion} is retained, given that retarded waves\index{retarded waves!evolution without constraint} are the only ones which are allowed to evolve without constraint, and this is why it is usually considered appropriate to take into account only the state vector\index{state vector!forward-in-time evolution} that evolves forward in time in order to obtain the probability of a whole process, even if this process may actually be most accurately described by a pair of state vectors\index{pair of state vectors!opposite chronological orders} evolving in opposite chronological orders.

But once it is understood that this time asymmetry\index{time asymmetry!boundary conditions} of the boundary conditions which are constraining the evolution of quantum systems is not a fundamental property of the wave function itself, but arises as a consequence of the requirement of diminishing entropy imposed on all past evolution by the global entanglement constraint that applies to the initial Big Bang state\index{initial Big Bang state}, then the two-state-vector formalism\index{two-state-vector formalism} becomes not only acceptable (as it does not require the existence of advanced waves), but actually essential to accommodate time symmetry\index{time symmetry!quantum-mechanical context} in a quantum-mechanical context.

An explanation of thermodynamic time asymmetry\index{thermodynamic time asymmetry!explanation} of the kind I have proposed in section \ref{sec:4.9} does not only render plausible the hypothesis that every quantum process\index{quantum process!backward-evolving counterpart} is complemented by a backward-evolving counterpart, but actually seems to require the existence of two histories\index{two histories!opposite chronological orders} evolving in opposite chronological orders, because, otherwise, it would be difficult to explain what enforces the then unique history\index{history!diminishing gravitational entropy in future}, to evolve with diminishing gravitational entropy in the future direction of time, before the hypothetical quantum bounce\index{quantum bounce}, for the portion of history that unfolds past the initial singularity\index{initial singularity!preceding history}, as required if gravitational entropy is to be minimum in the initial Big Bang state\index{initial Big Bang state!minimum gravitational entropy}, on our side in time of the singularity as well, given that under such conditions no history would be unfolding in the direction of time relative to which classical unidirectional causality\index{unidirectional causality!before past singularity} operates on the other side in time of the past singularity.

When it is recognized that there necessarily exists at least one history\index{history!unfolding from future to past} that unfolds from the future toward the past, then it becomes possible to explain the thermodynamic arrow of time\index{thermodynamic arrow of time!condition of low gravitational entropy} as being the consequence of the initial condition of low gravitational entropy imposed on the initial Big Bang state\index{initial Big Bang state} by the global entanglement constraint\index{global entanglement constraint}, because the evolution of at least one state vector\index{state vector!evolution determined by past conditions} is then determined by the conditions that existed in the singularity. In fact, this is a requirement that would apply to all processes in the context where historical consistency\index{historical consistency!quantum-mechanical viewpoint} must be observed, because, from a quantum-mechanical viewpoint, the consistency of past events with future events can only be obtained when those future events are also allowed to influence past events, as I will explain in section \ref{sec:5.12}.

Once this is understood, it is easy to see how a relativistically invariant model based on the sum-over-histories approach\index{sum-over-histories!relativistically invariant model} of quantum theory can be formulated that embodies the explicit time symmetry of the two-state-vector formalism\index{two-state-vector formalism!explicit time symmetry} by assuming that every quantum process involves both a conventional history (evolving without apparent constraint in the future direction of time) and a possibly distinct, time-reversed history\index{time-reversed history!causal independence} evolving independently (from the viewpoint of local causality) toward a state of lower entropy in the past direction of time.

This is an issue I will discuss more specifically in section \ref{sec:5.8}, but before I can do that, I must first explain why it is that a model involving two unique, but partly unobservable histories\index{histories!two unique histories}, unfolding in opposite directions of time (instead of two wave functions propagating in opposite directions of time), is not merely possible, but actually constitutes an essential requirement of a fully consistent, realistic interpretation of quantum theory\index{quantum theory!realistic interpretation}, despite the fact that what is usually assumed to be required in order to obtain the appropriate statistics is that all possible paths are followed all at once, in one single portion of history, for any given process.

\section{The constraint of scientific realism\label{sec:5.7}}

It has often been argued that the counter-intuitive aspect of quantum theory is not a real problem and merely indicates that there is a limit to what we are able to intuitively understand. It would then be incorrect to assume that the fact that there appears to be something incomprehensible with the current interpretation of the theory is due to the inadequacy of this interpretation. I believe, however, that this argument is invalid. In order to see what is wrong with this long-standing viewpoint, let's first suppose that we are, at the present state of our development, not sufficiently intelligent to understand quantum theory. The argument would then be that only some super-intelligence\index{super-intelligence} from the future (perhaps one that would run on a quantum\index{quantum!computer} computer) would eventually be able to overcome those limitations and to properly understand the significance of the empirically derived mathematical framework of quantum theory\index{quantum theory!mathematical framework}. Such a super-intelligence would, therefore, succeed at gaining a proper understanding of physical reality in a way that is simply impossible for us to achieve, due to the inherent limitations of our primitive intellect. But what does that mean in concrete terms?

When you carefully think about this question, it becomes obvious that the only thing that could happen is that this super-intelligence would then have developed a better interpretation of quantum theory\index{quantum theory!better interpretation}, because if the current mathematical framework is, in effect, appropriate to describe physical reality, then the only progress that \textit{could} be achieved would have to arise at the level of interpretation. You do not have to be super-intelligent to understand that and yet this is precisely what we fail to take into account when we suggest that the problem we experience while trying to make sense of quantum theory merely reflects the fact that it is not possible for us to understand the theory.

I believe that the lack of intelligibility of the standard interpretation of quantum theory\index{quantum theory!standard interpretation} is not a `fantastic new property of the physical world' which we happen to have discovered. It is a failure that originates in the inappropriateness of this interpretation and if this difficulty may be a consequence of the inadequacy of certain concepts we inherited from our experience of the world, it is also a problem that can be solved by making use of what we have learned from our experience of reality, as long as we do recognize that there is indeed a problem and that it deserves our attention. But those who still doubt the importance of a proper interpretation of quantum theory should take notice of the fact that, without interpretation, it would not even be clear that the theory is about probabilities of measurement outcomes, as this is an aspect that only came to be understood after the mathematics of the theory (regarding the Schr\"{o}dinger formulation\index{Schr\"{o}dinger formulation!quantum theory} in particular) had already been developed.

Now, it must be clear that quantum theory\index{quantum theory!counter-intuitiveness} \textit{is}, in effect, counter-intuitive and that it cannot be reduced to a classical view of the world by using the freedom we may have to interpret experimental facts and the current mathematical framework\index{mathematical framework!quantum theory} of the theory. Physical reality cannot be such as it was conceived at the epoch of Isaac Newton\index{Newton, Isaac}. Classical waves\index{classical waves} (as a means to explain quantum interference\index{quantum interferences}) and classical particles\index{classical particles} (not subjected to the constraints imposed by the uncertainty principle\index{uncertainty principle}) are gone and they will never form part of a consistent theory about the fundamental structure of reality ever again. But that does not mean that everything else is possible.

What is not allowed of a rational understanding of physical reality\index{physical reality!rational understanding} is inconsistency. The problem is that all known interpretations of quantum theory\index{quantum theory!known interpretations} do contain inconsistencies. Thus, either they contradict themselves, or else they do not agree with certain facts concerning that portion of reality which can be directly observed. This is usually understood by well-informed authors who recognize that the best that we can do in the present context is to pick as our necessarily inaccurate standpoint the interpretation which may be the least problematic for the kind of problem we are working on.

But while I have come to realize that some new conceptual elements, which had never been considered before, are necessary to formulate a fully consistent, yet straightforward interpretation of quantum theory\index{quantum theory!fully consistent interpretation} (which actually constitutes a more accurate physical theory), it also appears necessary to reject many of the outlandish concepts that came to be associated with a quantum-mechanical description of reality. Thus, I believe that the concept of history, or the concept of reality itself, must be simplified to once again be allowed to agree with the most basic empirical evidence, concerning in particular the uniqueness of facts\index{uniqueness of facts} and the particle nature of physical reality\index{particle nature of physical reality} (as a concept more elementary, but also more refined than its classical counterpart).

The problem here is that it is often believed that the notion of an elementary particle\index{elementary particle!unique trajectory} propagating along a unique trajectory is incompatible with the `complexity' which characterizes the quantum state of a system. But, as best understood by Richard Feynman\index{Feynman, Richard}, given the right formulation of quantum theory, not only is it unnecessary to reject the existence of elementary particles, or even to deny the relevance of the concept of trajectory, but it becomes imperative to recognize that those concepts actually form the substance of reality on the scale at which the most precise experimental data can be obtained.

I think that it is important to emphasize, therefore, that, even though common sense is not always a good guide for judging the validity of a physical theory, as the development of quantum mechanics itself illustrates, it would not be wise to conclude from this that more intuitive models are inappropriate and are necessarily ruled out by the apparent awkwardness of experimental facts, in the sense that our direct experience of reality would need to be considered irrelevant as a guide for elaborating a consistent interpretation of quantum theory\index{quantum theory!consistent interpretation}. We must keep in mind that classical physics\index{classical physics!counter-intuitive concepts} itself once involved counter-intuitive concepts which turned out to be inappropriate because of their awkwardness (like action at a distance\index{action at a distance}), or which only became fully understandable in the context of a more intuitive formulation of quantum theory (like the principle of least action\index{least action principle}).

Thus, I believe that, in the end, quantum reality will not be more difficult to visualize than classical reality, but will rather be more comprehensible, because it will be more consistent from a logical viewpoint. In any case, I believe that I'm justified in adopting a less counter-intuitive approach, given that the persistent problems which we are dealing with here have to do precisely with the apparent impossibility to provide a consistent, but also understandable representation of reality. However, instead of entering into a sterile debate about which of the ontological or the epistemological viewpoint constitutes a better approach to interpret quantum theory\index{quantum theory!ontological and epistemological viewpoints}\footnote{
The debate concerning interpretation has always centered around the problem of deciding whether the wave function\index{wave function!real entity or knowledge representation|nn} that allows to derive the quantum statistics of a process is a real `entity' or whether it is merely a representation of the knowledge that is available about what a (real) system is doing, which I believe is pointless, because even if the wave function is not physical reality itself it does provide the most accurate description of the state of a quantum system at any particular time and this description is a real \textit{aspect} of the system. The approach I will follow may actually be considered to allow a reconciliation of those two apparently incompatible viewpoints.},
 I will concentrate on explaining what the elements of an empirically accurate approach actually are that allow to reach consistency with the least amount of arbitrary hypotheses (I believe one does not need any).

To begin this discussion, it would be appropriate to point out that the most radical of those deficient approaches which were once proposed in order to make sense of quantum theory\index{quantum theory!most radical approach} is certainly that which is called quantum\index{quantum!logic} logic. It was suggested, in effect, that the logic that applies to physical reality may not be the ordinary Boolean logic\index{Boolean logic} with which we interpret ordinary facts, but some alternative logic, emerging from the apparently contradictory nature of certain conclusions made on the basis of a strict adherence to the rules which govern quantum reality\index{quantum reality!contradictory nature}. But while it is now recognized that such an approach goes too far as a tentative to adapt our mode of thinking to the reality of the quantum world, the fact that, at a certain epoch, quantum logic was considered to constitute a viable candidate for a solution to the problem of interpretation is quite indicative, I believe, of the extent to which we have departed from serving the true objective of science\index{science!true objective}, which is to explain facts by adapting and generalizing our physical laws and concepts to fit new experimental facts, in order precisely to avoid having to change the rules of logic\index{rules of logic} with which we analyze and understand reality.

The best example of such an adaptation is, of course, the shift to Riemannian spacetime\index{Riemannian spacetime} that was brought about by relativity theory\index{relativity theory} as a means to retain the validity of the concept of space in view of the equivalence of acceleration and gravitation\index{equivalence of acceleration and gravitation}. Indeed, if we were to reject Einstein's theory of gravitation\index{Einstein's theory of gravitation}, the only way we could retain the validity of the concept of physical space would be by altering the rules by which we formulate logical arguments\index{logical arguments!rules of formulation}, such as would be necessary to argue that despite all the evidence the Earth is flat. What the whole history of physics tells us is that it is always appropriate to use logical coherence\index{logical coherence} as a means to constrain our representations of reality and as a guide to assess the validity of our assumptions, while the rules of logic\index{rules of logic} themselves are rather like the rules of the game and can only be altered at the expense of invalidating most of everything else we have learned.

But the mere fact that quantum logicians were never able to dispense themselves from the need to use ordinary logic in order to reason about their own alternative system is quite indicative of the failure of their approach. I think that this is a particularly good example of the difficulties which the currently favored interpretations of quantum theory\index{quantum theory!currently favored interpretations} are facing as they stretch the notion of consistency, while trying to adapt to some perceived requirement of the mathematical framework\index{mathematical framework!quantum theory} of the theory, by going so far as actually allowing for contradictory accounts of factual aspects of reality\index{factual aspects of reality!contradictory accounts}. I will return to this question later in this section.

\bigskip

\noindent Not so long ago, it was suggested that certain difficulties that emerged as a result of the development of quantum field theory\index{quantum field theory} may indicate that the concept of an elementary particle\index{elementary particle!irrelevant concept} is no longer relevant to fundamental theoretical physics. One of those `difficulties' would have to do with the fact that, due to quantum uncertainty\index{quantum uncertainty!absence of localization in space}, particles can no longer be considered to be localized in space, as would seem to be necessary for the particle concept itself to make sense. Actually, in a relativistic context, it seems that the very fact that a particle is localized may depend on the state of motion of the observer which is assessing this fact, given that a particle's wavelength\index{wavelength of particle!velocity dependence} varies as a function of its relative velocity (actually its momentum).

Another aspect of the quantum-mechanical description of reality which would appear to constitute a serious challenge for the particle concept is quantum entanglement\index{quantum entanglement!challenge for particle concept} and the demonstration that what one particle does may, under certain conditions, depend on what another particle is doing at the exact same time in a remote location (relative to a given reference system), thereby apparently implying that only the ensemble, consisting of the two particles taken together, has physical significance. Finally, an additional difficulty arises from the fact that, due to the fluctuating nature of the quantum vacuum\index{quantum vacuum!reality of particle existence}, the very reality of a particle's existence may be called into question, because, even in empty space, particles would appear to be present. This problem is particularly severe in the context of a semi-classical approach\index{semi-classical approach}, in which the effects of acceleration and spacetime curvature\index{spacetime curvature!effect on quantum vacuum} on the quantum vacuum are taken into account and the presence of real (observable) particles becomes an observer-dependent property.

While I will not immediately address the issue of quantum entanglement and non-locality\index{quantum non-locality}, the conclusion I have reached is that, despite the difficulties mentioned here, the elementary particle concept is still viable in quantum field theory\index{quantum field theory!particle concept}. In the remainder of this section I will provide arguments to the effect that a realistic description of physical processes, based on the concept of particle trajectory\index{particle trajectories} is still desirable, even in the context where quantum interference\index{quantum interferences!multiple position states}, involving multiple distinct position states, must be assumed to constitute an essential aspect of reality. What emerges from this reflection is that it might be incorrect to suggest that particles cannot be localized in any way, because it may well be that particles in a pure momentum state\index{pure momentum state!unique unobservable particle trajectory} do follow unique, but unobservable trajectories in a certain sense which is merely incompatible with the \textit{classical} concept of trajectory\index{trajectory!classical concept}.

In such a context, the fact that the `wave packet'\index{wave packet!observer-dependent degree of localization} which is sometimes associated with the position state of a particle can be more or less localized in space, depending on the state of motion of the observer who measures this position, would not mean that a particle can actually be more or less `real', because such a variation would merely be a reflection of the dependence of observable macroscopic conditions\index{observable macroscopic conditions!reference system dependence} (here those which constrain the non-classical trajectory\index{non-classical trajectory} of the particle) on the choice of a particular reference system. But a detailed description of the realistic picture of quantum processes\index{quantum processes!realistic picture} that allows to articulate those considerations will only be provided in section \ref{sec:5.8}. In any case, I believe that the only real problem here is the general confusion that surrounds the question of deciding what it is exactly that remains acceptable about the particle concept\index{particle concept!acceptable elements} in a quantum field theoretical context, because all attempts at completely disposing of this essential concept have failed to provide a sensible, alternative conception of the nature of physical reality at the most-elementary level of description.

What I would like to immediately emphasize, though, is that, in light of the developments already introduced in chapter \ref{chap:2}, it is possible to conclude that vacuum fluctuations\index{vacuum fluctuations}, far from constituting a problem for the elementary particle\index{elementary particle!concept} concept, actually allow to provide a more consistent definition of what a matter particle really is. Indeed, what I previously explained is that positive-energy particles\index{positive-energy particles!missing vacuum energy and charge} must be considered to exist as a result of an absence of both negative energy and positive or negative charge in the fluctuating, electrically neutral vacuum\index{vacuum!electrical neutrality}, that is to say, from an absence of virtual particles\index{virtual particles} in the portion of zero-point vacuum fluctuations\index{zero-point vacuum fluctuations} that contributes a maximum negative value to the density of vacuum energy\index{vacuum energy density!maximum negative contribution} (while negative-energy particles\index{negative-energy particles!missing vacuum energy and charge} exist as a result of a similar absence of both positive energy and positive or negative charge in the vacuum). It therefore appears that the distinction between real particles\index{real particles!absence of virtual particles} and the virtual particles which are present in the vacuum is not as significant as one might imagine, given that the presence of real particles is actually equivalent to an absence of virtual particles in the quantum mechanical vacuum.

But it was also made very clear, in section \ref{sec:4.3} and then in section \ref{sec:4.7}, that, despite the fluctuating nature of the vacuum, there is a clear distinction between matter or radiation energy and vacuum energy, which is reflected in the electrical (or non-gravitational) neutrality of vacuum dark matter\index{vacuum dark matter!electrical neutrality} and in the absence of contribution to gravitational entropy\index{gravitational entropy} by the uniform distribution of vacuum energy\index{vacuum energy!uniform distribution} associated with the cosmological constant\index{cosmological constant}. On the basis of those developments, it becomes relatively straightforward to provide a clear and unambiguous definition of when it is that matter is present in a vacuum, that would also apply from the viewpoint of observers accelerating relative to a local inertial reference system\index{local inertial reference system!accelerating observer}, or in the presence of very strong gravitational fields (such as those which are present in the vicinity of a black hole\index{black hole!gravitational field}), and therefore the difficulties identified above would now appear to be rather insignificant.

But, in my opinion, one of the most powerful argument that can be used to support the idea that the elementary particle concept\index{elementary particle!concept} still constitutes a necessary and viable element of a consistent interpretation of quantum theory\index{quantum theory!consistent interpretation} (when it is allowed to obey the limitations imposed by the uncertainty principle\index{uncertainty principle}) is the observation that, even in the context where the hypothesis of the existence of elementary particles may seem to be the least appropriate, it nevertheless turns out that this assumption allows to explain, in a surprisingly simple way, certain key aspects of the processes involved. What I'm talking about is the use of virtual particles\index{virtual particles!elementary particle interactions} as the mediators of elementary particle interactions. The fact that it would be very difficult to explain certain properties of those interactions\index{interaction!range and strength}, like their range and their strength, without assuming that the interactions themselves are actually mediated by particles, even if those particles remain unobserved and cannot have classically well-defined energy and momentum states, is indicative of the usefulness and indeed of the necessity of assuming that, from a physical perspective, quantum fields\index{quantum fields!particles propagating between interaction events} actually are a manifestation of the existence of particles that propagate between interaction events\footnote{
Feynman\index{Feynman, Richard|nn} himself insisted that the concept of an external field\index{external field!concept|nn} becomes relevant merely in the context where the motion of a particle depends on a probability amplitude\index{probability amplitudes!particle interaction|nn} to interact with the particles mediating this field that varies only with the particle's position at a certain time, as may arise when a large number of such interactions are taking place over a relatively short period of time.}.

The problem we may have, in relation to this conclusion, is that, if particles do exist as real physical entities, it would seem that it is not possible to attribute a unique position state\index{position state!uniqueness} to those particles at all times, in the context where it is known that there are interferences between the probability amplitudes\index{probability amplitudes!multiple trajectories} associated with the many different trajectories that contribute to determine the transition probability for one single event involving the propagation of a particle in a given momentum eigenstate\index{momentum eigenstate}. This is why so many people prefer to assume that the wave function\index{wave function!immaterial nature}, despite its immaterial nature, may constitute reality itself; a hypothesis which raises difficulties of its own in the context where it must be recognized that this reality would be submitted, by the act of measurement, to discontinuous changes that may violate the spirit of relativity theory and the principle of local causality\index{principle of local causality!violation}, even if no information is communicated at faster-than-light velocities.

In any case, it must be clear that the wavelike nature of quantum processes\index{quantum processes!wavelike nature} is simply a consequence of the fact that the probability amplitudes\index{probability amplitudes!periodic evolution} that must be used in the calculation of transition probabilities\index{transition probabilities} are subject to periodic evolution and there is no sense in saying that a \textit{particle} may sometimes evolve as a wave, because the wavelike property is already well understood as being a property of processes which always involve particles and the problem really has to do with the apparent impossibility to attribute a definite location to those elementary particles\index{elementary particles!impossibility of definite location}, under general circumstances.

What I will explain, however, is that we have not yet exhausted all possibilities and that a realistic interpretation of quantum phenomena\index{quantum phenomena!realistic interpretation} that would involve elementary particles can still be formulated that would be compatible with the current mathematical framework of quantum field theory\index{quantum field theory!current mathematical framework} (if we allow for a slightly more elaborate concept of particle trajectory\index{particle trajectories!more elaborate concept}, while still rejecting the contradictory notion of wave-particle duality\index{wave-particle duality}). I believe that it is, indeed, possible to assume that a unique history\index{history!uniqueness} \textit{of some kind} is taking place, even for what regards the physical attribute of a quantum system (say its position in space) that is not under observation. This is a conclusion that would obviously contradict the orthodox interpretation of quantum theory\index{quantum theory!orthodox interpretation}, at least in its original form, given that, according to the conventional doctrine, there is no sense in speaking about the state of some physical attribute when no measurement has been effected to actually determine what this state is at a given time.

But if we recognize that the elementary particle\index{elementary particle!essential concept} concept is essential to a consistent interpretation of quantum theory\index{quantum theory!consistent interpretation}, then it seems that we have no choice but to conclude that the current interpretation\index{current interpretation!quantum theory} of the theory is incomplete, because it does not provide a clear and unambiguous description of what happens when the position of such an object is not under direct observation. Of course, certain modern interpretations, such as the consistent-histories interpretation of quantum theory\index{quantum theory!consistent-histories interpretation}, go some way into providing a more realistic picture of quantum phenomena\index{quantum phenomena!more realistic picture}, but they also appear to be incomplete, given, precisely, that they allow reality to be described only under particular circumstances, when a certain more or less arbitrary criterion of `consistency'\index{consistency criterion!arbitrariness} is met and given, also, that, despite their more appropriate handling of the quantum measurement problem\index{quantum measurement problem}, they still fail to explain the emergence and the persistence of a quasiclassical world\index{quasiclassical world!emergence and persistence}, as I will explain in section \ref{sec:5.10}.

In the introduction to this report I mentioned that I believe that it is essential to adopt a realistic interpretation of quantum phenomena\index{quantum phenomena!realistic interpretation} if we are to avoid deviating into a solipsistic and idealistic view\index{solipsistic and idealistic view of reality} of reality, according to which nothing consistent would really exist aside from our own mind (if that could ever be found possible). This criterion is particularly important in the context where the only thing that may be considered undeniable about reality is precisely that it is real. The problem is that the adjective `real' is usually assumed to be the characteristic of something that exists as a fact rather than as a mere possibility and therefore the characterization of \textit{quantum} reality as actually being real would appear to exclude the possibility that this reality may not always consist of \textit{observable} facts.

Thus, it is important to emphasize that what I have in mind here is the \textit{scientific} concept of realism\index{scientific realism!concept}, according to which it would be deemed appropriate to seek to describe the actual ways by which certain physical processes can occur, even when it is not possible to determine the specific paths which are followed by elementary particles in the course of any one particular process. In the context of the preceding discussion, it would, therefore, appear desirable to apply the criterion of physical reality\index{criterion of physical reality!elementary particle trajectories} not to the wave function specifically, as is usually proposed, but rather to the elementary particle trajectories that enter the sum-over-histories\index{sum-over-histories!formulation of quantum theory} formulation of quantum theory. The hypothesis would then be that it is appropriate to assume that, even in between position measurements\index{position measurement}, elementary particles follow real and to a certain extent unique (but not classically well-defined) trajectories in spacetime\index{spacetime trajectories!uniqueness}, despite the fact that those trajectories must, as a matter of principle, remain mere potentialities.

I understand, of course, that, despite being intuitively appealing, the hypothesis that some unobserved dynamic attribute of a quantum system could exist in a definite, unique, but unknown state, even as the various alternative paths\index{alternative paths!quantum interference} available to it interfere quantum mechanically with one another, appears to be ruled out by the fact that all possible histories must, in effect, be put to contribution in order to derive the right probability for a process to occur (that which is obtained by repeating the experiment a large number of times). Faced with this difficulty, one usually concludes that it is not possible to retain a realistic description of quantum phenomena\index{quantum phenomena!realistic description} that would involve elementary particle trajectories and that it is more reasonable to simply assume that reality cannot be unique in any way between measurements and that the question of what happens to unobserved attributes is meaningless from a scientific viewpoint, as originally proposed by Bohr\index{Bohr, Neils} and Heisenberg\index{Heisenberg, Werner}.

But, if one recognizes that the uniqueness of history\index{history!uniqueness} is a condition that cannot be ignored, one may alternatively propose that quantum interferences\index{quantum interferences!multiple trajectories} are not indicative of the fact that multiple trajectories must be taken into consideration simultaneously, but rather arise as a consequence of the existence of non-local, but otherwise classically well-behaved hidden variables\index{hidden variables!non-local} (occupying a unique classical state at all times) that would allow to causally determine the course of a conventional history involving otherwise ordinary objects.

Without entering into the details of each proposal, it is clear that they are both unsatisfactory, precisely because they both involve assumptions that contradict one key aspect of physical reality (either the uniqueness of history\index{history!observational requirement of uniqueness}, as an observational requirement, or the absence of instantly propagated effects\index{instantly propagated effects}, as a theoretical requirement). It must be clear, though, that, despite what is commonly believed, the first proposal is just as problematic as its alternative counterpart, even if it was favored by the originators of quantum mechanics on the basis of the fact that it involves fewer arbitrary assumptions.

It has always appeared preferable, in effect, to avoid postulating the existence of classical hidden variables\index{classical hidden variables!complex unobservable mechanisms}, given that any model based on such a requirement would necessarily involve complex mechanisms of an unobservable nature, whose validity could never be empirically confirmed. Yet, the argument that it is the non-locality of the hidden-variable models\index{hidden-variable models!non-locality} that makes them unacceptable is not very satisfactory. Indeed, if one recognizes that there must necessarily be a reality of some kind, then the only \textit{known} alternative to assuming the existence of hidden variables would be to consider the wave function itself as being the whole reality, which means that non-locality would also constitute an aspect of the orthodox interpretation\index{orthodox interpretation!non-locality}, because the wave function\index{wave function!non-local entity} is also a non-local entity, which is subject to non-local changes, as would occur in the course of certain measurements, because it does not concern one system in particular, but the ensemble of all processes (as I will further explain in section \ref{sec:5.12}).

Thus, it would appear that the only alternative to a non-local theory\index{non-local theory!alternative}, potentially involving complicated arbitrary constructs whose validity would remain unconfirmed, actually amounts to assume that reality is not real (when it is not subject to direct measurement). This is obviously a \textit{simple} assumption, but I'm not willing to accept that it would be mere scientific progress to consider it as a valid assumption about physical reality. One must come to recognize that such a position is not progress, but simple non-sense of the most scientifically objectionable kind. If a \textit{physical} reality exists, then I believe that what is certainly the most basic property that would need to characterize this reality is that it is, in effect, always real. This must be considered an essential consistency requirement and neglecting it would again amount to allow a logical contradiction to stand at the basis of our interpretation of the most fundamental of all physical theories.

Therefore, I suggest that one of the crucial points that cannot be neglected in trying to produce a consistent interpretation of quantum theory\index{quantum theory!consistent interpretation} is that the unique outcome of measurements\index{measurement!unique outcome} is indicative of the uniqueness of the history\index{history!uniqueness} that takes place in between measurements, even for what regards those dynamic attributes\index{dynamic attributes!unobserved} that are not subjected to direct observation. The existence of definite causal relationships\index{causal relationships!components of universe} between all components of the universe must be understood to actually require that every component of this physical reality be involved in only one such history at a time, in any one particular universe. The right interpretation of quantum theory\index{quantum theory!interpretation} must, therefore, emerge from a combination of two apparently incompatible requirements, which are provided, on the one hand, by this condition of uniqueness of history and on the other, by the necessity to allow quantum interferences\index{quantum interferences!distinct unobserved possibilities} to occur between the many distinct possibilities that may exist for the unobservable aspects of this unique history, even as may affect one single quantum process that need not be repeated many times.

It is the description of reality we are considering that must adapt to those two requirements if we are to avoid having to alter the rules of logic\index{rules of logic!alteration} to accommodate their simultaneous fulfillment. But I do agree with Copenhagenists that this must not be achieved by postulating the existence of hidden variables in a unique classical state\index{hidden variables!unique classical state}, whose effects would need to propagate at faster-than-light velocities\index{faster-than-light velocities!propagation of effects}, because, from all that we know, the principle of local causality\index{principle of local causality} provides as real a constraint on our description of physical reality as the existence of quantum interferences.

In fact, I have come to understand that the debate between Copenhagenists and classical hidden-variable\index{classical hidden variables!unobservable causes} theorists is not as meaningful as one might assume, because the only hidden-variable models that may allow to retain agreement with observational data are those that postulate that the hidden causes of the unique, classical evolution that takes place in between quantum measurements would remain unobservable as a matter of principle, even when they evolve deterministically (ignorance of the exact state does not arise from a practical limitation that could eventually be overcome, as in conventional statistical mechanics\index{statistical mechanics!conventional}). Thus, even though such classical hidden-variable models would contradict the postulate of objective indefiniteness\index{objective indefiniteness!postulate} of the orthodox interpretation of quantum theory\index{quantum theory!orthodox interpretation}, the fact that the hidden variables could never become part of experimental knowledge means that those models do not require a rejection of the concept of objective chance\index{objective chance} (associated with fundamental unpredictability\index{fundamental unpredictability}).

It would, therefore, appear that it is really just the naive, classical definiteness of the phenomenon which is assumed to determine the evolution of quantum particles that is problematic with those hidden-variable\index{hidden variables!naive classical definiteness} models, given that it necessarily requires the propagation of signals of a conspiratorial nature at faster-than-light velocities\index{faster-than-light velocities!propagation of signals} to achieve agreement with observational data. The real problem for current (classical) hidden-variable theories would then be that, instead of enhancing the domain of validity of the quantum-mechanical description of reality to the classical world, as an improved realistic interpretation of quantum theory\index{quantum theory!improved realistic interpretation} should enable to achieve, they just allow to perhaps reproduce the experimentally confirmed predictions of the theory through some unnatural and complicated contortion of classical reality that makes them even less appealing than the currently favored conventional approach.

Before I elaborate on what kind of physical reality might agree with the two basic requirements identified above (uniqueness of history and local causality), it is important to mention that the requirement that there exists a unique reality is different from Einstein's\index{Einstein, Albert} proposal that reality\index{reality!independent from observation} should be independent from whether or not a certain parameter is being observed, which assumes more than just a unique reality and which is irreconcilable with the mathematical framework of quantum theory\index{quantum theory!mathematical framework}.

We must recognize as an established fact that quantum reality\index{quantum reality!dependence on experimental conditions} is not independent from experimental conditions, even if it might be possible to assume that conjugate physical attributes\index{conjugate physical attributes!simultaneously unique states} like position and momentum can simultaneously be in unique (even though partly unobservable) states \textit{in a certain sense}, because, as I already explained, this unique reality must also give rise to quantum interferences between multiple states and it is only the physical attribute that is under direct observation at a given time, or in the course of a certain process, that is free of interferences. Assuming that reality is independent from experimental conditions would require that quantum interferences be absent altogether, which is certainly not compatible with any plausible interpretation of quantum theory in its present form.

If the state of conjugate observables\index{conjugate observables!simultaneous state determination} cannot be determined at the same time with an arbitrarily high degree of precision, it is precisely because the macroscopic constraints necessary to determine the exact state of those physical attributes cannot be realized together at the same time for the same process, while it is those macroscopic constraints (associated with the existence of records) that determine which physical observable\index{physical observable!quantum interferences} is not subject to quantum interferences (for reasons I will discuss in section \ref{sec:5.12}). Thus, even though I believe that it is necessary to assume that a unique reality\index{reality!dependence on macroscopic conditions} actually exists, regardless of whether it is being observed or not, I also think that it must be recognized that this reality does not evolve independently from the macroscopic physical conditions which determine what can be known, experimentally, of its actual state.

But it should be clear that the hypothesis that there exists a unique reality\index{reality!uniqueness} of some sort does not impose on quantum particles (say negatively-charged, positive-energy electrons propagating forward in time) that they be distinct individually, even when they possess the same \textit{static} attributes. What we must ask ourselves, therefore, is what the unique unobservable reality actually is if it does not conform to a classical representation in terms of simple, identifiable objects. Quantum theory\index{quantum theory!constraint on realistic description of reality}, from the viewpoint of its current interpretation, is not so much an answer to the problem of the fundamental nature of reality, as a constraint that must be obeyed by any realistic description of physical phenomena.

At this point, it is necessary to mention that I do know that from the viewpoint of someone who has been introduced to quantum mechanics\index{quantum mechanics!conventional introduction} in the conventional way, the requirements discussed above may appear irreconcilable, as it seems that the formalism of the theory itself cannot be dissociated from the Copenhagen interpretation\index{Copenhagen interpretation!quantum theory}, while the conventional definition of a quantum state\index{quantum state!conventional definition} would appear to be totally incompatible with a realistic interpretation\index{realistic interpretation!unique quantum mechanical history} that would involve a unique history. It is only when one begins studying relativistic quantum field theory\index{quantum field theory!sum-over-histories formalism}, that one is introduced to Feynman's\index{Feynman, Richard} approach and the sum-over-histories formalism (despite the fact that conventional quantum mechanics can also be formulated using path integrals\index{path integrals!conventional quantum mechanics}), at which point one has already been conditioned to believe that it is not possible to visualize quantum processes\index{quantum processes!unique histories} as involving unique histories of some sort, while, in fact, this is precisely what the sum-over-histories approach suggests and from a certain viewpoint even requires\footnote{
It is important to note that even a conventional formulation of quantum mechanics\index{quantum mechanics!conventional formulation|nn}, like Heisenberg's matrix mechanics\index{Heisenberg's matrix mechanics!unobserved virtual processes|nn}, can be interpreted as involving a summation over a series of intermediate, unobserved, `virtual' processes and it is significant that some of the originators of quantum theory were, in effect, open to such an interpretation (perhaps because they were not told by others how they should interpret their own theory), even though they did not see how it could be made truly viable.}.

In this particular sense, I was lucky, because I first learned of quantum theory\index{quantum theory!problem of interpretation} by reading about the problem of interpretation and Feynman's original approach, while I became familiar with the conventional formalism of quantum mechanics\index{quantum mechanics!conventional formalism} only later on, which means that rather than being critical of the reality of Feynman histories\index{Feynman histories}, I was rather critical of the conventional interpretation. I believe that this uncommon course is what allowed me to see more clearly how it can be that each independent elementary particle process\index{elementary particle processes!unique unobservable history} consists of a unique (even though partly unobservable) history, despite the fact that there always arise interference effects between the multiple histories\index{multiple histories!interference effects} which are allowed by the macroscopic, experimental conditions\index{macroscopic experimental conditions!quantum process} which are imposed on the process. What I would like to explain, therefore, is why it is necessary to assume that the multiple \textit{unique} histories depicted in Feynman diagrams\index{Feynman diagrams!multiple unique histories} correspond more than is usually recognized to the actual reality behind all quantum phenomena.

I believe that it is merely the fact that no truly acceptable realistic interpretation of quantum theory\index{quantum theory!realistic interpretation} has ever been proposed that motivates the widespread belief that the multiple histories depicted in Feynman diagrams do not correspond to anything actually occurring (must be considered purely fictitious) and merely constitute useful computational apparatus, despite the obvious similarity between the processes so described and the actual reality we experience. It has become very clear to me that what this formalism provides is nothing but a description of what is actually going on, which we are not able to directly observe, when some dynamic physical attribute\index{dynamic physical attributes!unobserved evolution} of a quantum system is evolving in between measurements.

Even ignoring the arguments provided so far in this section, concerning the relevance of the concept of elementary particle\index{elementary particle concept!quantum field theory} in quantum field theory, I think that one must recognize that it is very unlikely that the individual paths entering a sum-over-histories formulation of quantum theory\index{quantum theory!sum-over-histories formulation} could happen to be intuitively significant simply as a coincidence, without being related to what actually goes on in between measurements of the observable concerned. Perhaps that instead of insisting that our experience of reality is not a reliable guide for judging the value of certain hypotheses concerning unobservable aspects of this very same reality, we should instead try to figure out how the phenomena that cannot be directly observed can be described in a way that would better agree with what we do know about physical reality.

It is remarkable in this regard that, while Feynman\index{Feynman, Richard} himself believed that quantum reality involves particles and only particles, he also said that there is no way to explain or to understand what happens to those particles, even during the most simple of quantum processes, because it is not possible to assume that a particle in a given momentum eigenstate\index{momentum eigenstate!particle trajectory} goes along one trajectory or another in space, so that it may be preferable to give up trying to create a model of what is actually happening. I believe that this shows how deeply the philosophy behind the Copenhagen interpretation\index{Copenhagen interpretation!quantum theory} of quantum theory has become ingrained in our conception of reality, because if one person might have been allowed to understand what is the underlying reality behind all quantum phenomena, it should certainly have been Feynman and it is clear that his failure is in part attributable to the fact that, despite his remarkable insights, as all physicists of his generation he adhered to the notion that a realistic representation of quantum phenomena is not possible. But if those difficulties have been allowed to persist to this day, it is merely because we still do not understand the profound meaning of quantum strangeness\index{quantum strangeness} and remain ignorant of the fact that quantum phenomena \textit{can} be visualized.

What remains to explain, therefore, is how it can be that one and only one of the histories which can be depicted using Feynman diagrams\index{Feynman diagrams} corresponds to what really happens in the course of a specific quantum process\footnote{
I must mention that I'm aware that a method called `unitarity'\index{unitarity!modified Feynman approach|nn} is often used as a shortcut for the determination of quantum probabilities that constitutes a modification of Feynman's original approach, but this alternative technique does not require assuming that the original sum-over-histories\index{sum-over-histories!formulation of quantum theory|nn} formulation of quantum theory is not fundamentally the most accurate and it remains that the summation over all possible histories is more representative of what really goes on at a fundamental level, even if, from a practical viewpoint, it may be even less appropriate than the alternative approach for performing certain calculations under particular circumstances.},
 despite the fact that it is not possible to attribute to a quantum particle the physical properties of a classical object, in the sense that one cannot simultaneously determine both its momentum and its position with an arbitrarily high degree of precision.

For that purpose, it is necessary to point out that there is something highly problematic with the conventional viewpoint provided by Bohr's complementarity principle\index{complementarity principle}. What Bohr\index{Bohr, Neils} suggested, in effect, is not just that the conditions necessary for the measurement of a certain dynamic attribute\index{dynamic attributes!measurement conditions} is incompatible with those necessary for the measurement of its conjugate counterpart, but really that the concepts of momentum and position, for example, constitute mutually exclusive representations of a quantum system, so that it would not even be logically appropriate to conceive of a particle with a given momentum as being a localized entity. If one was to hold on to such a viewpoint, then, clearly, a realistic description of phenomena based on the sum-over-histories formulation of quantum theory\index{quantum theory!sum-over-histories formulation} would become impossible to achieve.

But, in fact, there is absolutely no reason to assume that the indefiniteness of the state of some unobserved attribute of a quantum system\index{quantum system!state indefiniteness} cannot be the consequence of a mere incompatibility between the macroscopic conditions necessary for the measurement of one dynamic attribute and those necessary for the measurement of its conjugate counterpart. When one understands the true nature of the constraints which allow decoherence\index{decoherence!elimination of quantum interferences} to take place and to rapidly eliminate quantum interferences for the physical attribute that is subjected to measurement (an issue I will address only in section \ref{sec:5.12}), it appears quite plausible that quantum\index{quantum!indefiniteness} indefiniteness actually arises as a consequence of this practical (but fundamental and unavoidable) limitation concerning the simultaneous realization of the macroscopic conditions necessary for a measurement of conjugate physical attributes\index{conjugate physical attributes!measurement}. Therefore, it is not \textit{a priori} impossible for a quantum particle which is known to be in a pure momentum state to follow a unique, but \textit{observationally} undetermined trajectory in space and only the existence of quantum interferences\index{quantum interferences!multiple distinct trajectories} involving multiple distinct trajectories would appear to contradict this conclusion.

There is certainly something true in Heisenberg's\index{Heisenberg, Werner} statement to the effect that ``\textit{the progress achieved (through the elaboration of quantum theory) was obtained at the price of having to abandon the possibility of visualizing natural phenomena in a way that is immediately and directly comprehensible to our conventional way of thinking}''. However, I would insist that what is inappropriate is not the requirement that it should be possible to visualize physical reality in a comprehensible way, but the requirement that this reality should, in effect, be similar in every respect to what it appeared to be before experiments began revealing that it is not possible to estimate the probability of certain processes without assuming the existence of interferences between the probability amplitudes\index{probability amplitudes!interferences} (a purely quantum mechanical concept) associated with the many possible, unobservable ways by which those processes can happen.

In order to progress toward this legitimate objective of visualizing quantum reality, we may once again consider the classical double slit experiment\index{double slit experiment}. What can be learned using this simple, but very general experimental arrangement is that, despite the fact that we are always dealing with discrete, localized particles\index{localized particles}, interferences, similar to those which can be observed when what is propagating is a classical wave\index{classical waves}, must be assumed to occur whenever a particle is allowed to propagate between a source and a detector through more that one possible path without giving rise to the formation of a permanent record\index{formation of records!particle trajectory} that would indicate through which trajectory the particle actually went.

Even though such interferences become apparent only in the statistical distribution of measurement results\index{statistical distribution of measurement results!interferences}, which is known to depend on the differences between the lengths of the possible paths along which a particle can propagate before its position is detected, the interferences must be considered to apply even for a single quantum process involving the propagation of one unique particle (because the observed interference patterns can be produced even when the particles are sent from source to detector only one at a time). The problem, therefore, consists in figuring out how it is possible for a localized particle in a given momentum eigenstate\index{momentum eigenstate} to give rise to those interferences involving many distinct possible paths if, as a particle, it must necessarily propagate in space by going through one definite, yet unobservable trajectory.

Stated in such a way, this aspect of the problem of the interpretation of quantum theory\index{quantum theory!problem of interpretation} appears at once very clear and quite unsolvable. But it took me a very considerable amount of time to simply realize that this is, in effect, how the problem must be stated, as this is not how most people see things. Indeed, it is \textit{not} usually assumed that the particle, as a particle, must necessarily go through a single trajectory or even through any trajectory at all, as this would immediately appear to give rise to an unavoidable contradiction, because, `obviously', a particle cannot go through one trajectory and produce interference effects\index{interference effects!multiple distinct particle trajectories} which involve multiple distinct trajectories. Anyone arguing that this is not necessarily the case would merely be a nostalgic of classical reality that does not accept the `undeniable' strangeness of reality unveiled by the observation of quantum mechanical phenomena. Such an approach to the problem of interpretation would necessarily have to deviate into classical hidden variables\index{classical hidden variables!non-local causality} and non-local causality.

But in fact, that is not the case. Not only is it possible to visualize what is going on when one acknowledges the validity of those premises, but this is the only way to arrive at an interpretation of all quantum phenomena that does not involve any arbitrary and undesirable assumptions that would either conflict with the observed uniqueness of experimental facts\index{uniqueness of experimental facts}, or else contradict one another (as when one speaks of a `probability wave'\index{probability wave} going through both slits all at once, which then `becomes' a particle when its position is detected), therefore implicitly or explicitly requiring an alteration of the conventional rules of logic\index{logic!alteration of conventional rules}.

It is important to understand that, while it is usually believed that logical contradictions may arise when one insists on requiring a realistic interpretation of quantum theory\index{quantum theory!realistic interpretation}, those contradictions are merely a consequence of holding on to a conventional, or naive conception of reality\index{reality!naive concept}, according to which it might be possible to obtain simultaneous experimental knowledge about the state of conjugate physical attributes of a quantum system.

It is usually believed, in effect, that one cannot assume that the conjugate dynamic attributes\index{conjugate dynamic attributes!unique states at all times} of a quantum system could be in unique states at all times without assuming that a precise knowledge of the state of those dynamic attributes would be available (which would violate the uncertainty principle\index{uncertainty principle!violation}). But, once we recognize that only the second assumption is inadequate and could give rise to factual contradictions\index{factual contradiction}, while one may be allowed to assume that a dynamic attribute of such a system still evolves along a unique path in any given portion of history (in a certain sense which will be clarified later) despite the fact that it is absolutely impossible to obtain knowledge concerning the state of this attribute, then it becomes possible for a realistic interpretation of quantum theory\index{quantum theory!realistic interpretation} to be formulated that is not logically inconsistent (even though, from a conventional viewpoint, such a proposal would appear to contain a contradiction).

What I will eventually explain is that the fact that a purely phenomenological model of quantum reality\index{quantum reality!phenomenological model} (such as that which constitutes the core of the orthodox interpretation of quantum theory\index{quantum theory!orthodox interpretation}) may appear to be better suited than a realistic model\index{realistic model!quantum reality} for explaining certain observations is merely a consequence of the fact that a realistic model cannot be applied to quantum phenomena\index{quantum phenomena!time-symmetric description} as they are conventionally described, but only becomes appropriate in the context of a time-symmetric description of those phenomena.

Following Einstein\index{Einstein, Albert}, I believe that one must be ready to take an intuitive leap and to derive, based on available experimental data, general postulates that may not always be immediately confirmed through direct observation, but which allow to better model the reality underlying those empirical facts. For what regards the problem of the interpretation of quantum theory\index{quantum theory!problem of interpretation}, this intuitive leap would actually consist in assuming that the particles involved in the description of elementary quantum processes are, in effect, real and that they are taking part in one unique history of some kind. Once this is recognized to be a legitimate and necessary requirement, the difficulty would then consist in understanding how such a realistic description of reality\index{realistic description of reality} could be made compatible with both the observational constraint imposed by the existence of quantum interferences\index{quantum interferences!observational constraint} and the theoretical constraint of a time-symmetric conception of causality\index{causality!time-symmetric conception}.

I think that one cannot be satisfied with assuming that what explains the existence of quantum interferences\index{quantum interferences} is the `fact' that a particle doesn't follow a unique path and actually propagates, from emission to detection, by simultaneously following, at once, \textit{all} possible trajectories. I believe that the notion that all the available paths are actually followed together, in the course of any single quantum process, constitutes one of those strange aspects of quantum reality\index{quantum reality!strange aspects} (as it is usually conceived) which are not merely unexpected, yet unavoidable, but which remain strange because they actually conflict with certain factual aspects of reality. What is quite amazing is that, even though such a notion is only slightly different from the usually rejected viewpoint according to which a particle may go partly through one slit and partly through the other (in a double slit experiment), it is often considered to provide an appropriate depiction of quantum reality.

But if one recognizes that such a representation is indeed incompatible with a realistic interpretation of quantum phenomena\index{quantum phenomena!realistic interpretation} that would not reject the empirical evidence for the uniqueness of historical facts\index{uniqueness of historical facts!empirical evidence}, it remains that one must take into account, in the determination of transition probabilities\index{transition probabilities}, any possible trajectory which is \textit{allowed} by the observable macroscopic conditions\index{observable macroscopic conditions} which are in effect while those transitions are taking place. In order to accommodate this fact, what is sometimes assumed (as I briefly mentioned in section \ref{sec:5.6}) is that any single quantum process\index{quantum process!two interfering histories} may actually always involve two interfering histories which, for some reason, would share the same observational conditions. But it remains to explain what justifies this assumption (which would still appear to conflict with the uniqueness of historical fact) and why it can be expected to give rise to the kind of classically well-defined reality we do experience.

It is certainly true that one of the criteria that may allow one to judge the validity of a conception of reality involving unobservable theoretical constructs\index{unobservable theoretical constructs!criterion of validity} is its usefulness for producing accurate predictions of experimental phenomena, but this is precisely why the currently favored interpretation must be rejected. I believe that if the notion that all histories occur all at once in the same universe is incompatible with the experimentally confirmed uniqueness of historical facts\index{uniqueness of historical facts!experimental confirmation} (in the context where the tentative solution to the quantum measurement problem\index{quantum measurement problem} that is provided by the many-worlds interpretation\index{many-worlds interpretation!quantum theory} of quantum theory is recognized to be ineffective, as I will argue in section \ref{sec:5.10}), then it must be rejected in favor of a conception of reality that does not require this uniqueness to be a mere illusion.

The problem, however, has always been that it would appear that the only realistic alternative to such an interpretation would require assuming that the wave function\index{wave function!as reality} itself constitute the reality, because, in the context where quantum interference\index{quantum interferences!unobserved attribute trajectories} is possible for the trajectories of unobserved attributes, this mathematical object (the state vector\index{state vector} more generally) does not merely provide a probability distribution\index{probability distribution} for the position of a particle which is in a momentum eigenstate\index{momentum eigenstate}, but may involve superpositions of position states\index{state superposition!complex-number weighting coefficients} with complex-number weighting coefficients, which means that position may sometimes appear to constitute an inappropriate concept for physical reality (while, of course, the same is true for momentum under distinct experimental conditions). This is not necessarily inadequate from a mathematical viewpoint, but it remains unsatisfactory from a physical viewpoint, especially in the context where this wave function\index{wave function!discontinuous changes} can be subjected to discontinuous changes that would violate the principle of local causality\index{principle of local causality!violation} whenever the quantum potentialities\index{quantum potentialities!actualization} involved are actualized, as I previously mentioned.

I believe that it is merely the fact that we fail to correctly visualize what is going on in between position measurements (for instance) that makes it look like physical reality cannot involve a unique history of some kind and needs to be replaced by some strange picture which deviates from the reality we experience to the point where reality itself looks unreal, in the sense that the currently favored concept of quantum reality\index{quantum reality!currently favored concept} is not only incompatible with observable aspects of reality, but also with the requirement of logical consistency\index{requirement of logical consistency} which is known to apply under more general circumstances (which would allow one to reject the possibility that a particle could be in a certain place and also in a different one, all at the same time, in one single portion of history).

What holds the key to a better understanding of quantum reality\index{quantum reality!better understanding} is the acknowledgment that what can be known about a quantum system does not allow one to tell everything about how it actually evolves, even though, as a matter of principle, there does not exist a more accurate description of the processes involved. This is the only potentially viable approach for a solution to this aspect of the problem of interpretation that does not involve contradictory statements about reality\index{reality!contradictory statements}, when one considers it inappropriate to assume that dynamic attributes\index{dynamic attributes!non-existence without observation} simply do not exist when they are not those about which direct experimental knowledge is available.

Although the approach I favor may, at first, seem problematic, it is actually much simpler to apply than its logical alternative, because the idea that a dynamic attribute does not exist when it is not subject to observation cannot be adapted to the case where such an attribute is only known to an intermediary level of precision, because, clearly, either an attribute exists or it doesn't, while it is undeniable that the state of any physical attribute can be determined with more or less precision by the appropriate measurement, as long as an inversely proportional uncertainty applies to its conjugate counterpart.

If, at least, it was possible to speak of certain quantum systems as definitely being observed, while other systems would not (under particular circumstances), then it might perhaps make sense to assume that what is measured is real and what is not measured doesn't exist, but in fact, there is always something that is known with arbitrarily high precision about a physical system, as long as the system remains causally related to the rest of the universe (this is what is implied by the linearity of Hilbert space\index{Hilbert space!linearity}) and it is merely the conjugate dynamic attribute\index{conjugate dynamic attributes!complete indetermination} of this observed attribute (however unnatural it is) which is completely undetermined, so that if one was to choose to follow the conventional approach, then, based on those considerations, one would be forced to somehow ascribe both reality and absence of reality to the same physical system (even though not to the same physical attribute), which again constitutes a logical contradiction.

Thus, despite what one is usually encouraged to believe, it seems necessary to assume (particularly if one wants to avoid having to consider the possibility of a reality\index{reality!created through observation} created through observation) that two systems prepared in the same quantum state may evolve differently at the level of the dynamic physical attribute\index{dynamic physical attributes!state indetermination} whose state is not determined by the macroscopic conditions of an experiment. I believe that this is what explains that a subsequent measurement of this originally undetermined attribute may produce outcomes that differ from one such measurement to another, even those which are performed under identical conditions, and if this is correct, then it would mean that it is inappropriate to assume that it is the act of measurement itself that introduces randomness into our description of quantum phenomena\index{quantum phenomena!randomness}. The more consistent approach I will propose, therefore, allows physical systems which are described by the same wave function to actually be different at a certain unfathomable level, even if the wave function still provides the most complete description of a quantum system and of how it will evolve.

From that perspective, it becomes apparent that something very problematic comes into play with the conventional interpretation of quantum theory\index{quantum theory!conventional interpretation} whenever post selection\index{post selection} is involved in the determination of which physical attribute of a system is actually measured (as would occur in the context of the delayed-choice experiments\index{delayed-choice experiments} discussed in section \ref{sec:5.6}). Indeed, if one assumes that only measured attributes are real, then it would mean that what is real at the present moment depends on what choice will be made in the future regarding which attributes are to be measured. This is so embarrassing that it is usually considered to support the view that quantum theory is not about reality at all, but about the outcome of measurements, while, in fact, what the possibility of post selection illustrates is rather the awkwardness of the conventional interpretation of quantum theory, in the context of which the reality of a physical attribute itself is dependent on what measurements are performed, either now or in the remote future.

Once the necessity of a realistic approach to quantum theory\index{quantum theory!realistic approach} (according to which all relevant physical attributes are assumed to be real regardless of experimental conditions) is recognized, then all that one must avoid is taking the easy way out and postulate that there exist hidden variables\index{hidden variables!classical} of a classical kind that would require violations of the principle of local causality\index{principle of local causality!violation} as a consequence of trying to explain in too simplistic (but actually quite complicated) a way how the multiple possible histories of unobserved dynamic attributes\index{unobserved dynamic attributes!interfering histories} are allowed to interfere, even in the course of one single quantum process.

In order to achieve a realistic description of quantum phenomena\index{quantum phenomena!realistic description} that does not contradict other essential aspects of reality, it is necessary to first understand that the most significant difference between the sum-over-histories formulation of quantum theory\index{quantum theory!sum-over-histories formulation} and the statistical mechanics\index{statistical mechanics} of classical systems (think about the phenomenon of Brownian motion\index{Brownian motion} in particular) has to do with the use of probability amplitudes\index{probability amplitudes!complex numbers} (complex numbers with both a magnitude and a periodic phase) instead of classical probabilities\index{classical probabilities!real numbers} (real numbers with a magnitude only) as elements of the summation process, which is what allows interferences\index{interferences!different possible histories} to arise between the different possible histories involved. From that viewpoint, what needs to be explained is how it is possible for a particle to follow a path along which all of its unobserved dynamic attributes\index{unobserved dynamic attributes!unique values} have unique values at all times, despite the fact that the probability amplitudes\index{probability amplitudes!interferences} associated with the many possible trajectories which can be followed by any one such attribute are allowed to interfere with one another, as if no definite trajectory was ever followed.

At this point, it may still appear justified to simply reject as implausible the hypothesis that there must, in effect, exist such a unique path. Once the requirement of a time-symmetric description\index{time-symmetric description!quantum reality} of quantum reality will be taken into consideration, however, it will become clear that it is just as inappropriate to reject the hypothesis of the existence of those unique trajectories, as it would be to reject the hypothesis of the existence of elementary particles\index{elementary particles!necessary hypothesis} themselves. John Von Neumann\index{Von Neumann, John} was certainly right when he claimed to have demonstrated that the ordinary reality\index{ordinary reality of everyday objects} of everyday objects cannot apply to quantum particles if those objects are to obey the principle of local causality\index{principle of local causality}. But, as I will explain, that does not necessarily mean that we need to reject the notion that quantum particles\index{quantum particles!unique trajectories of some kind} always follow unique trajectories of some kind (in the space of their unobserved attributes), which still constitutes a valid hypothesis, as long as those undetermined trajectories\index{undetermined trajectories!time-symmetric causality} adhere to the requirements of a time-symmetric conception of causality.

If the sum-over-histories formulation\index{sum-over-histories!formulation of quantum theory} really constitutes a fundamentally different formulation of quantum theory that cannot be derived from earlier formulations (namely Heisenberg's matrix mechanics\index{Heisenberg's matrix mechanics} and Schr\"{o}dinger's wave mechanics\index{Schr\"{o}dinger's wave mechanics}) by a simple mathematical transformation, as is usually understood, then one cannot reject the possibility that it is only by considering the reality it describes for what it is that we can begin to understand quantum theory. From that perspective, it is certainly incorrect to argue, as many authors do, that quantum theory\index{quantum theory!probability of measurement results} is only about the probability of measurement results and does not tell us anything about what goes on in between measurements. If the most adequate and general of quantum-mechanical formalisms\index{quantum-mechanical formalism!most adequate and general} does involve a certain description of what happens in between observations, then it would seem that it is merely our failure to understand why it is exactly that this description is relevant from a physical viewpoint that motivates our rejection of this realistic picture of quantum phenomena\index{quantum phenomena!realistic picture} and that justifies the commonplace belief that the formalism is not indicative of anything more profound.

In any case, one must keep in mind that the prevalent opinion that what the sum-over-histories formalism\index{sum-over-histories formalism!all paths followed all at once} indicates is that all paths are followed all at once in the course of any single process is not an unavoidable conclusion and that it cannot be claimed that no other choice exists for a realistic description of quantum phenomena. What I will explain is that it is still possible, in effect, to assume that a quantum particle must merely be \textit{allowed} to take any of the available paths, but that it does not actually go through all paths in the course of one single process. It is not true that we are confined to contradictory assessments of reality and that it is necessary to assume that quantum theory\index{quantum theory!contradictory assessments of reality} is about particles and yet that it is not about unique particle \textit{histories}.

It is not the hypothesis that there exists a unique and variable (but unobservable) history, which is incompatible with experimental facts, but rather the usually preferred hypothesis that similarly prepared systems always evolve in identical ways in between measurements. Indeed, measurement results\index{measurement results!uniqueness and variability} are clearly unique and variable (even under identical observational constraints) and it is merely our current assumptions regarding what remains unobservable which may turn out to be inappropriate. It must be clear, though, that I'm not claiming that the mathematical framework of quantum theory\index{quantum theory!mathematical framework} is incomplete, because I do recognize that it is impossible to provide a more accurate description of the state of a system than is allowed by the uncertainty principle\index{uncertainty principle}, so that, even if it is real, the unique history of an unobserved dynamic attribute\index{unobserved dynamic attribute!unique history} remains a mere potentiality for any specific process.

Experimental knowledge of both the exact momentum and the exact position of a particle is not allowed by the basic structure of quantum theory and I believe that this is a conclusion that cannot be overturned. In the language of the consistent-histories interpretation\index{consistent-histories interpretation!quantum mechanics} of quantum mechanics, one would say that the simultaneous determination of a particle's momentum and position can only take place on decoherent `branches' of history\index{decoherent branches of history}, which, from my viewpoint, actually means that it cannot occur at all, because this would require distinct observable macroscopic constraints\index{observable macroscopic constraints} to exist together simultaneously (for the same system) and if one wants to preserve the character of uniqueness of physical reality\index{uniqueness of physical reality}, then, obviously, one cannot argue that one set of mutually exclusive macroscopic constraints exist at the same time as a different one.

As a means to accommodate the uniqueness of measurement results\index{measurement results!uniqueness} in light of the existence of quantum interferences\index{quantum interferences!multiple possible histories} between the multiple possible histories of unobserved dynamic attributes\index{unobserved dynamic attributes}, what is usually proposed is that all histories\index{histories!different branches} actually occur all at once as different `branches', but that it is the decoherence effect\index{decoherence effect!vanishing of interferences} that allows \textit{observed} reality to appear unique, given that it requires the interferences that may exist between the different states of a dynamic attribute to vanish very rapidly upon a measurement of this physical attribute.

However, as I will explain in section \ref{sec:5.12}, it appears that decoherence\index{decoherence} can only achieve the goal of giving rise to a quasiclassical world\index{quasiclassical world} if we require the existence of a unique history\index{history!uniqueness} of some kind, in the sense that, in any given universe\index{universe!relationships of causality}, there should only exist relationships of causality among one set of mutually exclusive events. If those conclusions are valid, then it would follow that the only reason that we may still have to assume that all histories are followed at the same time, as different coexisting and interfering branches, is that this hypothesis may perhaps be required to solve the problem of quantum non-locality\index{quantum non-locality!problem}.

What is usually believed, in effect, is that when decoherence occurs and eliminates the interferences that previously existed between the many coexisting `branches' of history\index{coexisting branches of history!interferences}, one particular branch of history\index{branches of history!selection} is selected as that we experience and if this selection process is allowed to occur all at once, then non-local hidden variables\index{non-local hidden variables} may not be required to enforce the consistency of the outcome of measurements performed on entangled systems\index{entangled systems!consistency of measurement outcomes} at space-like separated events\index{space-like separated events}, because all those measurement results are just such as required to enforce the consistency of this whole history\index{history!consistency}, given that what is selected by the decoherence process is the entire branch of history in which we obtain those results.

But the only reason we may have to assume that an entire branch of history is selected all at once, is that this version of history is causally independent from all the other versions. If the multiple branches of history\index{multiple branches of history!causal independence} are not causally independent throughout the universe and at all times, then nothing can justify assuming that the particular branch of history which is selected by a measurement performed at a given moment, in a given location, necessarily determines what branch is selected as a result of a distinct, space-like separated measurement\index{space-like separated measurements}, because what is happening at this other location could just as well be determined by what happened in a different branch of history that is also causally related to this event locally.

The problem, however, is that a causally independent universe\index{causally independent universes!causally related events} actually consists of events which are causally related to one another and to nothing else, so as to form a unique history. Therefore, every single event in such a universe must be causally determined by the other events which are a part of it through the propagation of effects\index{effects!forward- or backward-in-time propagation} either forward or backward in time, without violating the future and past light-cone structure\index{light-cone structure!future and past} of spacetime. What this means is that if correlations exist between certain space-like separated events\index{space-like separated events!correlations} which form a causally independent history\index{causally independent history}, then non-local hidden variables\index{non-local hidden variables} may be required to give rise to those correlations.

I know that many people do not agree that this is a problem for the many-worlds interpretation\index{many-worlds interpretation!quantum theory} of quantum theory, even though they recognize that the multiple branches of history\index{multiple branches of history!causal independence} must, in effect, be causally independent from one another (as if they actually consisted of different universes), if it is to be possible for decoherence\index{decoherence!branch of history} to select an entire branch all at once. But the problem, once again, is that there is a logical contradiction here, because we cannot assume that we are dealing with causally independent branches, without assuming that what this causal independence means is that causal relationships are indeed restricted to events within that universe, which requires effects\index{effects!propagation between events} to propagate between those events to enforce correlations, because no external causes\index{external causes!universe} can be invoked to explain those regularities.

There should be no doubt that, if the multiple branches of history are causally independent from one another, so that a whole branch can be selected all at once following decoherence, independently from what is happening in other branches, then there must exist non-local hidden variables\index{non-local hidden variables} propagating effects faster than the relativistic speed limit, to enforce the historical consistency\index{historical consistency!measurement outcomes} of measurement outcomes at multiple remote locations, in the presence of quantum entanglement\index{quantum entanglement}; unless we embrace the kind of realistic time-symmetric interpretation of quantum theory\index{quantum theory!realistic time-symmetric interpretation} I will propose below, which does not require assuming that a quantum system goes through all available paths all at once in the course of every process occurring in the space of its unobserved dynamic attributes\index{unobserved dynamic attributes}.

Thus, by assimilating what I believe to be the only appropriate interpretation of quantum phenomena, we will go from a situation where it is necessary to assume that, either there is no reality at all between measurements, or else that all histories are followed all at once, to a situation where it is no longer possible or necessary to embrace such logically inconsistent viewpoints\index{logically inconsistent viewpoints} and where we are allowed to conceive of a universe\index{universe!one single history} as involving one single history of some kind, which, in effect, constitutes the most essential aspect of a physically adequate definition of what a universe actually is. It is not absolutely true, therefore, that the phenomenon of quantum non-locality\index{quantum non-locality} cannot be used to constrain our concept of physical reality in a way that would require it to better agree with certain properties of this reality which are known to apply under more general conditions, as Einstein once sought to achieve. But it is also paradoxical that what is usually assumed to be ruled out by the phenomenon of quantum entanglement is a realistic description of quantum phenomena.

What the violation of Bell's inequality\index{Bell's inequality!violation} by the results of multiple different experiments performed on pairs of entangled elementary particles\index{entangled particles} proves is merely that a naive concept of reality\index{reality!naive concept}, according to which all dynamic physical attributes\index{dynamic physical attributes!unique classical state} are in a unique \textit{classical} state at all times, could not be considered valid unless this reality involves non-local influences\index{non-local influences}. In other words it has become necessary to recognize that non-local correlations\index{non-local correlations} do arise at the most fundamental level of description of physical reality.

But this does not necessarily mean that any realistic description\index{realistic description!non-local hidden variables} of those phenomena would require the existence of non-local hidden variables that would propagate effects faster than the relativistic speed limit\index{relativistic speed limit}, because this non-locality\index{non-locality} may instead be a simple reflection of the fact that the basic structure of reality is richer than we usually assume, in the sense that it could be governed by a more general concept of causality that is not limited by the constraint of thermodynamic time asymmetry\index{thermodynamic time asymmetry}.

Given that quantum entanglement\index{quantum entanglement} is made manifest through quantum interference\index{quantum interferences}, the non-locality that is discussed here is not different from that I have already identified as emerging whenever one assumes that the wave function\index{wave function!as reality} itself constitutes physical reality. I believe that what this means is that it is not the hypothesis that there exists a unique history\index{history!uniqueness} of some kind which is problematic, but the notion that quantum non-locality\index{quantum non-locality} must necessarily involve a violation of the causal structure of spacetime\index{causal structure of spacetime} imposed by relativity theory\index{relativity theory}. I have already emphasized, in the discussion about time-symmetric causality\index{time-symmetric causality} from section \ref{sec:5.3}, that backward-in-time causation\index{backward-in-time causation} is not forbidden by relativity theory. But it should be clear that backward causation, even when it is restricted to operate in accordance with the principle of local causality, may actually give rise to non-local correlations\index{non-local correlations}.

The important point here is that the existence of such correlations would not allow faster-than-light communication\index{faster-than-light communication}, given that the backward propagated effects\index{backward propagated effects} are submitted to the condition of diminishing entropy in the past that is imposed by the constraint of global entanglement\index{global entanglement constraint} and in such a context information\index{information!flow from past to future} is only allowed to flow from the past toward the future and never in the opposite direction, while a flow of information\index{information!flow toward past} toward the past would be required for faster-than-light communication to occur. Amazingly, this is precisely the kind of limit that is observed to apply in the case of those non-local correlations which have been experimentally demonstrated to occur in the course of certain quantum phenomena, as a result of entanglement. I believe that this is not just a coincidence, but that it actually confirms what I have said above concerning the time-symmetric nature of causality and the crucial role played by this property in a quantum-mechanical context.

If this is the true origin of quantum non-locality\index{quantum non-locality}, then it would mean that what is actually ruled out is merely the existence of non-local hidden variables\index{non-local hidden variables!faster-than-light propagation} that would violate the principle of local causality\index{principle of local causality!violation} by propagating effects at faster-than-light velocities (which would allow faster-than-light communication and therefore also the flow of information from the future toward the past), while the non-local correlations\index{non-local correlations} that follow from backward-in-time causation\index{backward-in-time causation} would actually be a fact which we were traditionally allowed to ignore only because it does not allow signals or information\index{information!instant communication} to be communicated instantaneously (due precisely to the origin of those non-local correlations) and which can, therefore, only be revealed through subtle correlations of otherwise random outcomes of measurements, performed on carefully entangled quantum systems.

What should be clear, in any case, is that the observed absence of backward-in-time signaling\index{backward-in-time signaling} need not be a consequence of the inadequacy of a realistic time-symmetric interpretation of quantum theory\index{quantum theory!realistic time-symmetric interpretation}, as it can also be a consequence of the effectiveness of the constraint which was identified in section \ref{sec:4.9} and that gives rise to the thermodynamic arrow of time\index{thermodynamic arrow of time} under more general circumstances. Only if that was not the case, would the backward-in-time causation that may be involved in giving rise to quantum non-locality be allowed to violate the principle of local causality that is enforced by relativity theory\index{relativity theory}. It is not appropriate to conclude that the experiments which have confirmed that certain quantum phenomena require the existence of non-local correlations have proven that those phenomena are irreconcilable with any commonsense interpretation of the theory. What must be abandoned is not scientific realism\index{scientific realism}, but the conventional interpretation of quantum theory\index{quantum theory!conventional interpretation} which forces us to reject the principle of local causality and to return to a conception of reality that would involve instantaneous action at a distance\index{instantaneous action at a distance}.

It is important to note, in this regard, that it is the locality assumption\index{locality assumption} that would allow one to conclude, based on the results of certain recently performed experiments described in Ref. \cite{Proietti-1} which involve multiple entangled photons\index{multiple entangled photons}, that there may coexist many mutually incompatible accounts about what constitutes a known, or observationally confirmed fact. Those experimental results, which involve the violation of certain inequalities similar to, but distinct from the conventional Bell's inequality\index{Bell's inequality}, were initially assumed to support the claim that factual truth\index{factual truth!relative notion} is a relative notion (and therefore that reality\index{reality!lack of objectivity} may not be objective), a conclusion which would appear to confirm the relevance of the relational interpretation of quantum theory\index{quantum theory!relational interpretation}.

But once we recognize that quantum non-locality\index{quantum non-locality} is not optional and that it was actually shown, by even more straightforward methods, to itself constitute an unavoidable aspect of reality, then the inappropriateness of the radical conclusions which were drawn, based on the results of the above discussed experiments (regarding the lack of objectivity of observationally derived facts), becomes all the more obvious, even aside from the fact that they would (once again) have given rise to logical contradictions\index{logical contradictions}. Thus, it should be clear that the assumption that the experimental results obtained in one part of such an experimental setup cannot be correlated non-locally with those obtained in a remote part of the same setup is incorrect and it is only when we are not willing to take this aspect into consideration that those experiments seem to imply that reality is not objective and that the truth of certain experimentally established facts, which all happened in the same universe, may be an observer-dependent aspect of reality.

I believe that this only shows how important it is to recognize that non-local correlations\index{non-local correlations} do arise in the quantum realm, even if effects are always constrained to propagate at velocities no larger than that which is imposed by the relativistic speed limit\index{relativistic speed limit!light-cone structure of spacetime} associated with the light-cone structure of spacetime, either forward or backward in time.

What I have tried to make clear in this section is that it is highly preferable to adopt a realistic interpretation of quantum phenomena\index{quantum phenomena!realistic interpretation}, because all alternative proposals are known to involve logical contradictions\index{logical contradictions} at one point or another and those difficulties are always attributable precisely to a rejection of scientific realism\index{scientific realism!rejection}. What is unsatisfactory, however, is the absence of a realistic interpretation that would agree with the multiple specific constraints imposed by the mathematical structure of quantum theory\index{quantum theory!mathematical structure}, like non-locality\index{non-locality} or quantum interference\index{quantum interferences} (more generally). I believe that if the orthodox interpretation of quantum theory\index{quantum theory!orthodox interpretation} is still preferred by most researchers in the field, despite the fact that it requires rejecting scientific realism, it is because something essential is missing from all known realistic interpretations that could make one of them acceptable.

The problem to which I will now turn, therefore, is that of explaining in an intuitively satisfactory, but logically consistent way, without rejecting as mere illusion the uniqueness of historical facts\index{uniqueness of historical facts}, why it is that the probability amplitudes\index{probability amplitudes!multiple particle trajectories} associated with the many trajectories available to a quantum particle are allowed to interfere with one another when its position state is not under direct observation, as if the particle actually followed several different trajectories all at once in the course of one single process. It is here that it will finally be shown that, despite what is usually believed, this is not an impossible task.

\section{Time-symmetric quantum theory\label{sec:5.8}}

It is quite amazing that one single requirement allows to satisfy, all at once, both the condition of scientific realism\index{scientific realism} in face of quantum interference\index{quantum interferences} or state superposition\index{state superposition} and the principle of local causality\index{principle of local causality} in face of quantum entanglement\index{quantum entanglement}. This requirement is that of time-symmetric causality\index{time-symmetric causality!requirement}. There should be no doubt, indeed, that the only way one can avoid having to conclude that there exist non-local effects\index{non-local effects} propagating faster than the relativistic speed limit\index{relativistic speed limit}, in the context of a realistic description of quantum phenomena, is by assuming that certain effects actually propagate backward in time. But it is usually believed that such backward causation\index{backward causation} would be even more problematic than the existence of non-local hidden variables\index{non-local hidden variables}.

I must admit that I don't really understand what motivates that opinion, which to me appears even more arbitrary than the rejection of negative energy states\index{negative energy states}. What could be worse than an outright rejection of relativity theory\index{relativity theory} and the principle of local causality and what could be more difficult a task than rebuilding quantum physics from the ground up, while trying to provide a consistent classical hidden-variable theory\index{classical hidden-variable theories} that would allow to match all empirical constraints, by postulating explicitly non-local influences? But what is even more significant is that, as I previously explained, when special relativity is applied to a quantum mechanical description of elementary particle\index{elementary particle!interaction processes} interaction processes, it becomes an absolute necessity to embrace a time-symmetric concept of causality, according to which effects can propagate both forward and backward in time from causes that may originate from either the past or the future.

What must be understood is that backward-in-time causation\index{backward-in-time causation} is not necessarily problematic, even if the finality it involves may appear unnatural from the viewpoint of our conventional, unidirectional experience of time\index{unidirectional experience of time}. First of all, in a universe where entropy cannot grow in the past, backward-in-time causation would not allow us to tell the future in advance. But, as I already explained, it is also clear that backward causation\index{backward causation} does not allow one to change a known fact from the past. Classical causality\index{classical causality}, or the pairing of the distinction between causes and effects with the thermodynamic distinction between past and future, only comes into play at the macroscopic level where time asymmetry\index{time asymmetry!emergence} emerges from the constraint imposed by the presence of negative-energy matter on the initial Big Bang state\index{initial Big Bang state!global entanglement} at which global entanglement must take place. In other words, our experience of classical, unidirectional causality is not necessarily incompatible with backward causation, as long as the effects which are propagated backward in time do not give rise to the kind of backward-in-time signaling\index{backward-in-time signaling} that would require entropy\index{entropy!growth in past} to grow in the past.

Now, I previously mentioned that what quantum entanglement\index{quantum entanglement} appears to allow is precisely the kind of non-local correlations\index{non-local correlations} that would arise from such backward-in-time propagation of effects\index{backward-in-time propagation of effects}, which is required to occur with ever decreasing entropy\index{entropy!decrease toward past} in the past and which, for that reason, is not allowed to give rise to faster-than-light communication\index{faster-than-light communication}, as would be the case if classical hidden variables\index{classical hidden variables} were responsible for quantum non-locality\index{quantum non-locality}. A consistent interpretation of quantum theory\index{quantum theory!consistent interpretation} would be one that naturally agrees with this limitation in all situations, despite the fact that it would allow to explain non-local correlations. This must be considered an absolute requirement of any realistic approach in the context where no violation of the classical (unidirectional) principle of causality\index{unidirectional principle of causality!violation} has ever been observed to take place in the course of any measurement on entangled systems\index{entangled systems}.

If this is correct, then we need to ask how it is exactly that such backward causation\index{backward causation} is allowed to take place, in the context where the only particles we know about that do propagate backward in time are antimatter\index{antimatter!backward-in-time propagating particles} particles, while such particles are not always involved in the experiments which have revealed the existence of non-local correlations? What I have come to understand is that, in fact, such time symmetry\index{time symmetry!mathematical structure of quantum theory} is precisely what the mathematical structure of quantum theory naturally requires, as my discussion of the two-state-vector formalism\index{two-state-vector formalism} from section \ref{sec:5.6} emphasized.

As I previously explained, a mathematically equivalent version of quantum theory\index{quantum theory!mathematically equivalent formulation} can be formulated that involves two state vectors, one of which provides the state of a system as determined by past measurements, and the other the state of the same system as will be determined by future measurements. In between measurements, those two state vectors\index{two state vectors!unitary evolution} evolve in a conventional `unitary' manner, in the future following a past measurement, and in the past preceding a future measurement. Of course, this is not a realistic representation of quantum phenomena, as we are still dealing with wave functions, but at least, it shows that a formulation can be provided that allows to reproduce all the predictions of quantum theory (sometimes more naturally than even the standard theory) while satisfying the requirement of a time-symmetric description of quantum reality\index{quantum reality!time-symmetric description} (whatever this reality turns out to be).

One clear advantage of a time-symmetric formulation of quantum theory\index{quantum theory!time-symmetric formulation} is that it naturally accommodates the description of experiments in which the order in time of measurements\index{measurement!order in time} performed on a pair of entangled particles\index{entangled particles} is dependent on the state of motion of an observer. Indeed, when the chronological order of two measurements\index{chronological order of measurements!observer-dependent property} is an observer-dependent property (which would occur whenever those measurements\index{measurement!space-like separation} are separated by a space-like interval), a process of state vector reduction\index{state vector reduction!entangled particles} which may appear to be triggered by a measurement performed on one entangled particle, from the viewpoint of a certain observer, would appear to be triggered by the measurement performed on its entangled counterpart, for a different observer.

But it would be problematic to have to choose one or the other of two such measurements as being the cause of the outcome of the other measurement if it was not also possible to assume that it is this other measurement that is the cause of the outcome of the first one, because, in such a case at least, there is no objective criterion that would allow one to tell which event is the cause and which is the effect. Yet, from the viewpoint of the conventional approach, it would appear that it is necessarily the event that happens first that is the cause of the other event, even if this first event actually happens later from the viewpoint of a different observer.

In the context of a time-symmetric formulation of quantum theory\index{quantum theory!time-symmetric formulation}, however, the fact that future measurements are allowed to influence the present state of a system means that a certain reciprocity is allowed between the measurement that influences and the measurement that is influenced (both measurements exert an influence on the outcome of their remote counterpart). In other words, it is no longer necessary to assume that there exists an absolute distinction between a cause and its effect\index{causes and effects!absence of absolute distinction}, from a purely quantum-mechanical viewpoint, and this actually allows to avoid the contradiction that would otherwise emerge when we are dealing with measurements\index{measurement!space-like separated events} performed at space-like separated events on entangled systems\index{entangled systems}.

What one must retain is that if it was not for the existence of backward-in-time causation\index{backward-in-time causation}, then a clear distinction would need to exist between the causes of state vector reduction\index{state vector reduction!distinction between causes and effects} and their effects, even when we are dealing with entangled particles\index{entangled particles}, because reality itself is assumed to be different after a measurement, from a conventional viewpoint. But given that, in some cases, this distinction may be an observer-dependent property, then it would appear that the spirit of relativity theory\index{relativity theory!spirit violation} would be violated, even if it would be impossible to say exactly what distinguishes the cause from its effect, because, from a conventional viewpoint, this distinction would be required to exist. The fact that, in all known situations where non-local correlations\index{non-local correlations!absence of backward-in-time signaling} may arise, backward-in-time signaling is not allowed to occur, clearly shows that unidirectional causality\index{unidirectional causality} is not involved in the determination of the outcome of the second of two measurements performed on a pair of entangled particles, because if it was involved, then there would be no reason \textit{not} to expect backward-in-time signaling to occur, at least in some reference systems.

From such a viewpoint, it would appear that the prevalent belief that causality\index{causality!prevalent belief} must always operate forward in time is motivated by expectations similar in nature to those which \textit{originally} motivated the formulation of the Lorentz transformation\index{Lorentz transformation!original motivation} (the contraction of physical objects\index{contraction of objects} in motion relative to absolute space), because imposing a unidirectional conception of causality\index{causality!unidirectional conception}, in the context of quantum non-locality\index{quantum non-locality}, amounts to postulate a property of reality which, even if it did pertain to the physical world, would be required to have absolutely no distinguishable effect on it.

Now, even though the early time-symmetric formulations of quantum mechanics\index{quantum mechanics!early time-symmetric formulations} represented a step forward, the fact that they still didn't provide a realistic picture of quantum phenomena that would fully accommodate the particle concept and the requirement of a unique history of some kind means that, even aside from any other inadequacies, they cannot be the final answer to the problem of interpretation. Clearly, something essential is still missing and it is only after much questioning and while trying to figure out how the two-state-vector formalism\index{two-state-vector formalism} could be generalized to agree with the sum-over-histories formulation of quantum theory\index{quantum theory!sum-over-histories formulation} that I was able to obtain a truly consistent, realistic picture of quantum phenomena. I have become convinced, in effect, that the bold intuitive leap which I previously suggested one must be ready to take to achieve a more realistic interpretation of quantum theory\index{quantum theory!realistic interpretation} actually consists in recognizing that what we are dealing with here is a set of two unique histories\index{two unique histories!opposite directions of time} (involving unique particle trajectories) unfolding in opposite directions of time without directly interacting with one another in any way.

In such a context, what matters is not really the direction of propagation in time\index{direction of propagation in time!two versions of history} of the particles involved in those two versions of history, but an overall direction of time that only differs in a relationally defined way, such that, if the two versions of history\index{history!two versions} were to be otherwise identical, they would still differ in that the directions of propagation in time of all the particles involved in one of those histories would be opposite those in the other. But in fact, it is not possible to differentiate in any absolute way initial causes from final `causes'\index{final causes} and it is only the difference between the directions in which the two parallel histories\index{two parallel histories} unfold in time that has physical meaning and this relationship must be preserved even when the processes actually occurring in the course of those two versions of history differ in ways not forbidden by the macroscopic experimental conditions\index{macroscopic experimental conditions} imposed on those processes.

The important point, here, is that the paths followed by a quantum system in the space of its unobserved dynamic attributes\index{unobserved dynamic attributes!retarded and advanced paths} must, in effect, be allowed to differ for the retarded and the advanced portions of a process (the ordinary process and its time-reverse counterpart), even though the observed dynamic attributes\index{observed dynamic attributes!retarded and advanced paths} of the system must follow the exact same paths in both portions of the process.

What really happens, therefore, to a photon on its way to a detector in the double slit experiment\index{double slit experiment} (see Figure \ref{fig:5.1}) is not that it passes through both slits all at once, but that it has the \textit{possibility} to pass through any one of the two open slits in \textit{both} the retarded and the advanced portions of the same process (when the actual trajectory remains observationally undetermined), which therefore requires that both paths be taken into account in the determination of transition probabilities\index{transition probabilities!contribution of all paths} for any given process, even though a photon only ever goes through one particular slit in the retarded portion of history and then again through one particular (but possibly different) slit in the advanced portion of history. It is simple to verify that those assumptions allow to reproduce the predictions of the standard theory in any specific and possibly more complex situation (I will explain below why this should indeed be expected).

\begin{figure}
\begin{center}
\scalebox{2}[1]{\begin{picture}(200,360)
\put(20,320){\line(0,1){16}}
\put(20,342){\line(0,1){17}}
\scalebox{.5}[1]{\put(25,335){S}}
\put(20,220){\line(0,1){16}}
\put(20,242){\line(0,1){17}}
\scalebox{.5}[1]{\put(25,235){S}}
\put(20,120){\line(0,1){17}}
\put(20,143){\line(0,1){17}}
\scalebox{.5}[1]{\put(25,135){S}}
\put(20,20){\line(0,1){17}}
\put(20,43){\line(0,1){17}}
\scalebox{.5}[1]{\put(25,35){S}}
\put(180,322){\line(0,1){36}}
\scalebox{.5}[1]{\put(367,335){D}}
\put(180,222){\line(0,1){36}}
\scalebox{.5}[1]{\put(367,235){D}}
\put(180,122){\line(0,1){36}}
\scalebox{.5}[1]{\put(367,135){D}}
\put(180,22){\line(0,1){36}}
\scalebox{.5}[1]{\put(367,35){D}}
\put(100,300){\line(0,1){18}}
\put(100,322){\line(0,1){35}}
\put(100,362){\line(0,1){18}}
\scalebox{.5}[1]{\put(189,309){2}\put(189,363){1}}
\put(90,300){\vector(1,0){20}}
\scalebox{.5}[1]{\put(225,297){$t$}}
\put(100,200){\line(0,1){17}}
\put(100,222){\line(0,1){36}}
\put(100,262){\line(0,1){18}}
\scalebox{.5}[1]{\put(189,209){2}\put(189,263){1}}
\put(90,200){\vector(1,0){20}}
\scalebox{.5}[1]{\put(225,197){$t$}}
\put(100,100){\line(0,1){18}}
\put(100,122){\line(0,1){36}}
\put(100,162){\line(0,1){18}}
\scalebox{.5}[1]{\put(189,109){2}\put(189,163){1}}
\put(90,100){\vector(1,0){20}}
\scalebox{.5}[1]{\put(225,97){$t$}}
\put(100,0){\line(0,1){18}}
\put(100,22){\line(0,1){36}}
\put(100,62){\line(0,1){18}}
\scalebox{.5}[1]{\put(189,9){2}\put(189,63){1}}
\put(90,0){\vector(1,0){20}}
\scalebox{.5}[1]{\put(225,-3){$t$}}
\put(20,340){\vector(4,1){40}}
\put(60,350){\line(4,1){40}}
\put(100,360){\vector(4,-1){40}}
\put(140,350){\line(4,-1){40}}
\put(180,338){\vector(-4,1){40}}
\put(140,348){\line(-4,1){40}}
\put(100,358){\vector(-4,-1){40}}
\put(60,348){\line(-4,-1){40}}
\put(20,240){\vector(4,-1){40}}
\put(60,230){\line(4,-1){40}}
\put(100,220){\vector(4,1){40}}
\put(140,230){\line(4,1){40}}
\put(180,238){\vector(-4,-1){40}}
\put(140,228){\line(-4,-1){40}}
\put(100,218){\vector(-4,1){40}}
\put(60,228){\line(-4,1){40}}
\put(20,140){\vector(4,1){40}}
\put(60,150){\line(4,1){40}}
\put(100,160){\vector(4,-1){40}}
\put(140,150){\line(4,-1){40}}
\put(180,140){\vector(-4,-1){40}}
\put(140,130){\line(-4,-1){40}}
\put(100,120){\vector(-4,1){40}}
\put(60,130){\line(-4,1){40}}
\put(20,40){\vector(4,-1){40}}
\put(60,30){\line(4,-1){40}}
\put(100,20){\vector(4,1){40}}
\put(140,30){\line(4,1){40}}
\put(180,40){\vector(-4,1){40}}
\put(140,50){\line(-4,1){40}}
\put(100,60){\vector(-4,-1){40}}
\put(60,50){\line(-4,-1){40}}
\end{picture}}
\end{center}
\caption[The four possible, combined, retarded and advanced histories of a double slit or simple interferometer experiment]{The four possible, combined, retarded and advanced histories\index{retarded and advanced histories} of a double slit\index{double slit experiment} or simple interferometer experiment\index{interferometer experiment}, with one source $S$ and one position detector $D$, when the actual trajectory of the quantum particle remains unknown. The direction of the arrows corresponds to the flow of time. When the actual trajectory of the particle is subject to experimental determination, only the first two combined histories remain possible and the two trajectories no longer interfere, as the retarded and the advanced histories must be the same for any complete process.}\label{fig:5.1}
\end{figure}

It is merely the fact that we do not observe the actual trajectory followed by the photon that makes it necessary to consider both possibilities, all at once, for any single process, given that under such conditions this trajectory can be different for the retarded and the advanced portions of the process\index{retarded and advanced processes!differing trajectories}. But this does not necessarily mean that the trajectory actually is different for the two histories, only that it \textit{can} be and, as I will explain below, this is sufficient a motive for requiring that both trajectories be considered to contribute to the estimation of transition probabilities.

Any one history still involves a particle following a unique, unobservable trajectory\index{unobservable trajectory!uniqueness}, only, each process involves both a retarded history and an advanced history\index{retarded and advanced histories} (a pair of histories taking place in opposite directions of time) which are only required to share the same macroscopic experimental constraints\index{macroscopic experimental constraints!retarded and advanced histories}. Those histories are therefore allowed to differ in all aspects which are not constrained to a particular subset of possibilities by the observable `macroscopic' conditions (the retarded and advanced paths can differ as long as no permanent record of those differences ever becomes available) and this is why the probability amplitudes\index{probability amplitudes!interferences} associated with the many different, possible paths available to a quantum system are allowed to interfere with one another and must therefore be taken into account in the determination of the probability for the complete process (comprising those two histories) to occur.

Remarkably, if it was not for the fact that probability amplitudes\index{probability amplitudes!periodic evolution}, unlike conventional probabilities, are subjected to periodic evolution, which allows them to interfere constructively or destructively, then it would be impossible to deduce the existence of the advanced portion of a process (which may actually be any one of the two histories), because it is the periodic or wavelike aspect of the probability amplitudes associated with the retarded and advanced portions of history\index{retarded and advanced portions of history!interference} which allows them to interfere, when the dynamic attributes\index{dynamic attributes!experimental indetermination} involved are not subjected to direct experimental determination.

The greater consistency of the viewpoint proposed here is apparent in the fact that it is no longer necessary to assume that, when the path followed by a particle is not observed, the object actually behaves as if it was a different entity (a classical wave\index{classical waves}), because the interferences which are made conspicuous in the statistical distribution of measurement results\index{measurement results!statistical distribution} can be explained without requiring one to assume that the particle behaves differently when its position is not observed. What changes, when a different dynamic attribute is submitted to observation, is merely the macroscopic conditions imposed on a process, while the system involved still follows a unique, but unknown, and possibly different trajectory in the retarded and advanced portions of history\index{retarded and advanced portions of history!unique trajectories}, which unfold in the space of the unobserved attribute.

It is only when a particle is constrained by the experimental conditions to follow a certain definite path (when a record of the actual slit through which the particle went is available) that interferences are absent, for the dynamic attribute involved, because, in such a case, the particle must follow the same path during both the retarded and the advanced portions of history.

It would, therefore, be incorrect to maintain that it is not possible to visualize what occurs to a photon as it propagates from source to detector in a double slit experiment\index{double slit experiment!visualization of photon propagation}, when its trajectory is not observed. It is not nonsense to speak of the passage of the photon through one particular slit, even when this trajectory remains experimentally undetermined, as long as one recognizes that the actual trajectory followed by the particle can be different for the retarded and advanced portions of the process\index{retarded and advanced processes!differing trajectories}. From this viewpoint, what looks rather absurd is the conventional assumption that an elementary particle\index{elementary particle!undetermined trajectory} whose trajectory is observationally undetermined follows, at once, all possible paths. When it is properly understood, quantum theory\index{quantum theory!proper understanding} is no longer as unsettling as it used to be (this comment will become even more apposite when other essential aspects of this approach are discussed, which allow to justify its inevitability).

From the viewpoint of the interpretation of quantum theory proposed here, there would no longer arise logical contradictions in the description of the state of a system when a certain dynamic attribute of the system is in a state of superposition\index{state of superposition!dynamic attribute} (which is always the case for at least one physical attribute).

We may consider, for example, an electron whose spin\index{electron spin!measurement} has been measured to be up along the horizontal axis. Under such conditions, the spin of this electron along the vertical axis must be considered undetermined. But this cannot be understood to mean that the spin of the electron is either up along the horizontal axis and up along the vertical axis, or else up along the horizontal axis and down along the vertical axis, as one might consider appropriate from a classical perspective. Whenever one tries to experimentally confirm the apparently indisputable validity of this legitimate hypothesis, the results one obtains show that it cannot be true. It may therefore appear that whenever an electron is in a definite state of spin relative to the horizontal axis, its spin state\index{spin state!mutually exclusive directions} along the vertical axis, if it is real, must be such that it cannot be described without violating the conventional rules of logic\index{logic!violation of conventional rules}, because it would seem to be allowed to point along two mutually exclusive directions all at once, which, from a realistic viewpoint\index{realistic viewpoint!contradiction}, does, in effect, constitute a contradiction.

But once it is understood that two causally independent histories\index{two causally independent histories} are involved in any single process, then it becomes clear that what the results of the discussed experiments mean is not that the vertical spin of the electron\index{vertical spin of electron} is in no state at all (which would require rejecting the possibility of a realistic description of quantum phenomena), or that it is, at once, in all possible states (which would require rejecting the conventional rules of logic\index{logic!rejection of conventional rules}), but merely that while the vertical spin state of the electron\index{vertical spin of electron!retarded and advanced histories} in the retarded portion of history can be either up or down, the same vertical spin state can also be either up or down in the advanced portion of history, which means that four different combinations of states are allowed, thereby contradicting the hypothesis that this vertical state can only be either up or down and nothing else, for any single process, or at any single time (which actually consists of two different times that must simply correspond with one another for the retarded and advanced portions of history, as I will later explain).

One is therefore allowed to assume that the spin of an electron\index{electron spin!unique possibly unobservable state} along any axis is always in a unique, but possibly unobservable state in any one of the two parallel portions of history, even though it is not in a unique state for any \textit{process} (when a process is adequately considered to involve both a retarded and an advanced portion), as experiments confirm. Thus, if those experiments with electrons, as well as more decisive observations of the same kind, do show that quantum strangeness\index{quantum strangeness!inevitability} is unavoidable, it would be incorrect to assume that what they demonstrate is that a realistic interpretation of quantum theory\index{quantum theory!realistic interpretation} is impossible and that there cannot be a unique reality of some kind behind the observed phenomena.

The contradictions which are encountered in the context of a more conventional, realistic interpretation of quantum theory\index{quantum theory!conventional realistic interpretations} are only made apparent in the statistical distribution of measurement results\index{measurement results!statistical distribution} and always concern physical attributes which actually remain unobserved, while it is precisely at this level that the alternative interpretation proposed here differs from the conventional approach. But given that this realistic, time-symmetric interpretation of quantum theory\index{quantum theory!realistic time-symmetric interpretation} allows non-local correlations\index{non-local correlations!faster-than-light propagation} to arise, even when no effect is propagated at faster-than-light velocities (as I will show in the following section), then it also appears inappropriate to argue, as is often done, that only a rejection of scientific realism\index{scientific realism!rejection} (the idea that there must exist an `objective' reality in between measurements) may allow one to avoid the conclusion that quantum non-locality\index{quantum non-locality!action at a distance} arises from instantaneous action at a distance. Quantum non-locality is not an illusion, but action at a distance can be avoided, even in a realistic interpretation of the theory.

It is the fact that early time-symmetric interpretations of quantum theory\index{quantum theory!early time-symmetric interpretations} involve conventional wavelike phenomena\index{conventional wavelike phenomena} that makes them undesirable as realistic interpretations. But once the dual nature of the state vector is understood to be a consequence of the existence of two actual histories\index{histories!opposite time directions} in which particles propagate, once through any of the available paths and then again through any of those same available paths, but in the opposite direction of time, then the time-symmetric nature of quantum reality\index{quantum reality!time-symmetric nature} becomes a more significant asset, given that it allows to reproduce the statistics of quantum measurement results and to explain interference effects\index{interference effects!realistic picture of reality} involving distinct paths, while naturally providing a picture of quantum reality that satisfies the requirements of scientific realism.

Of course, the reality unveiled here is not classical, because it involves probability amplitudes\index{probability amplitudes} instead of classical probabilities and it requires the existence of an unobserved counterpart to every process (because we really experience only one of the two parallel portions of history\index{two parallel portions of history!one experienced history} at any single time). But then, what we are dealing with \textit{is} quantum reality and not classical reality and only logical consistency\index{logical consistency!representation of reality} provides an unavoidable criterion for judging the validity of any experimentally accurate representation of reality. If quantum strangeness\index{quantum strangeness!inevitability} itself cannot be avoided, then there must certainly remain some unexpected element in any empirically established model. In fact, it appears that it is the remaining, unexpected aspects of quantum reality\index{quantum reality!unexpected aspects} that make the theory truly consistent in a way that would be impossible classically and, as such, they are certainly not undesirable.

As I explained in the preceding section, consistency\index{consistency!unique non-classical reality} merely dictates that physical reality must, in effect, be real and therefore unique in \textit{some} particular way, but it does not \textit{a priori} constrain this reality to conform to some preconceived criteria of appropriateness we may believe should apply, that would be based on an experience of physical reality which is restricted to a subset of experimental conditions, namely those where quantum interference\index{quantum interferences} and entanglement\index{quantum entanglement} are usually unapparent. What's more, I'm not suggesting that two processes can be observed to take place in parallel that could differ from one another, in violation of the uniqueness of historical facts\index{uniqueness of historical facts}, but merely that there exists an unobservable counterpart to history\index{history!unobservable counterpart}, which is required in order to explain certain observable features of reality (the interferences) and which remains identical to its time-reversed version from the viewpoint of observable macroscopic properties, even if it may differ from it at the level of intricate details.

Therefore, what constitutes a decisive advantage of the time-symmetric interpretation of quantum theory\index{quantum theory!time-symmetric interpretation} proposed here is that it naturally agrees with the observation that all results of quantum measurements\index{quantum measurement results!uniqueness} are, in effect, unique, so that there is no longer a problem associated with the objectification of quantum measurement results\index{quantum measurement results!objectification}. In other words, it is no longer necessary to assume that all potentialities are actualized\index{actualization of potentialities} following measurement and that it is merely the splitting of history\index{history!splitting branches} into non-interfering branches (that would be attributable to decoherence\index{decoherence}), that explains the fact that we actually experience only one, non-interfering reality\index{reality!unique and non-interfering} for observed dynamic attributes\index{observed dynamic attributes}.

It seems that the error that is made, in the context of most conventional interpretations of quantum theory\index{quantum theory!conventional interpretation}, is that we fail to recognize that, if we were to take into account the existence of the advanced portion of every quantum process\index{advanced portion of quantum process}, it would simply no longer be necessary to assume that all paths from either the retarded or the advanced portion of history\index{retarded or advanced portion of history} are somehow being followed all at once, because the simple fact that the advanced portion of the process can be different from the retarded portion, while still obeying the macroscopic constraints\index{macroscopic constraints!experiment} of the experiment, is sufficient to guarantee that it is only when all possible paths are taken into consideration that the right predictions, concerning the probability of occurrence of the whole process, will be obtained.

Those considerations are also valid in the case where we are dealing with the probability of one single process (like the passage of a unique photon from source to detector in a double slit experiment\index{double slit experiment}), even if particles do not always follow all possible interfering paths in the course of one single time-symmetric process\index{time-symmetric processes!interfering paths} (but at most two of them), because, in the context where probability amplitudes\index{probability amplitudes} are involved, it is possible for one time-symmetric history\index{time-symmetric histories!negative probability contribution} (composed of a retarded and an advanced portion) to contribute negatively to the final probability of a process, and as I will explain below, this actually allows all the different alternative paths to contribute to the probability of one single time-symmetric process.

Another advantage of such a realistic, time-symmetric interpretation of quantum theory\index{quantum theory!realistic time-symmetric interpretation} (involving both forward and backward propagated effects\index{forward and backward propagated effects}) is that it allows to enforce the historical consistency\index{historical consistency!factual aspects of reality} of factual aspects of reality in a way that is particularly significant in the case of entangled systems\index{entangled systems}. Indeed, if one is to assume that the retarded and advanced portions of history\index{retarded and advanced histories!observable conditions} share the same observable, macroscopic conditions (a requirement whose validity will be justified in section \ref{sec:5.12}), then the result of a measurement performed on one of two entangled particles\index{entangled particles!compatibility of measurement results} must be compatible with both the experimental conditions of this measurement and those of any measurement that may eventually be performed on the other particle, because in any reference system (from the viewpoint of any observer) there is as much causal influence\index{causal influence!reciprocal effects of measurements} from the first measurement on the second, as there is from that second measurement on the first (especially when those two events are separated by space-like intervals).

What is apparent here, therefore, is that a quantum-mechanical description of reality involves some form of causal circularity\index{causal circularity} of the kind that would arise if time travel\index{time travel} was a possibility. But, as I mentioned in section \ref{sec:5.4}, the only problem that may arise in the context where such closed causal chains\index{closed causal chain} would be considered a possibility does \textit{not} have to do with the fact that, if historical consistency\index{historical consistency} is always preserved, this would seem to contradict our expectations regarding free-will\index{free-will!expectations} (a difficulty which is significant merely from the viewpoint of our conventional, unidirectional experience of time\index{unidirectional experience of time}), but with explaining how it is, in effect, that the consistency of history (the idea that all facts must agree with one another under all circumstances) can be preserved, regardless of what happens. What remains to understand is how it is that this requirement is enforced at the level of time-symmetric quantum-mechanical processes\index{time-symmetric quantum processes}.

It should be clear, first of all, that quantum theory does appear to be the appropriate framework for implementing historical consistency, as it already allows to appropriately handle the closed causal chains occurring as a result of the existence of antiparticles\index{antiparticles!particles propagating backward in time} as negative-energy particles propagating backward in time. Thus, the usual approach to estimating the probability for a process to occur, which amounts to sum up the probability amplitudes\index{probability amplitudes!summation} for all possible ways by which a process can occur and then to take the square of this complex number\index{complex number!square value}, appears to allow historical consistency to be satisfied, only, it is not completely clear why, in effect, such an annoying procedure allows to produce consistency, in the context where backward causation\index{backward causation} would be assumed to constitute an unavoidable aspect of a quantum-mechanical description of reality.

To understand what is going on, it is necessary to first recognize that a complete quantum process (one to which can be attributed a certain probability) actually consists in the combination of a retarded history\index{retarded history}, unfolding from a given past state toward a given future state through one particular unobservable path\index{unobservable paths} forward in time, followed by an advanced history\index{advanced history}, unfolding from the same future state toward the same past state through another particular and still unobservable path backward in time, or vice versa (as it may be the advanced history that is `followed' by the retarded history backward in time). Thus, it is essential that the two possible segments of history, which are unfolding in opposite directions of time, be combined to actually give rise to one complete time-symmetric history\index{time-symmetric histories}, to which can indeed be assigned a definite classical probability\index{classical probabilities} (instead of a mere probability amplitude\index{probability amplitudes}).

It would then be by adding the probabilities for all such time-symmetric histories, formed of a pair of histories which are both compatible with the observable past and future experimental conditions that we would obtain the final correlation probability\index{correlation probabilities}. Now, even though such a procedure can be shown to produce transition probabilities\index{transition probabilities} equivalent to those of the conventional approach under similar circumstances, the problem is that it is not always possible to obtain meaningful results from this procedure, unless one limits the scope of the questions that can be asked, concerning the history of a system and its environment, by adopting a suitable coarse-graining\index{coarse-graining}. It is only under such conditions (when certain details are left aside concerning the processes which are described) that classically meaningful probabilities can be obtained for various alternative pair of histories\index{pair of histories}.

In the context of the conventional, modern interpretation of quantum theory\index{quantum theory!modern interpretation} (the consistent-histories interpretation\index{consistent-histories interpretation!quantum theory}), what this would be assumed to mean is that, when described with a maximum level of details, certain histories are simply nonsense and cannot be considered to actually occur as real physical phenomena. This would be the case, for example, of the history of a photon as it goes from source to detector in the classical double slit experiment\index{classical double slit experiment}, when the particular path taken by the particle is not subjected to direct observation, because it seems that one cannot obtain a classically meaningful probability\index{classically meaningful probabilities} for a unique history of such a kind. But I believe that this self-imposed and somewhat arbitrary restriction, concerning what can be considered real of reality itself, is not appropriate and arises merely because we do not understand the profound significance of those apparently inconsistent probabilities\index{probabilities!apparent inconsistency}, which only emerges when they are considered in the context of a realistic and fully time-symmetric interpretation of quantum theory\index{quantum theory!realistic time-symmetric interpretation}.

It must be clear that what I find problematic about the formalism of consistent histories\index{consistent histories!formalism} is the restriction that is usually imposed regarding what can be \textit{meaningfully} described of quantum reality, not the logic of the conclusion, made in the context of the conventional interpretation of quantum theory\index{quantum theory!conventional interpretation}, concerning what can be \textit{classically} described of quantum reality\index{quantum reality!classical description} (which histories can be assigned classically meaningful probabilities) and under which circumstances.

What I'm trying to explain is that the criterion of decoherence\index{decoherence criterion!families of coarse-grained histories}, which is imposed on families of coarse-grained histories in the context of the consistent-histories interpretation of quantum theory\index{quantum theory!consistent-histories interpretation}, is not really a criterion for assessing what is consistent of reality, but merely a criterion for assessing what is \textit{classically} well-defined of this reality. I believe that it is incorrect to argue that common sense logic (conventional logic\index{conventional logic!increasing inadequacy}) is increasingly less adequate for describing reality, when we consider increasingly smaller scales, even though it is certainly true that the probability that various alternative histories\index{alternative histories!interference} interfere with one another rises as those histories are being described with an increasing amount of detail (using a finer coarse-graining\index{coarse-graining}). Clearly, either conventional logic applies or it doesn't and one cannot try to justify how nonsense could be made acceptable by relying on the confused principle of complementarity\index{complementarity principle} (the apparent freedom to describe the same reality with mutually incompatible concepts).

The problem with this conventional interpretation of quantum strangeness\index{quantum strangeness!conventional interpretation} is that it would cease to provide a logically consistent picture of reality\index{reality!logically consistent picture}\footnote{
It must be clear that my use of the term `consistency' does not have the meaning it has in the context of the consistent interpretation of quantum theory\index{quantum theory!consistent interpretation|nn}, where it refers merely to the classical definiteness of a history\index{history!classical definiteness|nn}.}
 precisely on the scale where the unconventional phenomena we may want to understand are occurring (the quantum\index{quantum!scale} scale). But this difficulty arises merely when we fail to understand that conventional logic\index{conventional logic!unobservable unique reality} applies \textit{not} to the observed phenomena themselves, but to the unobservable, unique reality, which consists in each of the two portions of history\index{histories!opposite time directions} that unfold in opposite directions of time for every process, on any scale. The fact that conventional logic still applies on the classical scale, even from a more conventional viewpoint, can therefore be understood to result, not from the fact that reality is only consistent on such a scale, but from the fact that the two time-reversed portions of history must always be the same on such a scale (for reasons I will explain in section \ref{sec:5.12}).

Anyhow, what is usually considered undesirable of the probabilities\index{probabilities!negative or larger than one} that may sometimes be obtained for a combined pair of histories\index{pair of histories} is that they can assume negative values, or normalized values larger than one. I believe that one can only begin to understand why the existence of negative probabilities\index{negative probabilities} in the intermediary stages of the estimation of a final transition probability\index{transition probabilities} is not catastrophic when one recognizes that it is precisely the circularity of all quantum causal chains\index{quantum causal chains!circularity} (that follows from the existence of both a retarded and an advanced portion to every quantum process\index{quantum process!retarded and advanced portions}) that enforces the historical consistency\index{historical consistency!enforcement} of the present with a given future.

In the context where one must take into account the existence of quantum interferences\index{quantum interferences}, it appears necessary, in effect, to impose on the phase of the wave function\index{phase of wave function} that it returns to a value that is as close to its initial value as possible, after a complete, time-symmetric process\index{time-symmetric processes} has occurred (once forward and once backward in time), if this unobservable parameter is to have any physical significance. Of course, this initial value can be any arbitrarily-chosen one, as only \textit{changes} to the phase and the amplitude of the wave function, occurring as a result of the evolution that takes place during the retarded and advanced portions of a quantum process\index{quantum process!retarded and advanced portions}, are significant. In other words, if the phase was originally $\pi$ radiant it cannot end up being $2\pi$ radiant (if the amplitude of the wave function remains unchanged) after a complete time-symmetric process has taken place, otherwise a contradiction would have occurred, as those two phases are the perfect opposite of one another and therefore correspond to two maximally distinct unobservable initial conditions\index{unobservable initial conditions!phase of wave function} (of the phase itself) which can only belong to two mutually exclusive instances of physical reality\index{reality!mutually exclusive instances}. But it is precisely under such conditions that negative probabilities would arise for the particular time-symmetric history involved.

Probabilities larger than one\index{probabilities!larger than one} merely constitute another facet of the same problem, because, as Feynman\index{Feynman, Richard} pointed out \cite{Feynman-3}, a greater than one probability for a given process to occur is equivalent to a negative probability\index{negative probabilities!of process not occurring} that the same process will \textit{not} occur. What I'm suggesting, then, is that, whenever the probability for a process to occur in one specific way is negative, one must assume that, if the process would occur in this specific way, it would diminish the chances that the observable macroscopic conditions\index{observable macroscopic conditions!probability of existence} which would have actually given rise to it existed in the first place, thereby making the sum of probabilities for all the possible ways the process could occur smaller than it would otherwise be, given that it would make the initial conditions\index{initial conditions!reduced likeliness of occurrence} themselves less likely to \textit{have} occurred (because the probability that the process would occur in such a way would decrease the likelihood that the process may occur in any possible way, instead of increasing it as is usually the case).

Thus, when a pair of minimally coarse-grained histories\index{minimally coarse-grained histories} (composed of both a retarded and an advanced process\index{retarded and advanced processes}) has a negative probability\index{negative probabilities} of occurrence, this can be interpreted as diminishing the chances that the process involved may occur by following any possible path (even those for which there is no destructive interference\index{destructive interference}). Likewise, when the probability for an individual pair of minimally coarse-grained histories to occur is larger than one, this can be interpreted as decreasing the likeliness that \textit{alternative} initial conditions\index{alternative initial conditions!likeliness of occurrence} have occurred, which is another way to say that it would actually increase the likeliness that the actual initial conditions\index{actual initial conditions!likeliness of occurrence} that gave rise to this history did indeed occur.

I'm therefore allowed to speak meaningfully not just about a reduction or an increase in the probability that some future outcome is realized when some past conditions are observed, but also about a reduction or an increase in the probability\index{probability reduction or increase!initial conditions} that a given set of initial conditions has actually been observed to occur, whenever a given set of future conditions\index{future conditions!probability of initial conditions} will be satisfied. Those additional contributions to the conventional measures of transition probability\index{transition probabilities!conventional measures} are dependent only on the degree of compatibility between the unobservable, initial phase of the wave function\index{phase of wave function} and that which is obtained as a result of the phase change that occurs in the course of the whole time-symmetric process\index{time-symmetric processes!retarded and advanced histories} (the combination of a retarded and an advanced history).

From the viewpoint developed here, therefore, a process is allowed to influence the very probability that certain boundary conditions\index{boundary conditions!probability of occurrence} necessary for its occurrence may be found to exist, not just in the future, but in the past as well. In the context of a time-symmetric interpretation of quantum theory\index{quantum theory!time-symmetric interpretation}, it should have been expected that such effects would arise, given that there is necessarily as much influence of the future on the past, as there is of the past on the future, which forbids the initial macroscopic conditions\index{initial macroscopic conditions!dependence on future events} from being determined independently from what happens in the unknown future.

What transpires, therefore, is that when the retarded and advanced portions of the history of a given unobserved physical attribute are such that they require changes to the phase of the wave function\index{phase of wave function} that would not allow it to return to its initial value (as would occur when the probability amplitudes\index{probability amplitudes!destructive interference} associated with the two possible paths in a double slit experiment\index{double slit experiment} interfere destructively), then one must assume that the probability that the very initial conditions\index{initial conditions of process!probability of occurrence} of the process (the emission of a photon with such an energy by this particular detector at this particular time) could themselves be observed to have occurred is reduced in proportion to the magnitude of this contradiction.

But if those initial conditions cannot be expected to have happened, then it means that the pair of minimally coarse-grained histories with which is associated a negative probability\index{negative probabilities!pair of minimally coarse-grained histories} would merely contribute to reduce the probability that the process is actually observed to take place following \textit{any} of the available paths, forward and backward in time, and this is why no single instance of a complete time-symmetric history\index{time-symmetric histories!two possibly different paths} (involving only two possibly different paths) contributes to the probability of a process independently from the other possible time-symmetric histories (involving all the other possible paths), despite the fact that only one such history actually happens.

As a result, even though the normalized probability that a given observable history\index{observable history!normalized probability} actually occurs must, in effect, be a number between zero and one, any \textit{unobserved} time-symmetric history\index{unobserved time-symmetric history!negative probability of occurrence} that contributes to determine this final probability could have a negative probability of occurrence, or a probability larger than one and still be describable as consistently as an observable history. The interpretation I'm proposing here, therefore, does not require one to reject as meaningless the histories with which are associated negative probabilities, as those time-symmetric processes\index{time-symmetric processes!realistic interpretation} can be interpreted in a realistic way and do not differ fundamentally from other time-symmetric histories (occurring once forward and then backward in time), given that they do contribute, in a meaningful way, to establish the final, positive probability for an observable (sufficiently coarse-grained) history\index{observable coarse-grained history!positive probability} to occur.

It is merely the fact that negative probabilities\index{negative probabilities} can only arise when quantum interferences\index{quantum interferences!absence of observation} are present, while, in general, interferences are only apparent when the actual path followed by a quantum system is not subjected to direct observation, that explains that we appear to be justified to assume that negative probabilities cannot arise and must be rejected as physically meaningless, because it is true that the probabilities of occurrence of such time-symmetric histories\index{time-symmetric histories!probabilities of occurrence} cannot be revealed through direct observations, as a matter of principle.

Once again, it is the fact that our experience of reality is limited to the portion of it that is directly accessible to our senses that explains that we do not directly experience negative probabilities\index{negative probabilities!indirect experience} and that we view them with suspicion, as if the histories they characterize could not be real. But the interpretation of negative probabilities I have proposed allows all of the histories that contribute to establish the statistics of time-symmetric quantum processes\index{time-symmetric quantum processes}, including those which must be assigned negative probabilities, to constitute potential instances of a unique, but partly unobservable physical reality, thereby allowing to fulfill the requirement of a realistic description of quantum phenomena.

Thus, the occurrence of negative probabilities\index{negative probabilities!clear meaning} in the context of the realistic, time-symmetric interpretation of quantum theory\index{quantum theory!realistic time-symmetric interpretation} I have proposed should not be considered a problem all by itself (that would justify rejecting as unreal the histories which give rise to those unconventional measures of probability), because I have shown how those negative measures of probability can be given a clear meaning in such a context (this is the originality of my approach), as long as the negative values involved do not show up in the final results of the estimation of a transition probability\index{transition probabilities!observed history} involving an \textit{observed} history. In fact, from a purely formal perspective, the proposed approach may be considered even more adequate than the conventional method, given that it always involves only the summation of real probabilities (one real, but possibly negative number for each time-symmetric process\index{time-symmetric processes!real but possibly negative probabilities}) instead of mere probability amplitudes\index{probability amplitudes!complex numbers} (complex numbers with no independent physical meaning).

In any case, it is now apparent that the most important weakness of early time-symmetric interpretations of quantum theory\index{quantum theory!early time-symmetric interpretations} (such as Cramer's\index{Cramer, John} transactional interpretation\index{transactional interpretation!quantum theory}, discussed in section \ref{sec:5.6}) is that they required assuming that the advanced waves\index{advanced waves!handshake process} which are part of a complete `handshake' process propagate backward in time in the same portion of history as that in which the retarded waves\index{retarded waves} propagate forward in time, instead of occurring as part of an independent segment of history\index{independent segments of history!retarded and advanced processes}, which would forbid any interaction with the processes taking place in the retarded segment\footnote{
For those reasons, the time-symmetric interpretation of quantum theory\index{quantum theory!proposed time-symmetric interpretation|nn} proposed here cannot form the basis of a solution to the problem of advanced waves\index{problem of advanced waves|nn}, because, in the present case, we are not dealing with advanced propagation as it could be observed to occur in the same portion of history and this shows, again, that the problem of the absence of advanced waves must be considered independently from the problem of the interpretation of quantum theory\index{quantum theory!problem of interpretation|nn}.}.
 It is important, therefore, to understand that, even though the retarded and advanced portions of history\index{retarded and advanced histories!sharing of macroscopic conditions} share macroscopic experimental conditions and even though their durations also correspond to a certain extent (for a given process), they do not take place simultaneously (even in the opposite order) in the same segment of history (how this can actually be made reasonable will be discussed in section \ref{sec:5.12}).

It is precisely the fact that we are dealing with two different portions of history (not really occurring at the same epoch) that allows the principle of local causality\index{principle of local causality} to be satisfied, despite the fact that the model proposed allows non-local correlations\index{non-local correlations} to arise, because the particles which are propagating in one of those two portions of history do not, in effect, interact with those which are propagating, at the corresponding moment, in the time-reversed portion of history\index{time-reversed portion of history}. As a result, this alternative approach allows to do away with advanced waves\index{advanced waves!classical waves} as real, \textit{classical} waves and this means that, contrarily to the early time-symmetric models, the alternative interpretation of quantum theory\index{quantum theory!alternative interpretation} proposed here is not a particular instance of classical hidden-variable theory\index{classical hidden-variable theories}.

This is certainly a suitable characteristic of the proposed model, because, as I previously mentioned, it is now understood that in order to reproduce the results of certain experiments in which quantum entanglement\index{quantum entanglement} is involved (the EPR-type experiments\index{EPR-type experiments}, which will be discussed in the following section), classical hidden variables would need to violate the principle of local causality through complex and highly unnatural mechanisms. I believe that a similar unnatural coordination of influences, now affecting experimental conditions, would be required if we were to assume, instead, that quantum non-locality\index{quantum non-locality} is an illusion attributable to what has been called `absolute determinism'\index{absolute determinism}, or the idea that every choice of measurement\index{measurement choices!predetermination} is determined in advance as a consequence of deterministic evolution. It is clear to me, indeed, that from a physical viewpoint, this latter proposal merely constitutes the same classical hidden-variable theory in disguise, because, given the sensibility to initial conditions\index{sensibility to initial conditions} that characterize deterministic evolution, the puzzling predetermination which it requires would remain unexplained, unless classical hidden variables\index{classical hidden variables} are involved.

What adds to the difficulties facing all such interpretations is that it was experimentally demonstrated, not so long ago \cite{Zeilinger-1}, that the classical hidden-variable hypothesis is in fact incompatible with the results of certain measurements that can be performed on a single quantum object, for which entanglement is irrelevant. Basically, what those experiments are designed to achieve is a measurement of five pairs of attributes of a photon that is in a state of superposition\index{state of superposition!three position states} of three position states. When the experiments are performed, it emerges that the statistical distribution of measurement results is incompatible with what is allowed in the case where classical hidden variables\index{hidden variables!naive realistic kind} (of the naive realistic kind) determine the outcome of those measurements, because the choice of which pairs of attributes are to be measured affects the outcome of the measurements.

What those results were immediately assumed to imply is that unmeasured attributes of a quantum system\index{quantum system!unmeasured attributes} cannot be considered to exist independently. But it must be clear that, in this particular case, just as in the cases where quantum entanglement\index{quantum entanglement} is involved, what has really been demonstrated is not that there cannot exist a unique reality\index{reality!uniqueness} of some kind in between measurements, but that this unique reality\index{reality!non-classical nature} cannot be classical in nature (it cannot involve a unique retarded history\index{retarded history!unique} and nothing else, for the unobserved physical attributes\index{unobserved physical attribute}). I have already explained, however, why reality cannot be uniquely characterized, in the classical sense, in between measurements and those developments clearly show that it is not necessary to reject the hypothesis of a unique reality for the retarded and advanced portions of a process\index{retarded and advanced processes} (independently), regardless of whether a physical attribute is being measured or not.

In the case at hand, it seems that what is happening is that the different possible measurements are affecting the macroscopic constraints\index{macroscopic constraints!unobserved histories} which are exerted on the unobserved retarded and advanced portions of history and are thus allowed to give rise to different patterns of interference\index{patterns of interference} for some correlated physical attributes, as also occurs in the case of entangled systems\index{entangled systems} (more about this in the following section). But the conclusion that there is no reality, independent of what is revealed by measurements, is not made unavoidable by those experiments, even if it is certainly true that this reality cannot be classical and must be conceived of in accordance with the requirements of time-symmetric causality\index{time-symmetric causality!requirements}.

It is also the fact that reality is not classical, even though it is unique in a certain sense, that allows to explain the otherwise puzzling thought experiments proposed by Yakir Aharonov, Jeff Tollaksen, Sandu Popescu, and their colleagues which are described in Ref. \cite{Aharonov-3}. Those experiments involve sending three electrons on two possible paths\index{three electrons on two possible paths experiment} in an interferometer and then effecting some post selection\index{post selection} (see section \ref{sec:5.6}) on some of the electrons to influence their past states backward in time and in the process give rise to quantum\index{quantum!correlations} correlations between the states of the electrons involved. What is remarkable, here, is that according to quantum theory, even if you send three electrons at a time in the interferometer, no two electrons will ever appear to have gone through the same arm of the interferometer during any single trial, as if it was possible for three electrons to simultaneously go through two possible paths without any two electrons ever going through the same path. But the paradox associated with such a thought experiment only arises when we fail to understand the fact that the trajectory of the electrons is only unique in the sense discussed above.

What would be proved by those experiments (if they were actually performed) is merely that, when none of the particles are directly observed to follow one path instead of another, then no pair of particles can, in effect, be determined to have followed the same path classically, that is, for both the retarded and the advanced portions of the process\index{retarded and advanced portions of process}. But this does not mean that a given pair of particles may not be following the same unobservable paths during either the retarded or the advanced portions of the process, as long as those trajectories actually remain unobserved, which is precisely the outcome of the condition imposed on the final state of the system in the experiments discussed here. No three particles can go through two different paths without two going through the same path, only, particles can go through no specific path during the complete time-symmetric process\index{time-symmetric processes} and obviously, if no particle goes through a single path from a classical viewpoint, then no \textit{pair} of particles can go through a single path either (from the same viewpoint), even if, from a realistic, time-symmetric viewpoint\index{realistic time-symmetric viewpoint}, there are always at least two particles following the same unique, but unobservable path, either forward or backward in time.

Finally, it is also important to mention that, even though the energy signs of the particles present in the advanced portion of history\index{advanced portion of history!particle energy signs} considered here are well-defined relative to the energy signs of the particles present in the retarded portion of history\index{retarded portion of history!particle energy signs}, in the sense that any positive-energy particle\index{positive-energy particles!propagating forward in time} that is observed to be propagating forward in time would be related by the observable macroscopic conditions to a negative-energy particle propagating backward in time\index{negative-energy particles!propagating backward in time} (those assumptions will be justified in section \ref{sec:5.12}), this does not mean that all the particles present in the retarded portion of history would have positive energy signs, while those present in the advanced portion of history would have negative energy signs.

In fact, in each of the two corresponding segments of history there may be both positive- and negative-energy particles propagating in any direction of time and all that we can assess is that the positive-energy particles which are \textit{observed} to propagate forward in time in one of the two portions of history must have negative energy (and positive action) as they propagate backward in time in the corresponding time-reversed portion of history. It should be clear, therefore, that there is no correspondence between the particles present in the advanced portion of history that is required to exist by the time-symmetric interpretation of quantum theory\index{quantum theory!time-symmetric interpretation} proposed here and the invisible negative-action particles\index{negative-action particles!invisibility}, whose physical properties were described in chapter \ref{chap:2} and which actually propagate in the same segment of history as ordinary positive-action particles, even though they also have their own counterparts in the advanced portion of history, as any other matter component.

\section{Quantum entanglement and non-locality\label{sec:5.9}}

Before I turn to the quantum measurement problem\index{quantum measurement problem} and share the most significant insights I have gained while working on the problem of the interpretation of quantum theory\index{quantum theory!problem of interpretation}, I would like to return to the important question of the viability of a realistic description of quantum phenomena in the context of the existence of non-local correlations\index{non-local correlations}. It is possible, in effect, to apply the interpretation which was developed in the preceding section to provide a realistic, yet locally causal description of the processes taking place in the course of an experiment of the Einstein-Podolsky-Rosen\index{Einstein-Podolsky-Rosen experiment} type, involving pairs of entangled photons\index{pair of entangled photons}.

What I will explain is that the experimentally confirmed violation of Bell's inequality\index{Bell's inequality!violation} does not make unavoidable the conclusion that instantaneous action at a distance\index{instantaneous action at a distance} must be an integral aspect of any realistic interpretation of quantum theory\index{quantum theory!realistic interpretation}, in the sense that we are still allowed to assume that no effect can propagate faster than the relativistic speed limit\index{relativistic speed limit!propagation of effects} in the course of the retarded and advanced portions of history\index{retarded and advanced portions of history}. The fact that I'm allowed to actually explain the existence of non-local correlations in such a way is significant, because, contrarily to what is often believed, quantum non-locality\index{quantum non-locality} is not only unexplained from a \textit{classical} perspective, it remains unexplained even in the context of the conventional interpretation of quantum theory\index{quantum theory!conventional interpretation}.

To help visualize the phenomenon of quantum non-locality, we may consider, for example, a simple interferometer experiment\index{interferometer experiment} in which the source, instead of emitting one photon in one direction (for which two possible trajectories would be allowed), would emit a pair of entangled photons\index{pair of entangled photons} which would be allowed to travel in opposite directions, each along one or another of two possible trajectories of equal lengths in which they would meet a mirror (one mirror for each path of a given photon) that would direct them toward a detector (one detector for each photon) that would allow to determine either the presence of interferences between the two paths available to a given photon (by simply detecting the arrival of the photon), or the exact path a photon took on its way to the detector (by measuring from which direction the photon arrived).

What's particular with such an interferometer experiment is that, when we choose to measure from which direction one of the two photons arrived (from which of the two mirrors it reflected) we also, inevitably, determine which path the other photon took in its own otherwise independent part of the experiment, so that, even if we try to measure interferences between the two paths available to this other photon, we do not observe any. This correspondence is made unavoidable by the fact that, when the direction of arrival of the first photon is determined, the direction of arrival of the second photon must be opposite that of the first photon in order that momentum be conserved in the initial state. Thus, it is only when we choose not to determine the exact path taken by \textit{any} of the two photons that we are, in effect, allowed to observe the presence of interferences between the two paths available to each of them.

From this perspective, it is apparent that, when we are dealing with entangled particles\index{entangled particles!entangled phases} (when the phases associated with the propagation of two otherwise independent particles have become entangled as a result of local contact), the choice of whether or not to observe interferences between the multiple paths available to the particles is not made locally, but constitutes a global property of the experiment, because, whenever the direction of arrival is measured for one of the entangled particles, it is no longer possible to measure interferences between the multiple paths available to the other particle.

To be more specific, the presence or the absence of quantum interferences between the multiple trajectories available to one of the two entangled photons\index{entangled photons!interference between multiple trajectories} as it propagates toward its detector is determined by the choice of which measurement is performed on the remotely-located photon in the future. Thus, the trajectory of the first photon must already be either classically well-defined or quantum-mechanically superposed right from the moment when the particle emerges from the initial entangled state\index{entangled state!initial state}, at the source, in order that any measurement that could be performed on that photon produces a result that actually agrees with the observational constraint set by the choice of which measurement will be performed on the second photon in the future.

I believe that what enforces the global character of this choice of measurement is the existence of an advanced portion of history\index{advanced portion of history}. Indeed, whenever we choose to determine the exact path taken by one of the two photons, the result of any measurement performed on the other photon must reflect that choice, because the effect of the measurement performed in the future on the first photon propagates backward in time along this photon's trajectory (in the advanced portion of history) to the initial entangled state (at the source) and then forward in time along the other photon's trajectory (in the retarded portion of history) to affect the measurement result performed on this second photon, even when this measurement is separated from the measurement performed on the first photon by a space-like interval\index{space-like interval}.

As a result, even if the experimental conditions that determine which attribute will be measured by the detectors are changed once the photons have already been emitted, the initial retarded and advanced states will already be such as to reflect that future change and will evolve in accordance with those altered conditions, because the initial retarded states of the two entangled photons\index{entangled photons!initial retarded states} are influenced by the choice of measurements performed at each detector in the future, through the advanced portion of the processes. Thus, the whole experimental setup with which the photons will interact in both the past and the future determines what is allowed to happen to both of them, even as they just leave the source in the forward-propagating version of history, before they interact with a detector. This means that the measurement that is performed on any one particular photon cannot \textit{alone} and under all circumstances determine what happens to both photons, because unidirectional causality\index{unidirectional causality} is not involved in conveying the influence that propagates along the advanced portions of the processes, from each future measurement back into the initial entangled state.

But it is still possible for the two entangled photons\index{entangled photons!unique corresponding trajectories} to have unique, corresponding trajectories in both the retarded and advanced portions of the process\index{retarded and advanced portions of process} under any conditions (such than when one photon goes through the upper path, the other photon always goes through the lower path), only, when those trajectories remain unobservable, they can be different, for the same photon, in the two portions of history. This is why, for any type of EPR experiment, it is the attribute of a particle which is correlated to that of the other particle by the conservation principles applying on the initial state (which in the example discussed here would be the direction of arrival of a photon\index{direction of arrival of photon!EPR-type experiment} associated with its momentum state) that must necessarily be classically determined for one particle, when it is so determined for the other particle.

It must be clear, also, that backward causation\index{backward causation} does not determine what the result of one particular measurement is whenever a measurement is performed on the other particle (this is the outcome of conservation laws), it only determines if interferences can actually be observed at any of the two detectors. This is why information\index{information!faster-than-light communication} cannot be communicated at faster-than-light velocities by making use of quantum entanglement\index{quantum entanglement}, because the outcomes of individual measurements are still random from a local viewpoint and to verify the presence of interferences one would need to repeat the experiment a large number of times, while using different configurations of the interferometer, but this would allow one to tell through which path a given photon went, by noting the time at which it arrived. Therefore, in practice, it is not possible to modulate a signal that would produce immediately recognizable effects in a remote location, by varying the type of measurement performed, even though more subtle non-local correlations\index{non-local correlations} can be observed to arise when one compares the statistical distribution of the outcomes of measurements performed on the two photons.

In all such experiments, interference effects observed at one location (on one particle) depend on the experimental conditions observed at a different location (on a different particle), simply because the changes to the phase of the wave function\index{wave function!phase changes} which arise in the course of those processes are occurring as a result of the boundary conditions\index{boundary conditions!complete time-symmetric process} applying on the complete time-symmetric process and not just on some portion of it associated only with one or the other particle. What those experiments, during which a violation of Bell's inequality\index{Bell's inequality!violation} is observed to occur, have revealed is that it is possible to demonstrate the existence of such non-local correlations\index{non-local correlations}, which cannot be attributed merely to the conditions imposed by conservation principles on the total momentum\index{total momentum!entangled photons} (or polarization state\index{polarization state!entangled photons}) of the two entangled photons, in the context where they are created by pair in the initial state (which merely requires that one photon goes through the upper path when the other goes through the lower path, in either the retarded or the advanced portion of the process\index{retarded or advanced portion of process}, even when those trajectories are not observed and remain classically undetermined).

Anyhow, as soon as the direction of arrival of a photon is measured at one or the other detector in the experiment described above, then it is no longer possible to measure interferences between the two possible trajectories for any of the two photons, because the retarded and the advanced trajectories are then exactly the same in all possible cases and for each photon, which means that there is no phase change for a complete time-symmetric process\index{time-symmetric processes!constructive or destructive interferences} and therefore no constructive or destructive interferences. If it seemed impossible, from a conventional viewpoint, to assume that the entangled photons\index{entangled photons!unique trajectories} propagated along a unique trajectory prior to such a measurement, it is because the measurement is what determines whether interferences will be observed or not, for both particles, and when interferences are indeed present the trajectories of the two photons are no longer well-defined from a classical viewpoint, which has always been interpreted to mean that there is nothing we can say of reality under such conditions.

But what emerges, from the more consistent perspective adopted here, is that it does appear possible to assume that each photon follows a unique, causally independent trajectory\index{causally independent trajectories!entangled photons} as it propagates toward its detector, from the moment when it is emitted by the source and right up to the moment when a measurement is performed on it, contrarily to what would appear to be allowed according to the orthodox interpretation of quantum theory\index{quantum theory!orthodox interpretation}, only, it turns out that in certain cases (when interferences are present) this unique trajectory can be different for the retarded and the advanced portions of the propagation process\index{retarded and advanced processes}, so that it cannot be argued that the photons are always in a \textit{classically} well-defined position state as they propagate toward their detectors. Once the validity of the interpretation proposed here is recognized, it follows that the idea that the concept of a localized particle\index{localized particle concept!quantum entanglement} may no longer be valid in the presence of quantum entanglement and that it should be replaced by a holistic concept of reality\index{holistic concept of reality} at a fundamental level is no longer justified and actually loses most of its appeal.

Those conclusions are certainly appropriate, given that it is not possible, in general, to tell which of two measurements (on one or the other photon) determines the time at which the intermediary states could be considered to no longer interfere and to actually become classically well-defined. I must acknowledge, however, that this is easier to understand in the context of EPR-type experiments\index{EPR-type experiments!linearly polarized photons} involving pairs of linearly polarized photons, in which case it is merely the \textit{difference} between the angles of polarization\index{angles of polarization!measurement} measured by the two detectors that determines if there are interferences or not. The fact that, in the experiment described above, one of the detectors may appear to be privileged in determining the presence or the absence of interferences at both detectors, when only one of the detectors measures the state of the correlated physical attribute\index{correlated physical attribute!entangled particles} of the entangled particles, should not be considered to undermine the validity of the hypothesis that the classical or superposed nature of the trajectories is, in general, determined by the configuration of both detectors.

The crucial point here is that the non-local correlations\index{non-local correlations!absence of backward-in-time signaling} revealed by such experiments can be established without any information carrying signal being sent backward in time. Each measurement unveils the state of a photon at the moment when this measurement takes place, but the choice of which measurement is performed influences the state of the photon as it reaches the source in the advanced portion of history\index{advanced portion of history}, just as when a past state influences a future state, only now backward in time and without entropy increase. But given that, in the above discussed experiments, this past state is an entangled state which results in the two photons sharing a common initial phase\index{common initial phase!entangled photons}, then it follows that the past state of the first photon is also causally influenced by the choice of measurement performed on the second photon in the future, in a way that is not that different from the usual manner by which effects are propagated forward in time, except that information cannot be carried by the effects so produced, given that entropy cannot rise as they propagate in the past direction of time.

This is all made unavoidable in the context where the retarded portion of history\index{retarded portion of history} experienced by the two photons must share the same observational constraints as apply to the advanced portion of history which is experienced by the same two photons (actually two identically prepared photons existing in a corresponding, but different portion of history), for reasons I will discuss in section \ref{sec:5.12}.

What this shows is that, instead of insisting that the wave function\index{wave function!state of knowledge of particular observer} may not be real, or that it merely represents the state of knowledge of one particular observer, which must be actualized on contact with information from another, previously independent observer (as one postulates in the context of an interpretation of quantum theory\index{quantum theory!QBism interpretation} such as `QBism'), we should instead recognize that the wave function does provide our best account of the exact quantum state of a system at any time, but that there are two such states (associated with two actual histories\index{two histories!forward- and backward-in-time evolutions} which both remain undetermined from the viewpoint of observation), one of which is evolving forward in time and the other of which is evolving backward in time.

In such a context, the fact that the wave function\index{wave function!subjective concept} may sometimes appear to be a subjective concept, dependent on whether information concerning the conditions of a future measurement to be performed on a system is available or not, can be seen to be a mere consequence of the fact that we cannot know in advance what the backward-in-time-evolving state\index{backward-in-time-evolving state!absence of advanced knowledge} is, before we obtain information about the outcome of that future measurement, even though it already affects the present state of the system. The adequacy of the latter viewpoint appears to have been confirmed by the fact that the process of actualization of quantum potentialities\index{quantum potentialities!actualization} is now understood to be a consequence of concrete changes that take place in the environment with which a system becomes entangled under very specific conditions (those responsible for triggering decoherence\index{decoherence!specific conditions}), thereby contradicting the hypothesis that it might be a subjective phenomenon.

At this point, it is necessary to mention that I'm aware of the fact that Murray Gell-Mann\index{Gell-Mann, Murray} (among others) once argued that the idea that EPR-type experiments\index{EPR-type experiments!non-locality} imply a certain form of non-locality is merely a distortion of reality, because (so he argued) what occurs when the direction of arrival is measured for one of the two photons is merely that we find ourselves in one particular `branch' of history\index{history!branch selection through measurement}, where both photons happen to follow a definite trajectory. What is problematic here, however, is not merely the fact that such an explanation would depend on the validity of the contradictory notion that a photon goes, at once, through all available paths in many `branches' of history, until a `splitting'\index{splitting of branches!history} takes place and those multiple histories no longer interfere with one another.

The real difficulty, as I explained in section \ref{sec:5.7}, has to do with the fact that, from such a viewpoint, unnatural coincidences would still be observed that would remain unexplained, because if the choice of which measurement is to be performed on one of the two photons affects the outcome of measurements performed on the other photon in a given `branch' of the universe's history and we require the principle of local causality\index{principle of local causality!branch of history} to apply within a particular branch of history, then it is not possible to explain how such a coordination of measurement results would occur, in one such branch of history where the global outcome would, in effect, be observed, unless we adopt the viewpoint I have proposed above, which makes the hypothesis of a multiplicity of branches of history\index{multiplicity of branches of history!unnecessary hypothesis} unnecessary for a solution to the problem of quantum non-locality\index{quantum non-locality!problem}.

The truth is that this rejection of the reality of quantum non-locality\index{quantum non-locality!rejection} is equivalent to the absolute deterministic viewpoint\index{absolute deterministic viewpoint} discussed in the preceding section, given that it requires one to assume that it is possible for pre-existing correlations\index{pre-existing correlations!non-local effects} to exist, which are not attributable to any locally propagated effect. It is quite ironical, therefore, that it was suggested that this viewpoint constitutes an alternative to classical action at a distance\index{action at a distance}, because, as I previously explained, in the context of a realistic interpretation of quantum theory\index{quantum theory!realistic interpretation}, absolute determinism\index{absolute determinism!classical hidden-variable theory} is actually a form of classical hidden-variable theory and therefore must involve faster-than-light signal propagation\index{faster-than-light signal propagation}.

It is now possible to be more specific regarding why it is that we would not be justified to assume that the principle of local causality\index{principle of local causality!violation} is violated when a state vector\index{state vector!reduction} is reduced, in a more general context, as when a photon is emitted by a source, whose propagation is described by an expanding spherical wave function\index{expanding spherical wave function}, after which its presence is detected in one particular location, thereby affecting the wave function over the entire volume. I believe that, if the conclusion that the principle of local causality is violated under such conditions is not unavoidable, even when we acknowledge the fact that the wave function\index{wave function!quantum state} always provides the most accurate description of the state of a quantum system, it is because this phenomenon can be described using the same realistic, time-symmetric approach which I used to explain the origin of quantum non-locality\index{quantum non-locality!realistic time-symmetric approach} as it arises in the case of entangled systems\index{entangled systems} and which involves two histories\index{two histories!opposite directions of time} (independent from the viewpoint of local causality) unfolding in opposite directions of time.

What happens is that the spreading wave function allows to accurately describe the results of any measurement that would reveal the existence of interferences between the multiple paths through which the photon might have traveled as its position remained unobserved and this requires that the wave function does indeed provide the most accurate account of the situation, before a measurement is performed that would reveal along which direction the photon propagated, but only as long as such a measurement is \textit{not}, in effect, performed, because, under such conditions, the retarded and advanced portions of the propagation process\index{retarded and advanced portions of propagation process} might take place along different paths.

However, if a measurement of the direction taken by the photon is effected at some point (before an alternative measurement is performed on the same photon that would allow to reveal an interference between two different trajectories it may have followed), then the photon can nevertheless be considered to have been constrained to follow a unique well-defined trajectory\index{trajectory!unique and well-defined}, all the way back to the emission process (when no macroscopic constraint seemed to apply on the wave function\index{wave function!absence of macroscopic constraint}), as a result of this measurement, given that in such a case the advanced portion of the process\index{advanced portion of process!enforcement of uniqueness condition} is allowed to enforce that condition of uniqueness (imposed on the photon's trajectory by the future measurement), through the effect it exerts on the source, backward in time.

Now, given that what I'm proposing is a time-symmetric interpretation of quantum theory\index{quantum theory!time-symmetric interpretation}, it is important to mention that, in such a context, there must exist a time-reverse analog to ordinary quantum entanglement\index{quantum entanglement!time-reverse analog}, which can be shown to actually give rise to non-local correlations arising from post selection\index{post selection!non-local correlations} (the phenomenon discussed in section \ref{sec:5.6} by which the choice of a measurement to be performed in the future is allowed to influence the evolution of a quantum system backward in time, as originally described in the context of the two-state-vector formulation of quantum theory\index{quantum theory!two-state-vector formulation}). Thus, even in the apparent absence of ordinary quantum entanglement established through local contact in the past, it should be possible to observe the existence of non-local correlations similar to those that may arise in a more conventional context, when it is a future state that is entangled in a certain way as a result of post selection.

From my viewpoint, the fact that those correlations do not allow faster-than-light communication\index{faster-than-light communication} can, once again, be explained as being a consequence of the fact that the constraint of global entanglement\index{global entanglement constraint} discussed in section \ref{sec:4.9} requires entropy to decrease in the past for all processes (occurring in a given universe), which means that no information-carrying causal signal can propagate toward the future and then backward in time to a distant location, as a result of post selection\index{post selection!absence of backward-in-time signaling}, even if causality does operate both forward and backward in time at a more fundamental level.

\bigskip

\noindent It is now possible to reflect on the traditional positions, regarding the significance of EPR-type experiments\index{EPR-type experiments!significance}. If we consider, first of all, the orthodox view\index{orthodox view!quantum entanglement} and Bohr's position\index{Bohr's position} concerning quantum entanglement, it amounted to assume that only the wave function\index{wave function!as reality} can be considered real, given that, when it is not the direction of arrival that is detected for one or another of the two entangled photons\index{entangled photons!direction of arrival}, then it seems to be impossible to say anything logically meaningful about the trajectories of both photons. The problem with this viewpoint is that, if the wave function is considered to be reality itself, then instantaneous action at a distance\index{instantaneous action at a distance} appears to be required, which is why the orthodox interpretation retreated into its idealistic position, according to which it simply doesn't make sense to speak about a reality behind observed phenomena (which may allow to avoid the conclusion that this reality\index{reality!non-locality} is non-local).

The position held by Einstein\index{Einstein's position} and the advocates of a realistic approach\index{realistic approach} was that this rejection of scientific realism\index{scientific realism!rejection} is not acceptable and that quantum theory\index{quantum theory!incompleteness} must simply be wrong or incomplete, given that it appears to require instantaneous action at a distance, when the consequence of a measurement on one of the two photons spreads to its entangled counterpart. Basically, then, one position required assuming that there is no reality, while the other required assuming that there is no entanglement. I believe that both positions were inappropriate in some way, but also accurate in a distinct way\footnote{
It has been argued by some of the originators of the consistent-histories interpretation of quantum theory\index{quantum theory!consistent-histories interpretation|nn} that Einstein\index{Einstein, Albert|nn} was misguided in trying to uphold a certain requirement of scientific realism\index{scientific realism!requirement|nn}, to which the \textit{conventional} interpretation\index{conventional interpretation!quantum theory|nn} of the theory does not conform, because it must be the theory that determines what is true of physical reality\index{reality!contradictory descriptions|nn}, even when it appears to require contradictory descriptions of it to be valid together at the same time. But I believe that it is rather this position which is misguided and this precisely because it constitutes an attempt at limiting what can be consistently described of reality in order to satisfy the perceived requirements of what is merely an inadequate or incomplete interpretation of the theory.}.
 Clearly, quantum theory and quantum entanglement\index{quantum entanglement!inevitability} are there to stay, but what I have tried to explain is that scientific realism\index{scientific realism!necessity} is not optional either and can be accommodated without requiring instantaneous action at a distance, when time-symmetric causality\index{time-symmetric causality!essential element} is recognized to be an essential element of this physical reality and the appropriate entropy-reducing constraint applies to the advanced portion of every quantum process\index{advanced process!entropy-reducing constraint}.

\section{The quantum measurement problem\label{sec:5.10}}

It is usually recognized that the two main conceptual difficulties with which we are faced when trying to formulate a consistent interpretation of quantum theory\index{quantum theory!consistent interpretation} are the existence of non-local correlations\index{non-local correlations} and the absence of an objective criterion for judging when it is that the multiple interfering potentialities\index{interfering potentialities!actualization} characterizing the state of some unobserved dynamic attribute\index{unobserved dynamic attribute} of a quantum system are actualized to a unique value, as happens when a measurement of this attribute is performed. In the preceding sections I have offered a viable solution to the problem of quantum non-locality\index{quantum non-locality!problem}, in the context of a realistic interpretation of quantum theory\index{quantum theory!realistic interpretation} based on the requirement of time-symmetric causality\index{time-symmetric causality}. But while progress was achieved in the last few decades in identifying the conditions necessary for the decoherence process\index{decoherence process} to occur, it remains that we haven't yet been able to determine exactly what is responsible for the persistence of quasiclassicality\index{quasiclassicality!persistence} that is observed to characterize the evolution of quantum systems\index{quantum system!entanglement with environment} when they become entangled with their environment, following a measurement.

The currently favored approach for a solution to the quantum measurement problem\index{quantum measurement problem} plays a role that is much the same as the late nineteenth-century approaches which were adopted in an attempt to solve the problem of the origin of thermodynamic time asymmetry\index{thermodynamic time asymmetry!problem of origin} through the use of statistical methods. Indeed, at some point, Boltzmann\index{Boltzmann, Ludwig} thought that he had solved the problem of the origin of the thermodynamic arrow of time\index{thermodynamic arrow of time!problem of origin}, because he had achieved significant progress in identifying its true origin. But as we now understand, it appears that he had not really provided a satisfactory explanation and that the remaining difficulties had not even been clearly identified.

Today, it is widely believed that the problem of quantum measurement has been solved by the recent advances achieved in identifying the conditions necessary for the phenomenon of decoherence\index{decoherence!phenomenon} to occur, while in fact this is not entirely correct, precisely because the consequences of thermodynamic time asymmetry on the evolution of quantum systems haven't yet been properly assimilated. This is the problem I will attempt to circumscribe in this section and to which I will be able to provide a satisfactory solution in section \ref{sec:5.12}. This will allow me to confirm, once again, that a realistic approach, according to which there must exist a unique reality of some kind, independently from whether or not a system is being observed, is not incompatible with the empirical evidence that singles out quantum measurement\index{quantum measurement!factual definiteness of reality} as the necessary condition for the factual definiteness of reality.

Originally, the quantum measurement problem\index{quantum measurement problem} had to do with the difficulty we were experiencing in trying to identify the exact nature of the conditions that give rise to the actualization of quantum potentialities\index{quantum potentialities!actualization}. In fact, the linearity of the equation that describes the evolution of the state vector\index{state vector!evolution} made it difficult to understand how it could be that quantum interferences\index{quantum interferences!vanishing} are, in effect, allowed to vanish for the dynamic attribute of a system under observation, so that they can give rise to a definite set of outcomes, to which meaningful probabilities\index{probabilities!meaningful} can be ascribed.

Thus, there appeared to be a conflict between observations, which indicate that quantum potentialities are actualized to definite non-interfering outcomes following what we call a measurement, and the theory itself, which seems to require quantum superposition of states\index{quantum state superposition!persistence} to persist indefinitely. From a conventional perspective, it would appear that when each possible final state of a quantum process becomes correlated with one possible state of a measuring device, if the quantum system was in a state of superposition of the observable concerned at the time when this correlation was established, then the whole measuring device\index{measuring device!state of superposition} should also be found in a state of superposition following the moment at which the measurement took place.

One of the earliest attempts at solving this quantum measurement problem\index{quantum measurement problem} became the actual justification for the conventional interpretation of quantum theory\index{quantum theory!conventional interpretation}, according to which all possible histories\index{histories!different branches} occur all at once as different interfering `branches'. What was proposed by Hugh Everett III\index{Everett III, Hugh} is that there is no actualization process\index{actualization process}, but that the superposed macroscopic states\index{superposed states!measuring apparatus} of a measuring apparatus, which become correlated with the multiple interfering states of a quantum system\index{quantum system!interfering states}, all exist simultaneously in parallel versions of history, while, for some reason, a `splitting' occurs following measurement\index{measurement!splitting of branches}, which is responsible for the fact that the multiple branches of history no longer interfere with one another.

The difficulty with this proposal, however, does not have to do only with the fact that it may allow logical contradictions to arise in the context of a realistic interpretation of quantum phenomena\index{quantum phenomena!realistic interpretation} (a particle could be in one location as well as in another, in the very same portion of history), it also has to do with the fact that, if all branches are followed together, then there is no \textit{a priori} reason why there could not be branches where a measuring device\index{measuring device!macroscopic state superposition} is in a state of superposition of macroscopic observables (at least in the context where decoherence\index{decoherence} alone would not be sufficient to allow one to expect that the absence of interferences that follows a measurement should persist indefinitely).

Nevertheless, the idea endured and was later revived when it was discovered that, under specific conditions, the phenomenon of decoherence\index{decoherence!phenomenon} must give rise to a diagonalization of the reduced density operator\index{diagonalization of reduced density operator} in the basis of the attribute under measurement (the probabilities of occurrence for superposed measurement results\index{superposed measurement results!probabilities of occurrence} must rapidly decrease to negligible values upon measurement of a given observable), which would appear to legitimate the splitting branches hypothesis\index{splitting branches!hypothesis}.

But if there should be no doubt that the discovery of decoherence itself was a step in the right direction, this does not mean that the validity of the hypothesis that there exist many continuously `splitting' branches\index{splitting branches!history} of history has been confirmed by those developments. Given that decoherence does not require the existence of those multiple branches of history, it appears that the only advantage that Everett's\index{Everett III, Hugh} original proposal \textit{might} provide does not really have to do with solving the quantum measurement problem\index{quantum measurement problem}, but with allowing one to avoid having to explain what determines the unique outcomes of quantum measurements\index{quantum measurement!determination of unique outcome} performed on a previously interfering physical attribute\index{interfering physical attribute!measurement}, in the context where the formalism of quantum theory appears to require that all possible histories occur all at once for the attributes of a quantum system which are not under measurement.

The actual purpose of the multiple-branches hypothesis\index{multiple-branches hypothesis}, therefore, is to allow one to avoid having to postulate the existence of distinct dynamical laws that would apply only during processes that can be qualified as measurements, given that, if it was possible to assume that all histories occur all at once, not only before, but also after a measurement has been performed, then it would no longer be necessary to explain what determines the one unique measurement result that appears to be actualized among the many different potentialities. Thus, it is argued that when all the superposed states\index{superposed states!actualization} of some physical attribute are actualized together in different splitting `branches' of the same history, there no longer needs to be a cause (of unknown origin) that would select from among all parallel histories\index{parallel histories!selection of measurement outcome} the one particular outcome that is actually observed following measurement.

But given that I have already argued, based on more general considerations, that it is not really necessary, in order to explain the existence of quantum interferences\index{quantum interferences}, to assume that all possible histories\index{histories!occurring all at once} are followed all at once in the course of each and every quantum process, then it would appear that it is preferable to recognize that the unique reality\index{reality!uniqueness} we do observe during measurements is a reflection of the uniqueness of the non-classical (time-symmetric) reality\index{time-symmetric reality!uniqueness} that exists in between measurements, instead of trying to argue that there must be a multiplicity of measurement results\index{measurement results!multiplicity}, which we do not observe, that would correspond with a multiplicity of histories\index{histories!multiplicity}, which we cannot observe either, but that would need to exist in between measurements.

What must be clear is that it is only when we ignore the requirement to adopt a suitable, realistic interpretation of quantum processes\index{quantum processes!realistic interpretation} that the uniqueness of history\index{history!uniqueness} actually constitutes an additional difficulty. But what's even more significant is that, as I will explain below, it appears that decoherence\index{decoherence} is not sufficient, all by itself, to predict that a physical attribute which was once measured should remain in a definite quasiclassical state\index{quasiclassical state} and should not give rise to quantum interferences\index{quantum interferences!macroscopic attributes} involving macroscopic attributes of a measuring device. Therefore, it seems that we should still expect that in some of the simultaneously occurring decoherent branches of history\index{decoherent branches of history}, macroscopic state superpositions\index{macroscopic state superposition} would develop at some point.

What I find most difficult to accept regarding the many-worlds interpretation\index{many-worlds interpretation!quantum theory}, however, is the fact that we are required to believe that the unique character of reality\index{reality!uniqueness} that we do observe on a classical scale is just an illusion, while we are also expected to assume that the hypothesis of a multiplicity of coexisting branches of history\index{coexisting branches of history}, which has never been directly confirmed by any observation, is valid under all circumstances. In other words, we are required to assume that what we see is not the true reality, while what is a mere hypothesis, that cannot yet be considered absolutely unavoidable, must be considered true, even though it is clearly incompatible with what we do know about reality.

The real problem, as I see it, is that even when we assume that, in the absence of measurement performed on a certain dynamic attribute, there coexist multiple branches of history\index{multiple branches of history!causal independence}, if those branches are really independent from the viewpoint of local causality, so that they may actually constitute many different universes, then their existence are of no use in trying to explain quantum interferences\index{quantum interferences!alternative versions of history} between alternative versions of history, because in such a case it becomes necessary to explain those interferences as being an outcome of constraints which would apply to only one single, independent history and which would merely require it to unfold in such a way as to allow for the existence of quantum interferences\index{quantum interferences} between all the possible ways it \textit{could} have happened.

It should be clear that there is nothing wrong with the hypothesis that there may exist a multiplicity of causally independent universes\index{causally independent universes}. In fact, there may be good reasons to recognize the validity of this particular hypothesis (which is not dependent on the validity of the many-worlds interpretation of quantum theory\index{quantum theory!many-worlds interpretation}), in the context where the weak anthropic principle\index{weak anthropic principle} appears to constitute the only possible explanation for certain otherwise unlikely properties of our universe. But while it may not be possible to reject the hypothesis that an infinity of causally independent universes exist in parallel, it must be clear that, all by itself, this hypothesis is not sufficient or adequate to solve the quantum measurement problem\index{quantum measurement problem} and to explain what determines the unique outcome that is obtained from any measurement\index{measurement!unique outcome} performed on a previously interfering physical attribute.

I'm aware, though, that it has been argued by Heinz Dieter Zeh\index{Zeh, Heinz Dieter} \cite{Joos-1} that the multiple-branches hypothesis\index{multiple-branches hypothesis!inevitability} may be unavoidable if one does not want to have to modify quantum theory, because the only way one can avoid having to assume that a unique state of an unobserved attribute existed before decoherence\index{decoherence} took place, that would merely have been revealed by the measurement, is by postulating that all histories occur all at once in the course of any one quantum process, which, therefore, allows all possible outcomes to themselves be actualized all at once as different branches, upon measurement.

It is known, in effect, that, for various reasons, a quantum measurement\index{quantum measurement} cannot be considered to simply consist in acquiring knowledge about a unique, pre-existing state of some unobserved attribute. But I have explained, in section \ref{sec:5.8}, that, if we cannot assume that a unique reality\index{reality!uniqueness} existed before a measurement was performed on some unobserved attribute of a quantum system, it is only because we usually assume a unique reality to be unique in the classical sense, while in fact the unique reality that would characterize a quantum process (in the absence of measurement on a certain dynamic attribute\index{dynamic attributes!absence of measurement}) is of a time-symmetric nature and involves both a unique retarded state\index{retarded state!uniqueness} and a unique and possibly different advanced state\index{advanced state!uniqueness} at all times, which guarantees the consistency of past evolution\index{past evolution!consistency with any future measurement} with \textit{any} future measurement and which requires all possible intermediary states to contribute to the final probability amplitude\index{probability amplitudes!contribution of intermediary states}, so that the future measurement does not allow to reveal a unique \textit{classical} path through which the system would have propagated.

This is why, despite what Zeh\index{Zeh, Heinz Dieter} suggests in another publication \cite{Zeh-1}, the fact that there would exist a unique, but unknown state prior to measurement\index{measurement!unique prior state} of an unobserved attribute does not violate the condition imposed by the Von Neumann equation\index{Von Neumann equation} (the quantum-mechanical generalization of Liouville's equation\index{Liouville's equation}) that ensemble entropy\index{ensemble entropy!absence of decrease} should not decrease during measurement. It is, therefore, simply the fact that future measurements\index{future measurement!influence on past states} also exert an influence on past evolution (as when a system is submitted to post selection\index{post selection}) that forbids one from assuming that an unmeasured physical attribute may have existed in a unique classical state\index{classical state!independence from future measurements} in the past, that would not depend on what measurement is performed on this attribute in the future. But it must be clear, once again, that this does not prevent a unique state from having actually existed at all times in the retarded and advanced portions of history\index{retarded and advanced histories} and therefore the conclusion that the uniqueness of measurement results\index{measurement results!uniqueness} can only be explained by postulating that all histories\index{histories!occurring all at once} are occurring all at once cannot be considered valid.

In any case, if quantum reality\index{quantum reality!unique time-symmetric type} was not of the unique time-symmetric type and the decoherent branches hypothesis\index{decoherent branches hypothesis} was assumed to alone provide a solution to the quantum measurement problem\index{quantum measurement problem}, then an alternative explanation of quantum non-locality\index{quantum non-locality!alternative explanation} would have to be found, as it cannot be provided by Everett's interpretation\index{Everett's interpretation} (as I explained in section \ref{sec:5.7}) and this is an additional difficulty for the conventional approach. Thus, I think that I have explained with enough clarity why it is that the frequently stated conclusion that it is just as difficult to provide decisive arguments in favor of the many-worlds interpretation\index{many-worlds interpretation!validity arguments}, as it is to provide arguments that would invalidate the idea, is not well founded, because the hypothetical, multiple branches of history\index{multiple branches of history} are not necessary, or even adequate to explain quantum strangeness\index{quantum strangeness}, while they also do not appear to be required to solve the quantum measurement problem (especially in the context where it is recognized that the splitting process\index{splitting process!multiple branches} would not, all by itself, allow one to avoid the difficulty associated with the non-locality of state vector reduction\index{state vector reduction!non-locality}).

Now, some theoreticians are worried about the fact that decoherence\index{decoherence!quantum cosmological context} would seem to be insufficient to solve the problem of the actualization of quantum potentialities\index{actualization of quantum potentialities!problem}, when we are considering the system under observation to be the universe as a whole (as becomes necessary in a quantum cosmological context). The problem they see is that, in such a case, there would be no outside environment degrees of freedom\index{environment degrees of freedom} to effect decoherence, while this is known to be a requirement under ordinary conditions. What constitutes a more serious difficulty, however, is that, from the viewpoint of the currently favored approach (the consistent-histories interpretation of quantum theory\index{quantum theory!consistent-histories interpretation}), decoherence is insufficient to explain the persistence of quasiclassicality\index{persistence of quasiclassicality!decoherence}, not just in the cosmological case, but even under more general circumstances and on a much smaller scale, as was first pointed out by Fay Dowker\index{Dowker, Fay} and Adrian Kent\index{Kent, Adrian} \cite{Dowker-1}.

This is not just a consequence of the fact that (ignoring my own contribution) we do not yet have a valid explanation for the irreversibility that characterizes the processes which give rise to decoherence\index{decoherence!irreversible processes}. It rather appears to be a basic insufficiency of the current approach, which is apparent in the fact that it fails to allow classical evolution to persist following decoherence\index{decoherence!persistence of classical evolution}, even when irreversibility is assumed to characterize the evolution of the environment degrees of freedom\index{environment degrees of freedom!irreversible evolution} without explanation, due to some boundary condition of low entropy that presumably applies to the initial Big Bang state\index{initial Big Bang state!low-entropy boundary condition} (I will return to this question below). An appropriate solution to the quantum measurement problem\index{quantum measurement problem}, therefore, must allow to predict the emergence of quasiclassicality\index{emergence of quasiclassicality!cosmological scale}, not just on the cosmological scale, but also on the much smaller scale of measuring devices, where the conventional approach is insufficient as well.

In any case, it is my intention to demonstrate that it is not necessary, in order to explain the nature of the outcomes of quantum measurement, to postulate the existence of distinct (perhaps fundamentally irreversible) dynamical laws that would apply only during a process that could, in effect, be characterized as a quantum measurement\index{quantum measurement!distinct dynamical laws}. Thus, while I do agree with the most knowledgeable experts that quantum theory, as it is currently interpreted, fails to explain the persistence of quasiclassicality\index{persistence of quasiclassicality!current interpretation} that is observed to follow any measurement, I do not believe that what is required in order to address this difficulty is a modification of the basic mathematical framework of the theory that would need to apply whenever measurements are performed, as was once proposed. It is at the level of interpretation that the appropriate solution will emerge that will allow us to solve the remaining difficulties surrounding quantum measurement\index{quantum measurement!remaining difficulties}. As I will explain in section \ref{sec:5.12}, what the current theory needs is not so much a modification of its structure, as an extension of its meaning.

It must be recognized, however, that the distinctive feature of all processes that can be characterized as giving rise to a measurement is indeed temporal irreversibility\index{temporal irreversibility!measurement processes}. A quantum measurement\index{quantum measurement!entanglement with environment} is nothing but the entanglement of a particular state of some attribute of a quantum system with some distinguishable, macroscopic degree of freedom of its environment whose future evolution is irreversibly influenced by this particular event. The fact that no quantum interference\index{quantum interferences!irreversibly evolving systems} is ever observed for irreversibly evolving systems indicates that the non-superposed nature of measurement results is related to the irreversible character of the measurement process. Thus, decoherence\index{decoherence!loss of phase coherence} itself (literally the loss of phase coherence) can only occur when a microscopic (quantum-mechanically evolving) system becomes entangled with some irreversibly evolving (entropy increasing) processes taking place in its environment (usually involving dissipation\index{dissipation}), so that the phase relations\index{phase relations!delocalization} that could have given rise to interferences become delocalized and are assumed to no longer be accessible to observation, as is already well understood. In fact, the ultimate manifestation of irreversibility appears to be decoherence\index{decoherence!irreversibility} itself.

It should not be unexpected, therefore, that all quantum measurements\index{quantum measurement!formation of record} involve the formation of a record, given that for a record of some past event to form, entropy must be growing in the future. Indeed, the formation of a record\index{formation of records!multiple effects from single cause} merely consists in the production of multiple persistent and somewhat independent effects in the future, which are all the outcomes of one single identifiable cause in the past and this is undoubtedly a process that is asymmetric with respect to the direction of time. What this means is that there is something very tangible occurring when a quantum measurement is performed and therefore, if it is true that our knowledge of a quantum system is altered when quantum potentialities\index{quantum potentialities!actualization} are actualized, it would not be appropriate to assume that the changes which are taking place in the course of a quantum measurement\index{quantum measurement!non-subjective changes} are subjective changes, because, following measurement, the observed physical attribute is no longer unique merely in the time-symmetric quantum mechanical sense, but acquires the same unique value in both the retarded and the advanced portions of history\index{retarded and advanced portions of history}.

It is not difficult to show that time irreversibility\index{time irreversibility!condition for measurement} is essential for a measurement to occur, while the mere complexity, or the large number of independent degrees of freedom of a macroscopic system with which a quantum system can become entangled, alone, is not sufficient a condition for triggering a measurement. It is, in effect, apparent in the formalism of quantum field theory\index{quantum field theory!formalism} that a near infinite amount of structure must be taken into account in estimating the probability of any process, given that additional fermion loops\index{fermion loops} and radiative correction terms\index{radiative correction terms} arise, at every level of approximation, on shorter scales. If we were to consider this small-scale complexity to provide the conditions for quantum measurement\index{quantum measurement!irrelevance of complexity} to take place, then it should be the case that the world would be quasiclassical down to a much smaller scale, given that all the complexity that is present at higher energies (and which can only be ignored as a result of the validity of the renormalization procedure\index{renormalization procedure}) would allow measurement to take place long before a quantum system even has the chance to become entangled with a macroscopic system.

I have provided strong arguments, in section \ref{sec:4.7}, to the effect that in the absence of matter there can be no persistent microscopic structure in the distribution of vacuum energy\index{vacuum energy!absence of persistent microscopic structure} and this means that no record of what takes place in the vacuum on smaller scales can exist, unless we directly reveal the existence of those processes by entangling them with irreversibly evolving degrees of freedom which leave persistent traces of their occurrence and this allows to confirm that the complexity of virtual processes\index{virtual processes!complexity} cannot be considered sufficient as a condition for quantum measurement to take place. When no irreversible change that could potentially carry information about a former state is allowed to take place, the predictions of quantum theory do not merely apply as much for the future as for the past, they do not apply at all, because there can be no measurement\index{measurement!irreversible amplification of effects}, that is to say, no irreversible process of amplification of the effects of one or another outcome of the unobserved evolution of a microscopic quantum system. From those considerations one can only conclude that the defining characteristic of the processes that allow quantum measurement\index{quantum measurement process!defining characteristic} to happen is not merely their complexity, but really their irreversibility.

This time asymmetry must not be confused with that which also characterizes the otherwise time-reversible `unitary' evolution\index{unitary evolution!time asymmetry} that takes place in between measurements and which is made conspicuous by the fact that the predictions of quantum theory\index{quantum theory!prediction of future evolution} are only valid for future evolution. The impossibility to accurately `predict' the past arises as a consequence of the fact that only a subset of states can be actualized in the past, due to the constraint of diminishing entropy\index{constraint of diminishing entropy!past direction of time} that exists for this direction of time and which also applies to classical evolution. It is the fact that no such a constraint applies on future evolution that allows predictions of future transition probabilities\index{transition probabilities} to be valid, while predictions of transition to past states do not apply in general. In section \ref{sec:4.9} I have explained that this constraint arises from the requirement that there exist relations of causality between all particles present in the initial state of maximum matter density of the Big Bang\index{Big Bang!initial state of maximum matter density}, which in the presence of negative-energy matter implies that gravitational entropy\index{gravitational entropy!minimum initial value} must have been minimal initially, thereby requiring the number of available microscopic states\index{microscopic states!availability} compatible with the known macroscopic conditions to be continuously decreasing in the past direction of time.

But while the time asymmetry that characterizes all measurement processes\index{measurement process!time asymmetry} has the same origin, it is a distinct phenomenon, that usually operates on a much shorter time scale and that does, in effect, give rise to a reduction of the state vector\index{state vector!reduction}. Yet, it is appropriate to remark that it is the global entanglement constraint\index{global entanglement constraint} defined in chapter \ref{chap:4} that actually explains the fact that decoherence\index{decoherence} is allowed to occur, which is necessary (even though not entirely sufficient) to explain the persistence of the quasiclassical nature of history\index{quasiclassical nature of history!persistence} that follows quantum measurements\index{quantum measurement}. In fact, this is the only explanation of time asymmetry\index{time asymmetry!explanation} that allows to deduce (rather than merely assume) that decoherence always occurs in one and the same direction of time for all measurement processes (as required for the logical consistency of history\index{logical consistency of history!requirements} according to Roland Omn\`{e}s\index{Omn\`{e}s, Roland} \cite{Omnes-1} (chap. 20, sec. 173)), as decoherence itself does not \textit{a priori} favor one direction of time over the other.

But when a measurement is performed on a dynamic physical attribute\index{dynamic physical attributes!interfering state} of a quantum system in an interfering state, what must also happen is a variation of the macroscopic constraints which apply on the wave function\index{wave function!macroscopic constraints} that describes the evolution of the system. When this is allowed to occur and the state vector\index{state vector!reduction} is reduced, an irreversible change is introduced in the evolution of the system itself and the outcome of this evolution is, in general, unpredictable (even if the wave function\index{wave function!deterministic evolution} itself always evolves deterministically in between measurements). But in the context of a realistic interpretation of quantum theory\index{quantum theory!realistic interpretation}, this cannot be understood to mean that it is the evolution that takes place in the course of a measurement\index{measurement!cause of quantum unpredictability} which is alone responsible for giving rise to the unpredictability of quantum phenomena, as is sometimes proposed.

If the state of some interfering dynamic attribute\index{interfering dynamic attribute!unique time-symmetric state} of a quantum system is unique, in the time-symmetric sense, before a measurement is performed, as I previously argued one must recognize, then it certainly cannot be assumed that the randomness of its evolution is merely a consequence of the events that take place during the subsequent measurement and it becomes necessary to admit that it is the unobserved paths\index{unobserved paths!random determination} followed forward and backward in time by the system as it approaches or emerges from the event at which a measurement is performed which are randomly determined and that this is what explains the unpredictability of the outcome of this measurement\index{measurement outcomes!unpredictability}.

It must be clear, in any case, that the randomness of quantum processes\index{quantum processes!randomness and uniqueness}, like their uniqueness, is not an illusion that emerges from the fact that an observer may be unable to perceive the evolution that supposedly takes place in all possible ways, all at once, in multiple branches of history\index{multiple branches of history}, as is sometimes suggested in the context of a many-worlds interpretation of quantum theory\index{quantum theory!many-worlds interpretation}. Randomness is a fact of the reality we experience that becomes perfectly acceptable in the context of a realistic time-symmetric interpretation of quantum theory\index{quantum theory!realistic time-symmetric interpretation}, where the deterministically-evolving wave function\index{wave function!deterministic evolution} is not reality itself and there exists a unique history\index{history!uniqueness} of some kind, even in between measurements.

Thus, if randomness becomes apparent only during quantum measurements\index{quantum measurement!randomness}, it is simply because it is only as a result of processes which can be characterized as measurements that the uniqueness of reality\index{uniqueness of reality!time-symmetric sense} (in the time-symmetric sense) is made apparent, while it is only at the level of individual histories\index{individual histories!unpredictability} that reality may be observed to vary unpredictably (given that the wave function itself evolves deterministically). But quantum evolution\index{quantum evolution!randomness} must be understood to always be random, even though in the absence of measurement, or when the macroscopic constraints applying on a system remain unchanged, this unpredictability\index{unpredictability!absence of observable consequences} is not apparent, because it has no observable consequences.

Once again, therefore, it seems that it is incorrect to assume that a fundamental distinction must exist between the `unitary' evolution\index{unitary evolution} that takes place in between measurements and the evolution that characterizes a process during which quantum potentialities\index{quantum potentialities!actualization} are actualized and this means that it should be possible to explain the quasiclassical nature of the evolution that follows a quantum measurement\index{quantum measurement!quasiclassical evolution} while remaining within the confines of the current mathematical framework of quantum theory\index{quantum theory!current mathematical framework}. The difference between observed and unobserved evolution is real, but only because the conditions that exist when there is an absence of knowledge provide a quantum system with more freedom regarding what it is allowed to do as it randomly evolves in the retarded and advanced portions of history\index{retarded and advanced histories!random evolution}.

It also transpires that the standard account, regarding the distinction between those situations in which a measurement takes place and those in which the usual `unitary' evolution law applies, is somewhat misleading, because, in fact, a quantum system is always in a state where at least one dynamic attribute\index{dynamic attributes!classically well-defined state} (as unnatural as it may be) is in a classically well-defined state, even though this means that the conjugate attribute\index{conjugate attributes!complete indetermination} is completely undetermined. This is a very important fact that is often overlooked and which actually holds the key to a solution to the remaining issues that prevent the formulation of a satisfactory explanation of the persistence of quasiclassicality\index{persistence of quasiclassicality}.

When a quantum measurement is performed, all that really happens is that the observable macroscopic constraints\index{observable macroscopic constraints} applying on the state of a system changes in such a way that an attribute (say position) which could have evolved differently in the retarded and advanced portions of history\index{retarded and advanced portions of history} before the measurement took place, must now follow the same evolution in both portions of history, while its conjugate attribute\index{conjugate attributes} (say momentum), whose value must have been the same in the retarded and advanced portions of history before the measurement was performed, can now have different unobservable values in the two portions of history. In such a context, it would certainly be inappropriate to argue that a \textit{fundamental} change occurs in the course of a process that can be qualified as a measurement, even though it is clear that some constraint, not present before the process took place, does, in effect, become significant for the future evolution of the attribute of the system which is subjected to measurement (I will have more to say concerning this issue in section \ref{sec:5.12}).

Now, the modern formulation of quantum theory\index{quantum theory!modern formulation}, the one which can most naturally accommodate the decoherence process\index{decoherence process}, is usually considered to be that of consistent histories\index{consistent histories!formulation of quantum theory}, which was developed in three steps by Robert Griffiths\index{Griffiths, Robert} \cite{Griffiths-1}, Roland Omn\`{e}s\index{Omn\`{e}s, Roland} \cite{Omnes-2}, and Murray Gell-Mann\index{Gell-Mann, Murray} and James Hartle\index{Hartle, James} \cite{Gell-Mann-1}. From this formalism emerges an interpretation of quantum theory\index{quantum theory!interpretation} according to which it is merely the fact that one may choose to ignore certain aspects of reality, and submit them to a summation process\index{summation process!ignored aspects of reality}, that allows one to obtain meaningful probabilities\index{meaningful probabilities} (which are positive and which add up to one) for the possible histories of a quantum system\index{quantum system!possible histories} which has become entangled with the summed-over portion of reality. It would then merely be the fact that one may choose to ignore what goes on in the environment with which a quantum system\index{quantum system!entanglement with environment} has become entangled that would allow one to find the system to be in a mixed quantum state\index{mixed quantum state}, instead of a pure quantum state\index{pure quantum state} for which interferences would be observed, following measurement.

More specifically, what the formalism of consistent histories\index{consistent histories!formalism} provides is a criterion for judging when it is that sufficiently coarse-grained histories\index{coarse-grained histories!absence of quantum interferences} are obtained (by ignoring certain details of the historical description of reality), which do not interfere with one another and which can therefore be attributed meaningful probabilities. Interestingly, the manner by which this is achieved is by considering pairs of coarse-grained histories\index{pairs of coarse-grained histories} (consisting of sets of alternative fine-grained histories\index{sets of alternative fine-grained histories} whose ignored details are allowed to differ in any possible way) which are subjected to decoherence\index{decoherence!pairs of coarse-grained histories} and between which there are virtually no interferences. When those conditions are satisfied, a meaningful probability for the process so described to occur can be obtained by applying the usual rule, which consists in multiplying the probability amplitude for a history\index{probability amplitudes!coarse-grained history} with the complex conjugate of the probability amplitude\index{probability amplitudes!complex conjugate} for the same coarse-grained history. But no interpretation is given for why it might be necessary to consider pairs of coarse-grained histories rather than single histories, even though this appears to be required from a mathematical viewpoint.

What the formalism of consistent histories provides is an improved definition of quantum measurement\index{quantum measurement!improved definition} as taking place continuously over the entire duration of a process, rather than at one particular event. This becomes possible as long as the \textit{local} environment degrees of freedom\index{environment degrees of freedom!irreversible evolution} which are left out of the description of the process evolve irreversibly, thereby allowing decoherence to arise. One of the advantages of such a viewpoint is that it is easier to see how it can be that the simple possibility for an event to happen allows a measurement to be performed, even if this event does not happen (as in the case of interaction-free measurements\index{interaction-free measurement}), because when something is, in effect, allowed to happen we simply are in a situation where one specific set of macroscopic experimental constraints\index{macroscopic experimental constraints} exists throughout the duration of a process, which would not exist otherwise, while different constraints mean a different measurement, not an absence of measurement.

But while the consistent-histories approach\index{consistent histories!approach} is certainly well-founded all by itself, given that it allows one to avoid having to refer to classical observers and classical measuring devices\index{classical observers and measuring devices} that would not be describable using the formalism of quantum theory\index{quantum theory!formalism}, it appears to be insufficient to predict the emergence of a classical world\index{emergence of classical world} (a maximum quasiclassical domain\index{maximum quasiclassical domain}). It is as if decoherence\index{decoherence!insufficiency of constraint} alone was not enough constraining a condition to guarantee an absence of quantum interferences between all the coarse-grained histories\index{coarse-grained histories!absence of quantum interferences} to which it may give rise, while no criterion currently exists to select as physically relevant only those future histories which actually describe a quasiclassical evolution\index{quasiclassical evolution!physically relevant histories}. As was the case with the original many-worlds interpretation of quantum theory\index{quantum theory!many-worlds interpretation}, it is not possible to avoid the conclusion that, in the course of certain otherwise `consistent' histories\index{consistent histories!superposition of macroscopic states}, a macroscopic measuring device may end up in a superposition of states, after becoming entangled with a quantum system\index{quantum system!entanglement with measuring device}.

There are, then, two problems affecting the consistent histories interpretation of quantum theory\index{quantum theory!consistent histories interpretation}. The first problem one must face has to do with the previously discussed lack of motive for justifying the application of the criterion of `consistency'\index{criterion of consistency} that is attributed to families of coarse-grained histories\index{families of coarse-grained histories} and according to which certain histories\index{histories!meaninglessness} would simply be meaningless, given that classically meaningful probabilities\index{classically meaningful probabilities} cannot be assigned to them. I have already mentioned that it appears preferable to allow our conception of reality to adapt to the fact that classical probability theory\index{classical probability theory} does not always apply, instead of trying to limit what may be consistently described of this reality through some arbitrary criterion that only serves to accommodate the limitations and the inadequacies of an interpretation that cannot fully satisfy the requirement of a realistic, time-symmetric description of reality\index{realistic time-symmetric description of reality}. Thus, I believe that it is important not to commit the error of enforcing consistency at the price of rejecting a realistic interpretation of facts, which would simply contribute to perpetuate the difficulties which are known to affect the original Copenhagen interpretation of quantum theory\index{quantum theory!Copenhagen interpretation}.

What should be recognized as nonsense is not the hypothesis that a photon follows a unique but unobservable trajectory of some kind in between measurements, but the decree that we should not even try to describe reality in situations where we do not yet know how to make sense of it. This reflection is especially relevant given that, in a quantum mechanical context, we are always dealing with probabilistic inferences, so that even histories which we may expect to be `consistent' might in some rare circumstances turn out to be `nonsense', which is certainly indicative of the arbitrariness of the restrictions imposed by the consistent-histories interpretation of quantum theory on our concept of reality. Therefore, to achieve further progress regarding the problem of quantum measurement\index{quantum measurement problem}, one must first realize that in face of the experimental evidence from which quantum theory has emerged, the desire to restrict the application of the criterion of logical consistency\index{criterion of logical consistency} to aspects of reality which behave in conformity with classical expectations is just as irrational as the desire to uphold determinism, that is to say, the predictability of future evolution.

The additional issue we need to consider, however, is more pragmatic. It has to do with the fact that, in the absence of a stronger and more specific constraint, there would be histories which could be characterized as `consistent' by the formalism of the theory, but which would not remain quasiclassical as time goes, following decoherence\index{decoherence!non-persistence of quasiclassicality}. This is the problem discussed in Ref. \cite{Dowker-1} and which I have mentioned earlier in this section. As Dowker\index{Dowker, Fay} and Kent\index{Kent, Adrian} explain, predictions only become possible, within the formalism of consistent histories\index{consistent histories!formalism}, once a set of histories, the physically relevant set\index{physically relevant set!histories}, which is based on a specific choice of dynamic attributes\index{dynamic attributes!choice} and a particular choice of coarse-graining\index{coarse-graining!choice}, has been selected, whose elements can then be attributed meaningful probabilities\index{meaningful probabilities}. But in a quantum-mechanical context, there appears to be total freedom over the choice of which dynamic attributes\index{dynamic attributes!freedom of choice} are used to specify the exact state of our physical systems and what elements of reality can be ignored and summed-over, and this is where the problem originates.

When no criterion exists to limit the choice of dynamic attributes, most `consistent' histories\index{consistent histories!non-persistence of quasiclassicality} do not remain quasiclassical in the future, even if they were so characterized in the past. Thus, the criterion of `consistency'\index{criterion of consistency!histories} appears to be insufficient to predict the persistence of the quasiclassical nature of history. In fact, it seems that the condition of `consistency'\index{consistency conditions!non-classical past} would not even allow one to assume that the past itself must have been classical up to the present moment, despite the fact that the existence of mutually consistent records\index{mutually consistent records!unique past} of a unique past appears to indicate that the whole observable universe evolved classically (without large-scale quantum interferences\index{quantum interferences!large scale}) as far back in time as one can tell. What remains problematic with the current approach, therefore, is the absence of a criterion, within the interpretation itself, for choosing the appropriate, physically relevant set of histories\index{physically relevant set of histories!choice} which would allow to describe the quasiclassical world\index{quasiclassical world} we do experience.

What was originally proposed by Murray Gell-Mann\index{Gell-Mann, Murray} and James Hartle\index{Hartle, James} is that, if we perceive a quasiclassical world\index{quasiclassical world!perception} it is because, as observers, we have evolved to take advantage of only those formulations of history\index{history!observer selection of quasiclassical formulations} according to which the world does, in effect, remain quasiclassical. The problem is that it appears that in the absence of a criterion for justifying the selection of the appropriate, physically relevant set of histories, the above mentioned results imply that the most likely explanation for the fact that one experiences a quasiclassical world would require one to reject all evidence of past quasiclassicality\index{past quasiclassicality!illusive evidence} and all expectations of future quasiclassicality\index{future quasiclassicality!illusive expectations} as being mere illusions and to satisfy oneself with having `explained' why it is that, at the present moment, one goes through a classical experience\index{classical experience!present moment} that is such as to give one the impression of living in a world that remained quasiclassical on a global scale during most of its history, even though that would not be the case.

I have already explained, however, why such solipsistic explanations\index{solipsistic explanation}, which require one to assume that one's current state of awareness\index{current state of awareness!exclusivity of classicality} is all that truly exists (or that evolves classically), are not acceptable in general, from the viewpoint of scientific realism\index{scientific realism}, and if there is one situation where this criticism would definitely need to apply it is certainly here. It seems to me that if such an approach is still often considered to constitute a valid explanation of the quasiclassical character of reality\index{quasiclassical character of reality!explanation}, it is merely because we cannot see how the remaining issues facing the current state-of-the-art interpretation of quantum theory\index{quantum theory!current state-of-the-art interpretation} could be resolved, so that we have come to believe that the solution may be that there is no problem after all, as long as we consider the world in the `appropriate' way.

But, if we really want to explain something, then, clearly, we must identify the constraint that allows to select the physically relevant set of histories\index{physically relevant set of histories!selection constraint} in which quasiclassicality is experienced by observers under all conditions, because the only alternative would be to retreat into a paranoid vision of reality\index{paranoid vision of reality!persistence of quasiclassicality}, where all that exists (in the classical sense) is the \textit{impression} of a persistent, large-scale, quasiclassical reality, despite the fact that there would be absolutely no reason for why such a deceptive state of consciousness should be experienced (which is the real problem). I believe that what those difficulties illustrate is the incorrectness of the basic assumption that no logically consistent interpretation exists for the interfering fine-grained histories\index{interfering fine-grained histories!logically consistent interpretation} which actually constitute the most fundamental elements of the consistent-histories formulation of quantum theory\index{quantum theory!consistent-histories formulation}.

It is significant, in this regard, that certain specialists have proposed a weaker and more general form of consistency conditions\index{consistency conditions!weaker form} \cite{Goldstein-1} that merely amounts to impose that the probabilities of coarse-grained histories\index{coarse-grained histories!positive probabilities} be positive, while still satisfying the usual probability sum rules\index{probability sum rules}. Those generalized `consistency' conditions\index{generalized consistency conditions!time reversal invariance} result in a formalism that is time reversal-invariant (which from my viewpoint is certainly a desirable property) and which selects sets of histories called \textit{linearly positive histories}\index{linearly positive histories} that include consistent histories\index{consistent histories!subset of possibilities} as a subset of possibilities. Once this is recognized to be a viable approach, however, one may be tempted to go one step further and simply allow negative probabilities\index{negative probabilities} as well, by considering the most complete sets of histories that would include all sets of linearly positive histories as a subset.

If such an even more complete generalization was never considered viable it is obviously due to the fact that negative probabilities\index{negative probabilities!absence of classical interpretation} cannot be classically interpreted (in such a context) and therefore appear meaningless and undesirable. Yet, Robert Griffiths\index{Griffiths, Robert}, suggested that it might be desirable to try to provide an interpretation of the probabilities which are known to arise when we consider quantum mechanical histories that do not satisfy the `consistency' criterion\index{consistency criterion!quantum mechanical histories}. Dowker\index{Dowker, Fay} and Kent\index{Kent, Adrian} themselves insist that there would be no logical contradiction in using an `inconsistent' set of histories\index{inconsistent set of histories!absence of contradiction} if a criterion existed that would allow one to select from it the physically relevant set\index{physically relevant set!selection criterion} and it was found that it allows a logically consistent description\index{logically consistent description!historical facts} of historical facts on a sufficiently `large' scale.

The problem is that, in the current context, the `consistency' criterion\index{consistency criterion!non-interfering sets of histories} appears to be necessary for selecting sets of coarse-grained histories that do not interfere with one another, as required by observations, while no satisfactory interpretation exists for negative probabilities\index{negative probabilities!interpretation}. But given that we still need to identify the constraint that allows one to choose the physically relevant set\index{physically relevant set!choice}, it cannot be ruled out that it might be this condition which enables to generate a historical description of reality that naturally satisfies both the criterion of `consistency' and that of persistent quasiclassicality\index{quasiclassicality!persistence}.

I have already suggested that, in the more appropriate context of a realistic, time-symmetric interpretation of quantum theory\index{quantum theory!realistic time-symmetric interpretation}, logical consistency\index{logical consistency!retarded and advanced histories} (in the general sense) would rather need to be satisfied by the unique retarded and advanced portions of history. But I also explained that, from such a perspective, an adequate interpretation of negative probabilities\index{negative probabilities!adequate interpretation} can be formulated that would confine them to unobservable aspects of physical processes. Thus, if a criterion can be found for the selection of a set of histories\index{set of histories!selection criterion} that is not \textit{a priori} `consistent', but that would nevertheless allow the quasiclassical character of reality\index{quasiclassical character of reality!emergence} to naturally emerge on the appropriate scale, then we may finally obtain a satisfactory extension of the current formalism that would allow to solve the quantum measurement problem\index{quantum measurement problem!solution}.

In fact, we may have another motive for recognizing that an additional constraint is necessary to explain the emergence of the quasiclassical nature of reality that is observed on a sufficiently irreversible scale. It was pointed out by Roger Penrose\index{Penrose, Roger} and apparently also by John Bell\index{Bell, John} and Bernard d'Espagnat\index{d'Espagnat, Bernard} that the current explanation for the reduction of the state vector through decoherence\index{decoherence!state vector reduction} is dependent on the hypothesis that it is impossible, in effect, to reveal the existence of quantum interferences\index{quantum interferences!environment degrees of freedom} involving the detailed configuration of the degrees of freedom of that part of the environment which has become entangled with a quantum system. But there is presently no valid reason to assume that such an unlikely procedure could not be carried out at some point in the future (even without deliberate intervention) and this means that the current explanation for the elimination of quantum interferences\index{quantum interferences!elimination} following measurement is only valid based on the assumption that the practical limitations that may prevent the observation of interferences between macroscopic states\index{macroscopic state interferences!limitations to observation} will \textit{never} be overturned.

Given that the existence of practical limitations to unveil superpositions of macroscopic states through a manipulation of the delocalized environment degrees of freedom\index{delocalized environment degrees of freedom} has been shown by Roland Omn\`{e}s\index{Omn\`{e}s, Roland} to be necessary for the validity of the factual definiteness of reality\index{factual definiteness of reality!necessary conditions} and the applicability of the conventional rules of logic\index{conventional rules of logic!conditions of applicability}, it is certainly significant that Omn\`{e}s himself has argued that one cannot definitely rule out the possibility that such an unlikely evolution could happen, but that given that it would mean that the world would no longer be `consistent', then he prefers to simply assume that the low probabilities involved imply that the decoherence process\index{decoherence process!definitive character} is definitive in principle.

In the context of a conventional many-worlds interpretation of quantum theory\index{quantum theory!many-worlds interpretation}, we would certainly be justified to assume that this condition needs to be fulfilled, as if it was not the case, then we should actually observe the multiple branches of history\index{multiple branches of history!macroscopic interferences} to interfere among themselves, even on the macroscopic level of measuring apparatuses, which does not only constitute an additional difficulty for this particular interpretation of quantum theory, but which also illustrates the necessity of providing a satisfactory explanation for the absolute irreversibility of the decoherence process\index{decoherence process!absolute irreversibility}.

Of course, we do observe an absence of interferences between alternative coarse-grained histories\index{alternative coarse-grained histories!absence of interferences} past a certain level of irreversibility of the ignored (summed-over) portions of a process\index{summed-over portions of process!irreversibility}\footnote{
It was once suggested that quantum interferences between alternative states are actually always allowed to occur, regardless of the size of the system under observation or the degree of irreversibility of its evolution, but that, if the existence of such interferences can be ignored, it is simply because they would be too difficult to reveal in the case of macroscopic systems\index{macroscopic systems!difficulty to reveal interferences|nn}. But it is usually recognized that this is not a valid proposal, because, in fact, nothing would be easier to distinguish than interferences between two different states of a pointer on a measuring device, given that this would necessarily be apparent in the statistical distribution of subsequent measurement results\index{measurement results!statistical distribution|nn}.}
 and this may appear to confirm the validity of the assumption that the practical limitations discussed here cannot be overcome. But we must recognize that we have, at present, no reason, from a theoretical viewpoint, to assume that such an unlikely reversal of fortune could not happen at some point in the future, because, even if there is only an infinitesimal chance that it does, given an infinite amount of time it should eventually happen and in such a case the consequences would be felt right now (this is made unavoidable in the context where the time-symmetric nature of quantum evolution\index{quantum evolution!time-symmetric nature} allows future measurements\index{future measurements!effect on past evolution} to exert an effect on past evolution).

Even if such a phenomenon was to occur only once, on a large scale, it would be possible to observe its consequences, because the usual assumption to the effect that there is no state superposition\index{state superposition!absence following measurement} following measurement would then no longer allow our estimation of transition probabilities\index{transition probabilities} to match observations, therefore indicating that the conventional hypothesis is incorrect. The fact that we usually do not observe such a disagreement means that the assumption that the decoherence process\index{decoherence process!true irreversibility} is in general truly irreversible is appropriate, even if it is not, at present, entirely justified. A satisfactory solution to the problem of quantum measurement\index{quantum measurement problem!satisfactory solution} should therefore allow one to gain confidence that, once decoherence has occurred, there is no chance that it may somehow be overturned at \textit{any} time in the future; which would allow to justify attributing the status of established facts to measurement results\index{measurement results!established facts}.

In any case, the often encountered statement to the effect that quantum theory\index{quantum theory!incompleteness of favored interpretation} has never been proven wrong, which would seem to invalidate the claim that the currently favored interpretation is incomplete, can no longer be considered accurate, given that, in the context of the developments discussed above, it seems that what the theory predicts is an absence of quasiclassicality\index{quasiclassicality!absence in past and future} in both the future and the past and this is clearly in conflict with what we do observe (for the past) and with what we have very good reasons to expect to observe (for the future). Therefore, a solution to the quantum measurement problem, the central problem of the interpretation of quantum theory\index{quantum theory!central problem of interpretation}, cannot merely consist in assuming that elementary particles\index{elementary particles!reality through interaction} acquire reality as a consequence of interaction with another part of reality (presumably a measuring device\index{measuring device}), as was originally proposed by some of the founders of quantum mechanics and as is still considered appropriate by advocates of the relational interpretation of quantum theory\index{quantum theory!relational interpretation}.

What I have tried to explain in this and the preceding sections of this chapter is that it is not necessary and not appropriate, or even possible to assume that no unique reality of any kind exists, for a quantum system\index{quantum system!unique reality between observations}, in between observations by a measuring device. The difficulty to explain the emergence of quasiclassicality\index{quasiclassicality!emergence} cannot be considered to mean that quantum theory\index{quantum theory!interaction with measuring devices} only allows to describe how microscopic systems interact with measuring devices, as if this was a requirement of a relational description of reality\index{relational description of reality}. In fact, as I will soon explain, it rather appears that a satisfactory solution to the quantum measurement problem\index{quantum measurement problem!satisfactory solution} actually requires considering that a well-defined and in some way unique, but unobservable reality does exist between measurements.

Particles\index{particles!no reality through interactions} do not become real through interactions, and the uniqueness of reality\index{uniqueness of reality}, which is observed during measurements, is not an effect that propagates as a result of further interaction, because, even if that was considered to be true, the emergence of quasiclassicality\index{quasiclassicality!emergence} would remain unexplained. It is not our intuition that such an explanation of quantum strangeness\index{quantum strangeness!wrong explanation} must be wrong that is at fault, but rather the orthodox interpretation of quantum theory\index{quantum theory!orthodox interpretation} and the insistence that we should not attempt to describe reality when it is not observed.

What emerges from those considerations is that, as undesirable as it may once have appeared, there seems to be something unavoidable with John Von Neumann's\index{Von Neumann, John} conclusion that something essential (although not necessarily fundamental) must differentiate a quantum system from the measuring device\index{measuring device!distinction from quantum system} and observer who effect a measurement on this system. Unless we are to allow for grossly inaccurate predictions, it is necessary to explain what justifies this distinction. But even though this difference can be recognized to have something to do with time irreversibility\index{time irreversibility} and even though it must come into effect following decoherence\index{decoherence}, its exact nature remains unidentified from the viewpoint of all known interpretations.

What explains that Von Neumann's conclusion\index{Von Neumann's conclusion} was never taken seriously is certainly his early proposal that the dividing line between superposed system and observing system may be determined by the level at which consciousness occurs, which could perhaps explain why it is that human observers never experience quantum interferences\index{quantum interferences!human experience}. Indeed, any reference to such qualitative aspects of physical reality as a degree of consciousness, or a level of cerebral development as possible causes of state vector reduction\index{state vector reduction!possible cause} is properly viewed with extreme suspicion by any physicist with a minimum level of cerebral development, while, in fact, such a reference is not necessary for the validity of Von Neumann's conclusion.

Once again, a perfectly valid deduction was ignored as a consequence of being associated with questionable assumptions which are not essential to its validity. But, if this is the truth, then it remains to identify the nature of this distinguishing physical property of measuring devices\index{measuring device!distinguishing property} and to explain why it has the decisive consequences it is observed to have, in the context where the basic mathematical framework of quantum theory\index{quantum theory!basic mathematical framework} is assumed to be valid under all circumstances. This is the task I will try to accomplish once I have clarified the role played by time in the most fundamental of quantum-mechanical frameworks.

\section{The emergence of time in quantum cosmology\label{sec:5.11}}

When searching for an adequate solution to the quantum measurement problem\index{quantum measurement problem} and a plausible explanation for the emergence of a quasiclassical world\index{emergence of quasiclassical world}, what one must first decide is whether quantum theory needs to be replaced by a better theory, or whether the current theory is appropriate to deal with those apparently insoluble difficulties. What I have been led to conclude is that quantum theory\index{quantum theory!incompleteness} is indeed incomplete and that it must be supplemented with new conceptual elements if it is to be made fully consistent with what we already know of physical reality that currently appears to conflict with its predictions. But, as I already mentioned, this does not mean that the current mathematical framework of quantum theory\index{quantum theory!current mathematical framework} (in its most appropriate form) must be rejected, or that the progress which was already achieved in trying to develop a better interpretation of the theory has become useless.

It is by building on earlier developments toward a time-symmetric formulation of quantum theory\index{quantum theory!time-symmetric formulation} that I will be able to address the remaining difficulties affecting the consistent-histories interpretation\index{consistent-histories interpretation!quantum theory} and to finally explain the quasiclassical nature of reality. For that purpose, however, it is necessary to first examine the extent to which time itself can still be assumed to constitute a meaningful concept in quantum cosmology\index{quantum cosmology!emergence of time} and to explain how it is allowed to emerge from a fundamental theory in which it may only be present in embryonic form. This has been made unavoidable by certain developments that took place in the field of quantum gravitation\index{quantum gravitation!irrelevance of universal time variable}, which appear to imply that the notion of a universal time variable may no longer be relevant to a fundamental description of reality, whether on the Planck scale\index{Planck scale} or on the cosmological scale.

Even though it is not before the first tentative quantum-mechanical descriptions of the universe\index{universe!quantum-mechanical description} as a whole were introduced that it was suggested that time may not be a fundamental concept for cosmology, the perceived difficulty is actually also present in classical cosmology\index{classical cosmology}. Indeed, it appears desirable, from both a practical and a theoretical viewpoint, to formulate relativity\index{relativity!dynamical theory} as a dynamical theory that would describe the evolution in time of the curvature of three-dimensional space\index{curvature of three-dimensional space!evolution in time}, given that such an approach can be more easily extended to a background-independent quantum gravitation theory\index{quantum gravitation theories!background-independence}.

But in a general-relativistic context, when it is recognized that all the meaningful physical attributes of the universe\index{physical attributes of universe!relational definition} must be defined in a purely relational way, without reference to any absolutely defined, external parameter, it transpires that any slicing of spacetime\index{slicing of spacetime} into three-dimensional space-like hypersurfaces\index{three-dimensional space-like hypersurfaces} and a time dimension (any particular choice of foliation\index{foliation!equivalence of choices}) is equivalent to any other. A general-relativistic description of the dynamics of the universe as a whole, therefore, does not allow to identify one particular dimension from among the four dimensions of spacetime as being that of time, given that the gravitational field equations\index{gravitational field equations!signature of spacetime metric} remain valid regardless of the choice of a particular signature for the metric of spacetime.

An additional difficulty also arises, due to the fact that the universe\index{universe!invariant total energy}, as a particular instance of isolated system, must have an invariant total energy which would appear to imply that no meaningful change can take place on the cosmological scale, as if time was, in effect, irrelevant. In fact, I have provided arguments in section \ref{sec:4.5} to the effect that the energy of the universe\index{energy of universe!null value}, just like its momentum, must actually be null (even when space is assumed to be flat on the largest scale) if no physical attribute of the universe\index{universe!physical attribute} is to be defined in reference to external, or metaphysical elements of reality, because if the universe\index{universe!absolute direction of time} as a whole had a positive or negative energy it would become possible to identify a particular direction in time as being of absolute (non-relational) significance. But this would appear to confirm the validity of the hypothesis that energy must be invariant on the global scale.

It seems that a similar conclusion would have to be drawn about the status of time in canonical quantum cosmology\index{canonical quantum cosmology}, where the same arbitrariness in the choice of a particular foliation\index{foliation!arbitrariness of choice} and the same absence of change to the energy content of the universe would now apply to the many different histories of extended three-dimensional space-like hypersurfaces\index{three-dimensional space-like hypersurfaces!histories}, which must be allowed to interfere with one another quantum mechanically. This is reflected in the fact that the most straightforward interpretation of the Wheeler-DeWitt equation\index{Wheeler-DeWitt equation!most straightforward interpretation} (the equation that would allow to determine the wave function of the universe) requires assuming that it is similar in form to the \textit{stationary} Schr\"{o}dinger equation\index{stationary Schr\"{o}dinger equation}, while time is notoriously absent from such an equation. It is sometimes suggested that what those difficulties demonstrate is that the hypothesis that time\index{time!existence hypothesis} exists as a unique dimension, distinct from the other three dimensions of space, is incorrect.

It should be clear, however, that the absence of change to the curvature of space\index{curvature of space!absence of global change} on a global scale, which is a consequence of the fact that the universe\index{universe!null gravitational energy} has a null value of gravitational energy, does not mean that time\index{time!local changes} is not a meaningful concept for relating the changes in space curvature which are taking place in one part of the universe with those occurring in another part of it, as long as we are actually dealing with different portions of the same universe, because it is not required of local subsystems\index{local subsystems!no energy invariance} that they have invariant energies as a consistency requirement and therefore change can certainly be observed to take place on an intermediary scale. In other words, even if we were to assume that time\index{time!irrelevance on global scale} is irrelevant on a global scale, this could not be understood to mean that it has no clear significance as a means to relate local measures of changes. What's important to recognize is precisely that, from a cosmological viewpoint, time\index{time!relationally defined physical property}, as a dimension distinct from space, has meaning only as a relationally defined physical property that allows multiple local measures of change to be compared, thereby enabling all observers to provide a unique description of the various processes taking place in the universe (or within their associated causal horizon\index{causal horizon}).

Thus, it is an exaggeration to suggest that time does not constitute a meaningful concept in quantum cosmology\index{quantum cosmology!meaningfulness of time} or that a null value of energy, for the universe\index{universe!null energy} as a whole, is indicative of the fact that time does not exist. Such a conclusion would be no more reasonable than the idea that a universe\index{universe!null momentum} with null momentum (relative to the global inertial reference system\index{global inertial reference system} determined by the average state of motion of all matter in the universe) would be indicative that space does not exist, which is so obviously inadequate a hypothesis that no one has ever suggested it could apply. To show that a conventional notion of time is not irrelevant to our description of reality on the cosmological scale, however, one must first explain how it is possible for time\index{time!differentiation from other spacetime dimensions} to differentiate itself from the other three dimensions of spacetime, despite the fact that all four dimensions are kept on an equal footing and are required to be equivalent, from a fundamental viewpoint, by relativity theory. It is this particular aspect of the problem of time in quantum cosmology\index{quantum cosmology!problem of time} that I will address in the remainder of this section.

Two points must be taken into account in order to explain the existence of a unique slicing of four-dimensional spacetime\index{slicing of four-dimensional spacetime!uniqueness} into three-dimensional space-like hypersurfaces\index{three-dimensional space-like hypersurfaces} that would consistently select one single dimension as being that of time. First, it needs to be recognized that there must exist unique relationships of causality\index{relationships of causality!ensemble of events in universe} between all local events comprising a four-dimensional universe. Second, it must be recognized that, at the fundamental quantum-gravitational level\index{quantum-gravitational level!principle of local causality}, it is possible for the principle of local causality to be enforced due to the existence of an embryonic element of time directionality\index{time directionality!embryonic element} in the causal structure of spin foams\index{spin foams!causal structure}.

Once this is recognized, then it becomes possible for a metric of spacetime\index{metric of spacetime!emergence of unique signature} with a unique signature to emerge that singles out one particular direction of four-dimensional spacetime as being that which is associated with the dimension of time across an entire space-like hypersurface\index{space-like hypersurfaces} (throughout the universe, on a given slice of spacetime). This is because, as I have explained in section \ref{sec:4.9}, the homogeneity of the initial matter distribution\index{initial matter distribution!homogeneity} at the Big Bang (which is responsible for the existence of a thermodynamic arrow of time\index{thermodynamic arrow of time}) arises precisely as a consequence of requiring a constraint of global entanglement\index{global entanglement constraint!spacetime slice} to apply uniformly over an entire slice of spacetime and this constraint is actually a condition for the existence of causal relationships\index{causal relationships!components of universe} between all components of the universe which are present in this initial Big Bang state\index{initial Big Bang state}.

It is important to understand that what distinguishes time\index{time!differentiation from other spacetime dimensions} from the other dimensions of spacetime, in a relativistic context, is merely the choice of a particular signature for the metric of spacetime\index{metric of spacetime!choice of signature} which is arbitrarily imposed on solutions of the gravitational field equations\index{gravitational field equations!solution} in order that they satisfy observational constraints. But what this distinction provides is merely a separation of spacetime into past and future light cones\index{past and future light cones!requirement of local causality} along one particular dimension, which is really a requirement of local causality. Thus, if the signature of the metric of spacetime\index{metric of spacetime!signature} was different and causality still operated uniformly, but along another dimension of spacetime, we would simply call this dimension time, while the other three dimensions would then all be analogous to conventional space. In fact, given that general relativity allows for local variations of the light-cone structure\index{light-cone structure!local variations}, one may say that what is produced as a result of spacetime curvature\index{spacetime curvature!local alterations of causal structure}, or due to the presence of local gravitational fields attributable to the presence of matter inhomogeneities, are merely smooth local alterations of the direction in which causality operates in spacetime.

Now, all that is required by the global entanglement constraint\index{global entanglement constraint!uniform matter density} is that at least one space-like hypersurface\index{space-like hypersurfaces} exists over which the matter density is sufficiently uniform, down to the quantum-gravitational scale\index{quantum-gravitational scale}, that no macroscopic event horizon\index{macroscopic event horizon} is present. But given that global entanglement\index{global entanglement!causal relationships} is a condition that is imposed in order that causal relationships be allowed to exist between all components of the universe, what is implied by this absence of macroscopic event horizon in the initial Big Bang state\index{initial Big Bang state!absence of macroscopic event horizon} is that the direction in which causality\index{causality!uniform direction in spacetime} operates actually is the same over all space. In other words, the embryonic, quantum-gravitational element of causal order\index{causal order!embryonic quantum-gravitational element of} must be found to operate in the same direction of spacetime in all locations, over at least one such hypersurface and right down to the quantum-gravitational scale\index{quantum-gravitational scale} (the Planck scale\index{Planck scale}), thereby consistently imparting on spacetime\index{spacetime!unique signature} a unique signature that is shared throughout the universe. I believe that this is what explains that the direction of spacetime in which time is flowing is still mostly the same over all of space today (except in the presence of strong local gravitational fields and macroscopic event horizons\index{macroscopic event horizon}), as necessary for the existence of a universal time variable\index{universal time variable}.

To put things a little differently, one could say that, if there were significant local differences in the alignment of light cones\index{light cones!local alignment differences} on scales larger than the quantum-gravitational scale, in the initial Big Bang state, this would be equivalent (for what regards causality) to the presence of macroscopic event horizons and the presence of event horizons on all but the shortest scale is precisely what is forbidden by the global entanglement constraint\index{global entanglement constraint}, in the presence of negative-energy matter. Therefore, if global entanglement is necessary for the existence of the universe\index{universe!ensemble of causally interrelated components} as an ensemble of causally interrelated components, then there must exist one space-like hypersurface\index{space-like hypersurfaces!unique light-cone orientation} over which the light cones and time itself are oriented in the same direction of spacetime in every location (this is easier to visualize when space is assumed to be two-dimensional).

It should be clear, however, that it is not merely the existence of causal or cosmological horizons\index{cosmological horizon} that imposes a condition of global entanglement\index{global entanglement condition!consistency requirement}, because global entanglement is an independent consistency requirement for the existence of causal relationships between all components of the universe\index{universe!causal relationships between all components} which, in the presence of negative-energy matter, actually allows the emergence of causal horizons\index{causal horizon!emergence as unidirectional phenomenon} as unidirectional phenomena (given that it allows the emergence of a thermodynamic arrow of time\index{thermodynamic arrow of time!emergence}). It is not logically inappropriate, therefore, to argue that it is when global entanglement is imposed that causal order\index{causal order!unique spacetime direction} must be found to apply in the same direction of \textit{spacetime}, uniformly, throughout the universe, as long as one recognizes that what is involved here is time-symmetric causality\index{time-symmetric causality}.

This is a significant result, because when a constraint of global entanglement is imposed on the initial Big Bang state\index{initial Big Bang state}, in the presence of negative-energy matter particles, the magnitude of early fluctuations in the densities of positive- and negative-energy matter is limited, to the extent that local variations of the light-cone structure\index{light-cone structure!absence of local variations} are virtually absent, so that time\index{time!uniform flow} flows uniformly over all space (proper time intervals\index{proper time intervals!uniform over space-like hypersurface} are the same across an entire space-like hypersurface), as would be the case by default in a Newtonian context. The above argument would therefore appear to provide the basis for a satisfactory solution to one of the last major unsolved problems still facing the most appropriate of current, tentative quantum theories of gravitation\index{quantum gravitation theories!last major unsolved problems}, which is the question of how it is possible for a universal time variable\index{universal time variable!emergence from timeless equations} to emerge from the timeless equations of the theory. Thus, it would no longer be necessary to appel to the weak anthropic principle\index{weak anthropic principle!explanation of time asymmetry} to explain, not only the observed time asymmetry and the unidirectional nature of causality, but really the very existence of a universal time variable\index{universal time variable!explanation of existence}.

Even though, from a classical perspective, relativity theory does not \textit{a priori} require that there is a preference for one particular dimension of four-dimensional space over any other, the condition that there should exist causal relationships between all parts of that four-dimensional reality\index{four-dimensional reality!causal relationships between events} (between all the events taking place in it) implies that one direction in four-dimensional space\index{four-dimensional space!direction of effects propagation} is singled out, \textit{uniformly}, as being that along which effects are propagated in the emerging spacetime and this is what gives rise to time\index{time!uniformly flowing variable} as the continuous and uniformly flowing variable we are accustomed to experience on a macroscopic scale.

The validity of the hypothesis that there does emerge such a singular dimension of time out of four-dimensional spacetime\index{spacetime!singular dimension of time} is what legitimizes a formulation of quantum cosmology\index{quantum cosmology!dynamics of space-like hypersurfaces} as having to do with the dynamics of extended three-dimensional space-like hypersurfaces, whose histories can be described as unique trajectories in superspace\index{superspace trajectories!histories of space-like hypersurfaces} (the configuration space of those three-dimensional objects). What is remarkable is that the viability of such a description is, in fact, a necessary condition for the elaboration of a consistent explanation of the quasiclassical nature of reality\index{quasiclassical nature of reality!consistent explanation} that emerges under conditions where time irreversibility is a characteristic of the processes involved, as I will explain in the following section.

The problem that there was, originally, with the proposal that quantum cosmology has to do with the dynamics of extended three-dimensional space-like hypersurfaces is that the introduction of a fundamental element of causality in quantum gravitation\index{quantum gravitation!fundamental element of causality} requires a decomposition into positive- and negative-energy solutions, as in conventional, relativistic quantum theory\index{relativistic quantum theory}, and it was not clear how this could be achieved in the context of such a model. But even though this difficulty appears to have been overcome, I still believe that significant progress could be achieved in developing the current covariant framework of spin-foam quantum gravity\index{spin-foam quantum gravity!covariant framework} into a fully satisfactory theory by taking into account the possibility for negative energy states\index{negative energy states!forward- and backward-in-time propagation} to propagate both forward and backward in time, which constitutes a necessary step in allowing a proper integration of the requirements imposed by the generalized, classical theory of gravitation\index{generalized gravitation theory} I have introduced in chapter \ref{chap:2}.

In any case, if local causality is, in effect, a decisive constraint on the quantum-gravitational scale\index{quantum-gravitational scale!local causality constraint}, then time itself necessarily constitutes a meaningful parameter in quantum cosmology\index{quantum cosmology!meaningfulness of time}, even on the global scale, because the separation of four-dimensional spacetime\index{four-dimensional spacetime!separation into space and time} into three dimensions of space and one uniformly-pointing dimension of time appears to be the defining character of a world that obeys the principle of local causality\index{principle of local causality} in the presence of negative-energy matter.

One must be careful, however, when considering a quantum-mechanical theory\index{quantum-mechanical theory!whole universe} that purports to describe the whole universe, because, from a realistic viewpoint, it would not be appropriate to describe the universe\index{universe!deterministically evolving wave function} by using a wave function evolving deterministically over its entire history. Indeed, by doing so, we would commit the same error we make in the classical theory of relating all past and future three-dimensional space-like hypersurfaces\index{space-like hypersurfaces!predetermined relationships} in a predetermined way to some arbitrarily-chosen present state, which makes it look like everything about history is resumed in one single stationary state. In a more realistic situation, the whole history would not be determined from knowledge of one particular global state and following each local measurement the state of the universe\index{state of universe!actualization following measurements} and its wave function would need to be actualized, which would reveal the random nature of the history that actually takes place and the absence of \textit{predetermined} relationships between the multiple extended three-dimensional spaces\index{extended three-dimensional spaces!relationships} forming a history, which in turn illustrates the relevance of time in characterizing the actual relationships.

Even in the context where a unique future\index{future!uniqueness} is assumed to exist in the same way a unique past\index{past!uniqueness} does, there is no rational motive to argue that time, as a measure of change, becomes an irrelevant notion, because such a conclusion would only be valid if we ignored the random aspect of quantum-mechanical processes\index{quantum-mechanical processes!randomness} (which is particularly unavoidable in the context of the existence of closed causal chains\index{closed causal chain}) and if we neglected the constraint imposed by the requirement that all components of the universe\index{components of universe!requirement of causal relationships} be causally related, which singles out the initial state of maximum matter density\index{initial state!maximum matter density} of the Big Bang as a state of minimum gravitational entropy\index{initial state!minimum gravitational entropy} from which all future evolution\index{future evolution!irreversibility} is taking place irreversibly, as I explained in section \ref{sec:4.9}.

Anyhow, it must be clear that, despite what is sometimes suggested, it is not true that time, or even space, do not exist at all in modern quantum gravity\index{quantum gravity!absence of time or space}. Indeed, a certain embryonic concept of space is clearly present in the structure of spin networks\index{spin network!embryonic concept of space}, which allows classical space\index{classical space!emergence} to emerge naturally when a sufficiently large number of fundamental, discrete elements of structure\index{discrete elements of structure!classical space} are combined according to purely quantum rules. Furthermore, even in such a context, we are still dealing with four-dimensional boundary conditions\index{four-dimensional boundary conditions} and this is certainly indicative of the relevance of time, even if this parameter may not explicitly appear in the equations which allow to determine the correlation probabilities associated with those four-dimensional boundary conditions. Actually, the mere fact that, even in a quantum-gravitational context, we are still speaking about `local' changes occurring in the configuration of spin networks\index{spin network!local changes to configuration} means that an additional degree of freedom \textit{must}, as a fundamental requirement, be allowed to emerge, which relates those local changes to one another.

The problem that there was, originally, is simply that, in the absence of a constraint of global entanglement\index{global entanglement constraint}, no universal time variable\index{universal time variable!emergence} was allowed to emerge, because no unique direction appeared to exist in spacetime\index{spacetime!absence of unique time direction} that would be associated with this degree of freedom and along which events could be sequentially ordered into some kind of \textit{universal causal chain}\index{universal causal chain}. Once it is recognized that causal relationships\index{causal relationships!components of universe} must exist among all components of the universe, however, then the most appropriate of the current fundamental theories do allow a certain concept of history\index{history!emergence of concept} to emerge given that, in the presence of negative-energy matter, this condition allows one particular dimension of four-dimensional spacetime\index{four-dimensional spacetime!dimension of causal order} to be selected uniformly, throughout one spin network configuration\index{spin network!configuration}, as being that along which causal order is established and for this reason alone, extended three-dimensional space-like hypersurfaces\index{space-like hypersurfaces!dynamic elements of quantum cosmology} may be considered to constitute the dynamic elements of a quantum theory of cosmology.

However, in my opinion, what would definitely invalidate a truly timeless quantum theory of gravitation\index{timeless quantum gravitation!invalidation} is precisely the fact that such a theory would be incompatible with the existence of a fundamental time-direction degree of freedom\index{fundamental time-direction degree of freedom} (such as revealed in particular by violations of the time reversal symmetry\index{time reversal symmetry!violation} operation $T$), while I have shown, in chapters \ref{chap:2} and \ref{chap:3}, that such a property is essential to a consistent description of physical reality, in a semi-classical context. Indeed, once it is recognized that, in quantum field theory\index{quantum field theory}, the propagation of elementary particles\index{elementary particles!propagation along any direction of time} can take place along any of two opposite directions of time, independently from the constraints imposed by thermodynamic irreversibility\index{thermodynamic irreversibility}, then a conflict emerges with the timeless viewpoint\index{timeless viewpoint}, given that if there is no time, then obviously there cannot be a fundamental direction in time, because any relationship of time directionality\index{time directionality} must necessarily involve a sequence of events causally related to one another following a definite and unique order, distinct from their spatial order, even when the classical spacetime structure\index{classical spacetime structure!emergence} in which those events are embedded is assumed to emerge from the combination of discrete elements.

Given the nature of the arguments which are usually proposed to support the conclusion that time is irrelevant in quantum cosmology\index{quantum cosmology!irrelevance of time} and therefore that it may not even exist, it would seem that solipsism\index{solipsism!rejection of time} is once again to blame for misleading even some of the most brilliant thinkers into this theoretical dead-end. Indeed, what a rejection of time would require us to assume is that there can be change and that all changes can be related to one another by the use of a reference system we call time, but that this is not enough to justify the conclusion that this reference system is the reflection of something real.

Thus, while we are allowed to recognize the emergence of a certain variable, distinct from spatial position, which is useful for comparing various local measures of change involving one or another physical attribute, and while the assumption that such a variable exists is undeniably useful and allows to simplify our description of reality, the fact that it is not possible to directly observe time\index{time!impossibility of direct observation} itself and the fact that this additional variable may no longer be globally significant under the most extreme conditions (on the shortest scale or in the presence of very strong gravitational fields) would mean that it cannot be considered a real physical property, even under more ordinary circumstances. All arguments against the existence of time as a meaningful concept in quantum cosmology involve such an element of solipsism. Time does not exist because it cannot be subjected to direct observation, or be the object of some measurement that would confirm that it is real. But that is just a perfect example of the kind of irrational conclusion one can draw based on such considerations, because what can be more obvious in fact, from our experience of physical reality, than the existence of change and the reality of time?

Now, it has been argued that it might be possible for time\index{time!thermodynamic phenomenon} to emerge as a mere thermodynamic phenomenon, despite the fact that it would not really exist from a fundamental viewpoint. What I'm talking about is the concept of `thermal time'\index{thermal time}, according to which the passage of time\index{passage of time!illusion of conscious experience} would actually be an illusion attributable to the fact that the irreversible time of our conscious experience appears to always be associated with heat\index{heat!dissipation} dissipation, which would appear to single out one particular physical variable as that relative to which energy remains unchanged, while in fact there would be nothing fundamental with such a variable. But the problem with this proposal is that there is, in fact, plenty of evidence for the relevance of a more conventional concept of time\index{conventional concept of time!elementary particle level} at the level of elementary particles, where irreversibility is not a defining characteristic.

The fact that we may be unable to perceive a preferred direction of time\index{preferred time direction!heat dissipation} in the absence of heat dissipation is certainly not altogether irrelevant to the problem of the existence of a classical spacetime continuum (given that dissipation\index{dissipation!decoherent space and time} appears to be necessary to explain the decoherent nature of space and time, as I will soon explain), but it is not that significant either, because we are not merely trying to decide whether unidirectional time\index{unidirectional time} is a valid concept, but with deciding if the whole concept of time is, in effect, relevant to a description of physical reality. However, if thermodynamics was the ultimate explanation for the existence of time, it would not be necessary to wait until we begin to explore reality on the quantum-gravitational scale\index{quantum-gravitational scale!absence of time} to witness an absence of time, because many phenomena are known to exist, on a much larger scale, that do not involve any irreversibility and yet they are still describable using space and time coordinates\footnote{
In the context where a satisfactory solution to the problem of the origin of thermodynamic time asymmetry\index{thermodynamic time asymmetry!problem of origin|nn} that is not based on the weak anthropic principle\index{weak anthropic principle|nn} is now available (this was the subject of section \ref{sec:4.9} of this report), the fact that the thermal-time\index{thermal time!explanation of thermodynamic time asymmetry|nn} hypothesis may appear suitable for an explanation of thermodynamic time asymmetry based on a certain interpretation of entropy growth\index{entropy growth!observer-dependent phenomenon|nn} as a purely subjective, observer-dependent phenomenon would no longer constitute a potential advantage of a timeless interpretation of quantum cosmology\index{quantum cosmology!timeless interpretation|nn}.}.

But if we were to insist that, despite all the evidence, time really doesn't exist, even under ordinary circumstances, we would then be left with having to conceive of the present as just one independent, momentary state among many possible states devoid of any causal relationships with one another. It was, in effect, suggested by Julian Barbour\index{Barbour, Julian} that such causally independent, momentary states may not be incompatible with our perception of the passage of time, if we assume that all that we really experience are momentary states of consciousness, which might be more appropriately described as memory states\index{memory states}.

But the problem here, again, is that even if such an explanation of consciousness as a state rather than as a process was possible (which I believe may not really be the case\footnote{
Memory\index{memory!formation process|nn}, as well as other basic mental faculties, are not really static events, but rather processes which require a certain duration to be experienced and if there is no duration, what one should expect to experience is not one everlasting memory, but nothing at all, which is certainly not compatible with my own experience of reality, at least.})
 we would then have no explanation for the fact that the present state of the universe, in which the state of our consciousness is contained, is one which is characterized by the existence of a large number of mutually consistent records\index{mutually consistent records!unique lower entropy past} of a unique lower entropy past, because such a configuration would not likely be chosen in a random trial, out of all the possibilities which would appear to exist for a momentary present state. The fact that what can be characterized as long-term records\index{long-term records!most stable structures} are usually preserved in what appears to be the most stable structures, while short-term memories\index{short-term memories!rapidly changing structures} are usually preserved in more rapidly changing structures, would also remain unexplained from a timeless universe perspective\index{timeless universe perspective}.

There were many attempts at trying to explain why such present states as revealed by our personal experience of reality may not really be unexplained, even when one assumes that all that exists in the universe\index{universe!space without time} is an extended space without any time. But in the end, one must recognize that those proposals are inadequate and that some observed properties of the present configuration of our universe would remain a complete mystery, unless one was ready to assume that those observations are not really indicative of the existence of a lower entropy past, even though there is absolutely no rational motive (even based on the weak anthropic principle\index{weak anthropic principle}) to legitimate the validity of such a conclusion.

Of course, if it had actually been demonstrated without doubt that time does not exist, then we may have no choice but to assume that all observational evidence of a low-entropy past\index{low-entropy past!observational evidence} is such a strange and deceptive illusion, but this is not true and the only reasonable conclusion we are allowed to draw from our observations is that the present state of the universe, regardless of its exact nature, must be related to one single past history through the existence of unique (but not predetermined) causal relationships unfolding back in time to the initial state\index{initial state!minimum gravitational entropy} of minimum gravitational entropy that allows to explain the existence, in the present state, of mutually consistent records\index{mutually consistent records!unique past} of a unique past.

It is usually recognized, in fact, that all that one may reasonably argue, concerning time\index{time!quantum-gravitational concept} as a quantum-gravitational concept, is that it is the continuity of its flow throughout space and the existence of a unique spacetime metric signature\index{spacetime metric signature!uniqueness} which do not apply at the most fundamental level. Thus, if, at some point, there was such a strong desire to do away with time, it is perhaps only due to the fact that we were unable to explain the singular character of time\index{time!singular character} as a dimension of spacetime, because, in the absence of guidance from the generalized theory of gravitation\index{generalized gravitation theory} I have introduced in the second chapter of this report, we couldn't understand the profound significance of the homogeneity of the initial matter distribution\index{initial matter distribution!homogeneity} at the Big Bang, which allows me to explain the near uniformity of the direction of propagation of effects in spacetime\index{propagation of effects in spacetime!uniform direction} and therefore of the flow of time\index{flow of time!uniform direction in spacetime} itself.

From a conventional perspective, it was rather convenient to simply assume that time does not exist at all, given that, like space itself, time is not present in its classical form at the most fundamental level. But it must be clear that if time, or more specifically causal order\index{time and causal order!quantum gravitation}, did not exist in any form at a fundamental quantum-gravitational level, then what we should definitely not experience is a dimension of time distinct from the other dimensions of space.

Now, despite the fact that I have criticized Julian Barbour's\index{Barbour, Julian} suggestion that our experience of the passage of time may not be incompatible with a timeless description of reality\index{timeless description of reality}, I must recognize that he, more than anybody else, is responsible for having convinced me of the validity of the concept of simultaneity hyperplanes\index{simultaneity hyperplanes}, or more generally of space-like hypersurfaces\index{space-like hypersurfaces!dynamical theory of space} as the basic building blocks of a dynamical theory of space that would be relevant to quantum cosmology\index{quantum cosmology}. The only problem I have with Barbour's interpretation has to do with his insistence that those global states\index{global states!absence of causal order} of the universe as a whole should all exist independently from one another and therefore cannot be causally related to one another following a unique and well-defined order (cannot be considered to form one single causal chain\index{causal chain!uniqueness} or to take part in one single history). But in fact, this need not be considered a requirement of a dynamical approach to quantum cosmology\index{quantum cosmology!dynamical approach} and as I have explained above, it would rather seem that there must exist unique causal relationships between those properly defined global states\index{global states!unique causal relationships}, despite the fact that there appears to be a lot of freedom in how spacetime\index{spacetime!slicing into space-like hypersurfaces} can be sliced into such space-like hypersurfaces.

We may, therefore, retain as valid the concept that the present state of the universe as a whole, regarding, in particular, its gravitational field or spacetime curvature, is provided by the current configuration of one such space-like hypersurface\index{space-like hypersurfaces!present state of universe}, which may be represented as a point in the appropriate configuration space (say the superspace\index{superspace} of canonical quantum cosmology\index{canonical quantum cosmology}), while the time\index{time!position along trajectory in configuration space} variable would enter the picture as the position along the actual trajectory followed by the global state in this configuration space. This becomes a valid proposal in the context where we now have a valid explanation for how it can be that one given spacetime dimension is uniformly selected as being that along which local causality is allowed to operate (as reflected in the uniqueness of the signature that must be assigned to the metric of spacetime\index{metric of spacetime!uniqueness of signature}) and to constitute a physically significant constraint that does not apply in the case of the other three dimensions of space, even in a general-relativistic context.

To be honest, I have to mention that the conclusion that the history of a universe's space curvature\index{history!space curvature of universe} can always be represented as a path in the configuration space of three-dimensional space-like hypersurfaces\index{three-dimensional space-like hypersurfaces} is dependent on the hypothesis that any solution of the gravitational field equations\index{gravitational field equations!closed time-like curves} that contains closed time-like curves (those hypothetical configurations of the curvature of space which would make conventional time travel\index{time travel!experience} experiences a reality) can be excluded. Usually, this is recognized to be possible merely if we assume without reason that the second law of thermodynamics\index{second law of thermodynamics} is valid under all conditions. But given the explanation I have provided in section \ref{sec:4.9} for the existence of the thermodynamic arrow of time\index{thermodynamic arrow of time}, the conclusion that closed time-like curves cannot naturally arise actually becomes unavoidable, because under such circumstances, the constraint that gives rise to thermodynamic time asymmetry\index{thermodynamic time asymmetry} must always operate in the same unique direction of time and invariably have as a consequence the diminution of entropy in the particular direction of time that points toward the initial state of minimum gravitational entropy\index{gravitational entropy!initial Big Bang state} of the Big Bang, as a requirement for the existence of causal relationships\index{causal relationships!requirement} between all components of the universe.

Therefore, a universe\index{universe!causally related space-like separated events} could not even exist, as a causally interrelated ensemble of space-like separated events, if it did not satisfy this unidirectionality constraint\index{unidirectionality constraint}, which would be the case if the direction in time of entropy decrease\index{entropy decrease!unique direction in time} could not be well-defined as a result of the curvature of spacetime\index{spacetime curvature} and this means that closed time-like curves\index{closed time-like curves} are actually forbidden. From my viewpoint it would, therefore, appear that it is \textit{always} possible to represent a particular history of the metric properties of space of the universe\index{history of metric properties of space!path in superspace} as some monotonic foliation of space-like hypersurfaces\index{monotonic foliation of space-like hypersurfaces}, that is to say, as a path in superspace.

It is, therefore, the existence of a unique direction in spacetime, along which effects must propagate, either forward or backward, that allows histories to be parameterized by a universal time variable\index{universal time variable!parametrization of histories} (associated with a particular slicing into space-like hypersurfaces\index{space-like hypersurfaces!particular slicing}) and that enables a description of space curvature\index{space curvature!evolution in time} as evolving with respect to this time variable, thereby legitimating the concept of history\index{history!ensemble of causally related global states} as consisting in an ensemble of causally related global states, that is to say, a universal causal chain\index{universal causal chain}. What I have shown is that the \textit{apparent} absence of a fundamental distinction between time and the other three dimensions of spacetime, which is an essential feature of relativity theory, does not constitute an insurmountable obstacle to achieving this objective, so that we are no longer justified to conclude that time is altogether absent in quantum cosmology\index{quantum cosmology!absence of time}.

This is certainly a significant result for the elaboration of a solution to the problem of the interpretation of quantum theory\index{interpretation of quantum theory!problem}, given that the existence of classical space and time\index{classical space and time!conventional quantum theory} is actually required by conventional quantum theory, for the description of histories, in the context where the various macroscopic experimental conditions which are shared by both the retarded and the advanced portions of a quantum process\index{quantum process!classical spacetime continuum} must be defined over one unique and classically well-defined spacetime continuum. Thus, spacetime\index{spacetime!decoherence} itself must be assumed to be decoherent, on a sufficiently large scale of distance and duration, under conditions where a history can be consistently defined, which means that quasiclassicality\index{quasiclassicality!curvature of space} must already apply to the gravitational field and the curvature of space in order that decoherence\index{decoherence!conventional quantum system} be observed at a higher level in the observed attributes of conventional quantum systems.

This, again, illustrates the fact that a uniformly oriented dimension of time\index{time dimension!uniform orientation in spacetime} must be allowed to emerge from a quantum theory of gravitation\index{quantum gravitation theories}\footnote{
Of course, even on an astronomical scale, the spatial uniformity of the flow of time\index{flow of time!spatial uniformity|nn} is only an approximation, because the metric properties of space and time\index{metric properties of space and time|nn} are influenced by the presence of positive-energy matter and by the inhomogeneities which are present in what remains of the negative-energy matter distribution, which means that, even from the viewpoint of the approach favored here, there is still no universally valid measure of the passage of time\index{passage of time!absence of universal measure|nn}.}
 before ordinary quantum processes can be appropriately described and conventional quantum theory\index{conventional quantum theory!semi-classical level} itself can be considered valid on a semi-classical level. The problem of the emergence of time\index{emergence of time!quantum cosmology} in quantum cosmology must, therefore, be recognized as constituting one particular aspect of the more general problem of the nature of the conditions necessary for the emergence of a quasiclassical world\index{emergence of quasiclassical world!necessary conditions}.

What this means is that in order to obtain a satisfactory interpretation of quantum theory\index{quantum theory!satisfactory interpretation}, one must first examine in which way gravitation and the curvature of space\index{space curvature!time-symmetric quantum mechanical description} could be subjected to the same time-symmetric, quantum-mechanical description as would apply to more conventional physical attributes under ordinary conditions. Achieving such an objective will allow me to identify additional constraints from which both the decoherent nature of spacetime\index{decoherent nature of spacetime} and the persistence of quasiclassicality\index{persistence of quasiclassicality} that characterizes all observed physical processes can be expected to arise, even in the context where quantum theory is assumed to be valid under all circumstances. What those considerations will demonstrate is that it is not just general relativity\index{general relativity!theory of universe as a whole} which really is a theory of the universe as a whole, as is usually recognized, but that quantum theory\index{quantum theory!cosmological theory}, from the viewpoint of its most accurate interpretation, is also essentially a cosmological theory.

\section{Universal causal chain and quasiclassicality\label{sec:5.12}}

We are now finally in position to examine how it is exactly that quantum theory can be extended, so as to become fully consistent from both a logical and an experimental viewpoint. It is here that all the breakthroughs achieved in the preceding chapters of this report, as well as in the preceding portions of the present chapter, while trying to provide a better understanding of so many aspects of physical theory associated with time directionality\index{time directionality} will converge to produce their most significant outcome: a logically consistent interpretation of quantum theory\index{quantum theory!logically consistent interpretation} that is valid at absolutely all levels of description.

It is certainly a positive development, already, that, in the preceding section, I have been able to conclude that time is still relevant to a description of our universe in a quantum-mechanical context. Under such conditions it becomes appropriate to define the intrinsic space curvature\index{intrinsic space curvature} over a particular three-dimensional slice of spacetime\index{spacetime!three-dimensional slice} at one particular moment as consisting of a single point in superspace\index{superspace}. The role of time then emerges quite straightforwardly as being that of relating those global states of the universe to one another into some kind of universal causal chain\index{universal causal chain}, while establishing the sequential (chronological) order of events.

What's remarkable is that the existing mathematical framework by which this particular approach can be formalized, which originates in the ADM formalism\index{ADM formalism}\footnote{
Despite the fact that the ADM formalism\index{ADM formalism|nn} of quantum gravity\index{quantum gravity|nn} (based on ADM variables\index{ADM variables|nn}) has been replaced by the more appropriate formalism of loop quantum gravity\index{loop quantum gravity|nn} based on a formulation of general relativity theory\index{general relativity theory!Ashtekar variables formulation|nn} in terms of connection, or Ashtekar variables, I think that it is still appropriate to describe the general concept of a dynamical theory of space\index{dynamical theory of space|nn} using the original approach to quantum cosmology\index{quantum cosmology!superspace|nn} which was based on superspace, as it allows to more intuitively visualize the phenomena involved and to more easily understand how the realistic interpretation of quantum theory\index{quantum theory!realistic interpretation|nn} developed here can be applied in a cosmological context, as far as the discreteness of space\index{space!discreteness|nn} can be neglected.}
 \cite{Arnowitt-1}, allows history\index{history!superspace trajectory} itself to be described as one particular trajectory in superspace \cite{Wheeler-1} \cite{Wheeler-2} \cite{Fischer-1}. Time, therefore, must be conceived of as the global variable to which are related the multiple local measures of change that take place as the curvature of space\index{curvature of space!superspace trajectory} evolves along such a trajectory in superspace. This allows to fulfill Reichenbach's\index{Reichenbach, Hans} vision of time\index{time!causal chain} as reducing, in its most essential form, to the general concept of a causal chain, that would allow to establish and maintain the invariant local topological structure of spacetime\index{spacetime!invariant local topological structure}, even when its metric properties\index{metric properties!local variations} are subject to local variations. From my viewpoint, however, it would not be appropriate to consider a conventional concept of causal chain\index{causal chain!conventional concept} that would involve time irreversibility\index{time irreversibility!fundamental} at a fundamental level, as Reichenbach contemplated, because irreversibility is a property that must rather emerge from the particular boundary conditions\index{boundary conditions!Big Bang} which existed at the Big Bang.

It must be clear, however, that it is the network of \textit{local} relationships\index{network of local relationships} that varies as we move along a trajectory in superspace\index{superspace trajectory}, because, from the viewpoint of its total energy content, the universe\index{universe!isolated system}, as the ultimate isolated system, would appear to remain in the same state without any change actually taking place (this is what motivates the unsubstantiated claim that time\index{time!irrelevance to quantum cosmology} may not be relevant to quantum cosmology, as I explained in section \ref{sec:5.11}). It must also be emphasized that what is provided by the concept of space-like hypersurface\index{space-like hypersurfaces!concept} is not a unique and absolutely defined characterization of reality, because, even when a universal time variable\index{universal time variable!emergence} is allowed to emerge, there are still many equivalent ways by which spacetime\index{spacetime!slicing into simultaneity hyperplanes} can be sliced into three-dimensional simultaneity hyperplanes (because simultaneity itself is a relative concept), which would appear to require a history of the universe's space curvature to consist, not in a unique trajectory in superspace, but rather in a given surface in the same infinite-dimensional configuration space, formed of the many equivalent trajectories which are associated with the same unique history of space curvature\index{history of space curvature!equivalent superspace trajectories}.

Thus, even if many equivalent possibilities exist for such a trajectory, they all provide alternative descriptions of the same causal chain\index{causal chain!alternative descriptions}, to which corresponds one unique history. Once again, the freedom that surrounds the choice of a suitable slicing of spacetime\index{slicing of spacetime} must not be considered to reflect the irrelevance of time for a description of the dynamics of space\index{dynamics of space} on the cosmological scale, as it is merely a reflection of its relational nature. The modern spin-foam quantum theory of gravitation\index{spin-foam quantum gravitation theory!covariance} allows to more appropriately deal with this freedom and to formulate the approach discussed here in a fully covariant way, with the additional benefit of providing a discrete, or quantized description of space and time\index{space and time!quantized description}.

Now, from the perspective of the developments introduced in the first portion of this chapter, it would appear that a quantum-mechanical description of the metric properties of space\index{metric properties of space!quantum-mechanical description}, concerning the universe as a whole, cannot merely involve adjoining a wave function\index{wave function!boundary conditions over superspace} to some boundary conditions defined over superspace, under the assumption that all possible histories compatible with those conditions happen, all at once, as different interfering branches\index{interfering branches!possible histories}. In a time-symmetric context, the purpose of a quantum cosmology\index{quantum cosmology!purpose} would rather be to estimate the probability of observing a global state of intrinsic space curvature\index{intrinsic space curvature!point in superspace} (represented as a point in superspace) when another such global state has been observed at a certain time in the past, by summing-up the (positive and negative) \textit{probabilities} associated with all the different ways by which those two points can be joined together as a result of the global state evolving, once forward and once backward in time, along two possibly distinct trajectories in superspace\index{superspace!two distinct trajectories} for which the local curvature of space\index{local curvature of space!unobservable differences} itself could differ, as long as those differences remain unobservable, that is to say, without irreversible consequences.

Here, again, we face the mystery of the existence of two interfering histories\index{two interfering histories} occurring in parallel, which would appear to merely complicate the causal chain picture of the universe's history by actually requiring bidirectional causality\index{bidirectional causality} to operate in opposite directions along two otherwise similar portions of history. But, even though this aspect of a quantum-mechanical description of the universe is certainly convenient, given that it allows to explain quantum non-locality\index{quantum non-locality!explanation}, it nevertheless remains unexplained. In order to begin to understand why this dual character of quantum reality is not as arbitrary and superfluous as it may seem, one must first examine how it is that causality would operate if there was no advanced portion to the history of the universe\index{history of universe!advanced portion}.

It only became clear to me what the organizing principle is that allows to clarify this situation when I began working on the problem of time travel\index{time travel!problem} and closed causal chains\index{closed causal chain}. It is at this point that I realized that, if the history of the universe was described by one universal causal chain\index{universal causal chain}, freely unfolding in the appropriate configuration space, along one particular direction corresponding to unidirectional time\index{unidirectional time} (say, that along which entropy is growing globally), there would need to be external causes\index{external causes!initial state of universe} that would determine how the universe began to get going along the particular trajectory over which it is found to have propagated in this configuration space (which, for now, may be assumed to be superspace\index{superspace!cause of particular trajectory}, even though, ultimately, one would need to consider a more general kind of configuration space\index{configuration space!non-gravitational degrees of freedom} that would also encode non-gravitational degrees of freedom).

This is a very important point, as an external cause is precisely what must be considered forbidden by the constraint of relational definition\index{constraint of relational definition!physical attributes of universe} of the physical attributes of the universe, which basically implies that there should be no `first cause'\index{first cause!external agent} which would not itself be the effect of an earlier cause and which may, therefore, need to be attributed to some external agent not governed by the causal structure of spacetime\index{causal structure of spacetime}. In fact, the same inconsistency would arise if the condition of continuity of the flow of time\index{condition of continuity of flow of time!particle world-line} along an elementary particle world-line, which was introduced in section \ref{sec:4.3}, was allowed to be violated and therefore this previously discussed constraint can be understood to really be a condition for the \textit{local} continuity of all causal processes\index{causal processes!condition of local continuity}.

The reader may recall the problem associated with so-called knowledge paradoxes\index{knowledge paradox}, that would arise from the viewpoint of unidirectional time\index{unidirectional time} when a time traveler would take a copy of some complex and highly valuable work of art, which happens to exist in the future, back to a time in the past before which it did not yet exist, thereby allowing it to be created instantaneously, without any apparent cause, so that the invention is allowed to exist in the future, which is necessary if it is to be brought back in time. I have explained in section \ref{sec:5.4} that such a phenomenon is not impossible in principle, but is simply very unlikely to occur, because it would actually require entropy to increase in the past direction of time, while the time traveler would be in the process of bringing back information from the future, which would constitute a violation of the second law of thermodynamics\index{second law of thermodynamics!violation}, given that it would involve a decrease of entropy in the future.

What can be learned from such a thought experiment is that, if the phenomenon described here is extremely unlikely, it would not, however, constitute a violation of the fundamental (time-symmetric) principle of local causality\index{time-symmetric principle of local causality}, because it would only involve a diminution of entropy that would be apparent from a unidirectional-time viewpoint\index{unidirectional-time viewpoint}, but would not require a real discontinuity in the flow of information\index{flow of information!absence of discontinuity} along the direction in which the time traveler would be progressing in time. Indeed, as I explained in section \ref{sec:5.3}, it must be recognized that there is no \textit{absolute} distinction between causes and effects\index{causes and effects!absence of absolute distinction} at a fundamental level and this means that the future can influence the past just as much as the past is allowed to influence the future, even in the same portion of history (as long as historical consistency\index{historical consistency!preservation} is preserved), which is what actually happens when an elementary particle\index{elementary particle!backward-in-time propagation} is propagating backward in time (in which case it has the physical attributes of an antiparticle, from a unidirectional-time viewpoint).

But if the present state of the universe\index{state of universe!causally determined cause} was determined by a certain cause (located either in the past or in the future) that is not itself determined by an earlier or later cause\index{causes!belonging to universe} that also belongs to the universe itself, but that would be necessary to set the universe on its course along one particular trajectory in superspace\index{superspace!cause of particular trajectory} (with which is associated one particular, initial state of the universe\index{initial state of universe} and one particular information content\index{information content!trajectory in superspace}), then a real problem would emerge, because, under such conditions, bidirectional causality\index{bidirectional causality!violation} would definitely be violated. Indeed, even if time\index{time!beginning at past singularity} was to actually begin at the moment when matter emerges from the past singularity, in the initial Big Bang state\index{initial Big Bang state!beginning of time}, there would still be a discontinuity in the causal chain trajectory\index{causal chain trajectory!discontinuity} and this is why I argued in section \ref{sec:4.5} that it is not desirable that matter\index{matter!creation out of nothing} be created out of truly nothing at the Big Bang.

But how could one avoid the conclusion that there needs to exist an external cause that would determine the initial (or the final) state of the universe\index{initial state of universe!external cause}, in the arbitrarily far past (or the arbitrarily far future), that is to say, how could one explain what determined the information contained in the extended three-dimensional space-like hypersurface\index{space-like hypersurfaces!information content} that constitutes the starting point along the universal causal chain\index{universal causal chain} that evolved into the present one? I believe that the truth is that we have no choice and that we must admit that a certain hypothesis, which may at first appear gratuitous and arbitrary, actually constitutes an absolutely essential condition that needs to be imposed in order that our quantum-mechanical description of reality\index{quantum-mechanical description of reality!absence of inconsistencies} be free of logical inconsistencies when it is applied to a description of the universe as a whole.

It is at this very precise point that quantum theory\index{quantum theory!most incomprehensible aspects} ceases to be baffling and that its most incomprehensible aspects become essential elements of a fully comprehensible representation of reality. What emerges from the original perspective developed in this chapter is that the history of the universe\index{history of universe!elongated closed causal chain} is nothing but an elongated, \textit{closed} causal chain that unfolds in superspace (or in some generalized configuration space\index{configuration space!matter degrees of freedom} where matter degrees of freedom would also be represented). There is no first cause\index{first cause!absence}. The initial impetus\index{initial impetus} that sets the universe on its course is provided by the universe itself, as all later states of the universe also constitute earlier states along this closed, universal causal chain\index{closed universal causal chain}. The universe\index{universe!cause of present as effect of present} truly brings itself into existence by providing the cause of its own present condition as being nothing but a remote effect of this very same present condition.

Perhaps that you remember my earlier discussion of the closed circuit analogy\index{closed circuit!analogy} from section \ref{sec:5.2}. What I explained is that most electrical circuits are really closed circuits and if they may not seem so under ordinary circumstances, it is simply because the circuits are usually extended in one particular direction and can only be recognized for what they really are by the fact that the cables in which they are confined are always composed of a pair of polarized wires\index{pair of polarized wires}, which betrays the fact that this unique path that seems to extend from source to sink is actually formed of the two branches of a closed circuit\index{closed circuit!two branches with opposite currents} in which the current flows in opposite directions. Well, I believe that one must come to accept as unavoidable that this is a very good analogy of what is described by the quantum-mechanical version of the history of our universe\index{history of universe!quantum-mechanical version}. This history\index{history!closed superspace trajectory} is a closed trajectory in superspace that is stretched to near infinite proportion along the direction relative to which unidirectional time\index{unidirectional time!stretching of superspace trajectory} unfolds, as allowed by the solution I have provided in the previous section to the problem of the origin of the differentiation between space and time\index{differentiation between space and time!origin}.

What I suggested, in the previous section, is that the existence of a time dimension distinct from the other three dimensions of space is an outcome of applying to the initial maximum-density state\index{initial maximum-density state!constraint of global entanglement} of the Big Bang a constraint of global entanglement, as a requirement for the existence of causal relationships\index{causal relationships!components of universe at Planck time} between all components of the universe which were present at the Planck time, which has for consequence that the same unique direction in spacetime is selected throughout the universe for the propagation of effects. But such a distinction between space and time (which is made apparent by the unique signature that must be attributed to the metric of spacetime\index{metric of spacetime!unique signature}) is what allows to consistently describe the history of the universe as consisting of a trajectory in superspace\index{history of universe!superspace trajectory}.

What a time-symmetric, quantum-mechanical description of the same reality allows, then, is for this trajectory to be a `polarized' version of history\index{polarized version of history!trajectory in superspace}, in the sense that it actually consists of two parallel histories\index{two parallel histories!same macroscopic conditions} which share the same observable macroscopic conditions, but whose corresponding segments are being propagated in opposite directions of time. Although this pairing of history and this polarization\index{polarization!history} would remain a complete mystery from a conventional viewpoint, in the context of the above discussion it becomes a natural and essential feature of physical reality that should actually have been expected all along, if only we had recognized that, from the viewpoint of logical consistency, causal self-determination\index{causal self-determination!consistency requirement} is not an optional requirement for the universe.

Indeed, if causality\index{causality!essential conditions for universe existence} is of any relevance to cosmology, it is certainly due to the fact that it imposes two essential conditions on the universe in order that it be allowed to simply exist in any possible way. The first of those two conditions is that all elementary particles\index{elementary particles!causal relationships} present in the universe must be causally related to one another as a result of having been in local contact with one another, over an entire space-like hypersurface\index{space-like hypersurfaces}, when the causal horizon\index{causal horizon!beginning of growth} began to grow in the first instant of the Big Bang. As I explained in section \ref{sec:4.9}, this must be considered necessary in order that all particles\index{particles!components of same universe} actually be components of the same universe. The existence of such a condition, which is responsible for the fact that gravitational entropy was minimum in the initial Big Bang state\index{initial Big Bang state!minimum gravitational entropy} of maximum matter density, is what allows me to assume that the history of the universe\index{history of universe!unique superspace trajectory} is, in effect, described by one unique trajectory in superspace, rather than by multiple, distinct (nonequivalent) and unrelated trajectories, which would really constitute the histories of many different universes, not causally related to one another.

But as I just mentioned, this global entanglement constraint\index{global entanglement constraint} is also responsible for the fact that time\index{time!distinct dimension} actually exists as a dimension distinct from the other three dimensions of space on a global scale; which is responsible for giving rise to the very causal structure of spacetime\index{causal structure of spacetime} (the near-uniform alignment of light-cones\index{light cones!alignment}). What's more, I will argue, below, that this condition is also necessary to explain the quasiclassical nature of reality\index{quasiclassical nature of reality!necessary condition}, under conditions where some dynamic attribute of a quantum system becomes entangled with irreversibly evolving degrees of freedom of the environment in which the system evolves (as must be the case following measurement\index{measurement!entanglement with environment}).

The second condition would then be that which I have just identified and which is that the universe\index{universe!causal self-determination} must be self-determined from the viewpoint of causality. This can be satisfied when the history of the universe\index{history of universe!closed causal chain} consists of a closed causal chain, represented by a closed trajectory in superspace\index{closed trajectory in superspace}, which requires the universe to eventually return to the exact same (but partly unobservable) state in which it currently is, as it evolves along this trajectory. This condition is what explains that it is necessary, in order to obtain the right correlation probabilities\index{correlation probabilities}, to take into account the existence of two otherwise independent histories\index{two independent histories!opposite time directions} evolving in opposite directions of time, which is the distinctive feature of the realistic, time-symmetric interpretation of quantum theory\index{quantum theory!realistic time-symmetric interpretation} developed in the preceding sections of this chapter.

What defines a universe, therefore, is not just the fact that all of its constituent elements (the particles\index{particles!causal relationships}) are causally related to one another despite the \textit{spatial} distances that separate them, but also the fact that the three-dimensional slices of spacetime\index{three-dimensional slices of spacetime!causal relationships} in which those constituent elements are located are all causally related to one another and to nothing else (they form one unique, closed causal chain\index{closed causal chain!unique}). When history\index{history!closed configuration space trajectory} consists of a closed trajectory in the space of all possible configurations, every single space-like hypersurface\index{space-like hypersurfaces!causal contact in time} can be in causal `contact', in time, with both a preceding and a succeeding such hypersurface from the same universal causal chain\index{universal causal chain} and this is what allows all those three-dimensional space-like hypersurfaces\index{space-like hypersurfaces!distances in time} to be causally related to one another, regardless of the `distances' that separate them \textit{in time}. Thus, the multiverse\index{multiverse!ensemble of closed universal causal chains} is not merely the ensemble of all possible, causally independent universes\index{causally independent universes} (those which may be characterized by distinct values of their global states of space curvature and other physical attributes at arbitrarily-chosen times), it is really the ensemble of all possible, universal causal chains which exist as nonequivalent, closed trajectories in superspace\index{superspace!closed trajectories}.

Despite what would seem to be implied by the progress that had already been achieved towards the elaboration of a consistent time-symmetric interpretation of quantum theory\index{quantum theory!consistent time-symmetric interpretation}, even though there appears to be two causally independent, but interfering histories\index{two causally independent interfering histories} to every process, from a cosmological viewpoint there is, in fact, only one history\index{history!uniqueness}, but it feeds back on itself, so as to form a closed causal chain. What happens is that, for some reason to be discussed below, the trajectory in superspace\index{superspace trajectory!causal chain} that corresponds to this causal chain goes through different, but observationally indistinguishable states, once by progressing in some direction of superspace as time goes, and then once again by progressing in the opposite direction (the state vectors\index{state vectors!two parallel portions of history} corresponding to those two parallel portions of history may differ, because the retarded state vector\index{retarded state vector!determination by past conditions} is determined by past conditions, while the advanced state vector\index{advanced state vector!determination by future conditions} is determined by future conditions, but the observable macroscopic constraints\index{observable macroscopic constraints} themselves do not differ).

Thus, there is no quantum system in a state of superposition\index{state of superposition}, going at once through all possible histories. There is one unique history\index{history!closed superspace trajectory}, the details of which remain in part unobservable to any observer, that unfolds as a closed trajectory in superspace (or some generalization of it), subject to the condition that this evolution happens along two mostly parallel trajectories\index{two parallel trajectories!joined together in superspace}, joined together at each extremity, as if two similar histories\index{two histories!opposite chronological orders} were occurring in opposite chronological orders, whose corresponding segments share all observable physical properties. But it is, in effect, only in this particular sense that we may assume history to be unique, because, even though we are always taking part in only one history, two independent histories\index{two independent histories!unfolding at same time}, which differ with respect to most unobservable physical attributes\index{unobservable physical attributes!differences}, are actually unfolding at the same time (in opposite chronological orders), from a \textit{unidirectional-time} viewpoint\index{unidirectional-time viewpoint}.

This interpretation allows to explain the fact that the interfering realities\index{interfering realities!absence of causal contact} are not in causal contact with one another locally, because, even if the two parallel portions of history\index{two parallel portions of history!shared macroscopic conditions} share the same macroscopic conditions, they do not really happen at the same epoch and therefore the particles present in the retarded portion of a process cannot interact with those which are present in the advanced portion of what only appears to be the same process. As a result, one can explain why it is, exactly, that the particles that take part in one possible history of a quantum process do not interact with those that take part in another such history, so that one can avoid the contradiction that would emerge, in the context of a more conventional approach, if it was assumed that all branches of history\index{branches of history!coexistence in same universe} coexist in the very same universe.

It must be clear that while this approach allows the state of the universe\index{state of universe!point along trajectory in superspace}, which is specified by a particular point along the trajectory in superspace, to be characterized by simultaneously well-defined values of conjugate attributes\index{conjugate attributes!simultaneously well-defined values}, observable data would still be subjected to quantum indeterminacy\index{quantum indeterminacy!states of conjugate attributes}, because no observer that is part of the universe can determine what the exact states of both a dynamic attribute and its conjugate counterpart actually are by imposing one particular set of experimental constraints.

Thus, when the extrinsic curvature\index{extrinsic space curvature} of space (associated with the rate of change of the intrinsic curvature of space\index{intrinsic space curvature} along a trajectory in superspace\index{superspace trajectory}) or the particle momenta\index{particle momenta} are determined with arbitrarily good accuracy, the intrinsic curvature of space or the particle positions\index{particle positions} become totally undetermined (from an observational viewpoint), even if there always exists a definite state of both intrinsic and extrinsic space curvature\index{intrinsic and extrinsic space curvature!definite state} at any point along the trajectory in superspace and this is reflected in the fact that the unobserved attribute is allowed to be in distinct, interfering states for the retarded and advanced portions of history\index{retarded and advanced portions of history}. The only difference between this situation and that which would appear to exist from the viewpoint of a more conventional interpretation of quantum theory\index{quantum theory!conventional interpretation} is that it can now be assumed that there does exist a unique state, in the two parallel portions of history\index{parallel portions of history!unique state}, for the unobserved attribute associated with a given set of observational constraints, even though this state can differ for the retarded and advanced portions of history and cannot, as a matter of principle, be subjected to direct experimental determination.

But despite the enormous clarification and simplification which are made possible by the adoption of such a viewpoint, it would remain to explain why it is, exactly, that we are allowed to expect that the states of those dynamic physical attributes\index{dynamic physical attributes!two portions of history} which are under observation do not differ much, most of the time, for those two portions of history, that is to say, we still need to explain why it is that, under such circumstances, the same unique path is shared by the processes which unfold in both the retarded and the advanced portions of the history\index{retarded and advanced histories!same unique path}, as required for the mathematical framework of quantum theory\index{quantum theory!mathematical framework} to be compatible with what is observed on a sufficiently large scale. As I previously mentioned, this is a particular aspect of the quantum measurement problem\index{quantum measurement problem}, or the problem of the origin of the quasiclassical nature of observed reality.

Actually, as I will explain below, it is the fact that the history of our universe consists of one single, closed trajectory in superspace\index{closed trajectory in superspace} that allows quasiclassicality\index{quasiclassicality!emergence} to naturally emerge as a property of the physical world, under appropriate conditions, and therefore it will be apparent that the fact that our world is, in effect, classical on a sufficiently large scale allows to confirm the validity of the hypothesis that the history of the universe constitutes a circular process that feeds back on itself. Thus, if it was not for the closed, or circular nature of quantum-mechanical history\index{quantum-mechanical history!circular nature}, not only would there be no reason to assume that two independent versions of history\index{history!two independent versions} are taking place at the same time, all the time, but it would also be impossible to explain why it is that the retarded and advanced portions of a process\index{retarded and advanced processes!sharing of observational constraints} share the same observational constraints or the same metric properties of spacetime\index{metric properties of spacetime} (because the criterion of consistency\index{criterion of consistency}, specified by the consistent-histories interpretation of quantum theory\index{quantum theory!consistent-histories interpretation} and enforced by decoherence\index{decoherence}, is insufficient to achieve such an outcome, as I explained in section \ref{sec:5.10}).

One thing should be clear already, though, and it is that if the history of the universe consists of a closed causal chain\index{closed causal chain!history of universe}, then the retarded and advanced trajectories in superspace\index{retarded and advanced superspace trajectories} must be smoothly joined at some point in what appears to be the future from the viewpoint of unidirectional time\index{unidirectional-time viewpoint} and also at a certain point in what appears to be the past from the same unidirectional-time viewpoint. As a result, no two points on the universal causal chain\index{universal causal chain} can be absolutely characterized as `earlier' or `later'. But it is also clear that the directions of propagation in time\index{direction of propagation in time!segments of closed causal chain} along two corresponding segments of the closed causal chain (those which appear to be in the same macroscopic state at a given instant of time) have significance merely as relationally defined physical properties (only the difference between those two directions has physical meaning), because no direction of propagation can be attributed absolute significance.

Thus, the direction of time associated with one or another portion of history along the universal causal chain is not a direction in configuration space\index{configuration space}, but a relationally defined property of the universal causal chain itself. Yet, the growth of (gravitational) entropy does allow for the existence of an objectively defined thermodynamic arrow of time\index{thermodynamic arrow of time!objective definition}, to which can be compared the direction of propagation in time along any given segment of the universal causal chain, and it is from the viewpoint of this unidirectional time parameter\index{unidirectional time!parameter} that the universal causal chain\index{universal causal chain!closure} would eventually appear to close, at which point time would come to an end\index{end of time}.

Now, it may appear that the hypothesis that the two parallel superspace trajectories\index{two parallel superspace trajectories!smooth meeting} associated with the retarded and advanced portions of history must be smoothly joined at a certain point in the future could never be proven right, given that, from the unidirectional-time viewpoint\index{unidirectional-time viewpoint}, the closure of the universal causal chain cannot be observed unless it has already occurred, in which case we would no longer be there to acknowledge this fact. But, as I mentioned above, the validity of the theoretical requirement of closure of the universal causal chain\index{universal causal chain!closure requirement} can actually be confirmed by the observation that reality is of a quasiclassical nature, for reasons I will soon explain. The existence of such an end of time, however, must be distinguished from that which would occur as a result of an interruption of the trajectory of the universal causal chain\index{universal causal chain!interruption of superspace trajectory} in superspace, which, despite what one might be tempted to assume, is not a real possibility, given that it would not constitute a simple bifurcation point in unidirectional time\index{bifurcation point!unidirectional time}, but would involve a causal discontinuity\index{causal discontinuity!bidirectional-time viewpoint}, even from a bidirectional-time viewpoint.

One important aspect that needs to be emphasized here is that the situation of a universe whose trajectory in superspace\index{superspace trajectory!closure condition} is submitted to a condition of closure is not the same as the situation of a universe which would evolve, as a result of Poincar\'{e} recurrence\index{Poincar\'{e} recurrence}, to the exact same \textit{observable} state in which it was at an earlier time (a state defined by the same observable macroscopic conditions\index{observable macroscopic conditions!different state vectors}, to which may still correspond two different state vectors), which, in principle, could be satisfied even if the parallel superspace trajectories\index{parallel superspace trajectories!absence of meeting} associated with the retarded and advanced portions of history\index{retarded and advanced portions of history} do not meet at any point along the shared, coarse-grained trajectory\index{shared coarse-grained trajectory} that would take the universe to its earlier observable state. In the present case, it must be assumed that, when the universal causal chain\index{universal causal chain!closure} closes in the future, this will be due to the fact that the retarded portion of history has, by chance, found itself in the exact same state as that in which the advanced portion of the process turned out to be, not just at an observable level, but even for what has to do with the unobservable state of those physical attributes which are the subject of quantum interference.

The evolution along the universal causal chain\index{universal causal chain}, therefore, will not merely take the universe to a state that is similar to that in which it once was, but eventually to the exact same point it once occupied in superspace, from which any further evolution would take the universe into the exact same history through which it once went, despite the random nature of this evolution (because this would actually be the same history). Yet, if an observer is present when the bifurcation point in superspace\index{bifurcation point in superspace!unidirectional time} is reached, she would not be able to experience the same history she once experienced, but in reverse, because what would happen is indeed a reversal of the direction along which the causal chain is unfolding with respect to \textit{unidirectional} time, which means that the thermodynamic arrow of time\index{thermodynamic arrow of time!reversal} would reverse along the causal chain trajectory, while reality can only be experienced in the opposite direction along this trajectory (the same direction as that in which the observer experiences reality in the retarded portion of history).

Under such conditions, both an observer that is part of the retarded process and its counterpart that evolves as part of the advanced process would experience the same reality and each of them would simply cease to exist at the bifurcation point, because consciousness\index{consciousness!thermodynamic process} is a thermodynamic process that necessarily takes place along the direction of time in which entropy is rising globally. But it must be understood that the closure of the universal causal chain\index{universal causal chain!closure in superspace} does not take place in position space, but really in superspace and therefore it does not involve an annihilation of the particles present in the retarded portion of history\index{retarded portion of history} by those present in the advanced portion of history\index{advanced portion of history} and it is not limited by the requirement of energy conservation\index{energy conservation!bidirectional-time viewpoint} that would otherwise need to apply with respect to unidirectional time, because, in such a case, continuity is only required to apply from the bidirectional-time viewpoint.

Thus, the point in the future at which the retarded and advanced trajectories of the universal causal chain\index{universal causal chain!meeting of retarded and advanced trajectories} would meet, from the unidirectional-time viewpoint\index{unidirectional-time viewpoint}, does not need to have extraordinary physical properties and could be any instant of time. Again, though, it must be clear that the closure of the universal causal chain is a phenomenon that takes place in superspace and therefore it may appear to violate the principle of local causality by occurring all at once in position space. Indeed, even if the bifurcation process\index{bifurcation process!universal causal chain} may appear to take place at different times, in distant regions of the universe, from the viewpoint of certain observers, once it happened in one region of the universe it would have to occur in all the other regions, as the condition that is responsible for the continuous decrease of entropy in the past direction of time does not allow for oppositely-directed thermodynamic arrows of time\index{thermodynamic arrow of time!opposite directions} to be present simultaneously in the same universe.

Also, if bidirectional time\index{bidirectional time!extension past initial singularity} can be extended to instants past the initial Big Bang singularity, then the moment in the past at which the universal causal chain\index{universal causal chain!closure past Big Bang singularity} would close would not necessarily need to be that at which the Big Bang itself occurs, but would likely be a time, arbitrarily distant in the past, prior to the Big Bang, when, by chance alone, the retarded and the advanced trajectories would meet in superspace. But it must be clear that the advanced portion of the known history is not the trajectory that unfolds prior to the Big Bang. Both the current history and that which may have taken place (with entropy growing in the opposite direction of time) before the Big Bang (as apparently allowed by certain quantum gravitation theories\index{quantum gravitation theories}) have their own retarded and advanced portions\index{retarded and advanced histories!times before Big Bang} of the same closed causal chain and could actually be very different histories, involving different sets of observable events.

If I believe that it is not \textit{a priori} necessary to assume that the retarded and advanced trajectories\index{retarded and advanced trajectories!meeting at Big Bang singularity} in superspace are likely to meet at the Big Bang singularity it is because, despite the uniformity of the matter distribution and the minimum gravitational entropy that characterizes the initial maximum-density state\index{initial maximum-density state!minimum gravitational entropy}, it remains that, even under such conditions, the unobserved, quantum-mechanically interfering dynamic attributes\index{unobserved dynamic attributes!retarded and advanced states} could be in different states for the retarded and advanced portions of history.

In fact, this is likely to be the case in the context where it must be assumed that, despite the fact that the amount of information associated with the number of elementary units of area\index{elementary units of area!space contraction} present in the vacuum diminishes when space is contracting, the amount of information contained in the microscopic state of the gravitational field\index{microscopic state of gravitational field!information growth} actually grows with the density of matter, for reasons I have explained in section \ref{sec:4.7}. Therefore, the probability of an exact correspondence of the retarded and advanced states\index{retarded and advanced states!probability of exact correspondence} is as small during the first instants of the Big Bang as it is at any other time, because the universe has the same information content\index{information content!invariance} and the same number of microscopic degrees of freedom\index{microscopic degrees of freedom!number invariance} during the first instants of the Big Bang as it has at any other time. It would only need to be assumed that a meeting of the retarded and advanced configuration space trajectories\index{retarded and advanced trajectories!meeting at Big Bang} occurs at the Big Bang if it turned out to be impossible for bidirectional time\index{bidirectional time!extension past initial singularity} to extend past the initial maximum-density state, as would be the case, from a classical viewpoint, in the presence of an initial spacetime singularity.

In any case, if the hypothesis that the universal causal chain\index{universal causal chain!closure requirement} must be closed is justified, then it becomes possible to confirm the validity of the conclusion stated at the end of section \ref{sec:5.8}, to the effect that the sign of energy of the elementary particles\index{elementary particle!backward-in-time propagation} which can be observed to propagate backward in time (with respect to unidirectional time\index{unidirectional time}) in the advanced portion of history\index{advanced portion of history} must be opposite that of the same particles which are propagating forward in time in the retarded portion of history\index{retarded portion of history}, while their sign of action remains unchanged.

Indeed, when the superspace trajectories\index{superspace trajectories!smooth meeting} associated with those two portions of history are smoothly joined at a remote point in the past, as well as in the future, then the particles which propagate forward in time in one portion of history must reverse their direction of propagation in time\index{direction of propagation in time!reversal}, from the unidirectional viewpoint, at both the past and the future bifurcation points\index{past and future bifurcation points}, due precisely to the fact that they do not reverse their direction of propagation from the viewpoint of \textit{bidirectional} time\index{bidirectional time viewpoint}. But this means that the energy signs, which necessarily remain unchanged relative to the direction of time in which those particles propagate (given that there is no reason to assume that their action signs would reverse), would appear to be reversed in comparison with those of the corresponding particles which propagate in the opposite direction of (unidirectional) time in the current portion of history, in agreement with the conventional description of advanced-wave phenomena\index{advanced-wave phenomena!conventional description}.

The only difference between the hypothetical phenomenon of advanced waves, as they are conventionally described, and that of the advanced portion of history\index{advanced portion of history} discussed above, has to do with the fact that the signs of all the non-gravitational charges\index{non-gravitational charge sign!reversal} carried by the particles present in the advanced portion of history would now appear to be reversed, from the unidirectional-time viewpoint\index{unidirectional-time viewpoint}, given that it is explicitly assumed that they remain unchanged, from the viewpoint of bidirectional time, when the trajectory of the universal causal chain\index{universal causal chain trajectory!bifurcation} bifurcates in the future and in the past (which should have been expected, given that the alternative definition of the time reversal operation\index{time reversal operation!alternative definition} I introduced in chapter \ref{chap:3} involves a reversal of charge from the unidirectional viewpoint). This is without consequences, however, because the relevant force fields\index{force fields!reversal of polarities} that provide the experimental conditions observed in the advanced portion of history all have their polarities reversed as well.

The circular nature of history\index{circular nature of history} is also what allows to explain that quantum interferences\index{quantum interferences} do occur, even in the context where we are assuming that the retarded and advanced portions of a quantum process\index{retarded and advanced portions of process} actually take place at two very distant epochs along the configuration space trajectory\index{configuration space trajectory}; which would appear to imply that they cannot exert any effect on one another (locally). I believe that if there are quantum interferences between the many possible paths allowed for the retarded and advanced portions of history, it is because the circular nature of history imposes a condition of continuity on the phase of the wave function\index{condition of continuity!phase of wave function} which is equivalent to that I have identified in section \ref{sec:5.8} when discussing the significance of the negative probabilities\index{negative probabilities} which occur in the context of the proposed realistic, time-symmetric interpretation of quantum theory\index{quantum theory!realistic time-symmetric interpretation}.

Given that what one would need to estimate, ultimately, is the probability of observing a certain history of the universe\index{history of universe!individual sub-processes} that comprises a detailed description of all the individual sub-processes (decoherent or not) which occur in the course of that history, then one must recognize that the phase of the wave function\index{phase of wave function!property of superspace trajectory} is actually a shared property of the unique superspace trajectory that provides the most accurate account of the history of the universe as a whole (or at least of its local space curvature\index{local space curvature!space-like hypersurface} over an arbitrarily large space-like hypersurface).

But, in the context where the entire history\index{history!closed causal chain} consists of a closed causal chain that feeds back on itself, there actually exists a constraint which imposes that all contributions, by intermediary sub-processes, to the variation of the phase of the wave function\index{phase of wave function!cosmological process} associated with the complete cosmological process, be such that they allow the phase to end up, after a complete turn, in the exact same state in which it was at the one point of the trajectory in superspace\index{superspace trajectory!boundary conditions} that constitutes both its initial and its final boundary conditions. If the universe\index{universe!phase of wave function} does, in effect, evolve back to the exact same state in which it once was, then it is certainly appropriate to assume that this state cannot itself be different from what it actually is, even for what regards unobservable physical properties like the phase of the quantum wave function, otherwise the concept would have no real significance.

It had, in fact, already been realized \cite{Wootters-1} that there is a single phase associated with the wave function of the whole universe\index{wave function of universe!single phase}, that is equivalent to one very rapidly rotating clock hand\index{rotating clock hand}. But in the context where time\index{time!periodic phenomenon} itself must be considered to constitute a periodic phenomenon, it follows that the cosmic wave function\index{cosmic wave function!periodic boundary conditions} must be similar to that which applies to quantum systems submitted to periodic boundary conditions (like an electron in orbit around a hydrogen nucleus, whose wave function\index{wave function!electron in orbit around nucleus} must necessarily involve an integer number of wavelengths). There may, thus, be something true to the previously discussed results from canonical quantum cosmology\index{canonical quantum cosmology}, which appear to indicate that the wave function of the universe\index{wave function of universe!stationary} is of the stationary kind, even if, in the present context, this no longer means that time is irrelevant to quantum cosmology\index{quantum cosmology!irrelevance of time}.

In any case, this cosmic wave function phase continuity condition\index{cosmic wave function!phase continuity condition} is what allows me to explain why it is, in effect, appropriate to impose on the unobservable phase of the wave function\index{wave function!unobservable phase} that it does not end up, in the course of an ordinary time-symmetric process\index{time-symmetric processes}, in a state that would be incompatible with that in which it initially was. I believe that such a requirement can be enforced in the context where certain time-symmetric histories\index{time-symmetric histories!negative probability contribution} (with which are associated negative probabilities\index{negative probabilities}) contribute to diminish the probability that a quantum process occurs by following any possible path, as a result of the influence they exert on the very conditions (both initial and final) necessary for their own occurrence.

It appears, in effect, that it is the requirement of continuity of the phase of the wave function\index{phase of wave function!requirement of continuity} along the closed, universal causal chain\index{universal causal chain} that explains that even individual time-symmetric processes\index{individual time-symmetric process!wave-function phase invariance} have a larger probability to occur when they leave the phase of the wave function invariant, because it is necessary to impose on the phase that it remains unchanged in the course of each individual time-symmetric process, even if the real constraint applies to the universal causal chain\index{universal causal chain!closure condition} that describes the evolution of the whole universe and on which is imposed the closure condition, because when one calculates the probability for an individual time-symmetric process\index{individual time-symmetric process!probability of occurrence} to happen, one implicitly assumes that the observable, macroscopic conditions\index{macroscopic conditions!ignored portion of history} imposed on the rest of history (the portion of it that is ignored) are such as to leave the phase of the wave function\index{wave function of universe!assumption of phase invariance} invariant, whenever those which are imposed on the individual process itself leave it unchanged, given that this is the only way to assess the likeliness that a certain quantum process\index{quantum process!probability of independent occurrence} will occur independently from the rest of the history of the universe (which must necessarily be assumed to occur if the process itself does).

Therefore, if the change of wave function phase\index{wave function phase change!individual time-symmetric process} associated with the observable, conditions imposed on an individual time-symmetric process is maximally destructive, then it means that the process cannot occur, because the whole history of the universe\index{whole universe history!destructive interference} in which the local process is embedded could not itself happen (given that it would not leave the phase invariant upon a complete run of the wave function around the closed causal chain\index{closed causal chain!wave function phase invariance}). This is why there are consequences to a variation of the phase that would take place in the course of an individual time-symmetric process\index{individual time-symmetric process!phase change consequences}, even though the condition of invariance actually applies on a global scale (of both space and time). Thus, it is now possible to understand why it is that there are quantum interferences between multiple different histories for ordinary quantum processes\index{quantum processes!origin of interferences}, even in the context where we assume that only one (circular) history actually takes place.

Those are very significant conclusions, given that, in the present context, observation of quantum interferences is the only way by which the advanced portion of history\index{advanced portion of history!observational evidence} can be deduced to exist by an observer that is present in the retarded portion of history\index{retarded portion of history}. Such an explanation of the origin of quantum interferences\index{quantum interferences!origin} would also appear to confirm that quantum non-locality\index{quantum non-locality!topology of configuration space trajectory} is a consequence of the non-trivial topology of the configuration space trajectory (which is here assumed to be the trajectory in superspace\index{superspace trajectory}). Indeed, according to Hans Reichenbach\index{Reichenbach, Hans} \cite{Reichenbach-1} (ch. 1, sec. 12), when faced with unexpected non-local correlations\index{non-local correlations}, one can either invoke `preestablished harmony'\index{preestablished harmony} in the form of instantaneous couplings of distant events that would violate the principle of local causality\index{principle of local causality!violation}, or else recognize that one is dealing with a compact topological structure\index{compact topological structure!periodicity} in which periodicity naturally arises. What I have tried to explain is that the history of the universe\index{history of universe!compact topological structure} is just such a structure and therefore its circular nature is what most naturally explains quantum non-locality\index{quantum non-locality!circular nature of history} as a phenomenon involving the entanglement of quantum phases\index{entanglement!quantum phases}.

Now, as I mentioned in section \ref{sec:5.6}, it has been argued by certain detractors of the earlier, more conventional time-symmetric interpretations of quantum theory\index{quantum theory!early time-symmetric interpretations} that the problem with any such interpretation is that it is not possible to distinguish between situations where interferences among different histories must be assumed to exist and situations where they can actually be ignored (under which conditions must the `handshake' process\index{handshake process!conditions of occurrence} be assume to occur). I have already explained that this erroneous conclusion arises merely when we fail to recognize that decoherence\index{decoherence} must occur, even from the viewpoint of a time-symmetric interpretation of quantum theory\index{quantum theory!time-symmetric interpretation}, under the same conditions in which it would be expected to happen according to a many-worlds interpretation\index{many-worlds interpretation!quantum theory}, despite the fact that the phenomenon has a different meaning in the context of a time-symmetric interpretation. But, as I mentioned in section \ref{sec:5.10}, there are two problems that one must face before one can conclude that decoherence does, in effect, constitute the cause of the quasiclassical character of macroscopic phenomena\index{macroscopic phenomena!quasiclassical character}, even in the context of a time-symmetric formulation of quantum theory.

The first of those problems has to do with the fact that it may never be possible to assume that decoherence\index{decoherence!uncertain irreversibility} itself constitutes a truly irreversible process. There is no reason, in effect, to reject the possibility that, given enough time, the processes giving rise to decoherence could eventually be reversed on an arbitrarily large scale, so that the many degrees of freedom of the environment\index{environment degrees of freedom!entanglement with superposed state} which have become entangled with the superposed state of a quantum system could be submitted to quantum interference\index{quantum interferences!environment degrees of freedom}, even without deliberate intervention and long after a measurement would normally be assumed to have occurred.

It appears that it is merely the improbability of such an evolution that explains that we do not feel compelled to recognize that quantum measurements\index{quantum measurement!non-definitive process} may not be definitive processes and could actually be overturned in the future, which would affect the validity of theoretical predictions concerning ongoing phenomena. One may be tempted to argue that this is not a real problem, because the potential for entropy growth\index{entropy growth!unlimited potential} may be unlimited in the future and this may allow one to expect that, as the effects of a measurement\index{measurement!irreversible spreading of effects} spread irreversibly into an ever larger portion of the environment, the possibility that quantum interferences involving all those correlated variables would occur becomes ever more insignificant.

Indeed, I have provided arguments in section \ref{sec:4.9} to the effect that the growth of gravitational entropy\index{gravitational entropy!unlimited growth} may be unlimited in our universe, due to the presence of negative-energy matter, which would appear to provide support for the conclusion that decoherence\index{decoherence!irreversibility} is truly irreversible, even in the context of a conventional interpretation of quantum theory\index{quantum theory!conventional interpretation}. The problem, however, is that, given an infinite amount of time, even such a continuously decreasing probability may not prevent fluctuations from eventually giving rise to quantum interference\index{quantum interferences!very large scale} on a very large scale. Therefore, it would seem that one cannot avoid the conclusion that decoherence is not definitive, which should have significant consequences at the present epoch.

Now, given that I have argued in section \ref{sec:4.7} that the expansion of the universe\index{expansion of universe!absence of information growth} does not take place with a real growth in the amount of microscopic structure or information (due to the variation of information associated with the diminishing strength of local gravitational fields\index{local gravitational fields!diminishing strength}), it would appear that the probability that the universal causal chain\index{universal causal chain!non-diminishing closure probability} closes at some point in the future (which would occur when the exact unobservable state of all the microscopic degrees of freedom\index{microscopic degrees of freedom!exact unobservable state} in the universe would happen to be the same in the retarded and advanced portions of history\index{retarded and advanced histories!exact same unobservable states}) is not diminishing with time. This conclusion may perhaps appear to be irrelevant to the problem discussed here, but that is not the case, because what it actually means is that one can expect that unidirectional time\index{unidirectional time!future end} will likely end at some point in the future (however far it might be), before a large enough fluctuation allows the ever growing entropy\index{growing entropy!fluctuation to lower value} to decrease sufficiently that it could lead to a reversal of the conditions necessary for the outcomes of quantum measurements\index{quantum measurement outcomes!objective facts} to constitute objective facts.

In other words, what we are now allowed to expect is that the universal causal chain\index{universal causal chain!closure before decoherence reversal} will eventually close in the future, before decoherence has the chance to be reversed on a global scale, which means that decoherence\index{decoherence} does not merely eliminate quantum interferences for all practical purpose, but must be assumed to give rise to classical outcomes of measurement\index{classical outcomes of measurement!decoherence} as a matter of principle, and this conclusion remains valid even in the context were we do not postulate that irreversibility\index{irreversibility!fundamental level} arises at a fundamental level. I believe that this constitutes the decisive argument that allows one to make sense, at long last, of the observation that quantum measurements\index{quantum measurement!definitive outcomes}, once effected, produce definitive outcomes which are never overturned.

The second problem one must confront is perhaps more unavoidable. I have explained in section \ref{sec:5.10} that certain relatively well-known developments \cite{Dowker-1} appear to indicate that the criterion of `consistency'\index{criterion of consistency!persistence of quasiclassicality} (in the sense of a consistent-histories interpretation of quantum theory\index{quantum theory!consistent-histories interpretation}) would not be constraining enough to allow one to expect that the quasiclassical nature of reality would persist following a quantum measurement\index{quantum measurement!irreversible process} (conceived as an irreversible process during which decoherence\index{decoherence!definitive process} is taking place), even when one is allowed to assume that decoherence itself is definitive and is not likely to ever be overturned.

I can now explain why it is that the realistic, time-symmetric interpretation of quantum theory\index{quantum theory!realistic time-symmetric interpretation} I have developed is more appropriate for predicting the emergence of a quasiclassical world\index{quasiclassical world!emergence and persistence} that remains classical once the effects of one or another outcome of a quantum measurement\index{quantum measurement!irreversible propagation of effects} irreversibly propagate into the environment. What holds the key to a complete and effective solution to this particular aspect of the quantum measurement problem\index{quantum measurement problem!effective solution} and to an explanation of the quasiclassical nature of `macroscopic' reality\index{quasiclassical nature of macroscopic reality!explanation} is, once again, the acknowledgment that the hypothesis of closure that applies to the universal causal chain\index{universal causal chain!closure hypothesis} is not optional and must, according to the arguments provided above, be imposed as an absolutely essential consistency requirement. It is only when I recognized the unavoidable nature of this condition that I was able to understand that, in the context where there must exist both a retarded and an advanced portion to every quantum process\index{quantum process!retarded and advanced portions}, additional constraints exist which only become apparent during processes which can be qualified as measurements.

So, what is it, in effect, that characterizes a process that can be described as a quantum measurement? The essential ingredient of decoherence\index{decoherence!irreversibility and dissipation} itself appears to be irreversibility (dissipation to be more specific), but as I mentioned above, decoherence can only be part of the solution. So, what happens, as a consequence of irreversibility, that does not take place under conditions where quantum interferences\index{quantum interferences} exist? To answer this question, it may help to consider what would be necessary for measurement \textit{not} to occur and quantum interference to exist, even after a quantum system\index{quantum system!entanglement with environment} becomes entangled with its environment. It is obvious that what would be required is that the state of the quantum system along with that of the immediate environment to which it has become correlated do not become entangled with an even larger portion of the environment. In other words, there would need to be no traces, in the larger portion of the environment, that would allow one to tell through which history the system and its immediate environment actually went.

Therefore, the point at which irreversibility enters the picture, for what regards quantum measurement\index{quantum measurement!record making}, is through the making of a record of the events involved (conceived precisely as the kind of process during which the effects of one or another of several alternative outcomes of the evolution of a microscopic system is amplified to macroscopic proportions). Only when the state of each physical attribute, whose determination would allow one to tell what the history of the system and its immediate environment was, is submitted to quantum interference\index{quantum interferences!condition for observation} before this information has the time to spread into the environment, can interferences actually be observed. It would, therefore, appear that, if temporal irreversibility\index{temporal irreversibility!elimination of interferences} is, in effect, necessary for the elimination of interferences, it is because the making of a record\index{making of record!irreversible evolution} can only occur when future evolution takes place irreversibly.

What is required for the making of a record\index{making of record!mutually consistent traces} is that one unique event in the past leaves multiple recognizable and mutually consistent traces of its occurrence in the future. A long-lasting record\index{long-lasting record} is one whose mutually consistent traces themselves each produce multiple recognizable and mutually consistent traces in the future that can all be traced back to the same unique original cause in the past. What happens when a quantum measurement\index{quantum measurement!record making process}, conceived as a particular, but general instance of such a record making process, comes into effect, therefore, is that a unique, particular outcome of the evolution of a quantum system produces a recognizable effect on a multitude of other events in the future, which would all have been affected in recognizably different ways had that original outcome been different or nonexistent. What must then be responsible for the elimination of quantum interferences\index{quantum interferences!cause of elimination} that follows decoherence is the fact that a growing number of observable variables become correlated with one unique, specific outcome of the evolution of a microscopic system, while \textit{all} of those variables would have evolved differently if another outcome had been obtained for the same measurement, in the past.

Now, the important point in all of this is that the spreading of effects\index{spreading of effects!position space} does not take place with respect to an arbitrarily-chosen dynamic attribute, but always relative to position space. Indeed, as I have emphasized in section \ref{sec:5.7}, at the most fundamental level, reality appears to consists of elementary particles\index{elementary particles!localization in position space}, which are objects that are localized in position space and which allow the propagation of effects\index{propagation of effects!position space} through local contact, again in position space. There is, thus, something very particular with position space\index{position space!unidirectional causality} for what has to do with unidirectional causality and the irreversible propagation of effects and this is apparent in the fact that the spreading of wave fronts\index{spreading of wave fronts!position space} always occurs in position space and not in configuration space.

The singular status of position space is made even clearer by the fact that the particular boundary condition\index{boundary conditions!cause of time asymmetry} which I have identified as being responsible for the asymmetry of the evolution in time of systems with a large number of independent, microscopic degrees of freedom\index{microscopic degrees of freedom} is a condition that is imposed on the \textit{spatial} distribution of matter in the first instants of the Big Bang. It is the homogeneity of the spatial distribution of positive- and negative-energy matter particles in the initial state of maximum matter density of the Big Bang\index{Big Bang!initial state of maximum matter density} that allows the universe to evolve irreversibly toward a state of larger gravitational entropy\index{gravitational entropy!irreversible growth}, characterized by a greater inhomogeneity of the two matter distributions, as space expands, in the future direction of time. As I explained in section \ref{sec:4.9}, this condition is what allows one to assume that the cosmological horizon\index{cosmological horizon!growth with time}, which limits the scale on which effects were allowed to propagate since the Big Bang, actually grows with time, from the minimum value it had in the initial singularity\index{initial singularity}.

What is allowed to happen, on a smaller scale, as a result of this particular condition, is for an irreversible spreading of effects\index{irreversible spreading of effects} into an ever larger volume of space to take place in the future direction of time, as elementary particles freely propagate in either the retarded or the advanced portion of history\index{retarded or advanced portion of history} (this is particularly apparent in those cases where dissipation\index{dissipation} is involved). In fact, as I explained in the previous section, the same constraint of global entanglement\index{global entanglement constraint} which gives rise to thermodynamic irreversibility\index{thermodynamic irreversibility} is also responsible for allowing time to differentiate from the other three dimensions of spacetime\index{time!differentiation from other dimensions} and therefore for giving rise to the causal structure of spacetime\index{causal structure of spacetime} which is governed by relativity theory and which is responsible for the fact that effects necessarily spread in position space, either forward or backward in time.

But what characterizes \textit{unidirectional} causality\index{unidirectional causality!irreversible spreading of effects} is not only the fact that it operates relative to a unique dimension of spacetime, but also the fact that it does, indeed, give rise to an irreversible spatial spreading of effects in the future direction of time, which is actually what the principle of local causality\index{principle of local causality} is usually considered to be all about. Thus, as time goes, a growing number of independent, microscopic degrees of freedom\index{microscopic degrees of freedom} can be influenced in recognizable ways by unique causes located in the past, while the reverse phenomenon is never observed to happen and this is really a property that is unique to the evolution of position states.

We are now very near a solution to a very old problem. What I have just explained is that the making of a record\index{making of record!quantum measurement} is essential for a quantum measurement to take place and that what it entices is the production of a multiplicity of correlated effects involving very many, otherwise independent variables which could all have evolved differently in the future, had the outcome of this measurement itself been different. A multitude of correlated effects as the outcome of one single quantum measurement\index{quantum measurement!multitude of correlated effects}. It is not very difficult to realize that, as time passes, the difference between the observable consequences of one single past measurement and what would have been the consequences of obtaining a different result for the same measurement becomes ever more significant. But in the context where one recognizes that the universal causal chain\index{universal causal chain!closed superspace trajectory} must, as a matter of principle, form a closed trajectory in superspace, then this remark becomes highly significant.

In a world that would have been quasiclassical on a macroscopic scale until now, if a measurement\index{measurement!retarded state of quantum system} performed on the retarded state of a quantum system was to give rise to an outcome that is different from that which was obtained as a result of a similar measurement\index{measurement!advanced state of quantum system} performed on the advanced state of the same system by a measuring device whose irreversible evolution actually also takes place in the future direction of time, then, as time goes (in the future), an exponentially growing number of independent variables from the environment of the system that takes part in the retarded portion of history\index{retarded portion of history} would be allowed to differ from those of the environment of the same system that takes part in the advanced portion of history\index{advanced portion of history}. This means that the two parallel trajectories in superspace\index{two parallel superspace trajectories!irreversible divergence}, which until now had always been very similar to one another, would begin to diverge in a way that would actually make it \textit{increasingly} less likely that they could ever meet at some point in the future, because of this property of the record making process\index{record making process} which is to produce an accumulation of recognizable changes in the states of an innumerable number of independently evolving microscopic degrees of freedom\index{microscopic degrees of freedom}, as a consequence of one little change in the past.

It is the requirement of closure, that applies to the universal causal chain\index{universal causal chain!closure requirement}, that constrains the future evolution of the retarded and advanced portions of history\index{retarded and advanced histories!absence of observable divergence} to not diverge in any \textit{observable} way, from the unidirectional-time viewpoint\index{unidirectional-time viewpoint}, because, if this condition was not obeyed, the number of independent variables, from both portions of history, that would need to change together in the same recognizable way at some point in the future, so as to allow a meeting of the parallel trajectories in superspace\index{two superspace trajectories!meeting conditions}, would become too large for the closure requirement to ever be fulfilled. As a result, the universal causal chain\index{universal causal chain!observationally identical trajectories} must be stretched into two observationally identical trajectories evolving side by side in superspace, along the unidirectional time direction\index{unidirectional time!direction}, for the whole duration of history, as if two indistinguishable versions of history\index{history!two indistinguishable versions} where taking place in parallel, all the time, without ever interacting with one another.

But the constraint of non-divergence of the retarded and advanced configuration space trajectories\index{retarded and advanced trajectories!non-divergence constraint} need not be any more restrictive than that, because the unobserved aspects of the configuration space trajectories\index{configuration space trajectory!unobserved aspects} do not give rise to the formation of a record and have no irreversible consequences and therefore are not required to correspond, for the two portions of history, by the requirement of closure of the universal causal chain. Quantum interferences\index{quantum interferences!growing unlikeliness} are not forbidden altogether, they merely become increasingly more unlikely as the entanglement of a quantum system with its environment becomes more significant and this is exactly what is required from an observational viewpoint.

It must be clear, however, that despite the unique role played by position space in giving rise to the formation of records\index{formation of records!unique role of position space}, the observed dynamic physical attribute of a quantum system\index{quantum system!observed dynamic attribute} (that which is known with perfect accuracy) is not necessarily always its position. The privileged status of position space\index{position space!privileged status} only means that even when the measured attribute is \textit{not} position, it is nevertheless a \textit{spatial} distribution of macroscopic constraints\index{spatial distribution of macroscopic constraints!quantum measurement} that allows such a measurement to be performed, because it is concerning those constraints that information is available in the form of records\index{records!spatial distribution of macroscopic constraints}.

This means that there is no freedom in deciding which dynamic attribute is classically well-defined in any particular situation where we have knowledge of a specific set of macroscopic conditions (while in fact the necessary conditions for classical definiteness\index{classical definiteness!necessary conditions} are always present for one and only one dynamic attribute, as I mentioned in section \ref{sec:5.10}). On the other hand, the dynamic attribute of a quantum system about which only a \textit{minimum} amount of information is available, as a result of the existence of records about the position states of various parts of a measuring device (the macroscopic observable conditions\index{macroscopic observable conditions}), is the attribute that may go through any possible trajectory (not necessarily in position space) during both the retarded and the advanced portions of history\index{retarded and advanced histories!unobserved attribute}, thereby giving rise to quantum interferences\index{quantum interferences!unconstrained dynamic attribute}.

What's important to understand is that, given that it is always for position space observables\index{position space observables} that the making of a record\index{making of record} of past events can take place, then it follows that the constraint of non-divergence of the retarded and advanced superspace trajectories\index{retarded and advanced superspace trajectories!non-divergence}, or more specifically of the observable retarded and advanced states\index{observable retarded and advanced states} of a time-symmetric quantum process\index{time-symmetric quantum processes}, is a constraint that always applies only to the dynamic attribute\index{dynamic attributes} of a system whose state is restricted to a subset of values as a result of being submitted to experimental conditions of such a nature. But such a constraint does not only give rise to non-interfering outcomes of measurement\index{measurement outcomes!non-interfering} following decoherence\index{decoherence}, but really to a quasiclassical evolution\index{quasiclassical evolution!persistence in time} that persists in time for the same family of consistent histories\index{family of consistent histories} (the physically relevant set of histories\index{physically relevant set!histories}).

It had already been remarked, in effect, that, from a phenomenological viewpoint, decoherence, even as it is usually conceived, appears to select position as the relevant collective observable\index{relevant collective observable!position} (that which becomes correlated with the microscopic system under study), at least for mechanical systems, in the presence of dissipation\index{dissipation}. It was conjectured that this is merely a consequence of the fact that the laws of physics (particularly in a quantum field theoretic context) are invariant under a change of reference system. In the present context, however, this could only be understood to mean that the selection of position as the relevant collective observable for decoherence is indeed a consequence of the fact that unidirectional causality\index{unidirectional causality!position space} (the irreversible spreading of effects) operates in position space, because what emerges, as a result of relativistic invariance\index{relativistic invariance}, is the causal structure of spacetime\index{causal structure of spacetime}, which, under appropriate conditions (when evolution is irreversible), gives rise to unidirectional causality and therefore to the existence of persistent records of past\index{persistent records of past} events.

The fact that the phenomenon of dissipation merely consists in one particular instance of irreversible spreading of effects\index{irreversible spreading of effects!position space} in position space would, therefore, appear to confirm that it is the closure requirement (that must be applied to the universal causal chain\index{universal causal chain!closure requirement}) that allows quasiclassicality\index{quasiclassicality!emergence and persistence} to emerge and to persist for those attributes of a quantum system whose states are restricted by macroscopic conditions\index{macroscopic conditions!spatial nature} of a spatial nature.

There should be no doubt that the existence of such an objectively defined, preferred basis is absolutely necessary from an observational viewpoint, because if none arose, it would be impossible to determine what causes the persistence of the quasiclassical nature of reality (even under the assumption that the universal causal chain\index{universal causal chain!closure requirement} must close at some point). Indeed, if reality was classical with respect to one family of consistent, coarse-grained histories\index{family of consistent coarse-grained histories} at a given time and then relative to another such family at a later time, as would be allowed in a more conventional context, then this reality would no longer appear classical from the first viewpoint after the transformation has occurred.

But when quasiclassicality is the outcome of imposing a requirement of closure to the universal causal chain and the spreading of effects in position space\index{spreading of effects in position space!irreversibility} occurs irreversibly, it follows that a preferred basis\index{preferred basis!representation of quantum states} (a preferred choice of dynamic attribute to represent quantum states) is naturally selected for the elimination of quantum interferences\index{quantum interferences!elimination} and it is from the viewpoint of the records which are available concerning the constraints (of a spatial nature) that select this dynamic physical attribute\index{dynamic physical attributes!selection constraints} that the world necessarily appears to remain classical following a measurement. I believe that those conditions, therefore, allow to satisfy Dowker\index{Dowker, Fay} and Kent's\index{Kent, Adrian} requirement for an additional, purely quantum-mechanical principle that would allow one to select a particular set of (consistent) histories\index{set of consistent histories!selection principle} as being of particular physical significance, without having to rely on solipsistic arguments\index{solipsistic arguments}.

So here we are, having actually explained why it is that, in practice, one never observes quantum superpositions involving macroscopic states of measuring apparatuses\index{states of measuring apparatuses!absence of superpositions}. If we never experience histories in which a cat is alive and dead all at once (following an experiment of the Schr\"{o}dinger's cat type\index{Schr\"{o}dinger's cat-type experiment}), it is because, if it was not the case that the cat was either alive or dead in the retarded and the advanced portions of history\index{retarded and advanced portions of history} alike, this would change the future in ways which would render impossible an eventual meeting of the retarded and advanced trajectories in (some extended version of) superspace\index{retarded and advanced superspace trajectories!meeting}, while this is necessary for the universe\index{universe!causal self-determination} to be self-determined from the viewpoint of causality. The identified constraint simply makes it extremely unlikely (as unlikely, in fact, as the growth of entropy that took place while the retarded and advanced states became distinct is important) that such an evolution could ever be experimentally deduced to have occurred.

The essential characteristic sought by Von Neumann\index{Von Neumann, John} and which would differentiate a measuring apparatus from the system it measures is simply the possibility that exists for the measuring device\index{measuring device!record of own history} to generate a record of the particular history it goes through, which has decisive consequences in the context where reality is a causal chain that must close at some point in the future. From that viewpoint, of course, quantum interference\index{quantum interferences!macroscopic states} of macroscopic states is not completely impossible, but even if such an unlikely phenomenon was to happen, then one would not \textit{see} a cat that is both alive and dead at the same time (despite the fact that one would then have to be in a state of superposition as well), because one is always confined to directly perceive only the portion of history (either the retarded or the advanced part) in which one happens to be located and in any such a history there is always a unique set of causally related facts\index{causally related facts!unique set}.

But this does not mean that a state of superposition involving a macroscopic portion of reality would have no apparent consequences, because if the advanced state was to become distinct from the retarded state on a large scale, then the estimation of transition probabilities\index{transition probabilities} for future processes would be affected in dramatic ways from the viewpoint of those observers which are part of the process, while it is under way, which means that their future would actually become unpredictable unless they assume that such a divergence from classicality\index{divergence from classicality} has indeed occurred and this is how they would actually gain knowledge of the existence of such a distinction between the current retarded and advanced states\index{retarded and advanced states!distinction}. But if the condition of closure of the universal causal chain\index{universal causal chain!closure condition} has the consequences I'm expecting it would have, then the observers which were part of such a process would not be allowed to remember through which history they went on either the retarded or the advanced portion of the process\index{retarded and advanced processes!memory of history}, after this macroscopic quantum interference\index{macroscopic quantum interference} is over, as otherwise this knowledge could spread into the environment\footnote{
This observation cannot constitute the basis of an alternative explanation of thermodynamic time asymmetry\index{thermodynamic time asymmetry!alternative explanation|nn}, because, if one does not assume that there exists a constraint for the retarded and advanced states\index{retarded and advanced states!constraint of non-divergence|nn} not to diverge that is made necessary by the independent condition of low initial gravitational entropy\index{low initial gravitational entropy!independent condition|nn}, which from my viewpoint is responsible for time irreversibility\index{time irreversibility|nn}, then one has no reason to expect that the retarded and advanced portions of history\index{retarded and advanced histories!macroscopic state convergence|nn} should converge back to the same macroscopic state after having diverged on a large scale and this means that our memory of the particular history that actually took place would not need to vanish and therefore its persistence would not need to be correlated with a history where entropy grows in the future.}.

The point that is perhaps the most difficult to understand concerning what I believe would qualify as an appropriate account of experiments of the Schr\"{o}dinger's cat type\index{Schr\"{o}dinger's cat-type experiment}, in which there would be quantum interferences\index{quantum interferences!macroscopic states} between macroscopic states, is that in the final state of such an experiment the cat would have to be neither in a live-with-no-poison-in-its-blood state, nor in a dead-with-poison-in-its-blood state, even though it is true that the animal may no longer exist in a recognizable form, because this is not the same as a cat that is dead due to having absorbed the poison released as a result of the measurement on the quantum particle having produced a negative result, even if it does mean that the cat may no longer be alive in the final state. What is required, therefore, is that it be impossible to tell, from the information that is present in the final state, whether the cat was killed by the poison or whether it might have been alive without any poison in its blood before the final measurement was performed that would have revealed the existence of quantum interferences, so that, even if the cat no longer exists in the final state, it would not be correct to say that it was killed as a result of the particular outcome of the particle disintegration experiment\index{particle disintegration experiment}.

In any case, given that no complex macroscopic system, such as a cat, was ever subjected to any reproducible experiment in which quantum interferences would have been observed, then it would appear that the requirement of closure, which I suggest must be imposed on the universal causal chain\index{universal causal chain!closure requirement}, is well-founded, because it does allow one to expect that macroscopic objects, which can never be completely isolated from their environment, should practically never be found in states of quantum superposition.

An additional advantage of the approach proposed here is that it allows one to understand how it is that historical consistency\index{historical consistency} would be enforced, in the context where a classical time travel experience\index{classical time travel experience} would occur and the course of history could potentially be altered so as to give rise to an alternate future. Indeed, when the effects of a future measurement\index{effects of future measurement!backward-in-time propagation} can be propagated backward in time (as a result of the existence of an advanced portion of history) and there is a condition for the retarded and advanced portions of history\index{retarded and advanced histories!sharing of observable conditions} to share the same observable macroscopic conditions in the future (so that the universal causal chain\index{universal causal chain} can close at some point), it follows that the present can only be influenced by the future to be such as to give rise (through forward-in-time causation) to classical outcomes of measurement, rather than to a retarded state that would differ from the advanced state.

In other words, the present cannot be influenced by the future in such a way that it would be likely to evolve toward a different future. Thus, even if the second law of thermodynamics\index{second law of thermodynamics!temporary violation} could be temporarily violated in a local region of space, perhaps as a result of a formidably improbable fluctuation, and information about the future would become available, no violation of historical consistency\index{historical consistency!absence of violation} could arise that would involve observable phenomena.

It is, therefore, the circularity of the causal process\index{causal process!circularity} and the existence of a thermodynamic arrow of time\index{thermodynamic arrow of time} arising from the requirement that all the elementary particles\index{elementary particles!causal relationships} in the universe be causally related to one another which allow the consistency of history to be preserved in the context of a time symmetric interpretation of quantum theory\index{quantum theory!time symmetric interpretation}. The conclusion that historical consistency would always be preserved in a quantum-mechanical context, therefore, need not depend on the hypothesis that all histories are followed all at once and that a `splitting of branches'\index{splitting of branches!alternative reality} occurs whenever an alternate reality is produced, as is often assumed, because it can be derived much more naturally by recognizing that for the universe\index{universe!causal self-determination} to be causally self-determined, its history must consist in a closed causal chain\index{closed causal chain}. Yet it does seem appropriate to assume that it is quantum theory that would ultimately be responsible for the impossibility of even a classical time travel paradox\index{classical time travel paradox!impossibility}, as I suggested in section \ref{sec:5.4}, because the limitation discussed here is made unavoidable as a result of the time-symmetric nature of quantum reality\index{quantum reality!time-symmetric nature}, which enforces consistency on a global scale (as necessary for the existence of non-local correlations\index{non-local correlations}) without violating the principle of local causality\index{principle of local causality}.

Now, it must be clear that, even though no record of the future can exist in the context where entropy only rises in this direction of time, reality would necessarily remain quasiclassical relative to the same family of consistent, coarse-grained histories\index{consistent coarse-grained histories!family} in the past direction of time as well, because if entropy rises continuously from as far back in time as the first instants of the Big Bang, then the condition imposed on future evolution by the requirement that the universal causal chain\index{universal causal chain!closure requirement} closes at some point in the future imposes that history\index{history!quasiclassicality} be quasiclassical right back to the initial singularity\index{initial singularity}.

Thus, if the property of quasiclassicality\index{quasiclassicality!entire history} is required to be observed for the entire duration of history, as a result of the constraint that applies on future evolution, then despite the fact that no record of the future\index{future!absence of records} exists which would constrain past evolution to remain quasiclassical (even in the context where the universal causal chain\index{universal causal chain!past closure} must also close at some point in the past), the condition will nevertheless also apply to past history, as a result of the condition that applies on the future, as long as entropy is actually growing in the future throughout this entire history. This is why we are allowed to expect that a unique past does exist that is compatible with our observation of the existence of mutually consistent records\index{mutually consistent records!unique past} of past events.

But if bidirectional time\index{bidirectional time!extension past initial singularity} extends past the initial singularity following a hypothetical quantum bounce\index{quantum bounce}, then it would be the condition that the universal causal chain\index{universal causal chain!closure requirement} closes at some point in the \textit{past}, on the other side in time of the initial singularity, that would require history to remain quasiclassical in the past (and therefore, again, also in the future), right from the instant at which matter emerges from the `initial' singularity, because entropy\index{entropy!growth in past} would then be growing in the past (for reasons I have discussed in section \ref{sec:4.9}) and records\index{records!about future} would only exist about the future (which would then be similar to our past, from a thermodynamic viewpoint).

Just as is the case for the future, it is not possible to say when it is exactly that a meeting of the retarded and advanced configuration space trajectories\index{retarded and advanced trajectories!past meeting} would occur in the past\footnote{
One should note that it is not possible to assume that the universal causal chain\index{universal causal chain!closure at Big Bang|nn} closes at the Big Bang and yet that there is a history taking place in reverse, prior to the initial singularity, otherwise the meeting of the retarded and advanced trajectories in superspace\index{retarded and advanced superspace trajectories!meeting|nn} would no longer have any meaning, even for the future, because, when the condition of closure of the universal causal chain\index{universal causal chain!closure condition|nn} would be met, history could nevertheless continue to take place as if nothing had actually happened.},
 because the only condition that must be imposed on the closure of the universal causal chain in the past is that it does not occur before the time at which the initial singularity\index{initial singularity} is formed (on our side in time of the Big Bang), because global entanglement\index{global entanglement!time of occurrence} must have had the time to occur, as otherwise the universe\index{universe!ensemble of causally interrelated components} would not have been allowed to exist as an ensemble of causally interrelated components.

\bigskip

\noindent Up to this point, I have only discussed the emergence of quasiclassicality\index{emergence of quasiclassicality} as it arises in a conventional quantum-mechanical context, where the metric properties of spacetime\index{metric properties of spacetime!unique background} constitute a common, unique background over which both the retarded and advanced portions of a process\index{retarded and advanced portions of process} unfold, either with or without quantum interference\index{quantum interferences!making of record}, depending on whether or not the particular history of the particles propagating over this background space gives rise to the making of a record. But what right do we have to assume that the metric properties of spacetime\index{metric properties of spacetime!retarded and advanced histories} themselves should always be shared by the retarded and advanced portions of history, if all other dynamic physical attributes can, under appropriate circumstances, differ and interfere for the two parallel trajectories of the universal causal chain\index{universal causal chain!two parallel trajectories}? If the other macroscopic conditions\index{macroscopic conditions!retarded and advanced histories} which are shared by both portions of history are so determined merely as a result of the fact that they give rise to an irreversible spreading of effects\index{irreversible spreading of effects}, then why would the metric properties of spacetime which are shared by both portions of history be simply given once and for all in their classical form, instead of being subjected to the same rules that govern the other physical attributes of our universe?

The truth, of course, is that the metric properties of space\index{metric properties of space!classical definiteness} are not always classically well-defined and that they may differ and interfere for the two parallel portions of history. It is well understood already that macroscopic changes to the gravitational field\index{gravitational field!decoherence} are a very potent way by which decoherence can be triggered, as confirmed by the fact that the motion of planets\index{motion of planets!absence of quantum interferences} is one of the phenomena for which the absence of quantum interferences is the most conclusive and the most persistent, while it was shown that this is not unrelated to the magnitude of the gravitational fields involved. Now, I have already mentioned that, in a quantum-gravitational context, what we would be dealing with are situations where the intrinsic curvature of space\index{intrinsic space curvature!retarded and advanced histories} would be allowed to differ in the retarded and advanced portions of history. I may now add that this would occur whenever information, in the form of records, would only exist about the extrinsic curvature of space\index{extrinsic space curvature!information} associated with the rate of change of intrinsic curvature along the actual trajectory that is followed in superspace\index{superspace trajectory}.

The intrinsic and extrinsic curvatures of some region of a space-like hypersurface\index{space-like hypersurfaces!intrinsic and extrinsic curvatures} are the quantum-gravitational equivalent of position and momentum and therefore they constitute conjugate physical attributes\index{conjugate attributes!intrinsic and extrinsic curvatures} whose states cannot be determined together with arbitrarily high precision using one unique set of experimental constraints. But this does not mean that all histories involving distinct intrinsic space curvatures\index{intrinsic space curvature} are followed all at once when the extrinsic curvature\index{extrinsic curvature!precise knowledge} is known with high precision, but merely that, under such conditions, the intrinsic curvature\index{intrinsic space curvature!retarded and advanced histories} may be different for the corresponding retarded and advanced portions of history, because information about the actual history is available (in the form of records) only for the extrinsic curvature.

The situation we normally experience (outside the quantum-gravitational regime\index{quantum-gravitational regime}) is one where the curvature of space\index{curvature of space!classical definiteness} in general is classically well-defined (information exists about both the intrinsic curvature\index{intrinsic space curvature!rate of change} and its rate of change) and there are no quantum interferences\index{quantum interferences!different states of space curvature} arising from the curvature of space being potentially different for the retarded and advanced portions of a process\index{retarded and advanced portions of process} (even when space is not flat locally), as is necessary for conventional quantum theory\index{conventional quantum theory} to provide a viable description of reality. But that need not always be the case and indeed, in situations where we would try to determine the extrinsic curvature of space with a very high degree of precision, by measuring the rate of change of the gravitational field\index{gravitational field!rate of change} over a very small time interval, then the intrinsic curvature of space\index{intrinsic space curvature!quantum interference} would be subjected to quantum interference, as its state would no longer be constrained to be the same in the retarded and advanced portions of history, for reasons I already mentioned.

Under such conditions, it would no longer be possible to estimate transition probabilities\index{transition probabilities!non-uniqueness of metric properties} while using one unique set of metric properties, that is to say, by assuming the existence of one single classical spacetime over which particles would propagate in both portions of a process and it would be necessary to take into account the possibility that the metric properties of spacetime, themselves, could evolve differently in the two portions of history, along with other unobserved dynamic attributes\index{unobserved dynamic attributes}. To determine which metric properties are likely to emerge upon observation, one would then need to take into account the existence of quantum interferences\index{quantum interferences!histories of space curvature} between the many possible histories of space curvature. When interference\index{interferences!constructive or destructive} would happen to be constructive, a given curvature would have more chances to be observed and when interference would be destructive, the very boundary conditions\index{boundary conditions!unlikeliness of existence} necessary for the observation of such a curvature would be unlikely to have existed in the past, or to eventually exist in the future.

From such considerations, it transpires that, if time\index{time!quantum superpositions} itself can be subjected to quantum interferences or superpositions, it is only in the sense that, on a sufficiently small scale, time may flow faster, or slower, locally, for the two portions of a quantum-gravitational process\index{quantum-gravitational process}, due to the fact that the curvature of space\index{curvature of space!two portions of history} may not be the same in both portions of history and may therefore give rise to differing durations for otherwise similar propagation processes. But given that the constraint of global entanglement\index{global entanglement constraint} that is responsible for selecting the particular signature of the metric of spacetime\index{metric of spacetime!particular signature} that gives rise to a universally valid distinction between time\index{time!differentiation from other spacetime dimensions} and the other three dimensions of spacetime only applies to the initial state at the Big Bang, it may be possible for the light-cone structure\index{light-cone structure!short-scale alteration} to be altered to such an extent that the causal order of events\index{causal order of events!reversal along time-like interval} would be reversed along a time-like interval, on a sufficiently short scale, in the context where the gravitational field\index{gravitational field!quantum indefiniteness} itself can be subjected to quantum indefiniteness, because the existence of closed time-like curves\index{closed time-like curves} is only forbidden on a time scale for which thermodynamic time asymmetry\index{thermodynamic time asymmetry} is required to apply.

It is not true, though, that there is no definite space and time in the quantum-gravitational regime\index{quantum-gravitational regime!space and time}. A unique curvature of space\index{curvature of space!uniqueness} does exist throughout history, only, it can differ for the two corresponding portions of history along the universal causal chain\index{universal causal chain!two corresponding portions of history}, to the extent that there may, in fact, no longer be a simple correspondence between those two portions of history\index{two portions of history!absence of correspondence} on a very small scale. Reality always remains a unique, closed causal chain\index{closed causal chain!reality}, even though, on a smaller scale, the regularity of the superspace trajectory\index{superspace trajectory!regularity} may be altered, given that the metric properties of spacetime\index{metric properties of spacetime!quantum randomness} and the gravitational field may no longer remain unaffected by the inherent randomness of quantum-mechanical evolution, which is then allowed to give rise to a divergence of the retarded and advanced trajectories in superspace\index{retarded and advanced superspace trajectories!divergence}, as long as no record exists regarding what those metric properties\index{metric properties!existence of record} actually are.

What is significant for a quantum-mechanical description of gravitation and space curvature\index{space curvature!quantum-mechanical description}, from the viewpoint of the developments introduced in the first part of this section, is that there must be a level at which the intrinsic space curvature\index{intrinsic space curvature!absence of superposition} cannot remain superposed (as it may be on the quantum-gravitational scale\index{quantum-gravitational scale}) and must give rise to a quasiclassical evolution\index{quasiclassical evolution!intrinsic curvature} and this turning point would be determined by the existence of information concerning the metric properties of spacetime\index{metric properties of spacetime!existence of information}.

It is, in effect, precisely when the effects on the propagation of elementary particles of a particular curvature of space\index{curvature of space!irreversible spreading of effects} irreversibly spread into the environment and gives rise to the formation of mutually consistent records\index{mutually consistent records!history} of a particular history, that the metric properties\index{metric properties!quasiclassical evolution} involved must begin to evolve quasiclassically, because the requirement of closure of the universal causal chain\index{universal causal chain!closure requirement} can only be satisfied when such an evolution is observed, just as is the case in a more conventional context and this means that temporal irreversibility\index{temporal irreversibility!emergence of classical spacetime} is an essential condition for a classical spacetime structure to emerge. It is only when the state of the gravitational field\index{gravitational field!observable state} becomes observable that it is no longer subjected to interference effects and that it is no longer allowed to affect the propagation of matter particles differently for the retarded and advanced portions of a process\index{retarded and advanced portions of process}.

It would, therefore, appear that the existence of a decoherent spacetime\index{decoherent spacetime} is itself dependent on the existence of unidirectional time\index{unidirectional time}, which emphasizes just how important it is that there exists an independent constraint, of the kind I have previously identified, for the emergence of temporal irreversibility\index{temporal irreversibility!emergence}, because, in a quantum-gravitational context, when the irreversible character of time itself does not emerge from the underlying theory, decoherence\index{decoherence!emergence of classical spacetime} cannot alone give rise to the classical spacetime structure.

What will be very important for the argument that will be developed in the concluding section of this chapter is the observation that, if random fluctuations of the metric properties of space\index{metric properties of space!random fluctuations} exist which have no observable effects of the kind that would require the gravitational field\index{gravitational field!retarded and advanced histories} to actually have the exact same configuration in both the retarded and the advanced portions of history, then those fluctuations might be allowed to exert an unexpected influence on the propagation of elementary particles\index{propagation of elementary particles!unexpected influence}, even on a scale well above that at which gravitation becomes as strong as the other interactions.

\section{A possible role for gravitation\label{sec:5.13}}

I must immediately warn the reader that the developments that will be the subject of this concluding section of the present report will probably be considered more speculative than other portions of my analysis and I would not myself consider such a judgment entirely inaccurate. Yet I believe that it is important to explain what I have learned concerning how it can be that, even when a quantum system is submitted to the same macroscopic boundary conditions\index{macroscopic boundary conditions} from one trial to another, many different possibilities are allowed for the one particular (unobserved) time-symmetric history\index{unobserved time-symmetric history} of the system that actually happens in the course of any given process. Despite the fact that this discussion comes last, it is actually based on results I had obtained in the earliest portion of my research program, while I was still working on the problem of elaborating a generalized, classical theory of gravitation\index{generalized gravitation theory} that would describe the interaction of positive- and negative-energy matter.

It is by pure chance that, while I was searching for a paper in the immense science and engineering library at McGill University\index{McGill University}, at the very beginning of my research career, I came upon an article in a very old volume of an obscure research journal that sought to explain the randomness of quantum measurement results\index{quantum measurement results!randomness} as being caused by perturbations attributable to the interaction of a quantum system with a background of gravitons\index{graviton!background} present in its environment. As I now understand, this was a particular instance of \textit{classical} hidden-variable theory\index{classical hidden-variable theories} which was inadequate mainly as a result of the fact that it was incompatible with the requirements imposed by quantum entanglement\index{quantum entanglement} and non-locality\index{quantum non-locality}.

Yet, for some reason, I had the strong intuition that the idea that gravitation was involved in explaining certain aspects of the quantum-mechanical description of reality was generally valid and should be further explored. This imperative remained in the back of my mind as a guiding principle as I explored other problems in fundamental theoretical physics\index{fundamental theoretical physics!problems} and even though I soon realized how such a proposal could be made viable, it is only much later that I came to understand that there is actually something unavoidable with the hypothesis that gravitation must become an integral element of a truly consistent formulation of quantum theory\index{quantum theory!consistent formulation}.

In the previous section I suggested that quantum theory\index{quantum theory!incompleteness}, as it is currently interpreted, is incomplete, given that it does not explicitly require history\index{history!closed universal causal chain} to be described by a closed, universal causal chain, while, as I have explained, such a concept is essential if we are to obtain a realistic theory\index{realistic theory!quantum theory} that allows for the emergence of a maximum quasiclassical domain\index{maximum quasiclassical domain!emergence}. But at this point, it was still possible to argue that the current formalism of quantum theory\index{quantum theory!current formalism} (in its most appropriate form) is compatible with this more complete version of the theory. However, given that the interpretation I have proposed is dependent on the assumption that there exists a unique reality\index{reality!uniqueness}, and in a certain sense, a unique history\index{history!uniqueness} behind all quantum-mechanical processes (a hypothesis which is necessary in order to maintain agreement with the uniqueness of the outcomes of quantum measurements\index{quantum measurement!uniqueness of outcome}), then it transpires that if our understanding of the theory is to be considered complete, one cannot avoid having to examine how this unique unobserved history\index{history!causal determination} may come to be determined from a causal viewpoint.

What should be clear, first of all, is that, while the closure requirement that must be imposed on the universal causal chain\index{universal causal chain!closure requirement} is constraining enough to imply that classical outcomes must follow measurements, the decoherence process\index{decoherence process!absence of unique measurement outcome} does not select one unique outcome of measurement, but rather leaves all potentialities on an equal footing. Thus, it should be clear that it is not decoherence or the closure requirement imposed on the universal causal chain that require that only one outcome be observed following a measurement on a previously interfering dynamic attribute of a quantum system\index{quantum system!interfering dynamic attribute}.

The uniqueness of the outcomes of quantum measurements\index{quantum measurement!uniqueness of outcome}, however, does impose on the reality\index{reality!uniqueness in between measurements} that must exist in between observations that it be unique. The problem is that if this unobserved reality is unique, while it is also allowed to vary from one process to another under unchanged experimental conditions, then it would seem that something essential remains unexplained by the theory. As a result, here again, one must face the possibility that quantum theory\index{quantum theory!incompleteness} is incomplete, but now in a way that would appear to require that it be reformulated. Indeed, even in the context of the realistic, time-symmetric interpretation of quantum theory\index{quantum theory!realistic time-symmetric interpretation} I have proposed, it would appear that the question of completeness can only be positively answered once one allows for a further extension of the formalism from the viewpoint of which the uniqueness of the unobserved portion of history\index{unobserved portion of history!uniqueness} would not constitute an additional problem, but would rather provide a hint as to what goes on when an elementary particle propagates in the space of its unobserved physical attributes.

Those remarks become particularly significant once one recognizes that it is not really possible to explain the uniqueness of measurement results\index{measurement results!uniqueness} themselves by simply postulating that the interfering branches of history\index{interfering branches!history}, which are usually assumed to occur all at once in the same universe, are all actualized together following measurement, but no longer interfere with one another, because this would not allow one to explain the persistence of quasiclassicality\index{persistence of quasiclassicality} that is observed to follow all measurements.

But I have also explained that it would be inappropriate to argue that a dynamic attribute\index{dynamic attributes!quantum indefiniteness} that is indefinite, in the quantum-mechanical sense of the word, could be objectively indefinite, in the sense that it would not satisfy the requirements of scientific realism\index{scientific realism!requirements} in any possible way. It would, therefore, appear necessary to conceive of a unique reality\index{reality!uniqueness} of some kind, such as that which emerges from the above discussed analysis, where, even in the absence of direct observation, particles always follow unique, but possibly different paths in the retarded and advanced portions of history\index{retarded and advanced histories!unique paths}. In such a context, however, the question necessarily arises as to what causally determines which unique unobserved path\index{unobserved paths!causal determination} is actually followed by a particle in between measurements?

You may recall that I have argued in section \ref{sec:5.10} that the unpredictability of quantum measurement results\index{quantum measurement results!unpredictability} cannot be a consequence of the measurement process itself. It rather seems necessary to assume that there is already randomness before a particle meets a detector, while it is still propagating in the two unobserved portions of history, and what remains unexplained is the variable nature of this evolution, which applies even for physical systems prepared in the exact same way. What is it that determines the particular history of a certain physical attribute\index{physical attributes!determination of unobserved history} that takes place in between measurements and which must merely be compatible with the outcomes of those measurements? What I have realized is that, in order to answer this question, it is necessary to recognize that, in its present form, quantum theory\index{quantum theory!idealization} is merely an idealization which must be reformulated to give rise to a more elaborate, but statistically equivalent model, in which the unique unobserved history\index{unobserved history!unobservable random causes} that takes place in the absence of observational constraints would be determined by fundamentally unobservable, random causes, whose existence is inevitable and does not have to be postulated on purpose in order to solve the above described problem.

A related question one may ask is whether the concept of objective chance\index{objective chance}, which is usually assumed to be implied by the fundamental unpredictability of quantum measurement results\index{quantum measurement results!fundamental unpredictability}, itself constitutes an appropriate notion, in the context of a realistic interpretation of quantum theory\index{quantum theory!realistic interpretation}? In other words, if objective indefiniteness\index{objective indefiniteness} is to be rejected, must one also reject the associated concept of objective chance? The conclusion to which I have arrived is that this depends on what we mean by objective chance.

If we are asking whether the unpredictability of measurement results can be circumvented given a more precise assessment of the microscopic state of a quantum system, then the answer would definitely be no. But if what we understand by objective chance is the idea that the unique unobserved path of a quantum system\index{quantum system!unique unobserved path} might be `determined' by nothing at all, instead of being the outcome of fundamentally unobservable causes\index{causes!fundamentally unobservable}, that is to say, if we are asking whether it is possible for an unobserved, variable feature of reality to have no identifiable cause, then the answer could only be provided in light of what we already know about reality at the level where it can be observed and by taking into account any possibility that there may be for such a variable feature to actually be causally determined (in the time-symmetric sense of the word). Only if we decide that an absence of causes\index{causes!acceptability of absence} is not physically unacceptable and if we can be confident that no influence exists that would provide such unobservable causes, can we argue that such a \textit{strong concept of objective chance}\index{objective chance!strong concept} is still applicable at the most fundamental level of description of physical phenomena.

It is often suggested that the concept of objective chance conflicts with common sense, but that this merely reflects another failure of our intellect to grasp the essentially distinct and counter-intuitive nature of quantum reality\index{quantum reality!counter-intuitive nature}. Again, however, I would like to argue that this is not all that there is and that, from the mere viewpoint of logical consistency, there is actually something problematic with assuming that a reality can differ and yet that such a difference need not be the result of any known cause, even of a fundamentally unobservable nature. What is easy to overlook is that, when we assume that a difference could exist that would have no `cause', then we actually allow for a violation of the requirement that the physical attributes of the objects which are present in our universe are to be describable by referring only to aspects of reality which are an integral part of this universe.

What I mean is that, if one assumes that it is acceptable for certain variable aspects of reality\index{variable aspects of reality!absence of identifiable causes}, which would exist beyond the observable portion of physical phenomena, to have no identifiable causes (even of a random nature) originating from within the universe in which those phenomena arise, then it may no longer be possible to avoid the conclusion that those variations actually are the product of external intervention, which would simply mean that our universe\index{universe!incomplete instance of reality} is an incomplete instance of reality. I believe that a physical model that would offer a complete account of what happens inside any given universe must, therefore, avoid postulating an absence of causes for variable aspects of that reality.

There is, thus, something rational in our aversion for a reality that would differ without any identifiable (even if potentially unobservable) causes, that is to say, there are good motives to doubt that a strong concept of objective chance\index{objective chance!strong concept} is relevant to our description of physical reality. No distinctive feature of our universe should have as a cause `nothing'. If events are, in general, related, in statistically significant ways, to other events of a similar nature through what we call causality\index{causality!relationships between events}, then we are justified to expect that there should be no event that would be related to something of an entirely different nature which we call nothing, but which could actually be anything at all.

That does not mean, however, that we have to reject the notion that physical reality\index{physical reality!fundamental unpredictability} is fundamentally unpredictable, as I already mentioned, because even a causally determined world would, in the context of the existence of closed causal chains\index{closed causal chain} and backward-in-time causation, involve an irreducible randomness\index{irreducible randomness!backward-in-time causation}, given that the cause of an event\index{cause of event!effect of same event} can be influenced backward in time by this very same event, despite the fact that no information is allowed to flow backward concerning that future event (so that it necessarily remains unpredictable), as I explained in section \ref{sec:5.4}. What this means is that, even if unobservable causes\index{unobservable causes!random but causally determined reality} were to be found to exert an influence on unobserved aspects of a quantum process, reality would remain fundamentally random and not just unpredictable, even if it is causally determined in every way. This is the exquisite beauty of time-symmetric causality\index{time-symmetric causality!non-deterministic}: it allows for causal determination without giving rise to complete determinism\footnote{
Such a conclusion would seem to confirm that a time reversal operation\index{time reversal operation!whole universe|nn} that would apply to the present state of the whole universe defined over a given space-like hypersurface\index{space-like hypersurfaces|nn} would not necessarily give rise to the exact same history in reverse, but could potentially give rise to an entirely different and genuinely unpredictable evolution, as I suggested in section \ref{sec:4.6}.}.

What allows the wave function\index{wave function!deterministic evolution} to evolve deterministically, but only until a measurement is performed, even in the context where one must assume that the underlying reality is randomly evolving, is the fact that we are dealing with a unique reality\index{reality!uniqueness} for which what happens in the future contributes to determine what happens in the past. In such a context, the outcome of a measurement performed on a quantum system at time $t_2$ can change what happens to the system as far back as the time $t_1$ when the system was prepared, which allows the evolution that takes place immediately after $t_1$ to agree with any possible outcome of that future measurement, despite the fact that this evolution is taking place randomly on a local level. Thus, the fact that the wave function evolved deterministically until time $t_2$ (at which decoherence\index{decoherence!state vector reduction} took place and the state vector was reduced), is not incompatible with the hypothesis that the system evolved randomly before that measurement, because this random evolution was influenced all along by what happened at a later time, when the measurement was performed, given that, from the viewpoint of time-symmetric causality\index{time-symmetric causality!constraints from future}, the system is required to obey constraints which may be determined by what happens in the future, as a result of that measurement.

But once the quantum potentialities\index{quantum potentialities!actualization} are actualized, at time $t_2$, the observed outcome of measurement is only required to be compatible with what \textit{actually} happened in the past and this is what explains that randomness becomes apparent. Quantum evolution\index{quantum evolution!perpetual randomness} is always random, but a real change is actually occurring when a measurement takes place, which makes it seem like this is where randomness originates, because right until the measurement is actually performed, multiple different outcomes are still possible and the system appears to evolve indifferently toward all those final states, all at once, and this is what makes this evolution appear deterministic, as it always happens in the same way from an observational viewpoint and must always be compatible with whatever could happen in the future.

In any case, as long as the unobservable causes\index{unobservable causes!variable paths of particles} which may explain the variable paths followed by quantum particles in between measurements remain unobservable, reality must remain unpredictable from the viewpoint of all observers. I would therefore object suggesting that the validity of a causal theory based on the realistic conception of reality\index{realistic conception of reality!causal theory} developed in the preceding sections of the present chapter would imply that the wave function\index{wave function!incomplete description of quantum state} provides an incomplete description of the state of a quantum system, because the wave function does provide the most complete account of how a system evolves as a result of the observable constraints exerted on it, only this still leaves us with a \textit{classically} indefinite state for the physical attributes which are left unconstrained by the macroscopic experimental conditions which apply to both the retarded and the advanced portions of a process\index{retarded and advanced portions of process}. I believe that this provides an important clue as to the nature of those unobservable random causes\index{unobservable random causes}.

What must be clear, also, is that the existence of such unobservable random causes, obeying the principle of local causality\index{principle of local causality}, is not ruled out by the phenomenon of quantum entanglement\index{quantum entanglement} in the context of the realistic, time-symmetric interpretation of quantum theory\index{quantum theory!realistic time-symmetric interpretation} I have proposed, because, even if the trajectories of two particles forming an entangled pair\index{entangled pair!unobservable causes} are separately influenced by those unobservable causes, when there is as much influence of the future on the past as there is of the past on the future, it is possible for the two entangled particles to evolve so as to enforce the non-local conditions imposed on the wave function\index{wave function!non-local conditions} as a result of their entanglement. This is why one must differentiate such an approach from the naive, realistic interpretations of quantum theory\index{quantum theory!naive realistic interpretations} which were proposed in the past and which can be appropriately called \textit{classical} hidden-variable theories\index{classical hidden-variable theories}.

Here it is the very concept of an objective reality\index{objective reality!different concept} that differs in essential ways, given that we are now dealing with a universal causal chain\index{universal causal chain} that feeds back on itself so as to give rise to two interfering, but otherwise causally independent versions of history\index{two interfering but causally independent histories} for each and every process, to which must be \textit{independently} applied the principle of local causality\index{principle of local causality!two versions of history}. Thus, the unobservable causes\index{unobservable causes} are allowed to exert different effects on the retarded and advanced portions of history\index{retarded and advanced histories!entangled particles} along the trajectories followed by any of two entangled particles, but given that the two portions of both processes interfere with one another quantum mechanically, as a result of being part of the same closed causal chain\index{closed causal chain}, then it becomes possible for non-local correlations\index{non-local correlations!independently evolving systems} to exist between the outcomes of measurements performed on the two otherwise independently evolving systems.

From my viewpoint, the reality that is causally determined\index{causally determined reality!non-classical uniqueness} is not unique in the classical sense and this is what allows even a causal theory\index{causal theory!compatibility with entanglement} to agree with the requirements imposed by the quantum entanglement of distant particles, without requiring complex and arbitrary non-local mechanisms of a conspiratorial nature, in contrast with all classical hidden-variable theories\index{classical hidden-variable theories}. The only difference between a causal theory involving unobservable causes of the kind I suggest may need to be considered and the orthodox interpretation of quantum theory\index{quantum theory!orthodox interpretation} would therefore be that, from my viewpoint, not only is it possible to assume that there can indeed exist a unique reality\index{reality!uniqueness}, even in between measurements of a certain physical attribute for which quantum interferences\index{quantum interferences} are observed, but it is also possible for this unique reality to be causally determined, as all observed phenomena. One of the advantages of this particular approach would, therefore, be that it naturally agrees with a much larger body of observational evidence, which clearly indicates that when there is an effect, there usually is a cause, even if its consequences may sometimes remain unpredictable.

\bigskip

\noindent In the second chapter of this report I have developed a generalized framework for relativity theory\index{relativity theory!generalized framework} that helped confirm the validity of the hypothesis that spacetime curvature\index{spacetime curvature!interaction} really is a consequence of the existence of an interaction. Indeed, once one recognizes that local inertial reference systems\index{local inertial reference system!observer dependence} and the curvature of space are dependent on the energy sign of the particles experiencing them, then one must accept that there is no such thing as a metric structure of spacetime\index{metric structure of spacetime!interaction} existing independently from the nature of the interaction that determines its physical properties. Thus, through an analysis of the \textit{quantum-mechanical} concepts of bidirectional time\index{bidirectional time!quantum-mechanical concept} and negative energy\index{negative energy!quantum-mechanical concept}, I was allowed to develop an improved, classical theory of gravitation\index{gravitation!improved classical theory}, which helped confirm the validity of the hypothesis that the metric properties of space and time really are the product of an interaction.

What I would like to discuss, now, is the possibility that a better understanding of the microscopic physical properties of \textit{classical} gravitational fields\index{classical gravitational field!microscopic properties}, which emerges from the fact that those fields are the outcome of a quantized interaction\index{quantized interaction!gravitation}, could provide the basis for a reformulation of quantum theory\index{quantum theory!reformulation} that would make it compatible with a causal interpretation of the randomly variable nature of the unobserved portion of quantum mechanical processes. It must be clear, however, that the approach I will propose does not constitute a replacement for current quantum gravitation theories\index{quantum gravitation theories} (such as loop quantum gravity\index{loop quantum gravity}), but merely provides a complementary contribution to the field, similar in scope to my derivation of the number of discrete degrees of freedom\index{discrete degrees of freedom!particle inside black hole} that characterize the state of matter particles inside a black hole (which was discussed in section \ref{sec:3.10}) or to my explanation of the emergence of a universal time variable\index{universal time variable!emergence} in the initial Big Bang state\index{initial Big Bang state} (which was discussed in section \ref{sec:5.11}).

The approach I have followed is actually the exact opposite of one that would be based on the many-worlds interpretation of quantum theory\index{quantum theory!many-worlds interpretation}, because, instead of positing a deterministic evolution\index{deterministic evolution!multiple simultaneous histories} involving multiple simultaneous histories, I'm assuming a random evolution involving one causally determined history\index{random evolution!unique causally determined history} (forming a closed causal chain\index{closed causal chain}). Thus, from my viewpoint, one no longer needs to assume that reality\index{reality!theoretical determinism and observational randomness} is deterministic from a theoretical viewpoint, but random from an observational viewpoint, which, all by itself, certainly constitutes significant progress.

What should be clear already is that, if quantum systems\index{quantum system!same conditions but different paths} do not always go through the same unobserved path when submitted to the same macroscopic conditions, this can only mean that, even when an optimal experimental characterization of the evolution of a physical system is available, it necessarily leaves aside fundamentally unobservable, but causally significant aspects of the process. It is only the fact that, from a conventional viewpoint, it appeared impossible to assume the existence of such causes without allowing violations of the principle of local causality\index{principle of local causality!violation} to occur that explains that we came to believe that such an otherwise more consistent viewpoint was no longer viable, even though a realistic, time-symmetric interpretation of quantum theory\index{quantum theory!realistic time-symmetric interpretation} of the kind I have proposed actually makes such an approach perfectly acceptable.

Once one recognizes that, as a matter of principle, no information could ever be obtained concerning the causes which may determine the random paths of unobserved dynamic attributes\index{random paths of unobserved attributes!unobservable causes}, then one must conclude that no violation of the uncertainty principle\index{uncertainty principle!violation} could occur as a result of the existence of such causes. It is also only under the incorrect assumption that additional information could be obtained about this unobserved layer of reality (that is not already accounted for by the quantum state of a system), that one would have to conclude that information\index{information!non-conservation} may no longer be conserved and that violations of the second law of thermodynamics\index{second law of thermodynamics!violation} may arise.

Now, even though it has long been my opinion that both the classical theory of gravitation\index{classical gravitation theory} and quantum field theory\index{quantum field theory} must be altered \textit{prior} to being integrated into a quantum theory\index{quantum theory!gravitational interaction} of the gravitational interaction, it is only after I realized that our understanding of \textit{classical} gravitation leaves aside important aspects which cannot be ignored in a quantum mechanical context that I began to appreciate the fact that the quantization issue does not concern merely the general theory of relativity\index{general relativity theory}, but that its resolution probably necessitates that quantum theory\index{quantum theory!necessity of reformulation} itself be reformulated so as to take into account those properties of the gravitational field\index{gravitational field!quantized nature} which arise as a consequence of its very quantized nature.

Thus, while I do recognize that the classical theory of gravitation\index{classical gravitation theory!quantization} must be subjected to a quantization procedure on the scale at which this interaction becomes as strong as the other known interactions, I also believe that the quantized nature of gravitation\index{gravitation!continuous force field approximation} would have consequences on a much larger scale where this interaction can still be appropriately described by using the approximation of a continuous force field associated with the curvature of spacetime\index{curvature of spacetime}. To be more specific, I believe that conventional quantum theory\index{conventional quantum theory!spacetime curvature} must come to integrate a certain element of spacetime curvature, even under those conditions where we currently assume the existence of a flat and invariant spacetime\index{flat and invariant spacetime}. For that purpose gravitation\index{gravitation!deterministic spacetime background} must no longer be assumed to merely be involved in defining an observable, deterministically evolving spacetime background, but must be understood to exert an unobservable random influence\index{unobservable random influence!gravitation} that contributes to causally determine the unobserved paths followed by elementary particles\index{unobserved paths of particles!causal determination}, even in the absence of local matter inhomogeneities, as long as it remains impossible to tell, based on the results of any measurements, what was the actual path taken by a particle as a consequence of those perturbations.

Those requirements can be fulfilled once one acknowledges that the trajectory of the universal causal chain\index{universal causal chain!superspace trajectory} in superspace, that describes the local evolution of the intrinsic or extrinsic curvature of space\index{intrinsic and extrinsic space curvature} for the whole universe, can differ for the retarded and advanced portions of history\index{retarded and advanced histories!different superspace trajectories}, as a result of the fact that unobservable random fluctuations of the classical gravitational field\index{classical gravitational field!unobservable random fluctuations}, which are attributable to the very quantum mechanical nature of this interaction, are affecting the trajectories of matter and radiation particles in the space of those dynamic physical attributes\index{dynamic attributes!unobserved trajectories} which are not the subject of direct observation.

What must emerge, therefore, is a theory where spacetime\index{spacetime!observable macroscopic constraints} does not merely provide an additional set of observable macroscopic constraints, as a result of the irreversibility of its effects, but where the local inertial reference systems\index{local inertial reference system!unobservable fluctuations} may be allowed to fluctuate in unobservable ways that may differ for the two time-reversed portions of a process, which are otherwise submitted to the same macroscopic conditions. The important point, here, is that, even though such fluctuations would indeed remain unobservable, they would nevertheless be physically significant, given that they would actually allow to explain what determines the unique and possibly distinct paths which are followed during the retarded and advanced portions of every quantum process\index{retarded and advanced processes!unique paths} by the dynamic attributes of a quantum system\index{quantum system!unobserved dynamic attributes} which are not submitted to observation.

From that viewpoint, even the classical spacetime continuum\index{classical spacetime continuum!uniform and static background} over which the unobservable paths\index{unobservable paths!quantum particles} of quantum particles are assumed to unfold in the absence of local matter inhomogeneities would no longer constitute a well-defined, uniform and static background, but would actually fluctuate as much as the particle trajectories themselves. This proposal is merely an extension of the general-relativistic idea according to which it is no longer possible to speak of a situation where there is an absence of gravitational field\index{gravitational field!absence}. Einstein\index{Einstein, Albert} himself reflected on the irrelevance of such a notion by noting that even in those situations where the metric of spacetime\index{metric of spacetime!Euclidean} is Euclidean and no mass is present nearby, there is still a gravitational field, only it is a field that does not vary with position (while an absence of gravitational field would require that there exist no metric properties at all). Here, the idea is that, even when it would appear, from a superficial, macroscopic viewpoint, that the gravitational field\index{gravitational field!invariance in space and time} does not vary with position or time, in fact it still exerts a decisive, random influence on the trajectories of elementary particles\index{elementary particle trajectories!random influence of gravitation} entering the sum-over-histories formulation of quantum theory\index{quantum theory!sum-over-histories formulation}.

Such a hypothesis is necessary due to the fact that even conventional quantum field theory\index{quantum field theory!integration of gravitation} implicitly takes into account the existence of gravitational interactions all along the unobservable trajectories of quantum particles, given that it assumes the relevance of a well-defined spacetime background\index{spacetime background!quantum field theory} over which the matter particles propagate. But, as Lee Smolin\index{Smolin, Lee} once remarked, it is difficult to imagine how a dynamical theory of spacetime\index{dynamical theory of spacetime} (such as a background-independent quantum theory of gravitation\index{quantum gravitation theories!background-independence}) could actually be derived from a theory where the geometry of space is assumed to be fixed (such as conventional quantum field theory\index{quantum field theory!fixed spatial geometry}).

What I'm suggesting is that, once we recognize that gravitation exerts a decisive influence, even on the scale of ordinary quantum theory\index{quantum theory!influence of gravitation}, then it is also necessary to recognize that, under such conditions, the gravitational field\index{gravitational field!absence of uniformity and constancy} is not constrained to have the properties of uniformity and constancy that we usually attribute to it in the absence of local matter inhomogeneities (when large measures of action are involved and classical physics is a suitable approximation). The gravitational field\index{gravitational field!omnipresence} definitely is omnipresent and does exert an effect at every `point' along the unobserved trajectories of elementary particles\index{elementary particles!unobserved trajectories} (including gravitons), but I believe that what the random nature of the paths followed in the unobserved retarded and advanced portions of any process\index{retarded and advanced processes!unobserved random paths} indicates is that this classical gravitational field\index{classical gravitational field!non-uniform and nondeterministic} cannot be required to be completely uniform and to evolve deterministically on a microscopic scale, but must rather be allowed to fluctuate in ways that could differ for the retarded and advanced portions of the process\index{retarded and advanced processes!gravitational field fluctuations}, when the existence of those random fluctuations would have no observational consequences.

If there is no valid motive to reject the possibility that the gravitational field\index{gravitational field!unobserved fluctuations} may so fluctuate in the absence of observations, then what one would have to recognize is that it is the local inertial reference systems\index{local inertial reference system!unpredictable variations} which are allowed to vary unpredictably with position and time. In fact, I believe that this should have been expected, even independently from any consideration of a quantum-mechanical nature, given that, from a Machian viewpoint\index{Machian viewpoint}, local inertial reference systems\index{local inertial reference system} are an effect of the gravitational interaction of the particles experiencing them with the ensemble of matter in the universe and such effects must necessarily vary unpredictably with both position and time, as the matter distribution itself is not perfectly unchanging and uniform over the entire universe and throughout history, even if, on the average, such fluctuations should necessarily cancel out, due to the large number of individual interactions involved.

What happens is that, when the trajectory of a particle\index{particle trajectories!multiple near simultaneous interactions} is the outcome of multiple, near simultaneous, quantized interactions, such as is the case with ordinary Brownian motion\index{Brownian motion}, then there necessarily arise fluctuations in the number of those interactions that produce a momentum variation in one direction, that are not necessarily matched by the fluctuations that simultaneously occur in the number of those interactions that produce a momentum change in the opposite direction and this must give rise to small random variations in the equilibrium of forces\index{equilibrium of gravitational forces!random variations} acting on the particle (which would here be the gravitational forces that determine the local inertial reference systems).

Thus, if the local inertial reference systems\index{inertial reference system!equilibrium of gravitational forces} which determine the trajectory of a particle with a given sign of energy must ultimately be conceived as being the outcome of an equilibrium in the sum of gravitational forces attributable to the interaction of this particle with all the matter in the universe with the same sign of energy, as I explained in section \ref{sec:2.4}, then one is certainly justified to assume that the unobservable trajectories of elementary particles are randomly influenced by the presence of fluctuations (also unobservable) in this equilibrium of gravitational forces, given that gravitational forces are themselves conveyed by elementary particles and must, therefore, fluctuate. The crucial point is that this would be true even in the context where the approximation of a classical spacetime continuum\index{classical spacetime continuum!approximation} would still be valid (and the metric would remain Euclidean locally), given that we are not concerned here with individual quantum interactions, but with fluctuations in a very large number of such interactions taking place nearly simultaneously.

What makes it possible for such unobservable fluctuations in the local equilibrium of inertial gravitational forces\index{gravitational forces!equilibrium} to have decisive consequences on the evolution of quantum systems, even outside the quantum-gravitational regime\index{quantum-gravitational regime}, is the fact that, even though the gravitational interaction\index{gravitational interaction!weakness} is very weak, inertia, as a gravitational phenomenon, constitutes a very significant influence for elementary particles, given that it is the outcome of the gravitational interactions which are taking place between a given particle and all the other matter particles present in the universe, whose number largely compensates the very small probability that the particle absorbs or emits a graviton\index{graviton!probability of absorption or emission} in the course of an ordinary quantum process. In such a context, it would appear that it is merely the fact that the random fluctuations of the classical gravitational field\index{classical gravitational field!random fluctuations} which determine the unobserved trajectories of elementary particles cancel out on the scale of action at which ordinary quantum theory itself becomes irrelevant that allows the metric properties of spacetime\index{metric properties of spacetime!deterministic evolution} to be described as deterministically-evolving under ordinary circumstances.

If those considerations are valid, it would then mean that what one needs to formulate is a realistic, time-symmetric version of stochastic gravitational field theory\index{stochastic gravitational field theory!time-symmetric version} (based on the generalized gravitational field equations\index{generalized gravitational field equations} introduced in section \ref{sec:2.14} and in accordance with the requirement of closure of the universal causal chain\index{universal causal chain!closure requirement}) that would apply to the retarded and advanced portions of every quantum mechanical process\index{quantum mechanical process!retarded and advanced portions}, independently. For this purpose, it is necessary to recognize that the classically well-defined, uniform and static spacetime background\index{uniform and static classical spacetime background} which is assumed to exist in the absence of local matter inhomogeneities merely constitutes an approximation that must emerge from a more accurate description where unobservable random fluctuations of the gravitational field\index{unobservable random gravitational field fluctuations} are present all along an unobserved particle trajectory.

The classical description can, therefore, be expected to break down on the action scale associated with ordinary quantum phenomena, where random fluctuations of the metric properties of spacetime are unavoidable. What explains that such fluctuations can usually be ignored is the fact that it is precisely on such a scale that they can be expected to remain unobservable, while they must cancel out, for the most part, when larger measures of action are involved. A more adequate formulation of quantum field theory\index{quantum field theory!more adequate formulation}, that would integrate the semi-classical description of gravitational fields\index{gravitational field!semi-classical description} envisaged here, would allow for random fluctuations in a medium that remains classically well-defined locally and would only break down on the quantum-gravitational scale\index{quantum-gravitational scale}, where it can be expected that the approximation of a classical spacetime continuum\index{classical spacetime continuum!approximation} is no longer valid.

This means that there are actually three levels of applicability to a theory of the gravitational field\index{gravitational field theory!three levels of applicability}, because the intermediary, semi-classical level, where gravitation\index{gravitation!semi-classical level} is usually assumed to be irrelevant, actually also involves this interaction in a decisive way. On such a scale, gravitation\index{gravitation!merger with quantum theory} may already be considered to merge with quantum theory, but merely in the sense that fluctuations of a quantum-mechanical origin must now apply to the classical gravitational field\index{classical gravitational field!quantum fluctuations}, while quantum evolution\index{quantum evolution!causal determination} becomes causally determined as a consequence of the very gravitational nature of the forces that determine the local inertial reference systems\index{local inertial reference system!gravitational forces} to which elementary particles are submitted, even in the absence of observable, local perturbations of the curvature of spacetime\index{curvature of spacetime!local perturbations}.

What makes this hypothesis significant is the universal nature of the gravitational interaction and the fact that it is allowed to affect not only the propagation of all matter particles, but also that of the particles which mediate their interactions, including the gravitational interaction itself, without having to refer to a pre-existing background spacetime\index{pre-existing background spacetime}, given that this is the interaction that determines the very metric properties of the spacetime\index{metric properties of spacetime} over which all particles propagate.

Now, if random local fluctuations of the metric properties of spacetime\index{metric properties of spacetime!unobservable fluctuations} actually occur, which remain unobservable, then they would have effects which would be indistinguishable from temporary violations of the conservation of momentum and energy\index{conservation of momentum and energy!temporary violations}, given that energy would be exchanged with the gravitational field, that would be unaccounted for classically. I believe that this is what explains that virtual processes\index{virtual processes!energy and momentum conservation violations}, like the ordinary processes of particle interaction, involve such violations of energy and momentum conservation, which are allowed to occur merely as long as they remain within the limits of quantum uncertainty\index{quantum uncertainty!violations of energy and momentum conservation}, that is to say, as long as they remain unobservable (even though they are absolutely necessary to explain the kind of phenomenon involved).

Even from a semi-classical viewpoint, the reality of a particle's existence\index{particle existence!dependence on local gravitational field} may depend on the presence of a local gravitational field or acceleration (think about the Unruh effect\index{Unruh effect} for instance) and in such a case all that matters is that once the presence of a particle is actually measured by a detector, even when this is made possible as a result of an exchange of energy with the gravitational field, then this event must become an established fact that is not dependent on the position or the state of acceleration of an observer. What's different, from the viewpoint of the approach advocated here, is that the virtual particles mediating an interaction\index{interaction mediating particles!real particles} can now be considered to be as real as ordinary matter particles, because what differentiates them is merely the fact that they do not exist permanently, with invariant energies, but merely as a result of energy exchanges with the randomly fluctuating classical gravitational field\index{fluctuating gravitational field!energy exchanges}.

Anyhow, if the non-measurable violations of energy and momentum conservation\index{energy and momentum conservation!non-measurable violations} which are allowed by quantum indeterminacy\index{quantum indeterminacy} are taking place as a result of undetectable exchanges of energy with the randomly fluctuating gravitational field, this would explain why it is that only the conservation of energy, momentum, and perhaps also angular momentum\index{energy and momentum conservation!explanation of violations} is allowed to be violated in such a way, while the electric and other non-gravitational charges of elementary particles (the static attributes\index{static attributes!rigorous conservation}) are always rigorously conserved, despite quantum uncertainty.

In this context, the fact that the quantum indefiniteness\index{quantum indefiniteness of position!momentum dependence} of a particle's position diminishes with the magnitude of its momentum would also appear all the more natural, given that a particle with a larger energy can be expected to interact with more gravitons\index{graviton!simultaneous interactions} simultaneously and therefore to be less affected by individual interactions, as if it was experiencing a reduced level of fluctuation in the equilibrium of gravitational forces\index{gravitational forces!equilibrium} attributable to all such interactions (which may help explain why the variation of the phase of the wave function\index{phase of wave function!energy and mass dependence} associated with the propagation of elementary particles is dependent not only on the energy of the particles involved, but also on their mass, even when gravitation is the only macroscopic constraint involved, as is the case in the context of the classical neutron interferometer experiment\index{classical neutron interferometer experiment} in a gravitational field).

From my viewpoint, the fact that quantum indefiniteness in momentum\index{quantum indefiniteness in momentum!cause of scale dependence} rises as we consider increasingly smaller distances would also appear all the more natural, given that it can be expected that random fluctuations of the classical gravitational field\index{classical gravitational field!random fluctuations} would rise as we consider shorter space intervals, over which the quantized nature of this interaction exerts a more significant influence.

To avoid confusion, however, it is necessary to understand that, despite the fact that the degree of randomness to which are submitted elementary particles as a result of the existence of unobservable fluctuations in the classical gravitational field\index{classical gravitational field!unobservable fluctuations} may depend on the magnitude of their energy (which determines the frequency of the wave function\index{wave function!frequency}), quantum interferences\index{quantum interferences} cannot be considered to be an aspect of the gravitational field itself, because random fluctuations in the metric properties of spacetime\index{metric properties of spacetime!random fluctuations} merely explain what determines the unique, unobserved trajectories which are followed in the retarded and advanced portions of history\index{retarded and advanced histories!determination of trajectories}, while it is still the constructive or destructive nature of the interferences associated with a complete time-symmetric process\index{time-symmetric processes!constructive or destructive interferences} which determines whether those trajectories are likely to be followed, when one takes into account the requirement of invariance imposed on the phase of the wave function\index{phase of wave function!invariance requirement}, in the context where the universal causal chain\index{universal causal chain!closure requirement} must be assumed to be closed.

An interesting outcome of this particular approach is that it allows one to more easily understand why it is that photons and other massless particles\index{massless particles!superluminal velocities and curved trajectories} are allowed to have non-measurable (but theoretically mandatory) velocities larger or smaller than the normal speed of light in a vacuum and to travel along curved trajectories on a small scale (as Feynman diagrams for radiative corrections\index{Feynman diagrams!radiative corrections} so appropriately illustrate), because when one takes into account the existence of unobservable, random fluctuations of the metric properties of spacetime, it is still possible to assume that massless particles in a given energy eigenstate\index{energy eigenstate} always travel along straight lines at their normal $c$ velocity, as long as one recognizes that this propagation takes place along the geodesics\index{geodesics!randomly curved spacetime} of a randomly curved spacetime.

Such an interpretation becomes possible in the context of the generalized gravitation theory\index{generalized gravitation theory} I proposed in chapter \ref{chap:2}, where matter configurations may exist which give rise, not to gravitational attraction and an apparent diminution of the speed of light (as a result of local space contraction\index{space contraction}), but to gravitational repulsion\index{gravitational repulsion} and an apparent increase of the limit velocity experienced by massless particles\index{massless particles!increase of limit velocity} (as a result of local space dilation\index{space dilation}). From such a perspective, the multiple possible trajectories of unobserved, dynamic quantum attributes\index{unobserved dynamic attributes!multiple possible trajectories} would simply be the causally determined geodesics\index{causally determined geodesics!randomly-fluctuating spacetime} of a randomly-fluctuating dynamical spacetime, rather than the random paths of particles evolving over a static and uniform spacetime background\index{static and uniform spacetime background!random paths}.

It must be emphasized that what I'm proposing is not that there arise stochastic perturbations of the Schr\"{o}dinger equation\index{stochastic perturbations of Schr\"{o}dinger equation} itself, when a quantum measurement takes place, as is sometimes proposed in order to try to explain the emergence of classicality\index{emergence of classicality} and the random nature of quantum measurement results\index{quantum measurement results!randomness}. Once again, it should be clear that the irreducible randomness of quantum processes\index{quantum processes!irreducible randomness} cannot be assumed to be a consequence of what goes on during measurement and the improved framework, that would allow to reproduce the statistical predictions of the current theory, would differ merely in that it would allow to explain what causally determines the particular, unobservable trajectories\index{unobservable trajectories!causal determination} followed by elementary particles in between measurements, while the absence of interferences that follows quantum measurement\index{quantum measurement!elimination of interferences} would still be a mere consequence of decoherence\index{decoherence}, now enforced by the requirement of closure of the universal causal chain\index{universal causal chain!closure requirement}.

From that viewpoint, it would also be decoherence and the closure requirement that would trigger the process of state vector reduction\index{state vector!reduction} which can be expected to arise before a quantum superposition of randomly fluctuating spacetime curvatures\index{fluctuating spacetime curvature!quantum superposition} becomes so significant that it would have consequences of an observable nature. The randomly fluctuating spacetime curvature\index{fluctuating spacetime curvature!observable consequences} that exists in a certain retarded portion of history\index{retarded portion of history} and that which exists in the related, advanced portion of history\index{advanced portion of history} are only allowed to differ as long as they remain without observable consequences, just as is the case with any other dynamic attribute of a quantum system. Such differences in the curvature of spacetime may trigger decoherence, like other divergences between the retarded and the advanced states\index{retarded and advanced states!divergence} of a quantum system, but it is precisely the fact that they are not required to do so when they do not give rise to any observationally recognizable effect, that explains how it is possible for the unique history\index{history!uniqueness} that takes place in between measurements of a dynamic attribute of a quantum system to vary from one virtual process to another, even under identical experimental conditions.

To sum up, I believe that, instead of simply adapting the current classical theory of gravitation\index{classical gravitation theory!quantum-mechanical reality} to accommodate the quantum-mechanical nature of reality, we should first redefine the foundations of quantum theory\index{foundations of quantum theory!redefinition} to take into account a certain overlooked, but unavoidable aspect of a consistent semi-classical theory of gravitation\index{semi-classical theory of gravitation}, that would allow to explain how the unique histories, which constitute the true fundamental elements of the formalism of quantum field theory\index{quantum field theory!fundamental elements}, are randomly determined from the perspective of time-symmetric causality\index{time-symmetric causality}.

It is important to mention, however, that the idea that the unique retarded and advanced portions of history\index{retarded and advanced histories!causal determination} which take part in every quantum process are causally determined by unobservable random fluctuations of the classical gravitational field\index{classical gravitational field!unobservable random fluctuations} is not absolutely necessary for the validity of the solutions I have provided to other aspects of the problem of the interpretation of quantum theory\index{quantum theory!problem of interpretation}. Thus, as I came to realize, this proposal is not even necessary to solve the problem of the objectification of quantum measurement results\index{objectification of quantum measurement results}, because even if the interfering quantum-mechanical histories\index{interfering quantum-mechanical histories!uniqueness} were not causally determined, they would nevertheless remain unique from a time-symmetric viewpoint\index{time-symmetric viewpoint!uniqueness of history}, which is sufficient to make them compatible with the uniqueness of the outcomes of quantum measurements\index{quantum measurement!uniqueness of outcome}.

But while we may never be able to directly confirm that the unobserved paths\index{unobserved paths!causal determination} of elementary particles, despite their absolutely unpredictable nature, are nevertheless causally determined in every way, the fact that it is already possible to envisage the exact form of a theory that would satisfy those consistency requirements should encourage us to recognize that the only reasonable conclusion is that reality is not fundamentally without causes.

\pagestyle{headings}

\chapter{Conclusion\label{chap:6}}

\section{Main results\label{sec:6.1}}

I have come a long way since first asking what would happen to a negative-mass object dropped in the gravitational field of the Earth. Yet I was able to confirm that my early intuition was right and that consistency dictates that the negative mass\index{negative mass!current expectations} would need to `fall' upward, despite the fact that this goes against current expectations. This is a conclusion for which I have provided ample justification and even if that was all I had been able to establish, I would already be very satisfied with the outcome of my undertaking.

But several other developments were introduced in this report which are all related to the issue of time directionality\index{time directionality} as a concept independent from the thermodynamic arrow of time\index{thermodynamic arrow of time}. In fact, the hypothesis of the existence of a fundamental time-direction degree of freedom\index{fundamental time-direction degree of freedom} has become the vital lead that allowed me to better understand many aspects of gravitational and quantum physics. Thus, despite the fact that, originally, the main objective of this report was to provide a consistent account of the way by which the alternative concept of negative energy that emerges from those considerations can be integrated into a classical theory of gravitation\index{classical gravitation theory!alternative negative-energy concept}, I have also made use of those theoretical developments to provide solutions to various specific problems in cosmology\index{cosmology!solutions to specific problems} and to develop a more adequate interpretation of quantum theory\index{quantum theory!more adequate interpretation}.

First of all, using the proposed description of negative-energy matter particles\index{negative-energy matter particles!voids in positive vacuum energy} as consisting of voids in the positive-energy portion of zero-point vacuum fluctuations\index{zero-point vacuum fluctuations}, I was allowed to show that the existence of negative energy-matter would not give rise to catastrophic vacuum decay\index{vacuum decay} and to the creation of energy out of nothing\index{creation of energy out of nothing}, even in the case of those negative energy states which are already predicted to occur under exceptional circumstances by quantum field theory\index{quantum field theory!negative energy states}. The prediction of an absence of vacuum decay is one of the most significant result of the alternative approach to classical gravitation which was developed in this report, because this potential problem is usually considered to be the most serious affecting a conventional theory of negative-energy matter. But I also provided solutions to several other problems which are usually believed to be associated with the concept of negative-energy matter. This allowed me to demonstrate the viability of a generalized gravitation theory\index{generalized gravitation theory} based on the proposed, alternative interpretation of negative energy states\index{negative energy states!alternative interpretation}.

An important outcome of my reformulation of the discrete symmetry operations\index{discrete symmetry operations!reformulation}, on the other hand, was the derivation of an exact binary measure for the number of microscopic degrees of freedom\index{microscopic degrees of freedom!matter in black hole} of matter under the influence of a black hole. This result is particularly noteworthy in that it allows to explain why it is that only one binary unit of information\index{binary unit of information} is encoded in every elementary unit of area\index{elementary units of area!black hole event horizon} on the event horizon of a black hole under all conditions, even though the area of the event horizon of an extremal black hole\index{extremal black hole} is only a fraction of that of an ordinary Schwarzschild black hole\index{Schwarzschild black hole} with the same mass (or the same number of elementary particles). The possibility that is allowed, in the context of the proposed interpretation of negative energy states, to generalize the analogy between classical, thermal equilibrium states and stationary black holes\index{stationary black holes!thermal equilibrium states} through an application of the concept of negative temperature\index{negative temperature}, on the other hand, allows to confirm the relevance of the concept of negative-energy matter for gravitation theory.

But while the most significant result derived in this report will probably remain the elaboration of a quantitative framework which generalizes the already elegant gravitational field equations\index{gravitational field equations!generalization} of relativity theory in a way that increases their simplicity, whenever negative-energy matter is present and the vacuum is allowed to contribute a non-vanishing measure of energy, the most concrete results are those which were obtained by applying the lessons learned while solving the problem of negative energy states\index{negative energy states!problem} to address several long standing issues in theoretical cosmology\index{theoretical cosmology!long standing issues}.

I believe that what this shows is that a cosmological model based on a consistent theory of negative-energy matter provides a fertile ground for understanding all sorts of astronomical phenomena in which the gravitational interaction plays a crucial role. First of all, using the proposed, alternative formulation of the gravitational field equations\index{gravitational field equations!alternative formulation}, I was able to show that the cosmological constant\index{cosmological constant!vacuum energy density}, must be conceived as an average, residual value of vacuum energy density, which is determined by the difference that may exist between the scale factor\index{scale factor!positive-energy observers} associated with the metric properties of spacetime\index{metric properties of spacetime!positive-energy observers} which are experienced by positive-energy observers and the scale factor\index{scale factor!negative-energy observers} associated with the metric properties of spacetime\index{metric properties of spacetime!negative-energy observers} which are experienced by negative-energy observers, while this allows one to expect that the value of this cosmological parameter should be as small as required by the weak anthropic principle\index{weak anthropic principle}.

What makes this possible is the fact that additional contributions to vacuum energy density\index{vacuum energy density!additional contributions}, arising from those zero-point vacuum fluctuations\index{zero-point vacuum fluctuations!negative-energy observers} which are directly experienced only by negative-energy observers, must be taken into account, which allow the natural value of the cosmological constant\index{cosmological constant!natural zero value} that would exist in the absence of a divergence between the scale factors experienced by opposite-energy observers to actually be zero, rather than the very large number associated with the energy scale of quantum-gravitational phenomena\index{quantum-gravitational phenomena!energy scale} which is produced by more conventional models.

But given that, in the context where energy must be assumed to be null for the universe\index{universe!null energy} as a whole, a non-zero initial value for the total energy of matter\index{total energy of matter!non-zero initial value} would give rise to a non-zero value for the gravitational energy of the universe\index{universe!non-zero gravitational energy} that would require space\index{space!global positive or negative curvature} to be positively or negatively curved on a global scale, while this would make the cosmological constant\index{cosmological constant!arbitrarily large initial value} arbitrarily large in the very first instants of the Big Bang, then it becomes possible to explain not only the fact that the current value of the cosmological constant\index{cosmological constant!explanation of smallness} is relatively small, but also the observation that space is perfectly flat on a global scale, as being unavoidable requirements of the weak anthropic principle\index{weak anthropic principle}.

In such a context, it appears that the presence in the primordial universe of negative-energy matter\index{negative-energy matter!universe energy budget}, governed by the proposed generalized gravitational field equations\index{generalized gravitational field equations}, is observationally confirmed, given that the presence of such matter is required to balance the initial matter energy budget, while still allowing the gravitational potential energy of matter\index{gravitational potential energy of matter!global scale} not to be null on a global scale, from the viewpoint of positive- and negative-energy observers, so that the rate of expansion\index{expansion rate!critical value} can be set to its critical value in the initial Big Bang state\index{initial Big Bang state}. This constitutes a further proof that the alternative concept of negative-energy matter\index{negative-energy matter!alternative concept} which I proposed, based on independent motives, is fully justified, even from a purely empirical viewpoint.

I must admit, however, that, for a while, I was not fully convinced that a solution as technically (although not necessarily conceptually) simple as that which I had derived (based on the hypothesis of the existence of negative-energy matter) could alone solve such a difficult and intricate problem as that of flatness\index{flatness problem}. What I had realized, of course, was that, if I was right, then it probably meant that inflation theory\index{inflation theory!dominant paradigm for cosmology} could no longer be invoked to solve other aspects of the inflation problem\index{inflation!problem} either and this was difficult to believe, given that inflation theory was the dominant paradigm for cosmology at the time when I obtained my first results.

But I came to recognize that this is the only appropriate conclusion and that there actually exists a more natural solution to the flatness problem that merely requires one to acknowledge the reality of negative-energy matter. Thus, even though I may have preferred arriving at a different conclusion, there is no longer any doubt in my mind that it is really the condition of null energy\index{null energy condition!universe} (imposed on the universe as a whole by the constraint of relational definition\index{constraint of relational definition!physical attributes} of physical attributes) and the balancing effect of negative-energy matter\index{negative-energy matter!balancing effect} which allow to explain the global flatness of space\index{global flatness of space!explanation} in the context where one recognizes that an observer must be present in the universe to measure a value for this parameter.

One amazing consequence of the particular approach to the problem of dark energy\index{dark energy!problem} which was proposed in the present report is that, despite the fact that it relies on the existence of a previously ignored symmetry principle, it nevertheless allows one to understand why it is that the current value of the cosmological constant\index{cosmological constant!explanation of non-zero value} is not perfectly null. But it also allows the density of the uniform portion of vacuum energy\index{uniform portion of vacuum energy!variable density} to vary with time, because the rate of expansion of space\index{expansion rate!observer dependence} measured by positive-energy observers can differ from that which is measured by negative-energy observers, under conditions where a lesser proportion of negative-energy matter survives the early annihilation of matter with antimatter\index{early matter-antimatter annihilation}.

This is certainly a result that will have decisive consequences for our understanding of the evolution of the universe, especially in the context where it is also possible to predict that the average density of vacuum energy\index{average density of vacuum energy!early universe} was significantly larger than it currently is in the early universe, before the decoupling of matter and radiation\index{decoupling of matter and radiation}, which allows to ease the Hubble tension\index{Hubble tension}, that is to say, the difficulty that emerges as a result of the fact that the current value of the Hubble constant\index{Hubble constant!present value} derived from measurements of cosmic microwave background\index{cosmic microwave background!temperature fluctuations} temperature fluctuations (based on more conventional assumptions) appears to differ from that which is obtained from low-redshift measurements\index{low-redshift measurements!Hubble constant}.

Also of importance is the conclusion that there must have existed additional gravitational attraction around visible, positive-energy matter overdensities\index{positive-energy matter overdensities!gravitational attraction} in the early universe, due to the presence of underdensities in the distribution of negative vacuum-dark-matter energy\index{negative vacuum-dark-matter energy!underdensities}. What makes this conclusion particularly significant is the fact that those forces can be expected to have accelerated the formation of large-scale structures\index{formation of large-scale structures!acceleration} in the primordial distribution of positive-energy matter in a way that actually helps explain the presence of well-developed galaxies at an epoch when they should not yet exist according to the currently favored cosmological models\index{cosmological models!currently favored}.

The conclusion that no such a contribution to structure formation should exist for negative-energy matter on stellar and galactic scales, in the context where it is both theoretically possible and observationally unavoidable to assume that the average density of baryonic negative-energy matter\index{baryonic negative-energy matter!average density} was negligible during the matter-dominated era\index{matter-dominated era} (while the presence of baryonic negative-energy matter overdensities would appear to be required for positive-energy matter underdensities to form on such a scale), is also of particular importance, given that it allows to explain why no observation that would provide an unmistakable evidence for the presence of gravitationally repulsive structures\index{gravitationally repulsive structures!observational evidence} has ever been performed.

It is quite remarkable, as well, that the additional source of gravitational attraction which arises from the presence of negative-energy matter underdensities\index{negative-energy matter underdensities!gravitational attraction} in the early universe is allowed to so adequately complement the contribution to gravitational instability\index{gravitational instability!ordinary dark matter} which is provided by ordinary dark matter, once the missing-mass effect\index{missing-mass effect!local variations in vacuum energy density}, which is usually believed to arise solely from the presence of weakly-interacting massive particles\index{weakly-interacting massive particles}, is understood to actually be a consequence of local variations in the density of vacuum energy, whose presence is attributable to the fact that opposite-energy observers experience different metric properties of spacetime\index{metric properties of spacetime!observer dependence} locally, in the presence of matter inhomogeneities.

But given that a clear distinction nevertheless exists between ordinary matter and vacuum dark matter\index{vacuum dark matter!distinction from ordinary matter}, due to the fact that the presence of ordinary matter\index{ordinary matter!missing vacuum energy and charge} is equivalent to both missing vacuum energy and missing vacuum charge, while vacuum dark matter\index{vacuum dark matter} arises merely from local differences between the maximum positive and negative contributions to the density of vacuum energy which are directly experienced by negative- and positive-energy observers (respectively), while this difference does not affect the electrical or non-gravitational neutrality of the vacuum\index{vacuum!electrical neutrality}, then it is possible for those two concepts to coexist unambiguously.

The conclusion that a much smaller portion of the missing-mass effect\index{missing-mass effect!baryonic dark-matter} can be attributed to the existence of baryonic dark-matter\index{baryonic dark-matter!reversed-bidirectional-charge particles} particles carrying reversed bidirectional charges is also significant, since it allows to explain exactly how the imbalance which exists in our universe between the number of ordinary matter particles and that of ordinary antimatter particles is allowed to arise. This can be achieved once we recognize that the condition of continuity of the flow of time\index{condition of continuity of flow of time!particle world-line} along an elementary particle world-line, which is responsible for the absence of non-gravitational interactions between particles with opposite bidirectional charges\index{opposite-bidirectional-charge particles!interactions}, does not prevent those same particles from interacting gravitationally with one another.

As a result, particles propagating opposite bidirectional charges\index{opposite-bidirectional-charge particles!pair creation} in the same direction of time can be produced by pairs out of gravitational radiation\index{gravitational radiation}, under the extreme conditions which existed in the first instants of the Big Bang\index{first instants of Big Bang} and when a violation of time reversal symmetry\index{time reversal symmetry!violation} exists, this allows more positive-action particles than positive-action antiparticles to be produced that can survive the early annihilation of matter with antimatter\index{early matter-antimatter annihilation}. Thus, it is possible to avoid the conclusion that there would exist an absolute lopsidedness of the universe\index{universe!absolute lopsidedness} with respect to the direction of time as a result of the violation of time reversal symmetry that is responsible for the presence of baryonic positive-energy matter, because, if a quantum bounce\index{quantum bounce} occurs, a similar asymmetry can be expected to arise relative to the opposite direction of time, in the portion of history that precedes the initial state\index{initial state!maximum matter density} of maximum matter density.

It is while I was trying to solve the mystery of the thermodynamic arrow of time\index{thermodynamic arrow of time}, however, that I was led to derive the most surprising results regarding classical cosmology\index{classical cosmology} and to gain the essential insights which allowed me to solve virtually all remaining aspects of the inflation problem\index{inflation!problem}.

First of all, I explained that there exists a usually ignored measure of information, concerning the microscopic state of the gravitational field\index{microscopic state of gravitational field!uniform matter distribution} produced by a uniform matter distribution, and that its value diminishes when the density of matter decreases below its average cosmic value locally. It is this variation that allows the total measure of information\index{information!invariance} in the universe to remain constant, even in the context where the amount of \textit{missing} information\index{missing information!microscopic state of gravitational field} required to describe the microscopic state of the gravitational field must be assumed to rise as a result of the growing strength of gravitational fields that takes place when the density of matter is increasing locally. As a result, it becomes possible to conclude that information\index{information!conservation} is always conserved, despite the growing inhomogeneity of the matter distribution, while a similar conclusion applies in the context where the expansion of space\index{expansion of space!information increase} itself contributes to increase the amount of information in the universe by continuously creating new quantum-gravitational units of area\index{quantum-gravitational unit of area!creation} in the vacuum.

Based on the notion that the thermodynamic arrow of time\index{thermodynamic arrow of time!origin} originates from the smoothness of the initial matter distribution\index{initial matter distribution!smoothness} at the Big Bang, I was then led to propose that it is the requirement that there must exist causal relationships between all the elementary particles\index{elementary particles!causal relationships} which were present in that initial state\index{initial state!maximum matter density} of maximum matter density that explains the asymmetric character of the growth of gravitational entropy\index{gravitational entropy!asymmetric growth}. 

Most people no longer hesitate to recognize that the physical properties of our universe\index{physical properties of universe!constraint of observer existence} are constrained to a very small subset of potentialities by the requirement that those properties allow for the spontaneous development of a conscious observer. What I have proposed is that solving that oldest of all physics problems, the mystery of the origin of the thermodynamic arrow of time, requires taking into consideration the similarly obvious requirement that, for the universe\index{universe!constraint on existence} itself to exist as a consistent whole, a certain requirement must be met, which can only be satisfied when the universe goes through a state of maximum matter density and minimum gravitational entropy\index{minimum gravitational entropy!initial state of universe} at least once during its history, because it is only under such conditions that all of its components can actually be causally related to one another.

Thus was solved that long-lasting puzzle. I believe that this unexpected outcome illustrates better than anything else the fact that serious consequences may follow when we choose to uphold, without good reasons, the validity of certain commonly held hypotheses, such as the positivity of energy\index{positivity of energy} and the absolute nature of gravitational attraction and repulsion\index{gravitational attraction and repulsion!absolute nature}, because it is as a consequence of not having being held by such a prejudice that I was allowed to solve the problem of the origin of the thermodynamic arrow of time\index{thermodynamic arrow of time!problem of origin}.

What I'm most satisfied with having achieved, however, is having been able to actually understand quantum theory. When I began doing research in fundamental theoretical physics\index{fundamental theoretical physics!research}, more than 30 years ago, I did not suspect that some of the early ideas and insights I was trying to develop would eventually become essential for producing a consistent interpretation of quantum theory\index{quantum theory!consistent interpretation}. But the hypothesis that the gravitational interaction is symmetric under exchange of positive- and negative-energy matter\index{positive- and negative-energy matter!symmetry under exchange} turned out to be essential to the formulation of an interpretation of quantum mechanics in which no implicit or explicit assumptions contradict one another or some observable aspects of reality, because this idea is what allows one to understand how it is possible for thermodynamic time asymmetry\index{thermodynamic time asymmetry!emergence} to emerge despite the time-symmetric nature of causality\index{causality!time-symmetric nature}.

Outside the context of a generalized gravitation theory\index{generalized gravitation theory} compatible with this condition, it would not be possible to assume that there exists a constraint on the emergence of a maximum quasiclassical domain\index{maximum quasiclassical domain!constraint on emergence} imposed by a requirement of closure of the universal causal chain\index{universal causal chain!closure requirement}. Actually, it wouldn't even be possible to assume that there exists a universal time variable\index{universal time variable} along which the causal chain can unfold. As a matter of fact, if we were to ignore those theoretical developments, it wouldn't be possible to assume that there is a reality at all in the absence of measurement, unless we were willing to reject some equally unavoidable theoretical requirements derived from observation, like the principle of local causality\index{principle of local causality}.

Now, the most significant aspect of a quantum-mechanical description of reality is certainly the use of interfering probability amplitudes\index{interfering probability amplitudes} in place of conventional probabilities\index{conventional probabilities} (or equivalently the appearance of negative probabilities\index{negative probabilities!time-symmetric histories} for time-symmetric histories). But in the context of an interpretation of quantum theory that satisfies the requirement of scientific realism\index{scientific realism!requirement}, the existence of interference effects can be understood to be a consequence of the circular nature of history\index{history!circular nature} that is associated with the requirement of closure of the universal causal chain\index{universal causal chain!closure requirement}. I believe that this, again, serves to demonstrate how crucially important it is to acquire a proper understanding of certain purely cosmological aspects of reality in order to develop a consistent interpretation of quantum theory\index{quantum theory!consistent interpretation}.

This dependence is further emphasized by the fact that gravitation may ultimately be involved in explaining what causally determines the one unique history\index{history!causal determination} which is followed in the course of any particular time-symmetric process\index{time-symmetric processes}, given that a more accurate understanding of the phenomenon of inertia\index{inertia}, which also arises from considerations of a cosmological nature, may require that quantum field theory\index{quantum field theory!reformulation} be reformulated so as to accommodate the randomly fluctuating nature of the classical gravitational field\index{classical gravitational field!randomly fluctuating nature} and in such a way relieve the theory from its dependence on the concept of a static and uniform spacetime background\index{static and uniform spacetime background}.

It is also the notion that effects cannot be restricted to propagate only in the future direction of time, as the classical principle of causality\index{classical principle of causality} would appear to require, that made unavoidable a picture of quantum reality\index{quantum reality!two causally independent histories} involving two corresponding, causally independent histories, unfolding in opposite directions of time, for each process. The understanding that this is made necessary when all causes\index{causes!origin within universe} are required to originate from within our universe then allowed me to elaborate the first complete solution to the quantum measurement problem\index{quantum measurement problem!complete solution}.

I have explained that it is the requirement of a relational description of reality\index{requirement of relational description of reality} that imposes a condition of continuity on the universal causal chain\index{universal causal chain!condition of continuity} which can be most naturally satisfied when causality\index{causality!circular phenomenon} is appropriately conceived of as a circular phenomenon in which time plays a role similar to that which would be played by space in a universe\index{universe!closed geometry} with closed geometry, while such a closure requirement is what allows to explain the persistence of quasiclassicality\index{persistence of quasiclassicality} following decoherence\index{decoherence}. In such a context, it becomes clear that there is no real difficulty associated with the assumption that there does exist a unique reality\index{reality!uniqueness} at all times, as long as one recognizes that this reality does not consist of one single classical history\index{classical history} propagating in one single direction of time at all times. Once this is understood, it is no longer necessary to retreat into complicated and confused philosophy in order to try to explain the simplest and most-elementary phenomena which are taking place right in front of us, all the time.

While I was progressing toward a better understanding of quantum mechanics, I realized that my position concerning the many-worlds interpretation of quantum theory\index{quantum theory!many-worlds interpretation} is somewhat similar to my position regarding the weak anthropic principle\index{weak anthropic principle!author's position}. Indeed, while I do believe that both anthropic selection\index{anthropic selection} and the existence of a multiplicity of causally independent universes\index{multiplicity of causally independent universes} are necessary concepts, I also provided decisive arguments to the effect that the many-worlds interpretation, which is often considered to be a multiverse theory\index{multiverse!theory}, is not viable as a realistic interpretation of quantum theory\index{quantum theory!realistic interpretation}, from both a theoretical and an observational viewpoint. But I have also explained why the quantum measurement problem\index{quantum measurement problem!idealistic issue} is not to be considered a mere idealistic issue in a world where the emergence of quasiclassicality\index{emergence of quasiclassicality!subjective notion} would be a subjective notion, associated with the biased nature of the perception of reality that would be characteristic of our conscious experience. Such a subjective approach, however, could only be made legitimate on the basis of the validity of the weak anthropic principle, whose relevance is therefore diminished by the developments I have introduced in the last portion of this report.

It is somewhat ironical, therefore, that the weak anthropic principle\index{weak anthropic principle!multiplicity of universes} was once considered to be bad science on the basis of the fact that it would require the existence of multiple universes, whose existence could not be confirmed by any \textit{other} means, because, as I previously mentioned, this stubbornness is actually a form of solipsism\index{solipsism} which, in the above described context, would be supported by the weak anthropic principle, which would therefore require the existence of a multiplicity of universes.

Concerning the realistic conception of quantum reality\index{quantum reality!realistic conception} developed in this report, it is perhaps appropriate to note that, while it can be expected that the most virulent objections to such an interpretation would probably have to do with the `hypothesis' of a unique reality\index{reality!uniqueness} behind interfering quantum-mechanical histories\index{quantum-mechanical histories!interferences}, I think that this resistance is not merely an undesirable by-product of the long tradition of instrumentalism\index{instrumentalism!tradition} that emerged from the Copenhagen interpretation of quantum theory\index{quantum theory!Copenhagen interpretation}, but also constitutes an unfortunate consequence of the more profound inadequacy of a philosophical position that originates from Descartes'\index{Descartes, Ren\'{e}} desire to free himself from the `superfluous' hypothesis that his mind may not be all that there is in the world.

I must emphasize, once again, that it is my strong belief that the most significant challenge currently facing fundamental theoretical physics\index{fundamental theoretical physics!most significant challenge} and the development of a consistent philosophy of the natural world is that of overcoming the psychological barrier associated with the reluctance to accept as real what one cannot perceive directly and to realize the sterility and the inadequacy of the opposite viewpoint, when what one wants to assess is the nature of reality itself. Here it may not be consistency alone which is at stake, but the very meaningfulness of the whole exercise, that which embodies the quest for the ultimate representation of reality.

\section{Historical perspective\label{sec:6.2}}

The significance of the developments introduced in this report can be better appreciated by describing the progress achieved from a historical perspective. If we start with general relativity\index{general relativity!relativity of all motion}, I think that what the theory allowed us to understand is that all motion, including acceleration, is relative and that the state of motion of an object must be defined in relation to the state of motion of the rest of the matter in the universe. Thus, relativity theory embodied in its structure the requirement of a relational definition of physical properties\index{relational definition of physical properties!relativity theory}. But it also failed to integrate the requirement of the relativity\index{relativity!of sign of energy} of the sign of energy. The common belief which existed, since the creation of the general theory of relativity, is that energy\index{energy!positive-definiteness} must be considered positive-definite, because, otherwise, apparently insurmountable problems would arise.

What quantum field theory\index{quantum field theory!negative energies} allowed us to understand, on the other hand, is that negative energies are unavoidable for properly estimating the probability of all possible transitions involving particles and antiparticles\index{particles and antiparticles!transition probabilities}. But the current interpretation of this theory also failed to accommodate the fact that no constraint exists that would justify assuming that those negative energies are only useful for computational purposes and do not show up as properties of real matter particles, distinct from ordinary particles and antiparticles, when gravitation comes into play.

What I have tried to achieve in the first portion of this report is to generalize relativity theory\index{relativity theory!generalization} to produce a fully relativistic theory, compatible with the requirement that the sign of energy\index{sign of energy!relativity} should also be a relative physical property. What motivated those developments was a better understanding of the relationship between the sign of energy\index{sign of energy!relationship with direction of propagation in time} of a particle and its direction of propagation in time, which again arose from applying the requirement of relational description\index{requirement of relational description!physical attributes} of physical attributes. In such a context, it appeared, in effect, that a concept of negative energy\index{negative energy!alternative concept} distinct from that which is usually assumed to be relevant to quantum field theory was not only allowed, but was required by a truly consistent classical theory of gravitation\index{classical gravitation theory!consistency}.

Once it had been shown that the difficulties usually associated with negative-energy matter\index{negative-energy matter!difficulties} can be solved without rejecting the physical relevance of the whole concept of negative energy, there appeared to no longer be any rational motive for rejecting the possibility that negative energies\index{negative energy!propagation forward in time} can propagate forward in time and give rise to gravitational phenomena distinct from those involving exclusively positive-energy matter. It thus became clearly inappropriate to attribute a preferred status to positive-energy matter\index{positive-energy matter!preferred status} and this, in turn, meant that we are no longer justified to assume that, even as it becomes integrated with general relativity, the concept of gravitationally repulsive matter\index{gravitationally repulsive matter} cannot involve negative energy, but must merely give rise to the notion of an observer-dependent metric of spacetime\index{metric of spacetime!observer-dependence} devoid of any theoretical justification.

It has been clearly emphasized in this report that it is only when the concept of negative energy\index{negative energy concept!classical gravitation theory} is well integrated to classical gravitation theory, by considering the equivalence between negative-energy matter\index{negative-energy matter!missing positive vacuum energy} and missing positive vacuum energy that a consistent theory (for which all measures of energy are relative) emerges which agrees with all experimental and observational constraints. The original approach which was developed in the preceding chapters is thus unique in that it actually allows to account for the very existence of the phenomenon of inertia\index{inertia}, despite the fact that both positive- and negative-mass matter must be present on the largest scale. It alone also enables the success of the standard model of cosmology\index{standard model of cosmology!prediction of expansion rate} at predicting the rate of expansion experienced by positive-energy observers to be reproduced in a bi-metric theory\index{bi-metric theory}.

It must be clear that the concept of negative energy already existed before the developments I proposed in order to make it a consistent notion were introduced. But negative energy\index{negative energy!absolute non-relational definition} was always defined in an absolute or non-relational manner which, as I have shown, leads to serious difficulties, in particular because it would give rise to violations of the principle of inertia\index{principle of inertia!violation}. Indeed, the idea that energy could be negative in an absolutely defined way and should therefore gravitationally repel all matter, regardless of its energy sign, as if this repulsion was a distinctive physical property of negative-energy matter itself, was here shown to give rise to undesirable effects, even aside from the plain logical inconsistency it would involve. The alternative interpretation of negative energy states\index{negative energy states!alternative interpretation} which I have proposed has allowed to avoid those problems, while also making unnecessary the hypothesis that only positive-energy matter can exist in stable form, because it explains why matter in a negative energy state is unobservable from the viewpoint of observers made of positive-energy matter and why even negative vacuum-dark-matter energy\index{negative vacuum-dark-matter energy} is mostly absent, at the present epoch, in regions of the universe occupied by positive-energy stars and galaxies.

It has also become possible to explain why it is that weak gravitational lensing\index{weak gravitational lensing!experiments} experiments have not revealed the presence of gravitationally repelling negative-energy matter\index{negative-energy matter!overdensities} overdensities, because it is now possible to assume that the average density of baryonic negative-energy matter\index{baryonic negative-energy matter!average density} is currently much smaller than that of baryonic positive-energy matter\index{baryonic positive-energy matter}, while the absence of such matter implies that overdensities in the distribution of negative vacuum-dark-matter energy\index{negative vacuum-dark-matter energy!overdensities} must be virtually absent on the scale of individual stars and galaxies and can only exist in smoothly distributed form in the largest voids in the distribution of positive-energy galaxies\index{distribution of positive-energy galaxies!largest voids}. Thus, it was actually explained why negative-energy matter has remained mostly out of reach of astronomical observations, so that this fact no longer constitutes a profound mystery.

Concerning quantum reality\index{quantum reality!non-local character} now, the problem that there was from a conventional viewpoint is that we regarded its distinctive non-local character as a mere curiosity and we were convinced that it did not constitute a challenge to our conventional understanding of causality\index{conventional understanding of causality!challenge}, simply because we could not see how the difficulty could be resolved if it is, in effect, real. The fact that the mathematical framework of quantum theory\index{quantum theory!mathematical framework} nevertheless allowed to produce accurate predictions, while the kind of non-locality\index{non-locality!no instantaneous communication} involved did not allow information to be transmitted instantaneously, appeared to legitimize this position and this is what explains that people stopped searching for a solution to the problem of the apparent incompatibility between quantum entanglement\index{quantum entanglement!incompatibility with relativity theory} and the constraint imposed by relativity theory on the propagation of effects.

But all along, we continued searching, with more and more sophisticated experiments, for possible loopholes that could explain quantum non-locality\index{quantum non-locality!outcome of unidirectional causality} as being an outcome of conventional unidirectional causality, just like people kept searching for experimental evidence of our motion relative to absolute space\index{absolute space} over a century ago. This happened because we were not willing to accept the conclusion that our concept of causality\index{causality!inadequate concept} is inadequate in some ways, given that it does not allow to explain facts without requiring the propagation of effects\index{propagation of effects!faster-than-light velocities} at faster-than-light velocities.

It is the fact that the notion of time directionality\index{time directionality!poor understanding} remained so poorly understood, even after the progress which was achieved in this area by the creation of quantum field theory\index{quantum field theory}, that explains that there was so much confusion over what constitutes an appropriate definition of the discrete symmetry operations\index{discrete symmetry operations} (from the viewpoint of both clarity and consistency) when I began studying the subject. It is indeed the stubbornness to consider time\index{time!conventional viewpoint} from a conventional viewpoint, where only one direction is allowed for this degree of freedom, that explains that the time reversal operation\index{time reversal operation!inappropriate definition} was never appropriately defined and that the time-symmetric nature of causality\index{causality!time-symmetric nature}, which allows one to make sense of quantum non-locality\index{quantum non-locality!proper understanding}, was never properly assimilated. This was allowed to occur despite the clues arising from the discovery of antimatter and the successful description of antiparticles\index{antiparticles!propagation backward in time} as particles propagating backward in time.

The commonsense feeling inherited from our experience of thermodynamic time\index{thermodynamic time} is so strong that it is still commonly believed that antiparticles are merely identical particles which happen to have opposite charges, rather than being the same particle propagating backward in time, as seems to be required from a mathematical viewpoint. This is what explains that time reversal\index{time reversal!reversal of charge} was never considered to involve a reversal of charge, as I have shown to actually be required. But once this was recognized, the possibility opened up to explain other facts. It is, in effect, by using this insight that I was able to propose an explanation for the fact that a finite number of discrete degrees of freedom\index{discrete degrees of freedom!particle inside black hole}, which is proportional merely to the area of a black hole, allows to completely specify the microscopic state of all the elementary particles which were captured by the gravitational field of any such an object. Therefore, it can no longer be argued that the notion of backward-in-time propagation\index{backward-in-time propagation} is merely an expedient for facilitating the calculations of probability amplitudes. Our notion of time direction has been irretrievably altered and there is no going back.

\chapter*{Acknowledgements}
\addcontentsline{toc}{chapter}{\protect{Acknowledgements}}
\markboth{}{}

I would like to thank the government of Qu\'{e}bec and its taxpayers which, through the generous social programs they offer, have allowed me to benefit from a steady source of income during the years in which I was working on the present project without any support from academia or the industry. They have not only allowed me to benefit from the conditions necessary to achieve the depth of knowledge required to conduct this research, but they have also saved my life at times when studying physics was the only activity that still gave enough meaning to my existence that it actually felt endurable.


\newpage
\pagestyle{headings}

\addcontentsline{toc}{chapter}{\protect{Bibliography}}
\bibliographystyle{unsrtnat}
\bibliography{References}

\addcontentsline{toc}{chapter}{\protect{Index}}
\small
\printindex
\normalsize

\end{document}